\documentclass{15719_aa} 

\usepackage{latexsym, graphicx, natbib, longtable}
\usepackage{longtable}
\usepackage{lscape}
\usepackage{txfonts}
\def\HI{\hbox{\rm H\,{\sc i}}}
\begin{document}
   \title{The WSRT Virgo {\HI} filament survey II}

   \subtitle{Cross Correlation Data}

   \author{A. Popping
          \inst{1} \inst{2} 
          \and
          R. Braun\inst{3}
          }

   \offprints{A. Popping \email{attila.popping@oamp.fr} }

   \institute{Laboratoire d'Astrophysique de Marseille, 38 Rue Fr\'{e}d\'{e}rique Joliot-Curie, 13388 Marseille Cedex 13, France
    \and 
    Kapteyn Astronomical Institute, P.O. Box 800, 9700 AV Groningen, the Netherlands
    \and
    CSIRO -- Astronomy and Space Science, P.O. Box 76, Epping, NSW 1710, Australia}

   \date{}

 
  \abstract
  {The extended environment of
  galaxies contains a wealth of information about the formation and
  life cycle of galaxies which are regulated by accretion and feedback
  processes. Observations of neutral hydrogen are routinely used to
  image the high brightness disks of galaxies and to study their
  kinematics. Deeper observations will give more insight into the
  distribution of diffuse gas in the extended halo of the galaxies and
  the inter-galactic medium, where numerical simulations predict a
  cosmic web of extended structures and gaseous filaments.}
   {To observe the extended environment of galaxies, column
  density sensitivities have to be achieved that probe the regime of Lyman limit
  systems. {\HI} observations are typically limited to a brightness
  sensitivity of $N_{HI} \sim 10^{19}$ cm$^{-2}$, but this must be
  improved upon by $\sim2$ orders of magnitude.}
  {In this paper we present the
   interferometric data of the Westerbork Virgo {\HI} Filament Survey
   (WVFS) -- the total power product of this survey has been published
   in an earlier paper. By observing at extreme hour angles, a
   filled aperture is simulated of $300\times25$ meters in size, that
   has the typical collecting power and sensitivity of a single dish
   telescope, but the well defined bandpass characteristics of an
   interferometer. With the very good surface brightness sensitivity
   of the data, we hope to make new {\HI} detections of diffuse
   systems with moderate angular resolution.}
  {The survey maps 135 degrees in Right
     Ascension between 8 and 17 hours and 11 degrees in Declination
     between $-$1 and 10 degrees, including the galaxy filament
     connecting the Local Group with the Virgo Cluster. Only positive
     declinations could be completely processed and analysed due to
     projection effects. A typical flux sensitivity of
     6~mJy~beam$^{-1}$ over 16 km~s$^{-1}$ is achieved, that
     corresponds to a brightness sensitivity of $N_{HI} \sim
     10^{18}$~cm$^{-2}$. An unbiased search has been done with a high
     significance threshold as well a search with a lower significance
     limit but requiring an optical counterpart. In total, 199 objects
     have been detected, of which 17 are new {\HI} detections.}
  {By
       observing at extreme hour angles with the WSRT, a filled
       aperture can be simulated in projection, with a very good
       brightness sensitivity, comparable to that of a single dish
       telescope. Despite some technical challenges, the data provide
       valuable constraints on faint, circum-galactic {\HI} features. }


   \maketitle
%

\section{Introduction}
In the current epoch, numerical simulations predict that most of the
baryons are not in galaxies, but in extended gaseous filaments,
forming a Cosmic Web (e.g. \citealp{1999ApJ...511..521D, 1999ApJ...514....1C}) Galaxies are just the brightest pearls in
this web, as the baryons are almost equally distributed amongst three
components: (1) galactic concentrations, (2) a warm-hot intergalactic
medium (WHIM) and (3) a diffuse intergalactic medium.  Direct
detection of the intergalactic gas is very difficult at UV, EUV or
X-ray wavelengths \citep{1999ApJ...514....1C} and so far the clearest
detections have been made in absorption
(e.g. \citealp{2007ApJ...658..680L, 2008ApJS..177...39T}). In
this and previous papers in this series, we make an effort to detect
traces of the intergalactic medium in emission, using the 21-cm line
of neutral hydrogen. Most of the gas in the Cosmic Web will be highly
ionised, due to the moderately high temperatures above $10^4$ Kelvin,
resulting in a low neutral fraction and relatively low neutral column
densities. A more detailed background and introduction on this topic
is outlined in \cite{2010PhDT..APOPPING} and \cite{2010arXiv1012.3236P}

To investigate column densities that probe the Lyman Limit System
regime, very deep {\HI} observations are required with a brightness
sensitivity significantly better than $N_{HI} \sim 10^{19}$
cm$^{-2}$. Reaching these column densities is important to learn more
about the distribution of neutral hydrogen in the inter-galactic
medium and to have a better understanding of feedback processes that
fuel star formation in galaxies. In \cite{2010arXiv1012.3236P} and
\cite{2010A&A...HIPASS} two {\HI} surveys have been presented that
reach these low column densities in a region of $\sim 1500$ square
degrees. The first data product described in \cite{2010arXiv1012.3236P}
is the total power data of the Westerbork Virgo Filament Survey, an
{\HI} survey mapping the galaxy filament connecting the Virgo Cluster
with the Local Group. The survey spans 11 degrees in Declination from
$-$1 to +10 degrees and 135 degrees in Right Ascension between 8 and
17 hours. This survey has a point source sensitivity of 16 mJy
beam$^{-1}$ over 16 km s$^{-1}$ corresponding to a column density of
$N_{HI} \sim 3.5 \cdot 10^{16}$ cm$^{-2}$.

The second data product presented in \cite{2010A&A...HIPASS} is
reprocessed data, using original data that has been observed for the
{\HI} Parkes All Sky Survey \citep{2001MNRAS.322..486B, 2006MNRAS.371.1855W}. The 1500 square degree region overlapping the WVFS
was reprocessed to permit comparison between these data products and
detections. The point source sensitivity of the reprocessed HIPASS
data is 10 mJy beam$^{-1}$ over 26 km s$^{-1}$, corresponding to a
column density of $N_{HI} \sim 3.5 \cdot 10^{17}$ cm$^{-2}$.

In this paper, a third data product is presented: the
cross-correlation data of the Westerbork Virgo Filament Survey. As
explained in \cite{2010arXiv1012.3236P}, the aim of the WVFS was to achieve very high
brightness sensitivity in a large region of the sky, to permit
detection of {\HI} features that probe the neutral component of the
Cosmic Web. The configuration of the array was chosen such that
the dishes of the interferometer form a filled aperture of $\sim 300$
meters in projection by observing at extreme hour angles. Because of
the much smaller beam size compared to the WVFS total-power or HIPASS
observations, we will be able to identify brighter clumps within
diffuse features if these are present.

The special observing configuration creates some
technical challenges itself. This novel observing strategy requires
non-standard data-reduction procedures, which will be explained in
section 2. In section 3 we will present the results, starting with a
list of detected features. Objects are sought both blindly, by using a
high signal-to-noise threshold, and in conjunction with a known
optical counterpart by using a lower threshold. New {\HI} detections
and diffuse structures are briefly discussed, however detailed analysis
of these features will be presented in a follow up paper, also
discussing new and tentative detections obtained from the WVFS
total-power data and the re-processed HIPASS data as described in
\cite{2010arXiv1012.3236P} and \cite{2010A&A...HIPASS}. We will end with
a short discussion and conclusion, summarizing the main results.

\section{Observations and data Reduction}
The basic observations have already been described
in \cite{2010arXiv1012.3236P}, where the total power product of the
Westerbork Virgo Filament Survey is presented. Cross-correlation data
were acquired simultaneously with that total power data. We will only
summarise the observational parameters as these have been discussed
previously and concentrate more on the observing technique and data
reduction as these are very non-standard for this data set.

\subsection{Observations}
The galaxy filament connecting the Virgo Cluster with the Local Group
has been observed using the Westerbork Synthesis Radio Telescope
(WSRT) in drift scan mode. Data was acquired in two 20 MHz IF bands
centred at 1416 and 1398 MHz. In total $\sim 1500$ degrees has been
observed from $-$1 to 10 degrees in Declination and from 8 to 17 hours
in Right Ascension. Forty-five strips have been observed at fixed
Declinations of Dec~=~$-1, -0.75, -0.5 \dots$ 10 degrees. The
correlated data was averaged in Right Ascension every 60 seconds,
corresponding to an angular drift of about 15 arcmin, to yield Nyquist
sampling in the scan direction. All regions have been observed twice,
once when the sources were rising and once when they were setting.

\begin{figure}[t]
  \includegraphics[width=0.5\textwidth]{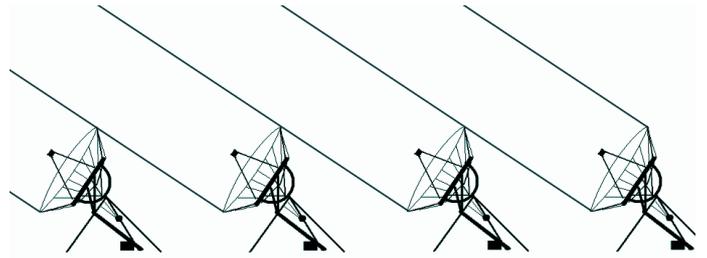}
  \caption{Observing mode of the WSRT dishes: 12 of the 14 dishes are
    placed at a regular interval of 144 meters. When observing at
    large hour angles, an approximately filled aperture of 300 by 25
    meters can be simulated in projection.}
  \label{config}
\end{figure}

\subsection{Observing technique}
\begin{figure}[t]
  \includegraphics[width=0.5\textwidth]{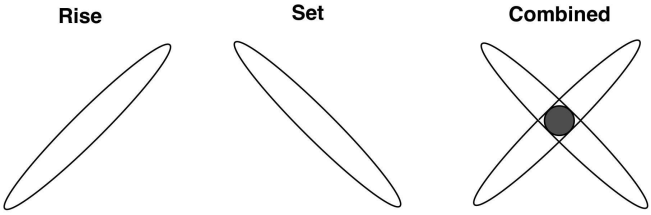}
  \caption{Due to the filled aperture of $\sim 300 \times 25$ meters,
    a very elongated beam is created. By observing each pointing twice
    at complementary orientations, the combined beam of the two
    observations is nearly circular, although with a significant
    side-lobe level.}
  \label{beam}
\end{figure}

The data has been obtained at very extreme hour angles between $\pm
80$ and $\pm 90$ degrees, to be able to achieve a filled-aperture
projected geometry. The WSRT has 14 dishes of 25 meter diameter, of
which 12 can be placed at regular intervals of 144 meter. At these
extreme hour angles, the dishes do not shadow each other, however the
separation is so small that there are no gaps in the UV-plane. Using
the 12 dishes at regular intervals, a filled aperture is created in
projection of $300\times25$ meter in extent as demonstrated in
Fig.~\ref{config}. When observing in this mode, we can achieve the
high sensitivity of a single dish telescope, but benefit from the
excellent spectral baseline properties and PSF of an
interferometer. Although each pointing is only observed one minute at
a time and two minutes in total, the expected sensitivity after one
minute of observing is $\Delta N_{HI} \sim 5\times 10^{17}$ cm$^{-2}$
over 20 km s$^{-1}$ in a $\sim 35 \times 3$ arcmin beam over a $\sim
35\times35$ arcmin instantaneous field-of-view. It is important to
note that the shape of the beam in a single snap-shot is extremely
elongated. The high column density sensitivity is only achieved in
practise for sources which completely fill the beam, which is only
likely for the nearest sources, or when they are fortuitously aligned
with the elliptical beam. However, each pointing is observed at
two complementary hour-angles; one positive and the other negative,
implying that the orientation of the snap-shot beams is also
complimentary. When combining the observations, the resulting beam has
a well defined circular/elliptical shape, even though the orientation
of each of the two beams is only a few degrees on either side of
vertical. The concept is demonstrated in Fig.~\ref{beam} where the
combination of the individual elongated beams forms a symmetric,
approximately circular response.

\subsection{Data Reduction}
Both the auto and cross-correlation data have been obtained
simultaneously, but were separated before importing them into Classic
AIPS \citep{1981NRAON...3....3F}. The reduction process of the
auto-correlations or total power data has been described in detail in
\cite{2010arXiv1012.3236P}. Here we will describe the cross-correlation data product.

A typical observation consisted of a calibrator source (3C48 or
3C286), the actual drift scan at a fixed Declination and another
calibrator source. After importing the data into AIPS, the calibration
and drift scan data were concatenated into a single file.

The total dataset consists of 90 drift-scans at fixed declination,
containing 91 baselines in two polarisations; half observed at an
extreme negative hour angle and the other half at an extreme positive
hour angle. Each baseline was inspected manually in Classic AIPS,
using the SPFLG utility. Suspicious features appearing in the
frequency and time domain of each baseline were inspected
critically. Features that were not confirmed in spectra acquired
simultaneously were flagged as radio frequency interference (RFI).

The calibrator sources were used to determine the bandpass calibration
and the bandpass solutions were inspected by eye before
application. Continuum data files were created from the line data by
averaging the central 75\% of the frequency bandwidth. These were used
for determining the gain calibrations. As we are using the {\it AIPS}
package which was originally developed for VLA data, a re-definition
of the polarisation products is necessary for correct gain
calibration. The VLA measures right and left circular polarisations
($RR=I+V$, $LL=I-V$, $RL=Q+iU$ and $LR=Q-iU$), while the WSRT data
consists of the two perpendicular linear polarisation products
($XX=I-Q$, $YY=I+Q$, $XY=-U+iV$ and $YX=-U-iV$). Both definitions are
in terms of the same true Stokes parameters ({\it I, Q, U, V}). By
redefining a calibrator's parameters as ({\it I$'$, Q$'$, U$'$, V$'$})
= ({\it I, -U, V, -Q}) it is possible to successfully calibrate the
$(XX, YY, XY$ and $YX)$ data by treating it as if it were $(RR, LL,
RL$ and $LR)$. Redefinition of polarization products has no effect for
sources that are not polarised. However the calibrator, 3C286, is
known to be about 10\% linearly polarised. The ``apparent Stokes''
values that have been used for this calibrator are (14.75, -1.27, 0,
-0.56) Jy for the first IF and (14.83, -1.28, 0. -0.57) Jy for the
second IF. Gain solutions determined with the continuum data are
applied to the line data as well. The calibrated data is exported from
{\it AIPS} into the {\it uvfits} format and imported into the {\it
  Miriad} \citep{1995ASPC...77..433S} software package. {\it Miriad}
has been used to further reduce the data. The mosaic scans are split
into individual pointings. Continuum emission is subtracted from the
line data using a first order polynomial, excluding the edge of the
bandpass and regions containing galactic emission.

\begin{figure}
  \includegraphics[angle=270, width=0.5\textwidth]{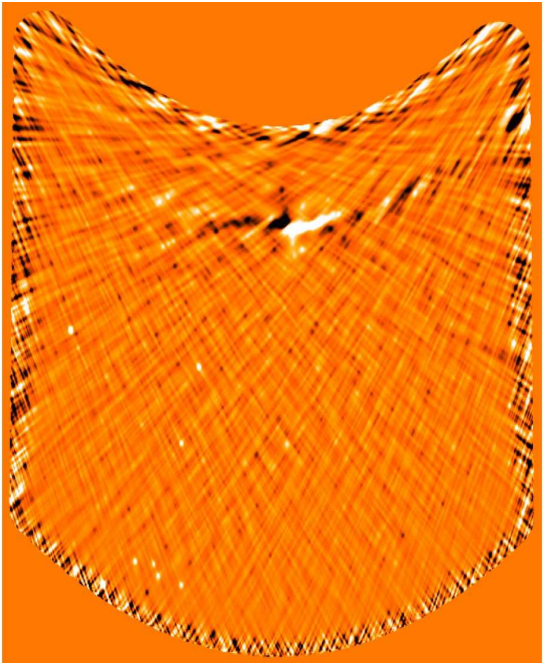}  
  \caption{Example of a channel map from one of the processed sub-region
    cubes (Cube 9) from the WVFS cross-correlation data. The inner
    region is well sampled by the pointings and the noise is uniform,
    while emission at the edges is noisier. There is 25\% overlap in
    Right Ascension between adjacent cubes. Unfortunately the
    coordinates could not be shown in this plot due to projection
    effects, as explained in the text.}
  \label{field}
\end{figure}

\begin{figure}
  \includegraphics[angle=270, width=0.5\textwidth]{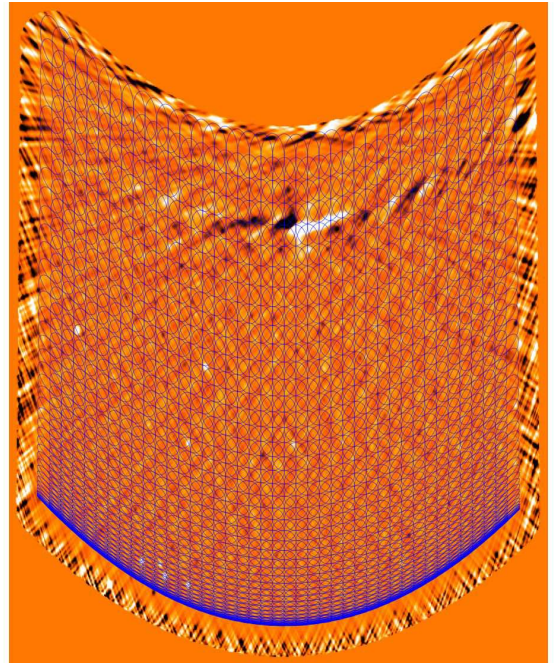}
  \caption{Same image as Fig.~\ref{field}, but now with the positions
    of the pointings overlaid. Each cube contains data from 45
    pointings in Declination, by 40 pointings in Right Ascension
    covering 11 by 10 degrees. Due to the NCP projection of
    the data, the Declinations close to zero degrees are increasingly
    distorted, as can be seen in the shape of the pointings that are
    squeezed toward the lowest declinations. Unfortunately, pointings
    with both positive and negative Declinations could not gridded
    simultaneously due to limitations of the NCP projection.}
  \label{fieldann}
\end{figure}

Although there are only scans at 45 different Declinations, each scan
contains 540 pointings in Right Ascension. Because the complete survey
cannot be imaged simultaneously due to computer memory and image
projection limitations, all the scans have been split into individual
pointings. The complete survey is separated in 18 blocks of 40
pointings in Right Ascension and 45 pointings in Declination, with a
ten pointing overlap in Right Ascension between neighbouring
blocks. This corresponds to sub-regions of $\sim 10 \times 11$ degrees
in size. The central positions of each of the sub-region cubes is
listed in table~\ref{cubes}. When inverting the data from the $UV$ to
the image domain a uniform weighting scheme was applied, as this is
the most optimal weighting for the 12 inner antennas of the
array. Because of the regular antenna spacing, many baselines have the
same length but the $uv$-plane is fully sampled at all spacings
between one and eleven dish diameters. To suppress the side-lobes due
to incomplete sampling of the longer baselines involving antennas 13
and 14, a Gaussian taper has been applied to the visibility data with
a FWHM of 200 arcsec; as this is approximately the size of the final
beam. The 250 individual velocity channels were imaged between 150 and
2000 km s$^{-1}$ with a sampling of 8.24 km s$^{-1}$.  An example of a
single channel in a sub-region is shown in Fig.~\ref{field}, the inner
10 degrees in Right Ascension and Declination have uniform mosaic sampling
and an approximately constant noise level. Fig.~\ref{fieldann} shows
the same field, overlaid with all the individual pointing patterns
that have been used to mosaic this field.

Unfortunately the data from pointings with negative Declinations are
missing from the cubes, as well as the axis labels in the two plots
that are shown. The data has been inverted using the {\it Miriad}
software package, which grids and inverts the data using the natural
NCP (North Celestial Pole) projection that applies to data obtained
with an East-West interferometer. In most cases this is not a problem,
however the NCP definition breaks down at a Declination of zero degrees. The
NCP projection is defined in \cite{1971PhDT.......153B} by:

\begin{equation}
L = \cos \delta \sin \Delta \alpha
\end{equation}
and
\begin{equation}
M = (\cos \delta_0 - \cos \delta \cos \Delta \alpha) / \sin \delta_0
\end{equation}

At Declinations approaching zero, the $M$-value diverges to infinity,
making any projection impossible. The commonly used imaging tasks in
both {\it Miriad} and {\it AIPS} have been found to work effectively
when given either all positive or all negative Declinations, but
failed when given both simultaneously.  Various methods of
circumventing this problem have been tested, but none of them was
successful in re-projecting all of the data to a more useful grid. We
have therefore chosen to neglect the negative Declination pointings of
the cross-correlation data, as they contain less than 10\% percent of
the total. Other projections for the northern part of the data have
been considered, but these are not favorable since they yield a highly
position dependant PSF. The data is gridded in NCP-projection, for
which a well defined beam applies. When re-projecting the $\sim
10\times 10$ degree field to e.g. a SIN-projection, the response to
point sources, particularly at the lowest Declinations, is severely
distorted. The effect of the NCP-projection can also be seen in the
shape and distribution of the pointings in Fig.~\ref{fieldann} as the
pointings at low Declination are squeezed into narrow ellipses. Quite
apart from the NCP-projection, it is an inherent property of an
east-west interferometer, that the north-south spatial resolution that
can be achieved at Dec=0 is only as good as the primary beam, which is
$\sim 35$ arcmin.\\

\begin{figure*}
  \includegraphics[width=0.5\textwidth]{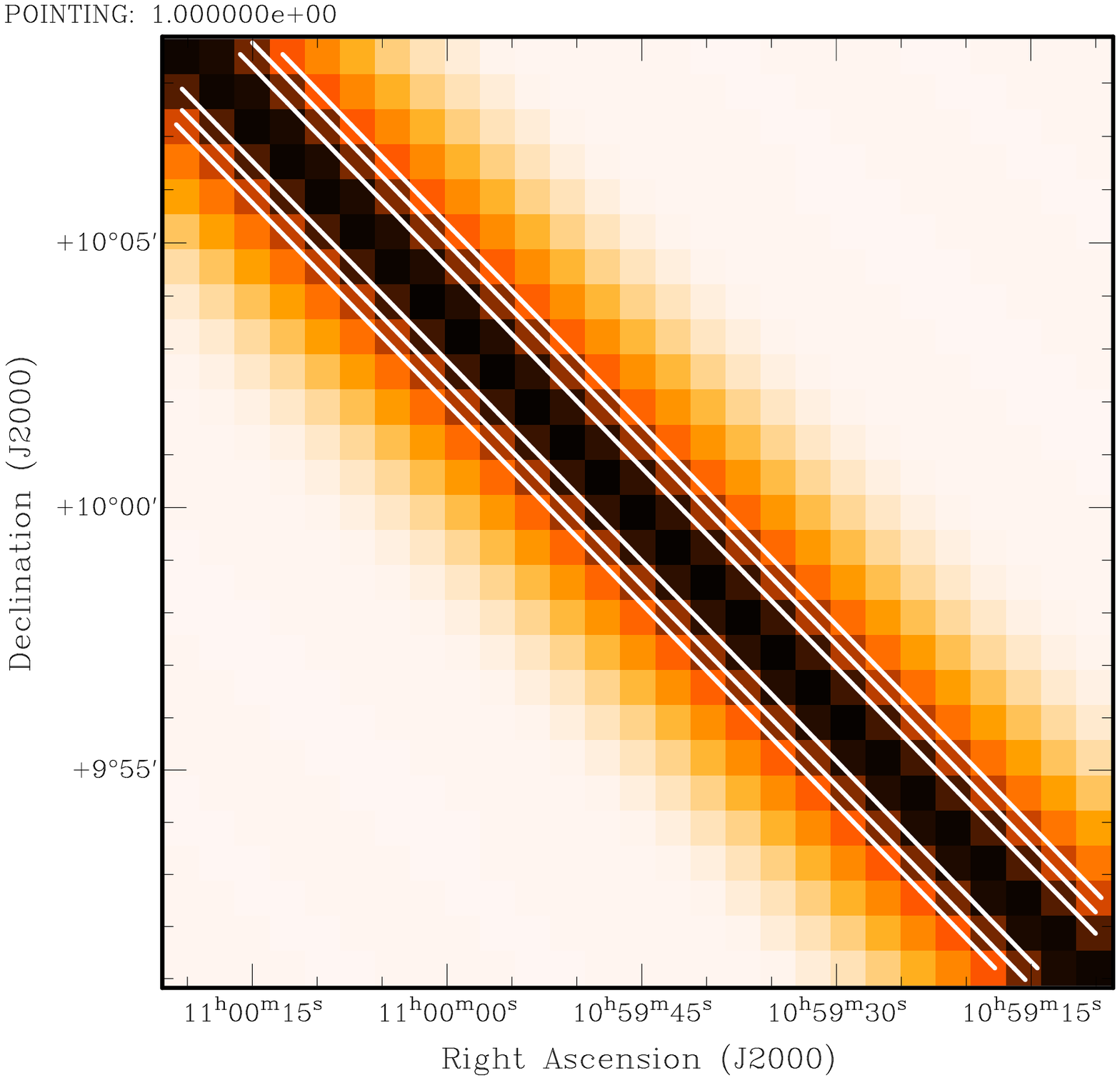}
  \includegraphics[width=0.5\textwidth]{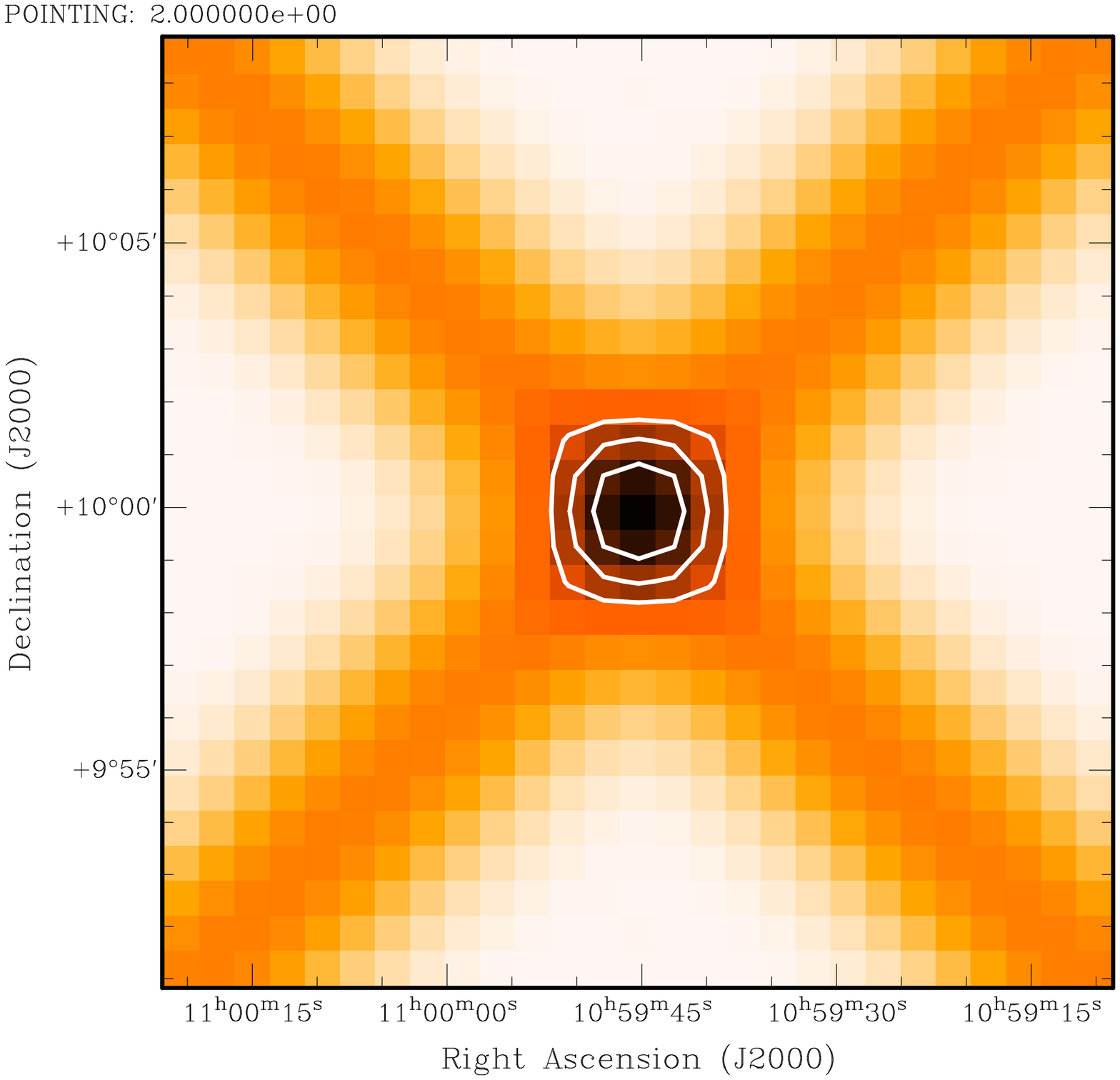}
  \caption{Beam shape of a central pointing in the strip at Dec=10
    degrees. The left panel shows the synthesized beam of a single
    observation, while the right panel shows the beam for the
    combination of 
    two complimentary scans. Both panels have the same intensity scale and
    contours are drawn at 70, 80 and 90 percent of the peak.}
  \label{beamplot}
\end{figure*}

Fig.~\ref{beamplot} shows the synthesized beam for
one of the central pointings in the strip at Dec=10 degrees. The left
panel shows the beam due to a single observation, which is extremely
elongated. The right panel displays the beam after combining both
complimentary observations. When averaging the two elongated beams, a 
circular main-lobe is formed, although substantial X-shaped side-lobes
are also apparent.\\

After the cubes have been imaged, the data were Hanning smoothed using
a width of 3 channels, to eliminate spectral sidelobes and lower the
RMS noise. Although the channel sampling of the cubes is unchanged, the
velocity resolution of the data is decreased to $\sim 16$ km
s$^{-1}$.\\

\begin{figure*}
  \includegraphics[width=0.5\textwidth]{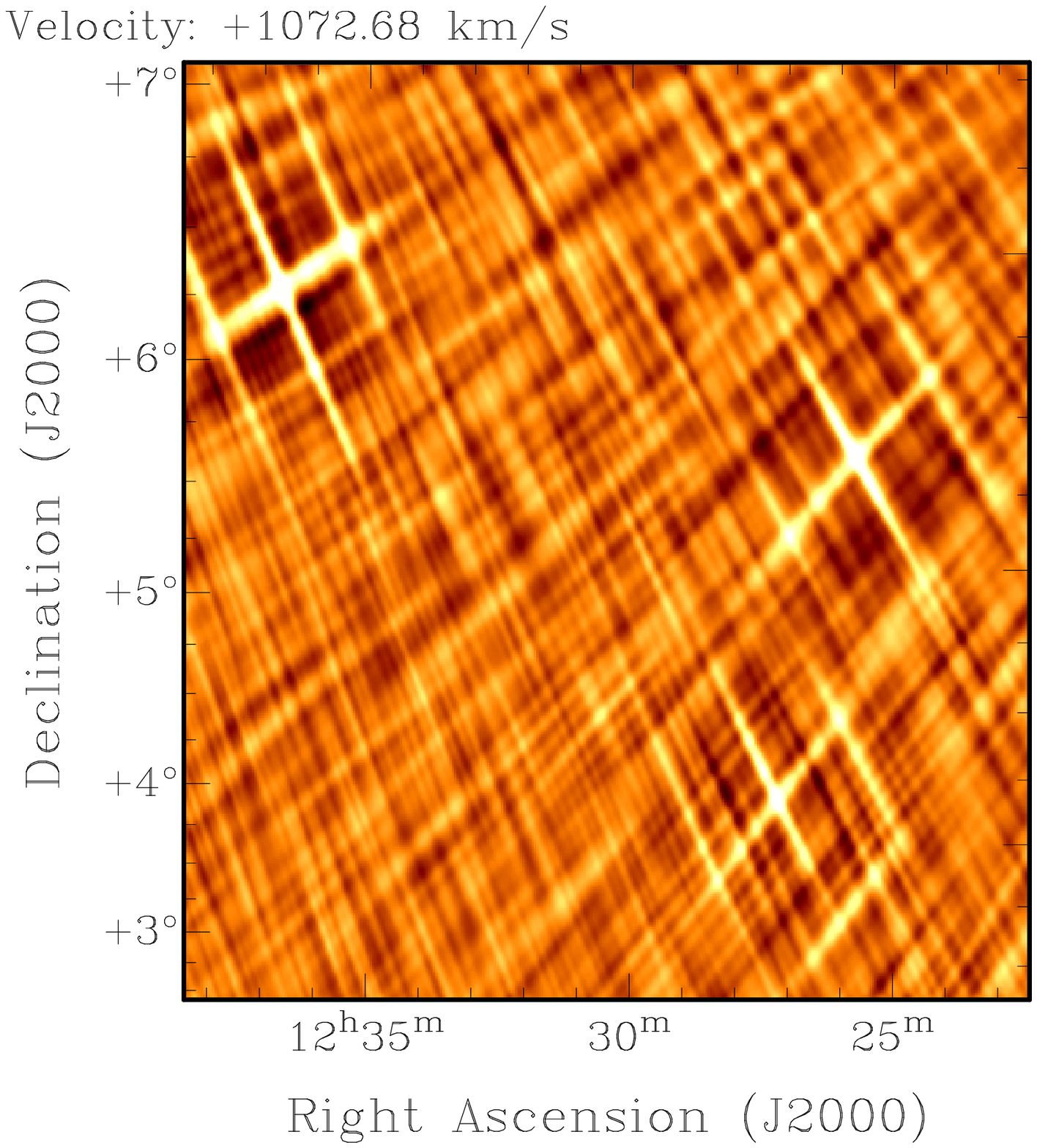}
  \includegraphics[width=0.5\textwidth]{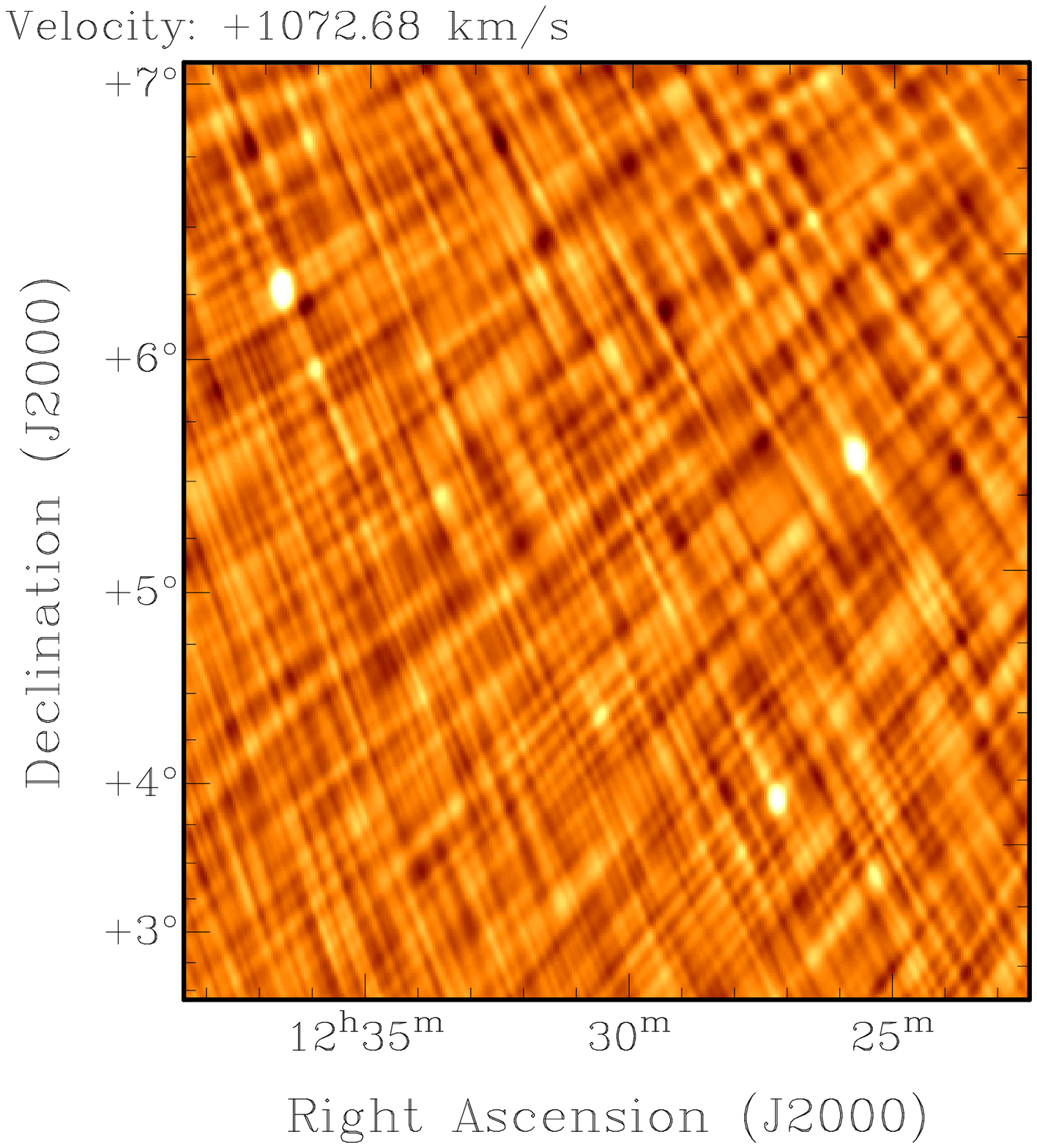}
  \caption{The left panel shows a region of the {\it dirty} map; data
    that has only been inverted from the {\it u,v} plane to the image
    plane. The right panel shows exactly the same region after one pass
    of {\it cleaning}. While in the left panel the cross
    pattern of the beam is very apparent, in the right panel the
    side-lobes are almost completely subtracted.}
  \label{clean}
\end{figure*}

After Hanning smoothing has been applied, the beam and smoothed {\it
  dirty} maps were deconvolved using the {\it MOSSDI} task within {\it
  Miriad} to create {\it clean} cubes. This task is similar to the
single-field {\it CLEAN} algorithm, but can be applied to
mosaic-data. Only one pass of {\it CLEAN} deconvolution has been done
without any masking. Because of the relatively large size of the 18
cubes, the cleaning step takes significant processing power. For the
cleaning step a cutoff level of 50 mJy beam$^{-1}$ was used. This
cutoff level was determined empirically to be optimal in cleaning the
data as deeply as possible while not creating false
components. Although on the whole the data quality was significantly
improved by deconvolution, the brightest sources are still suffering from some
residual side-lobe artifacts. A second cleaning pass using a clean mask was not
undertaken, since residual sidelobes prevented effective mask
definition in an automatic way. The survey volume was deemed too
large, to determine a mask manually. Nevertheless the improvement in
dynamic range is
significant as can be seen in the example shown in
Fig.~\ref{clean}. In the left panel a channel is shown before cleaning
and in the right panel the same field is shown after cleaning. In the
left panel a very strong X-shaped sidelobe pattern can be seen at the
location of bright sources. In the right panel the side-lobes have almost
completely disappeared.\\

\subsection{Sensitivity}

We reach an almost uniform noise level throughout the survey area of 6
mJy beam$^{-1}$ over 16 km s$^{-1}$ which corresponds to a column
density sensitivity of $N_{HI} \sim 1.1 \cdot 10^{18}$ cm$^{-2}$.  The
noise level in each of the individual 18 cubes is listed in
table~\ref{cubes}. Although we reach very high sensitivities, we note
that the data is affected by some residual side-lobe contamination in
the vicinity of bright sources.

Typical features that occur in the data cubes are illustrated in
Fig.~\ref{field}. At the edge of the field there is an increase in the
noise due to the finite number of mosaic pointings included in each
cube. The large dark and light structure in the upper part of the
field is caused by solar interference during the observation. This
instance of solar interference is very extreme and certainly not
typical. Distributed over the field are many galaxies that appear as
point sources. The X-shaped residual side-lobe pattern is still
apparent within the noise around the brightest of these sources in the
field.

\begin{table}
\begin{center}
\begin{tabular}{lccc}
\hline
\hline
Cube &  RA$_{\textrm{min}}$ &  RA$_{\textrm{max}}$  & rms [Jy beam$^{-1}$] \\
\hline
1  & 07:57:04 & 08:41:35  & 6.4 \\
2  & 08:26:10 & 09:11:42  & 5.5 \\
3  & 08:56:19 & 09:41:44  & 5.3 \\
4  & 09:26:25 & 10:11:49  & 6.3 \\
5  & 09:56:29 & 10:41:53  & 5.8 \\
6  & 10:26:34 & 11:11:59  & 5.1 \\       
7  & 10:46:39 & 11:42:04  & 5.8 \\
8  & 11:26:45 & 12:12:08  & 5.5 \\
9  & 11:56:49 & 12:42:15  & 5.5 \\
10 & 12:26:56 & 13:12:18  & 6.5 \\       
11 & 12:56:59 & 13:42:24  & 6.4 \\
12 & 13:27:04 & 14:12:28  & 7.0 \\
13 & 13:57:09 & 14:42:34  & 6.6 \\
14 & 14:27:14 & 15:12:38  & 6.2 \\ 
15 & 14:57:19 & 15:42:43  & 6.6 \\
16 & 15:27:24 & 16:12:48  & 6.1 \\
17 & 15:57:29 & 16:42:53  & 6.1 \\
18 & 16:27:35 & 17:01:56  & 5.9 \\
\hline
\hline

\end{tabular}
\end{center}
\caption{Right Ascension range and noise levels of the 18 individual
  sub-cubes of the WVFS cross-correlation data. All cubes are centred
  at a declination of 5 degrees.}
\label{cubes}

\end{table}

\subsection{False Positives}
\label{false_pos}
Although the reduced data has a good flux sensitivity, there are
many artifacts in the data, that are important to understand. Because
the "cleaning" step during the data reduction has not been perfect,
there are artifacts in the vicinity of bright groups of sources that
are caused by side-lobes in the beam. In some cases the residual side-lobes
of a bright source are similar in strength to fainter sources in it's
vicinity.

At some locations, the quality of the data was not optimal and parts
of several Declination strips had to be flagged. As a result, the
restoring beam is not well matched to the data at these locations and
the sensitivity is $\sqrt{2}$ worse. Due to the extended nature of the
mosaic pointing pattern, such instances are often compensated by
adjacent pointings. The effect of poor data quality is most apparent
around a declination of 9.25 degrees (due to solar interference),
where the noise is enhanced throughout the entire survey.

Although an attempt was made to observe only during night time hours,
portions of scans are suffering from solar interference. The regions where
this occurs are relatively isolated, however the data quality in these
regions is significantly impaired. 

Galactic {\HI} emission is another potential cause of false detections at radial
velocities below $\sim +400$ km s$^{-2}$.

\subsection{Flux determination}
In the case of single dish observations with a large beam, the flux
density of an object can be determined by integrating the spectrum
over the line-profile of an object. When multiplied with the velocity
resolution of the observations, this gives the line strength in
[Jy~km~s$^{-1}$].  This method can be used when the telescope beam is
more extended than the spatial size of an object implying that the
sources are unresolved. This assumption has been used for the HIPASS
data and for the WVFS total power data. Although the WVFS
cross-correlation data has a relatively large beam size compared to
typical interferometric observations, some objects are apparently
resolved.

The flux of each object is therefore determined in two different
ways. The first method simply employs the integral of the line
strength in the spectrum with the highest peak brightness. The error
in this estimate is given by:
\begin{equation}
\sigma = \sqrt{\frac{1.5 \cdot W_{20}}{v_{res}}} \delta v \times \textrm{rms } 
\end{equation}
where $v_{res}$ is the velocity resolution of the data, while $\delta
v$ is the channel separation in [km~s$^{-1}$].

The second method is a more sophisticated one, that is better suited
to extended sources. Ideally the flux would be determined
interactively for each object, by selecting the regions that contain
significant emission in a moment map integrated over the velocity
extent of the source. While this is possible when observing a modest
number of individual galaxies, the total area of the WVFS survey and
the number of objects is too large to treat each object manually. An
automatic method is used, which consists of integrating the flux of
the moment maps in the vicinity of each peak out to a certain
radius. Ideally, beyond a certain radius the integrated flux density
remains approximately constant apart from fluctuations due to the
noise and local background. Because of side-lobe and large-scale
background effects this is often not the case; the integrated flux
drops again or keeps increasing. A radius has to be determined that
yields the best estimate of the integral, while restricting the
effects of confusing features as much as possible

For each object a zeroth moment map is created and the radial profile
of the flux values is determined. A Gaussian function is fitted to the
radial profile, the $\sigma$ value of the fit is an indication of how
extended the source is. The pixel brightness is integrated for all
pixels within a radius of $3.5\sigma$ and converted from
[Jy~beam$^{-1}$~km~s$^{-1}$] to [Jy~km~s$^{-1}$] by division with the
beam area. The error of the integrated flux is calculated from:
\begin{equation}
\textrm{error} = \sqrt{\frac{1.5 \cdot W_{20}}{v_{res}}} \times
\sqrt{\frac{1.5 \cdot A_{s}}{A_{\textrm{beam}}}} \delta v \times \textrm{rms }
\end{equation}
where $A_{s}$ is the surface area of the source and
$A_{\textrm{beam}}$ is the surface area of the synthesised beam. The
integrated value of the pixel values is used, rather than the integral
of the Gaussian fit, to be more sensitive to possible extended
emission which is likely to be suppressed by the wings of the
Gaussian.\\

\subsection{Flux correction}
\begin{figure*}[t]
  \includegraphics[width=1.0\textwidth]{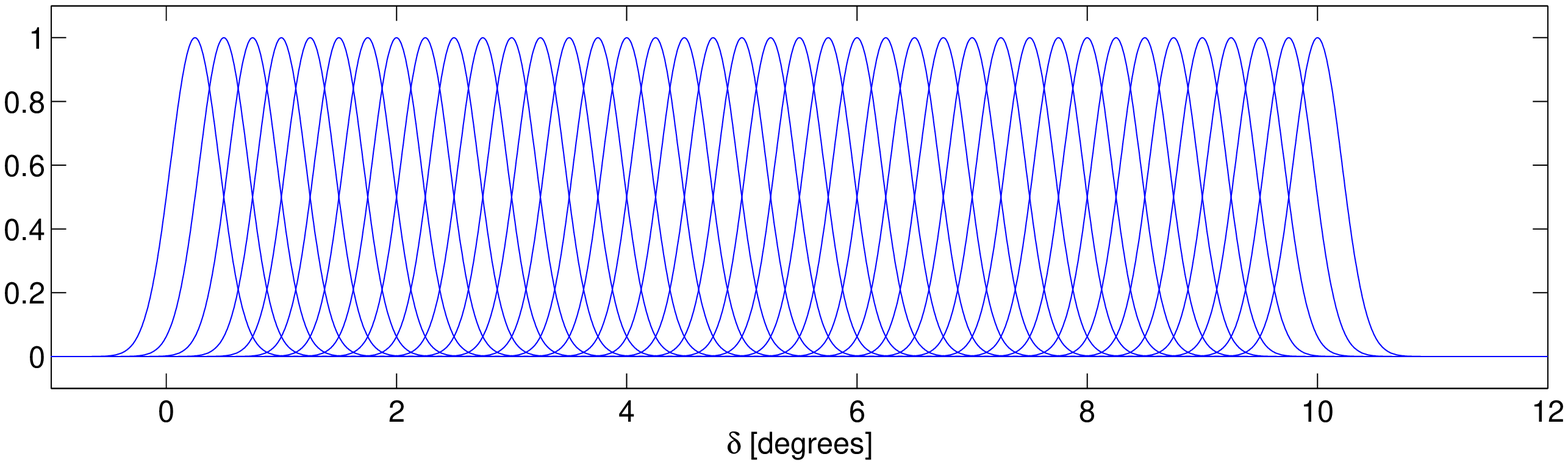}
   \includegraphics[width=1.0\textwidth]{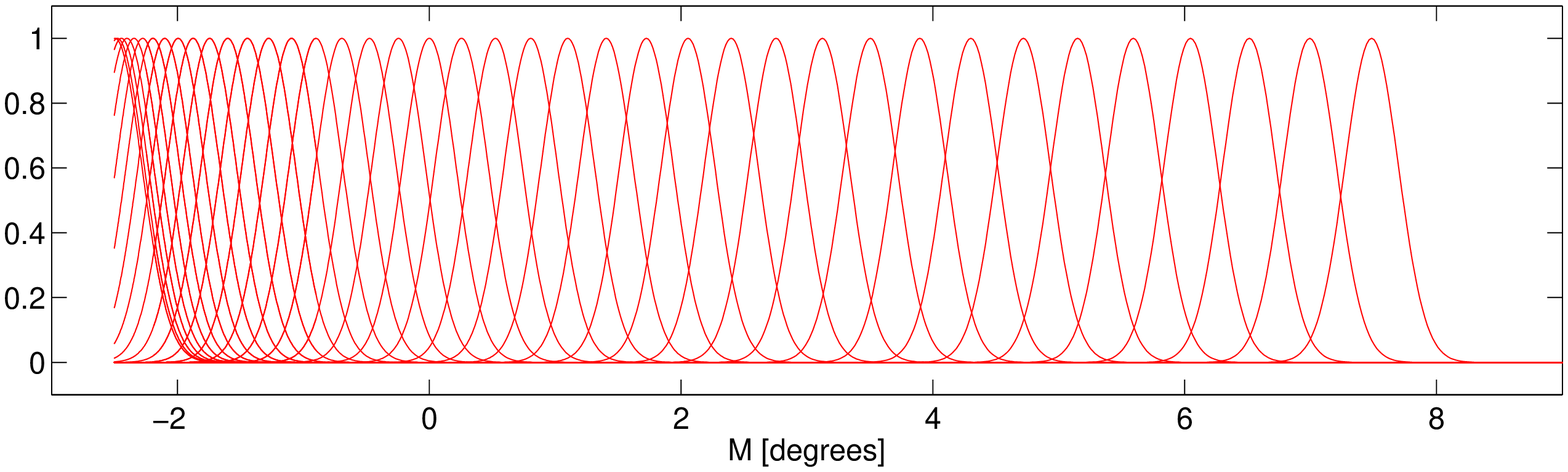}
\caption{Beam response of pointings in the mosaic as function of the
  observed declination $(\delta)$(top panel) and the {\it NCP}
  projected declination $(M)$. The observed mosaic is Nyquist sampled,
  with a regular interval between beams of 15 arcmin. When projecting
  the data, the sampling becomes different.}
  \label{observed_pointings}
\end{figure*}

We have noticed a shortcoming in the {\it invert} task within {\it
  miriad} when the data is converted from the {\it u,v} plane to the
{\it image} plane. When creating the mosaiced image, each individual
pointing is inverted and then the set of images are combined using the
primary beam model for the relative weightings.

When gridding the data using the {\it NCP} projection the
offset in declination with respect to the tangential point is
calculated using equation 2. In this function $\delta_0$ is the
central declination, $\delta$ is the observed declination that has to
be gridded and $\Delta \alpha$ is the difference between the central
Right Ascension and the Right Ascension to be gridded. At declinations
close to zero, the differences in $M$ in the projected frame become
very small: $\delta = 0+\epsilon$ and $\delta = 0+2\epsilon$ are
gridded to the same pixel in the projected frame and for $\delta = 0$
there is a singularity as there is no solution at all.

In the observed frame, the mosaic is Nyquist sampled and there is a
pointing every 15 arcmin between 0.25 and 10 degrees in
declinations. This is shown in the top panel of
Fig.~\ref{observed_pointings} where the response of each beam is
plotted as function of declination. At each position the sum of all
the beam responses is equal and the weighted sum is always unity. 

The separation between the pointings with respect to each other
changes, when they are converted to the projected {\it NCP} frame. Not
only the position of pointings changes, the complete shape of the
primary beam becomes different, and a distorted beam should be applied
when doing the weighting. We suspect that this is not happening and
that the undistorted beam is used. This is demonstrated in the bottom
panel of Fig.~\ref{observed_pointings} where the undistorted beam
response of the pointings is plotted as function of the projected $M$
value. The $M$ value gives the offset with respect to the reference
pixel, so $M=0$ corresponds to $\delta=5$ degrees.

In an image the flux is determined by the weighted sum of the beam
response of all the contributing pointings.:

\begin{equation}
S_{\delta} = \frac{\Sigma S_i(\delta) \cdot B_i(\delta)}{\Sigma B_i(\delta)}
\end{equation}

and 
\begin{equation}
S_{M} = \frac{\Sigma S_i(\delta) \cdot B_i(\delta)}{\Sigma B_i(M)}
\end{equation}

Where $S_i(\delta)$ is the measured flux by a pointing, $B_i(\delta)$
is the beam response in the observed frame, while $B_i(M)$ is the beam
response in the projected frame.

In the projected frame the sum of the beam responses changes
dramatically with declination if the projection is not applied to the
shape of the primary beam.

As a result, fluxes appear lower in the projected frame at low
declinations, as the sum of the beam responses by which the flux is
weighted is larger. At high declinations the opposite is the case where
the flux values in the projected frame become enhanced.

\begin{figure*}[t]
  \includegraphics[width=0.5\textwidth]{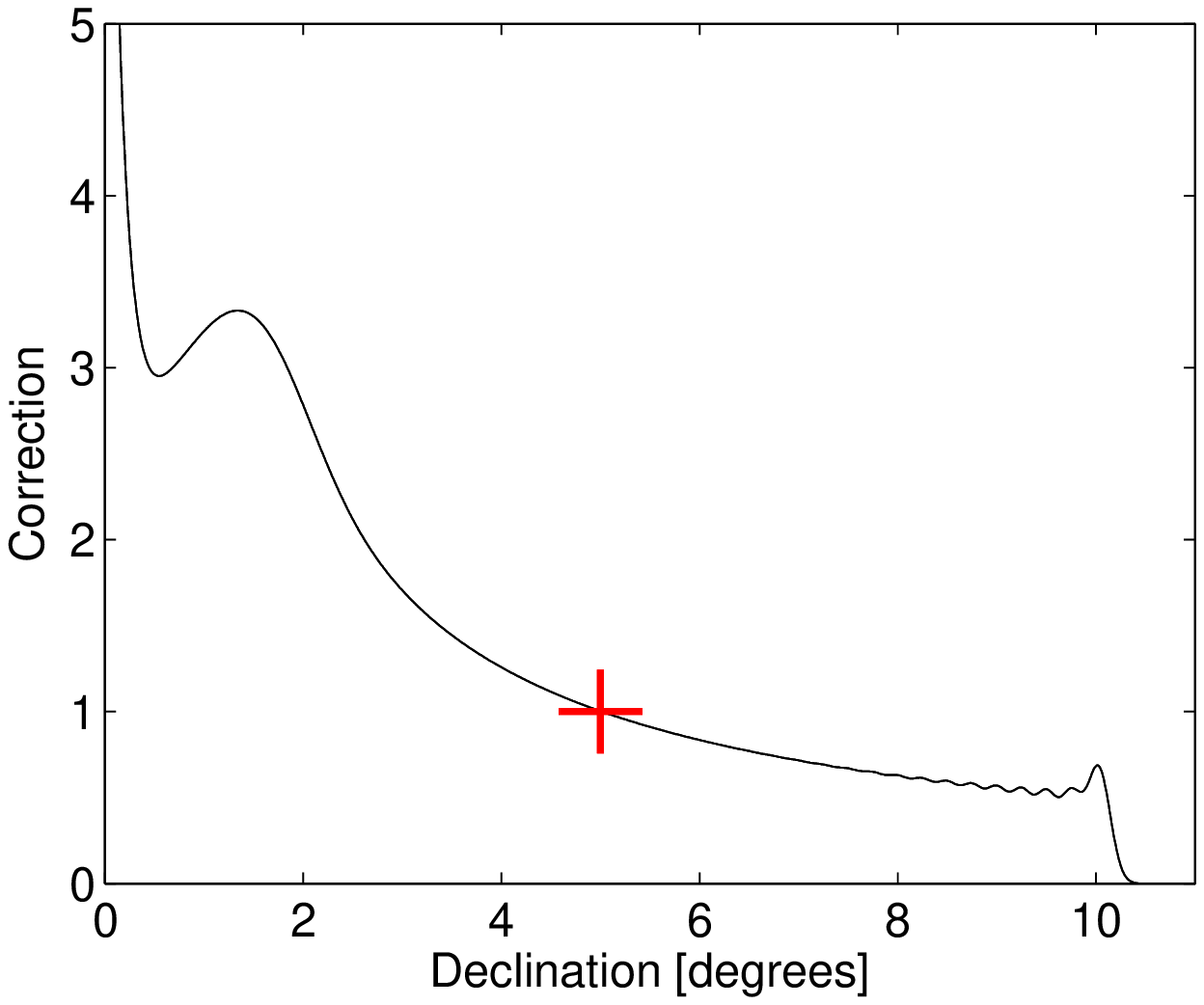}
  \includegraphics[width=0.5\textwidth]{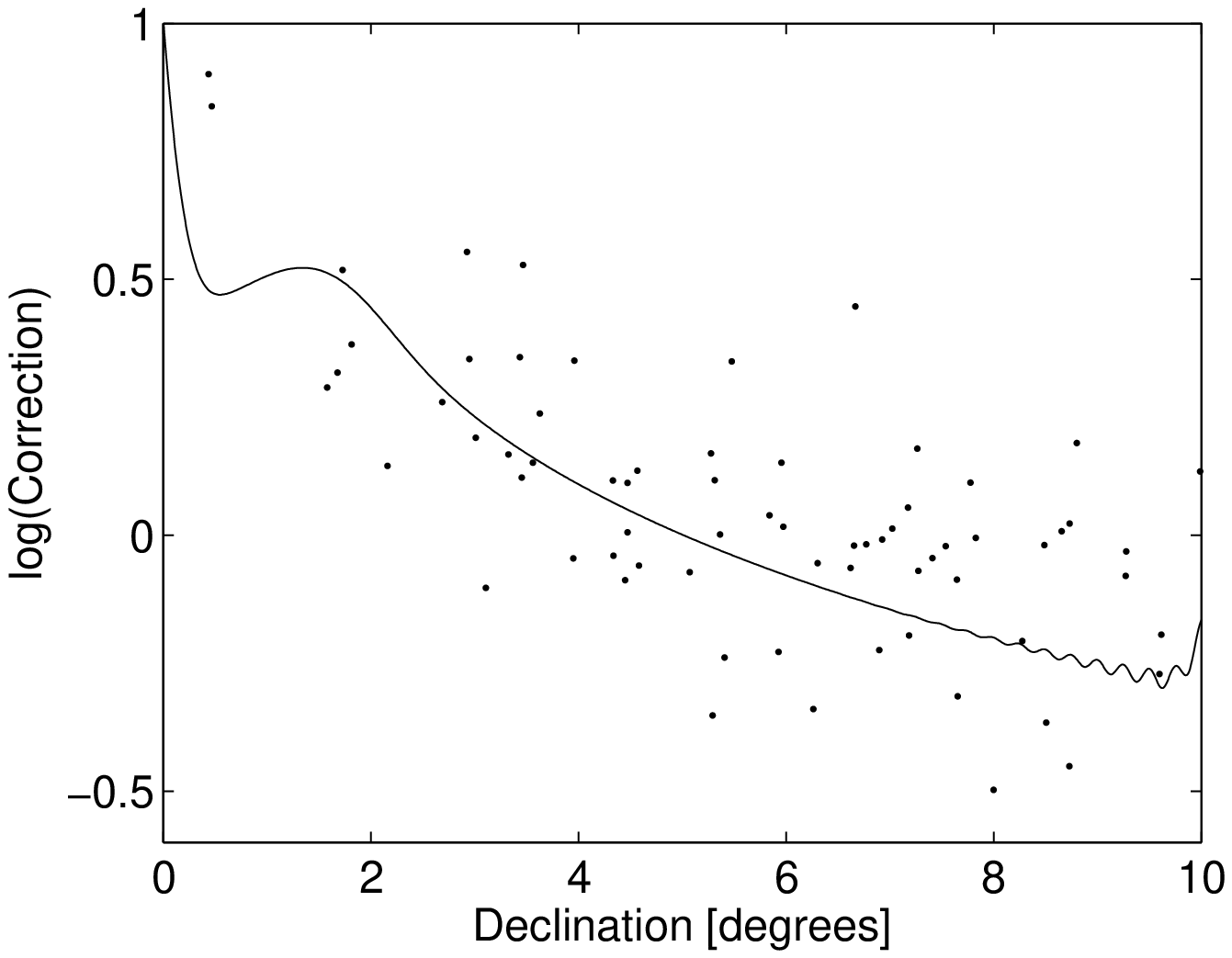}
\caption{Left panel: Correction factor as function of declination, to
  be applied on the obtained fluxes. Right panel: Correcting factor on
  a logarithmic scales with data points representing the flux ratio
  between WVFS total-power and cross-correlation data, before applying
  the correction on the cross-correlation data.}
  \label{dec_correct}
\end{figure*}

This effect is clearly visible in the data, as the measured flux of
objects at low declination is systematically to low. This is
demonstrated for a sample of objects in the right panel of
Fig.~\ref{dec_correct}. The data points show the ratio between the
fluxes obtained from the WVFS total power data as described in chapter
4 and the fluxes from the cross-correlation data, without a flux
correction. The flux ratios are plotted on a logarithmic scale against
declination. At low declinations the total power fluxes are much
higher, while at high declinations the opposite is the case.

The correction that has to be applied is given by:

\begin{equation}
C(\delta) = \frac{\Sigma B_i(M)}{\Sigma B_i(\delta)}
\end{equation}

This ratio is plotted in the left panel of Fig.~\ref{dec_correct} as
function of declination. The cross indicates the central position
of the projected grid at 5 degrees, where the ratio is exactly 1 as no
correction has to be applied. For declinations below 5 degrees the
fluxes have to be scaled up, while for high declinations the fluxes
have to be scaled down. At both ends of the correction function there
is a bump, as the edges of the mosaic are not Nyquist sampled in the
observations, so the integrated beam responses are different here. For
declinations approaching zero degrees the correction goes to infinity,
because of the singularity in the {\it NCP} projection here.

In the right panel of Fig.~\ref{dec_correct} the same correcting ratio
is plotted on a logarithmic scale, together with a sample of data
points as described before. Although the scatter is large the data
points follow the correcting function reasonably well. The fluxes
obtained from the total power and cross-correlation data actually can
be different as the total power data is more sensitive to extended
emission, but to get the cross-correlation data on the right level,
the correction has to be applied.

\section{Results}

\subsection{Source Detection}
\label{source_detect}
Because of the large extent of the WVFS, and the large number of independent
pixels, an automated source finding algorithm is essential to obtain a
list of candidate detections. Although the sensitivity of the data is
good, source detection is not straightforward because of the artifacts
that are apparent in the data as described in section
\ref{false_pos}. Two strategies have been employed to circumvent these
complications and to obtain a list of candidate detections that is as
complete as possible. The first method is a blind search that uses a
clipping level of 8$\sigma$. This conservative clipping level will
provide a reliable list of bright features. The second method uses a
less conservative clipping level of 5$\sigma$ that will also yield
many false positives. An extra constraint on these features is that an
optical counterpart is required within a suitable search radius and at a
comparable radial velocity. The search radius is variable, depending
on the declination of the candidate feature, as will be explained below.

For both approaches the {\it Duchamp} \citep{2008glv..book..343W} source
finding algorithm has been used, with different control parameters. The two
methods and their results are described below. When searching for
objects, {\it Duchamp} uses the processed, three dimensional data
cubes. The source finder has been run on each of the 18 cubes
individually, although for Cubes 8 and 9 every pixel above a
declination of 9 degrees has been blanked. This region could not be
used due to solar interference, which heavily affected the quality of
the data. The interference is so strong here that it increases the
noise and causes many false and unreliable detections. As a
consequence, any objects between $\sim 172$ and $\sim 190$ degrees in
Right Ascension and above 9 degrees in Declination will not be
detected.

\begin{figure}[t]
  \includegraphics[width=0.5\textwidth]{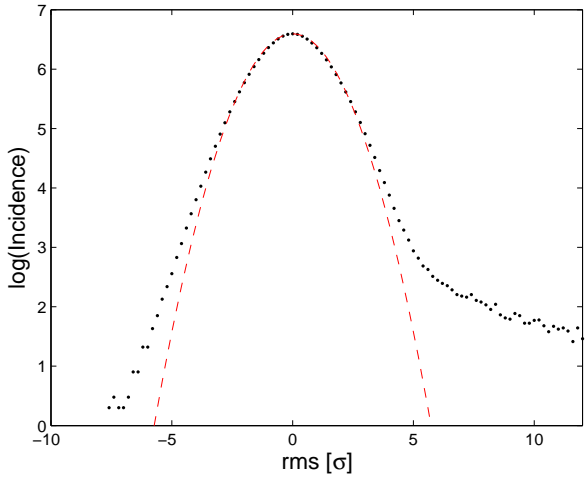}
  \caption{Histogram of incidence of brightness in units of $\sigma$
    in Cube 13 of the WVFS. At a level of $-5\sigma$ there are still
    many pixels, indicating that the number of false positives at the
    $+5\sigma$ level is also likely to be relatively high. At a level
    of $-8\sigma$ there are no negative pixels, so above a clipping
    level of $+8\sigma$ no false detections are expected.}
  \label{noise_histo}
\end{figure}

\subsection{Blind Detections}
Although the noise in the data is approximately Gaussian, it is clear
from an inspection of the histogram of brightnesses that there are
excess negative outliers from this ideal distribution. Due to the
purely relative nature of interferometric data (the absence of the
auto-correlation), artifacts caused by e.g. side-lobes, continuum
sources and solar interference are symmetric in their positive and
negative excursions with respect to zero. For this reason, the number
of noise and artifact pixels within a negative brightness interval is
equal to the number of likely false positive pixels within the same
brightness interval on the positive side of the histogram. This is
illustrated in Fig.~\ref{noise_histo} where a histogram of the observed
brightness is plotted from the central portion of Cube 13. The
incidence of each brightness scaled by the local RMS fluctuation level
is plotted. At 4 or 5 $\sigma$ the number of negative pixels is still
significant, but these rapidly drop to zero for a brightness below
$-$8$\sigma$. A clipping level of +8$\sigma$ was therefore chosen
since it should yield no false positive detections due to either noise
or artifacts.\\

For the actual source finding {\it Duchamp} has been used with four
different settings, that are summarised in table~\ref{duchamp8}. The
cubes were first Hanning smoothed to velocity resolutions of 16,
32, 64 and 128 km s$^{-1}$, to be more sensitive to sources with
different line widths. Although an initial clipping level of 8$\sigma$
has been used for peak identification, the detected features are
"grown" to a level of 3$\sigma$. Furthermore a certain number of
pixels in the velocity domain is required, that is representative for
the velocity resolution.

\begin{table}[t]
\begin{center}
\begin{tabular}{lccccc}
\hline
\hline

       &  $V_{res}$ & rms  & clip & grow  & size \\
       &  [km s$^{-1}$] &  [mJy beam$^{-1}$] & [$\sigma$] &  [$\sigma$] &  [pixels] \\
\hline
Setup1 & 16 & 6.7 & 8 & 3 & 3\\
Setup2 & 32 & 5.0 & 8 & 3 & 5\\
Setup3 & 64 & 4.1 & 8 & 3 & 9\\
Setup4 & 128 & 3.1 & 8 & 3 & 17\\

\hline
\hline
\end{tabular}
\caption{{\it Duchamp} parameters for the blind {\HI} search in the WVFS cross-correlations.} 
\label{duchamp8}
\end{center}
\end{table}

\begin{figure}
  \includegraphics[angle=270,width=0.5\textwidth]{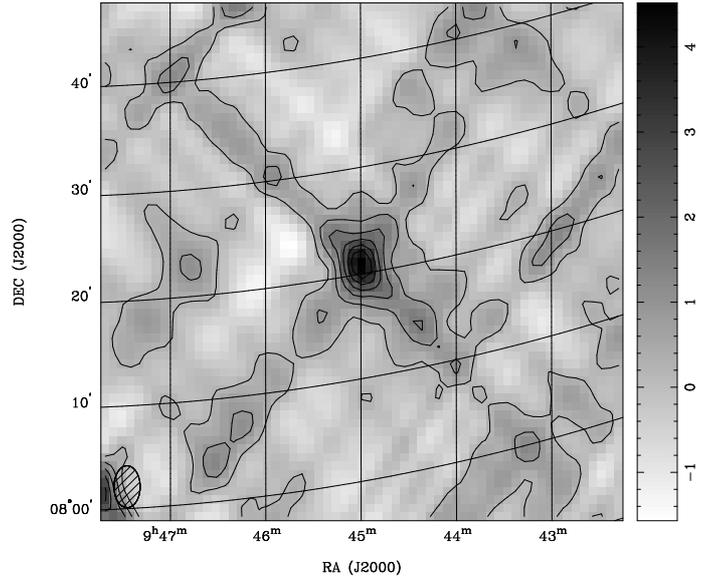}
  \caption{Example of a source with a bright residual side-lobe
    pattern. An {\it X} pattern is clearly visible and those regions
    within the side-lobes with multiple contours might be
    mis-identified as individual objects by the {\it Duchamp}
    source-finding algorithm.}
  \label{side_fig}
\end{figure}

For each candidate detection, the spectrum was determined and a moment
map has been created by integrating the data cube over the velocity
width of the detection. Furthermore, an optical counterpart was sought
for each source within a radius of 7 arcmin. A radius of 7 arcmin was
chosen both to accomodate some intrinsic offset of the {\HI} and optical
centroid as well as those instances of limited positional accuracy, as
will be explained below.

All spectra and moment maps were inspected visually, to look for
artifacts. Continuum artifacts, or solar interference can result in a
false detection, but are easily recognised by eye.

The moment maps were used to eliminate those candidate sources coincident
with residual side-lobe artifacts. When candidate sources are
coincident with one of the arms of the X-shaped residual side-lobe of
a bright nearby source, then they are very likely unreliable.
An example of a clear residual side-lobe structure is shown
in Fig.~\ref{side_fig}.

Furthermore, all multiple detections of the same source were excluded
from the list of detections. Features that are at the edge of a cube
are likely to be detected twice in two adjacent cubes. In the cubes
with different velocity resolution, the peak flux of the candidate is
not always at the same spatial position, resulting in a slightly
different apparent centroid. Finally, sources are sometimes counted twice when
there are two bright spectral components that do not connect. For
example in the case of a double horned profile, where the region between
the two peaks does not exceed the 3$\sigma$ level which was set as the
``growing'' limit.

A total of 135 sources have been detected using the blind detection
approach, which have a peak brightness exceeding 8$\sigma$. The
properties of each detection are listed in the online appendix, where the spectrum of each detection is shown as well.

The first column gives the name of the
source, consisting of the characters "WVFSCC" ({\it W}esterbork {\it
  V}irgo {\it F}ilament {\it S}urvey {\it C}ross-{\it C}orrelation),
followed by six plus six digits for the Right Ascension and
Declination respectively in hh:mm:ss and dd:mm:ss format. The second
column gives the optical ID if present, followed by the Right
Ascension, Declination and Velocity. The sixth and seventh column give
the line-width at 50 and 20\% of the peak ($W_{50}$ and $W_{20}$). The
following three columns give the integrated line-strength $(S_l)$, the
integrated moment map flux density $(S_i$) and the error in the
integrated flux density $(\sigma_S)$. The differences between the
different flux measurements and the method of estimating the
uncertainty are explained in section~\ref{flux_section}. The last column in the table indicates whether the object has been found in the blind search $(B)$ or in the search where an optical identification is required $(O)$. When objects appear in both search methods, this is indicated with two letters.

Of the 135 detected sources in the blind search method, only WVFSCC~120929+080730 does not have an optical counterpart, the properties of this objects are given in table~\ref{sources}, which has the same column description as the online appendix, but only showing new {\HI} detections in the survey.

\onecolumn


\begin{landscape}
\begin{center}
\begin{longtable}{lllrrccccccc}

\caption{Physical properties of new {\HI} detections in the Westerbork Virgo Filament Survey.\\
{\it (a,)}: Units of  $V_{Hel}$ are given in [km s$^{-1}$].\\
{\it (b)}: Units of  $W_{50}$ are given in [km s$^{-1}$].\\
{\it (c)}: Units of  $W_{20}$ are given in [km s$^{-1}$].\\
{\it (d)}: Line strength $(S_l)$ in units of [Jy km s$^{-1}$].\\
{\it (e)}: Integrated flux $(S_i)$ in units of [Jy km s$^{-1}$].\\
{\it (f)}: Error in $(S_i)$ in units of [Jy km s$^{-1}$]. \\ 
{\it (g)}:  Detection method, blind $(B)$ or with optical identification $(O)$.\\
} 
\label{sources}\\


\hline
\hline
 \small Nr.     &    \small   Name           &   \small Optical ID.              &    \small   RA [hh:mm:ss] &  \small Dec [dd:mm:ss] &  \small $V_{Hel}^a$  &  \small $W_{50}^b$ &  \small $W_{20}^c$  &  \small $S_l^d$  &  \small $S_i^e$ &  \small $\sigma_S^f$ & \small Detection$^g$ \\
\hline   
\endfirsthead

\hline
\hline
 \small  Nr.     &    \small   Name           &   \small Optical ID.              &      \small RA [hh:mm:ss] &  \small Dec [dd:mm:ss] &  \small $V_{Hel}$ &  \small $W_{50}$ &  \small $W_{20}$ &  \small $S_l$  &  \small $S_i$ &  \small $\sigma_S$  & \small  Detection\\
\hline
\endhead

\hline
  \multicolumn{10}{c}{{Continued on Next Page\ldots}} \\
\endfoot

\hline \hline
\endlastfoot


  \small    1    &  \small       WVFSCC 091006+070245    &  \small    SDSS J091019.53+070141.2          &  \small     09:10:06.89    &  \small    07:02:45.74    &  \small     1471      &  \small        38       &  \small        98       &  \small     2.9        &  \small       2.5       &  \small     0.1  & \small O \\ 
  \small    2   &  \small       WVFSCC 091244+085446    &  \small    SDSS J091246.59+085620.1          &  \small     09:12:44.40    &  \small    08:54:46.51    &  \small     1254      &  \small        26       &  \small        39       &  \small     0.5        &  \small       0.8       &  \small     0.1   & \small O\\ 
  \small    3   &  \small       WVFSCC 092116+093914    &  \small    SDSS J092114.98+094352.3          &  \small     09:21:16.13    &  \small    09:39:14.37    &  \small     1368      &  \small        49       &  \small        76       &  \small     0.5        &  \small       1.4       &  \small     0.1   & \small O\\ 
  \small    4   &  \small       WVFSCC 093450+062749    &  \small    CGCG 035-007                      &  \small     09:34:50.79    &  \small    06:27:49.11    &  \small      545      &  \small        72       &  \small        95       &  \small     0.9        &  \small       1.2       &  \small     0.2   & \small O\\ 
  \small    5   &  \small       WVFSCC 094456+082003    &  \small    2MASX J09445892+0822116           &  \small     09:44:56.14    &  \small    08:20:03.72    &  \small     1734      &  \small       161       &  \small       185       &  \small     2.7        &  \small       6.4        &  \small    0.2  & \small O, B \\ 
  \small    6   &  \small       WVFSCC 095531+082217    &  \small    UGCA 188                          &  \small     09:55:31.31    &  \small    08:22:17.84    &  \small     1263      &  \small        62       &  \small       164       &  \small     1.4        &  \small       3.9       &  \small     0.2   & \small O\\ 
  \small    7   &  \small       WVFSCC 100407+063139    &  \small    MRK 0714                          &  \small     10:04:07.61    &  \small    06:31:39.15    &  \small     1286      &  \small        28       &  \small        66       &  \small     0.8        &  \small       0.9       &  \small     0.1   & \small O\\ 
  \small    8   &  \small       WVFSCC 102542+053843    &  \small    CGCG 037-033                      &  \small     10:25:42.21    &  \small    05:38:43.58    &  \small     1161      &  \small        91       &  \small       106       &  \small     1.8        &  \small       2.8       &  \small     0.2   & \small O\\ 
  \small    9   &  \small       WVFSCC 111057+093634    &  \small    TOLOLO 1108+098                   &  \small     11:10:57.95    &  \small    09:36:34.50    &  \small     1582      &  \small        84       &  \small       156       &  \small     1.4        &  \small       3.0       &  \small     0.1   & \small O\\ 
  \small    10   &  \small       WVFSCC 120929+080730    &  \small     		                    &  \small     12:09:29.73   &  \small    08:07:30.13   &  \small     1175     &  \small        24      &  \small        55      &  \small   0.5       &  \small     1.6           &  \small     0.1      &  \small  B \\ 
  \small    11   &  \small       WVFSCC 121225+025023    &  \small    LEDA 135791                       &  \small     12:12:25.60    &  \small    02:50:23.41    &  \small      873      &  \small        51       &  \small        69       &  \small     2.6        &  \small       3.9       &  \small     0.3   & \small O\\ 
  \small    12   &  \small       WVFSCC 122357+081730    &  \small    SDSS J122405.10+081737.8          &  \small     12:23:57.79    &  \small    08:17:30.18    &  \small     1346      &  \small        28       &  \small        64       &  \small     0.5        &  \small       0.7       &  \small     0.1   & \small O\\ 
  \small     13   &  \small         WVFSCC 122711+085019   &  \small  VCC 0976                          &  \small     12:27:11.92   &  \small    08:50:19.56   &  \small     1220     &  \small        84      &  \small       302      &  \small   4.3       &  \small    13.5          &  \small      0.2     &  \small  O, B  \\ 
   \small   14   &  \small       WVFSCC 123310+090816    &  \small   VCC 1496                           &  \small     12:33:10.55    &  \small    09:08:16.52    &  \small     1114      &  \small        83       &  \small       120       &  \small    1.6         &  \small      7.5       &  \small     0.5    & \small O, B\\ 
   \small   15   &  \small       WVFSCC 123720+044705    &  \small   VCC 1713                           &  \small     12:37:20.25    &  \small    04:47:05.82    &  \small     1642      &  \small        19       &  \small        37       &  \small    0.7         &  \small      0.9       &  \small     0.2   & \small O \\ 
   \small   16   &  \small       WVFSCC 143926+090655    &  \small   SDSS J143912.44+090805.8           &  \small     14:39:26.66    &  \small    09:06:55.44    &  \small     1581      &  \small        64       &  \small        74       &  \small    0.6         &  \small      2.1       &  \small     0.1  & \small O  \\ 
   \small   17   &  \small       WVFSCC 160650+063353    &  \small   CGCG 051-043                       &  \small     16:06:50.30    &  \small    06:33:53.19    &  \small     1757      &  \small        21       &  \small        60       &  \small    0.8         &  \small      1.3       &  \small     0.2   & \small O \\ 


\end{longtable}
\end{center}

\end{landscape}


\twocolumn

\subsection{Confirmed Detections}
Because of the high clipping limit in the blind search, many objects
have been missed that are at a lower significance, which might still
be real sources. In a second approach, all features are sought with a peak
brightness of at least 5$\sigma$. This limit was chosen to permit the
deepest possible search for the {\HI} counterparts to known objects while
keeping the number of likely false positives manageable. An optical
identification was sought for each candidate within a variable search
radius, using the NASA Extragalactic Database (NED) \footnote{The
  NASA/IPAC Extragalactic Database (NED) is operated by the Jet
  Propulsion Laboratory, California Institute of Technology, under
  contract with the National Aeronautics and Space
  Administration.}. Only those features which have an optical
counterpart at a radial velocity that is within 100 km s$^{-1}$ of the
detected feature are accepted. Because the survey has been done using
an east-west array, the synthesized beam size is increasing towards
lower Declinations. For features above 8 degrees in Declination, a
search radius of 6 arcmin was employed, which is on the order of the
synthesized beam FWHM. For lower declinations, the search radius is
scaled as a function of the Declination:

\begin{equation}
r(\delta) = 8 \frac{\sin8}{\sin\delta} \textrm{   for $2<\delta<8^\circ$}
\end{equation}
For Declinations below 2 degrees a constant search radius of 24 arcmin
is used, which is of the same order as the primary beam of the
telescope. This scaling is chosen to track the distortions of the NCP
projection at these low Declinations. Allowing even larger search radii is
not necessary, as the uncertainty in position cannot be larger than
the primary beam.

A very similar method has been used to identify counterpart features
as for the blind search, employing {\it Duchamp} with two different
settings, as given in Table~\ref{duchamp5}. Only two velocity
resolutions have been used, as using more combinations did not appear
to be practical. The list of candidate features was visually inspected
in a very similar way as the blind search detections. Multiple
detections were omitted, as well as false positives caused by
artifacts in the data.

\begin{table*}
\begin{center}
\begin{tabular}{lccccc}
\hline
\hline

       &  $V_{res}$ [km s$^{-1}$] & rms [mJy beam$^{-1}$] & clip [$\sigma$] & grow [$\sigma$] & size [pixels] \\
\hline
Setup1 & 16 & 6.7 & 5 & 3 & 3\\
Setup2 & 32 & 5.0 & 5 & 3 & 5\\
\hline
\hline
\end{tabular}
\caption{Duchamp parameters for the counterpart search in the WVFS} 
\label{duchamp5}
\end{center}
\end{table*}

In a similar fashion as for the blind search, all the spectra and
moment maps of the candidate features have been inspected visually to
look for artifacts. All detections with their properties are
listed in the appendix, together with the sources that have been detected through the blind detection.

A total of 198 {\HI} features with a peak brightness exceeding 5$\sigma$
could be identified with an optical counterpart within the search
radius. We have compared the list of detections with the HIPASS
catalogue. In total 53 objects are not listed in the HIPASS database,
of which 16 are completely new {\HI} detections. On the other hand 51
objects have not been detected in the current data, which are listed
in the HIPASS catalogue. The properties of the new {\HI} detections are listed in table~\ref{sources}.

The WVFS has a better point source sensitivity than HIPASS, which
explains the sources that are detected in WVFS and not in HIPASS. For
the HIPASS detections, a clipping level of $4\sigma$ has been used
which corresponds to approximately 60 mJy beam$^{-1}$ over 18 km
s$^{-1}$ taking into account the point source sensitivity of the original
HIPASS cubes. The $5\sigma$ clipping limit that has been used in this
survey corresponds to $\sim 30$ mJy beam$^{-1}$ at a very comparable
velocity resolution of 16 km s$^{-1}$.

Although the point source sensitivity of the WVFS is almost a factor
two better than HIPASS, the latter has a much better brightness
sensitivity due to the larger beam. The 14 arcmin beam of the Parkes
telescope is approximately six times larger in area than the
synthesised beam of the WVFS, resulting in a three times better
brightness sensitivity. This may explain the relatively high number of
sources that have not been detected in the WVFS. Diffuse sources will
be more easily detected in the larger beam of a single dish like the
Parkes telescope.

\begin{figure}
  \includegraphics[width=0.5\textwidth]{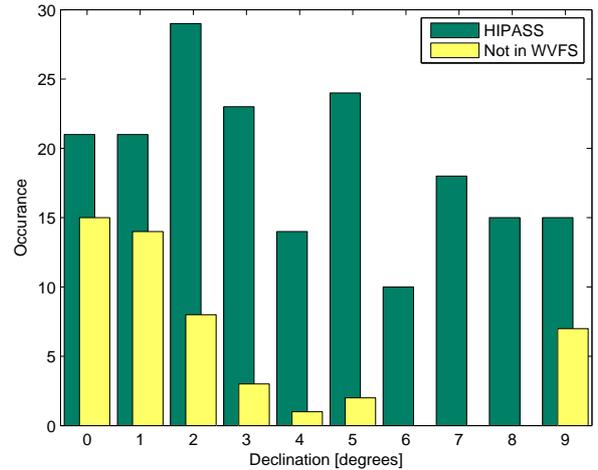}
  \caption{The dark histogram shows the number of HIPASS sources as
    function of declination between 0 and 10 degrees, binned in
    intervals of 1 degree. The light histogram shows the number of
    HIPASS objects that have not been found in the WVFS source
    detection procedures. The distribution of undetected sources is not
    uniform, as their proportion increases toward lower Declination.}
  \label{no_hipass}
\end{figure}

The incidence of all the 51 HIPASS sources that are missed in the
current survey are plotted as a histogram of Declination with a bin
size of 1 degree in Fig.~\ref{no_hipass}. The dark bars of the
histogram represent the distribution of sources catalogued by HIPASS,
while the light coloured bars show the number of HIPASS sources that
have not been detected by WVFS. The histogram does not show a uniform
distribution for the missed sources, several sources that have not
been identified in WVFS are in the region that has been excluded from
the search due to solar interferences, as explained in
section~\ref{source_detect}. This explains the high number of
non-detections in the last bin, above a declination of 9 degrees. What
is striking is that the vast majority of the sources that have not
been detected in the WVFS cross-correlation data have a low
Declination. This is again an effect of the shortcoming in the
gridding, as has been described earlier. At declinations below 2
degrees the projected fluxes become diluted by a factor more than two
and in many cases end up below the detection limit.

\subsection{Flux comparison}
\label{flux_section}
\begin{figure}
\begin{center}
  \includegraphics[width=0.5\textwidth]{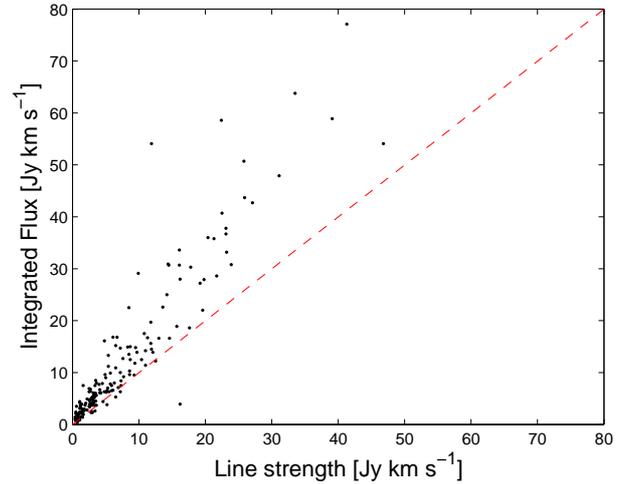}
\end{center}
  \caption{Integrated flux of the WVFS cross correlation data plotted
    against the line-strength. For many objects, the integrated flux
    is significantly larger than the line strength implying that
    detections are spatially resolved.}
  \label{flux1}
\end{figure}

\begin{figure}
\begin{center}
  \includegraphics[width=0.5\textwidth]{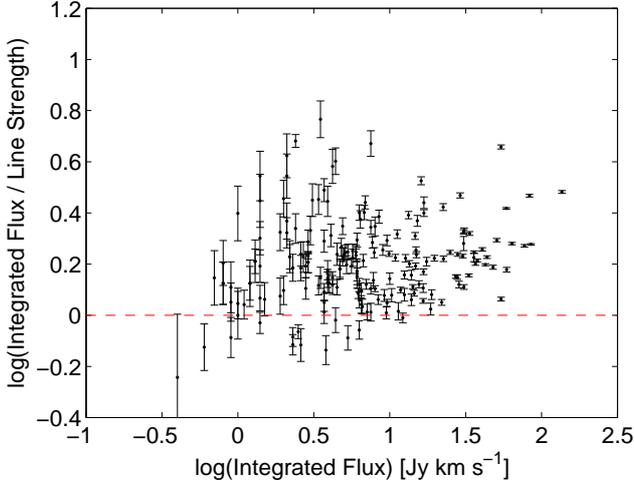}
\end{center}  
\caption{Ratio of integrated flux and line-strength is plotted on a
  logarithmic scale as function of the integrated flux. For almost all
  sources the emission is more extended than the synthesised beam and
  the integrated flux is much higher than the integrated line-strength.}
  \label{flux2}
\end{figure}

The two different methods by which the flux was determined are
compared in Fig.~\ref{flux1} where the integrated flux is plotted as
function of the line-strength. The dashed line through the origin of
the plot indicates where the fluxes are equal. For low flux values,
there is reasonable agreement between the two methods. For high fluxes
the differences are increasing dramatically, with most bright sources
being much more extended than the synthesised beam. This is shown in a
different way in Fig.~\ref{flux2}, where the ratio of the two fluxes
with their error bars is plotted on a logarithmic scale against the
integrated flux on a logarithmic scale. Here the differences become even
more apparent. In addition to a systematically higher flux in all the
integrated moment maps, the ratio also increases with source
brightness.  This result implies that most objects do have extended
emission on the scale of the $\sim5$ arcmin synthesised beam.\\

\begin{figure}[h]
\begin{center}
  \includegraphics[width=0.5\textwidth]{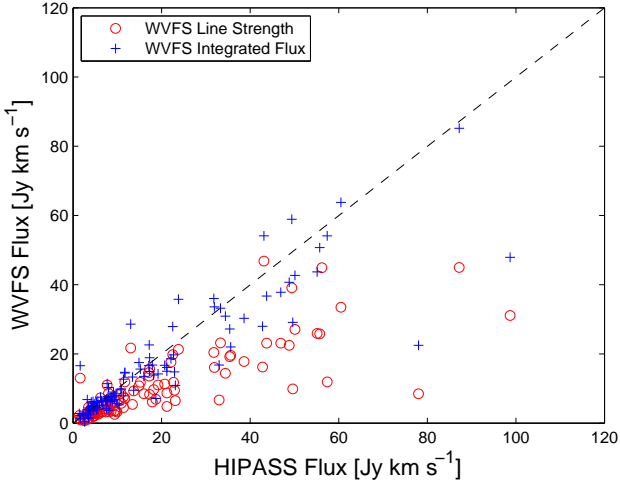}
\end{center}  
\caption{WVFS line-strength (circles) and integrated flux (crosses)
  are plotted as function of HIPASS flux. Although the line-strength
  is very low for many objects with respect to the HIPASS fluxes,
  there is good correspondence between the HIPASS data and the WVFS
  integrated fluxes.}
  \label{flux3}
\end{figure}

\begin{figure}[h]
\begin{center}
  \includegraphics[width=0.5\textwidth]{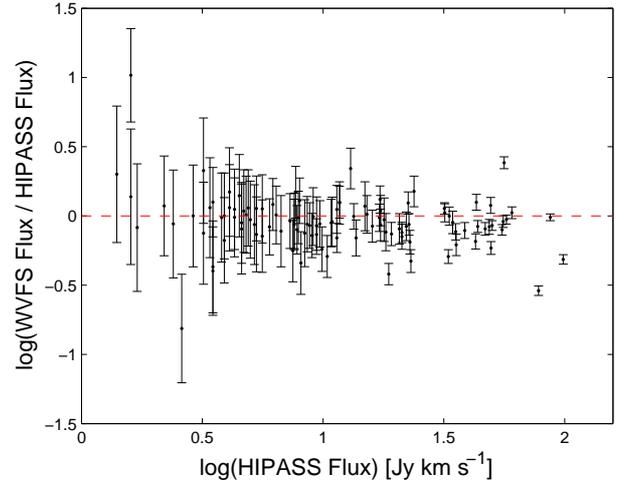}
\end{center}
  \caption{Ratio of WVFS integrated flux and HIPASS fluxes as function
    of HIPASS fluxes, plotted on a logarithmic scale.}
  \label{flux4}
\end{figure}

As a second test, all measured fluxes have been compared with fluxes
obtained from the HIPASS catalogue, where available. Fig.~\ref{flux3}
shows both the WVFS line-strength (circles) as the integrated flux
(crosses) plotted against HIPASS fluxes. The dashed line indicates
where the plotted fluxes are equal. Fluxes determined from the
line-strength measurement are much too low; while there is reasonable
correspondence between the integrated WVFS fluxes and the HIPASS
fluxes. This is also shown in Fig.~\ref{flux4}, where the ratio
between the integrated WVFS fluxes and the HIPASS fluxes is plotted on
a logarithmic scale against the HIPASS fluxes. At low flux values the
scatter is very large, but overall there is very good
correspondence. To be less influenced by the scatter, for all objects
with a HIPASS flux stronger than 5 Jy km s$^{-1}$ the mean and median
values have been determined. The mean of the ratios is 0.91 with a
standard deviation of 0.33, while the median of the ratios is 0.87,
these numbers indicate that there is an excess in the catalogued
HIPASS fluxes of $\sim 10$\% with respect to the WVFS
cross-correlation data.

\begin{figure*}[t]
\begin{center}
  \includegraphics[width=0.45\textwidth]{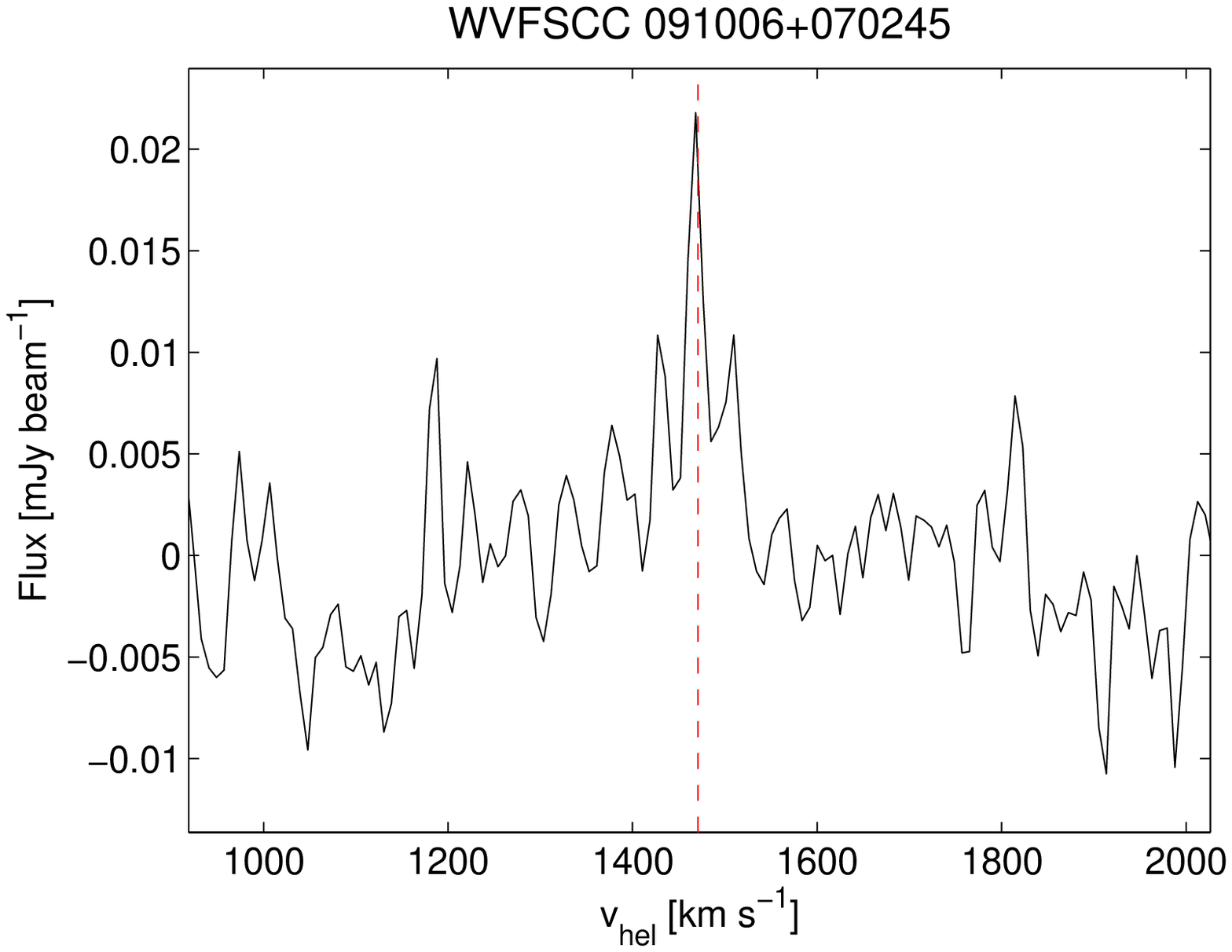}
  \includegraphics[width=0.45\textwidth]{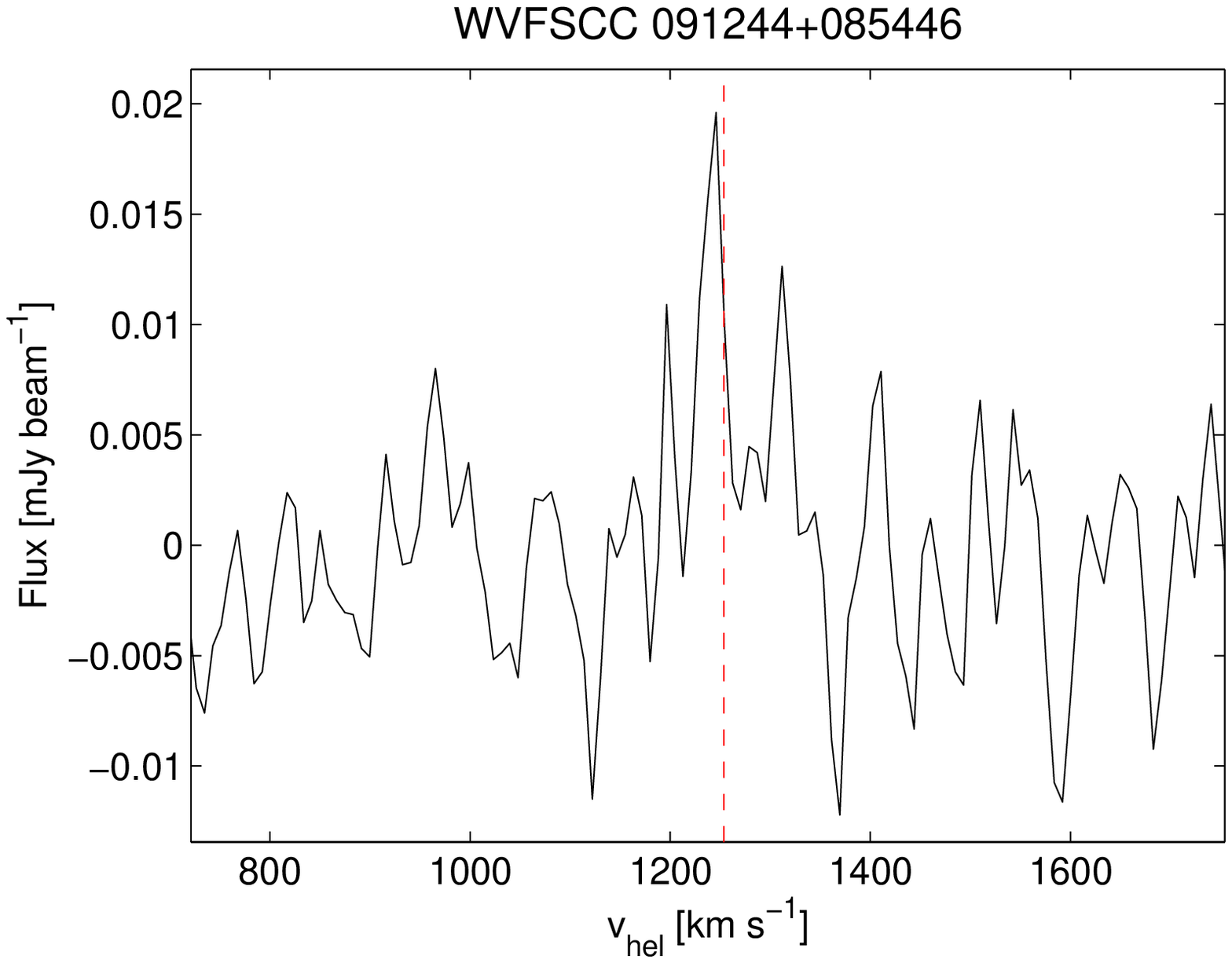}
  \includegraphics[width=0.45\textwidth]{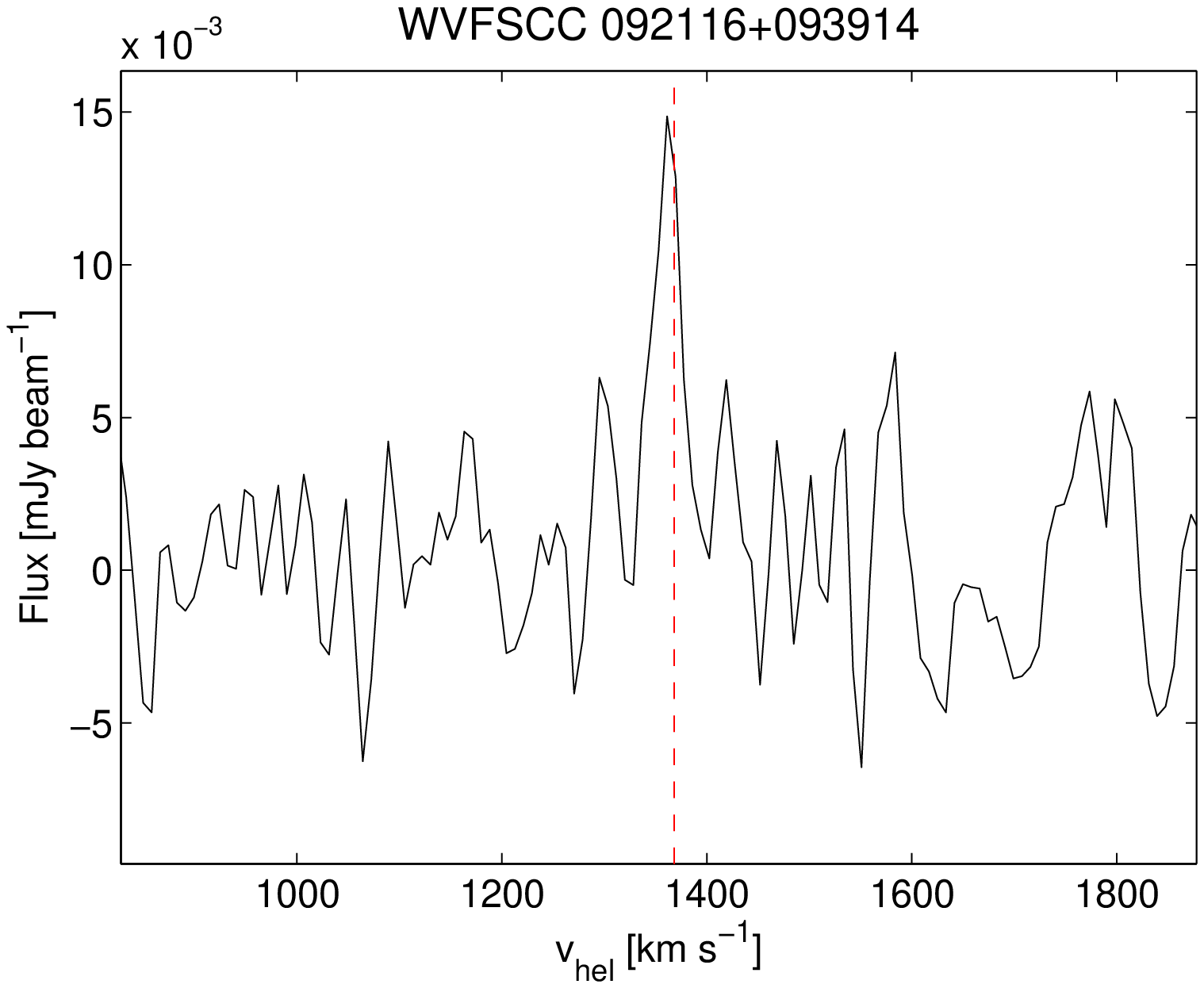}
  \includegraphics[width=0.45\textwidth]{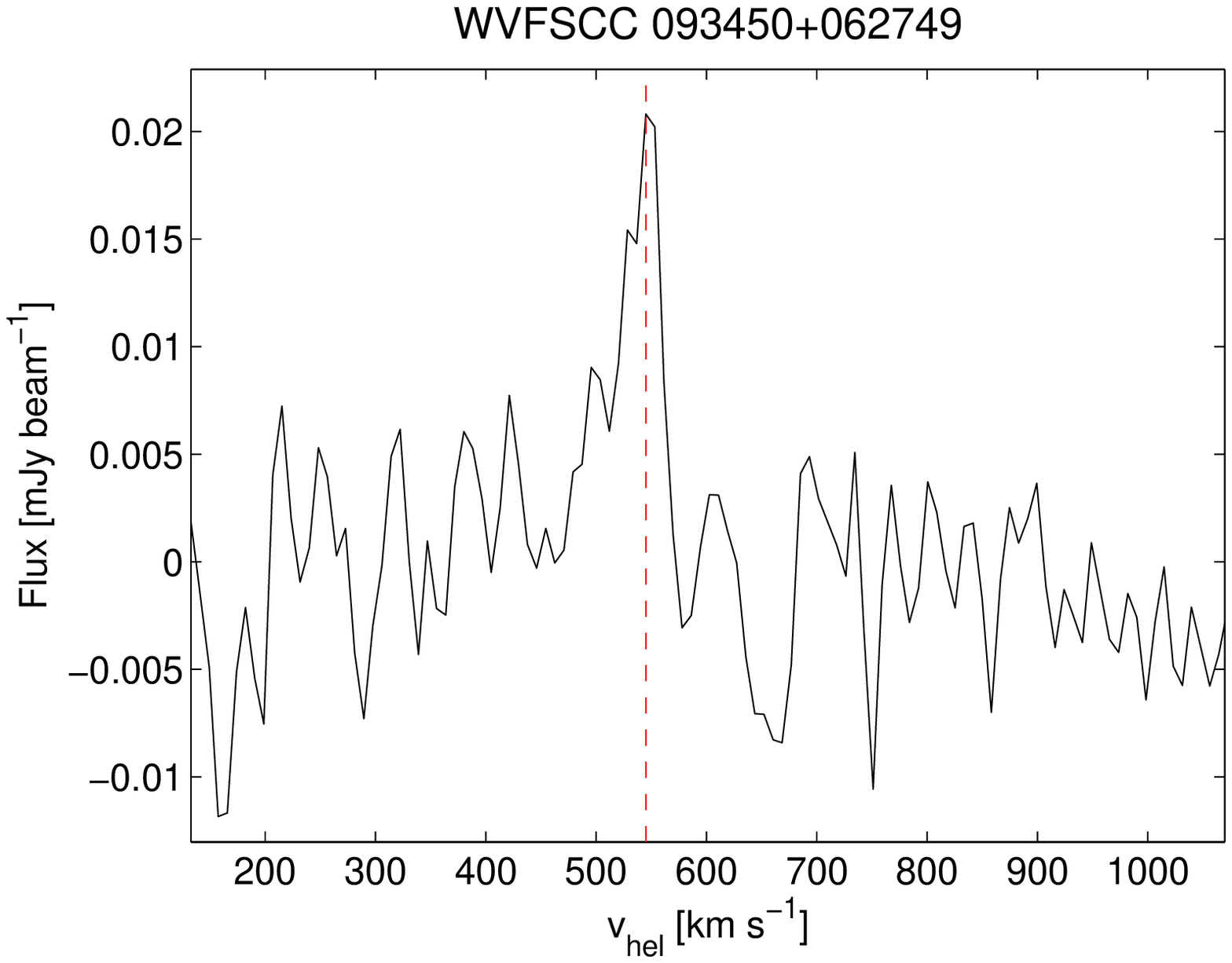}
  \includegraphics[width=0.45\textwidth]{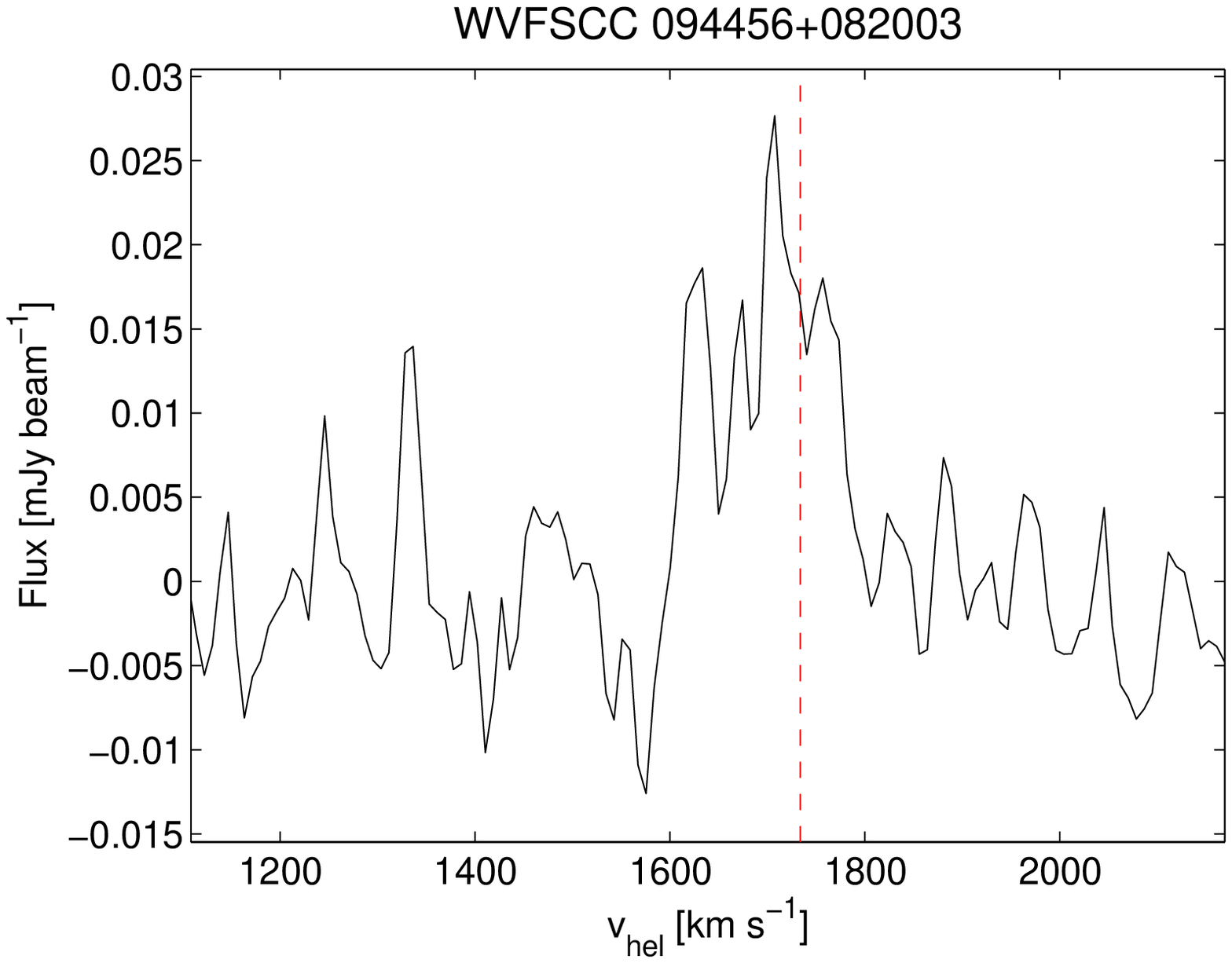}
  \includegraphics[width=0.45\textwidth]{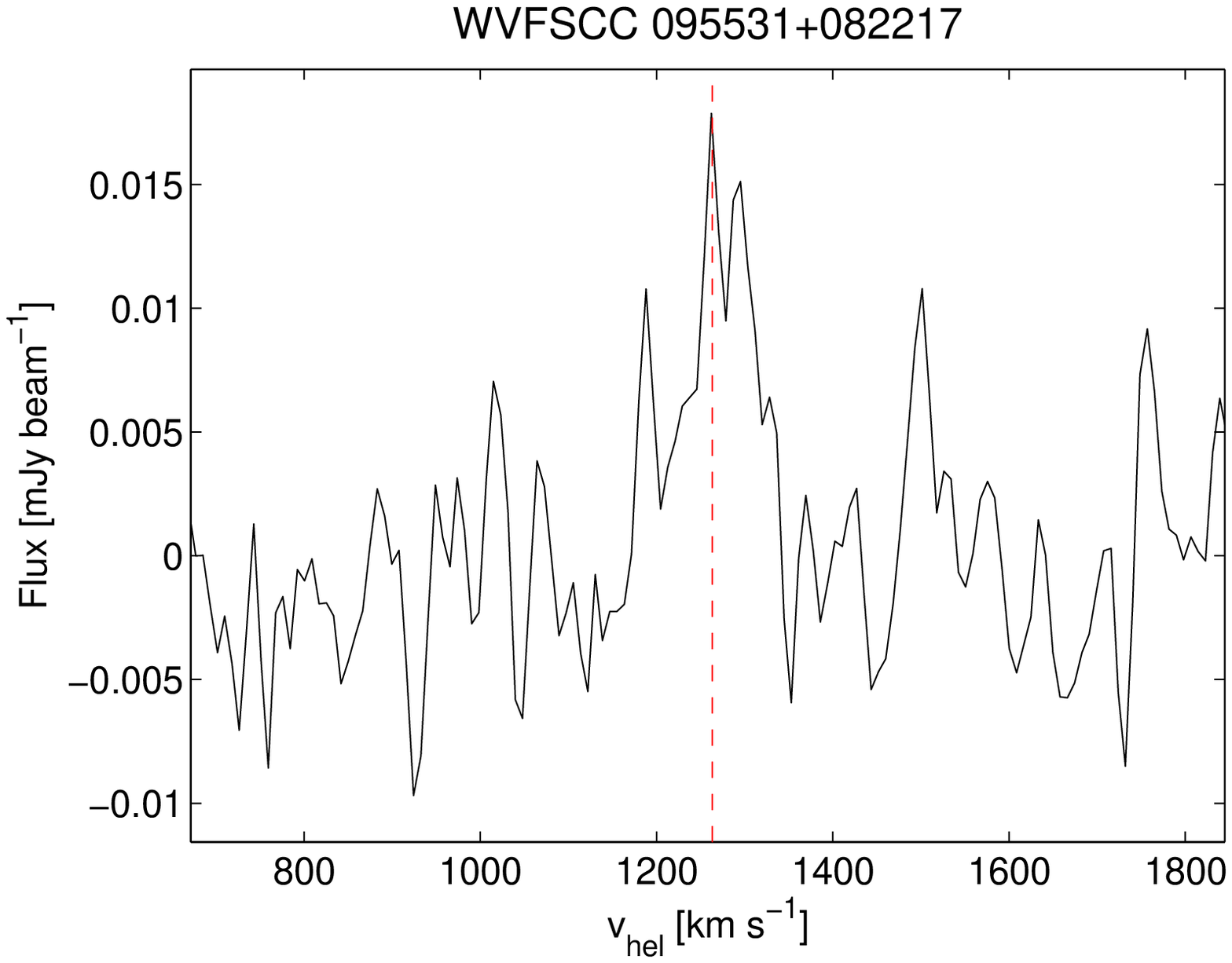}
\end{center}

  \caption{Spectra of new {\HI} detections is the WVFS cross correlation
  data. The central velocity of each object is indicated by the dashed
  vertical line.}
  \label{new_HI}
\end{figure*}

\begin{figure*}[t]
\begin{center}
  \includegraphics[width=0.45\textwidth]{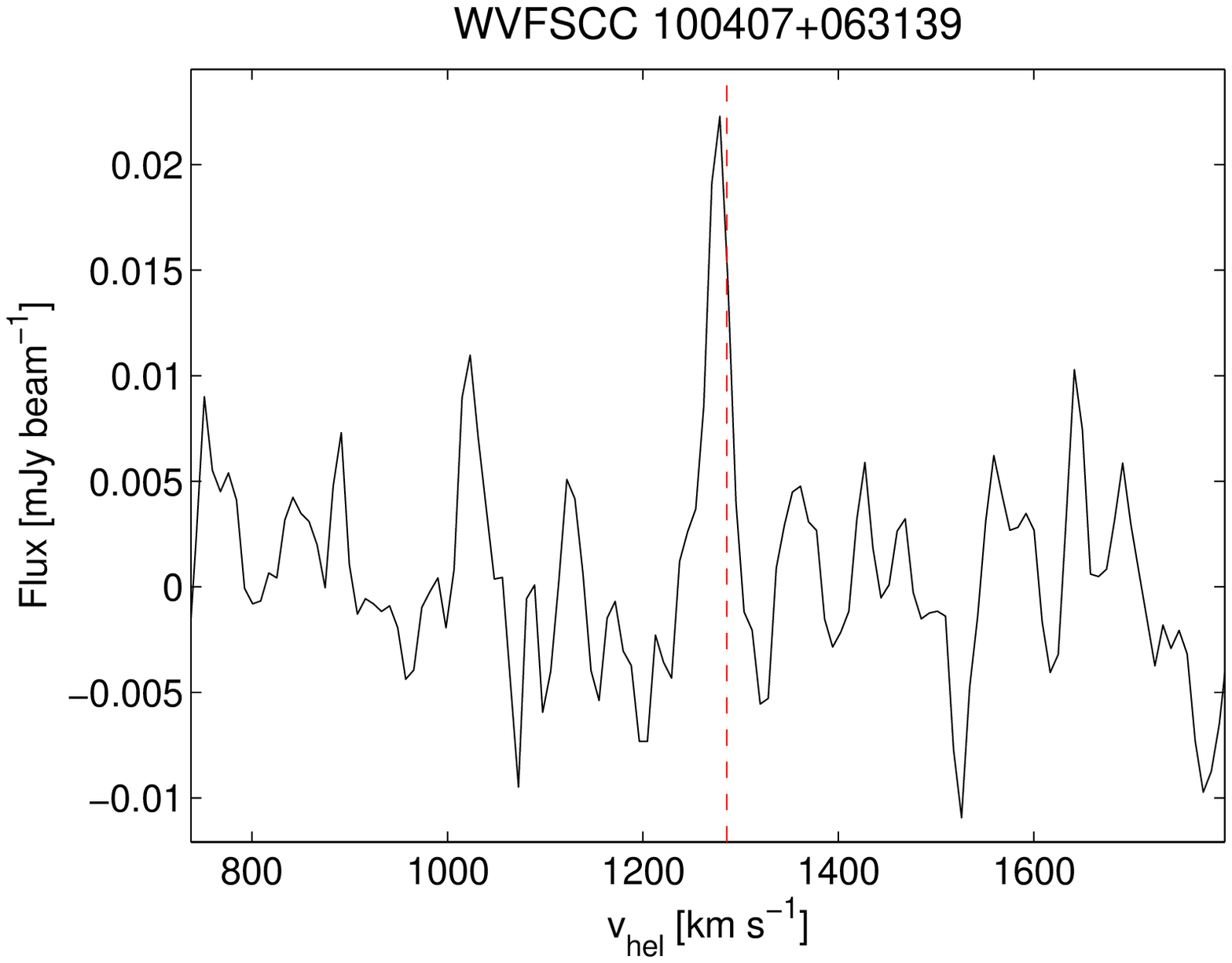}
  \includegraphics[width=0.45\textwidth]{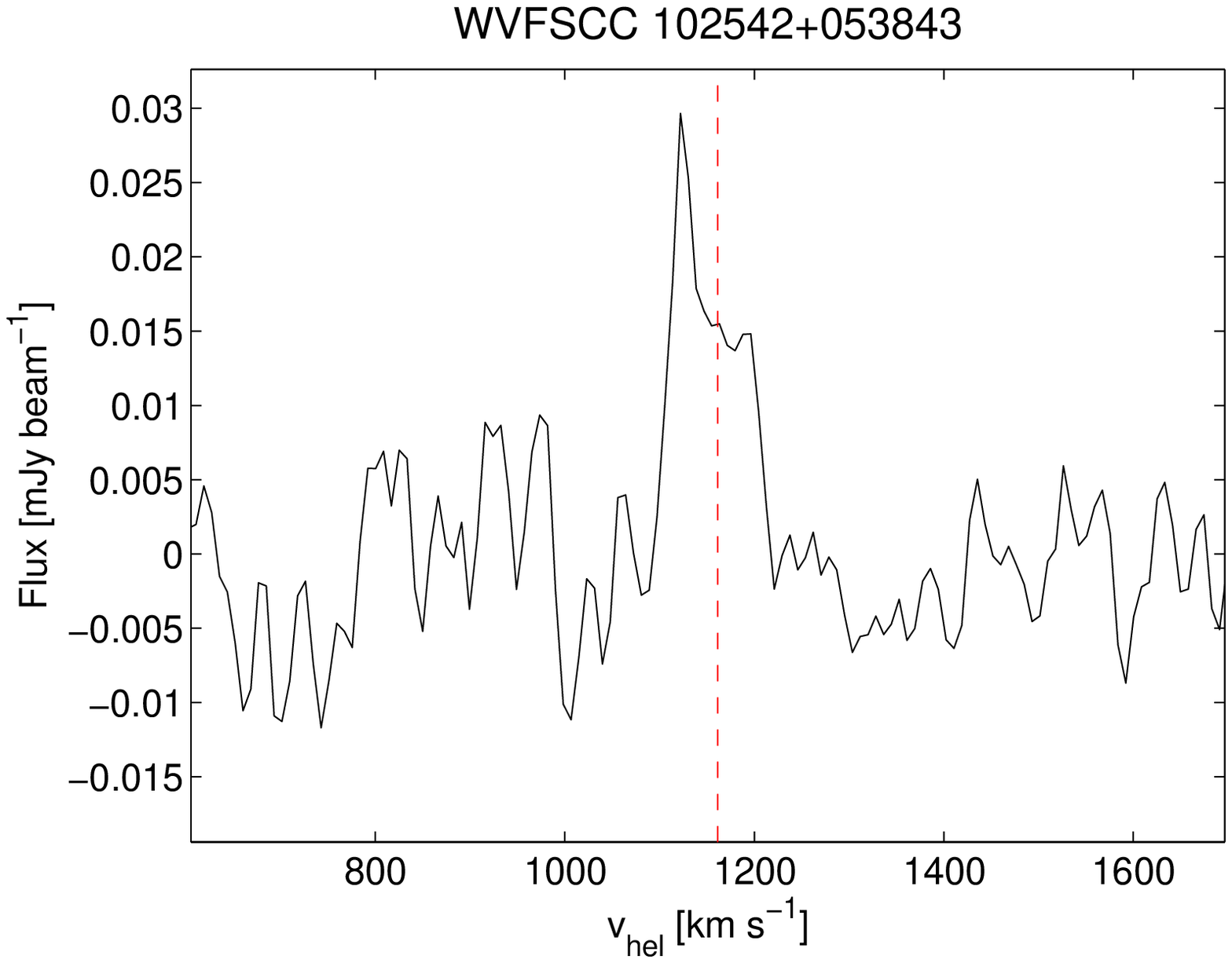}
  \includegraphics[width=0.45\textwidth]{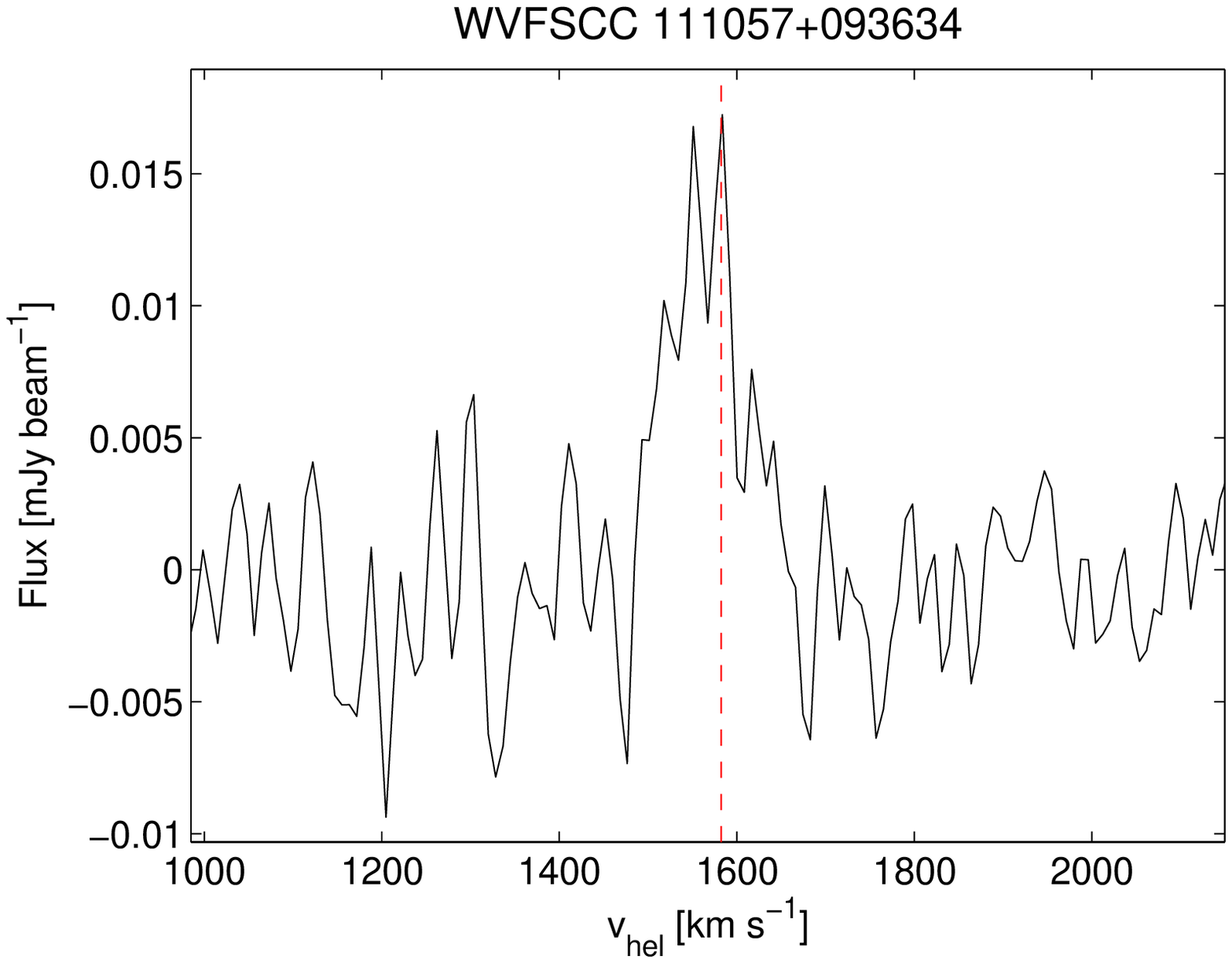}
  \includegraphics[width=0.45\textwidth]{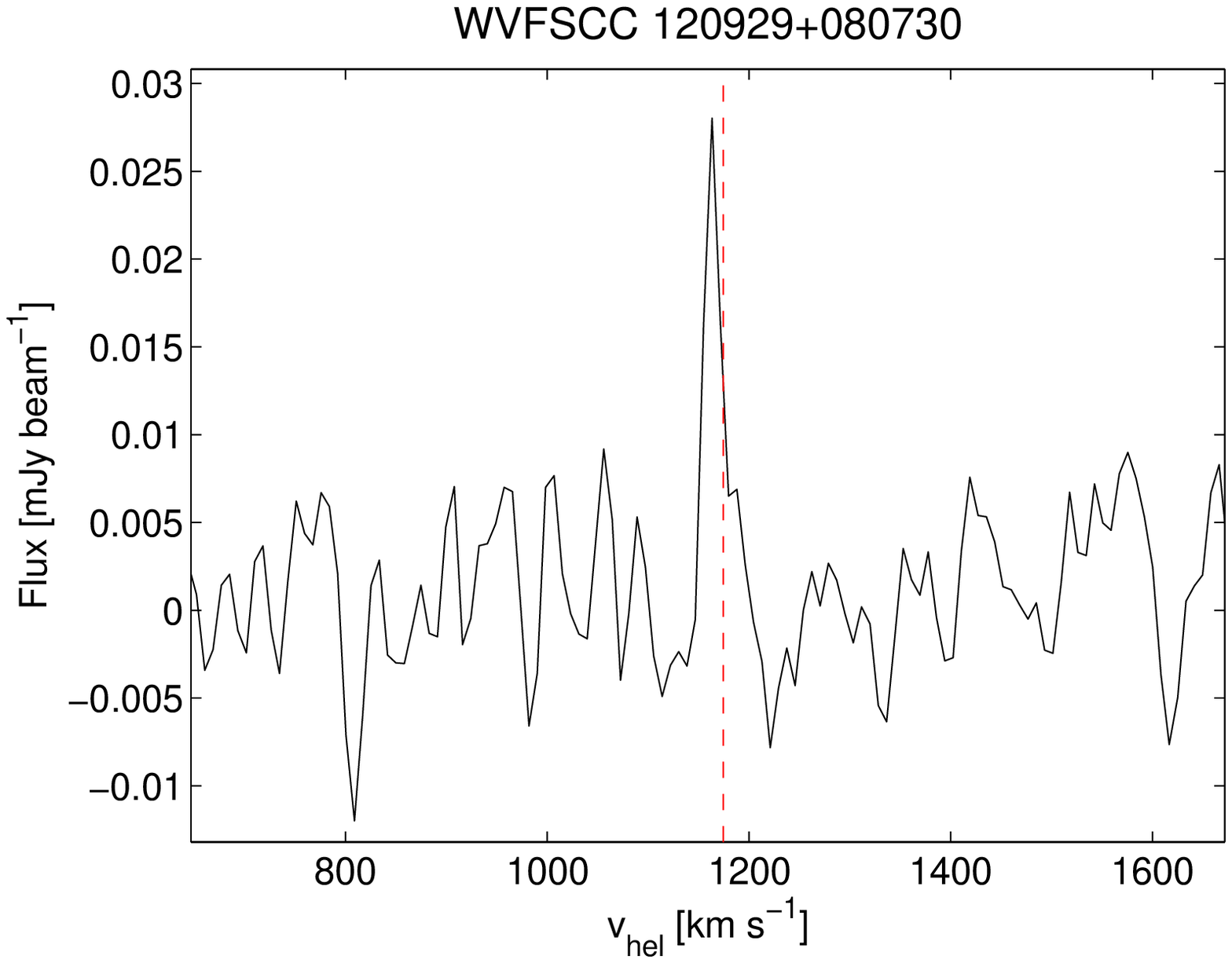}
  \includegraphics[width=0.45\textwidth]{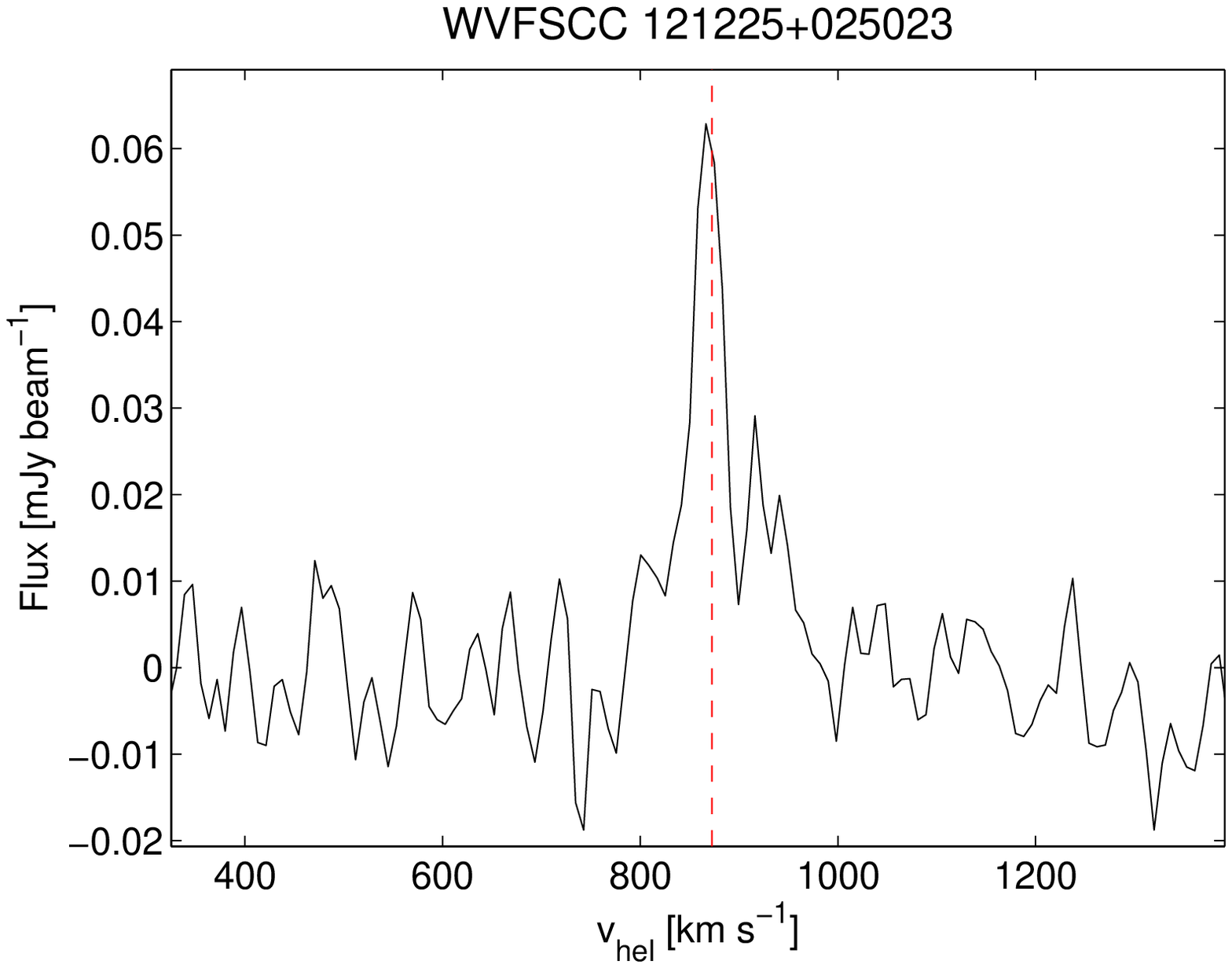}
  \includegraphics[width=0.45\textwidth]{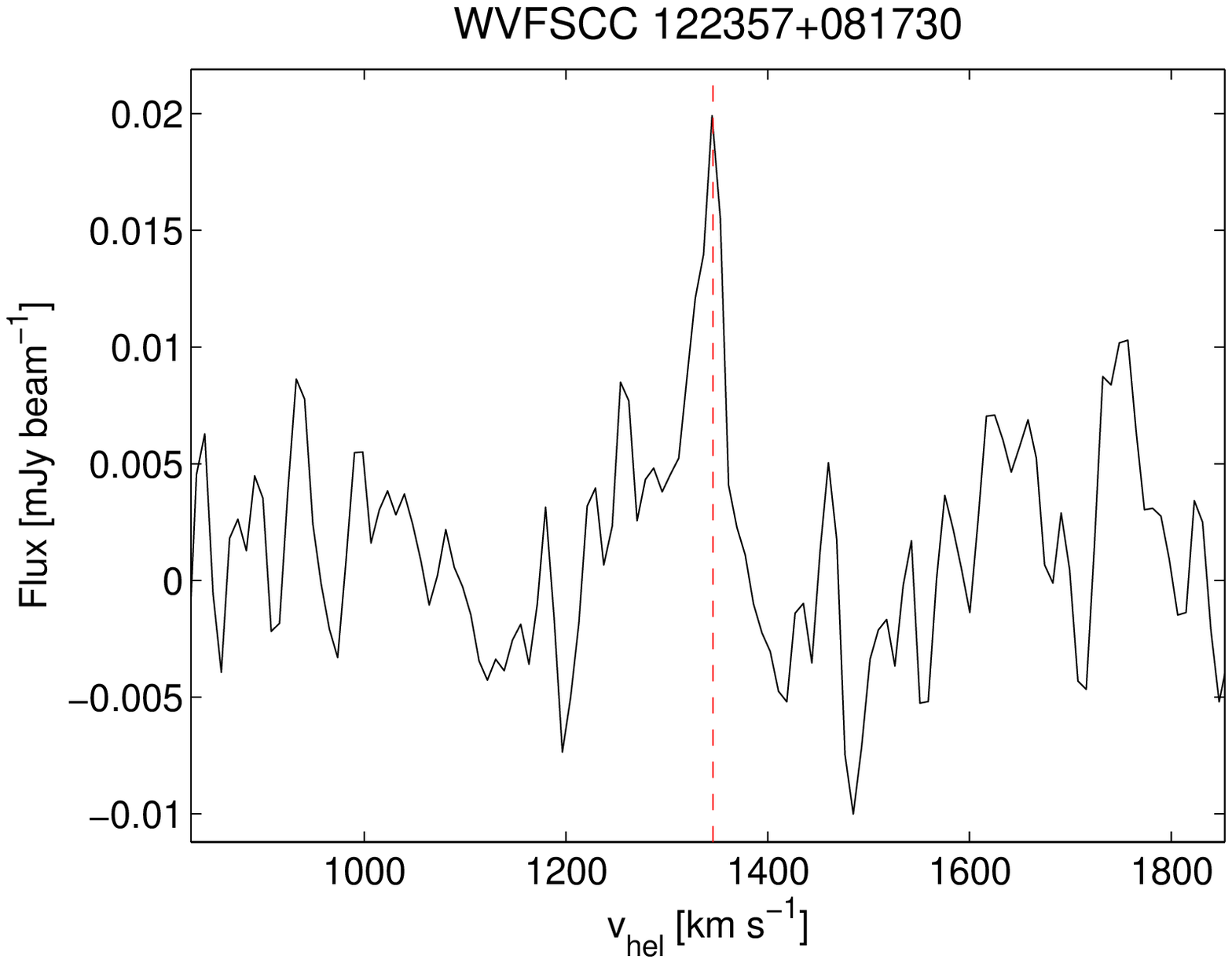}
 \end{center} 
  {\bf Figure ~\ref{new_HI}} continued.
  
\end{figure*}

\begin{figure*}[t]
\begin{center}

  \includegraphics[width=0.45\textwidth]{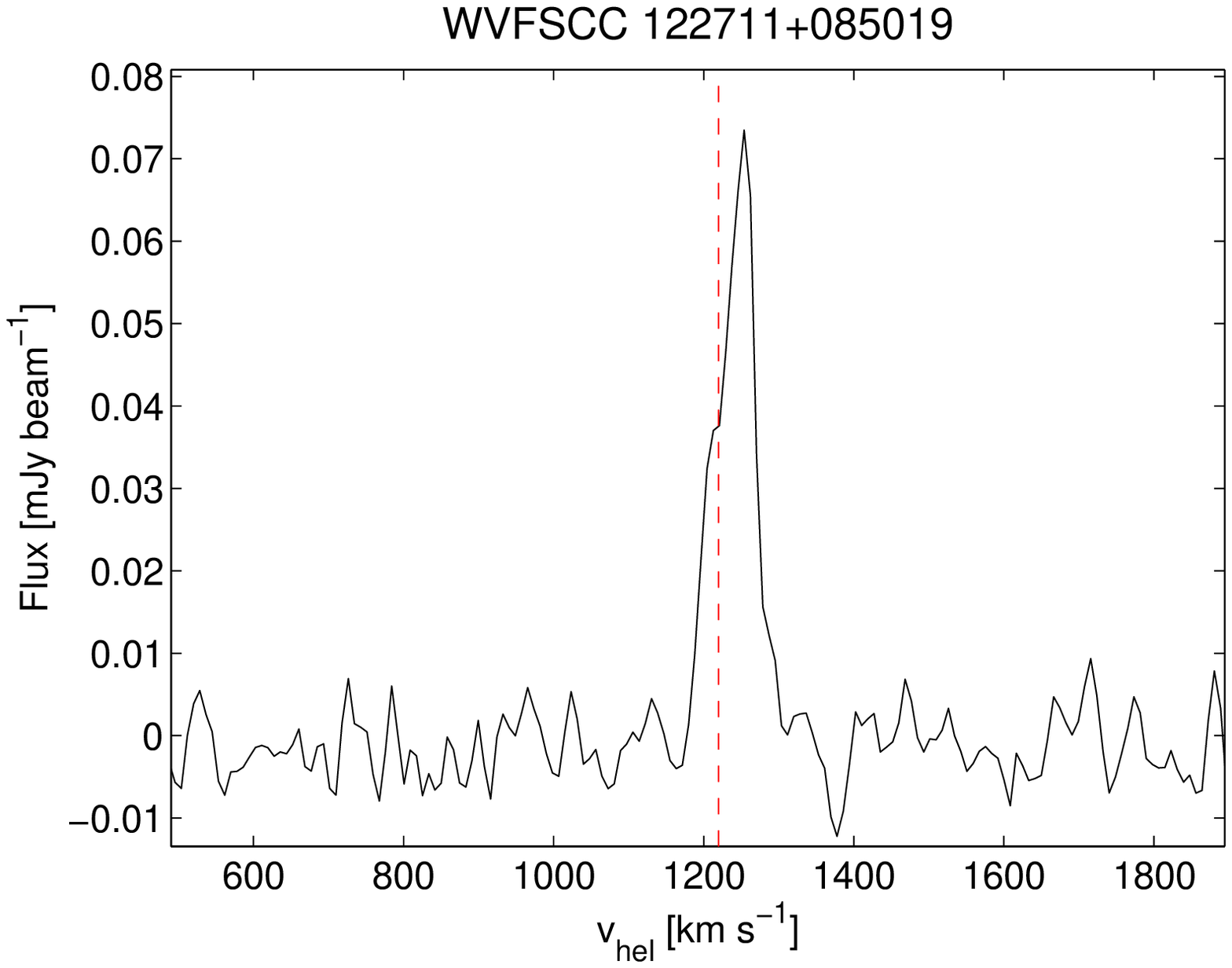}
   \includegraphics[width=0.45\textwidth]{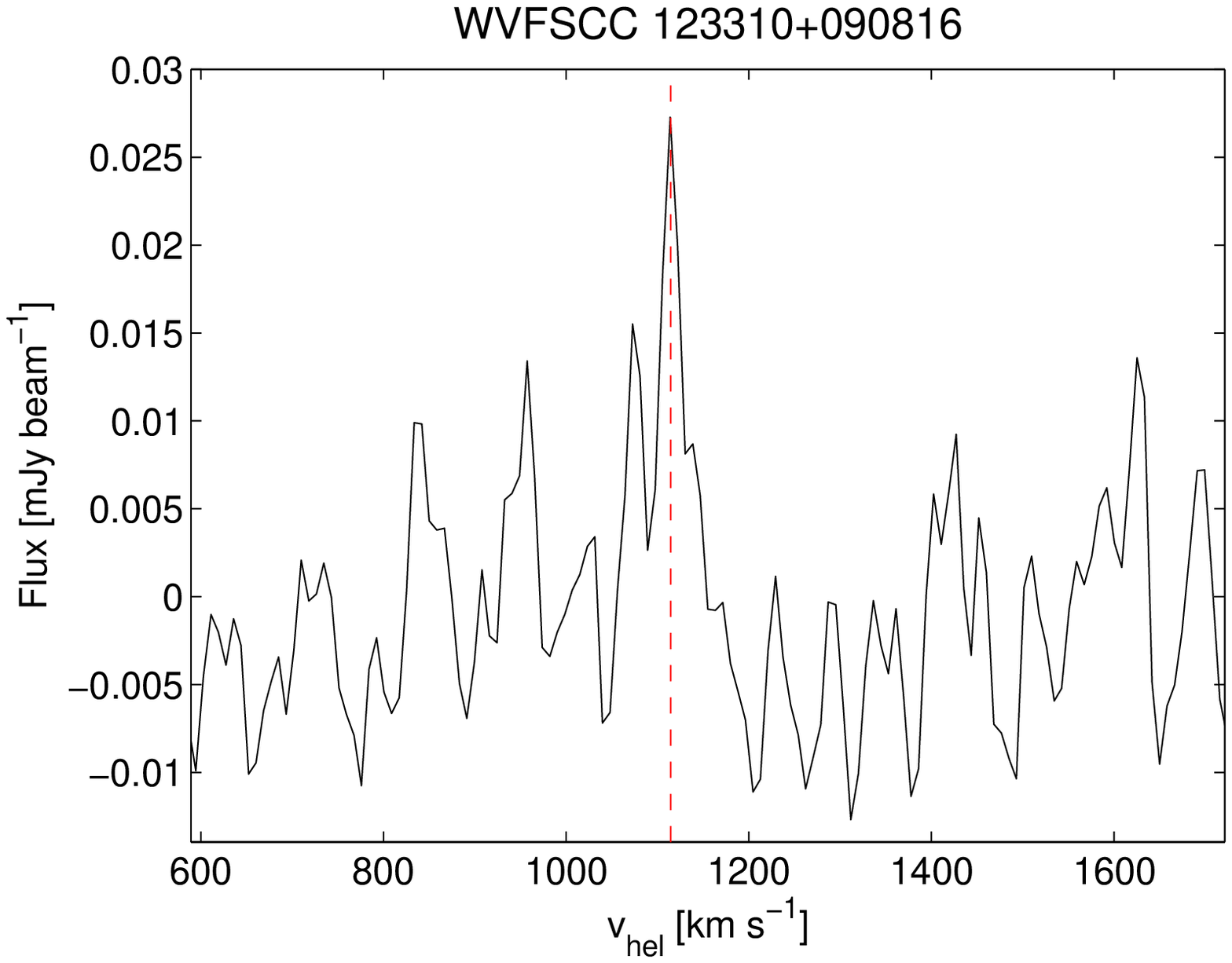}
  \includegraphics[width=0.45\textwidth]{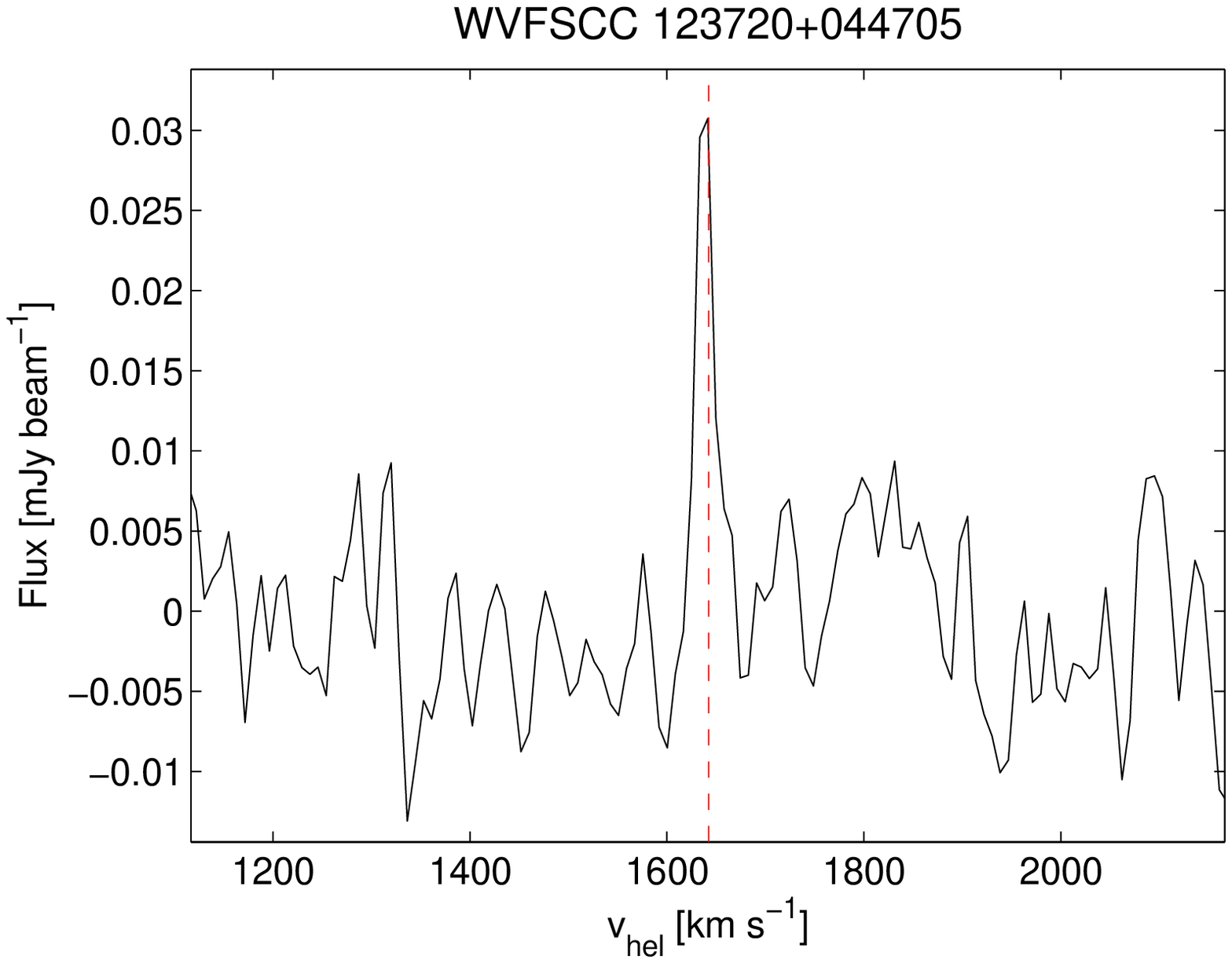}
  \includegraphics[width=0.45\textwidth]{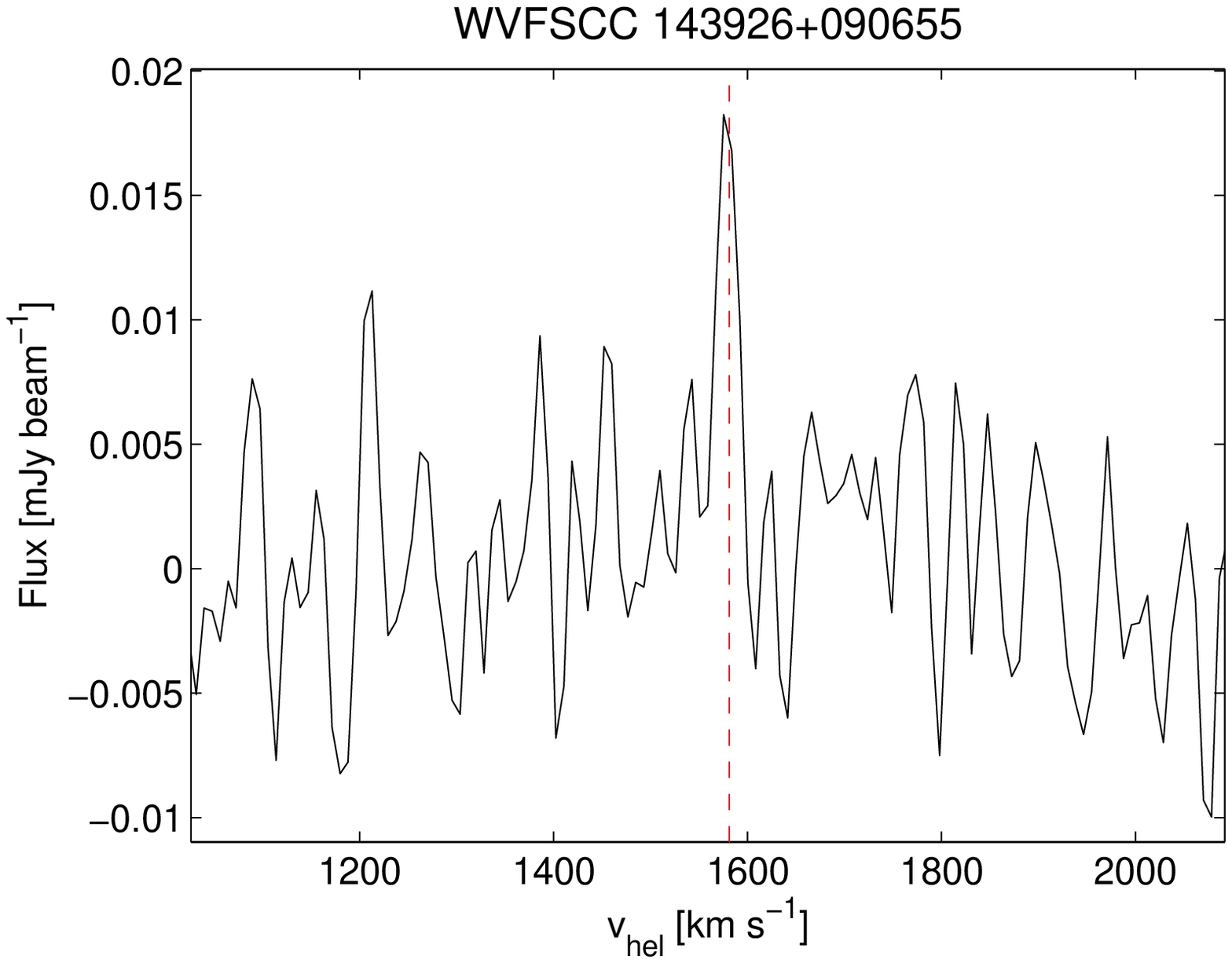}
  \includegraphics[width=0.45\textwidth]{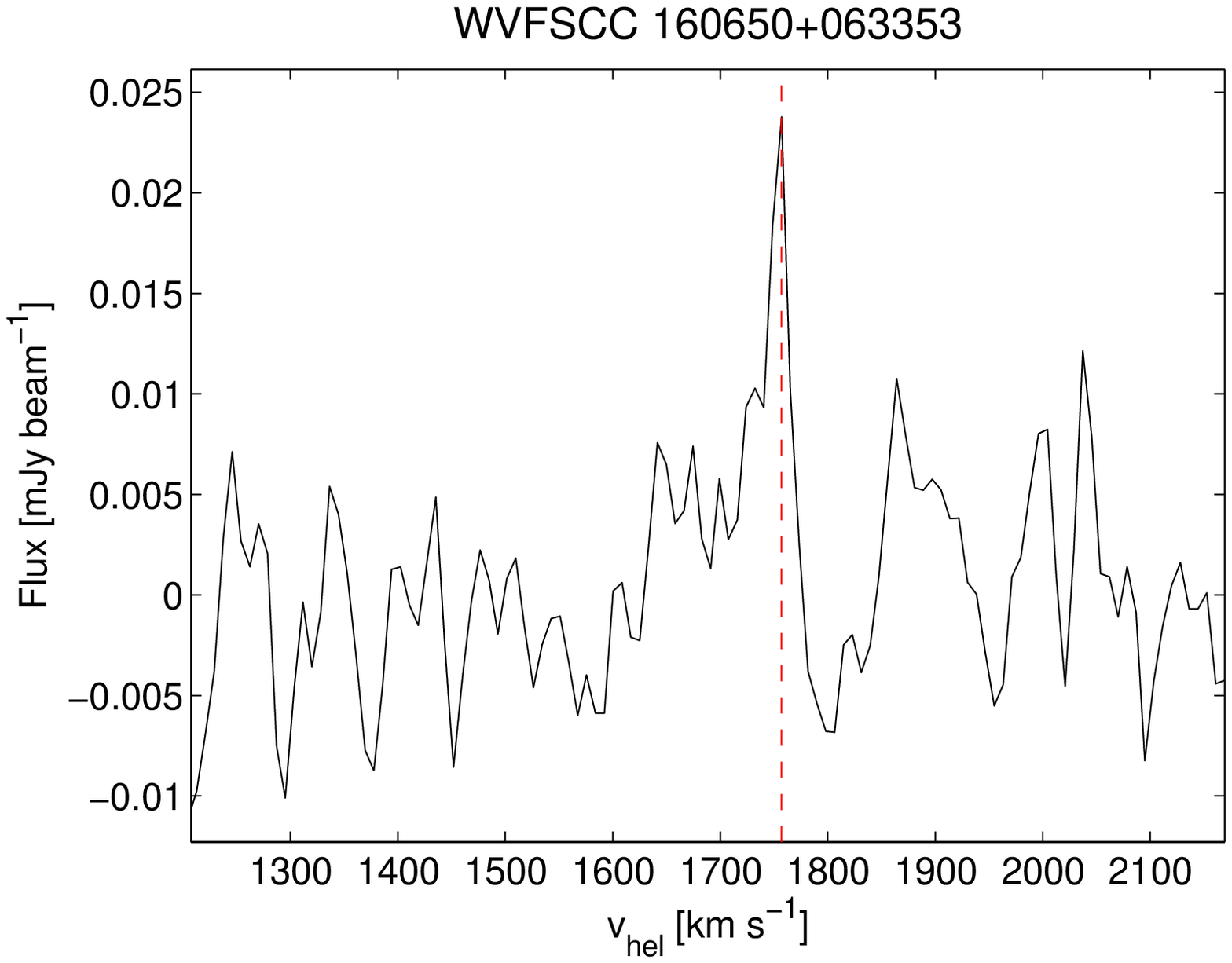}

   \end{center} 
  {\bf Figure ~\ref{new_HI}} continued.
  
\end{figure*}

\subsection{New {\HI} detections}
Both the blind search for neutral hydrogen and the search based on
identification of an optical counterpart have resulted in a number of
new {\HI} detections. Spectra for all these detections are shown in
Fig.~\ref{new_HI}; {\it WVFSCC~120929+080730} is the only one without
an optical counterpart. Typically all detections have a peak brightness
of between 25 and 50 mJy beam$^{-1}$, which is quite faint. Based on the
fluxes derived from WVFS, none of these sources could have been
detected by HIPASS, as the peak flux is below the clipping level that
has been utilised for the HIPASS source finding. The new {\HI} detections
do not share any particular attributes, and the spectra have a variety 
of shapes. The line-widths vary from a few tens of km
s$^{-1}$ to several hundreds of km s$^{-1}$ and both single peaks
as well as ``double-horned'' profiles are encountered.

\subsection{{\HI} without an optical counterpart}
The blind {\HI} search resulted in one source, for which no optical
counterpart is found. {\it WVFSCC~120929+080730} is just at the $8\sigma$ detection limit of
the blind {\HI} search. This feature has one strong peak, but has a
relatively narrow line-width of only 22 km s$^{-1}$ at FWHM as can be
seen from the spectrum in Fig.~\ref{new_HI}. The peak column density
of this feature is $N_{HI} \sim 1.5\times10^{19}$ km s$^{-1}$. No cataloged
optical counterpart is found within a radius of one degree, at the
relevant velocity. The nature of this detection is not clear, as the
DSS image of this position also shows no likely counterpart. Possibly
this is an example of an intergalactic gas filament that is a 
component of the Cosmic Web. Comparison with other data products will
have to reveal whether this detection can be confirmed in independent
data.

\subsection{Extended Neutral Hydrogen}
For each detection a moment map has been created by integrating the
cubes in the velocity domain over the line-width of the detection. All
integrated maps were visually inspected to search for extended emission
or companions. For several galaxies there is clear evidence of
extended {\HI} emission; we will discuss these cases in more detail
individually.\\

\begin{figure}[h]
\begin{center}
  \includegraphics[angle=270,width=0.5\textwidth]{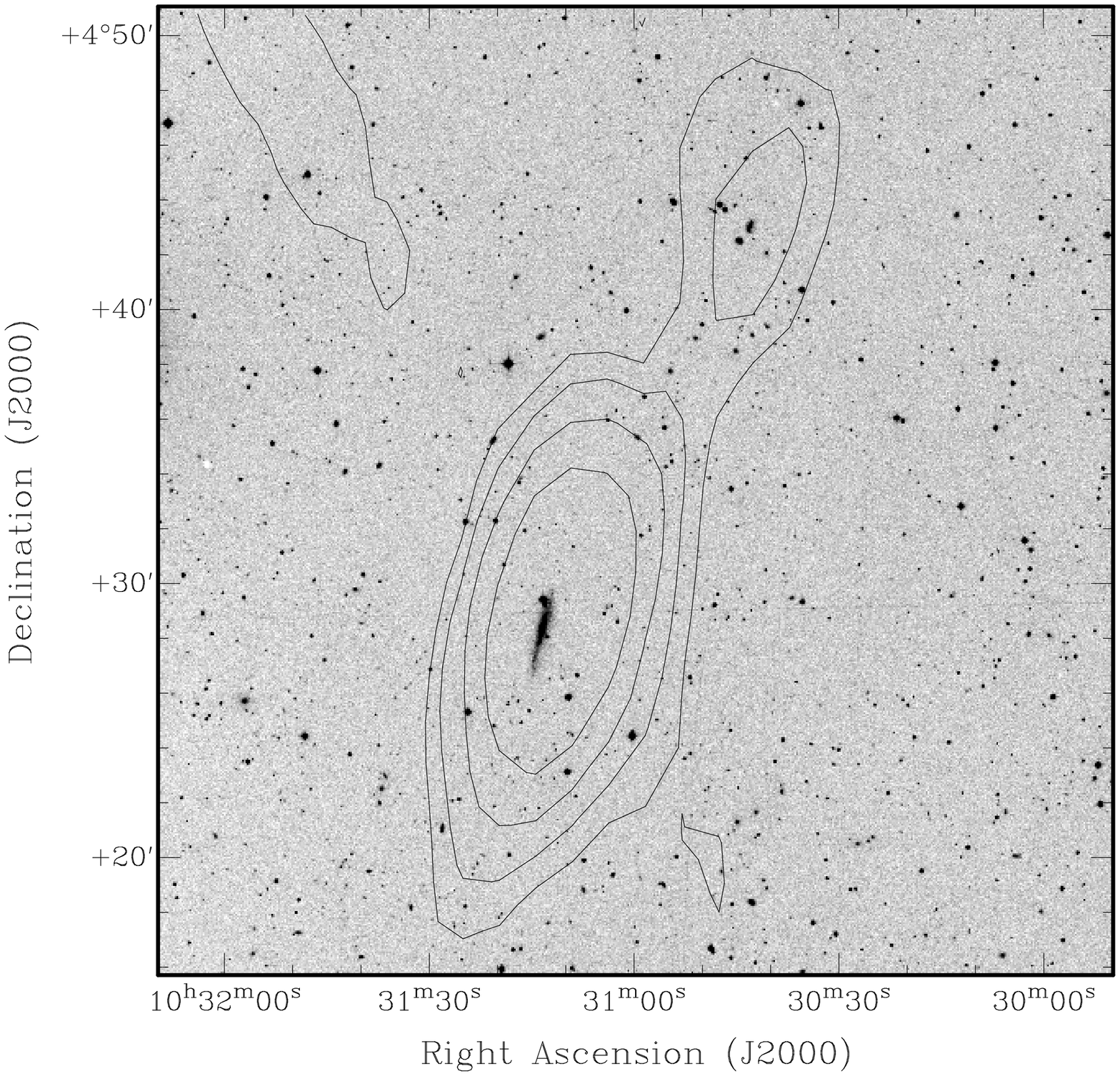}
\end{center}
  
\caption{A second generation red band DSS image, overlaid with {\HI}
  contours of WVFSCC~103109+042831. Contour levels are drawn at
  1, 2, 4 and 8 Jy~km~s$^{-1}$. The main galaxy is
  UGC~5708, the small companion in the north west is most likely
  CGCG~037-059}
  \label{fil_1}
\end{figure}

{\bf WVFSCC~103109+042831}: The main galaxy UGC~5708 has a clear
extension toward a nearby companion to the north west. The companion
is separated from UGC 5708 by approximately 20 arcmin, which
corresponds to 100 kpc. Both the radial velocity and the line width of
the companion are very similar to that of UGC~5708 as can be seen from
the spectra in Fig.~\ref{fil_1}. The peak column density of the
companion is $N_{HI} \sim 3\cdot 10^{19}$ cm$^{-2}$. The contours are
overlaid on a second generation DSS image. An optical counterpart can
be identified exactly at the peak of the contours. This optical
counterpart is probably CGCG~037-059 which is classified as a galaxy
triplet. Only one radial velocity, of 11944
km s$^{-1}$ is known for this triplet . If CGCG~037-059 is indeed a galaxy triplet that is
gravitationally bound, then a relation with the {\HI} gas is not likely
because of the very large deviation in radial velocity and the
similarity in spatial position is purely coincidental. The other
possibility is that the classification of CGCG~037-059 is incorrect
and that the {\HI} gas is associated with an object with an
unknown redshift.\\

\begin{figure}[h]
\begin{center}
  \includegraphics[angle=270,width=0.5\textwidth]{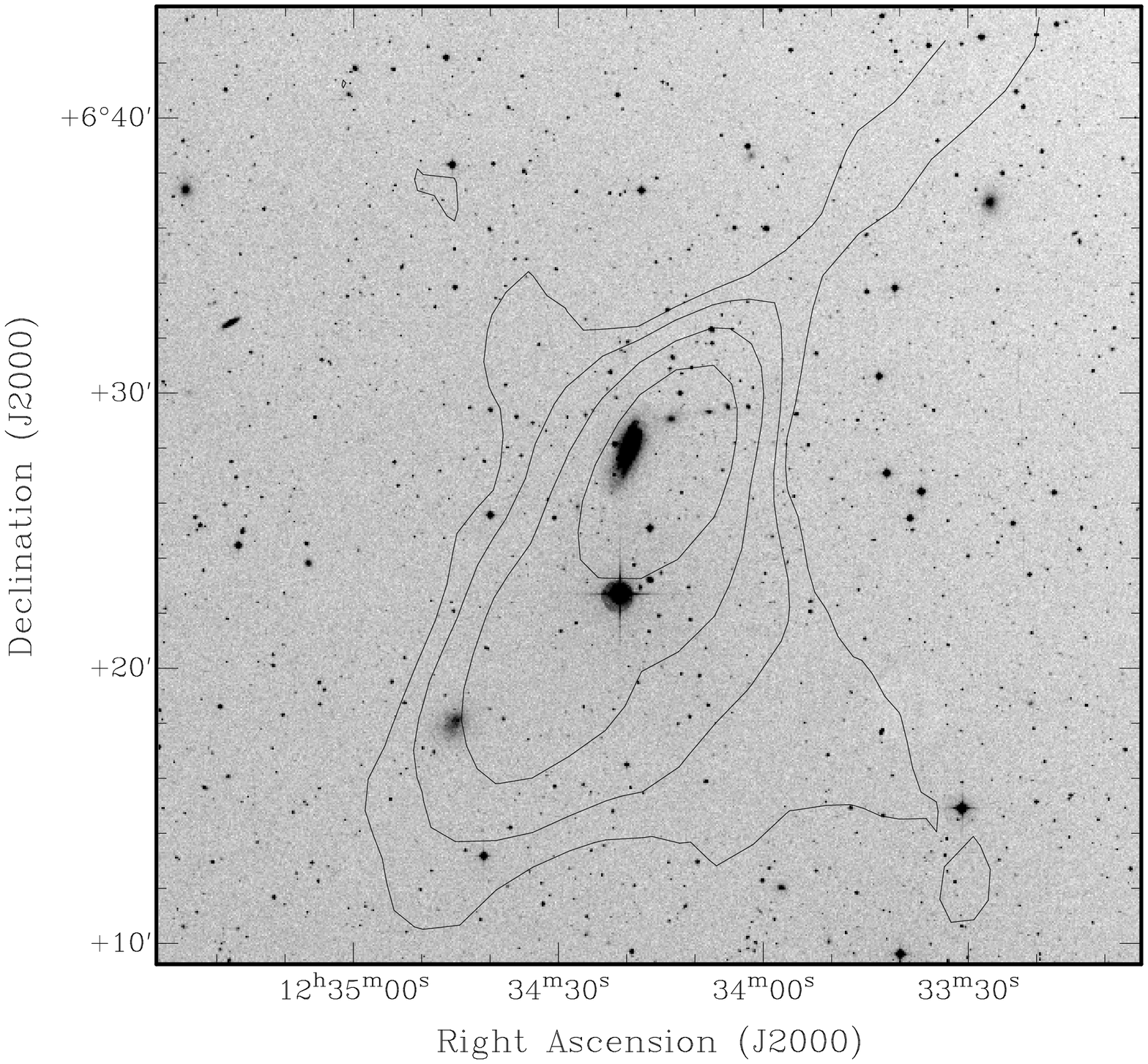}
\end{center}
  \caption{{\HI} contours of WVFSCC~123428+062753 at 1, 2, 4
    and 8 Jy~km~s$^{-1}$ are overlaid on a second generation DSS
    images. The main galaxy is NGC~4532, while the diffuse galaxy in the
    South-East that is attached is Holmberg~VII.}
  \label{fil_2}
\end{figure}

{\bf WVFSCC~123428+062753}: The main galaxy is NGC 4532, that has an
extended {\HI} companion towards the south east as can be seen in
Fig.~\ref{fil_2}. Although the gas looks very extended, it belongs to
the irregular galaxy Holmberg~VII, at a radial velocity of 2039 km
s$^{-1}$. Due to the relatively large synthesised beam it has not been
identified as an individual object, but this is not the first
detection of neutral hydrogen in this galaxy.\\

\begin{figure}[h]
\begin{center}
  \includegraphics[angle=270,width=0.5\textwidth]{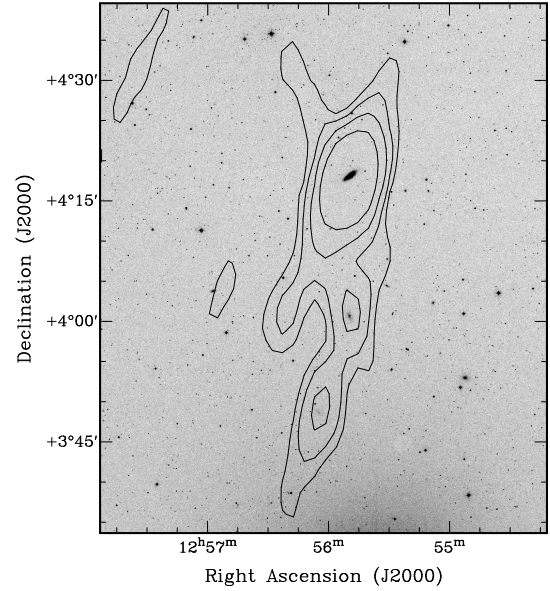}
\end{center}
  
\caption{{\HI} contours of WVFSCC~125550+041348 at levels of 1, 2, 4 and
  8 Jy~km~s$^{-1}$ overlaid on a red band, second generation DSS
  images. The main galaxy NGC~4808 has a long tail of {\HI} towards the
  south, connecting two other galaxies. South-East of the main galaxy
  is a very faint signal possibly associated with another galaxy.}
  \label{fil_3}
\end{figure}

{\bf WVFSCC~125550+041348}: This is a very extended structure with at
least 4 concentraions as can be seen in Fig.~\ref{fil_3}. The main
galaxy is NGC 4808, which has a long tail of {\HI} gas towards the
south. The apparent extension to the south-east is most likely caused
by a residual side-lobe. When looking at the
optical DSS image with the {\HI} contours overlaid, an optical
counterpart can be identified for all {\HI} peaks. The first optical
galaxy south-east of NGC 4808 is CGCG 043-077 and has only one {\HI}
contour. The peak column density of this detection is $N_{HI} \sim
1.4\cdot 10^{19}$ cm$^{-2}$, this is the first {\HI} detection of this
galaxy. The two galaxies to the south are UGC~8053 and
UGC~8055, both with a peak column density of $N_{HI} \sim 4\cdot
10^{19}$ cm$^{-2}$, which have been previously detected in {\HI}.

\section{Discussion and Conclusion}
The WSRT has been used in a very novel observing mode to simulate a
filled aperture in projection of $300 \times 25$ meters by observing
at very extreme hour angles. Because of the very short observing times
per pointing it is a technical challenge to observe and reduce this
data, while still achieving useful results. In total 22,000 pointings
have been observed that cover a total area of $\sim 1500$ square
degrees. Each pointing has been observed two times for a period of one
minute. Normally an integration time of one minute with an
interferometer is not sufficient to fill the {\it u,v} plane, however
as there are essentially no gaps between the individual antennas in
projection, and the two scans have a complimentary orientation, a well
defined synthesized beam could be formed. The observing method is limited to a
narrow range in declination, but has been very successful.\\

In the reduced data we reach a flux sensitivity of $\sim
6$~mJy~beam$^{-1}$ over 16 km~s$^{-1}$. The synthesised beam has an
average size of $395 \times 245$ arcsec, which results in a
brightness sensitivity of $N_{HI} \sim 10^{18}$ cm$^{-2}$. Such a
brightness sensitivity can typically only be achieved with
single-dish observations. Because the WSRT is an interferometer, the
calibration and stability of the bandpass is significantly superior to
that of a single dish.

The drawback of interferometric observations is that a {\it cleaning}
step has to be applied to correct for the synthesised beam shape,
especially for bright objects. Although the synthesised beam is well
defined, the side-lobes in our observations are very strong, making
the {\it cleaning} step a critical one. Because of the large size of
the survey, only one pass of {\it cleaning} deconvolution could be
applied. Improved deconvolution results would require detailed masking
of real emission features during component identification. Because of
the relatively high side-lobe level, any automatic masking procedure
is unlikely to be reliable. Each emission peak would have to be
inspected visually, to be able to distinguish a real
emission feature from a side-lobe artifact. Because of the size of the
survey this approach was not considered practical. As a result of
these limitations, there are low level artifacts in the data,
although the flux sensitivity in the reduced data is very good.\\

An extra complication in processing the data is that the north-south
synthesized beamsize increases towards low Declinations since the WSRT
is an east-west array. Furthermore, the natural image plane projection
of an east-west array is the North Celestial Pole ({\it NCP})
projection \citep{1971PhDT.......153B} which becomes undefined at zero
degrees. As a result, only positive declinations could be analysed in
our mosaiced images. Although this did not dramatically affect our
results, it is still a major point of concern.

We found another serious complication in using an {\it NCP} projection
in a wide field survey close to a declination of zero degrees. This
complication is not caused by a shortcoming in the projection, but
more likely a shortcoming in the imaging software that has been
used. When gridding the observed data or {\it u,v} coordinates to the
projected plane, the observed flux is not being conserved due to an
incorrect wighting when combining pointings in a mosaic. At positive
declinations,the imaged flux of data below the declination of the
reference pixel is diluted while for data above the reference pixel
the imaged flux is enhanced compared to the observed flux. This effect
is very apparent when using a {\it NCP} projection but probably also
happens when using other projections. When observing small fields, or
fields between $\sim 20$ and $\sim 70$ degrees in declination this
effect is negligible, however in our case a significant correction to
the derived fluxes had to be applied.

In general, other image plane projections are required at Declinations
near zero degrees to enable both positive and negative Declinations to
be imaged simultaneously. However several major interferometers
are East-West arrays, including the WSRT and the Australia Telescope
Compact Array (ATCA). All these telescopes are being upgraded, partly
to serve as a pathfinders for the SKA (Square Kilometre Array). A
general aim of future telescopes, especially the ones that use a FPA
(Focal Plane Array) is to conduct large surveys of the entire
sky. Ideally these surveys will have significant overlap with deep
optical surveys, however several major optical surveys are
concentrated at Declinations near zero degrees.\\

Two search methods have been applied to the reduced WVFS data, both
using the {\it Duchamp} source finding algorithm. The first method is
a blind search at a peak brightness limit of $8\sigma$. The second
method uses an initial peak brightness limit of $5\sigma$, but has the
additional requirement that all detected features need to have an
optical counterpart. In the blind search 138 objects have been
detected, while the second search resulted in 198 {\HI} counterparts to
cataloged optical galaxies. Of all the detections 16 are new {\HI}
detections and only 1 detection does not have an optical
counterpart. On average, the interferometric total fluxes of detections are
$\sim 10$\% lower than the catalogued fluxes in the HIPASS archive.

There are many features in the cube with a peak of 3 or 4$\sigma$ and
an integrated flux that probably exceeds 8$\sigma$. It is very likely
that many of these features are real, however they cannot be
identified reliably by automated source finders. In a subsequent
paper, the WVFS cross-correlation data will be compared with the WVFS
total-power data and the re-reduced HIPASS data as described in
\cite{2010arXiv1012.3236P} and \cite{2010A&A...HIPASS}. Both surveys
contain several new {\HI} detections and tentative filaments. As this
is a limited number of sources, we can do a targeted search in the
WVFS cross-correlation data. Although the brightness sensitivity of
each of the three surveys is almost an order of magnitude different,
comparison of the data will be useful to distinguish extended and
diffuse emission from denser {\HI} clumps.

\begin{acknowledgements}
 The Westerbork Synthesis Radio Telescope is oper- 
ated by the ASTRON (Netherlands Foundation for Research in Astronomy) with 
support from the Netherlands Foundation for ScientiÞc Research (NWO)

\end{acknowledgements}

\bibliographystyle{15719_aa}
\bibliography{names,thesisbibliography}

\newpage

\onecolumn
\begin{appendix}
\section{Physical properties of {\HI} detections in the Westerbork Virgo Filament Survey}


\begin{landscape}
\begin{center}
\begin{longtable}{lllrrccccccc}

\caption{Physical properties of {\HI} detections in the Westerbork Virgo Filament Survey.\\
{\it (a,)}: Units of  $V_{Hel}$ are given in [km s$^{-1}$].\\
{\it (b)}: Units of  $W_{50}$ are given in [km s$^{-1}$].\\
{\it (c)}: Units of  $W_{20}$ are given in [km s$^{-1}$].\\
{\it (d)}: Line strength $(S_l)$ in units of [Jy km s$^{-1}$].\\
{\it (e)}: Integrated flux $(S_i)$ in units of [Jy km s$^{-1}$].\\
{\it (f)}: Error in $(S_i)$ in units of [Jy km s$^{-1}$]. \\ 
{\it (g)}:  Detection method, blind $(B)$ or with optical identification $(O)$.\\
} 
\label{sources}\\


\hline
\hline
 \small Nr.     &    \small   Name           &   \small Optical ID.              &    \small   RA [hh:mm:ss] &  \small Dec [dd:mm:ss] &  \small $V_{Hel}^a$  &  \small $W_{50}^b$ &  \small $W_{20}^c$  &  \small $S_l^d$  &  \small $S_i^e$ &  \small $\sigma_S^f$ & \small Detection$^g$ \\
\hline   
\endfirsthead

\hline
\hline
 \small  Nr.     &    \small   Name           &   \small Optical ID.              &      \small RA [hh:mm:ss] &  \small Dec [dd:mm:ss] &  \small $V_{Hel}$ &  \small $W_{50}$ &  \small $W_{20}$ &  \small $S_l$  &  \small $S_i$ &  \small $\sigma_S$  & \small  Detection\\
\hline
\endhead

\hline
  \multicolumn{10}{c}{{Continued on Next Page\ldots}} \\
\endfoot

\hline \hline
\endlastfoot


  \small    1    &  \small       WVFSCC 083122+065754    &  \small    CGCG 032-035                      &  \small     08:31:22.24    &  \small    06:57:54.32    &  \small     1840      &  \small       125       &  \small       207       &  \small     2.9        &  \small       4.9       &  \small     0.2  & \small O, B  \\ 
  \small    2    &  \small       WVFSCC 083628+051108    &  \small    HIPASS J0836+05                   &  \small     08:36:28.06    &  \small    05:11:08.70    &  \small     1857      &  \small        35       &  \small       100       &  \small     1.6        &  \small       2.0       &  \small     0.2  & \small O \\ 
  \small    3    &  \small       WVFSCC 084014+053812    &  \small    UGC 04524                         &  \small     08:40:14.27    &  \small    05:38:12.27    &  \small     1985      &  \small        70       &  \small       138       &  \small     2.6        &  \small       3.4       &  \small     0.2   & \small O, B\\ 
  \small    4    &  \small       WVFSCC 084135+045857    &  \small    NGC 2644                          &  \small     08:41:35.12    &  \small    04:58:57.69    &  \small     1932      &  \small       106       &  \small       125       &  \small     3.0        &  \small       3.7       &  \small     0.2   & \small O, B\\ 
  \small    5    &  \small       WVFSCC 090629+061848    &  \small    UGC 04781                         &  \small     09:06:29.08    &  \small    06:18:48.54    &  \small     1451      &  \small       181       &  \small       202       &  \small     8.5        &  \small      13.5        &  \small    0.2   & \small O, B\\
\end{longtable}
\end{center}

\end{landscape}


\section{Spectra of all {\HI} detections in the WVFS cross-correlation data}

\begin{figure*}[!h]
  \begin{center}
 
\includegraphics[width=0.32\textwidth]{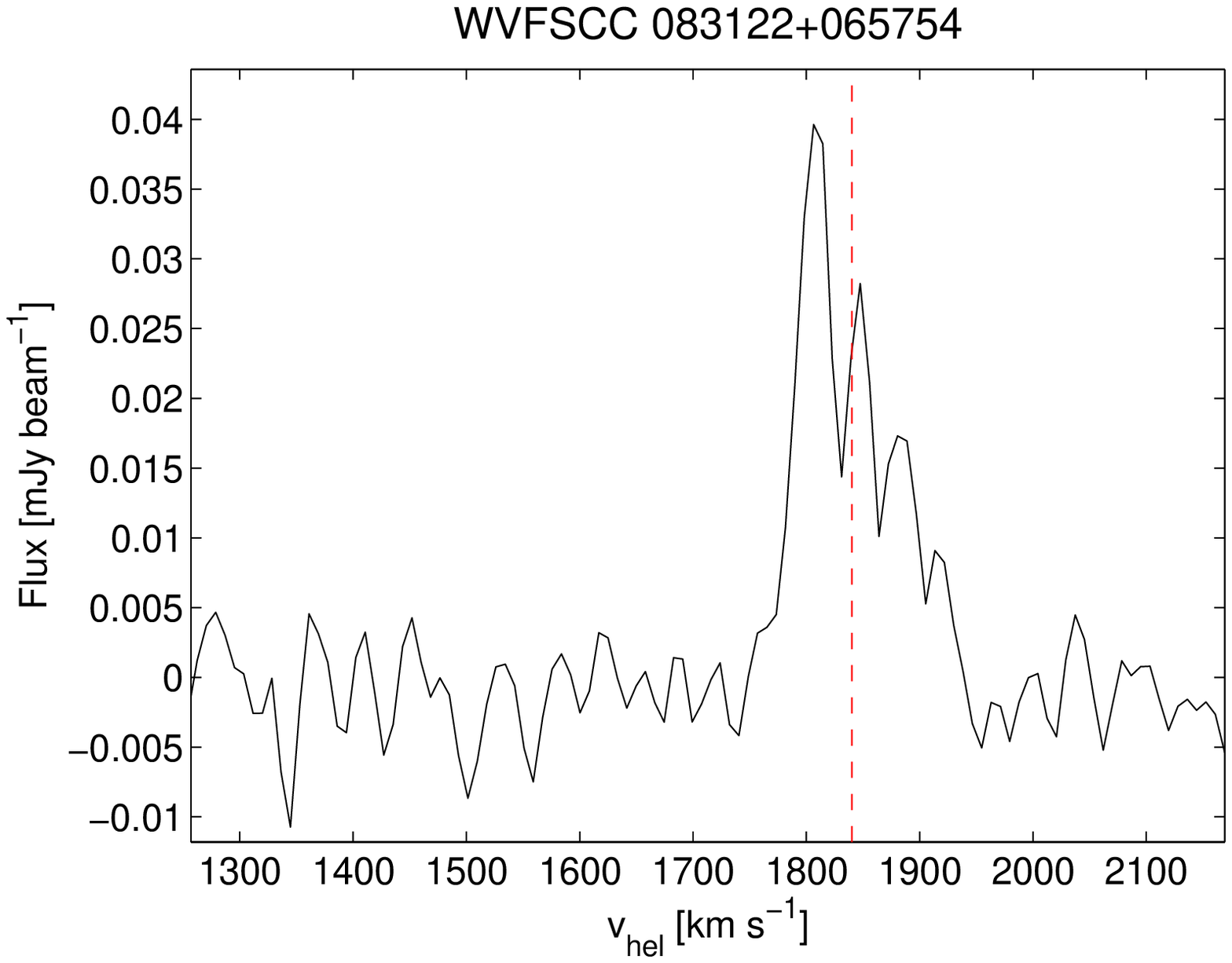}
\includegraphics[width=0.32\textwidth]{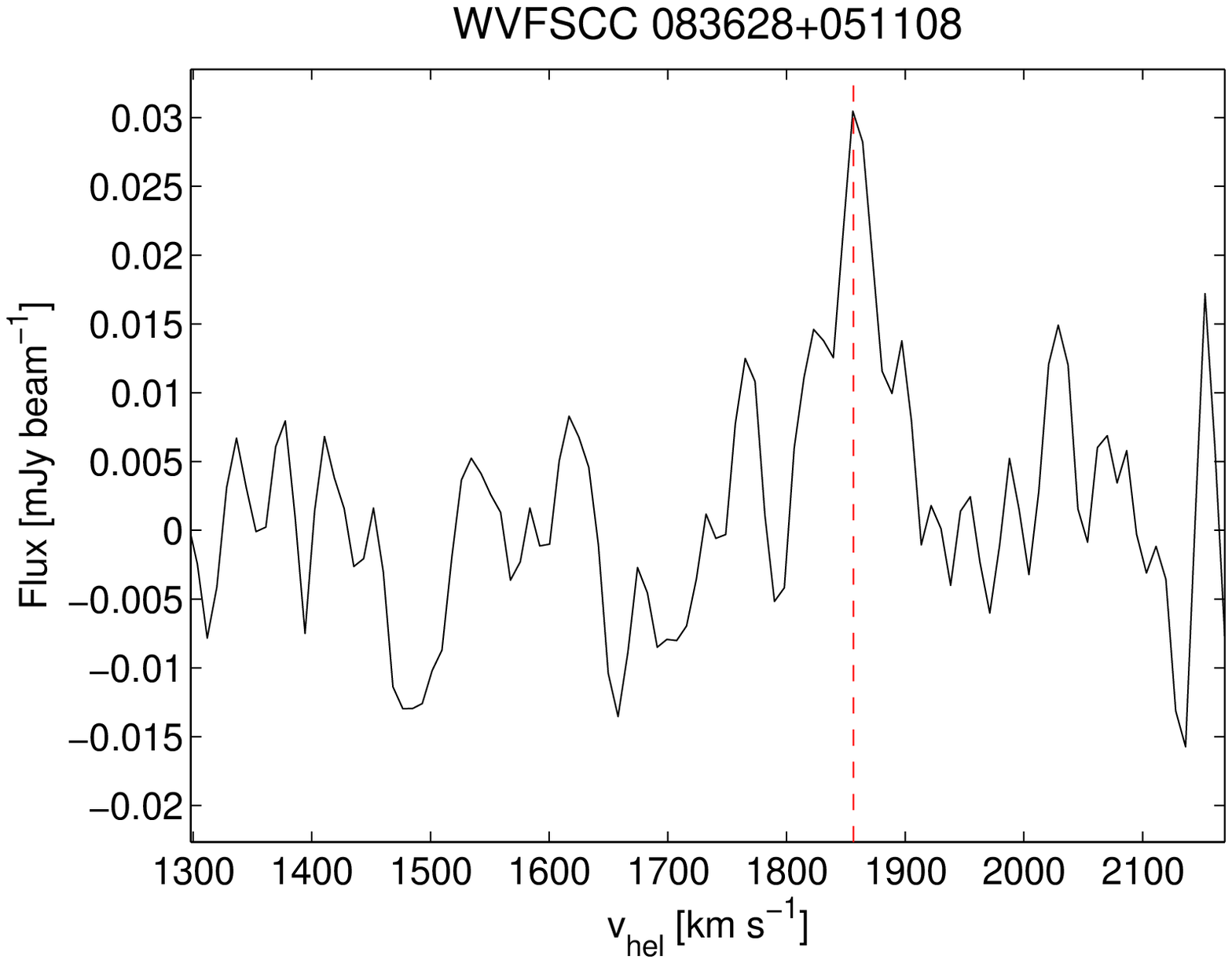}
\includegraphics[width=0.32\textwidth]{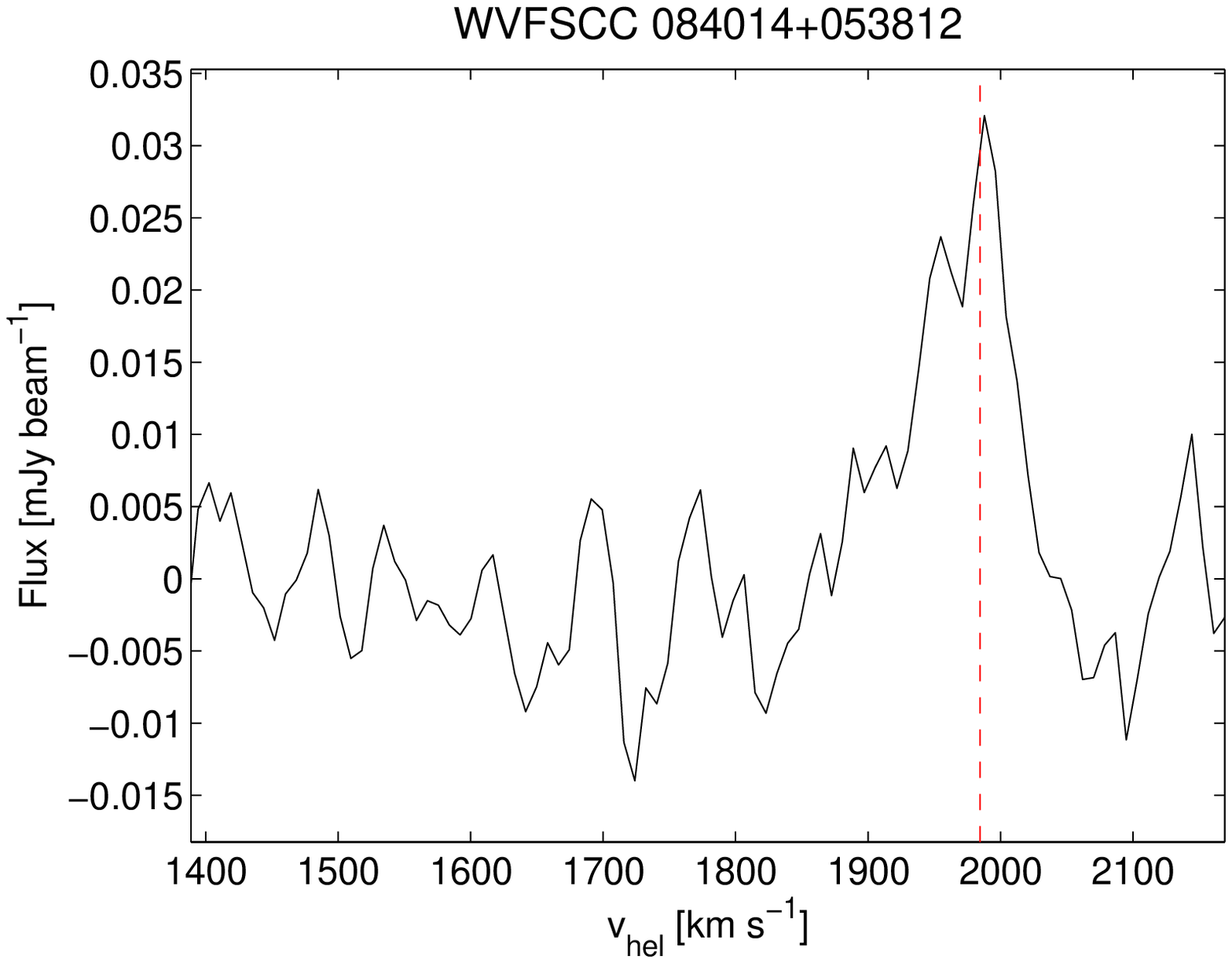}
\includegraphics[width=0.32\textwidth]{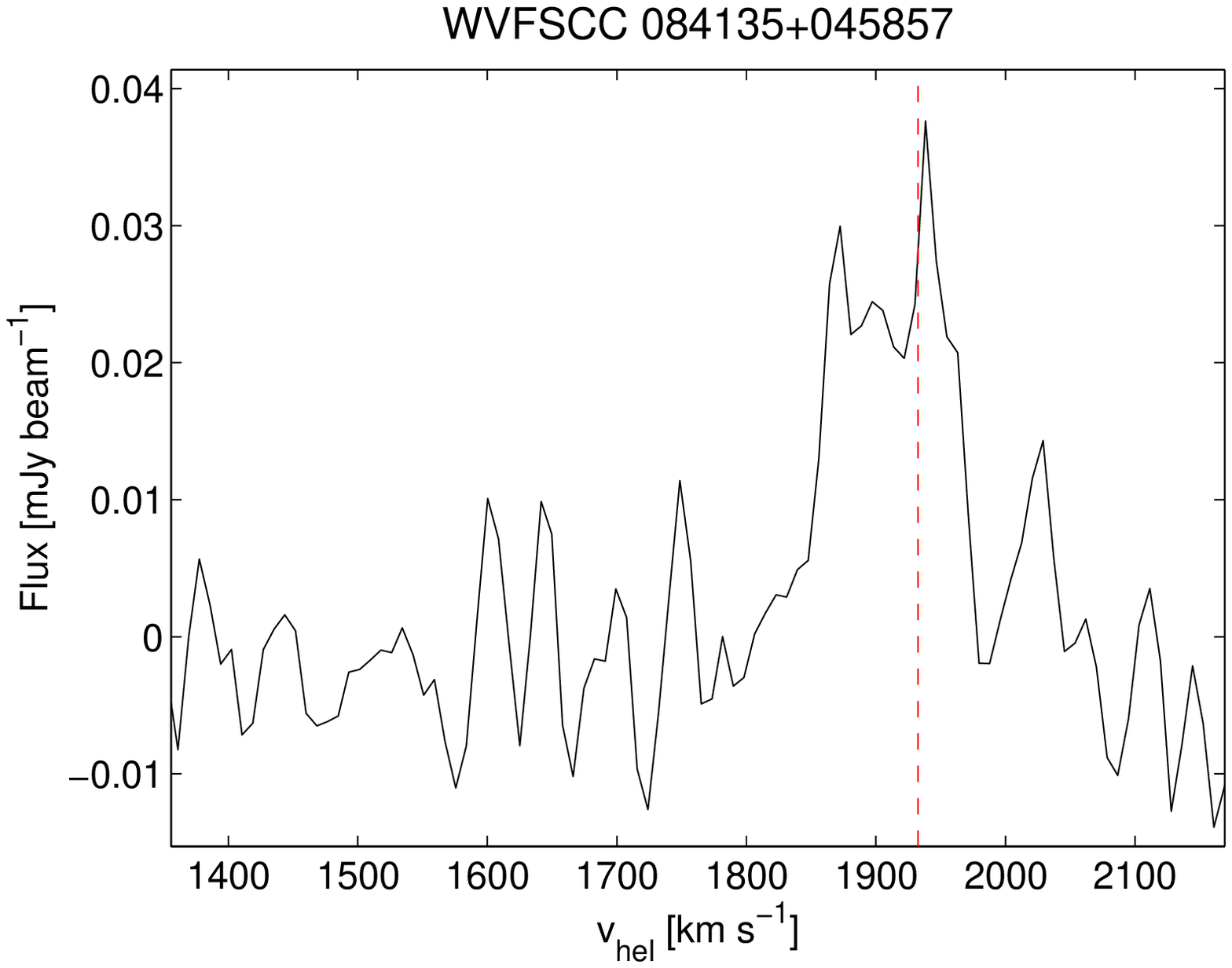}
\includegraphics[width=0.32\textwidth]{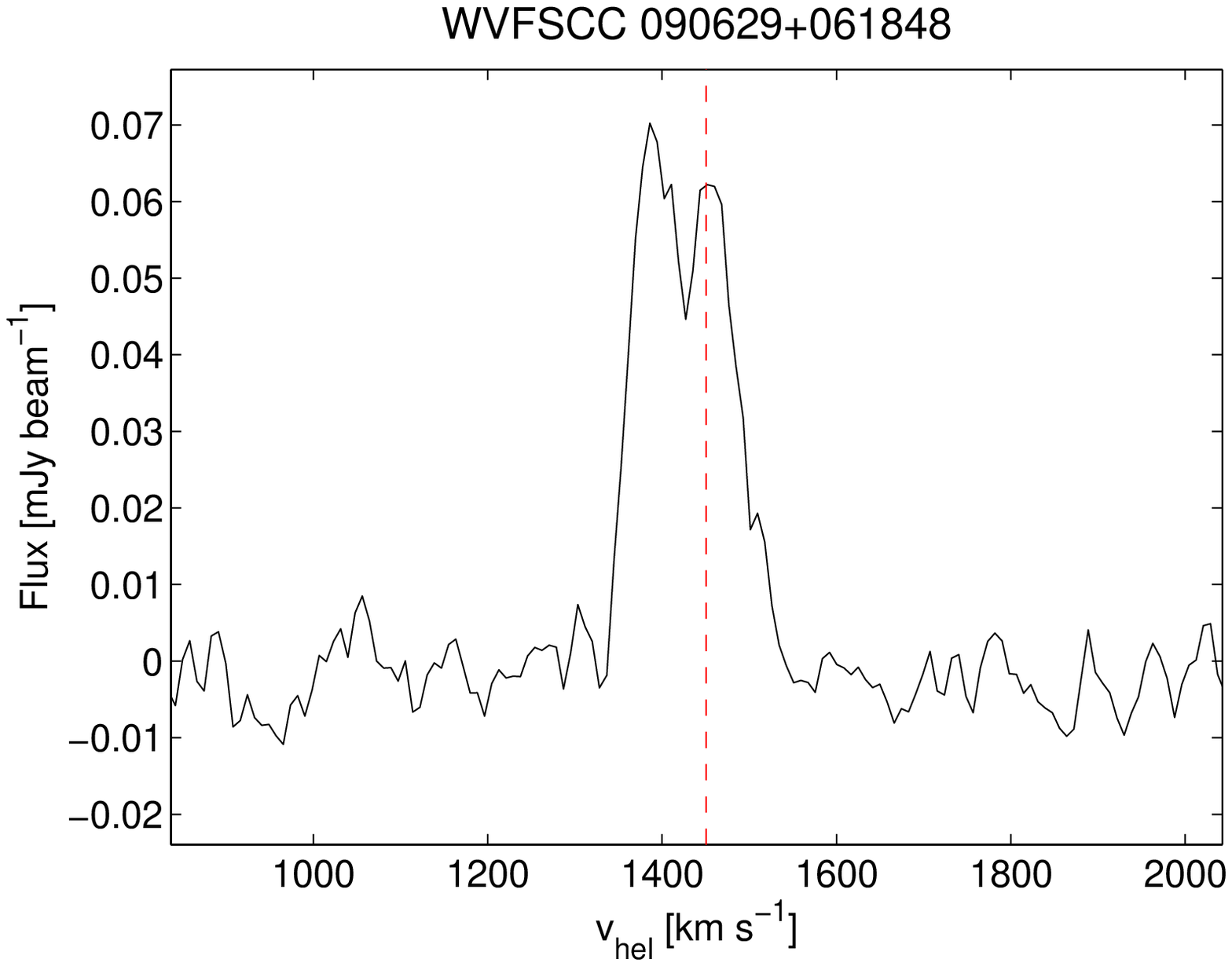}
\includegraphics[width=0.32\textwidth]{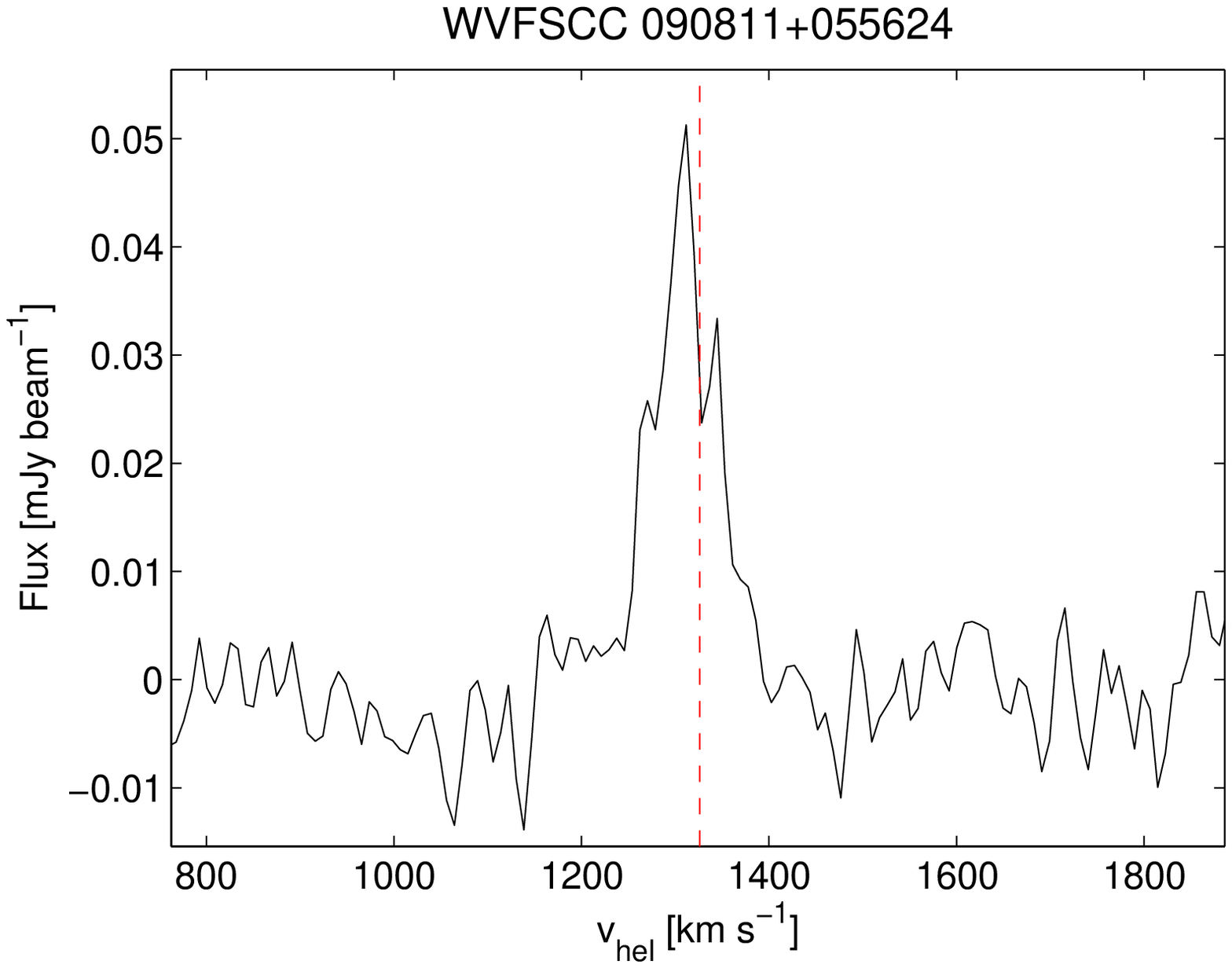}
\includegraphics[width=0.32\textwidth]{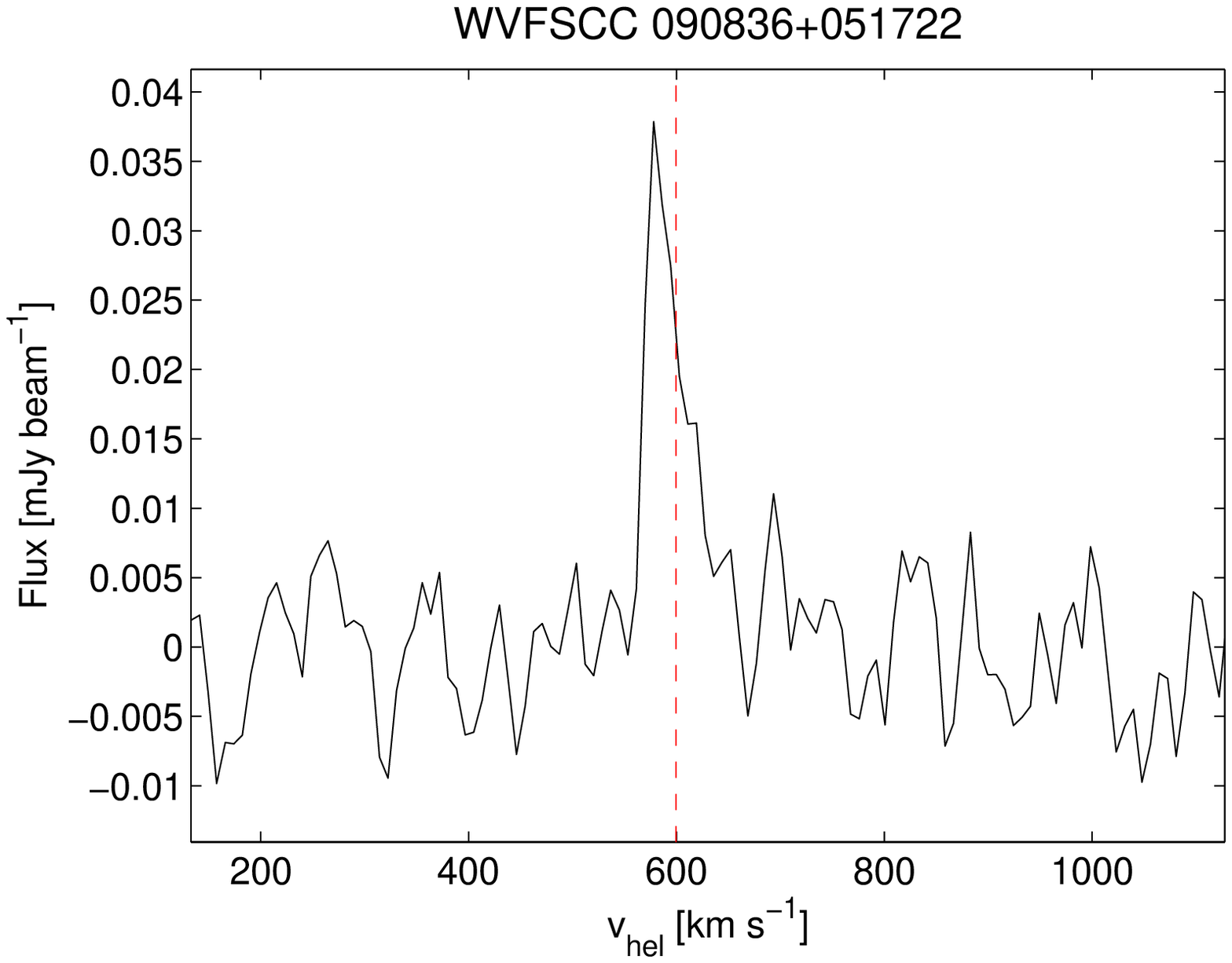}
\includegraphics[width=0.32\textwidth]{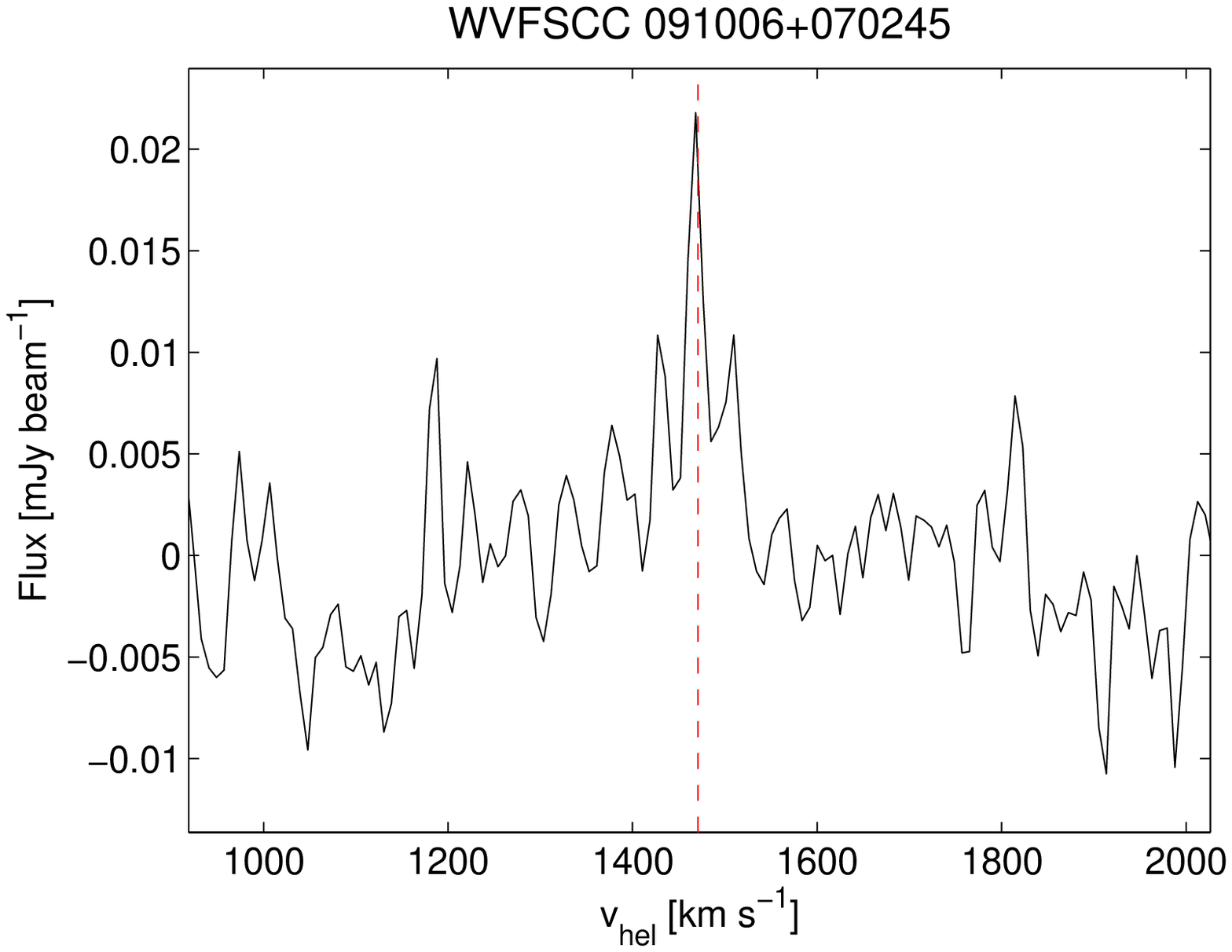}
\includegraphics[width=0.32\textwidth]{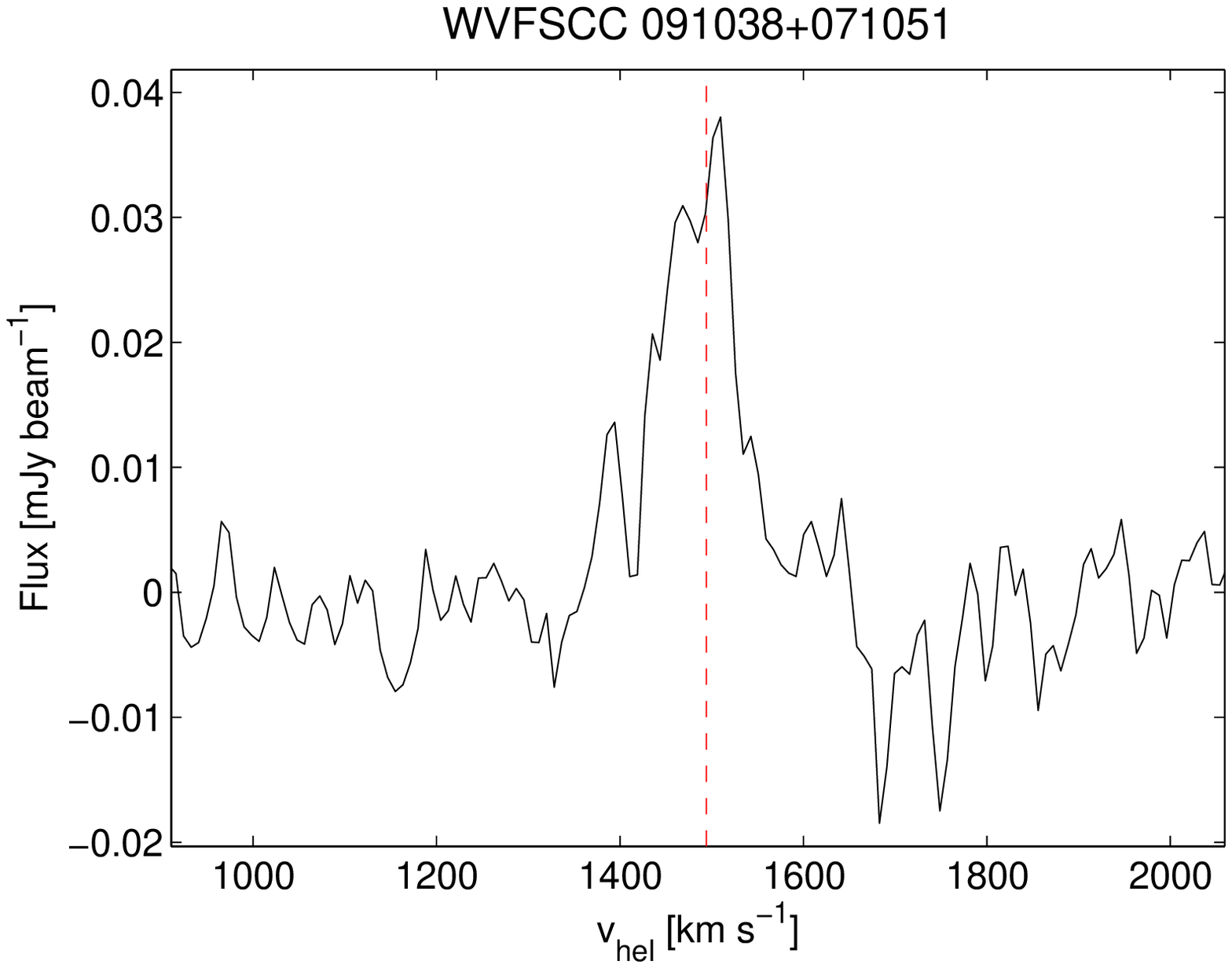}
\includegraphics[width=0.32\textwidth]{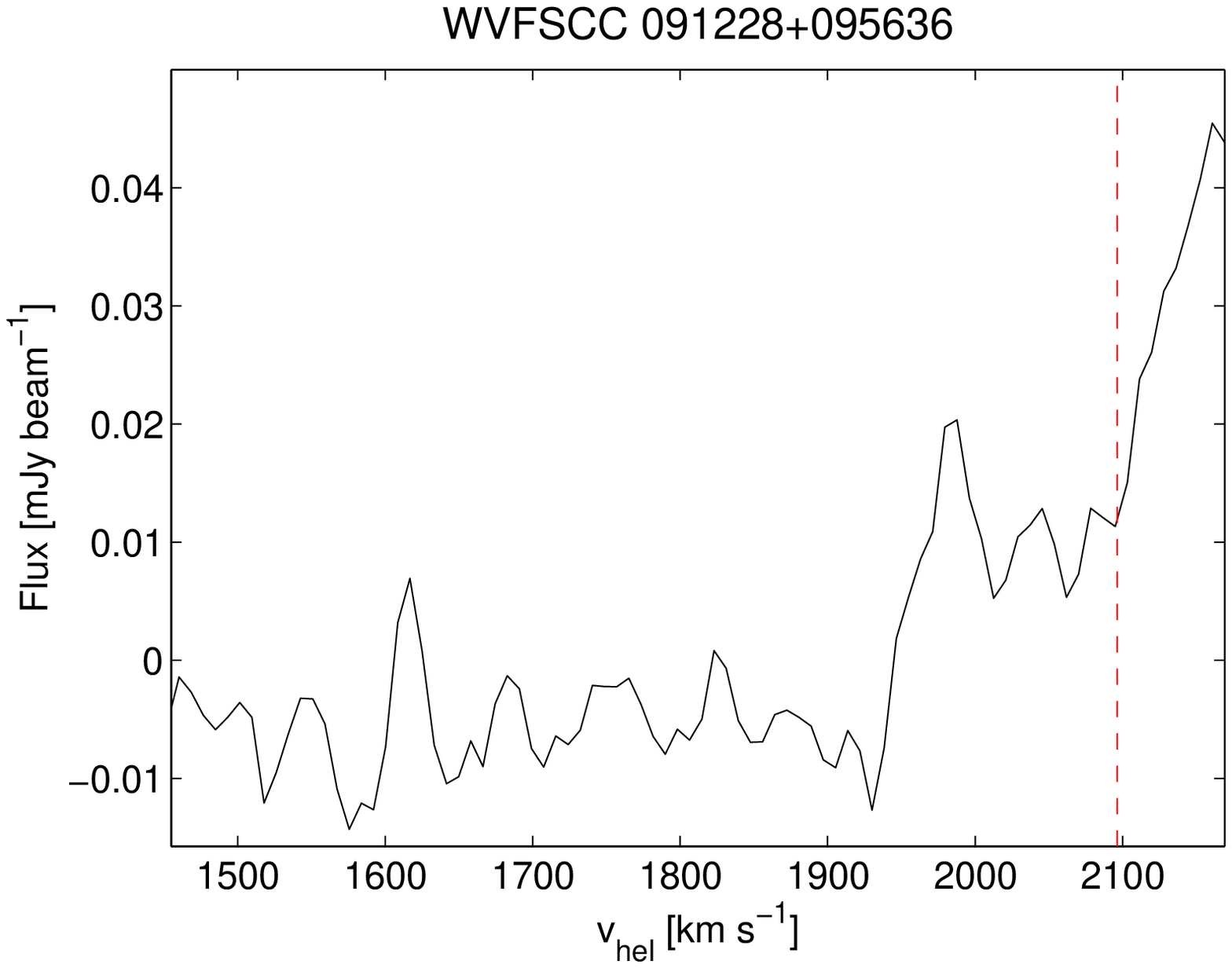}
\includegraphics[width=0.32\textwidth]{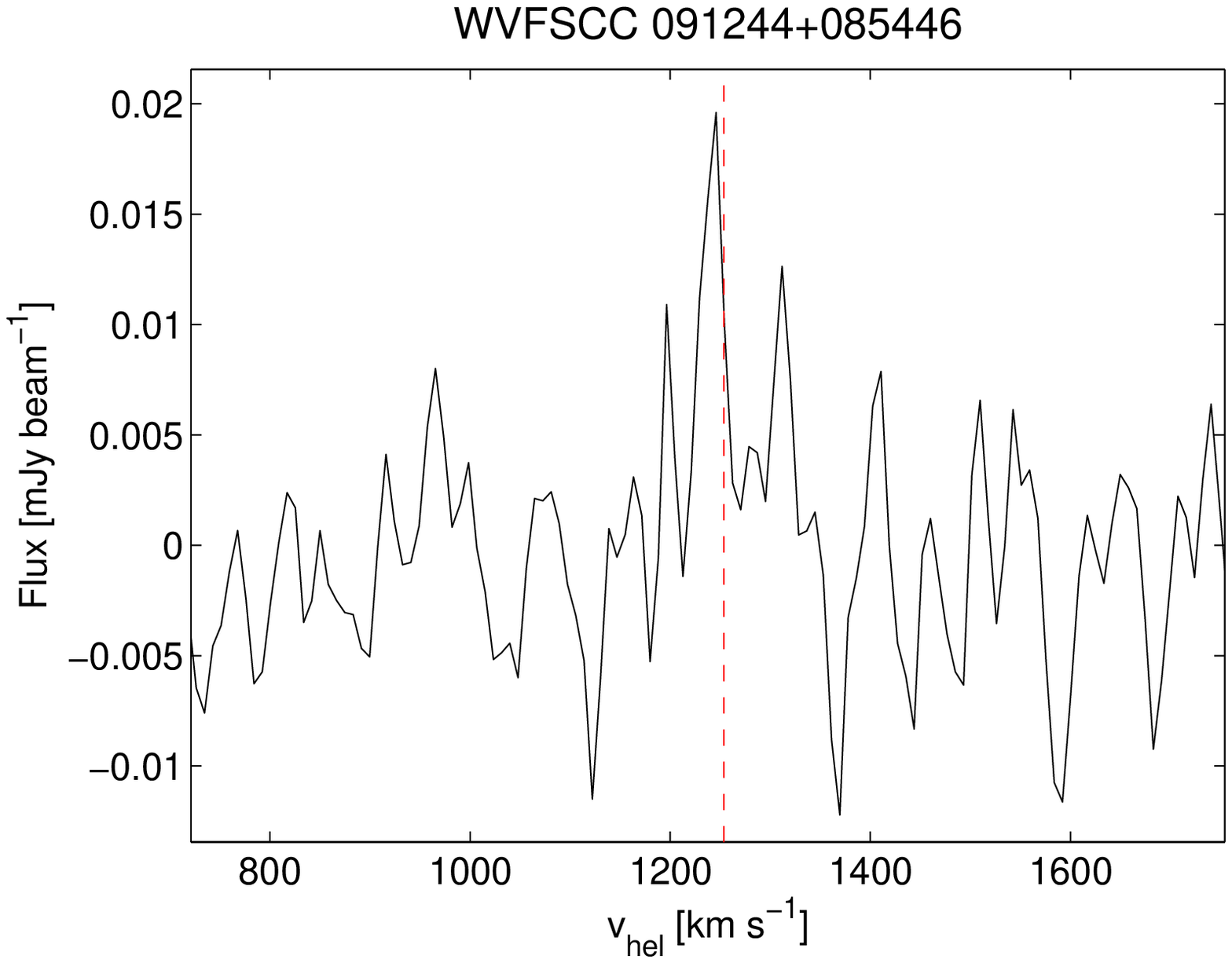}
\includegraphics[width=0.32\textwidth]{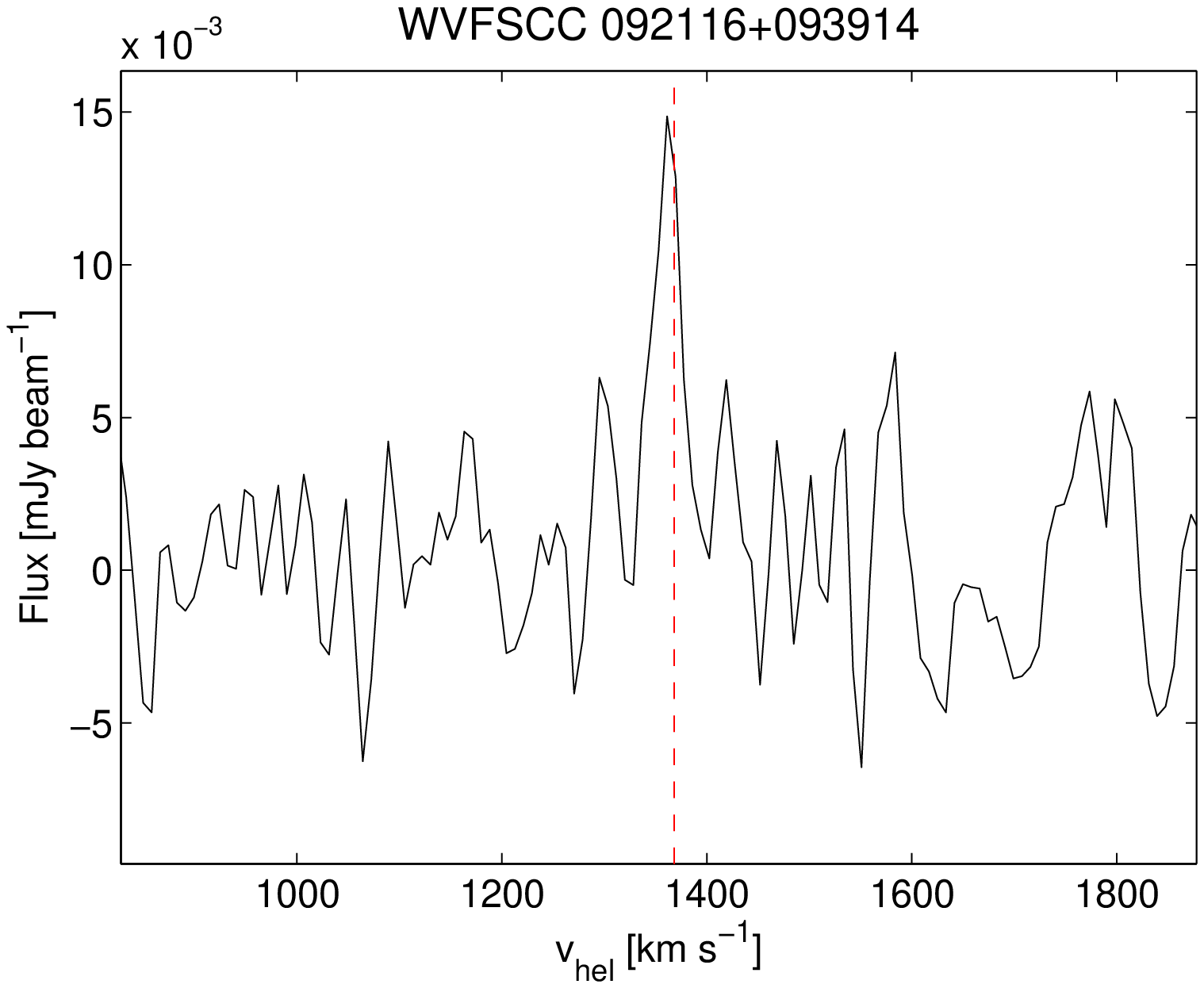}

 \end{center}                                            
  \caption{Spectra of detections of neutral hydrogen in the WVFS
    cross-correlation with a known optical counterpart. Detections
    were found using an 5$\sigma$ threshold and are only accepted if
    an optical counterpart at a corresponding redshift is cataloged.  The
    central velocity of each object is indicated by the dashed line.}
  \label{all_spectra2}                                    
\end{figure*}


\begin{figure*}
  \begin{center}

\includegraphics[width=0.32\textwidth]{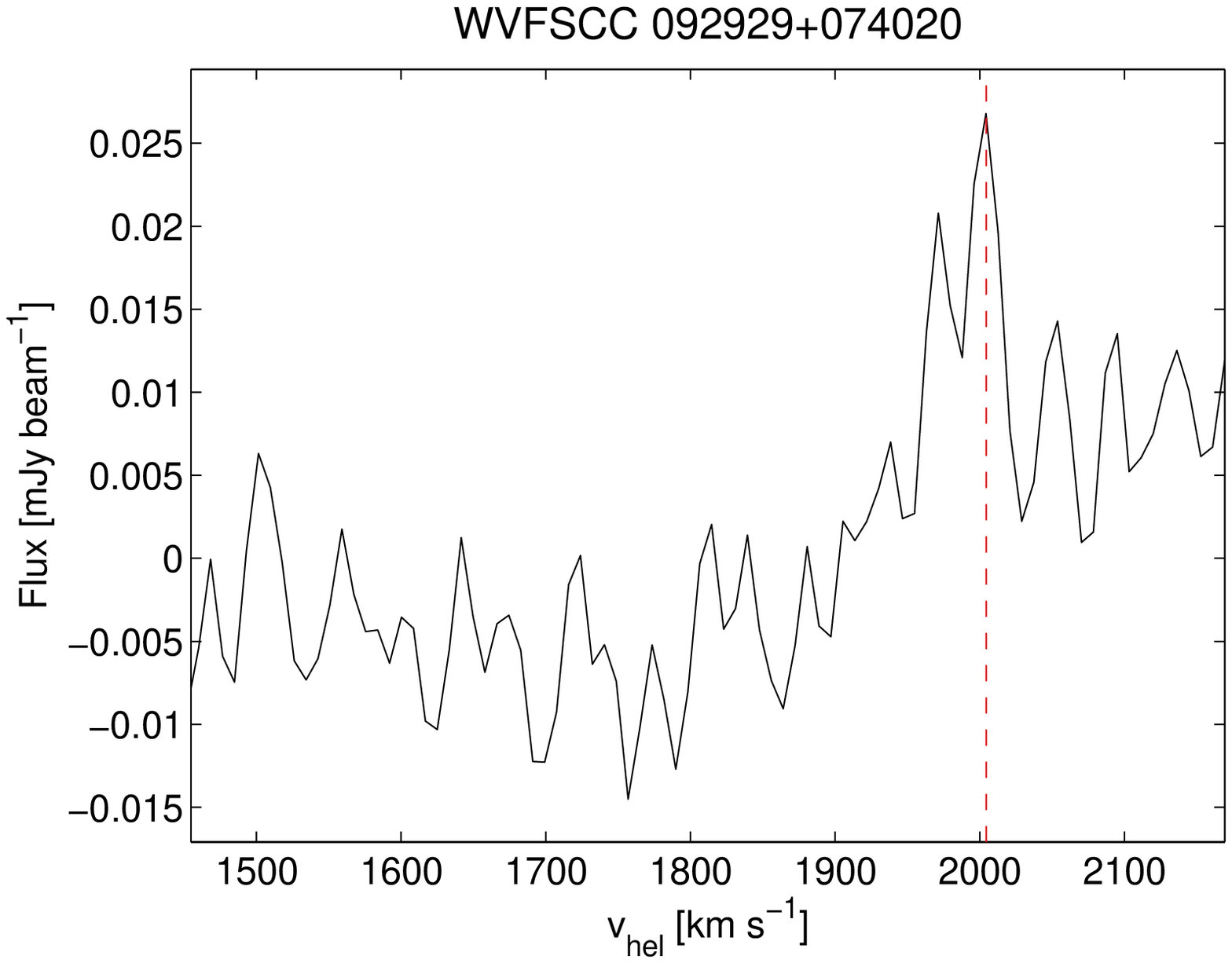}
\includegraphics[width=0.32\textwidth]{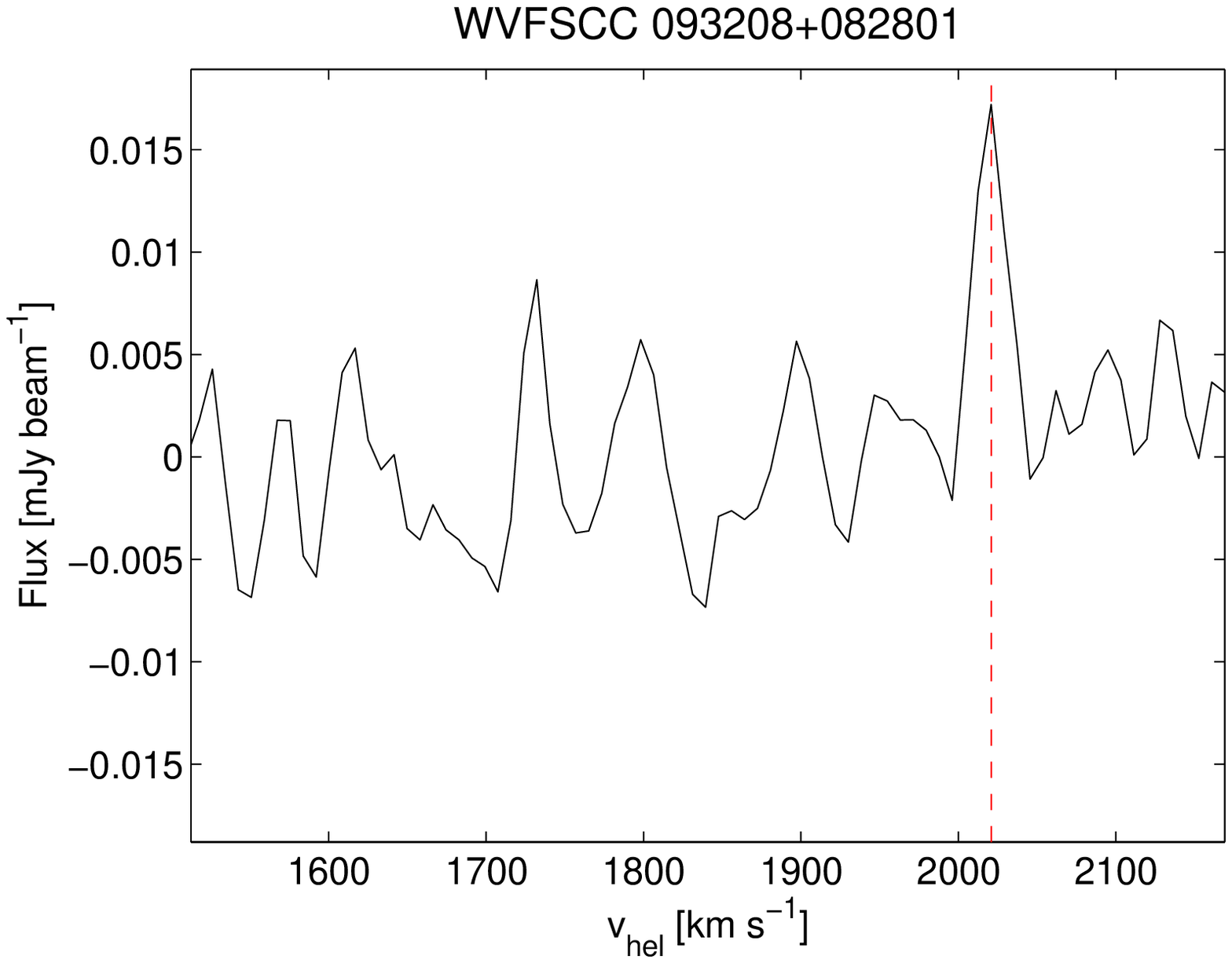}
\includegraphics[width=0.32\textwidth]{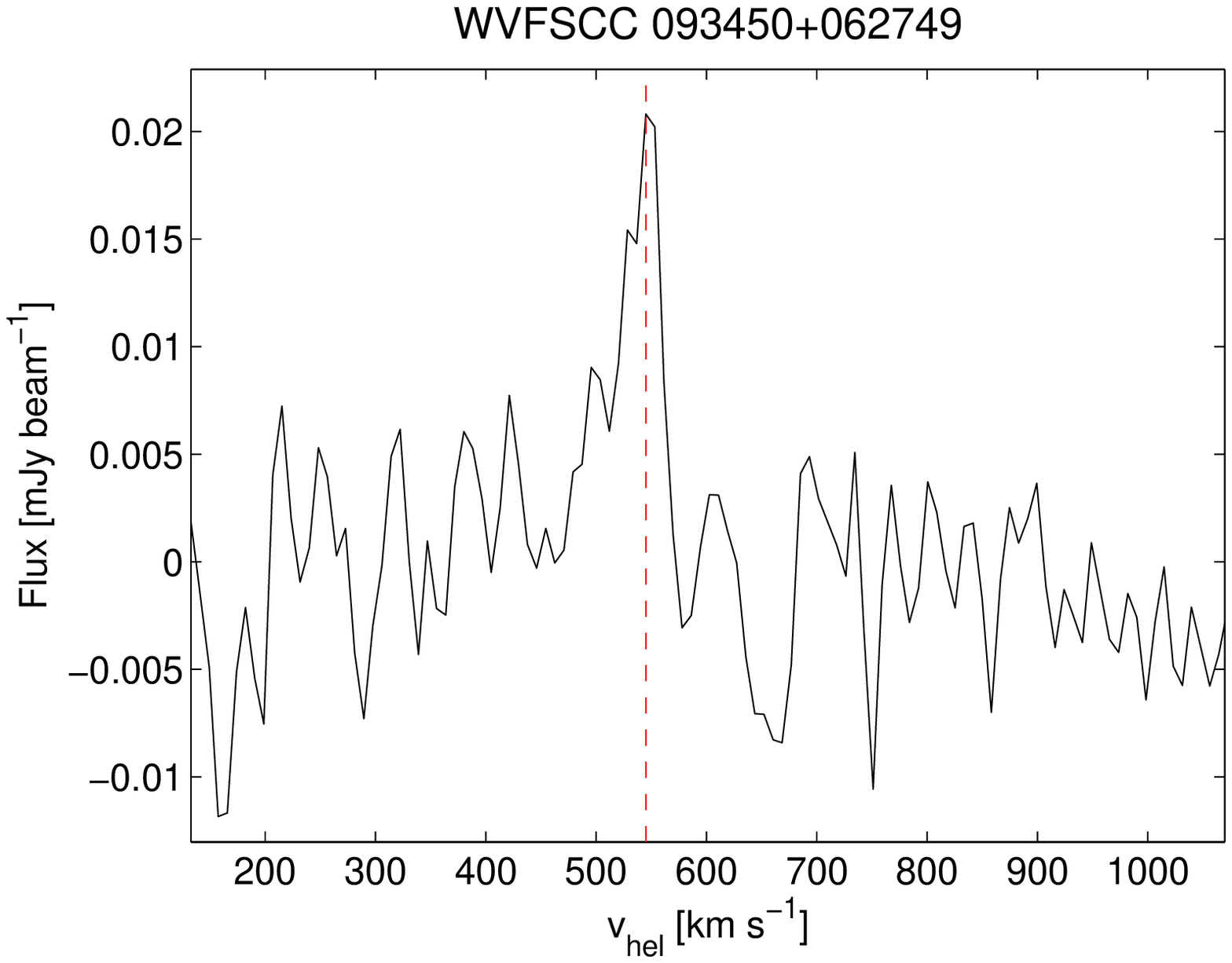}
\includegraphics[width=0.32\textwidth]{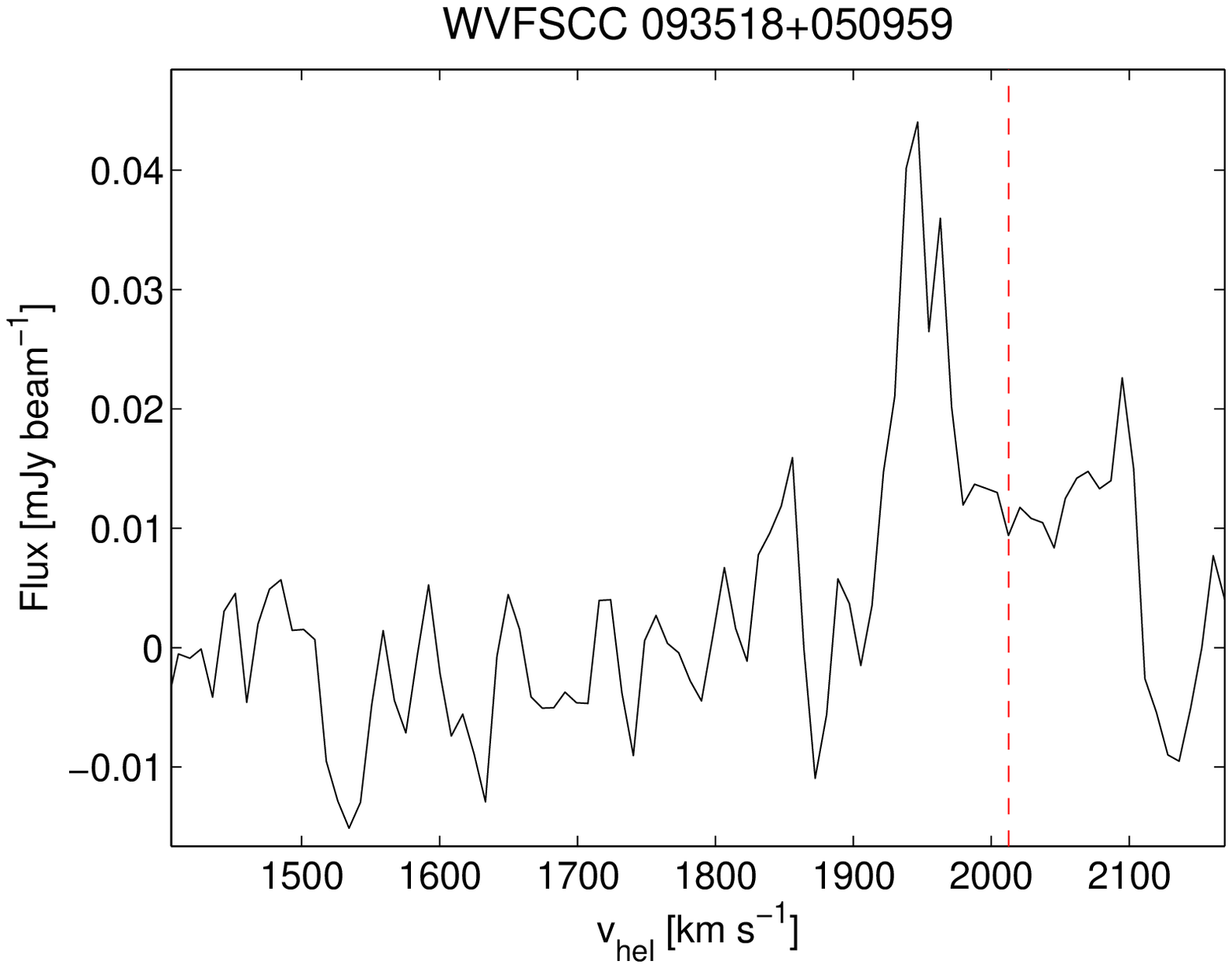}
\includegraphics[width=0.32\textwidth]{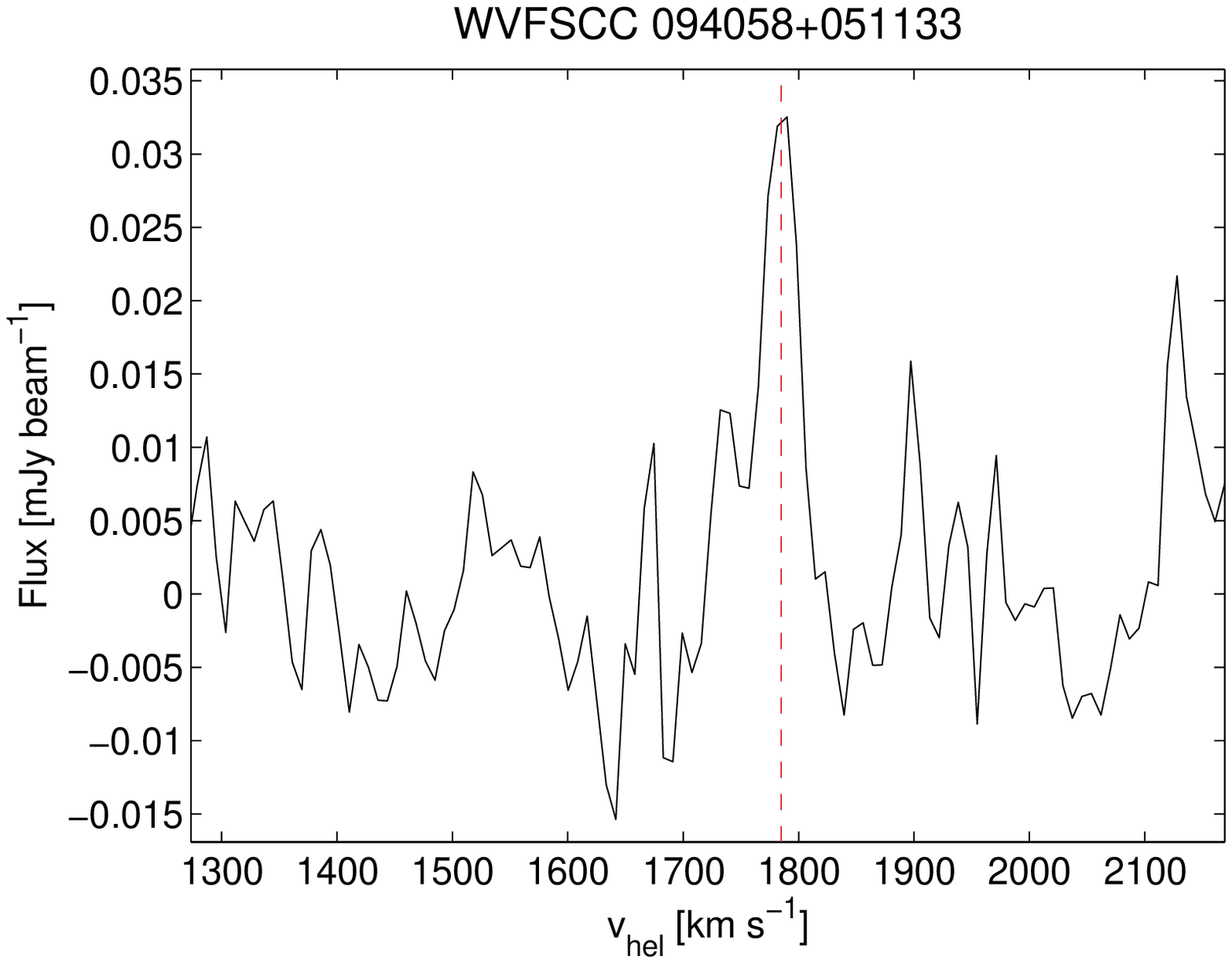}
\includegraphics[width=0.32\textwidth]{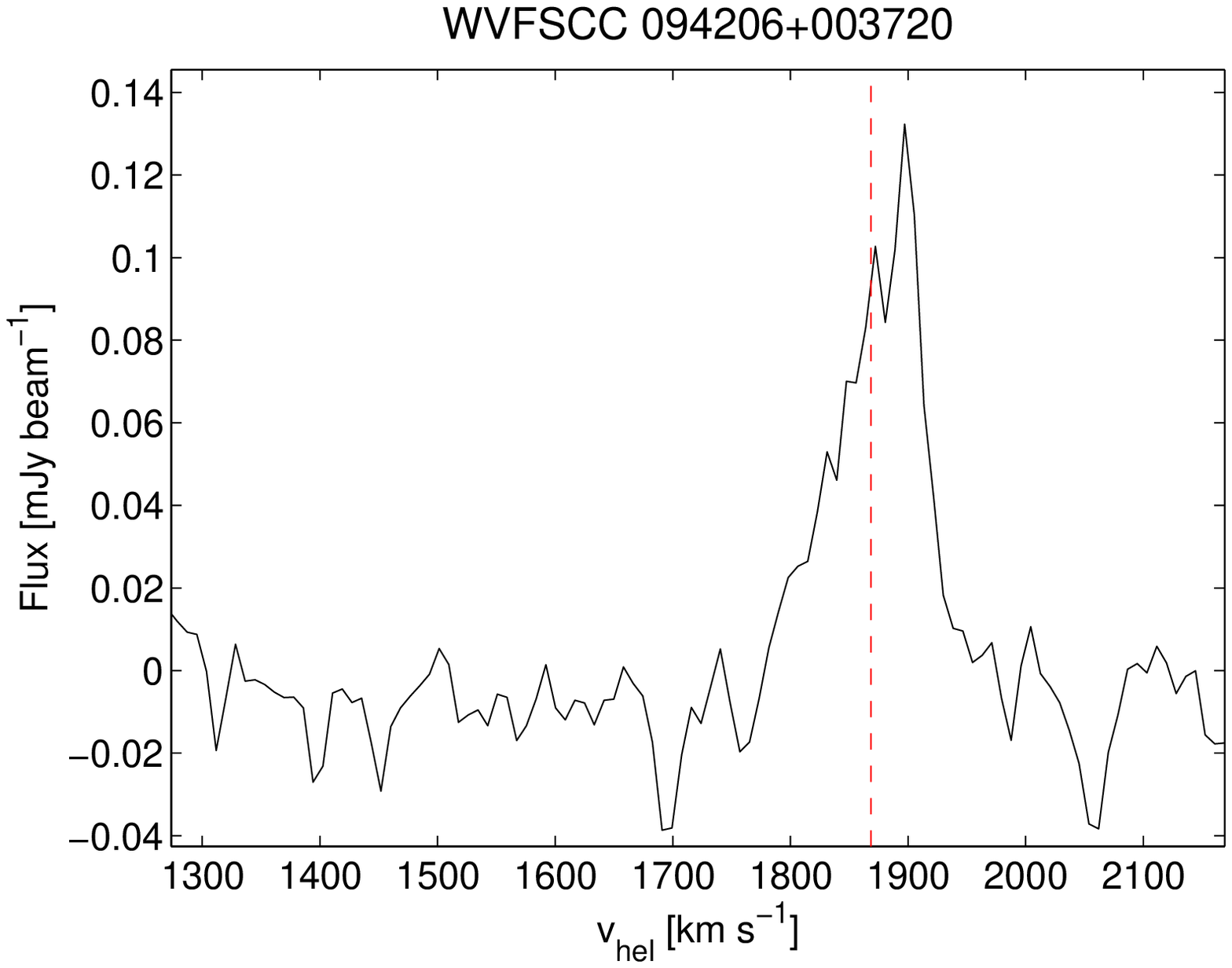}
\includegraphics[width=0.32\textwidth]{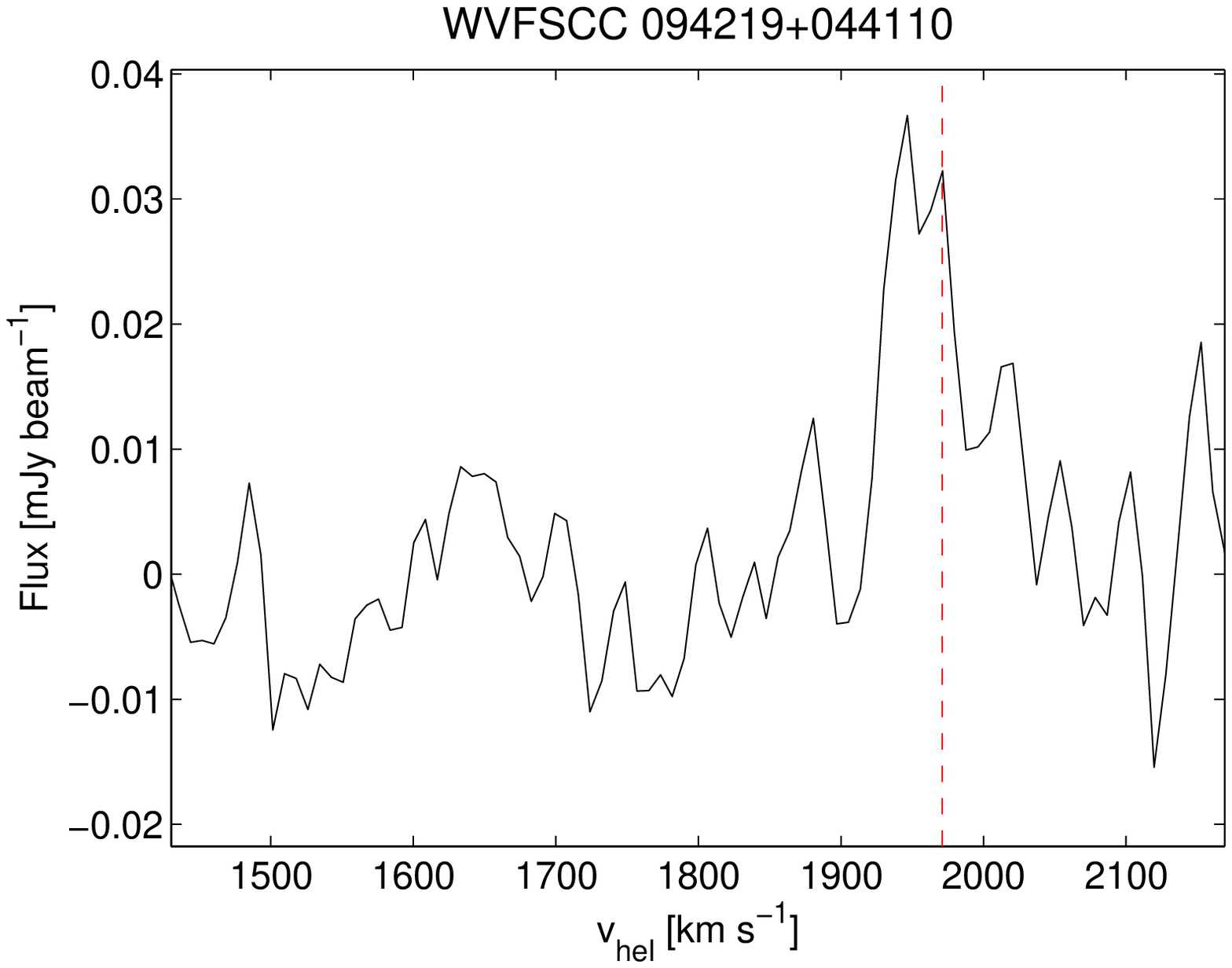}
\includegraphics[width=0.32\textwidth]{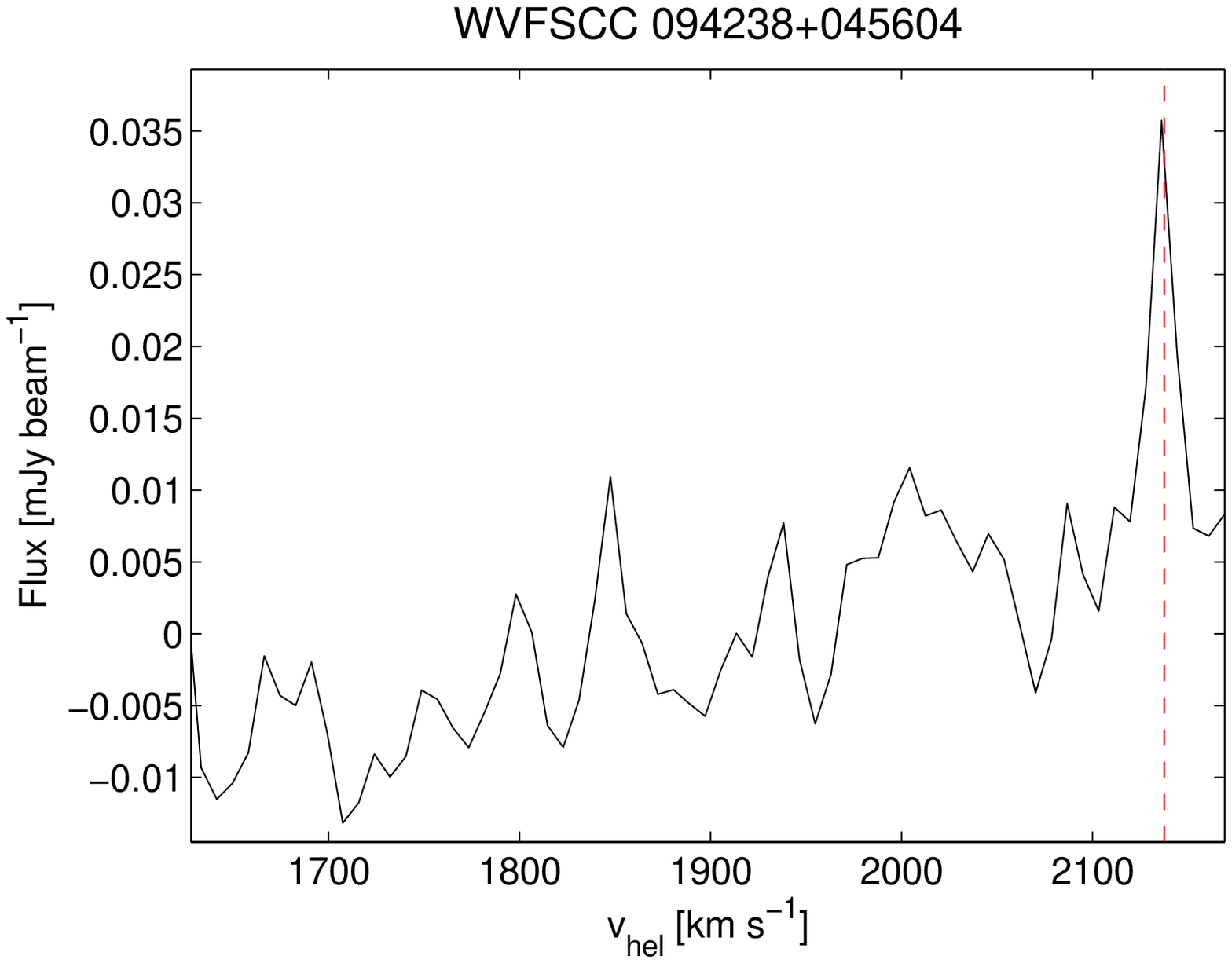}
\includegraphics[width=0.32\textwidth]{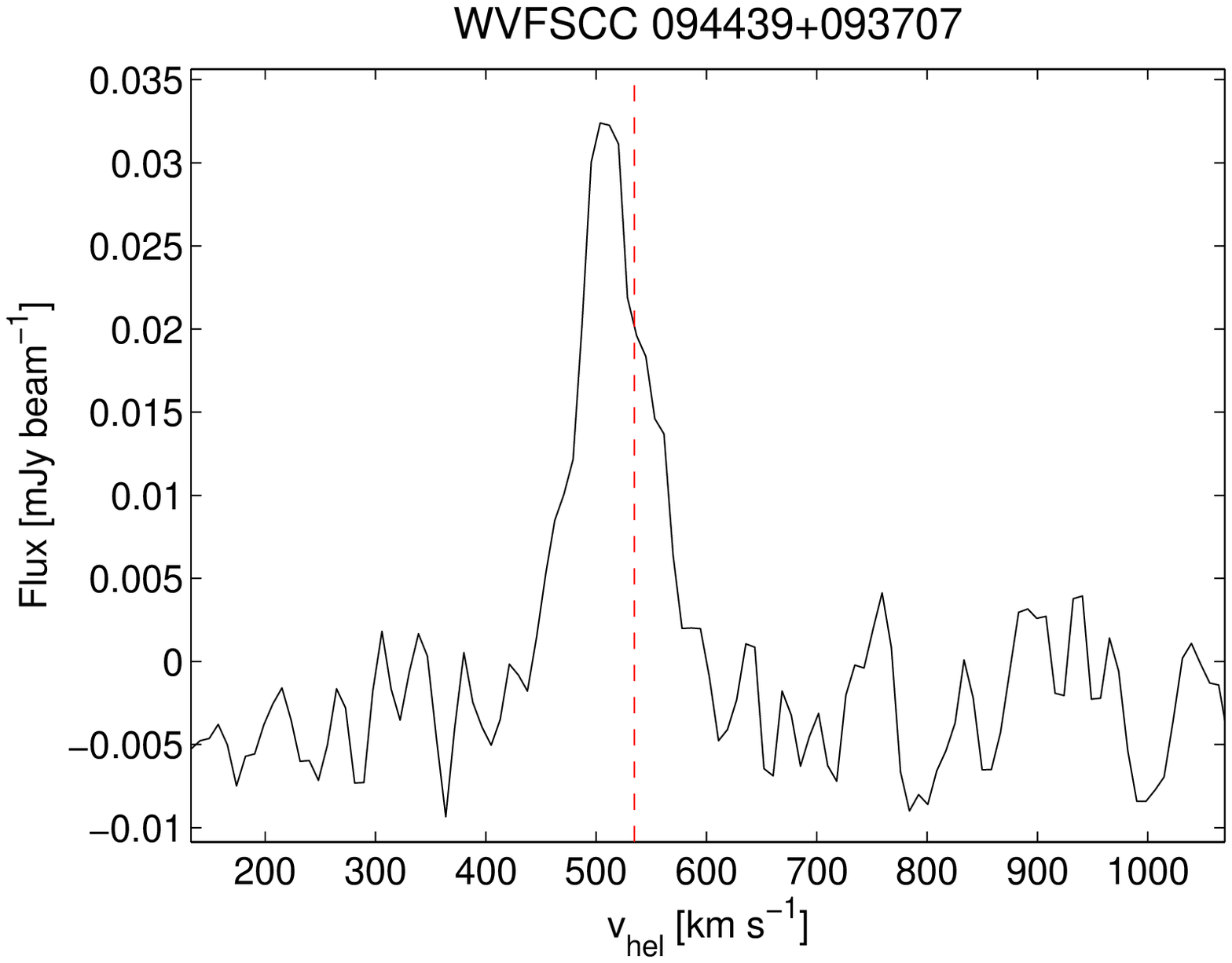}
\includegraphics[width=0.32\textwidth]{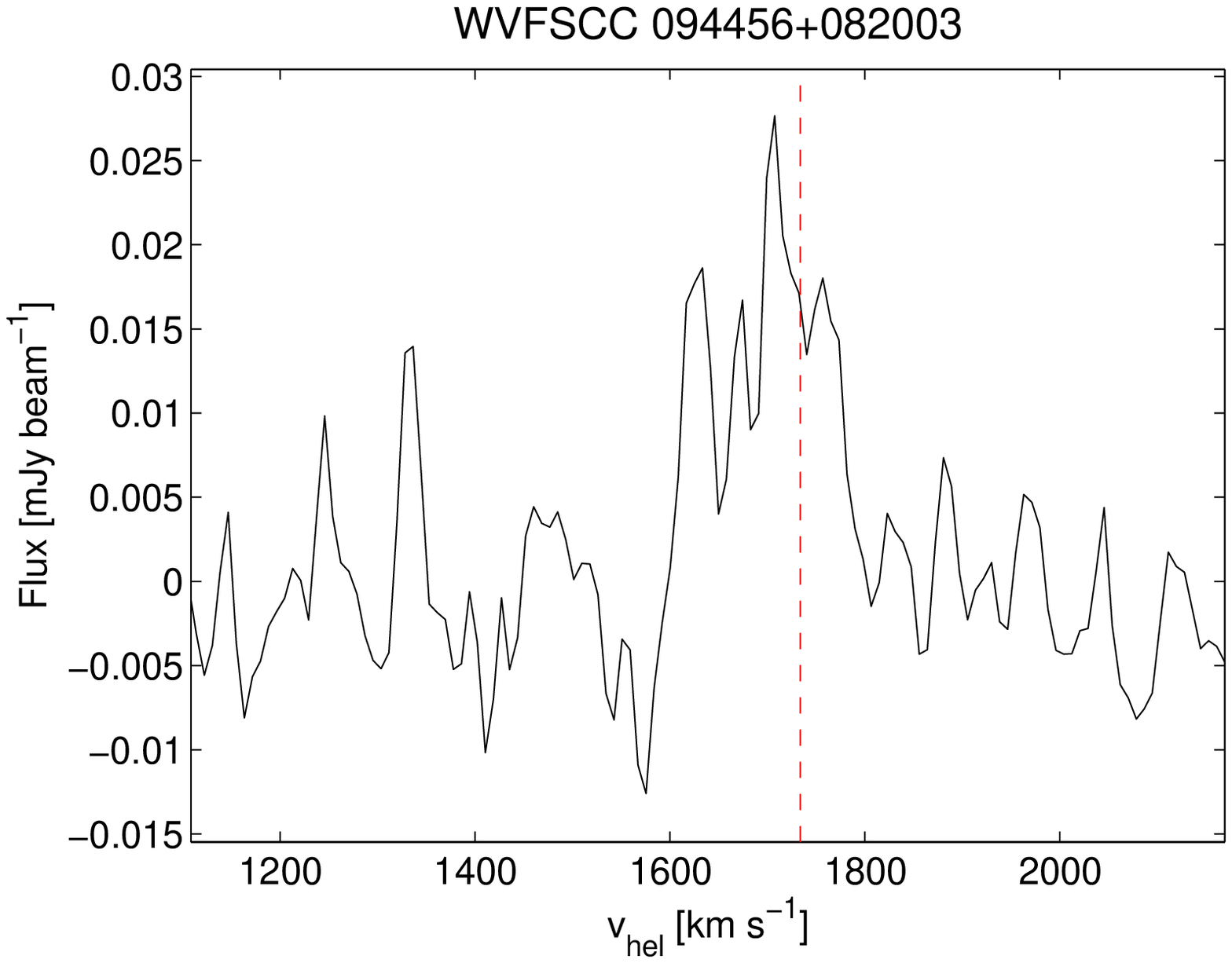}
\includegraphics[width=0.32\textwidth]{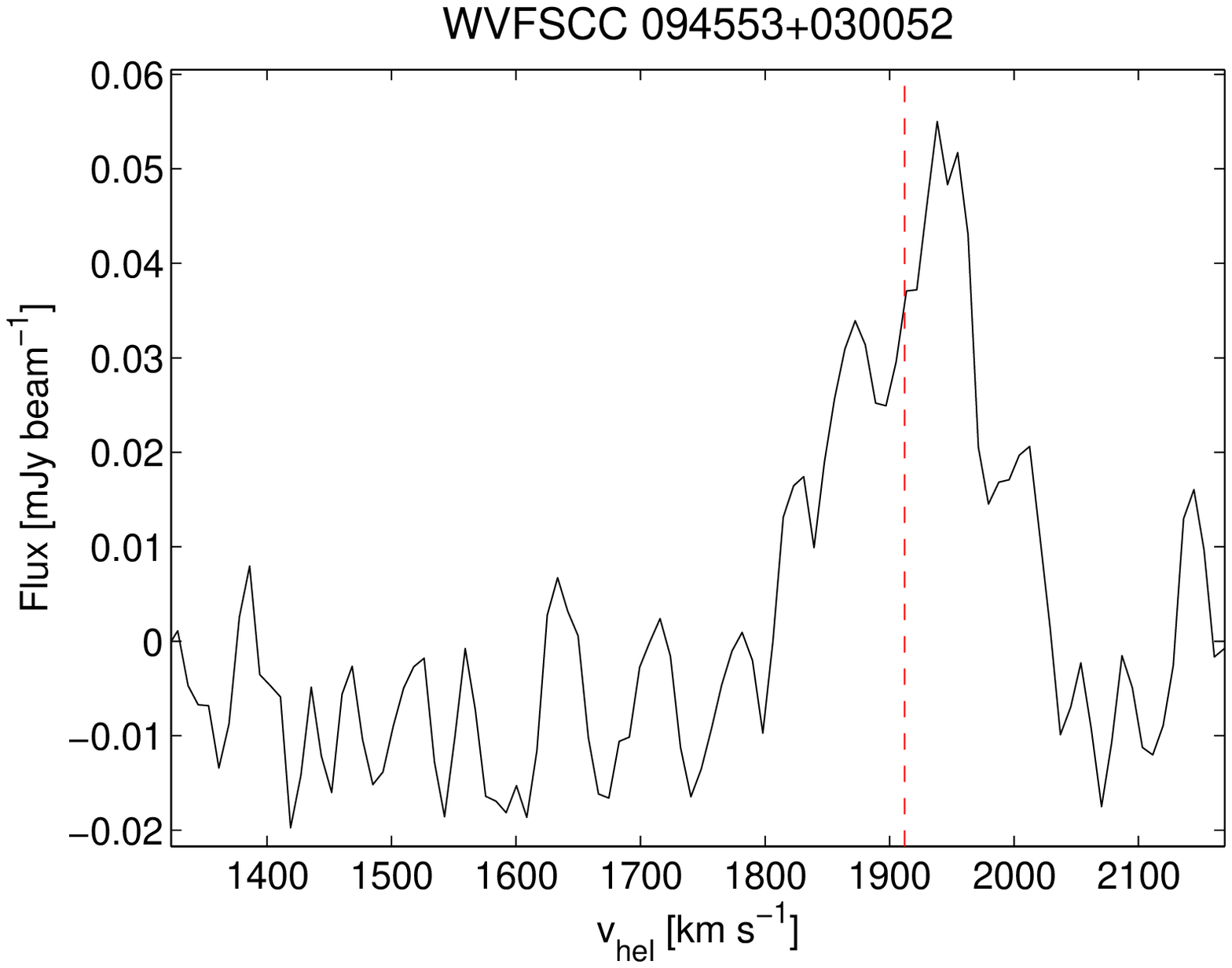}
\includegraphics[width=0.32\textwidth]{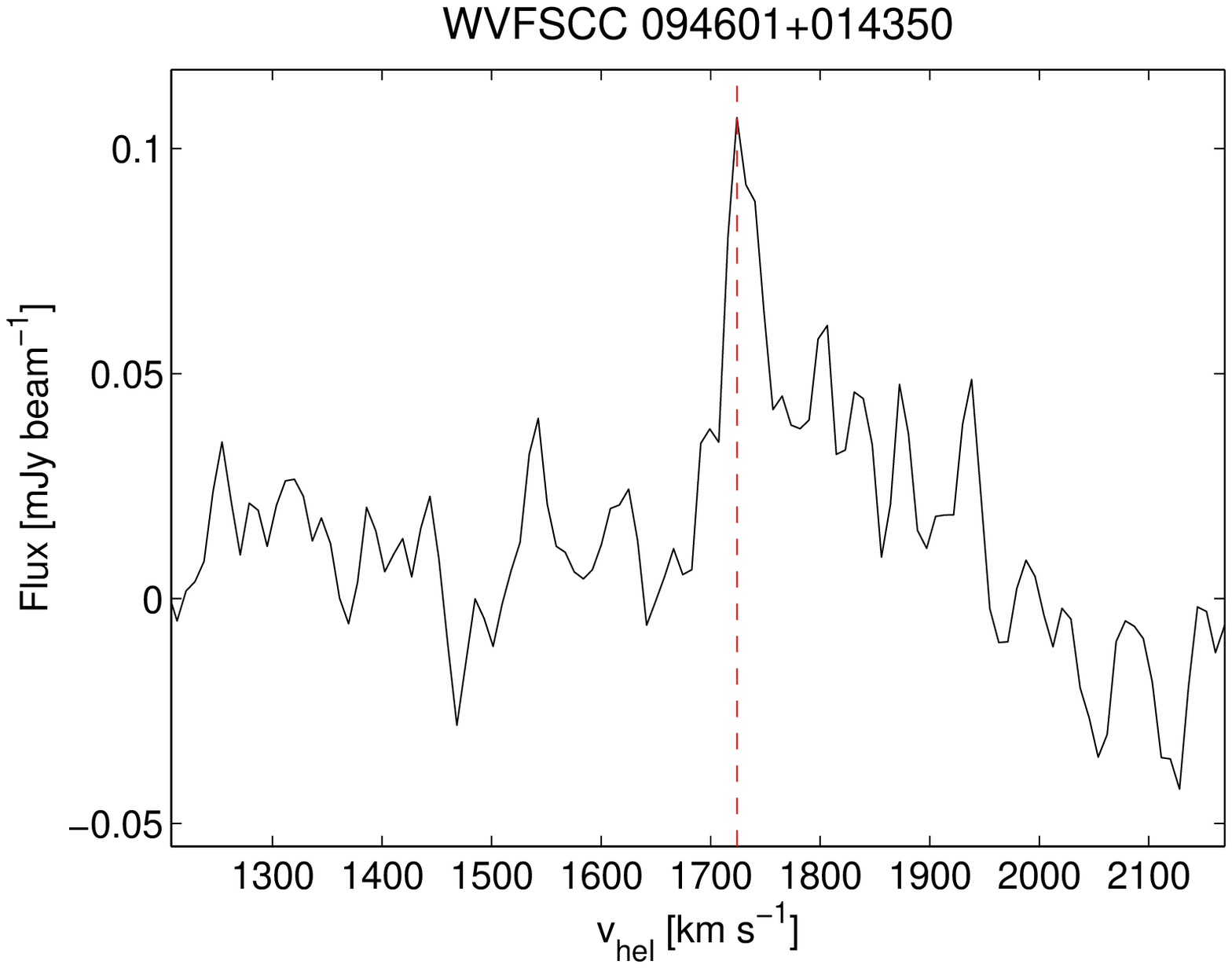}
\includegraphics[width=0.32\textwidth]{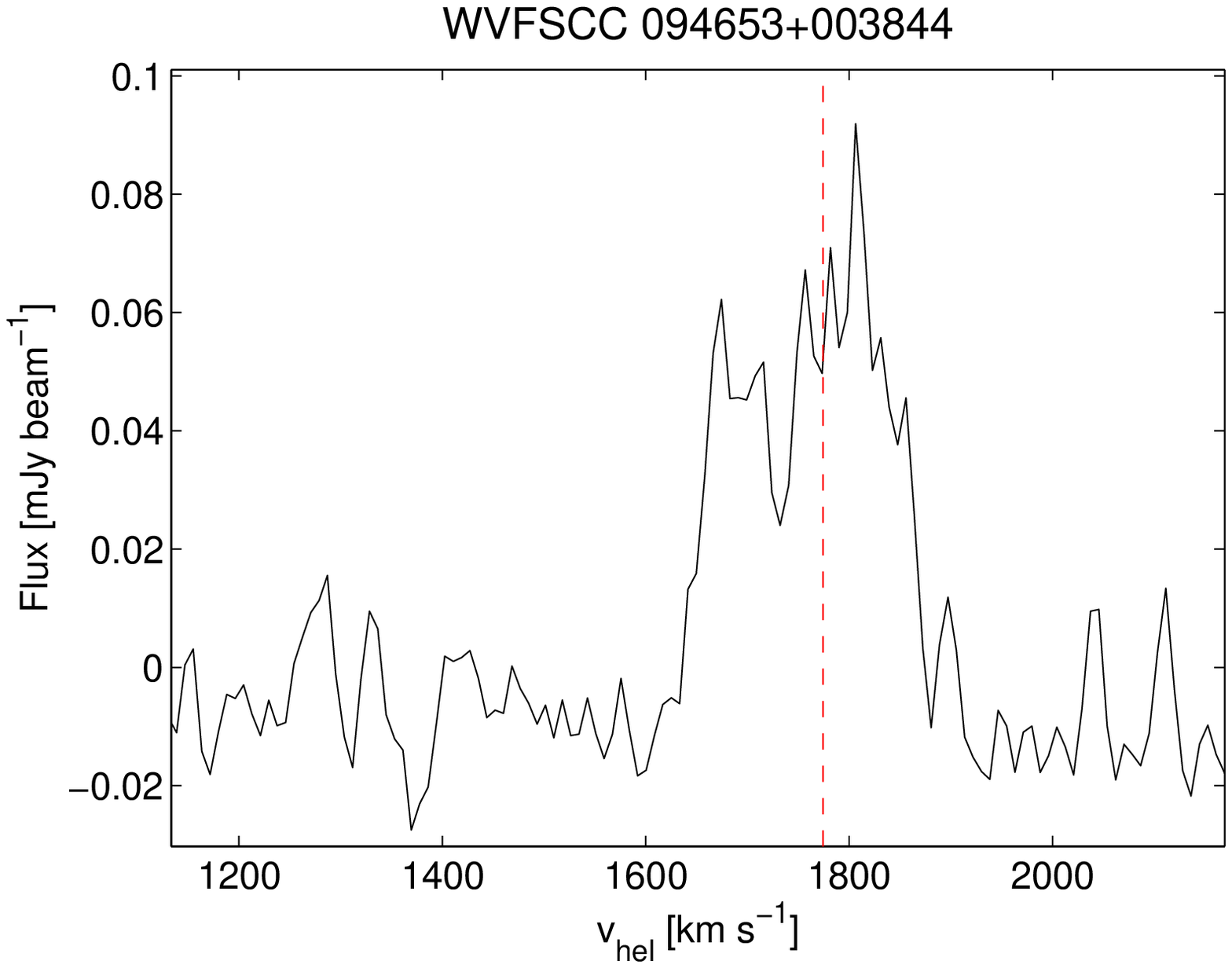}
\includegraphics[width=0.32\textwidth]{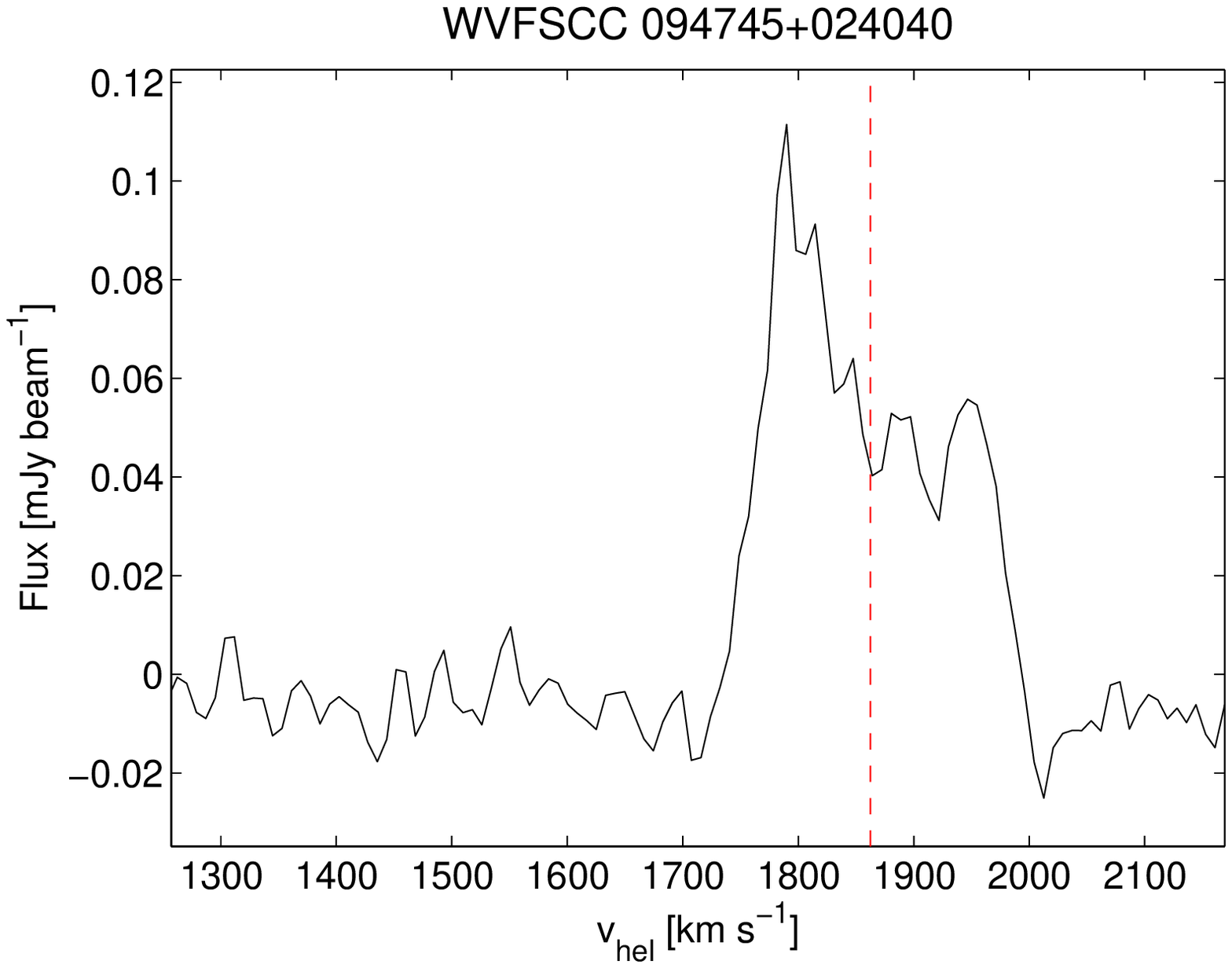}
\includegraphics[width=0.32\textwidth]{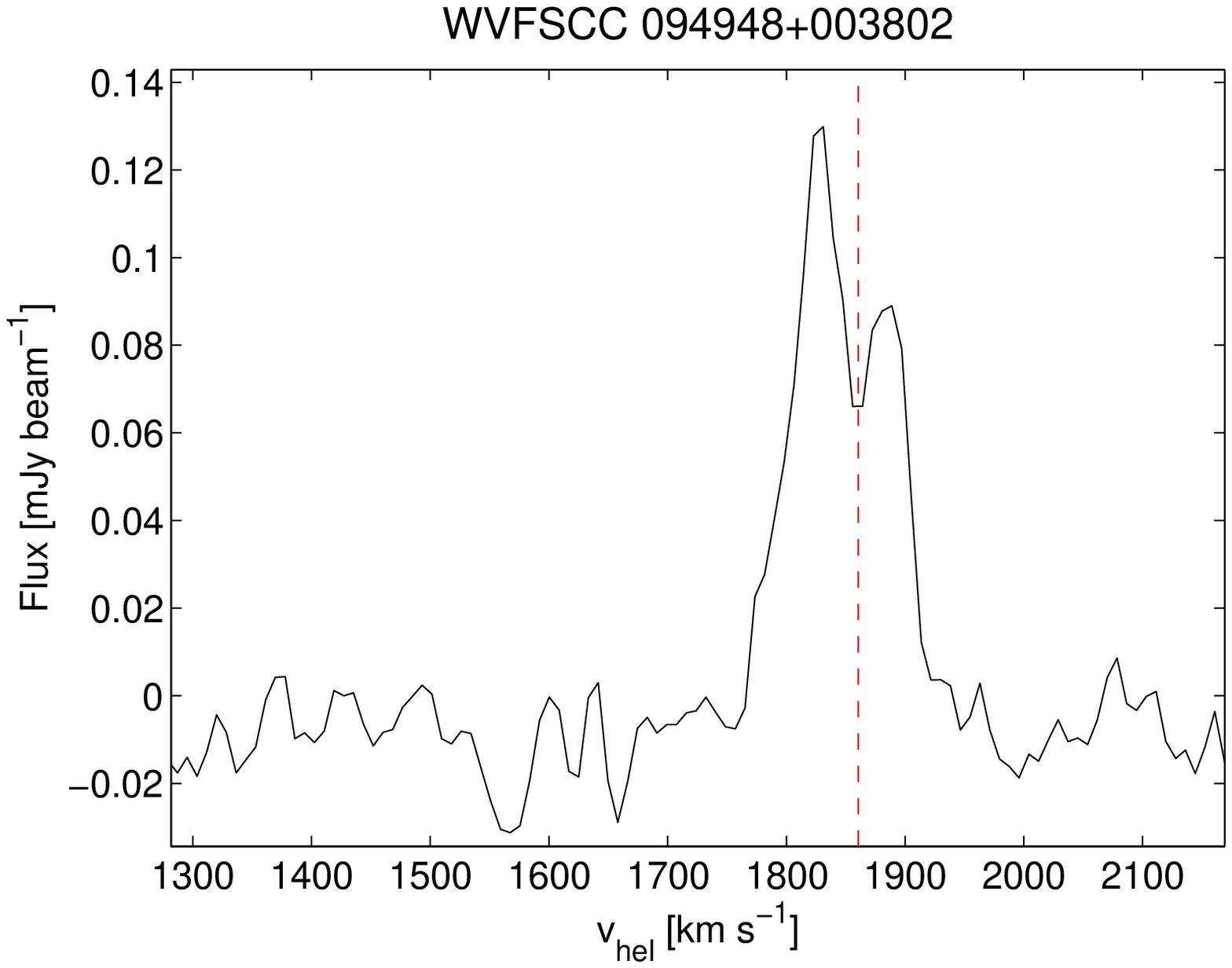} 
                                                         
\end{center}                                            
{\bf Fig~\ref{all_spectra2}.} (continued)                                        
 
\end{figure*}


\begin{figure*}
  \begin{center}

\includegraphics[width=0.32\textwidth]{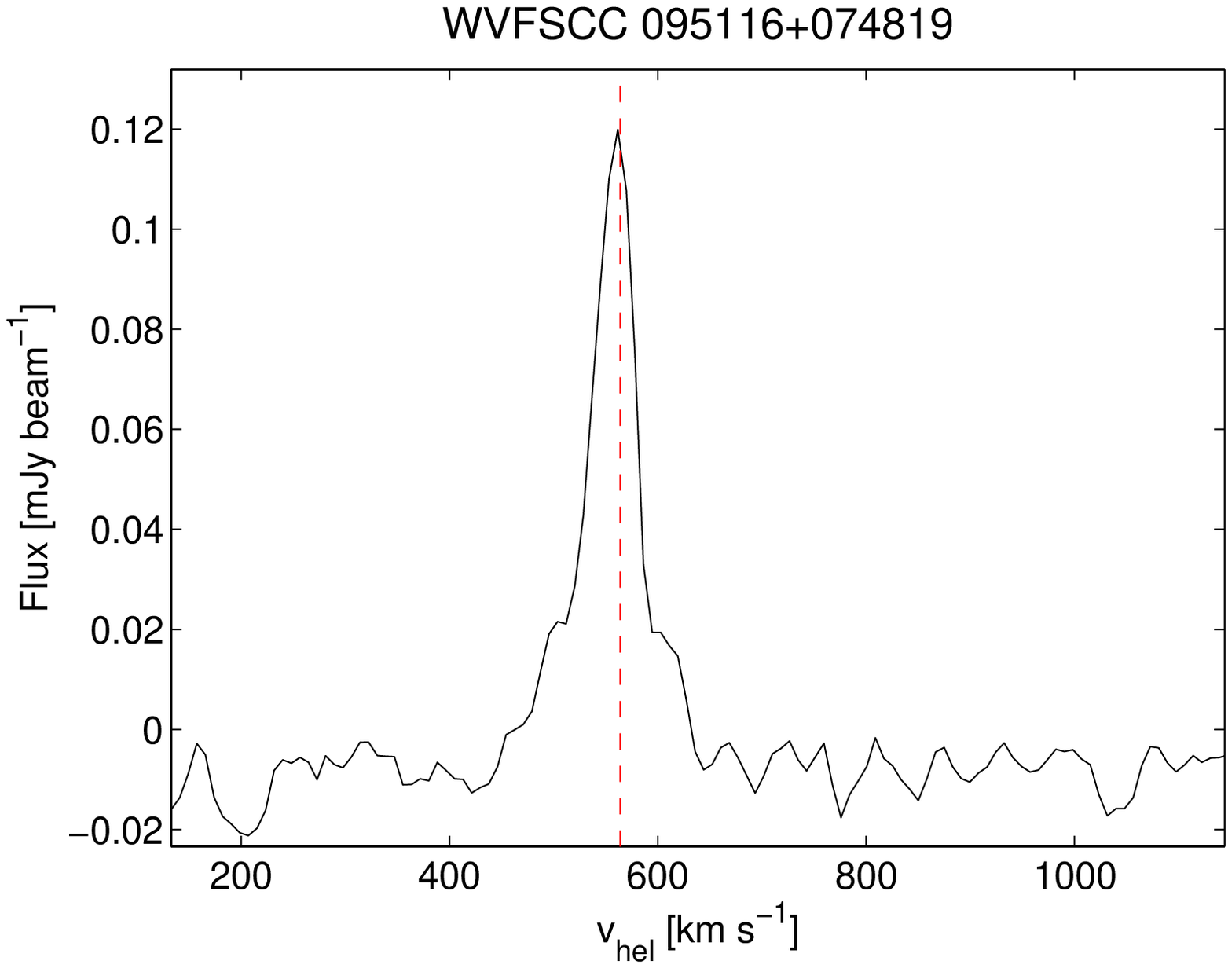}
\includegraphics[width=0.32\textwidth]{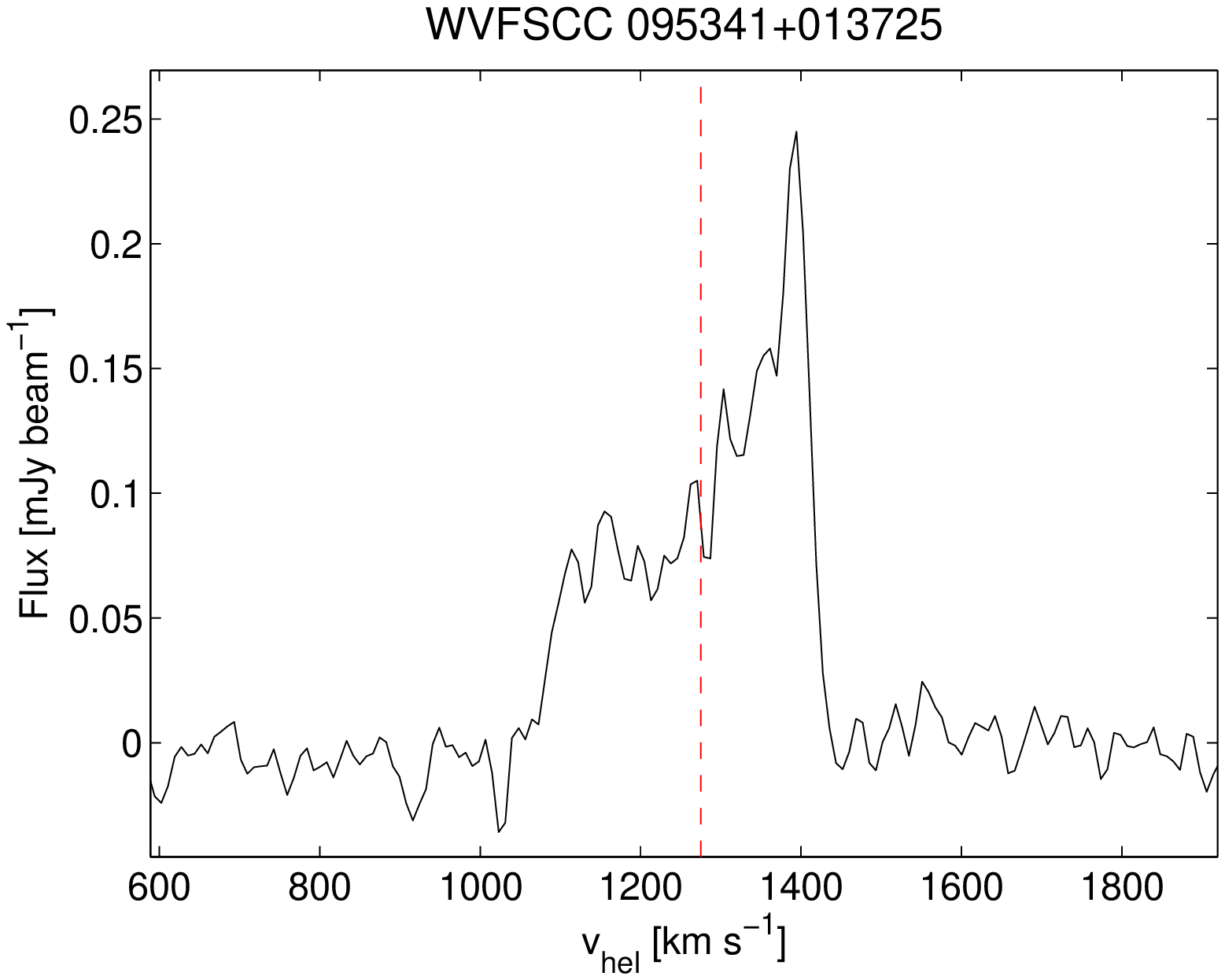}
\includegraphics[width=0.32\textwidth]{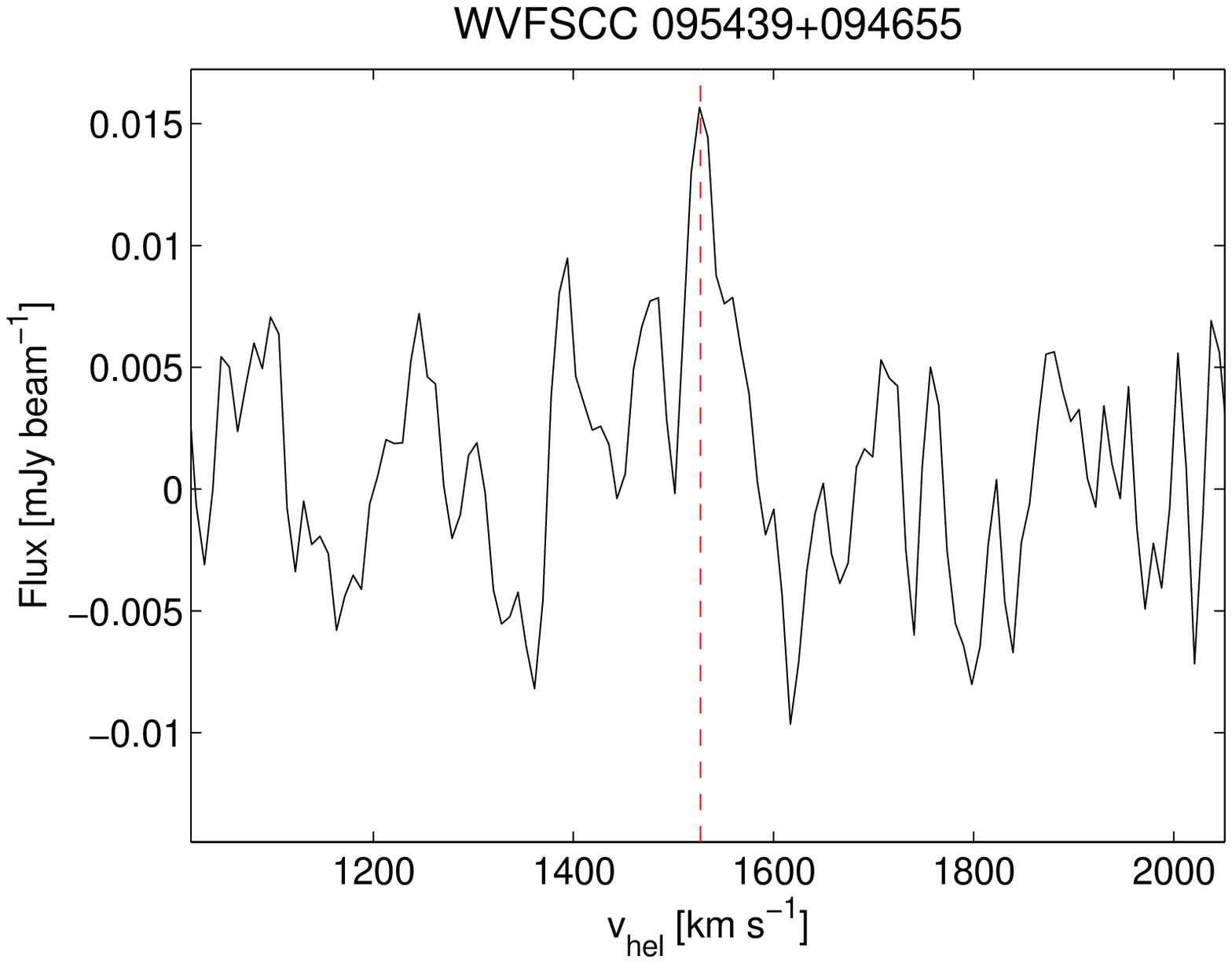}
\includegraphics[width=0.32\textwidth]{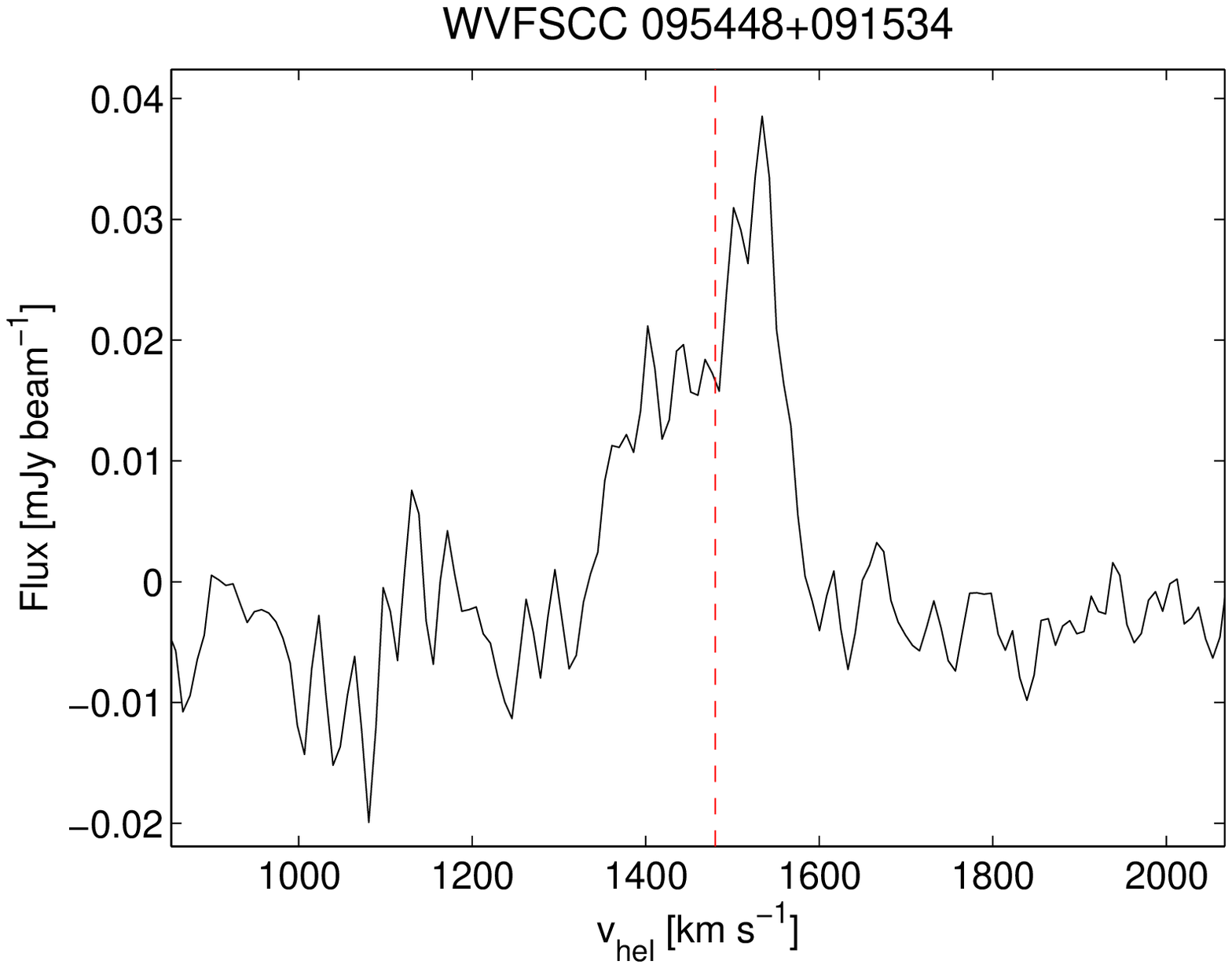}
\includegraphics[width=0.32\textwidth]{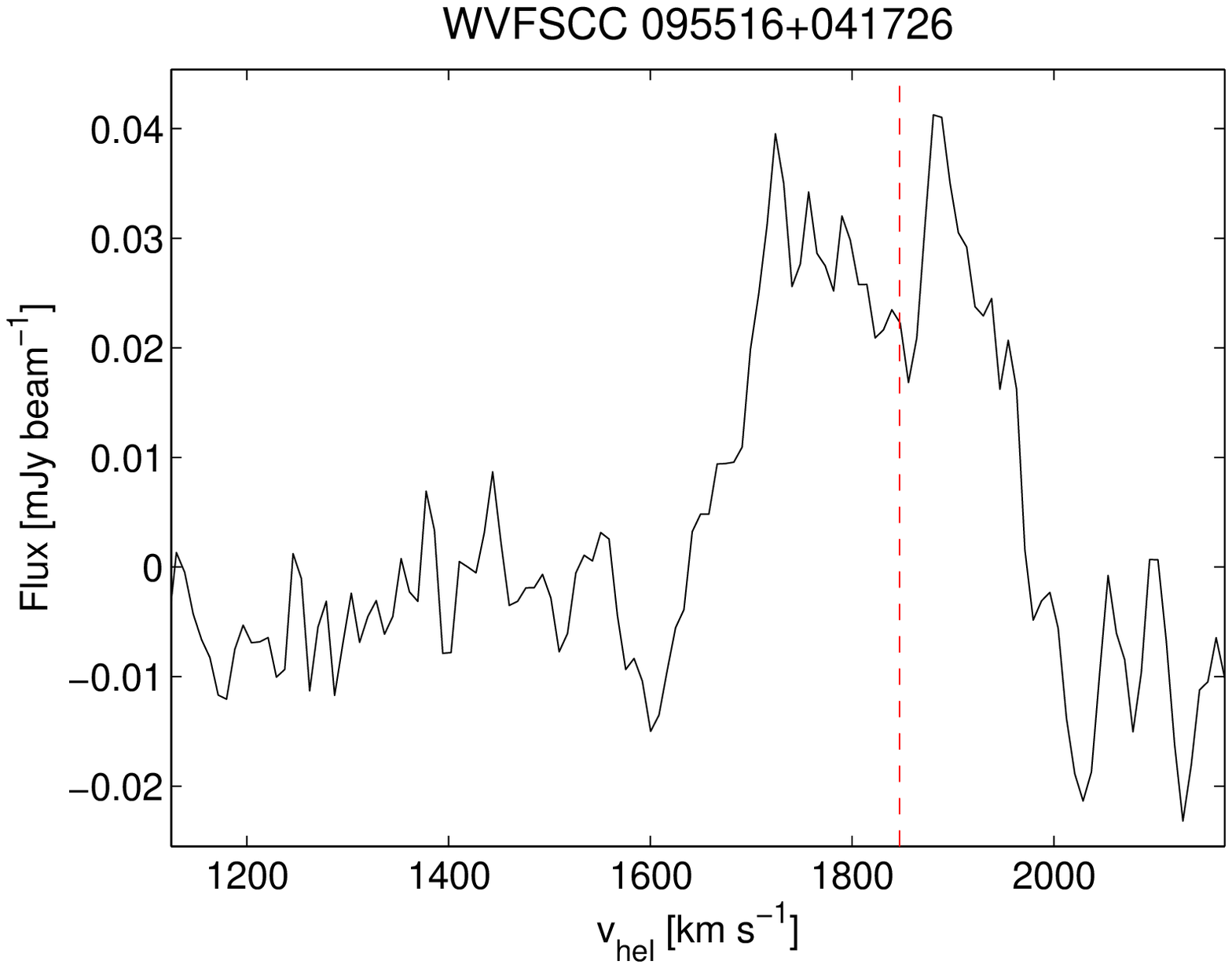}
\includegraphics[width=0.32\textwidth]{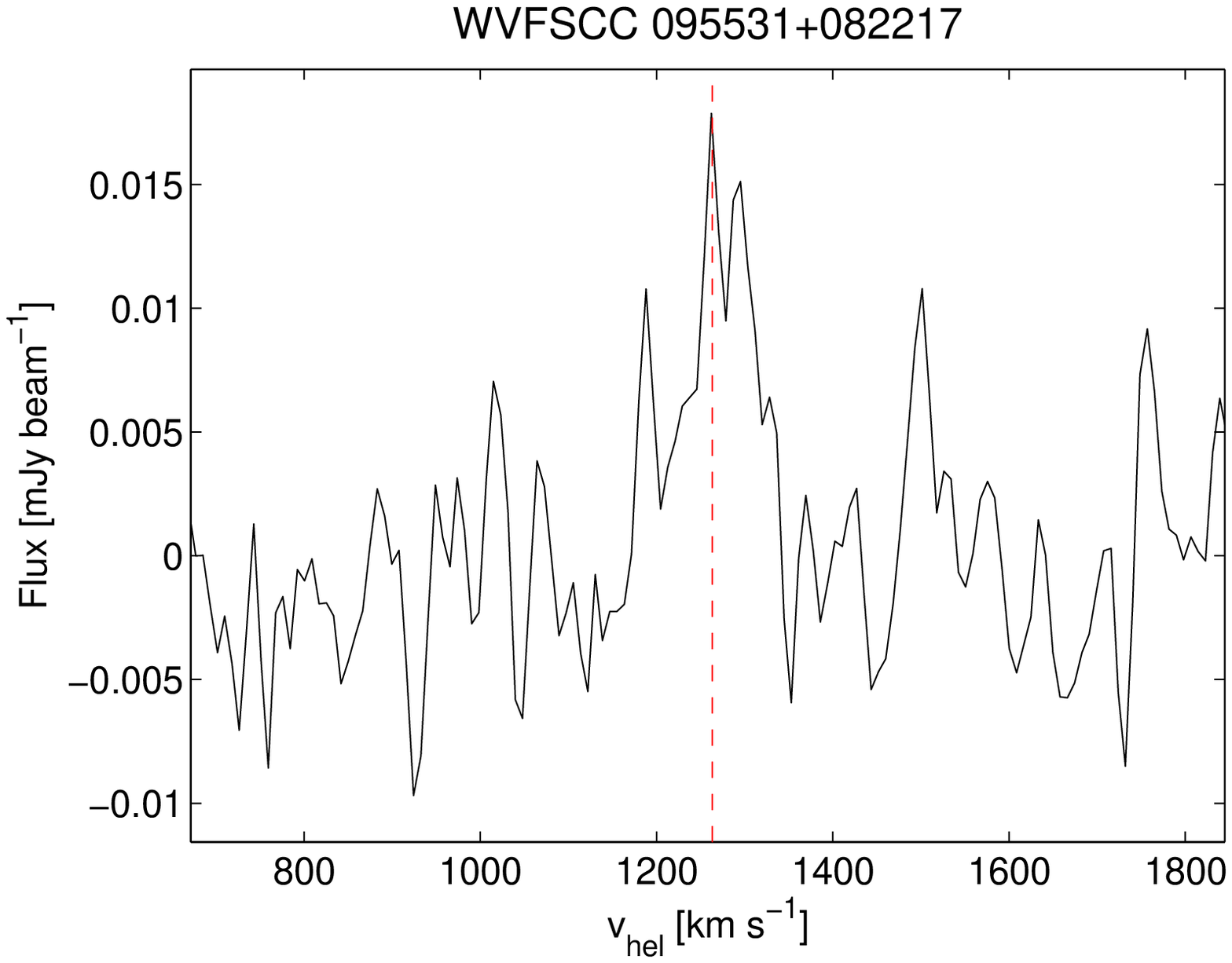}
\includegraphics[width=0.32\textwidth]{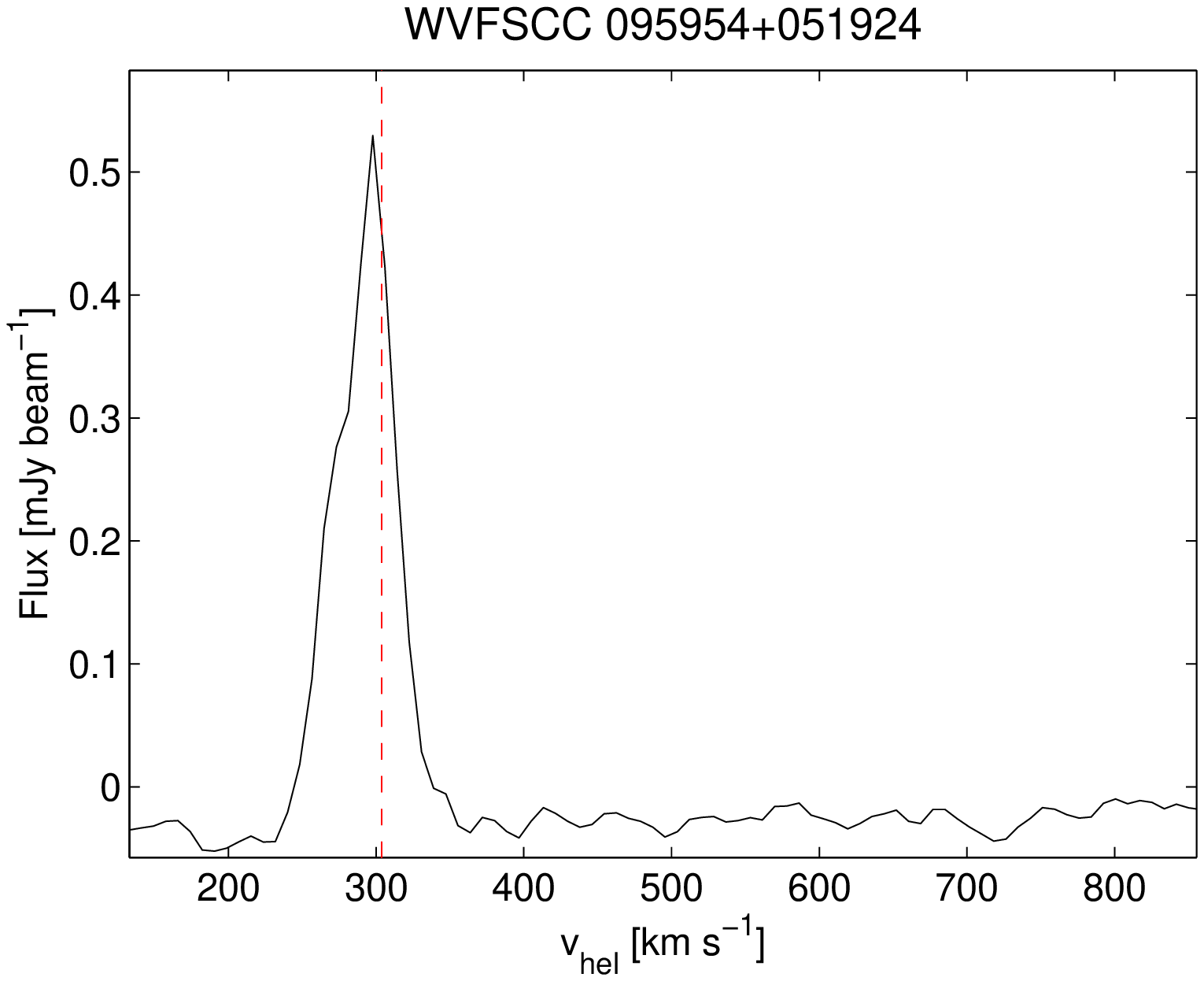}
\includegraphics[width=0.32\textwidth]{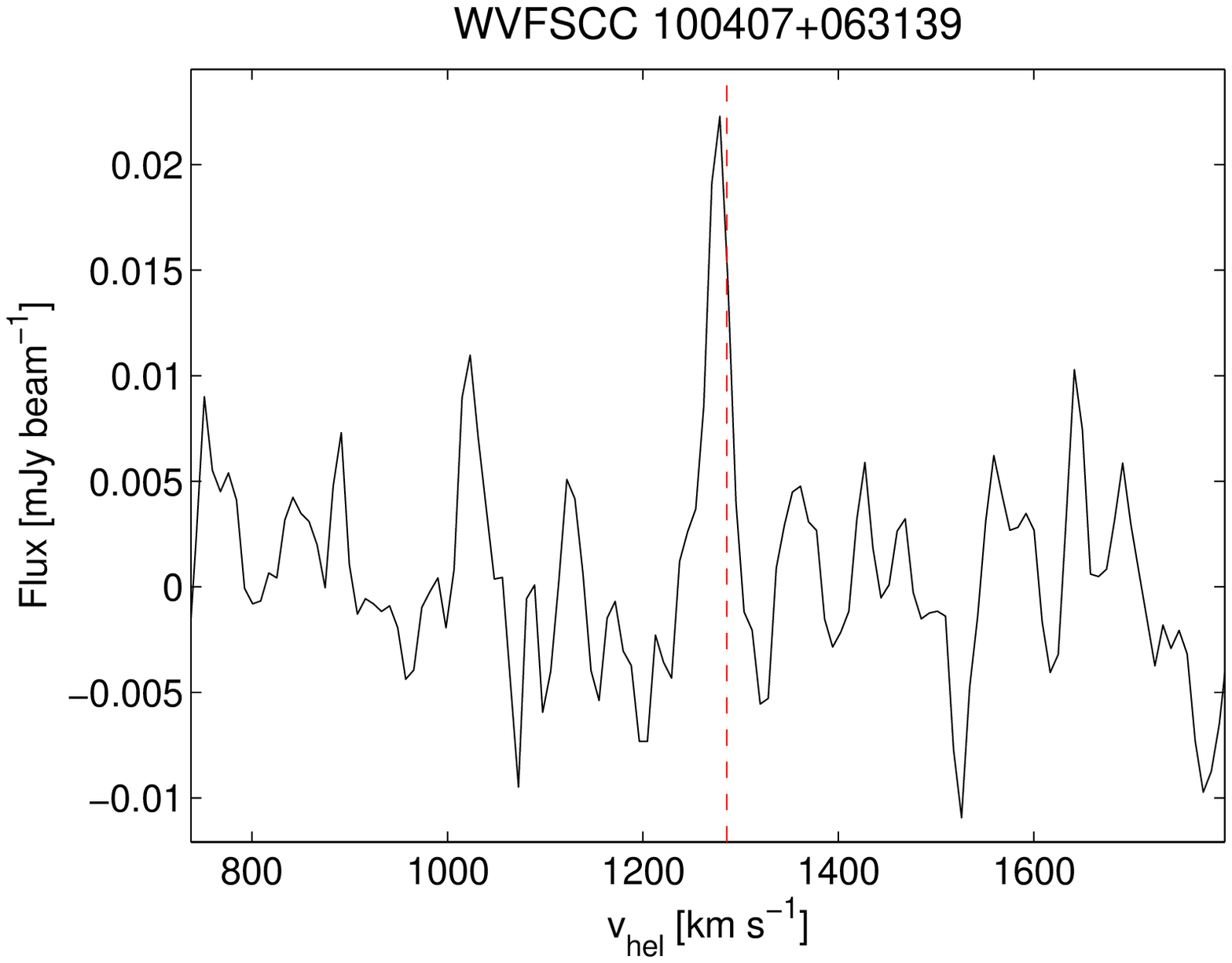}
\includegraphics[width=0.32\textwidth]{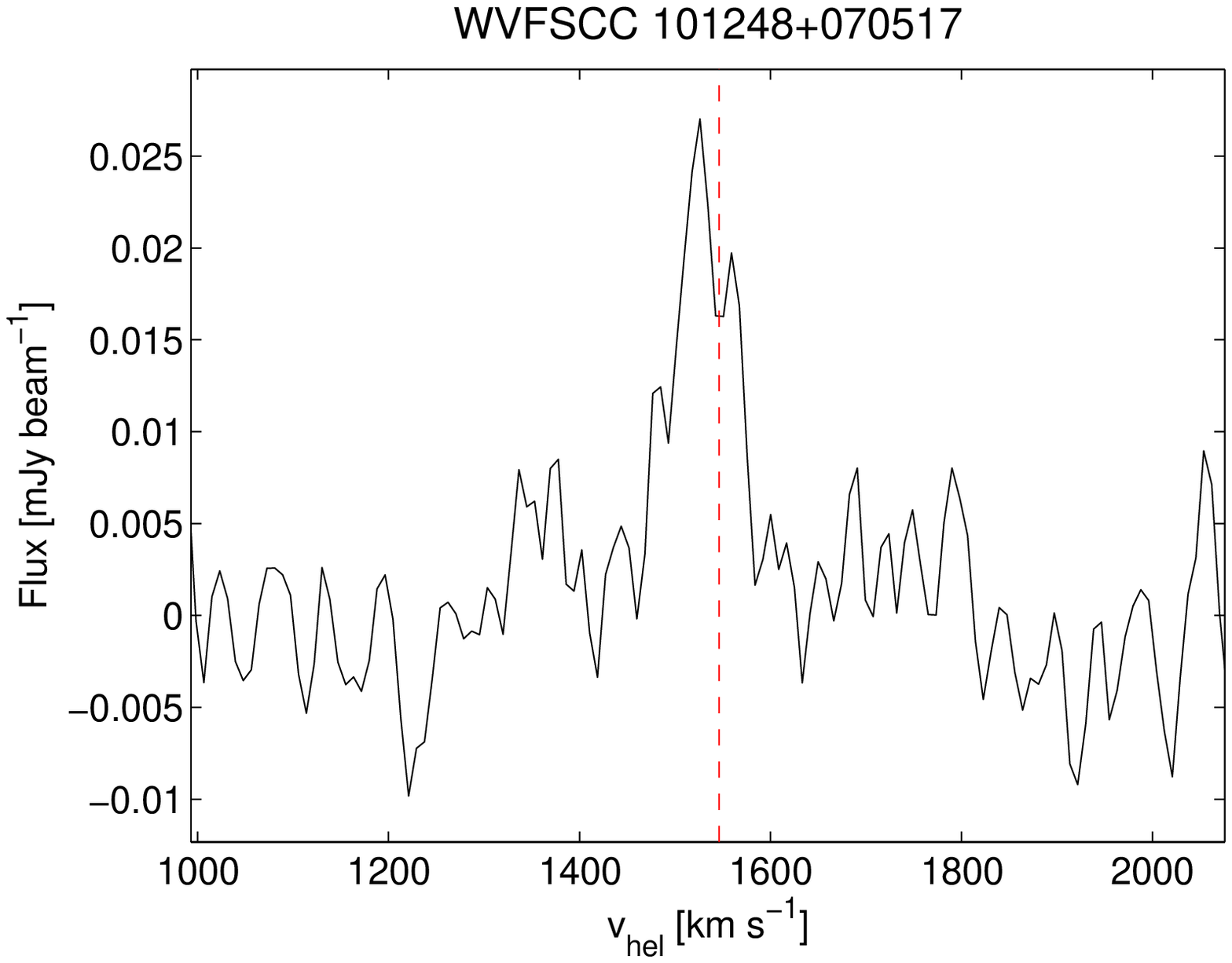}
\includegraphics[width=0.32\textwidth]{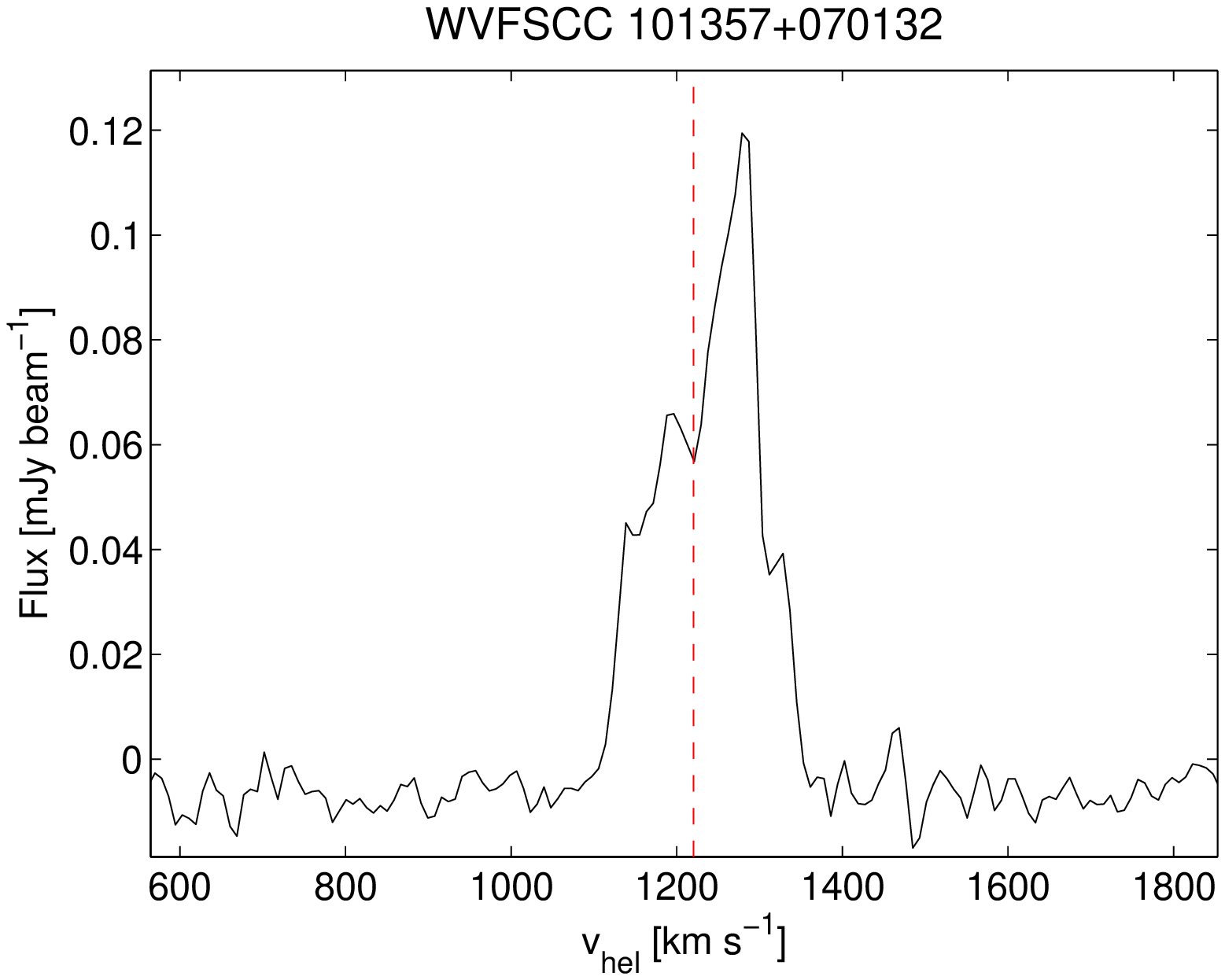}
\includegraphics[width=0.32\textwidth]{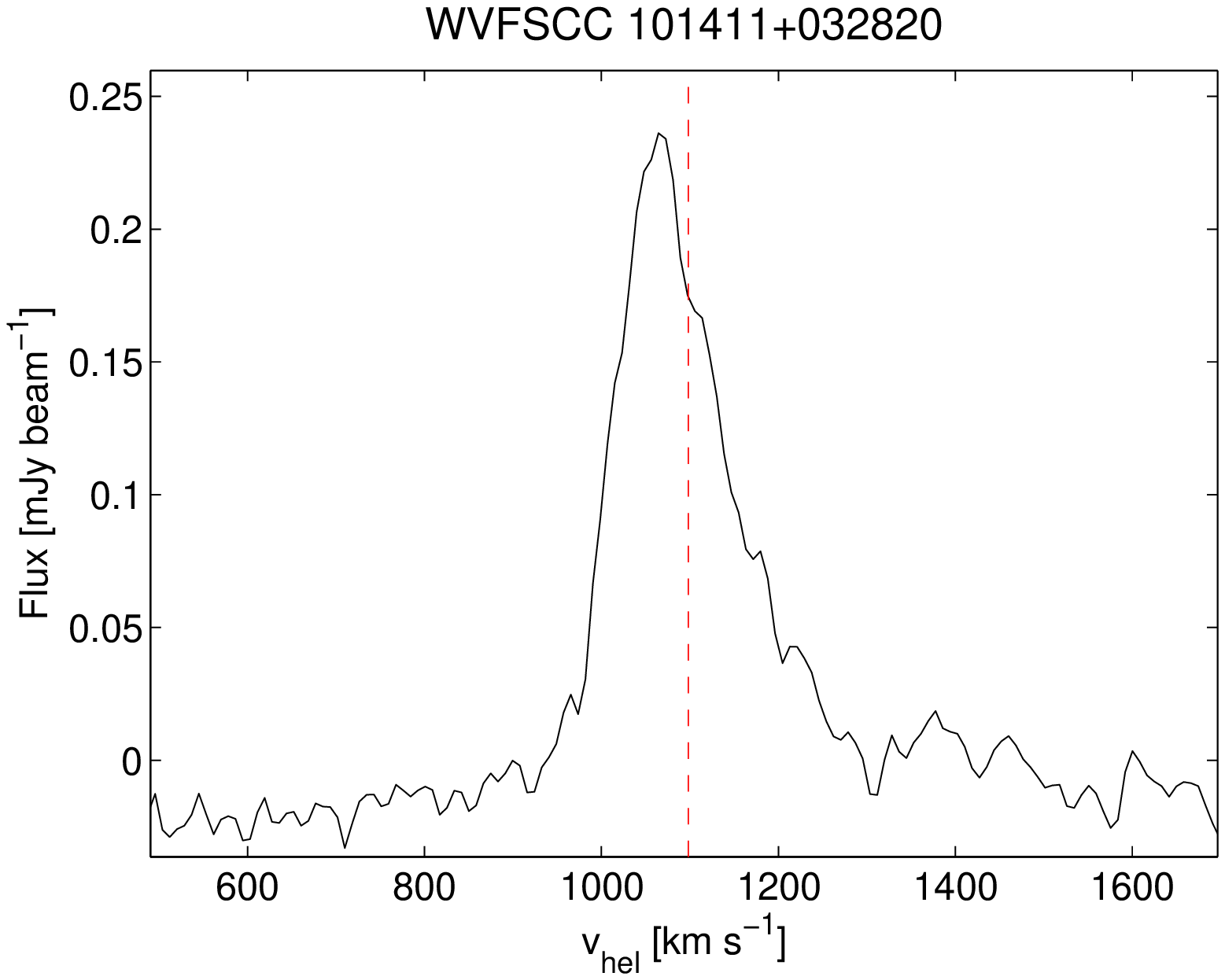}
\includegraphics[width=0.32\textwidth]{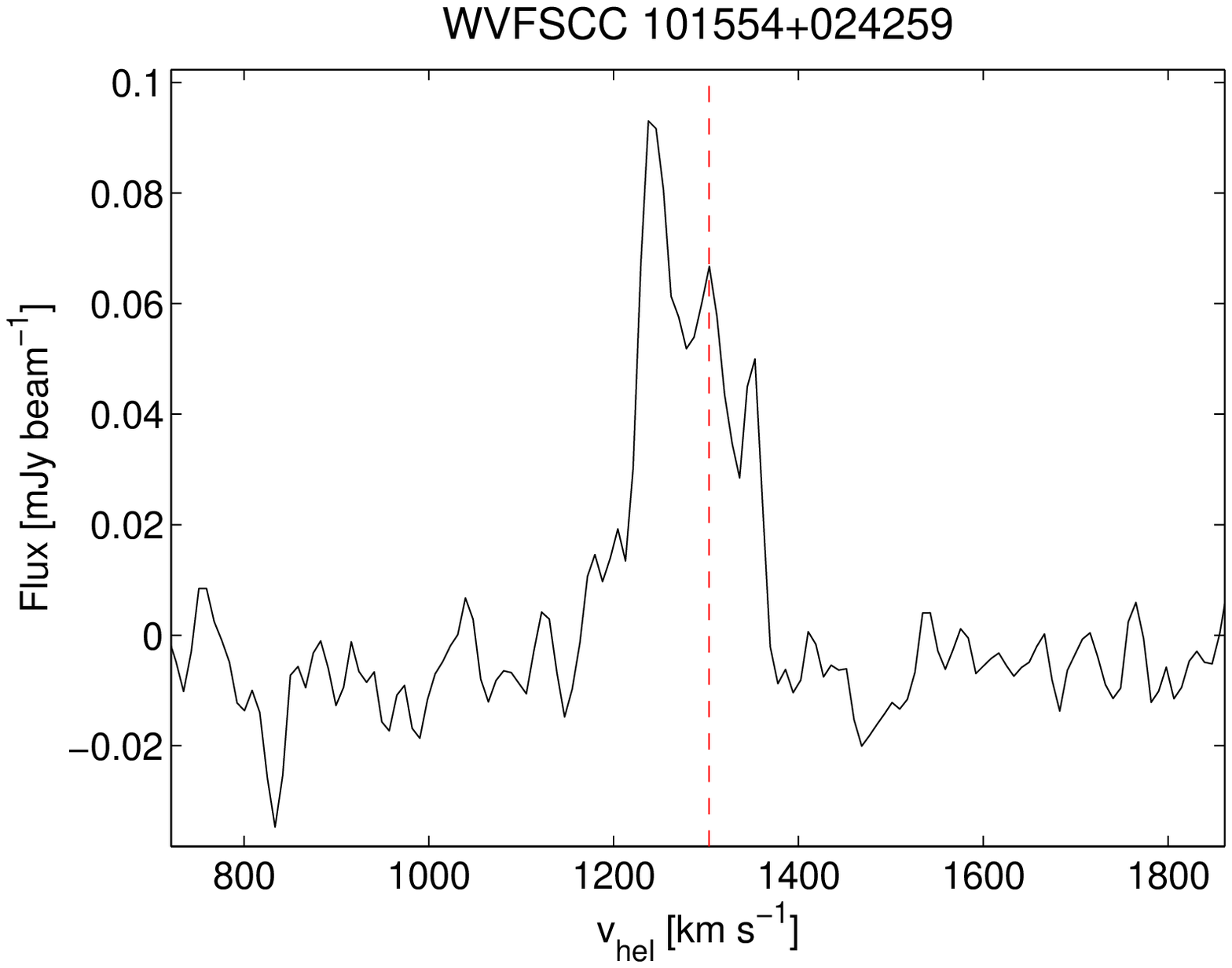}
\includegraphics[width=0.32\textwidth]{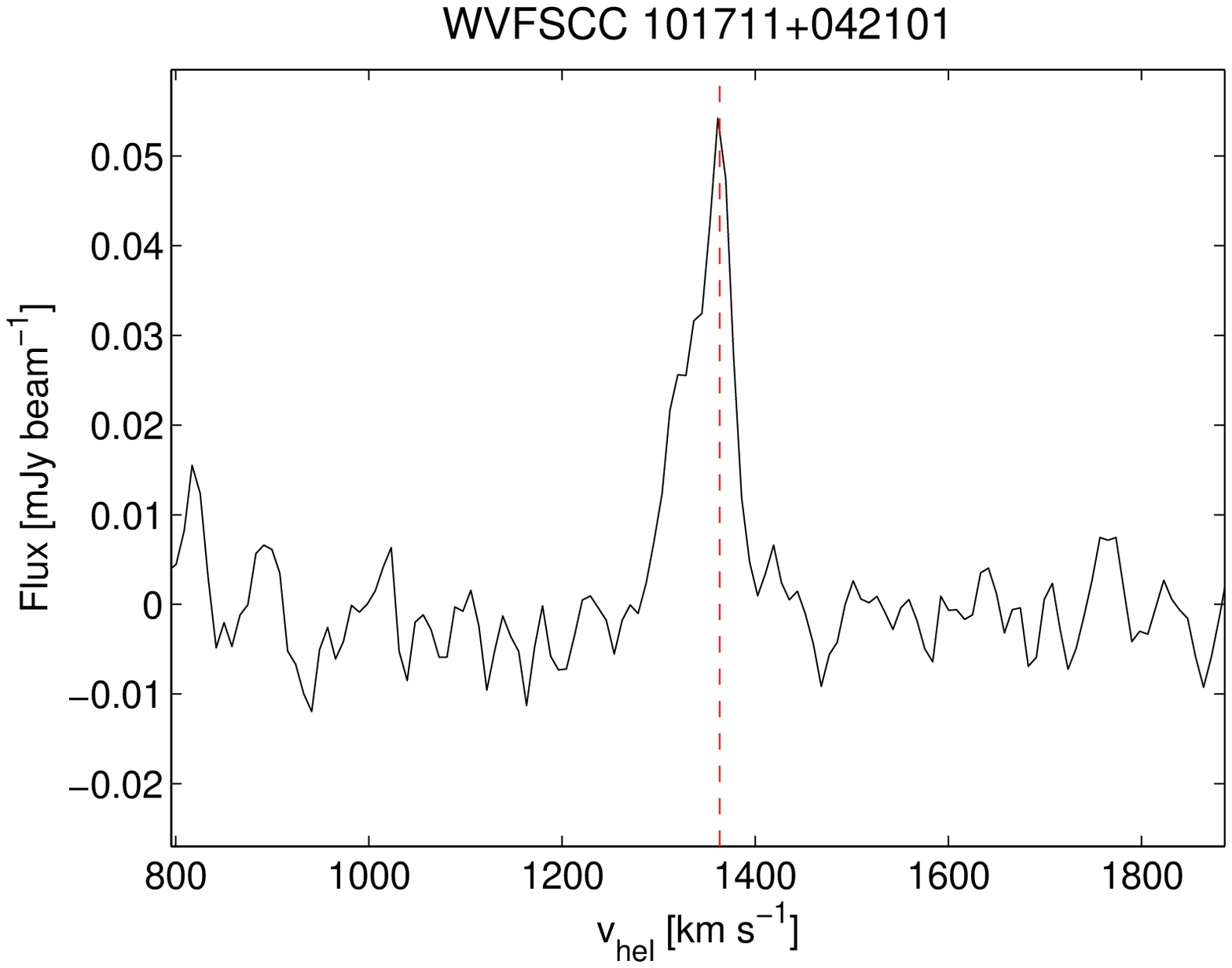}
\includegraphics[width=0.32\textwidth]{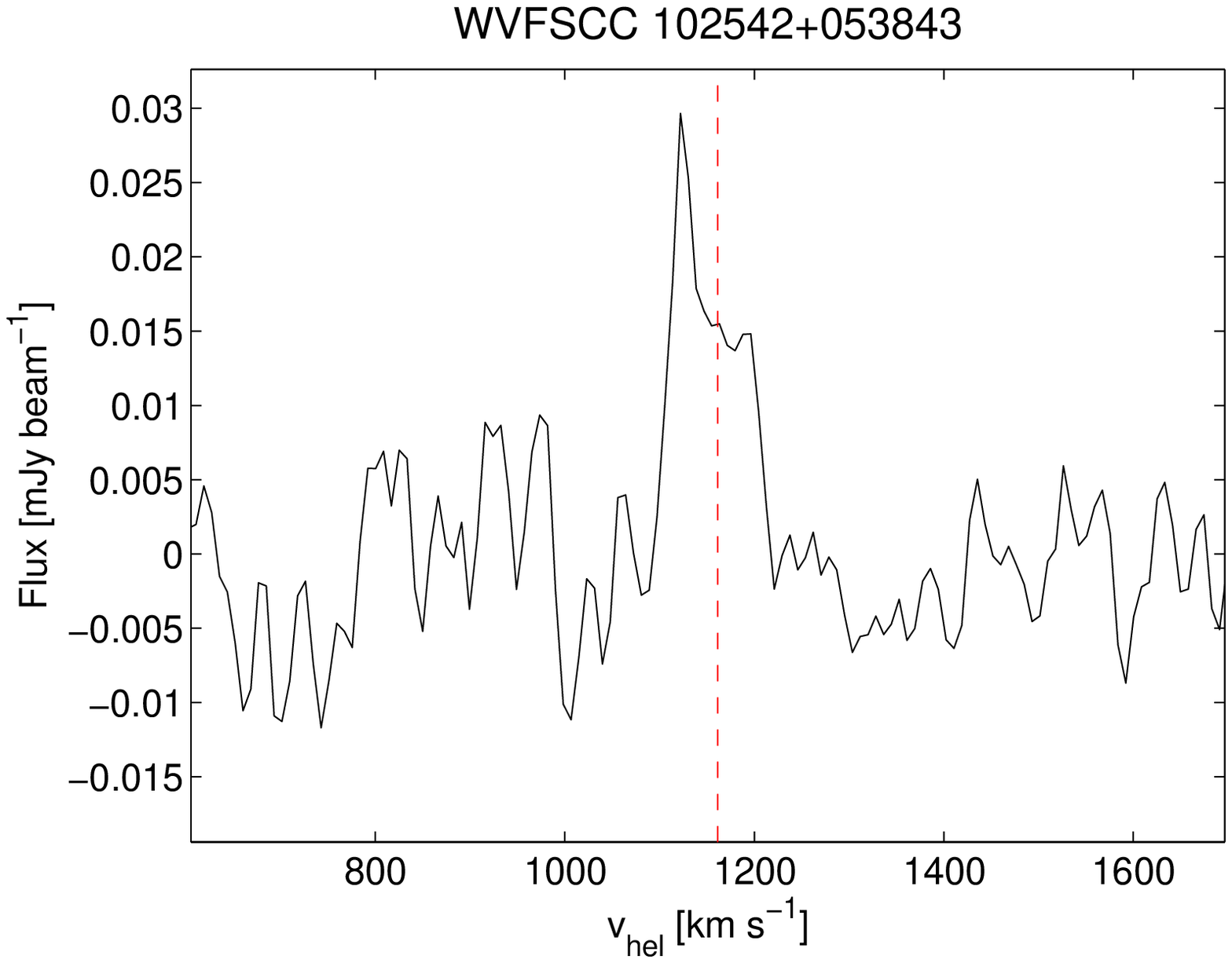}
\includegraphics[width=0.32\textwidth]{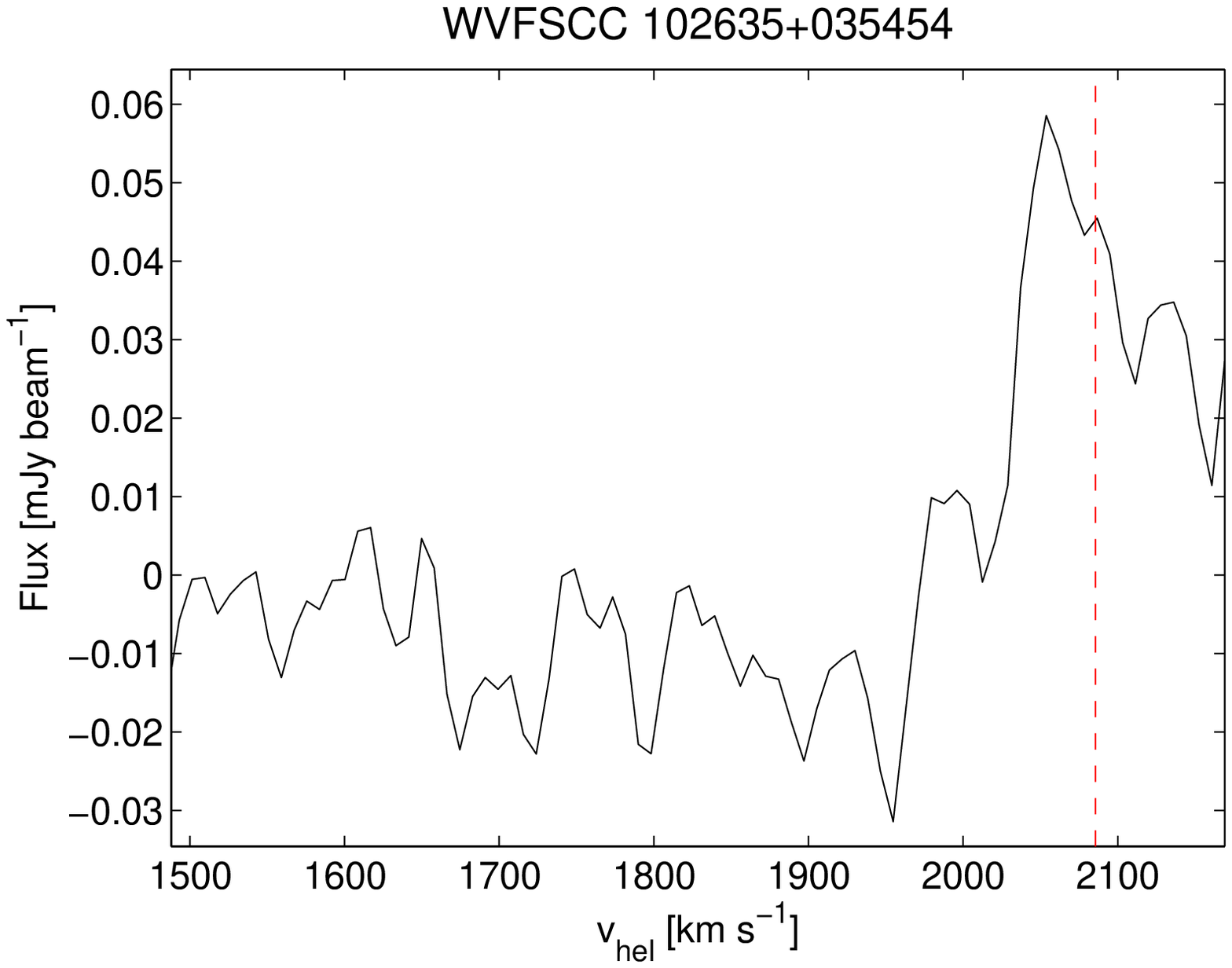}

\end{center}                                            
{\bf Fig~\ref{all_spectra2}.} (continued)                                        
 
\end{figure*}


\begin{figure*}
  \begin{center}

\includegraphics[width=0.32\textwidth]{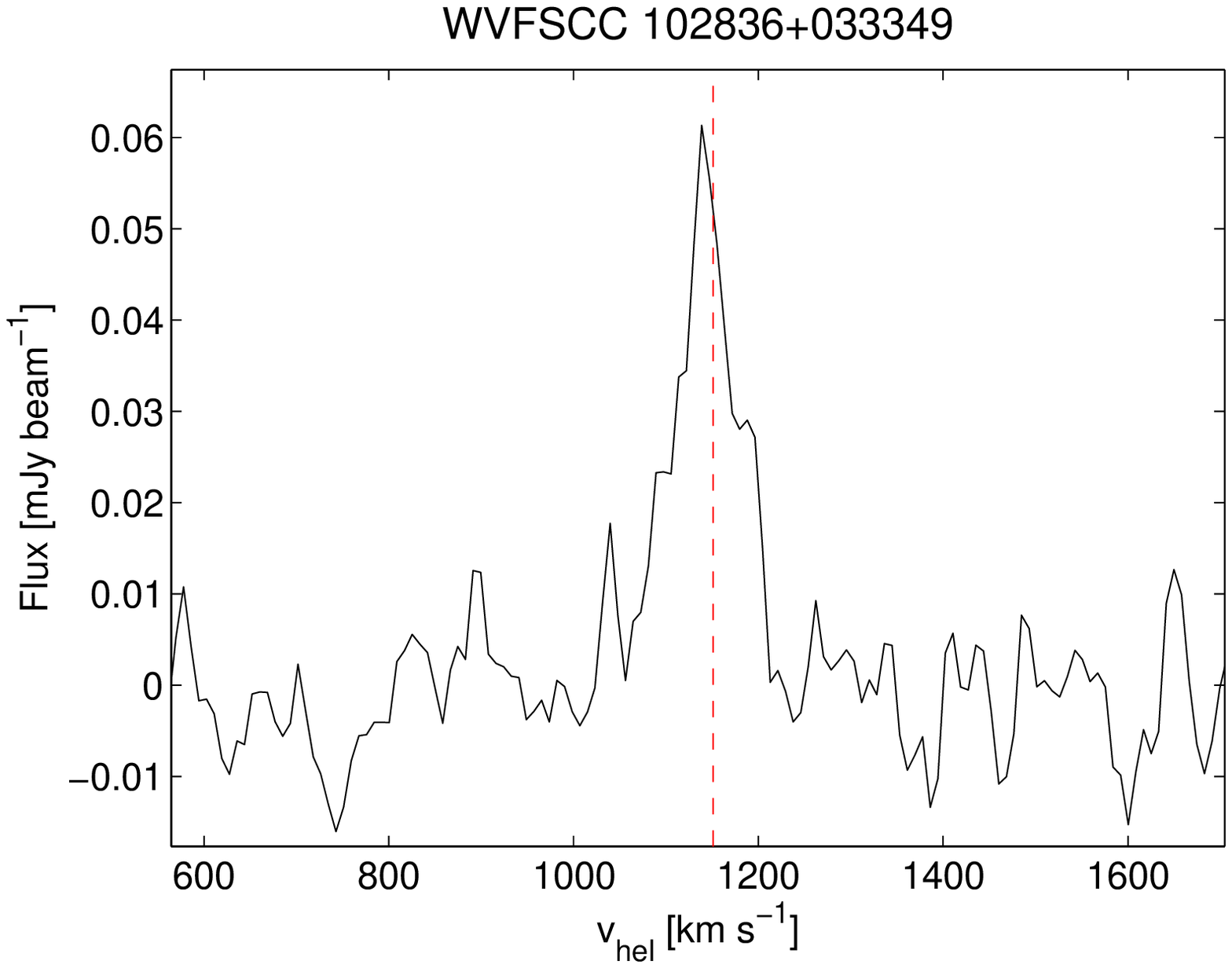}
\includegraphics[width=0.32\textwidth]{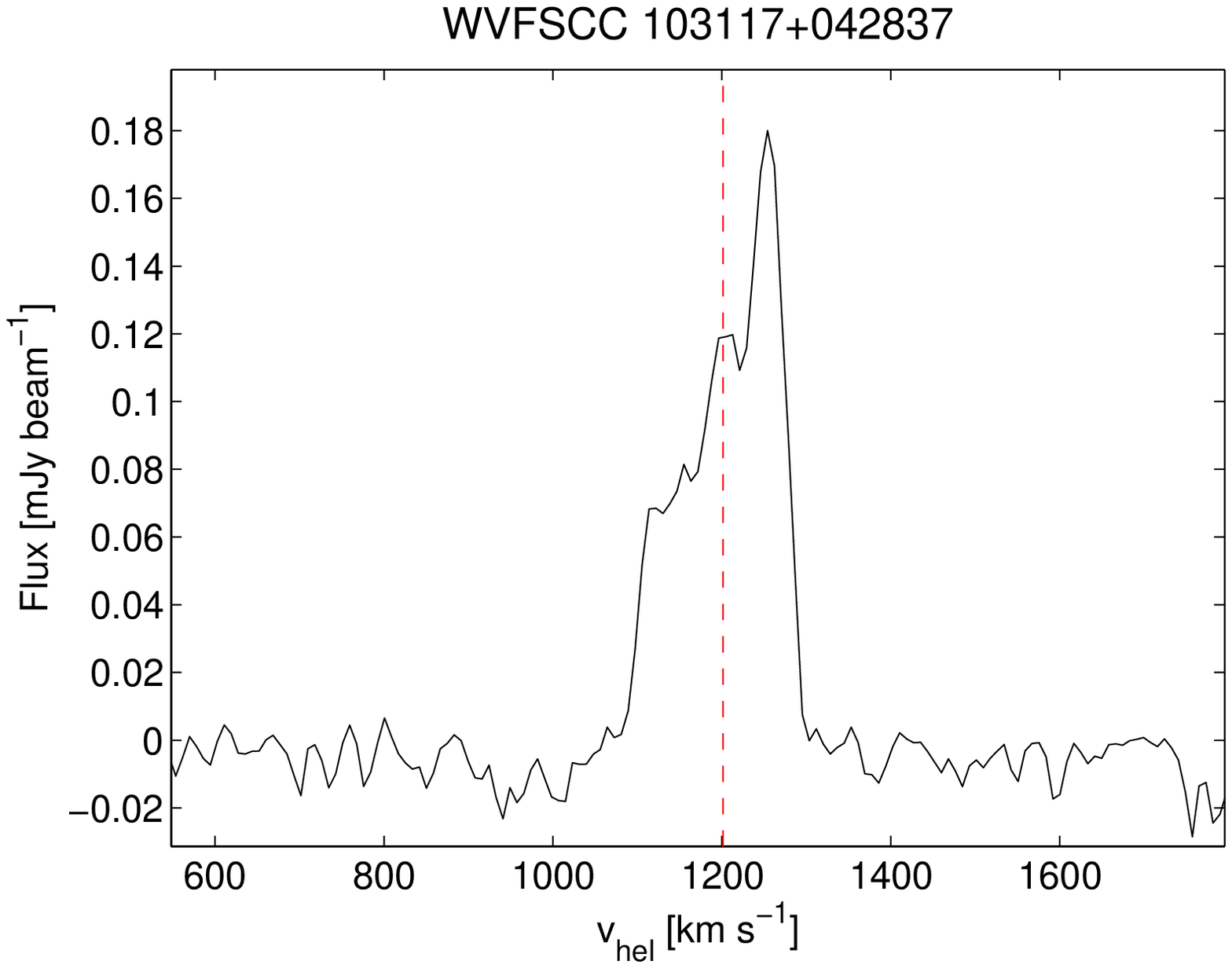}
\includegraphics[width=0.32\textwidth]{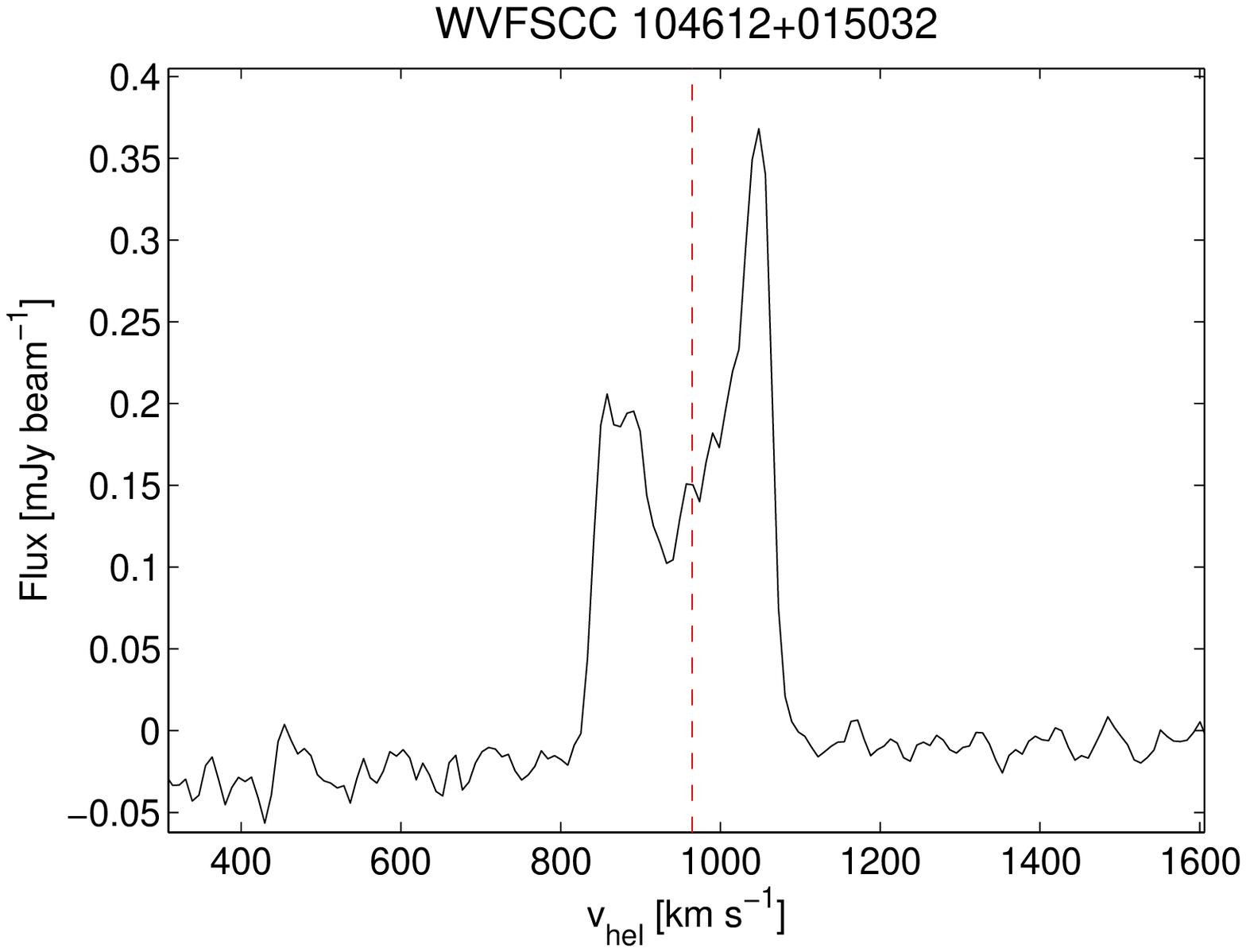}
\includegraphics[width=0.32\textwidth]{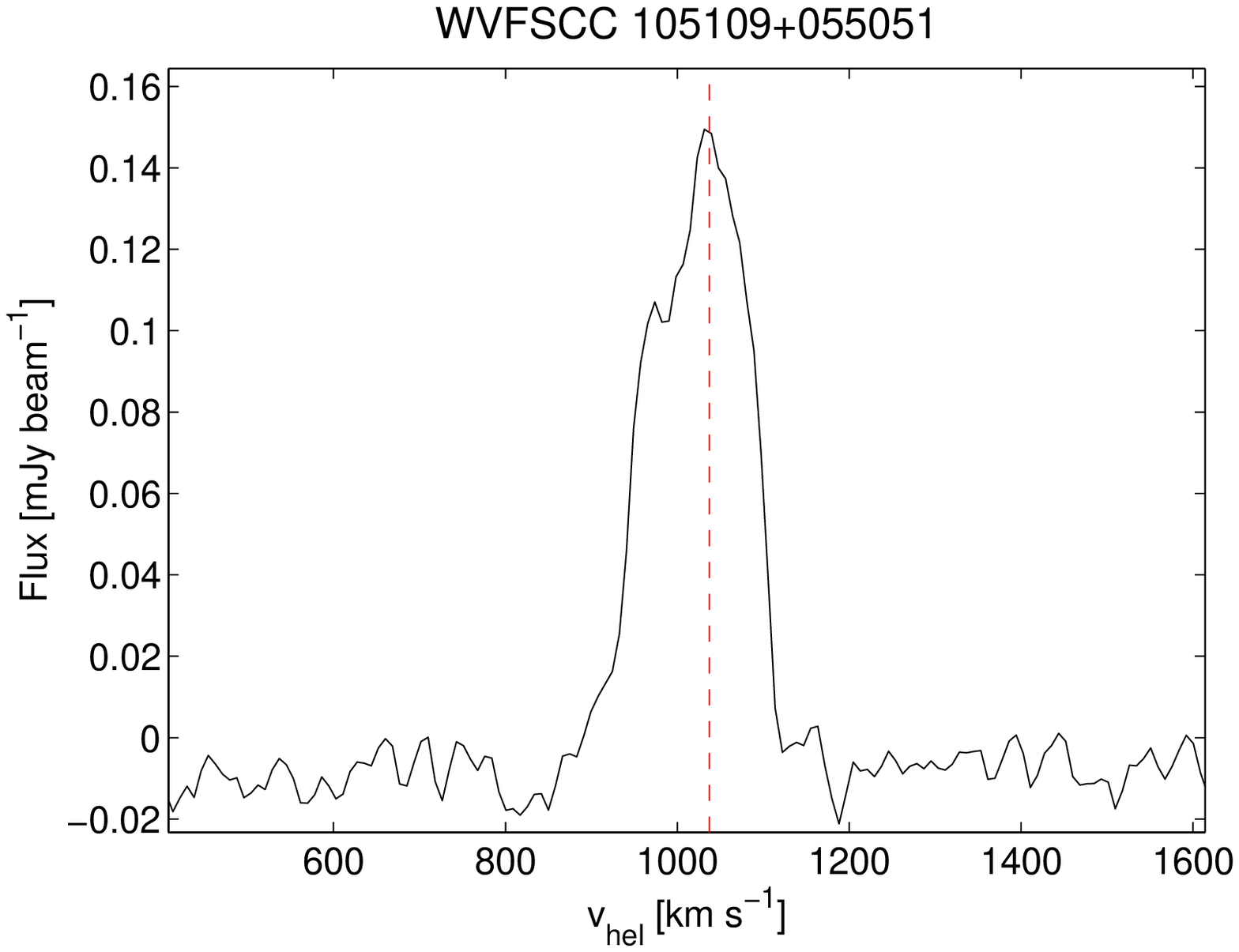}
\includegraphics[width=0.32\textwidth]{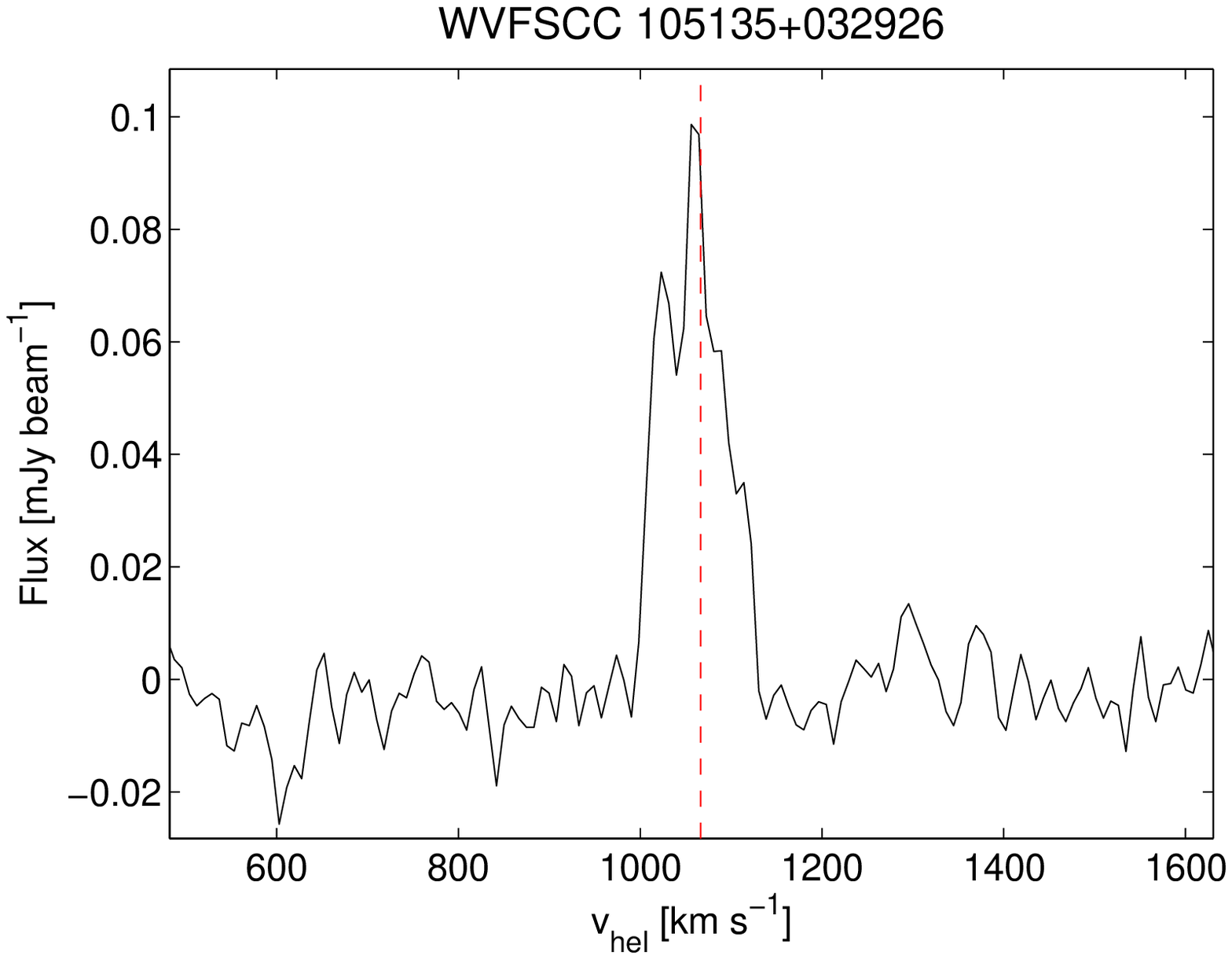}
\includegraphics[width=0.32\textwidth]{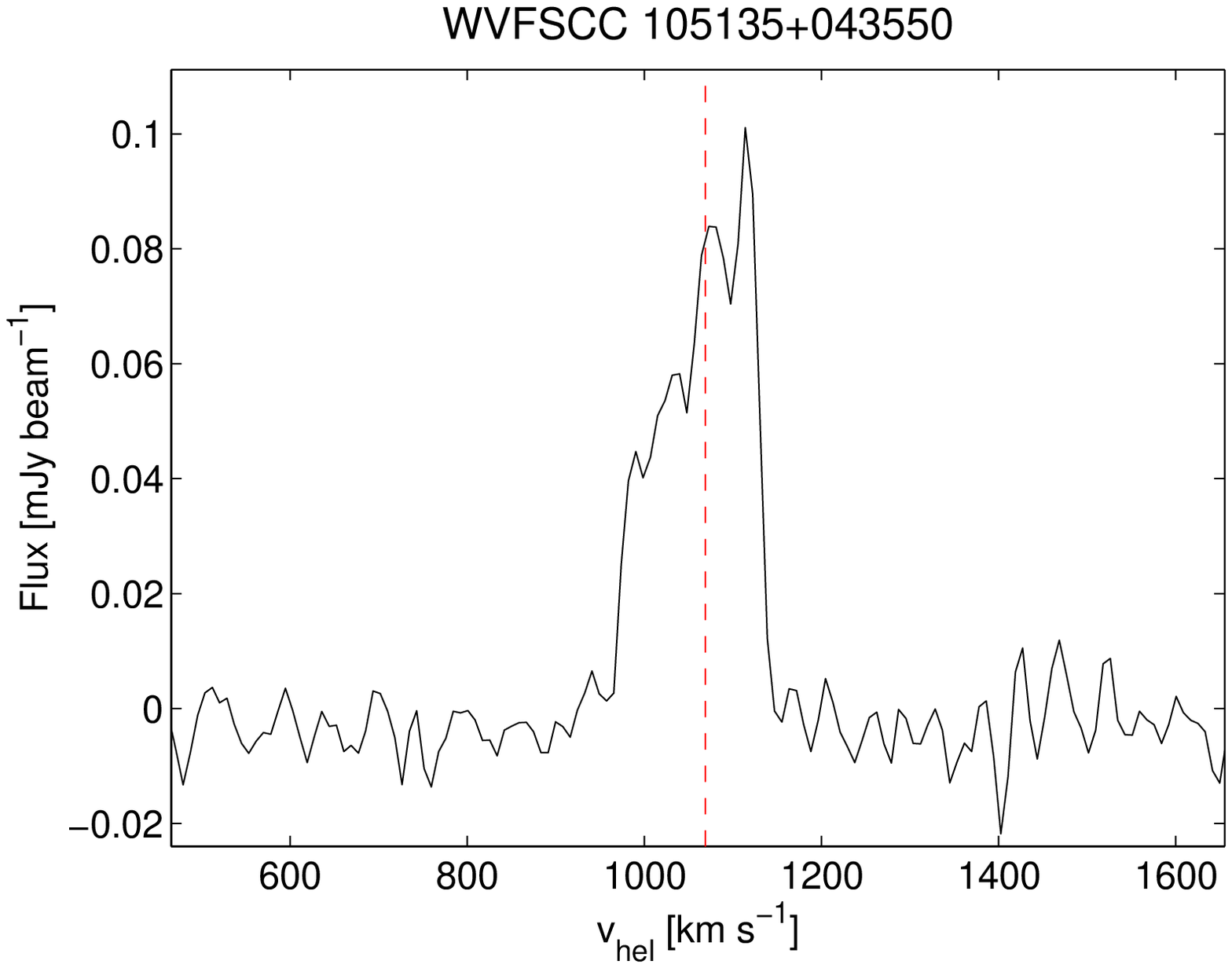}
\includegraphics[width=0.32\textwidth]{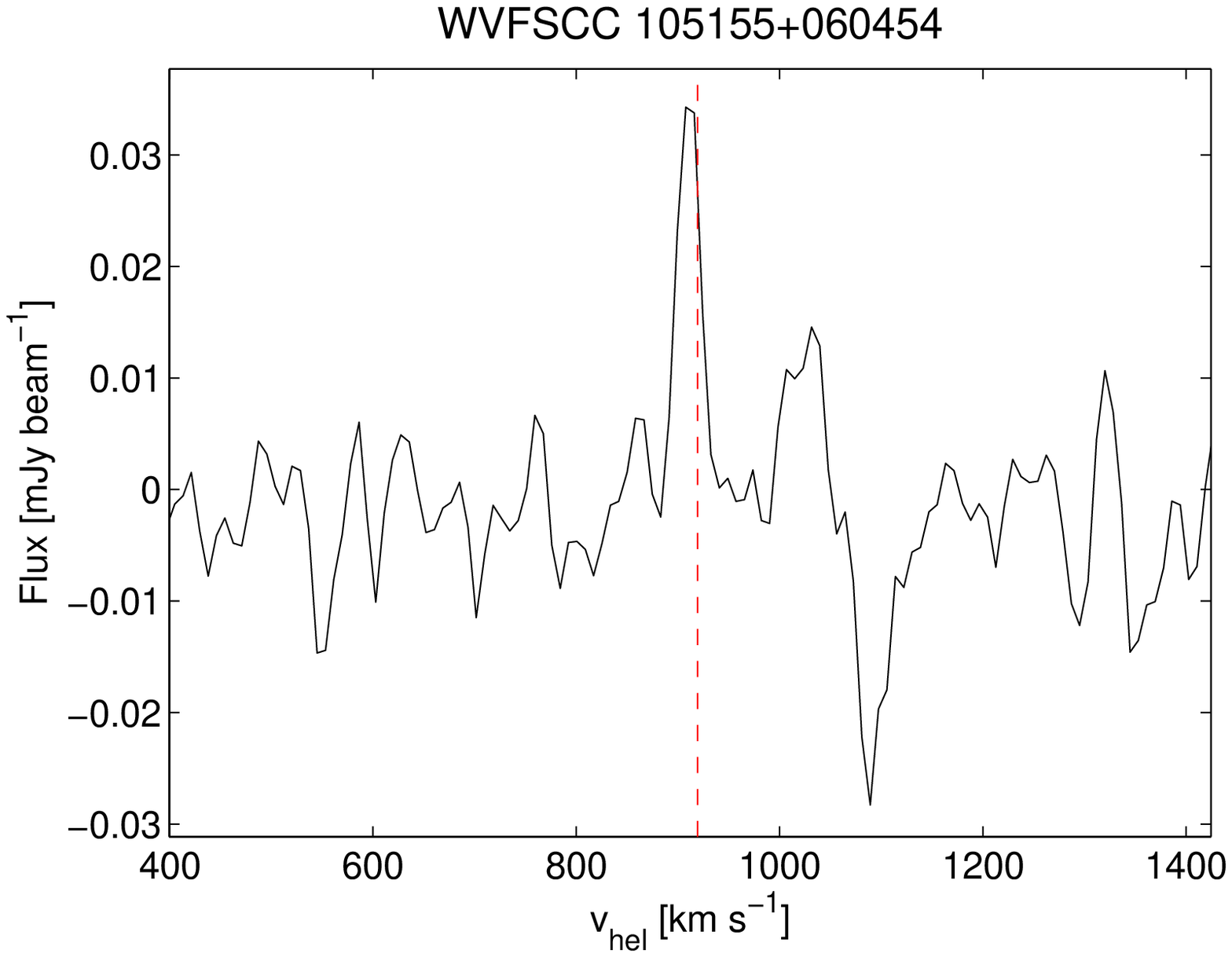}
\includegraphics[width=0.32\textwidth]{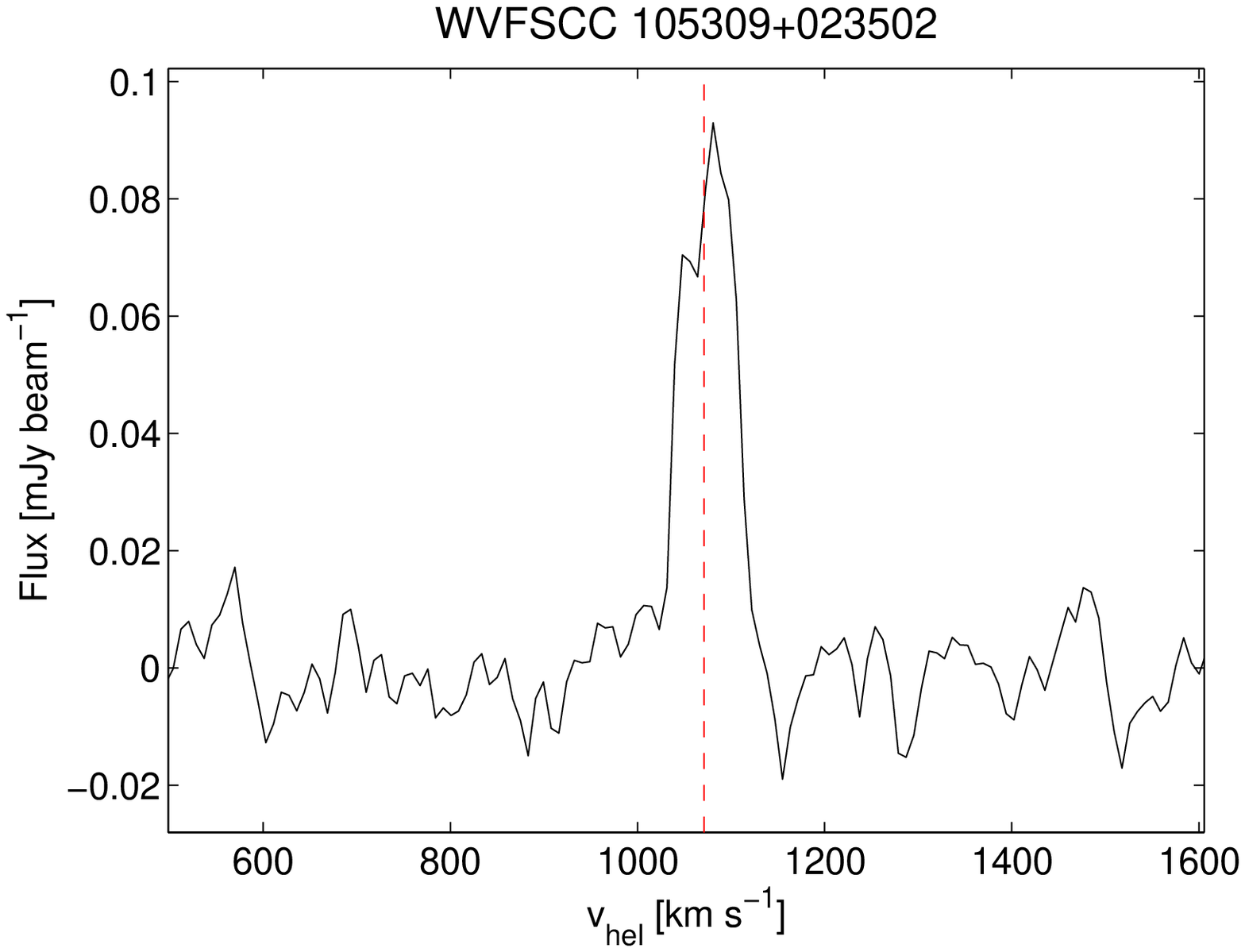}
\includegraphics[width=0.32\textwidth]{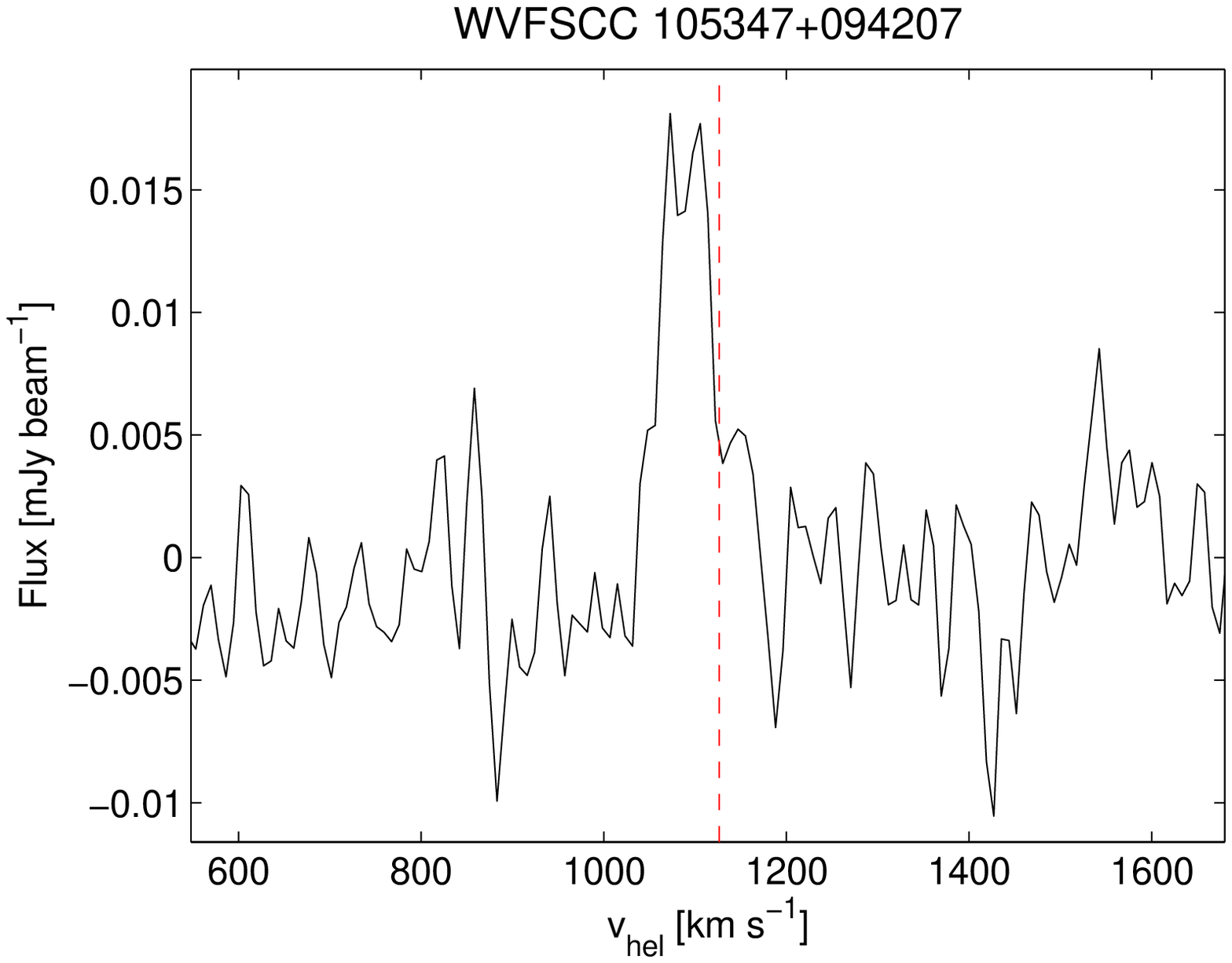}
\includegraphics[width=0.32\textwidth]{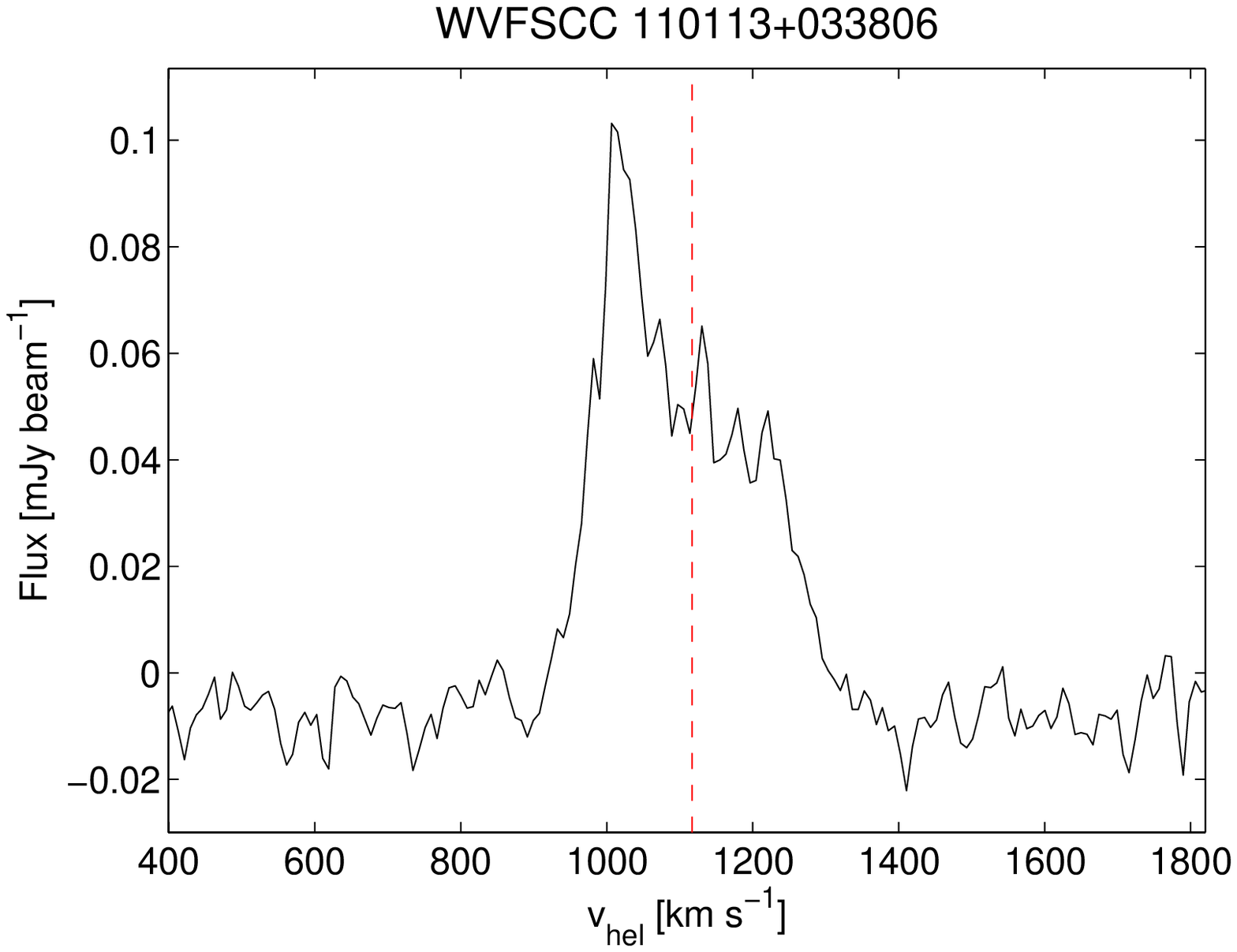}
\includegraphics[width=0.32\textwidth]{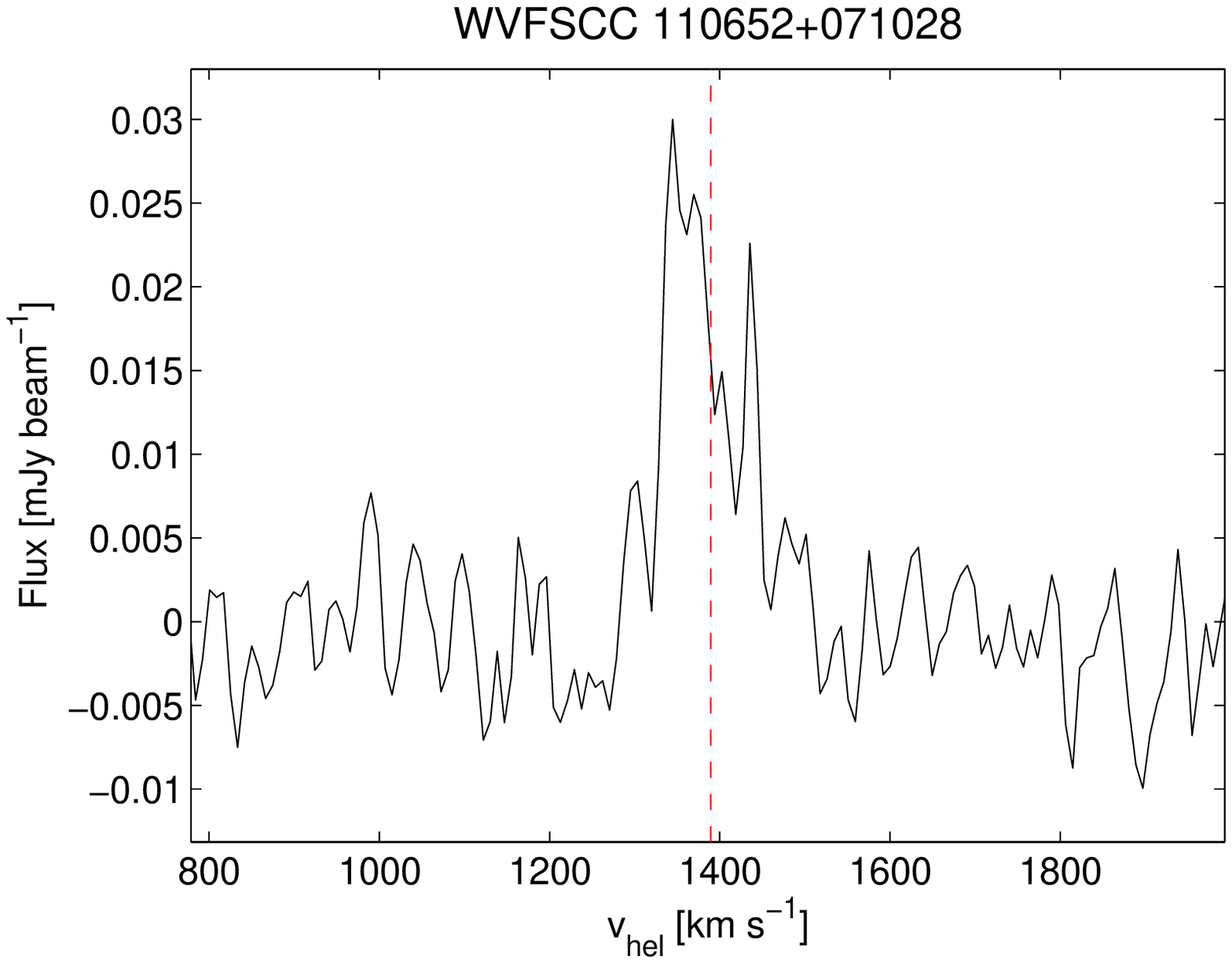}
\includegraphics[width=0.32\textwidth]{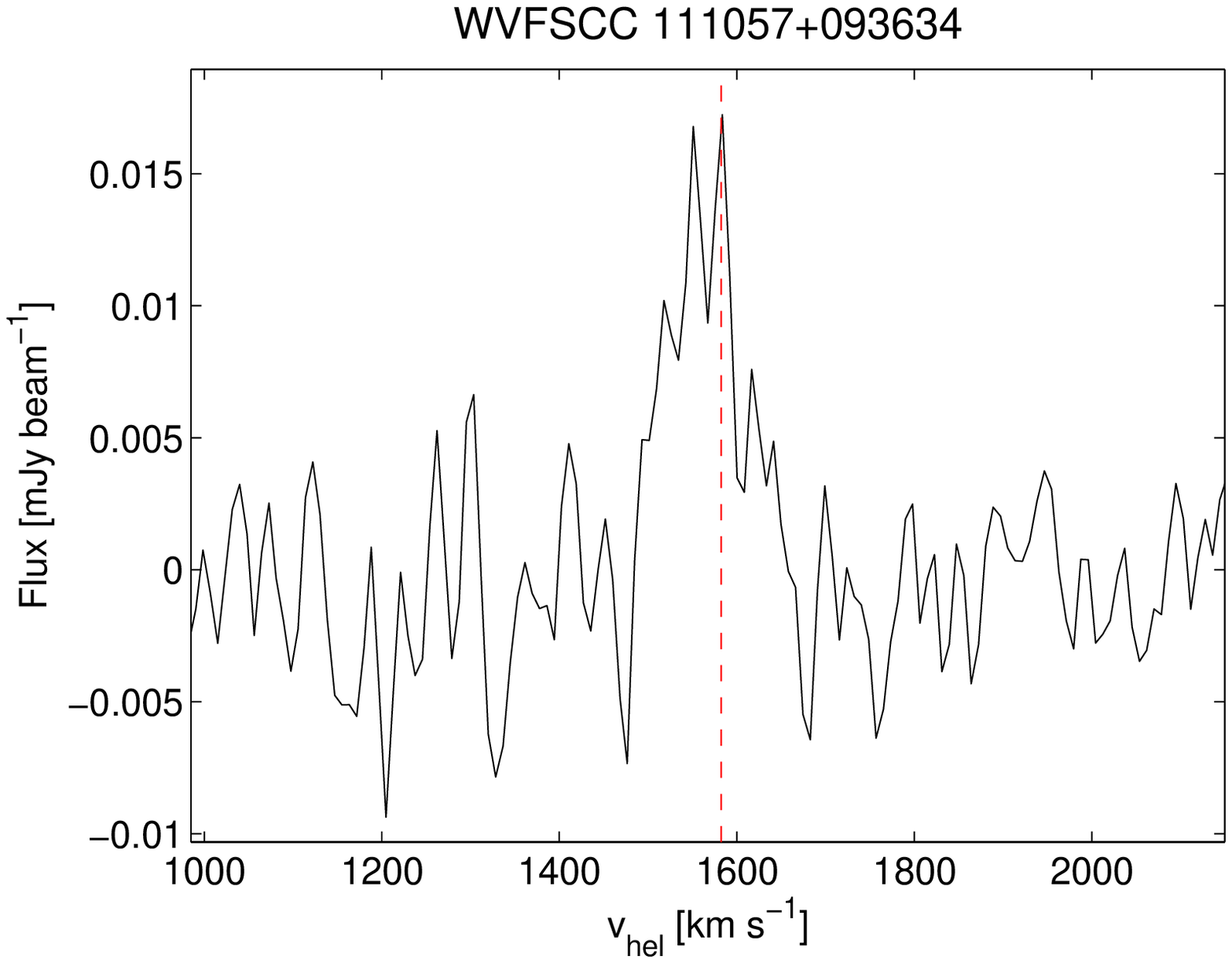}
\includegraphics[width=0.32\textwidth]{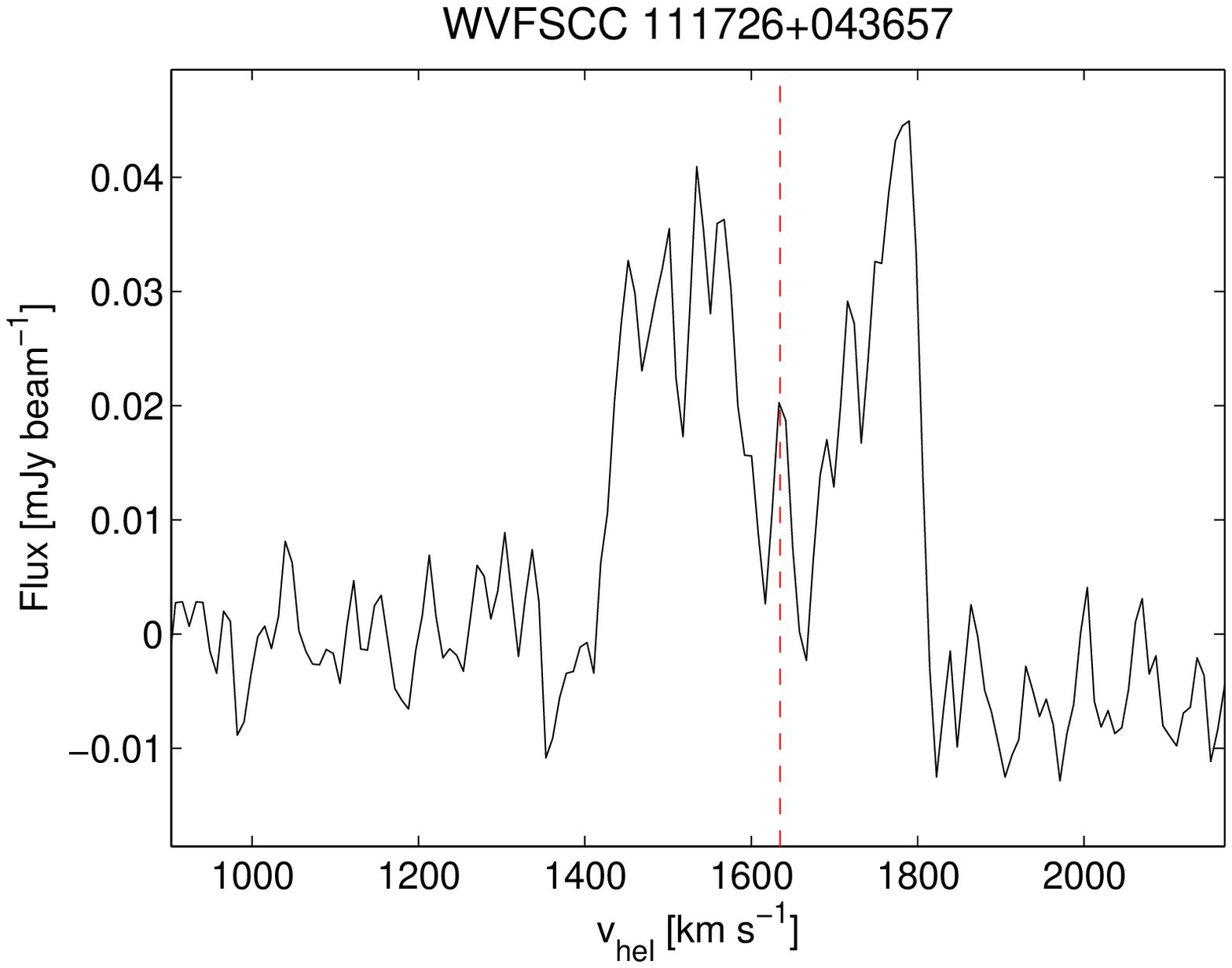}
\includegraphics[width=0.32\textwidth]{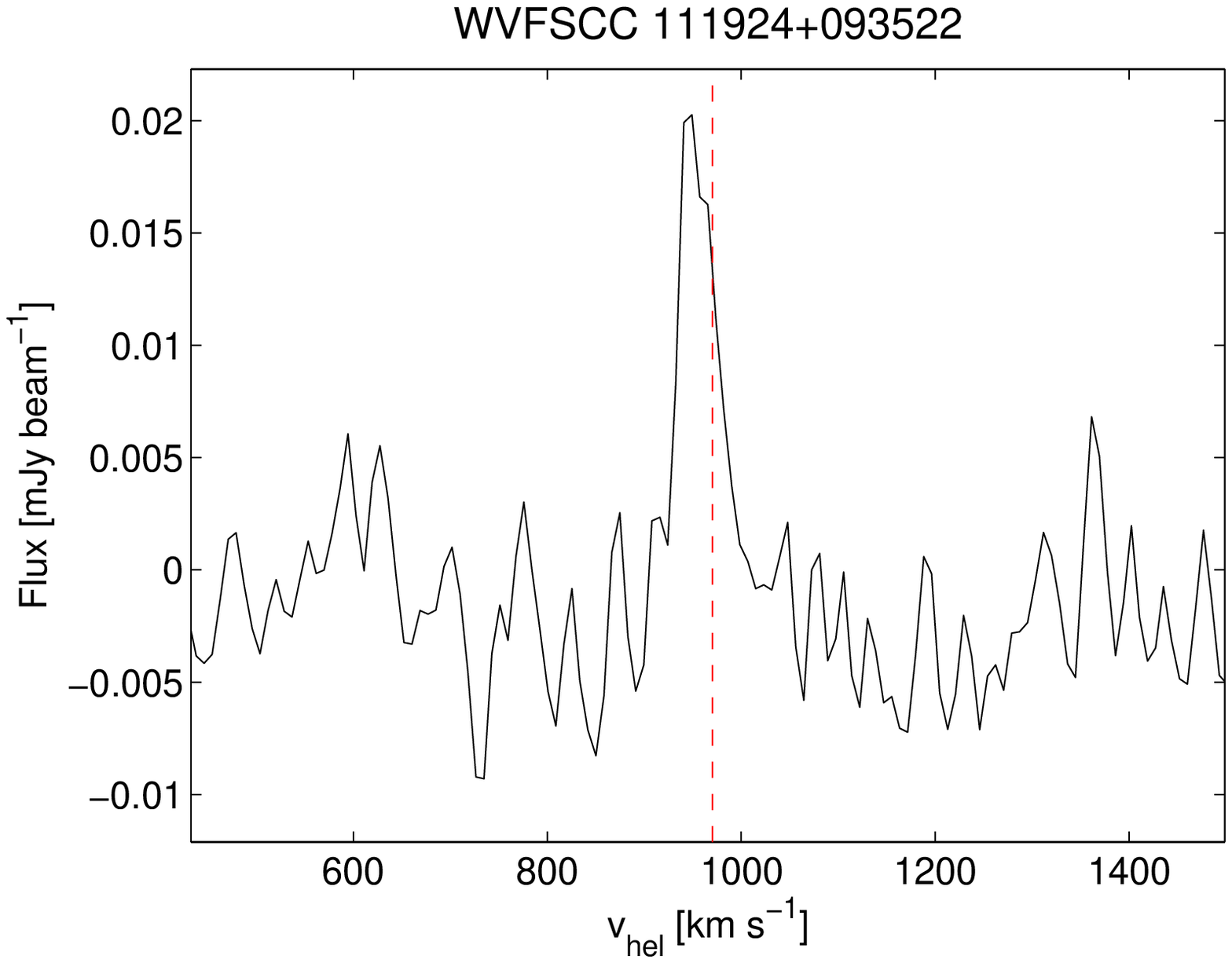}
\includegraphics[width=0.32\textwidth]{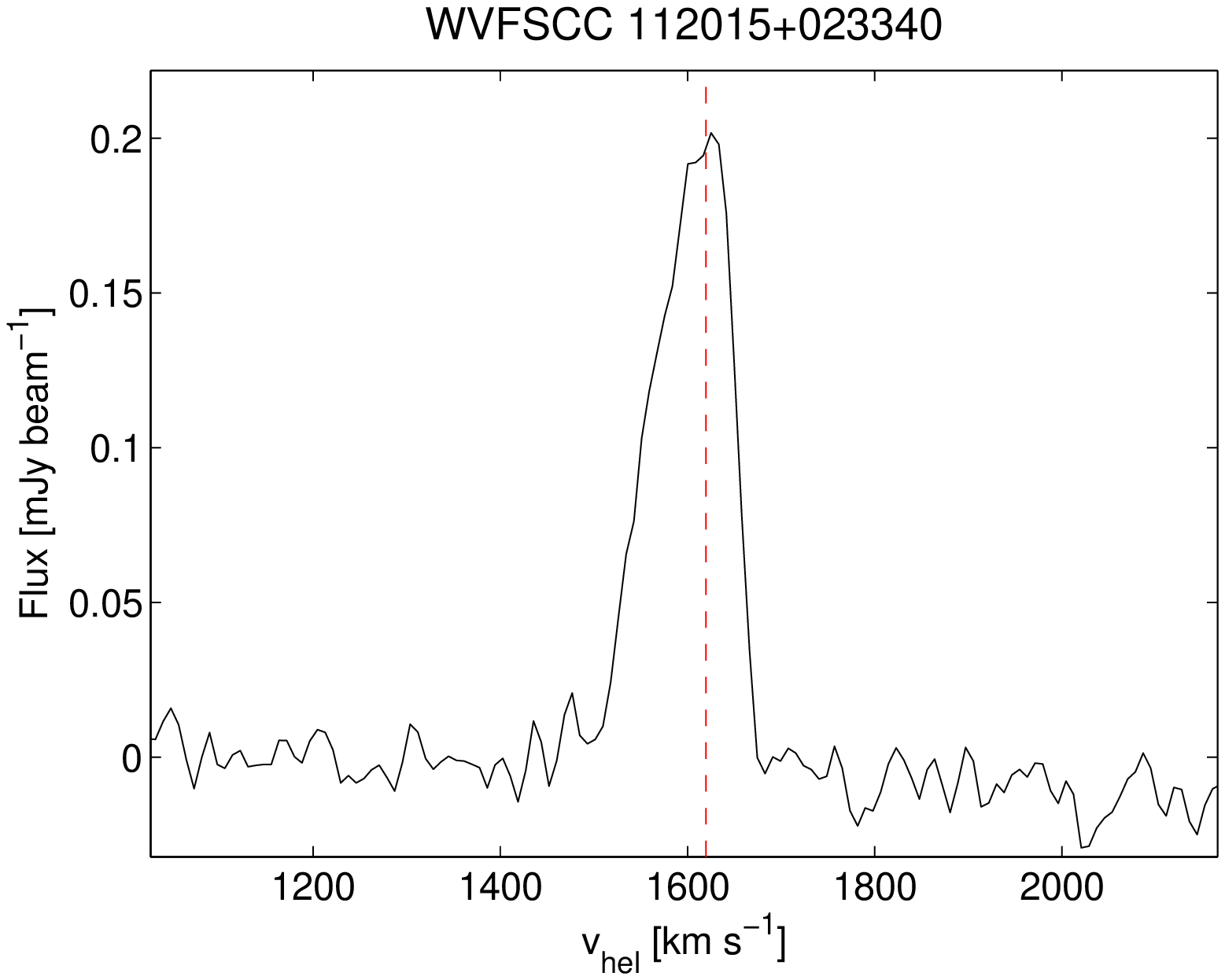}

\end{center}                                            
{\bf Fig~\ref{all_spectra2}.} (continued)                                        
 
\end{figure*}


\begin{figure*}
  \begin{center}

\includegraphics[width=0.32\textwidth]{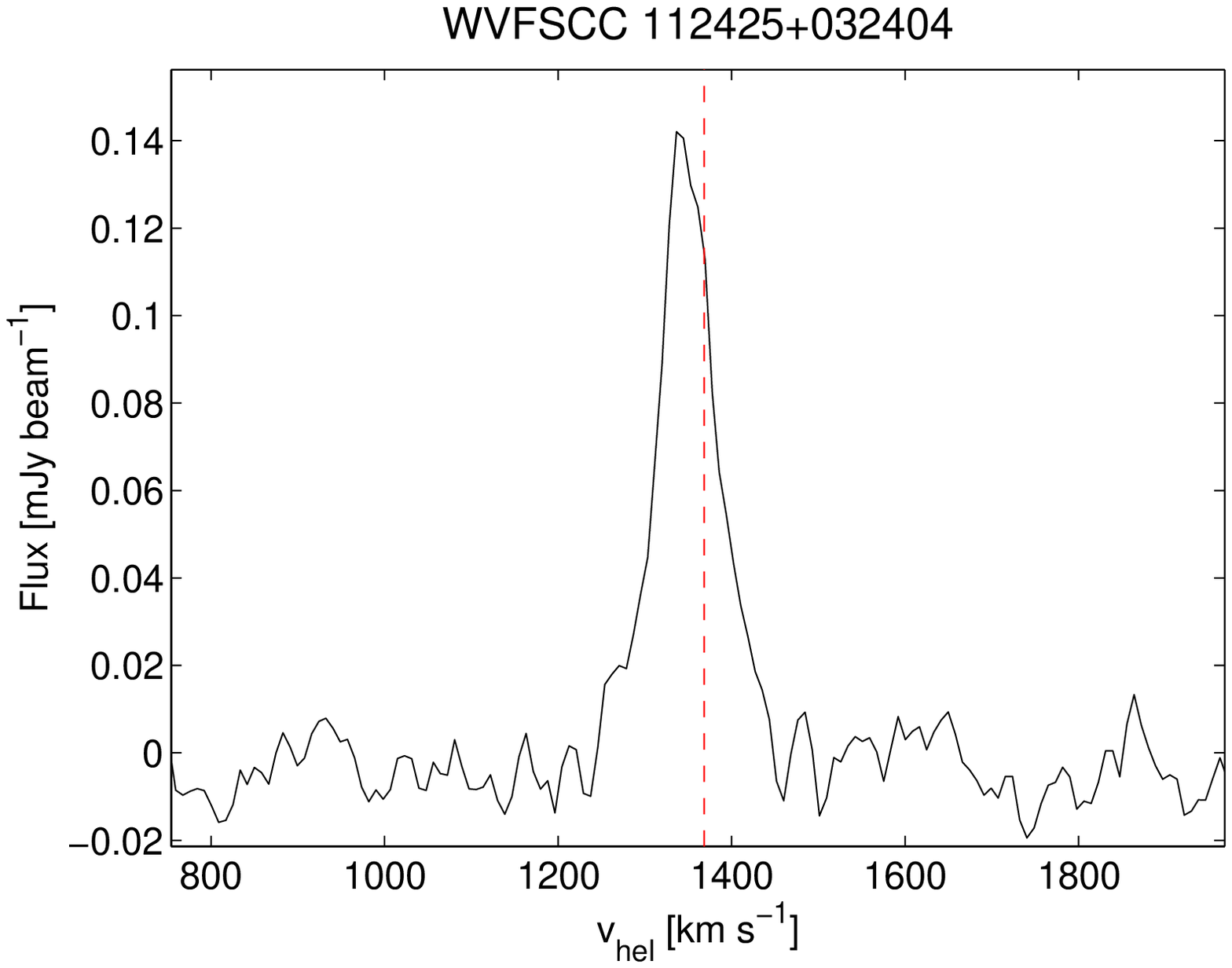}
\includegraphics[width=0.32\textwidth]{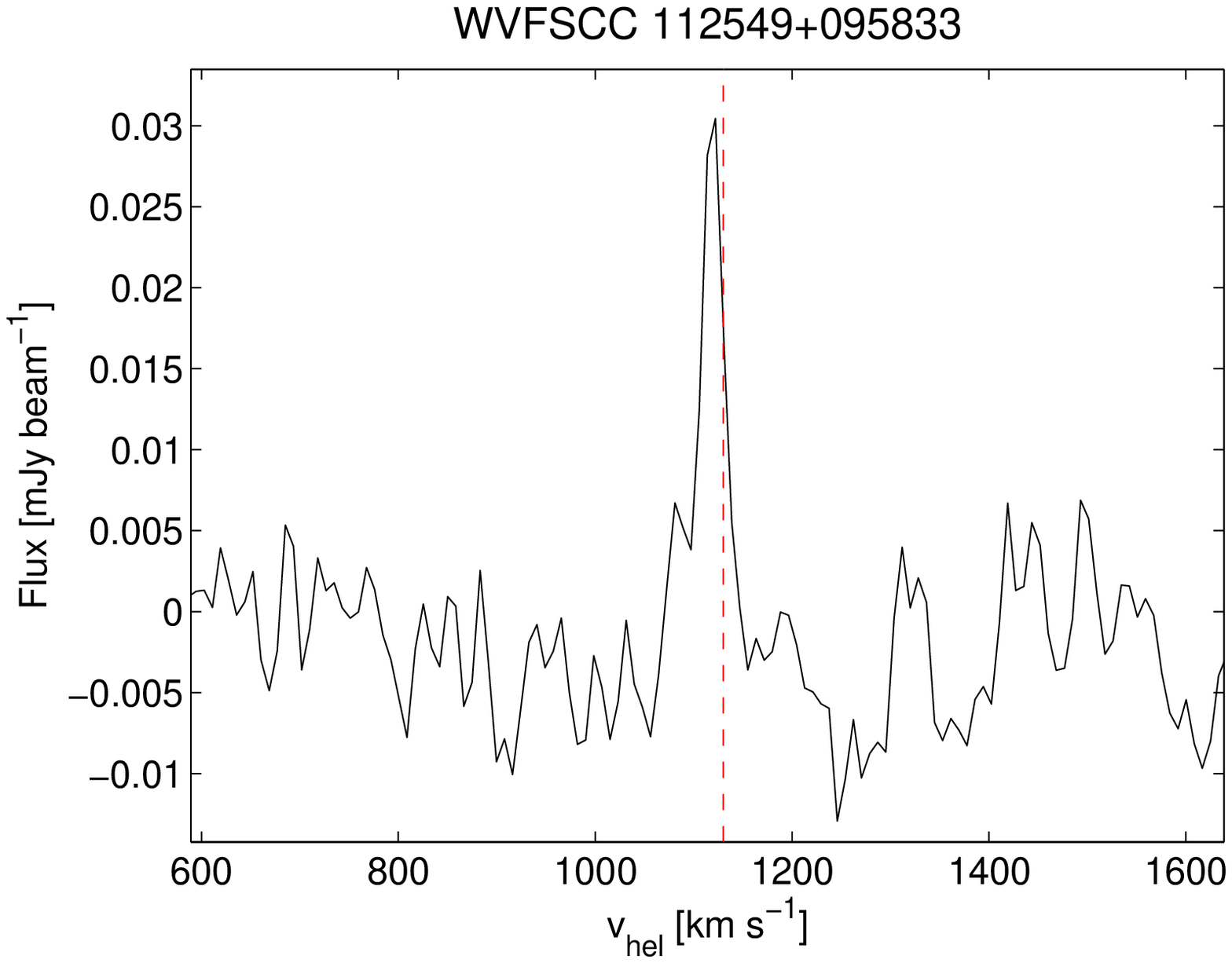}
\includegraphics[width=0.32\textwidth]{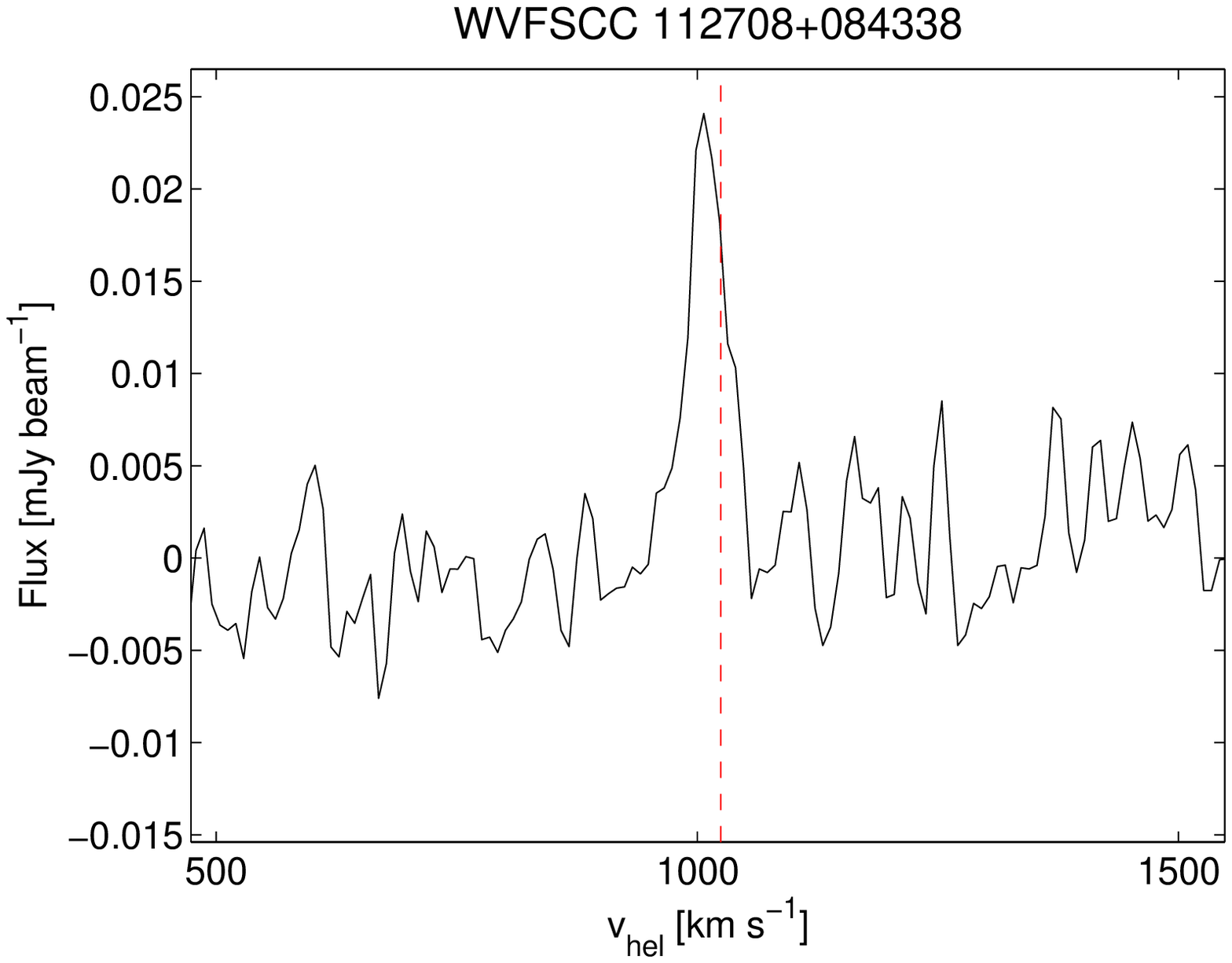}
\includegraphics[width=0.32\textwidth]{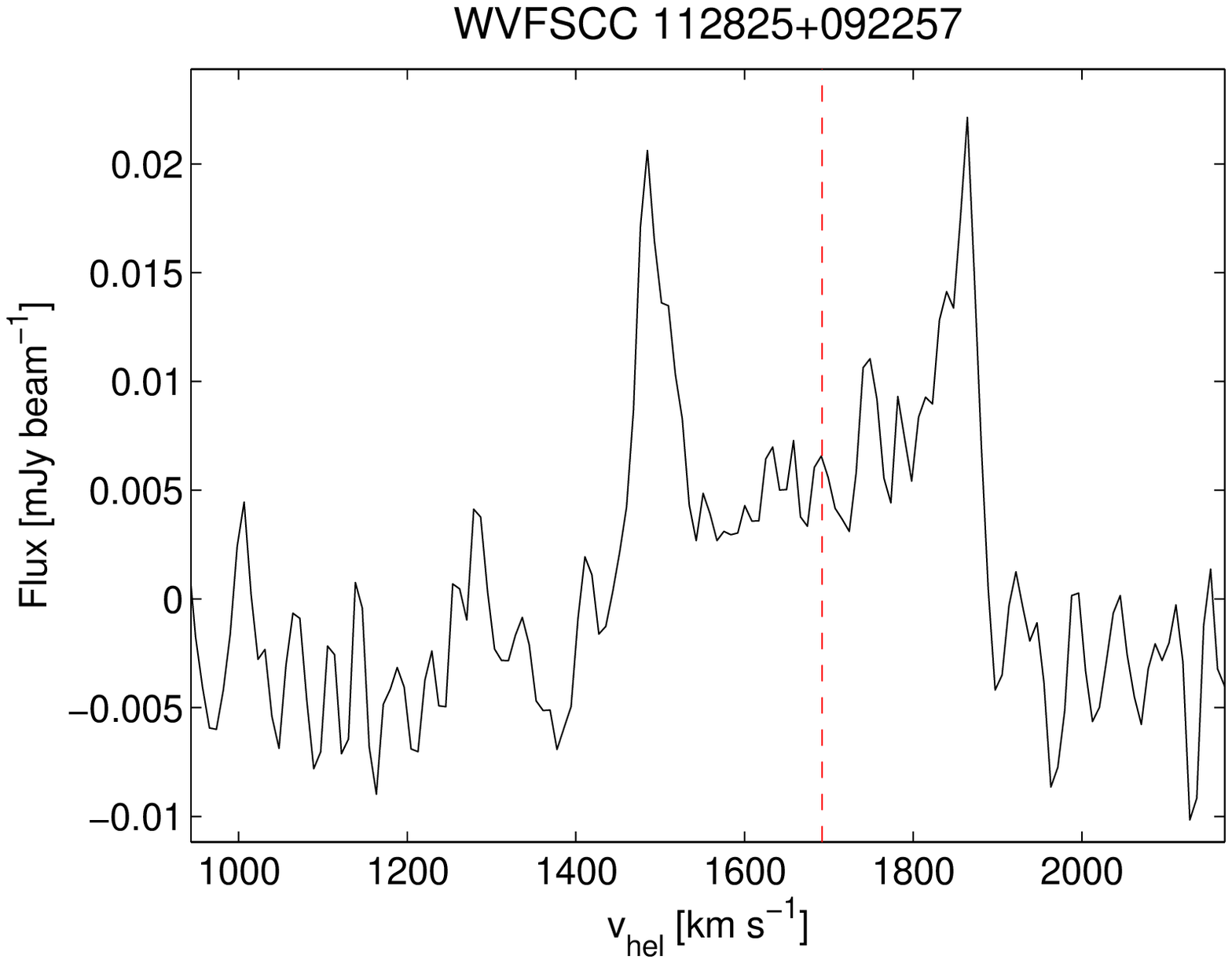}
\includegraphics[width=0.32\textwidth]{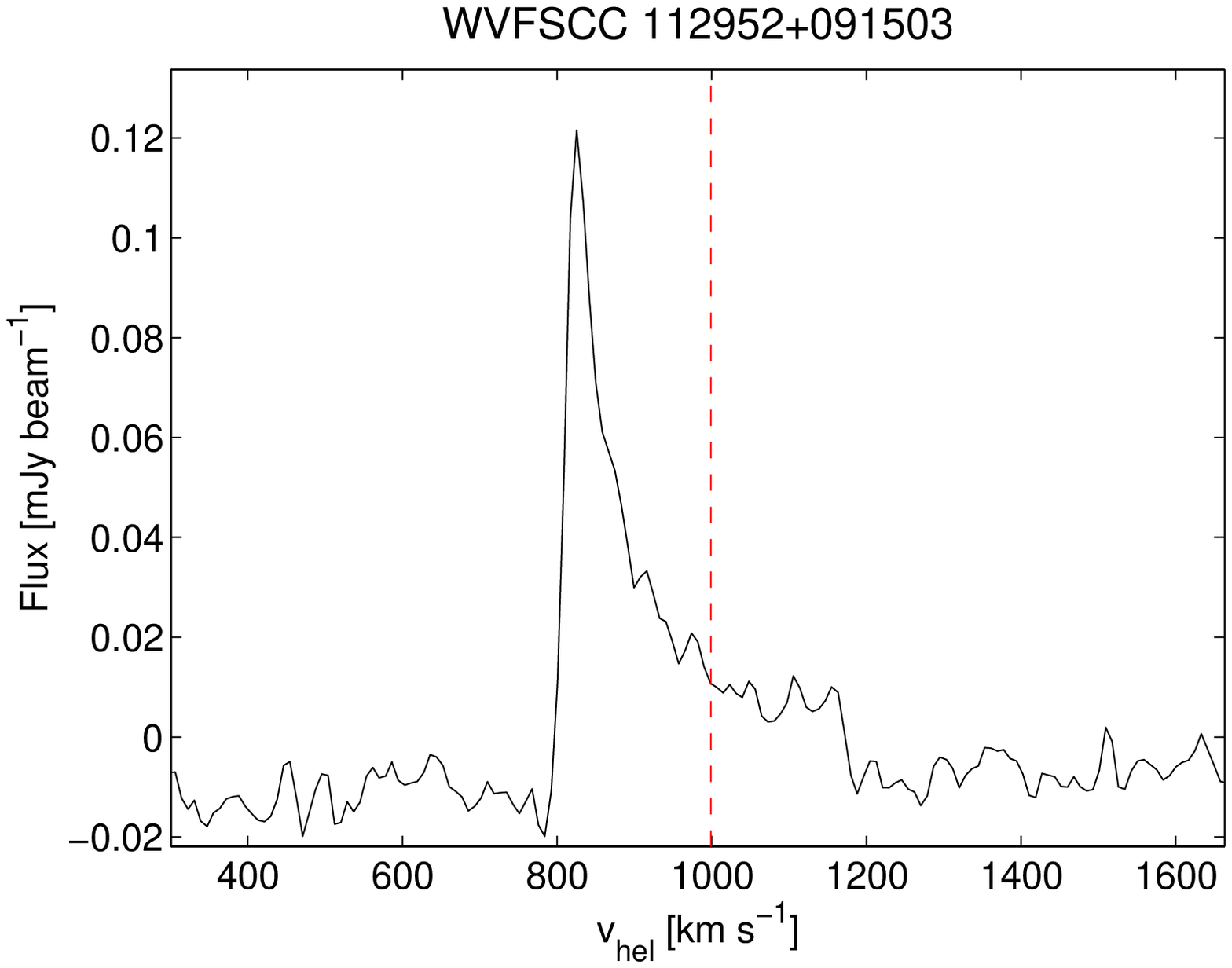}
\includegraphics[width=0.32\textwidth]{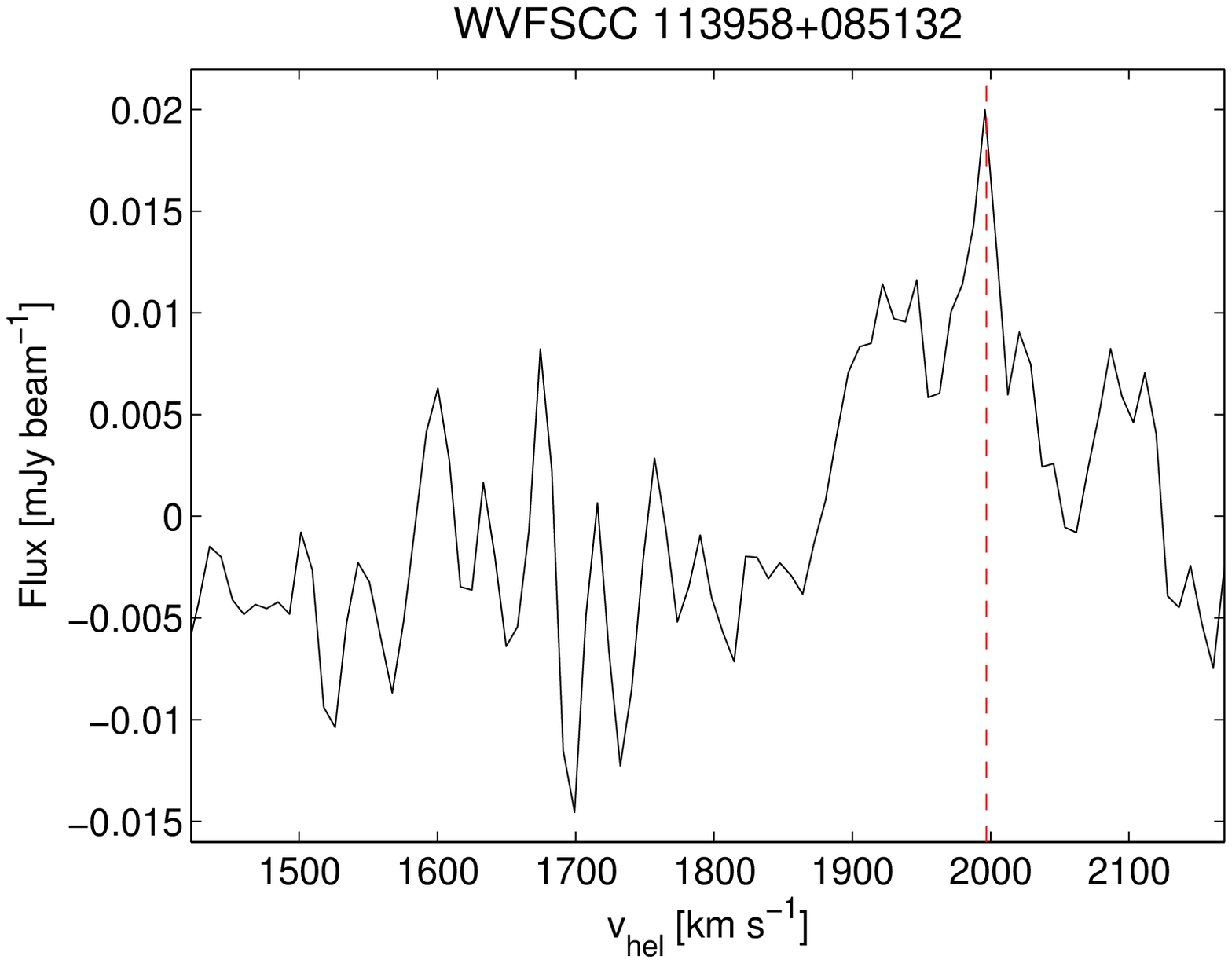}
\includegraphics[width=0.32\textwidth]{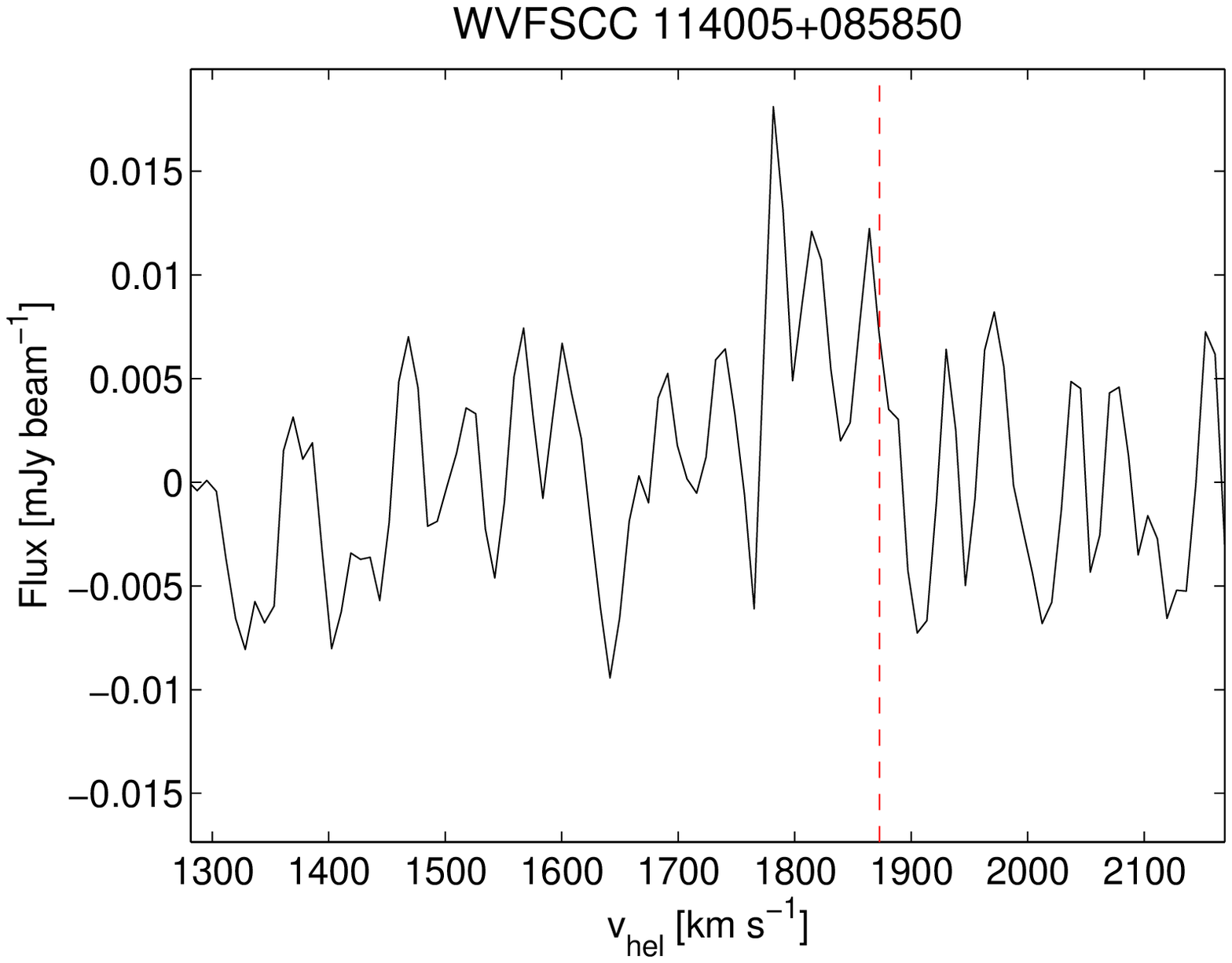}
\includegraphics[width=0.32\textwidth]{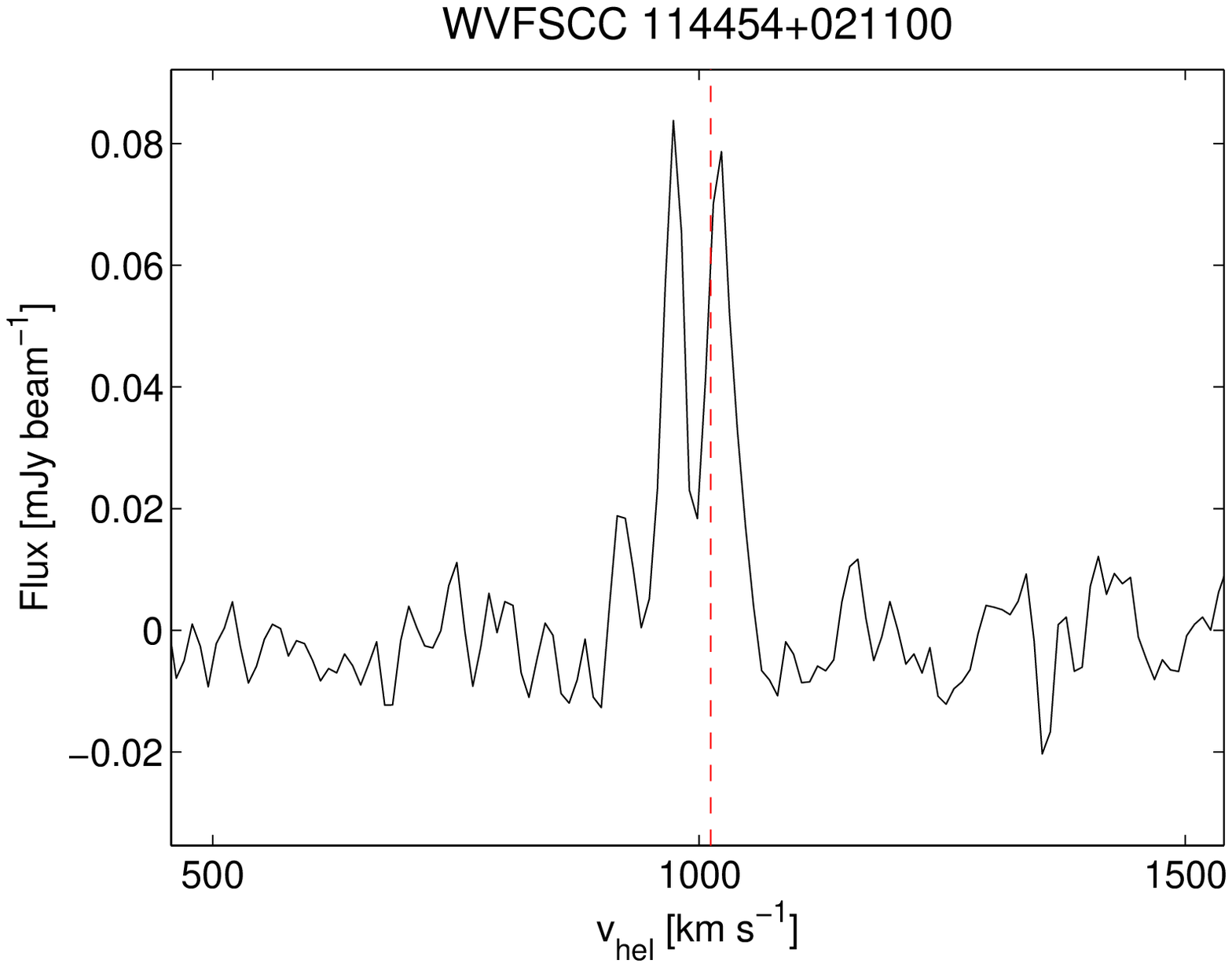}
\includegraphics[width=0.32\textwidth]{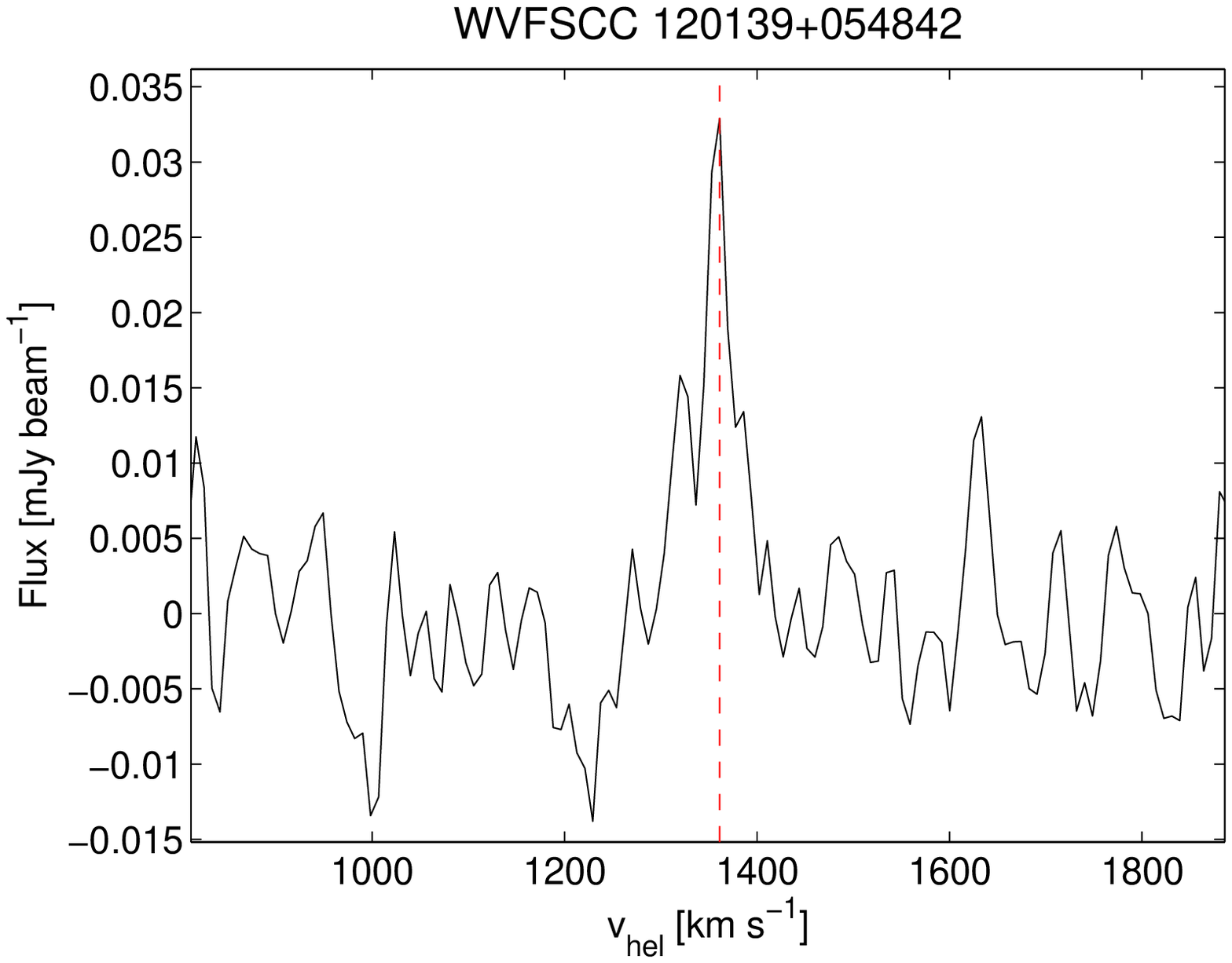}
\includegraphics[width=0.32\textwidth]{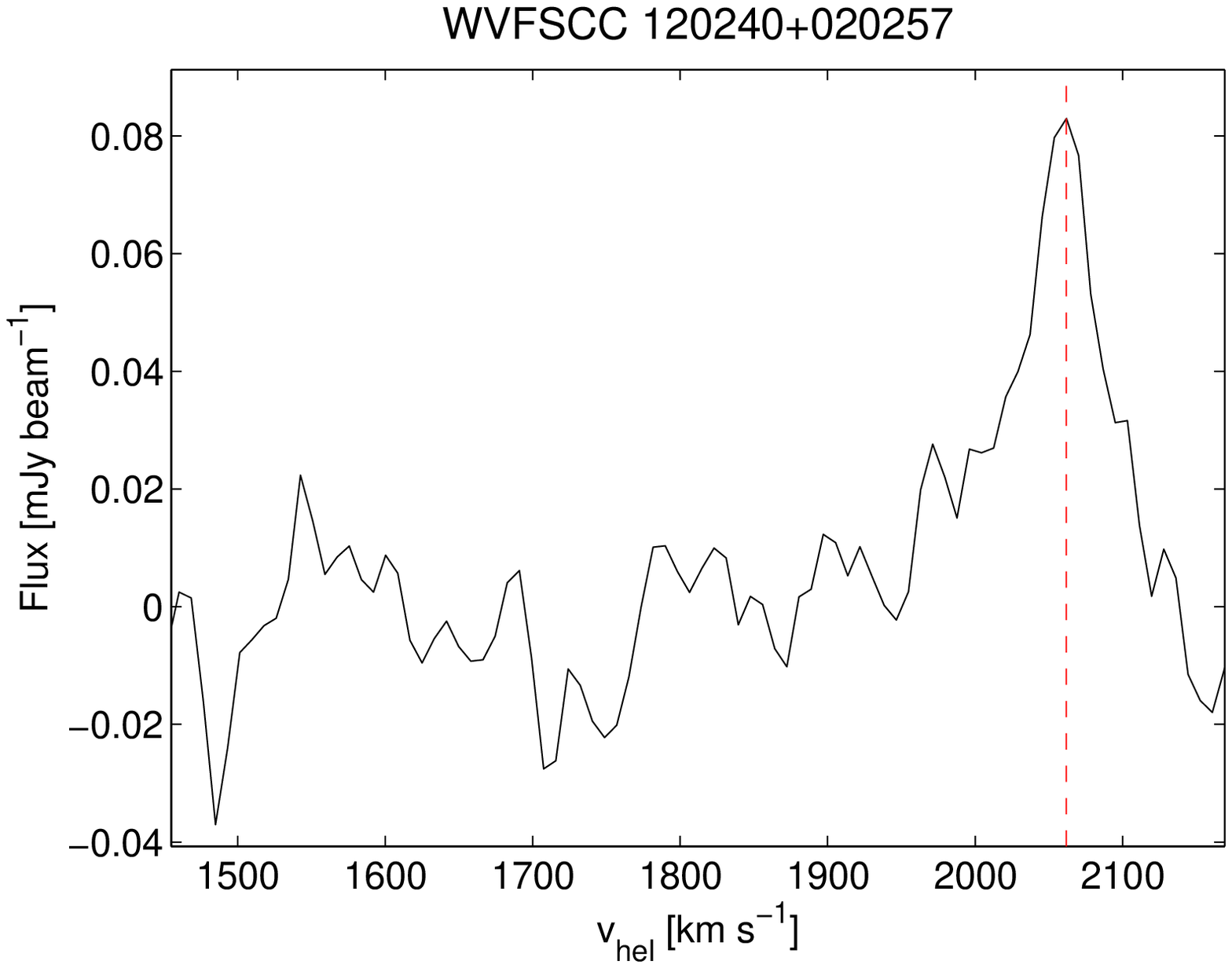}
\includegraphics[width=0.32\textwidth]{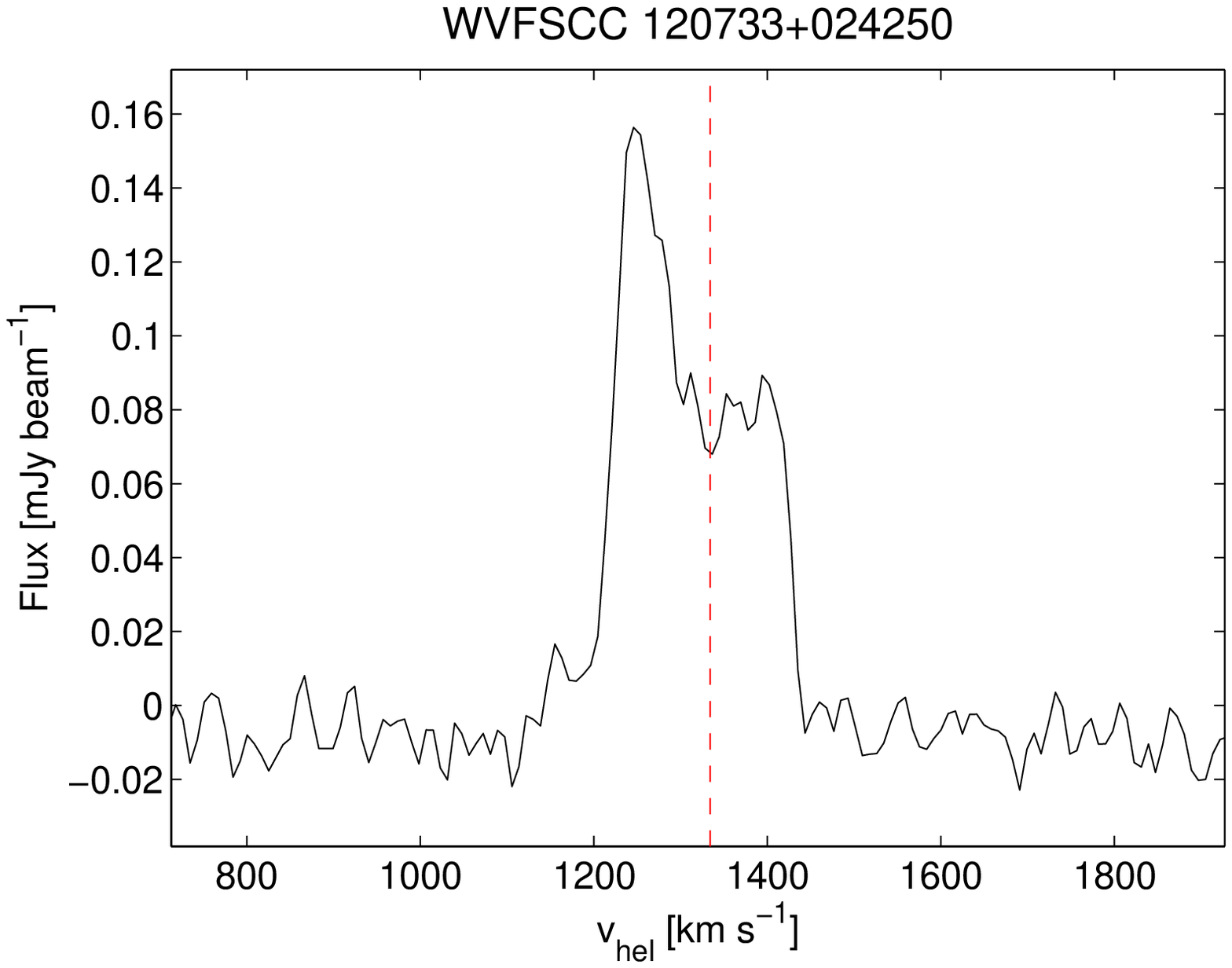}
\includegraphics[width=0.32\textwidth]{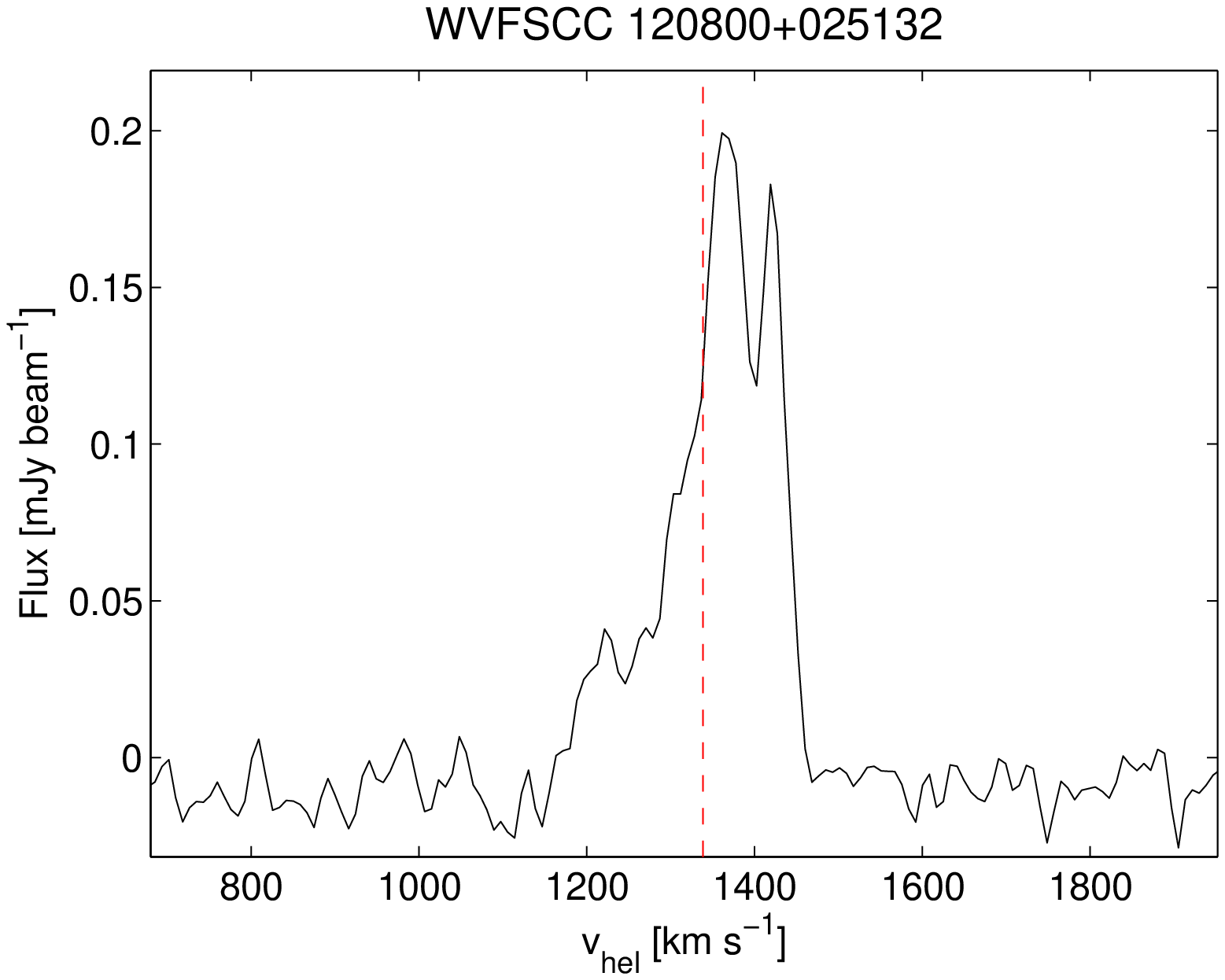}
\includegraphics[width=0.32\textwidth]{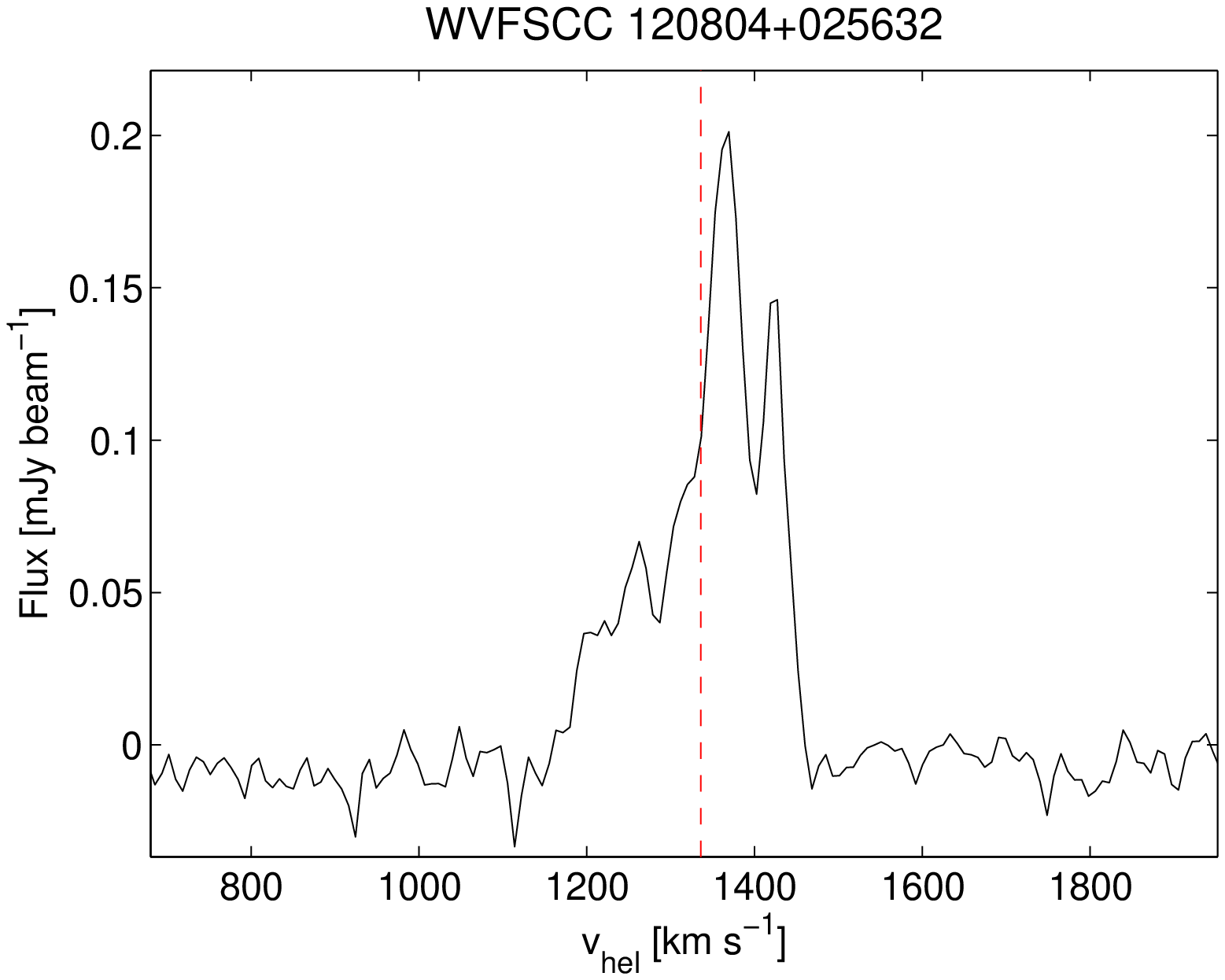}
\includegraphics[width=0.32\textwidth]{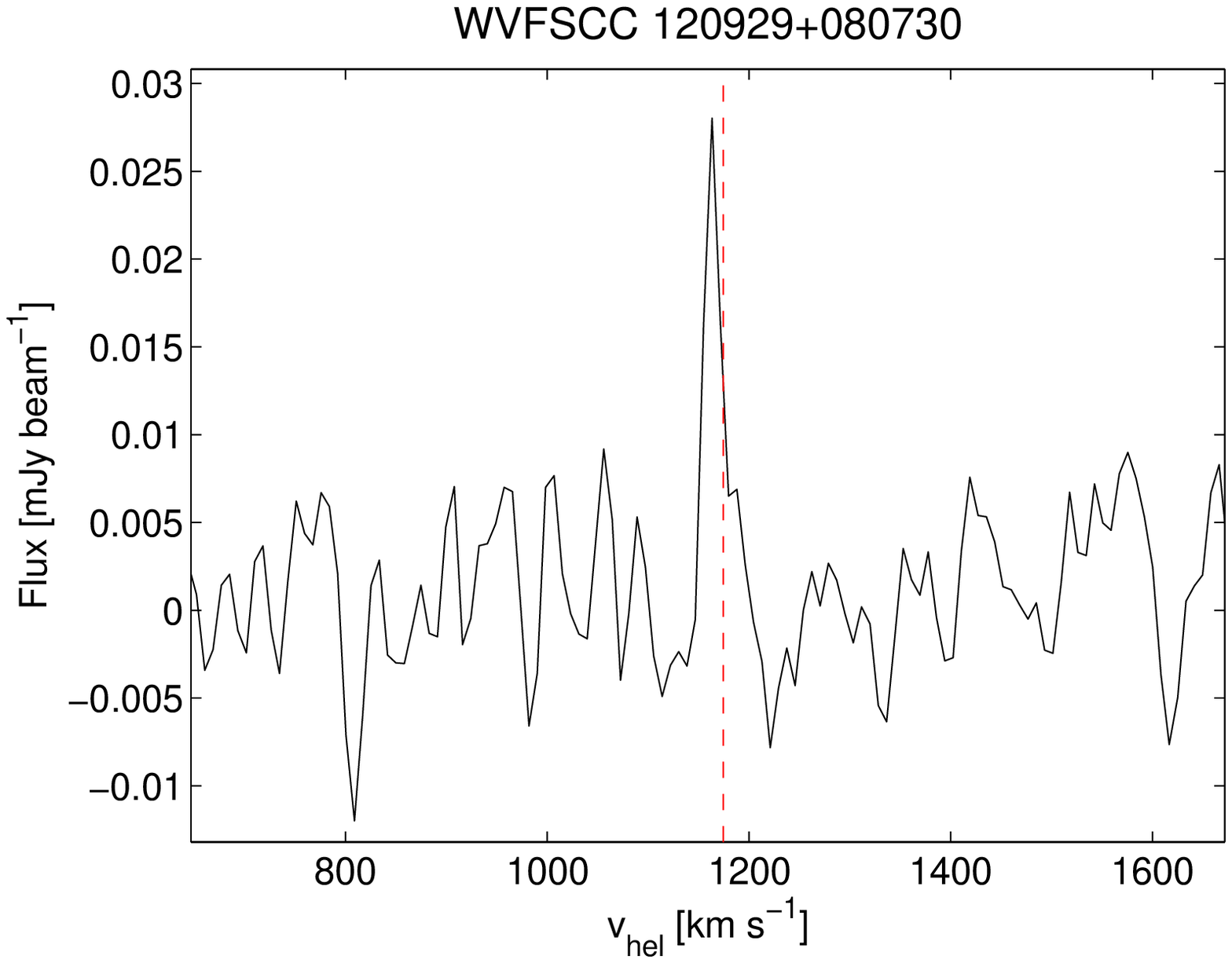}
\includegraphics[width=0.32\textwidth]{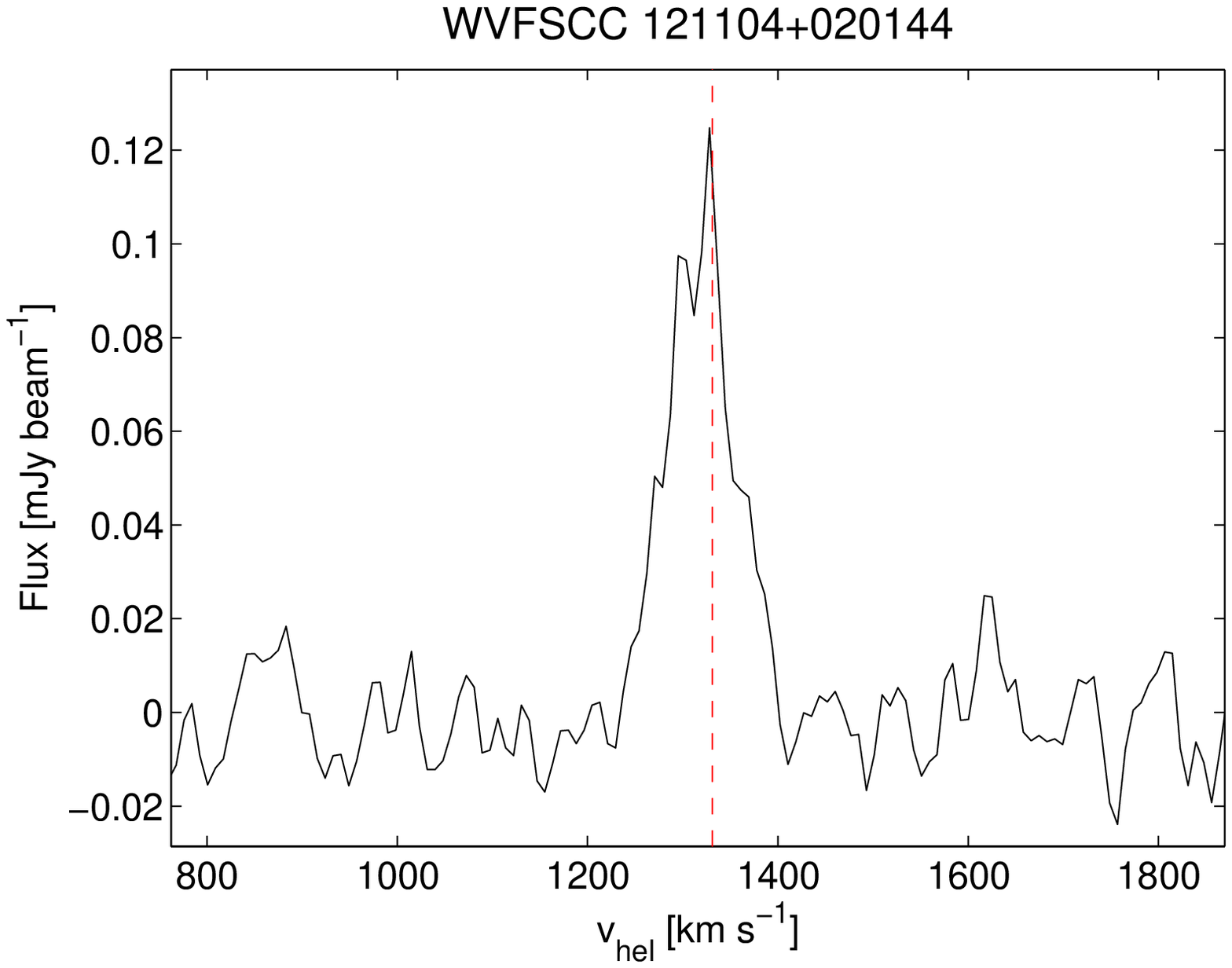}

\end{center}                                            
{\bf Fig~\ref{all_spectra2}.} (continued)                                        
 
\end{figure*}


\begin{figure*}
  \begin{center}

\includegraphics[width=0.32\textwidth]{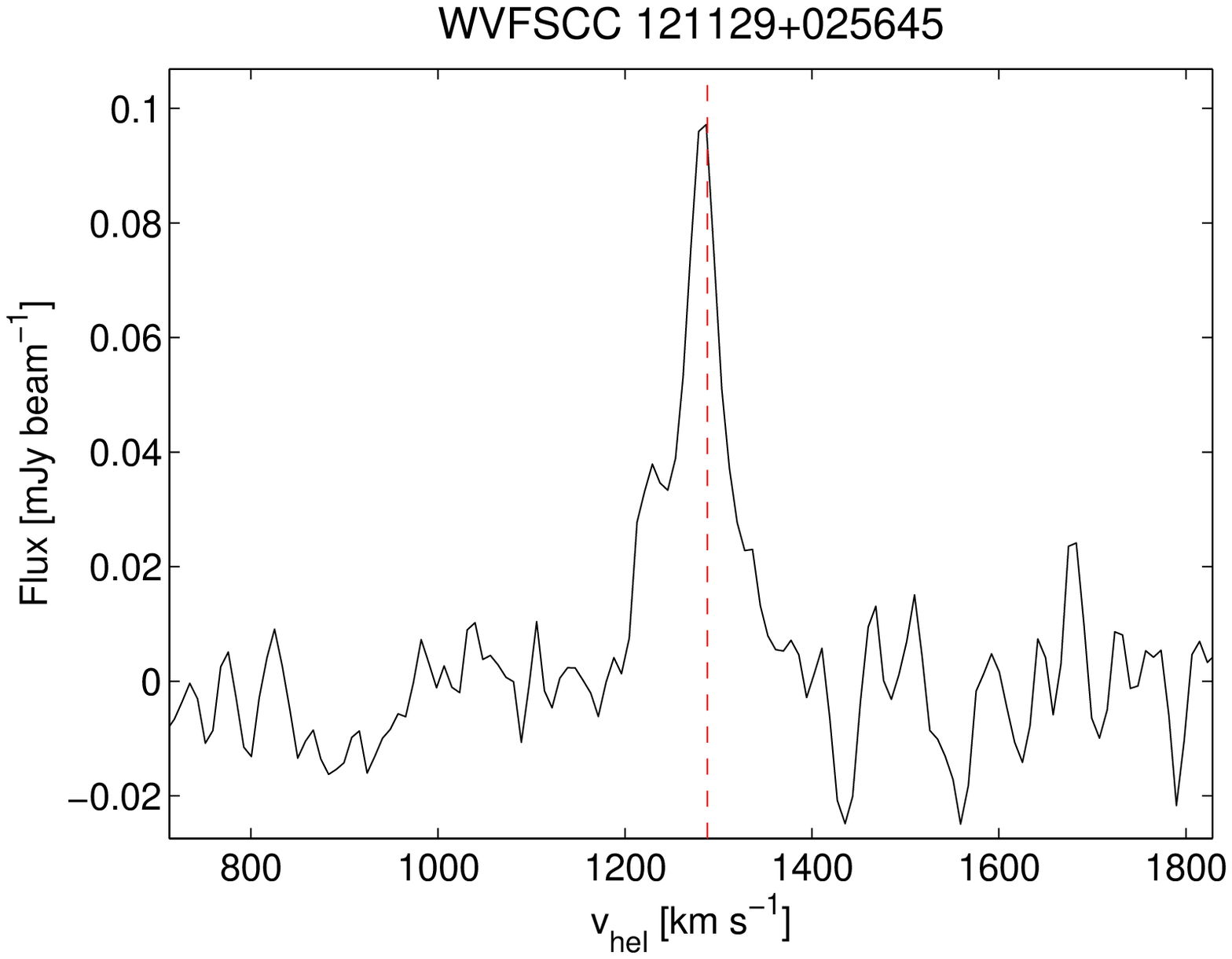}
\includegraphics[width=0.32\textwidth]{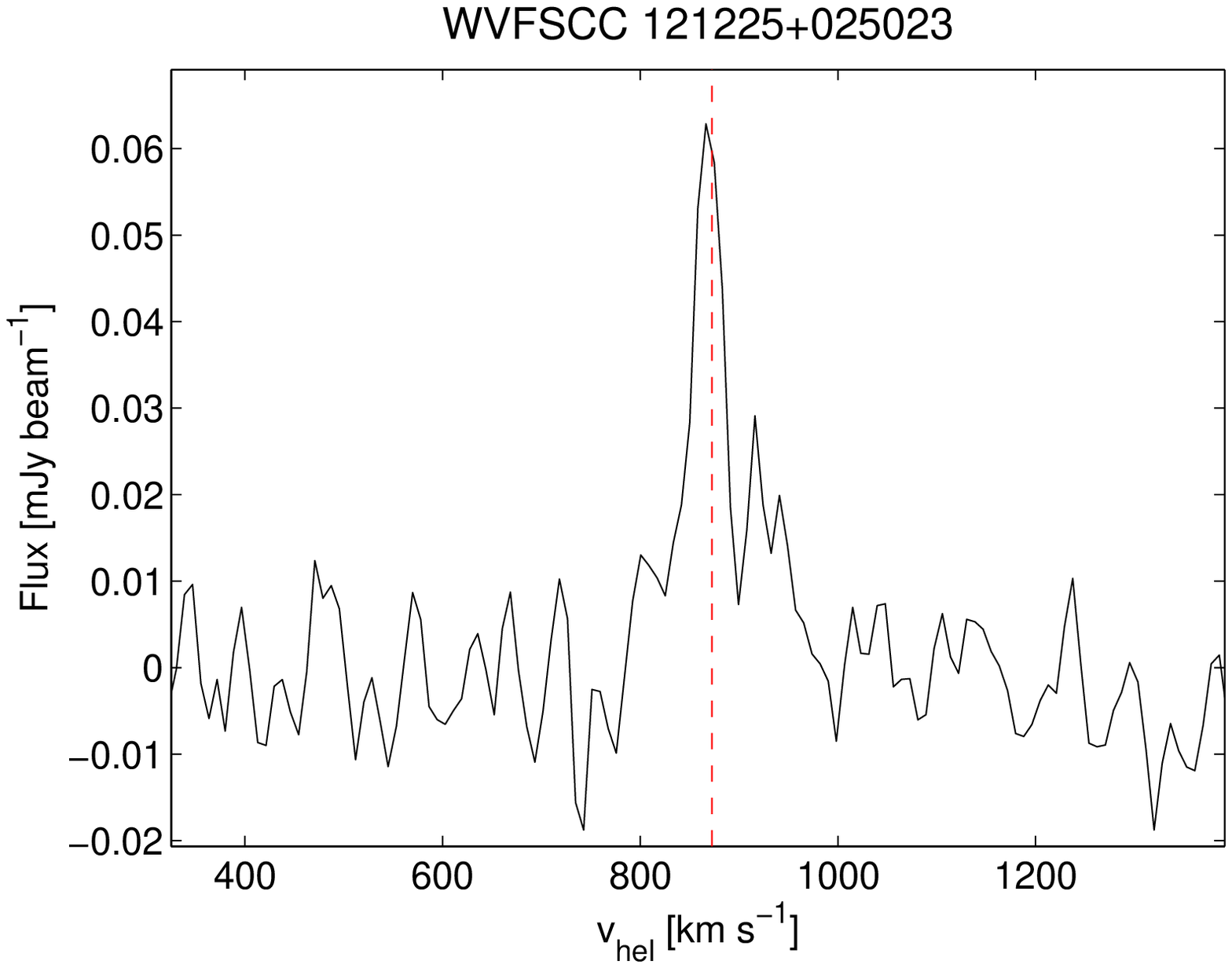}
\includegraphics[width=0.32\textwidth]{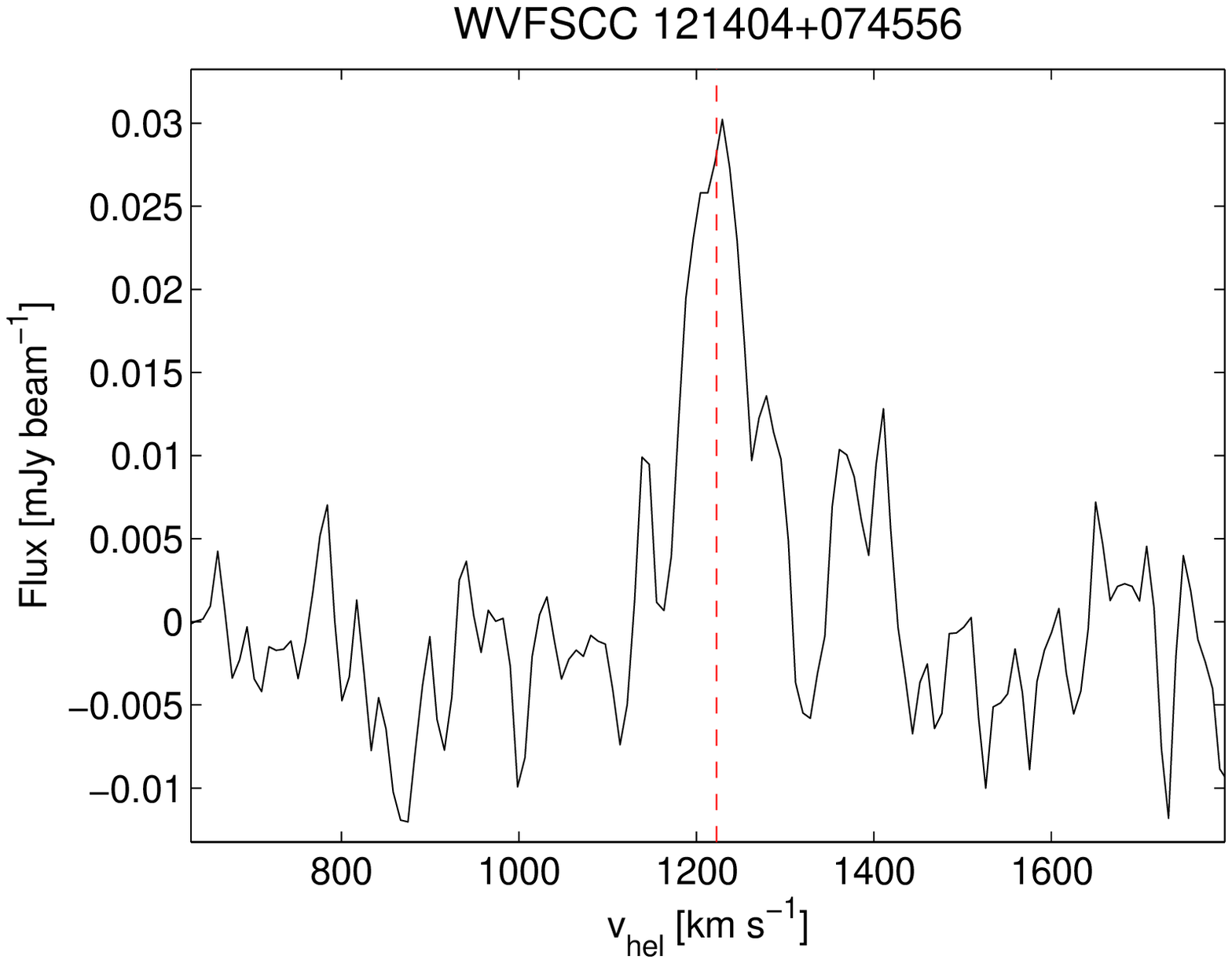}
\includegraphics[width=0.32\textwidth]{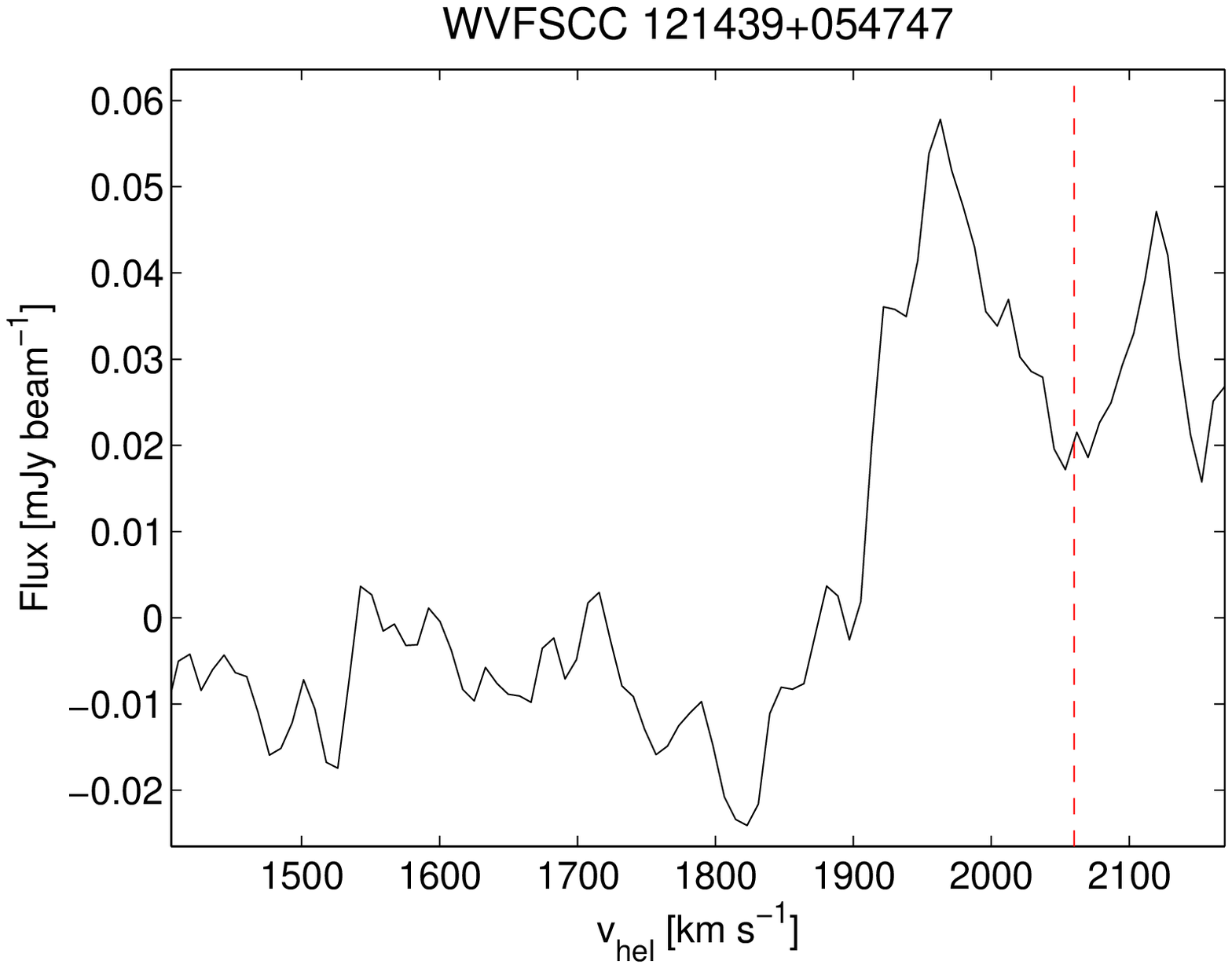}
\includegraphics[width=0.32\textwidth]{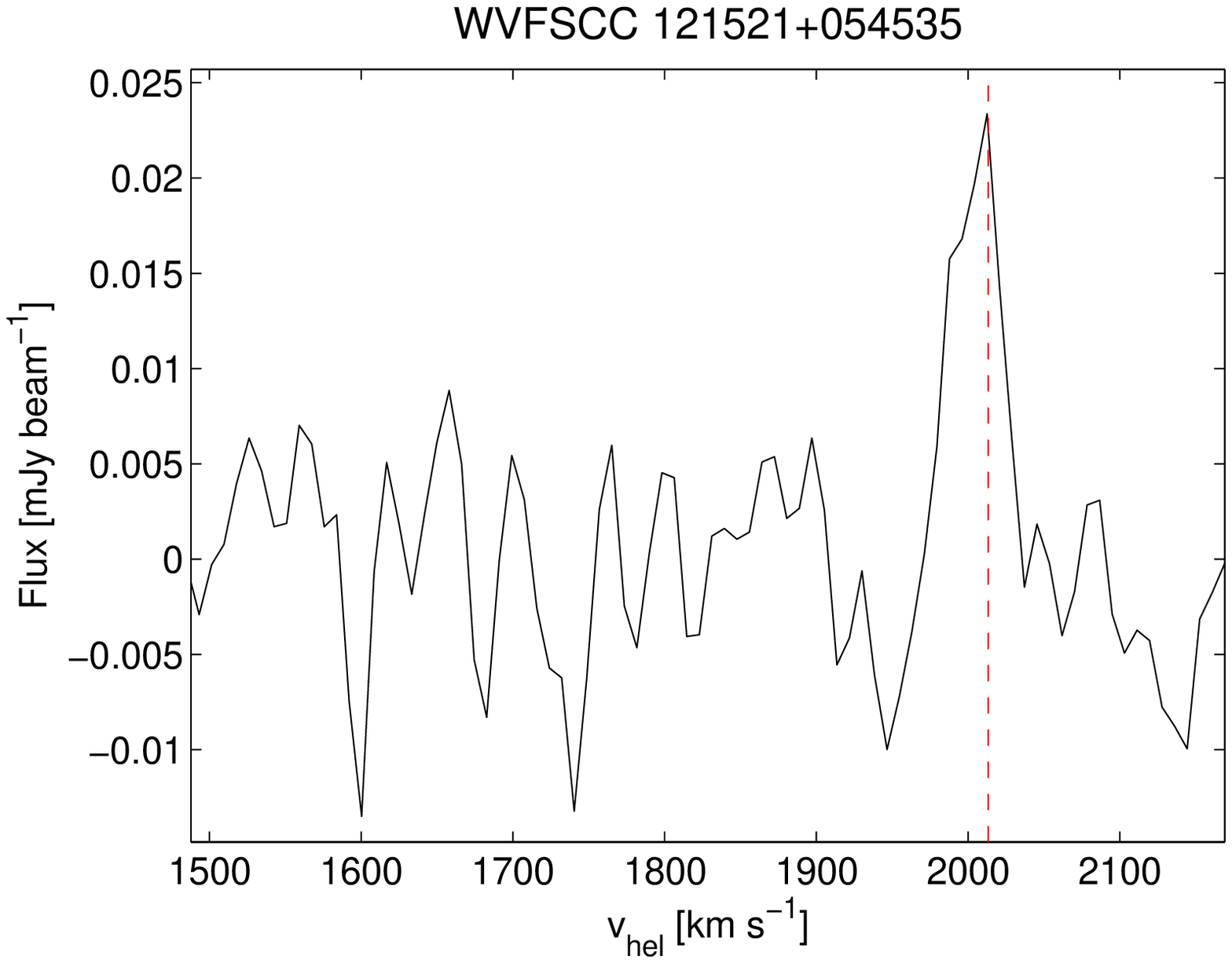}
\includegraphics[width=0.32\textwidth]{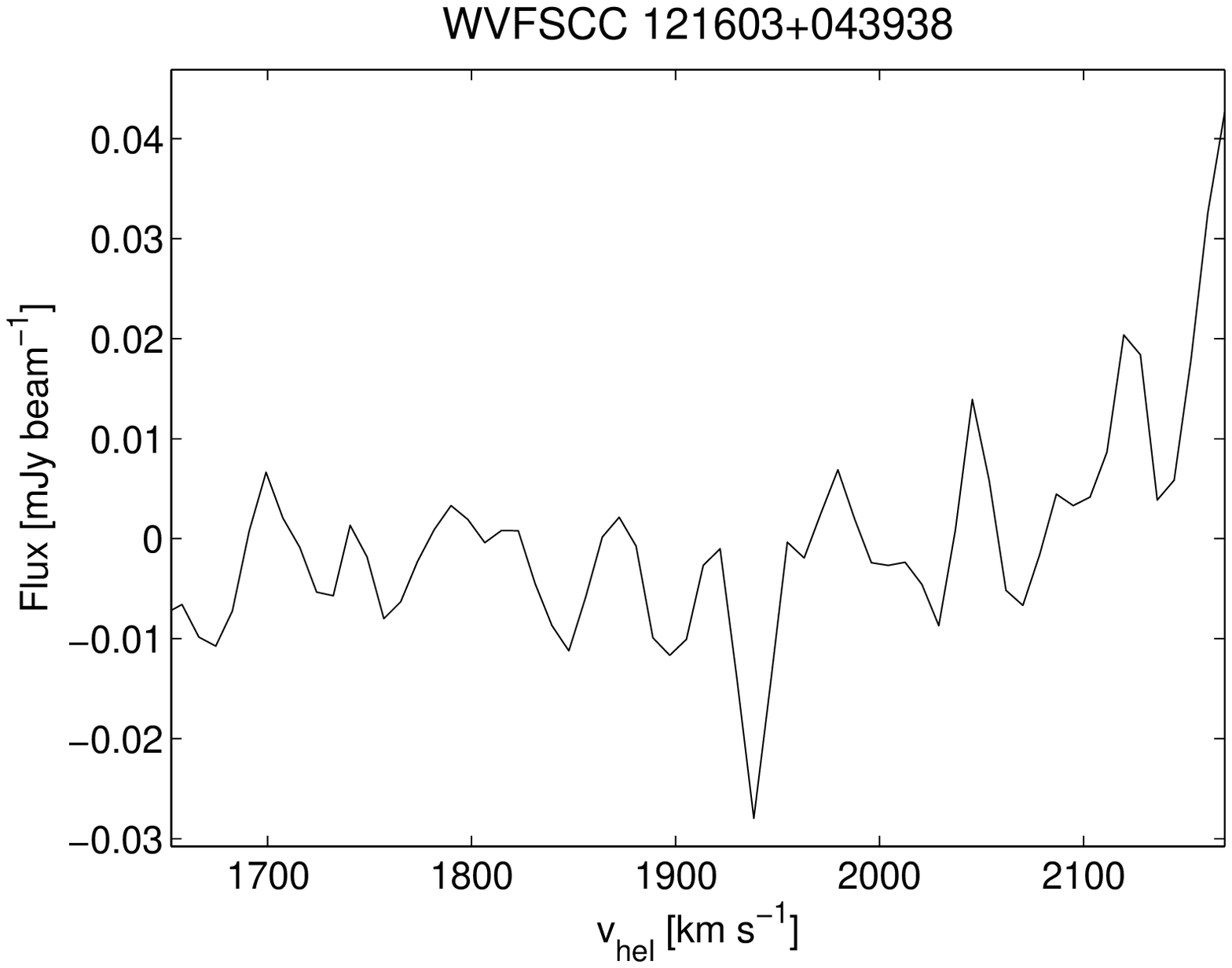}
\includegraphics[width=0.32\textwidth]{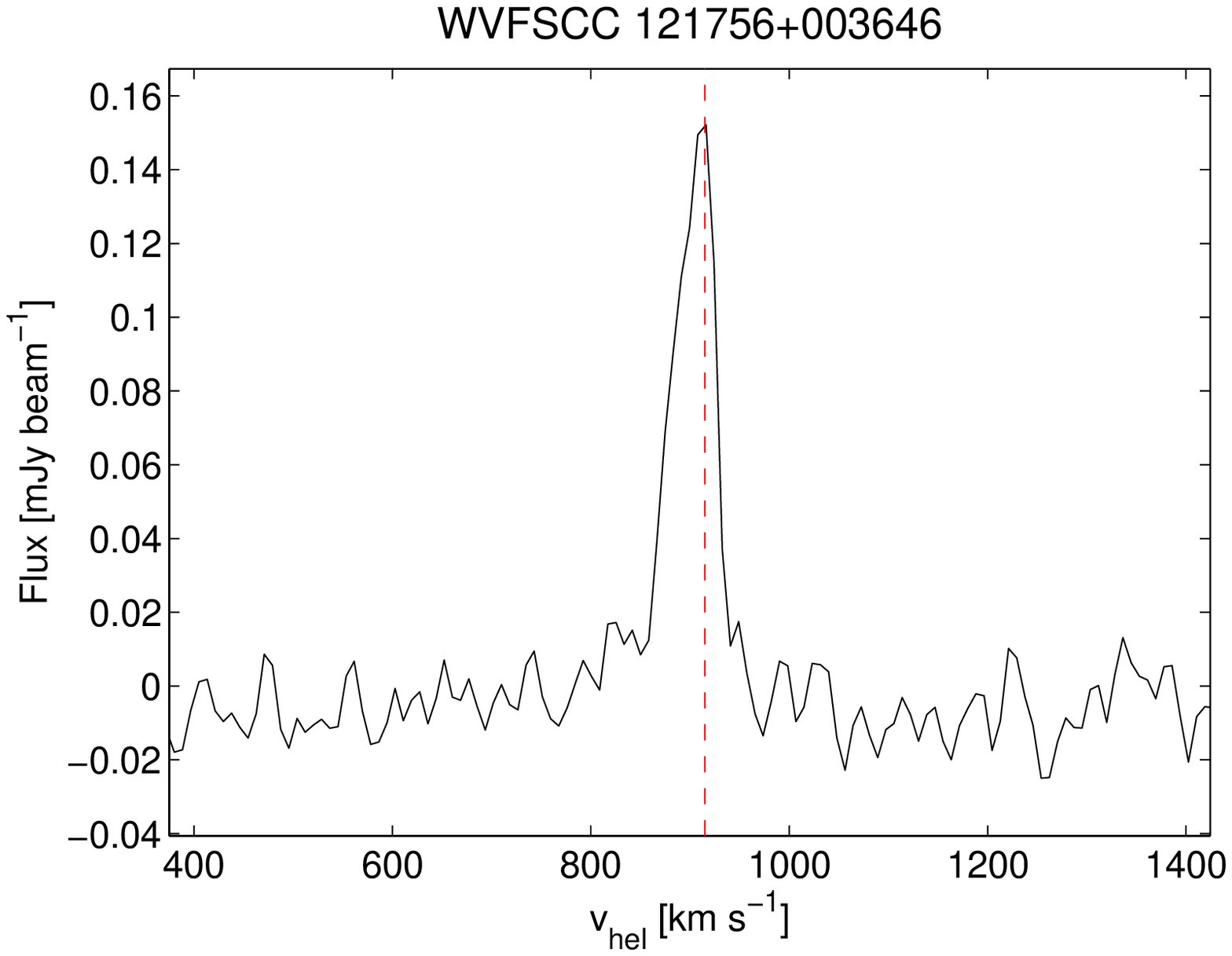}
\includegraphics[width=0.32\textwidth]{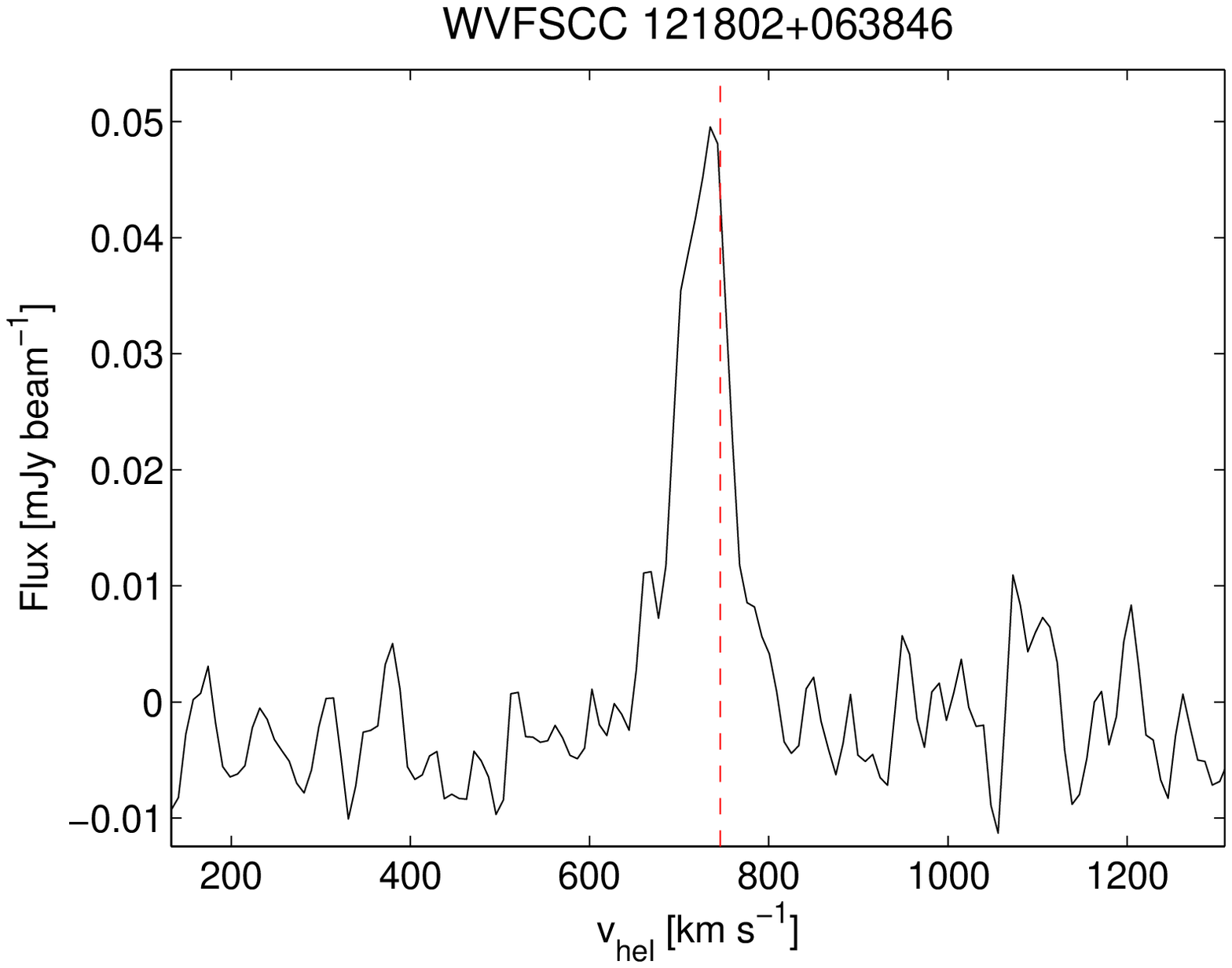}
\includegraphics[width=0.32\textwidth]{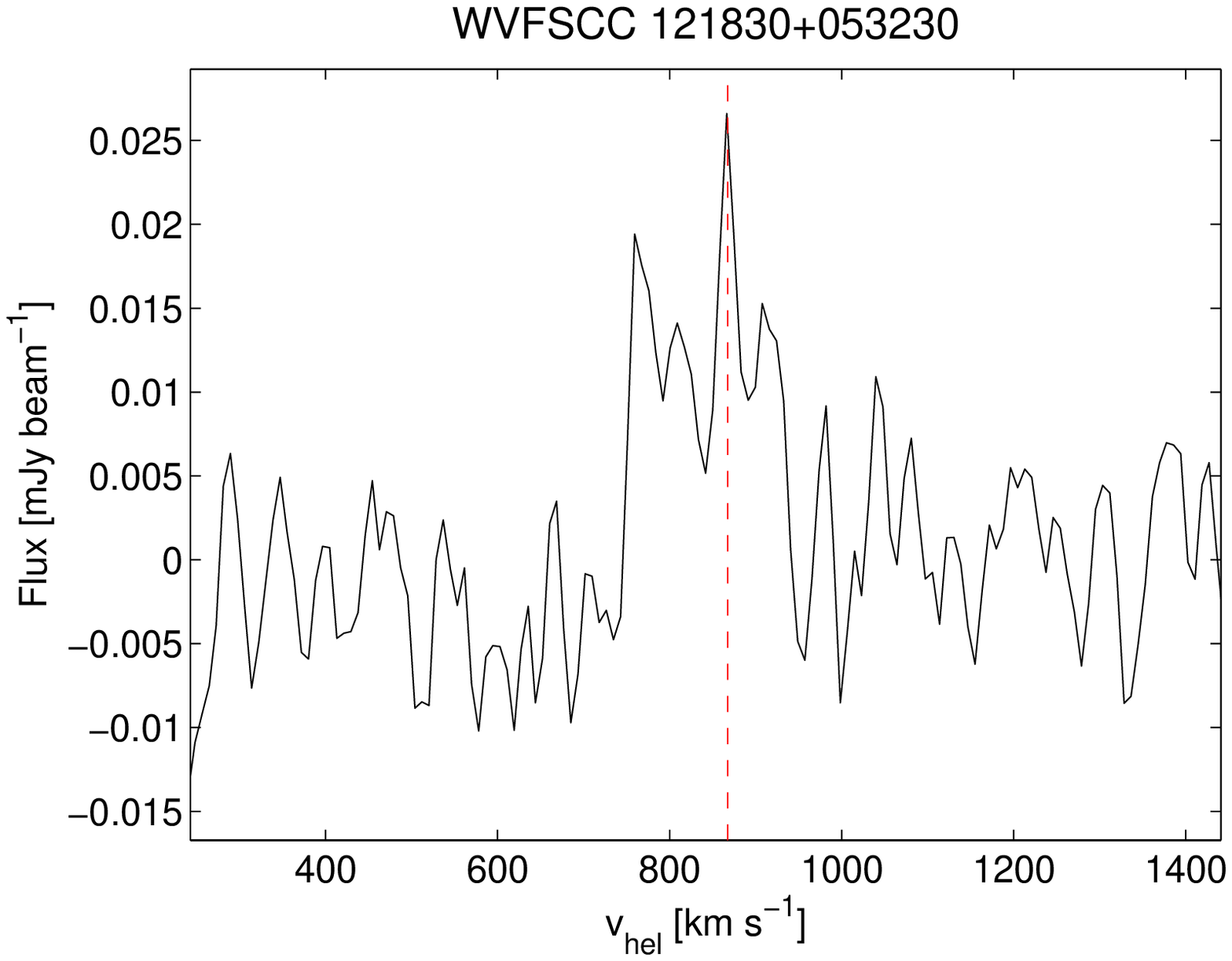}
\includegraphics[width=0.32\textwidth]{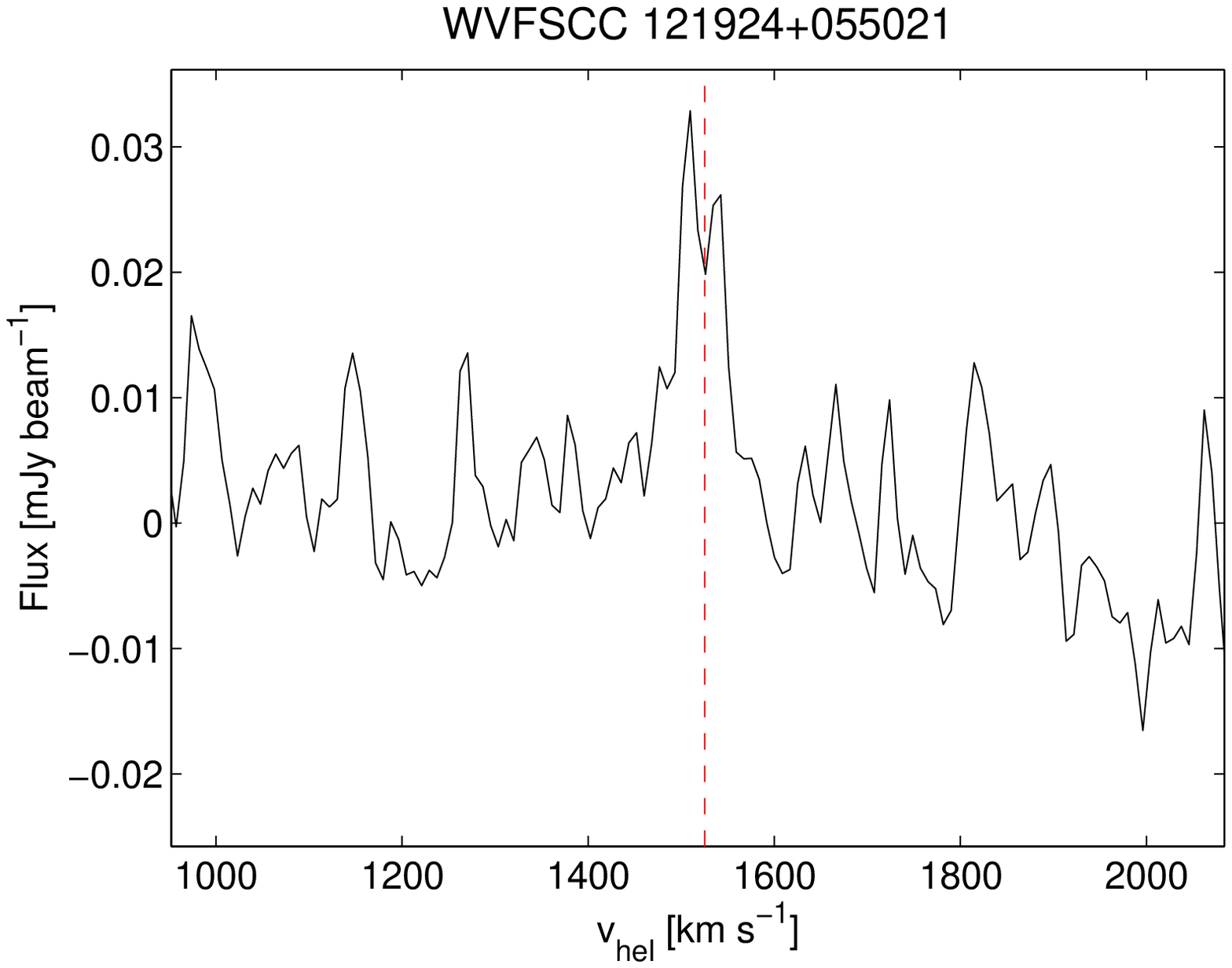}
\includegraphics[width=0.32\textwidth]{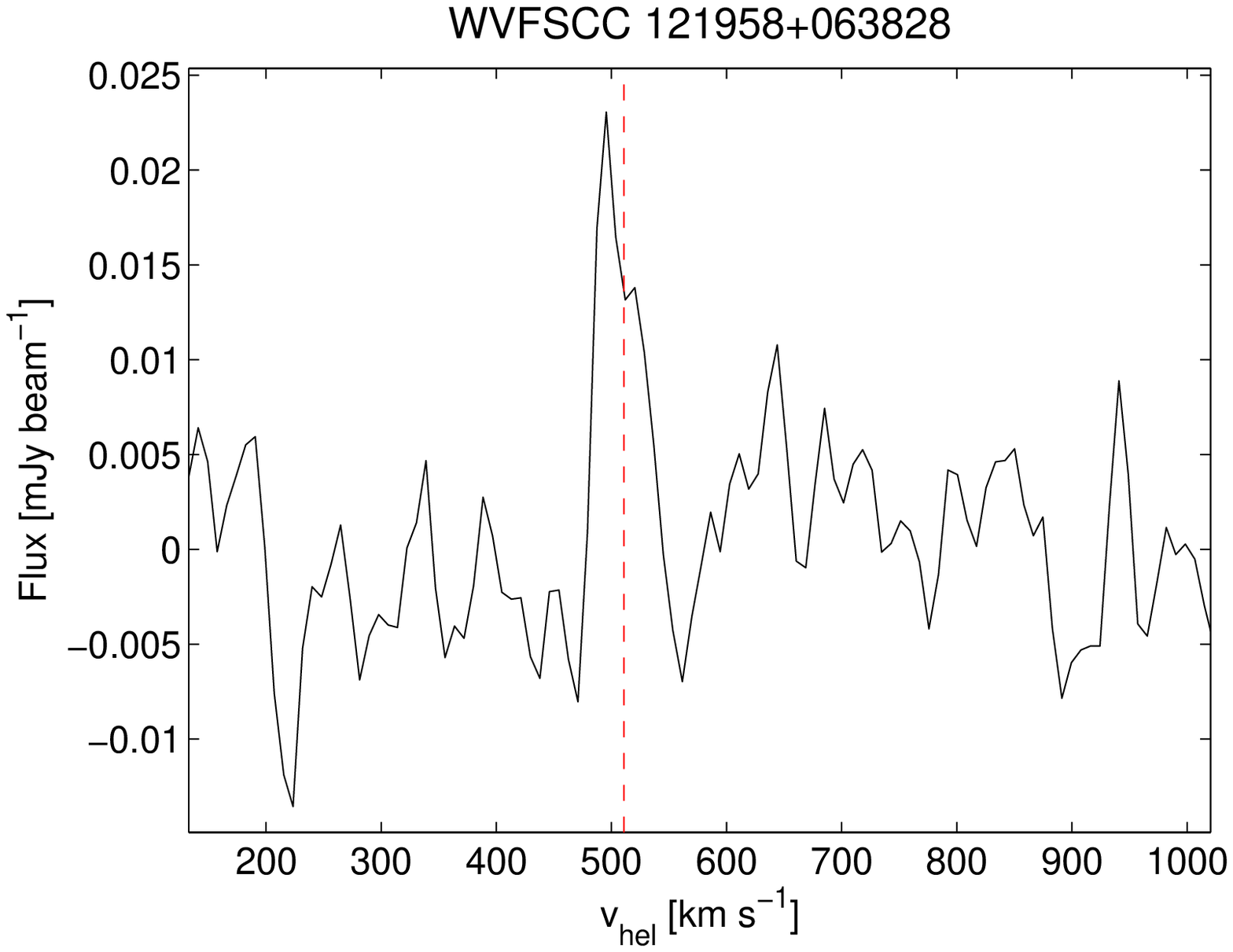}
\includegraphics[width=0.32\textwidth]{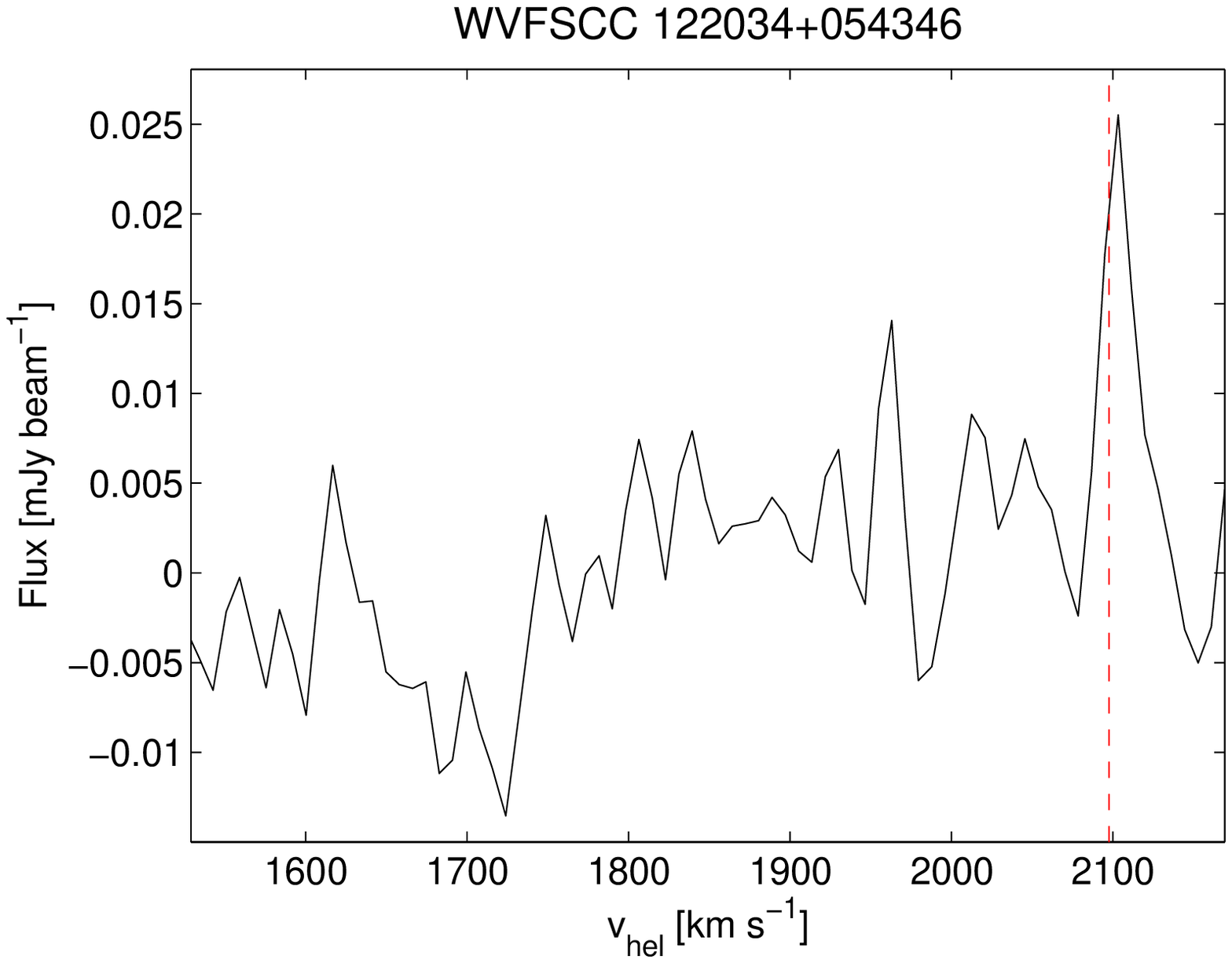}
\includegraphics[width=0.32\textwidth]{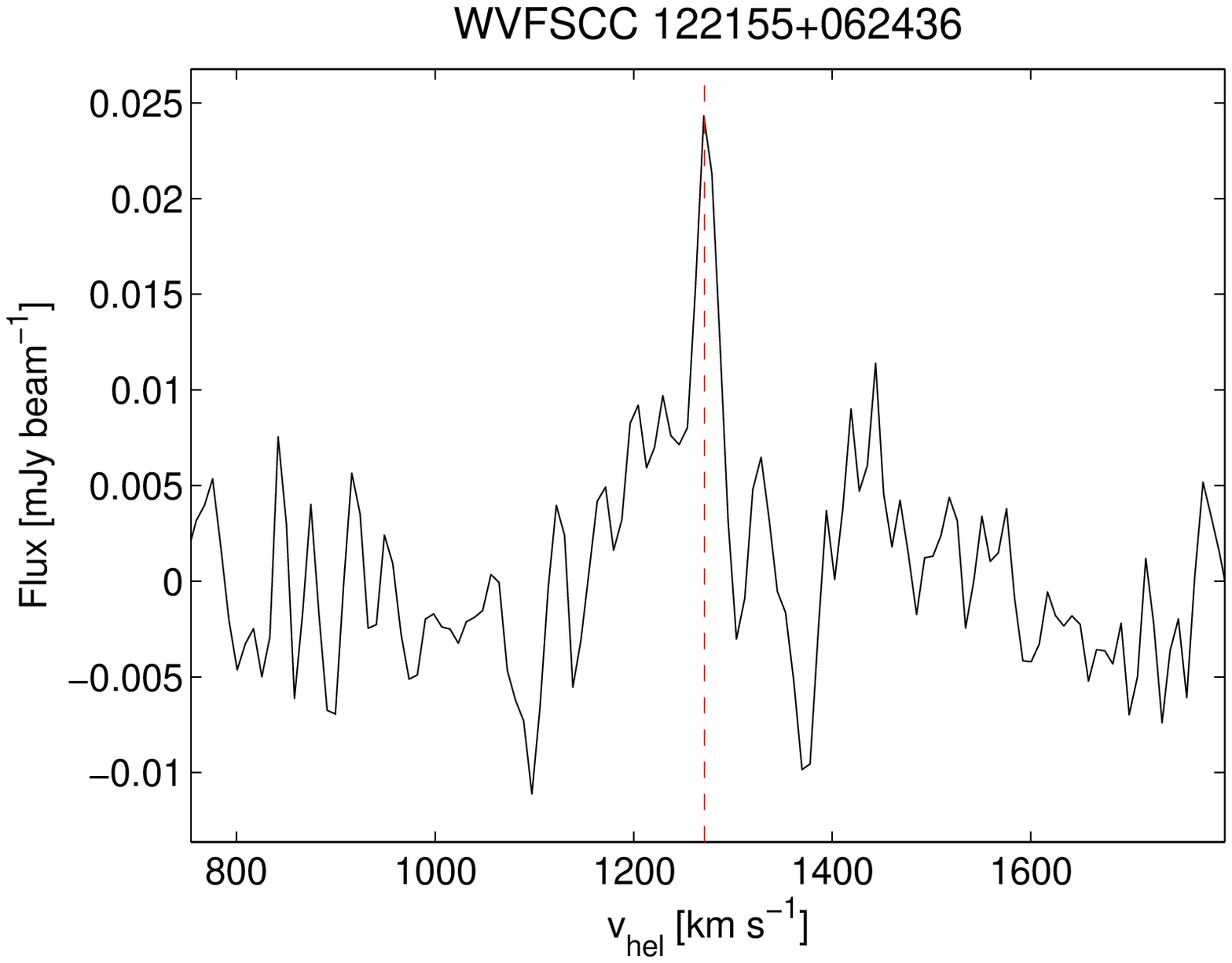}
\includegraphics[width=0.32\textwidth]{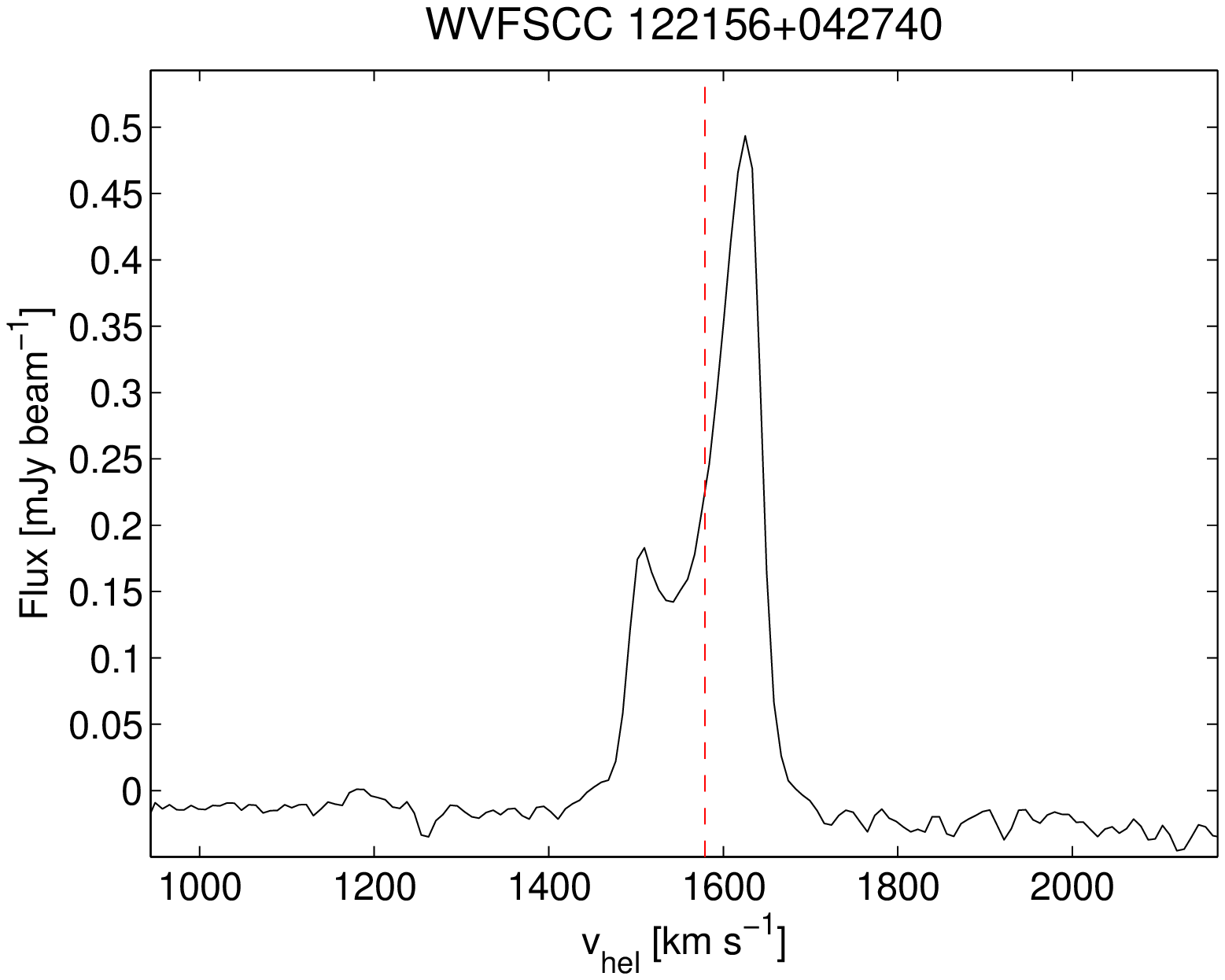}
\includegraphics[width=0.32\textwidth]{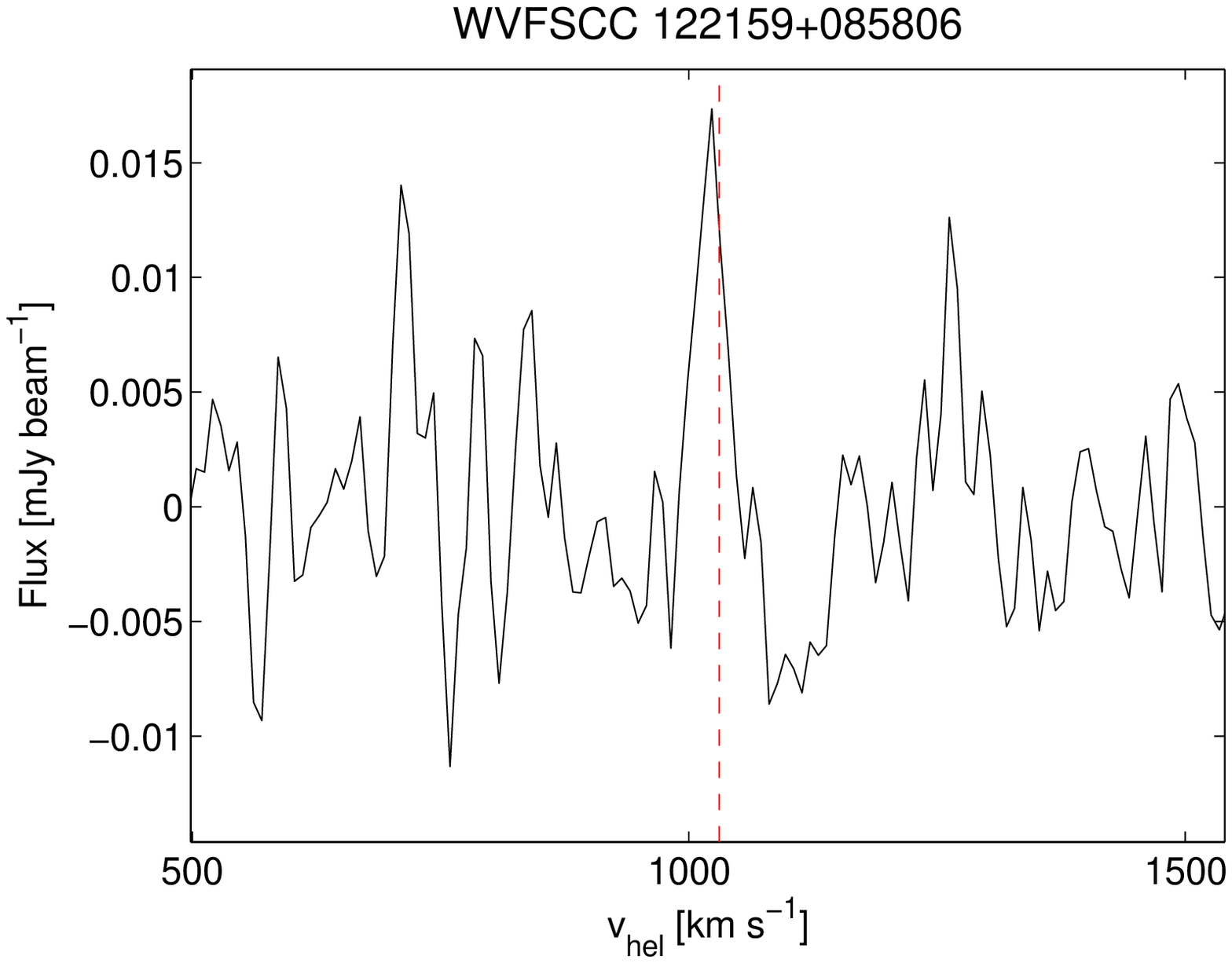}

\end{center}                                            
{\bf Fig~\ref{all_spectra2}.} (continued)                                        
 
\end{figure*}


\begin{figure*}
  \begin{center}

\includegraphics[width=0.32\textwidth]{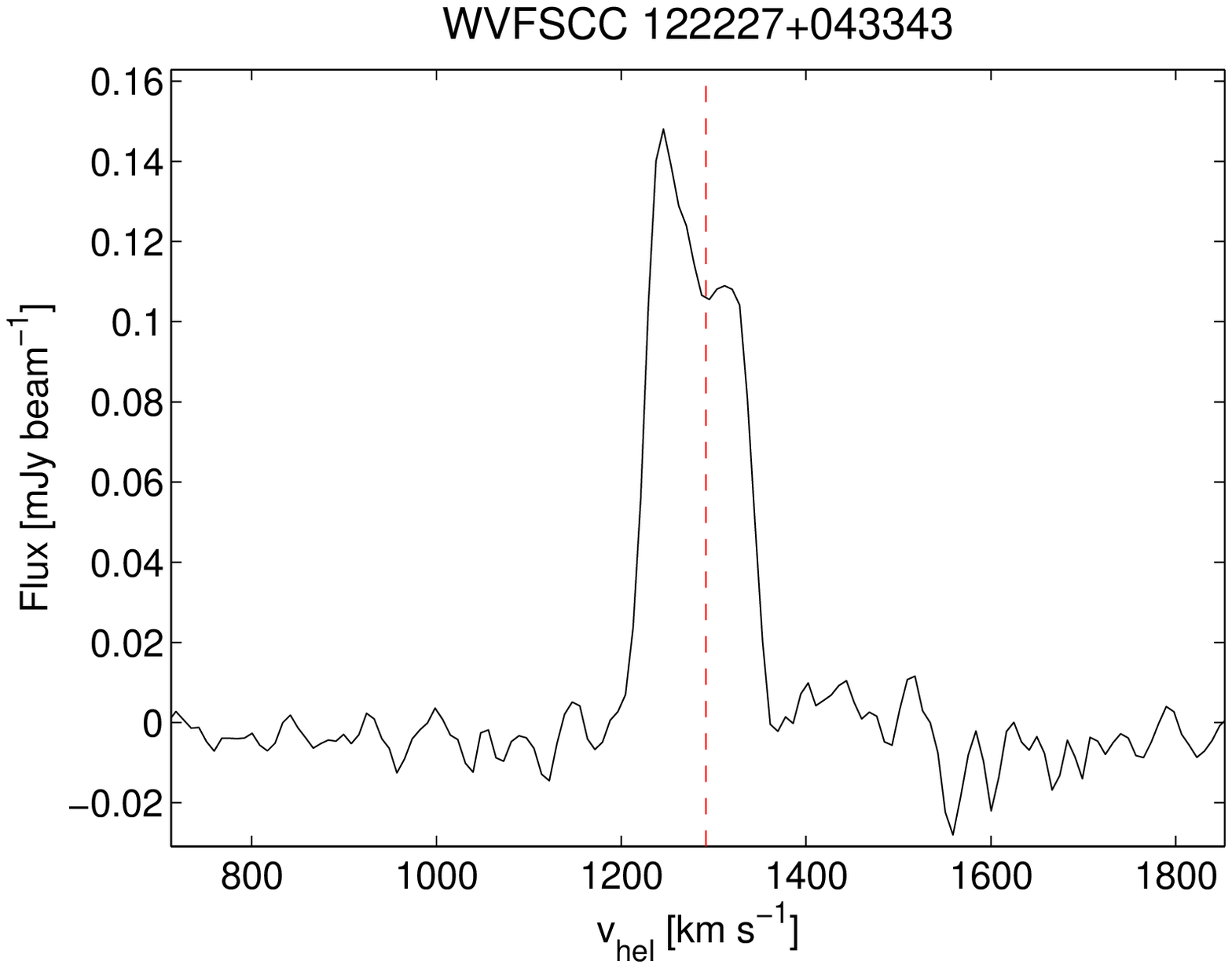} 
\includegraphics[width=0.32\textwidth]{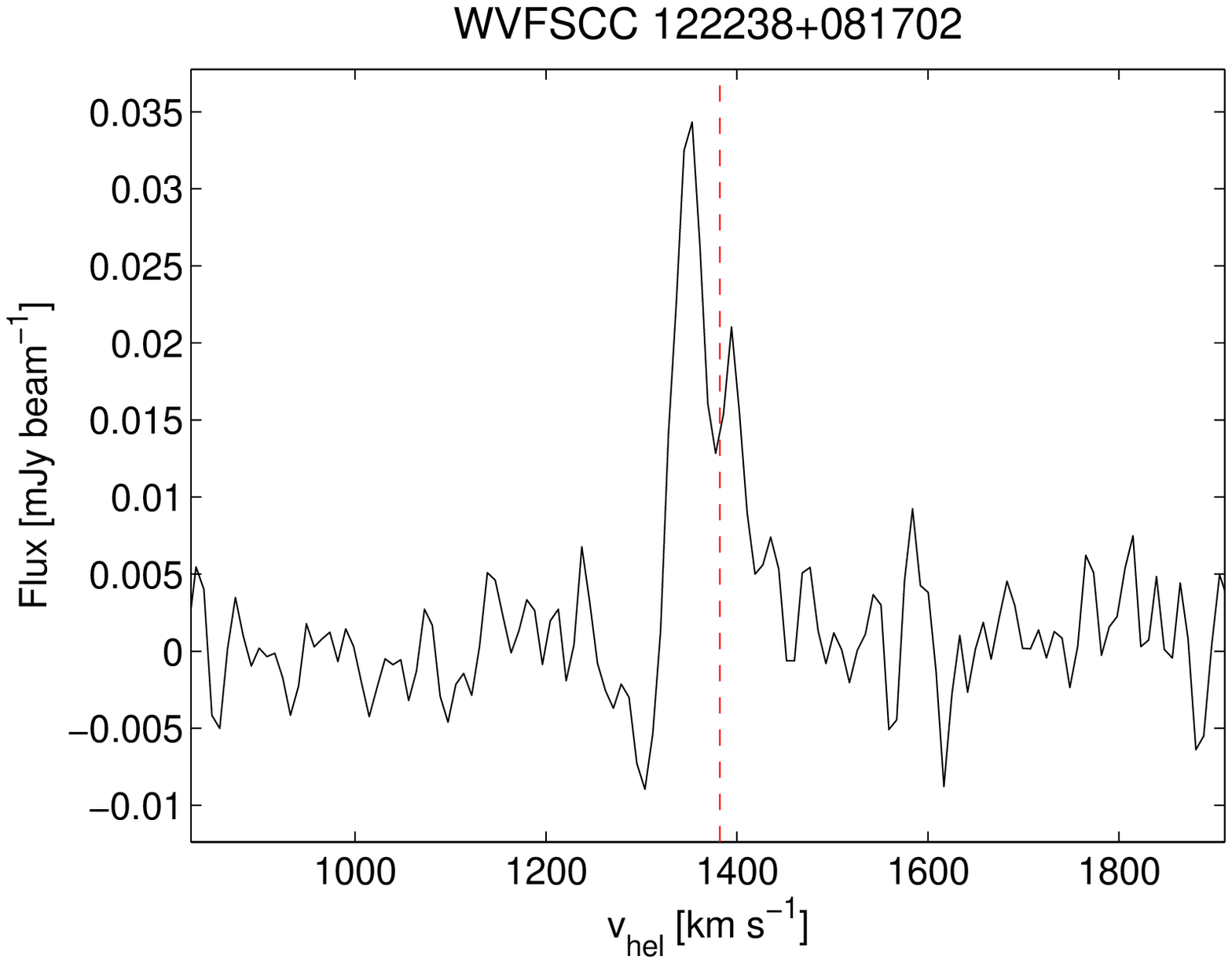}
\includegraphics[width=0.32\textwidth]{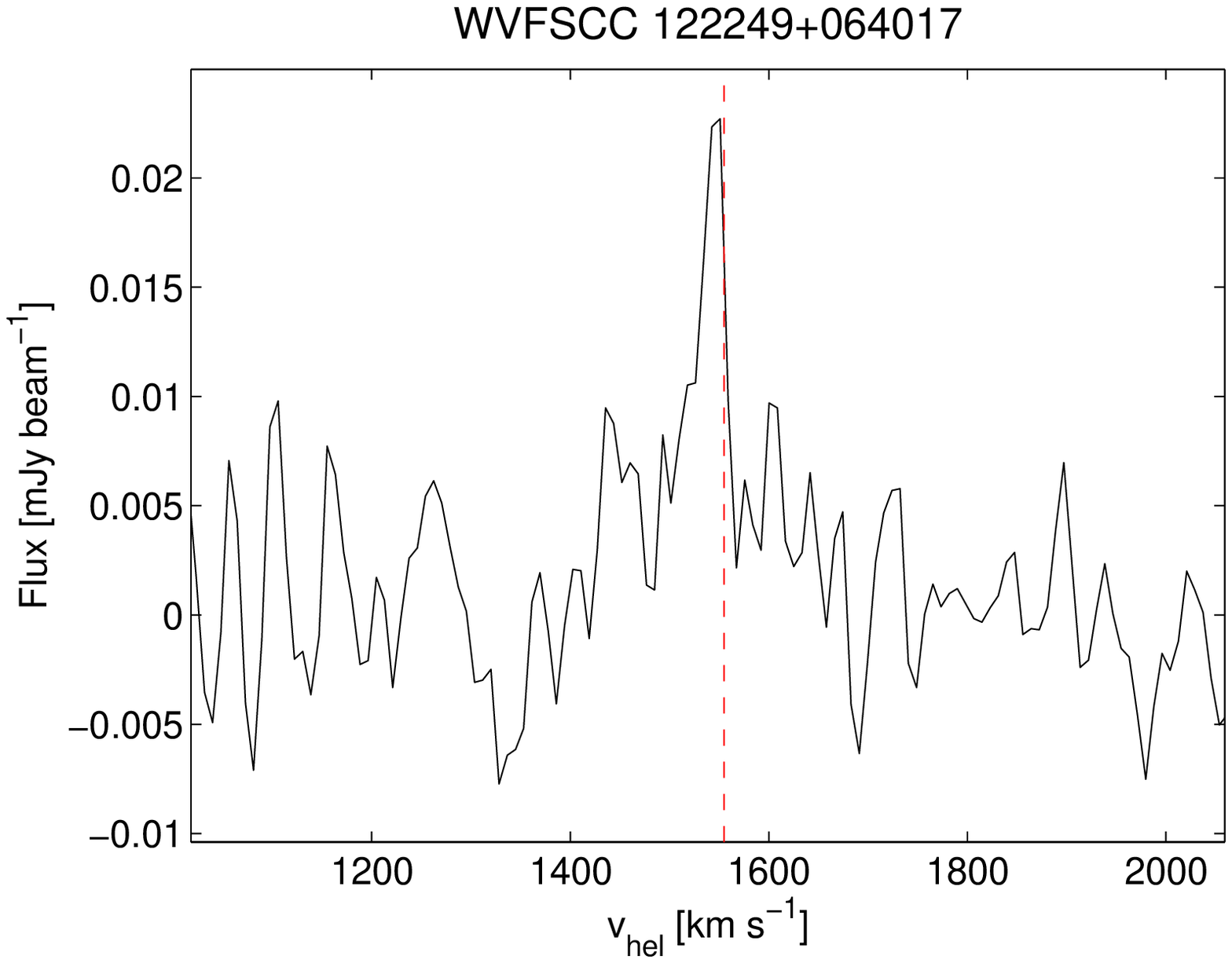}
\includegraphics[width=0.32\textwidth]{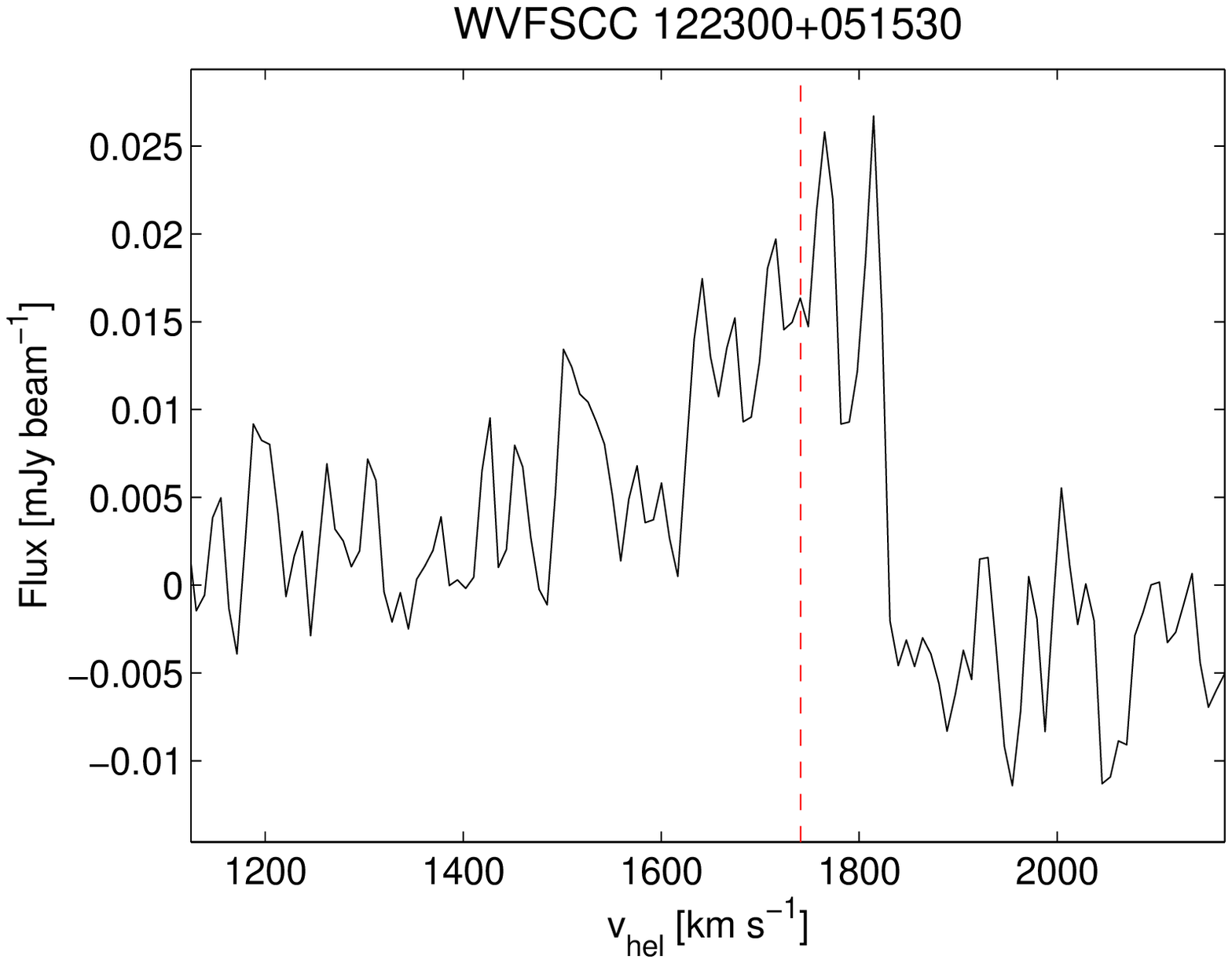}
\includegraphics[width=0.32\textwidth]{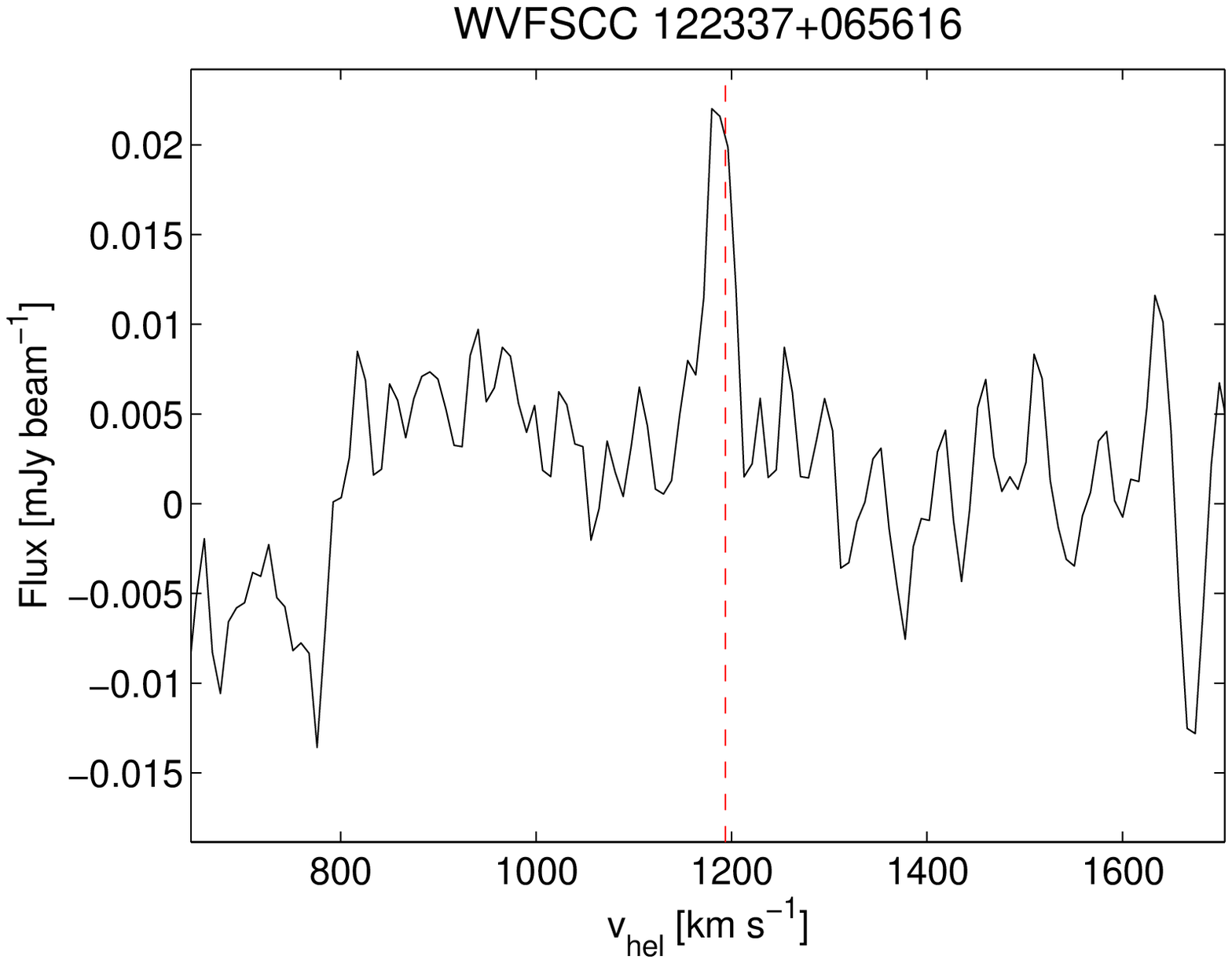}
\includegraphics[width=0.32\textwidth]{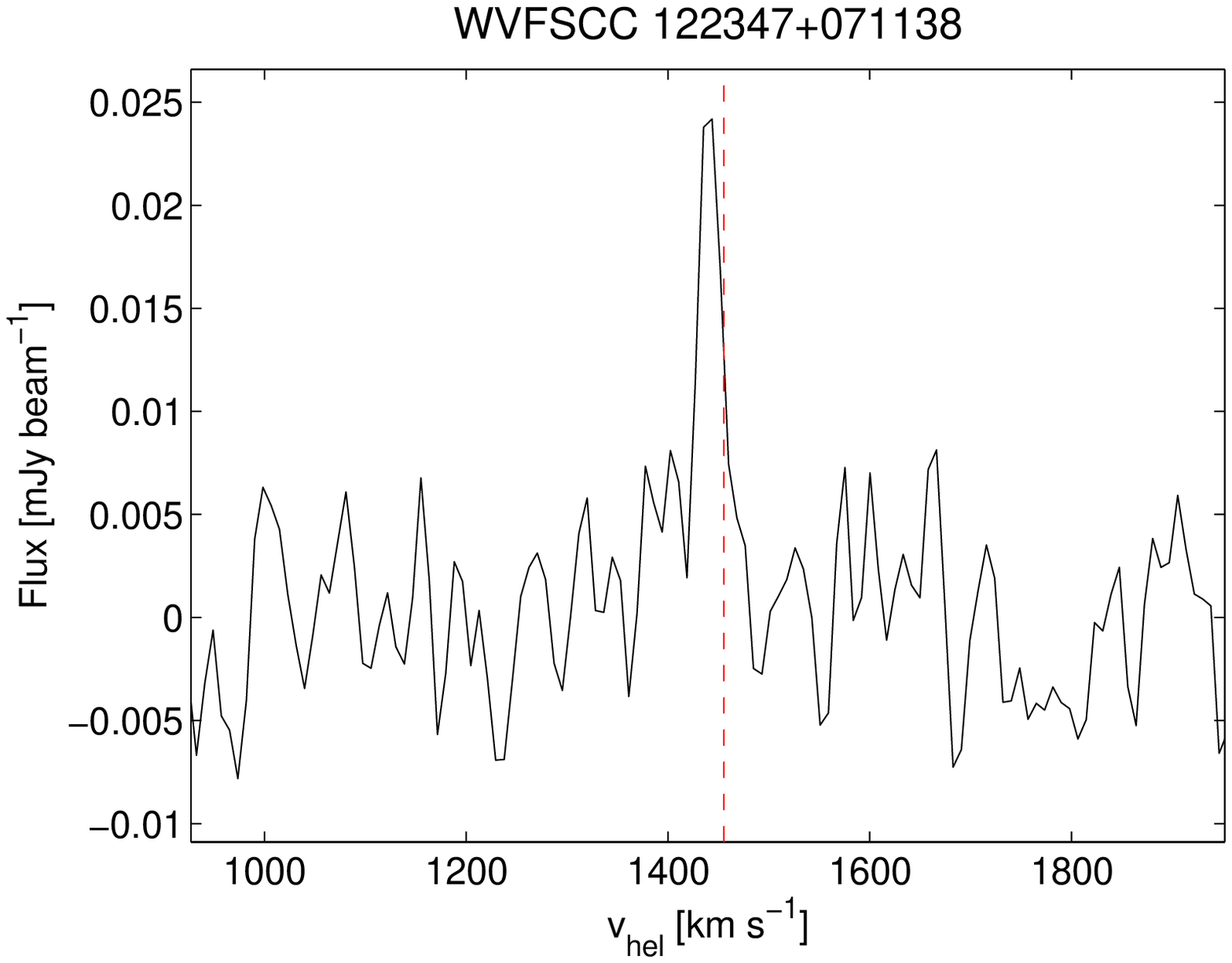}
\includegraphics[width=0.32\textwidth]{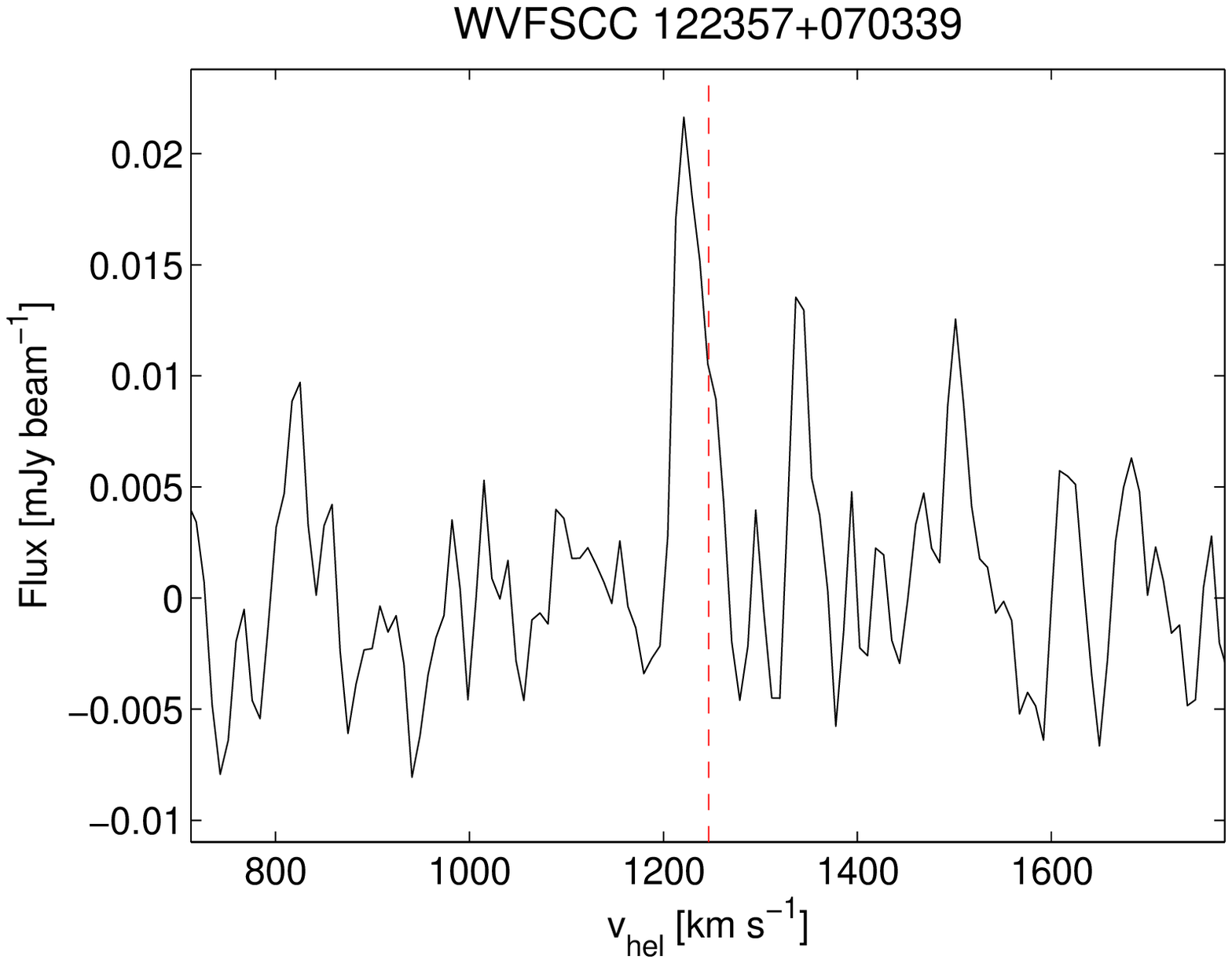}
\includegraphics[width=0.32\textwidth]{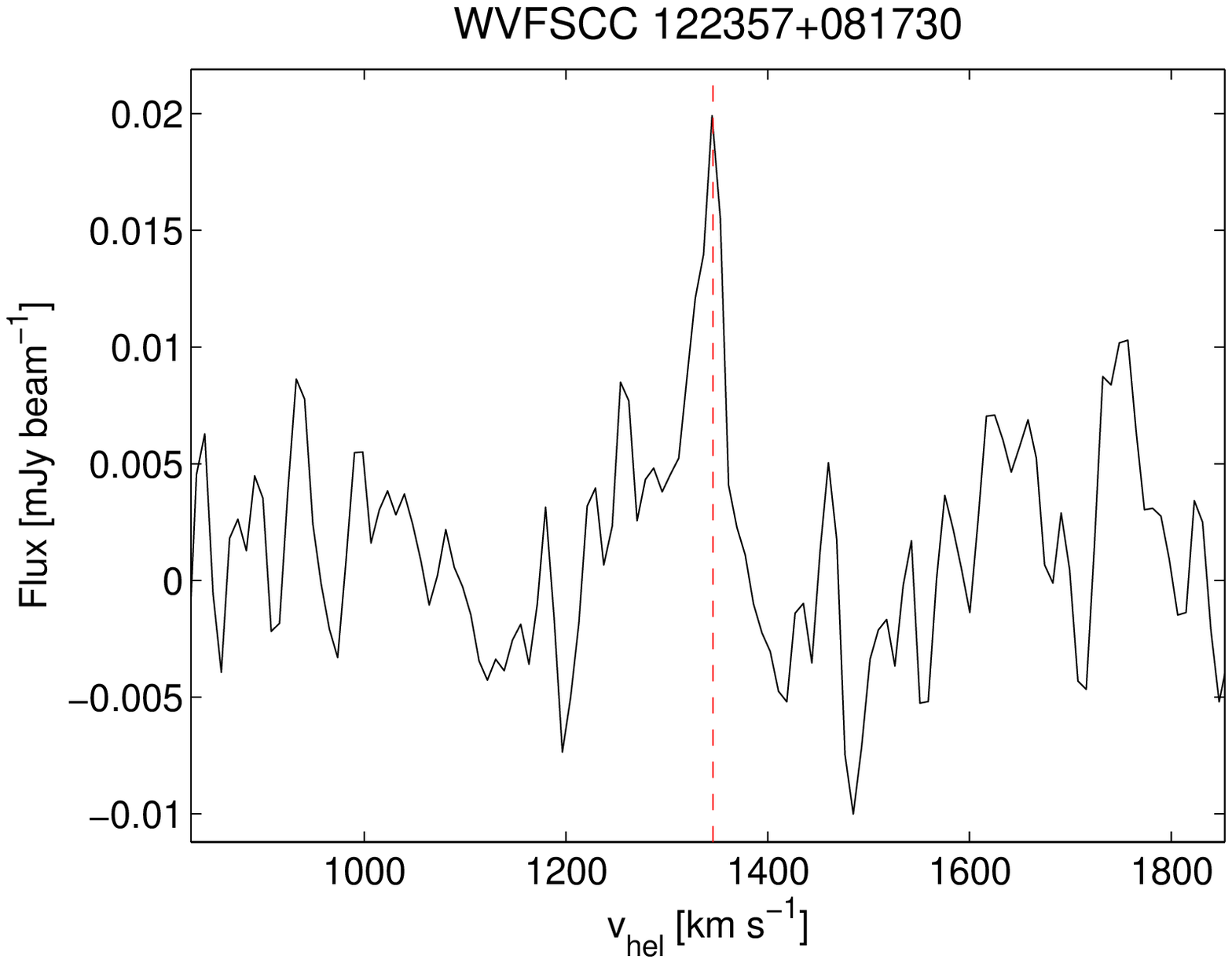}
\includegraphics[width=0.32\textwidth]{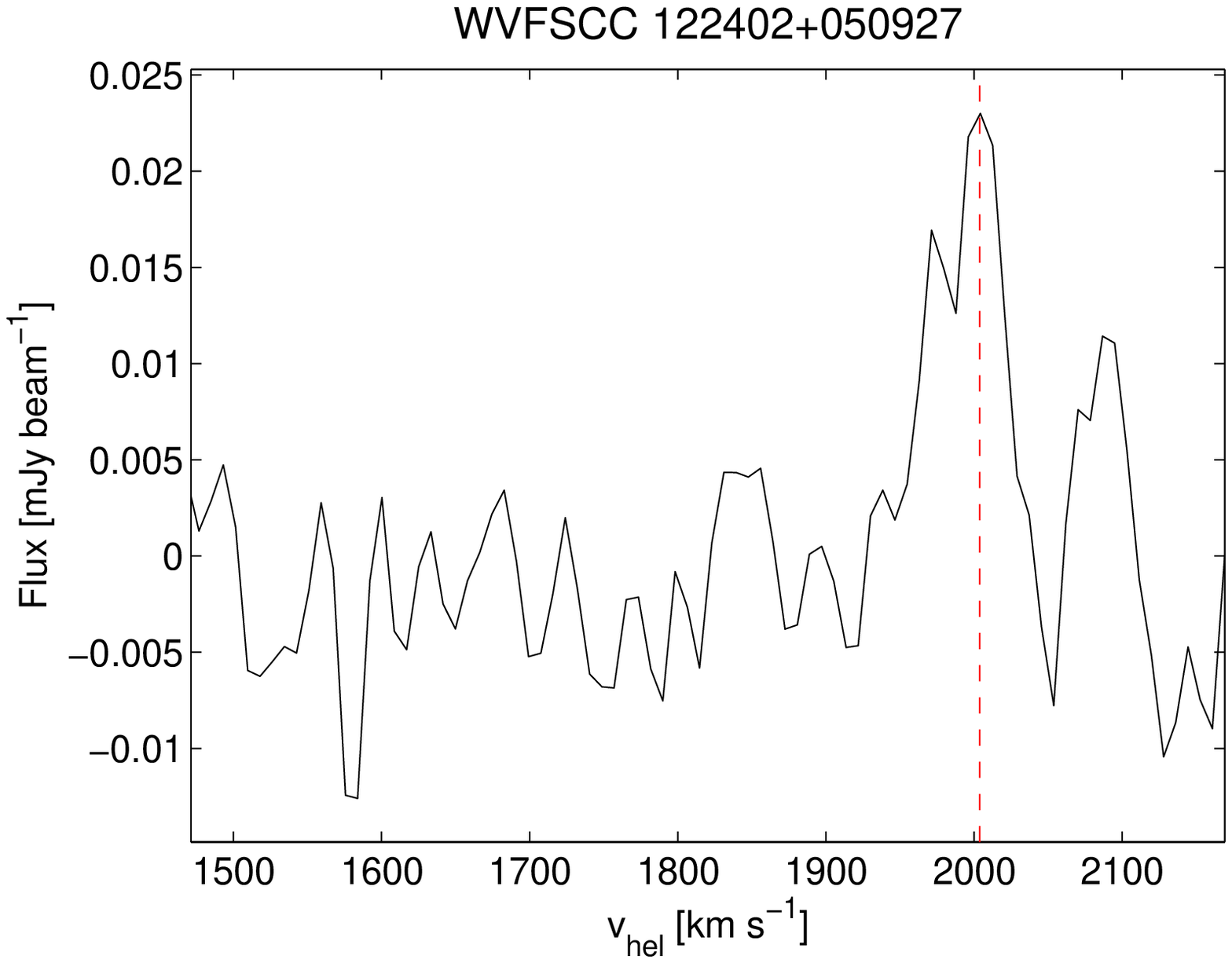}
\includegraphics[width=0.32\textwidth]{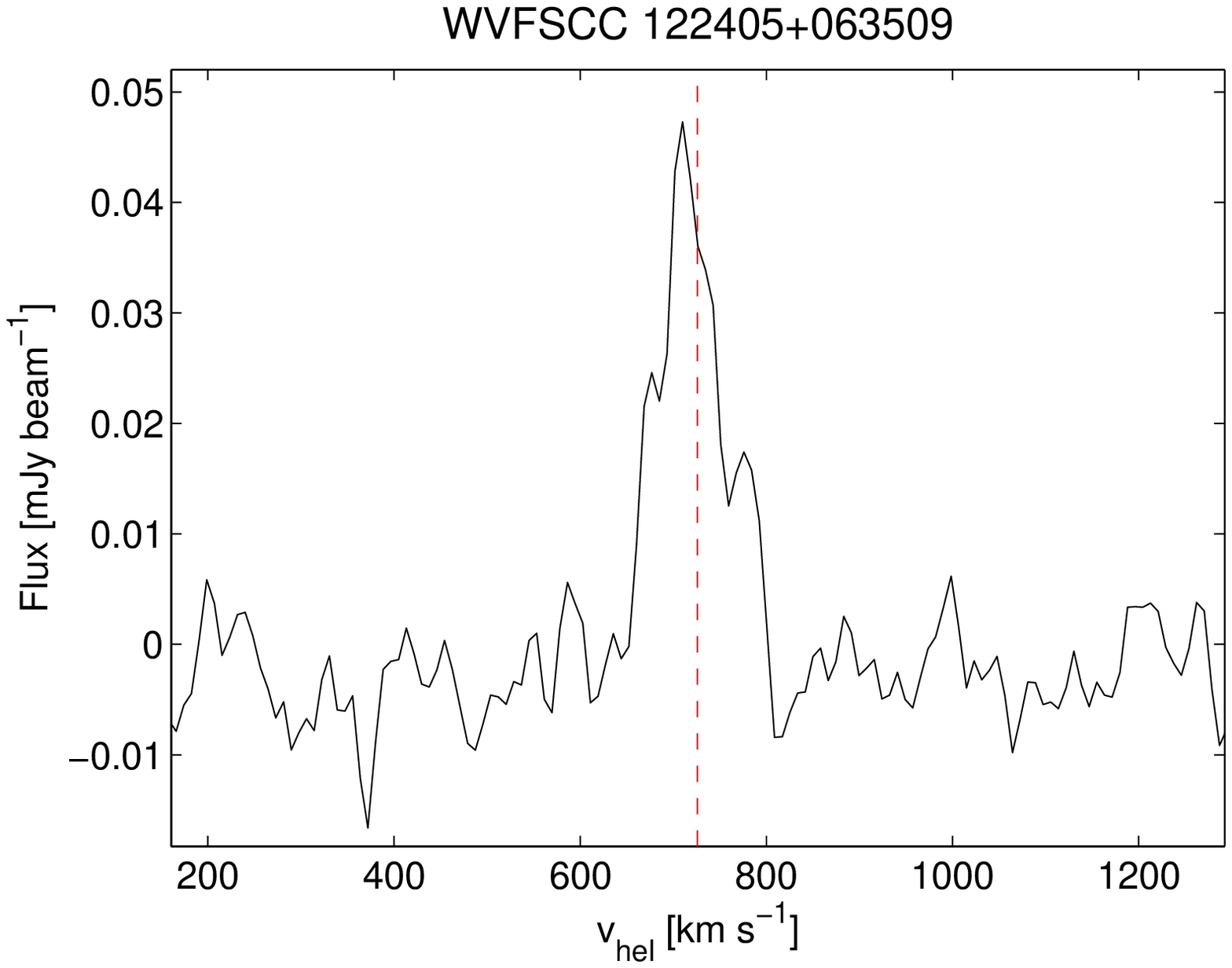}
\includegraphics[width=0.32\textwidth]{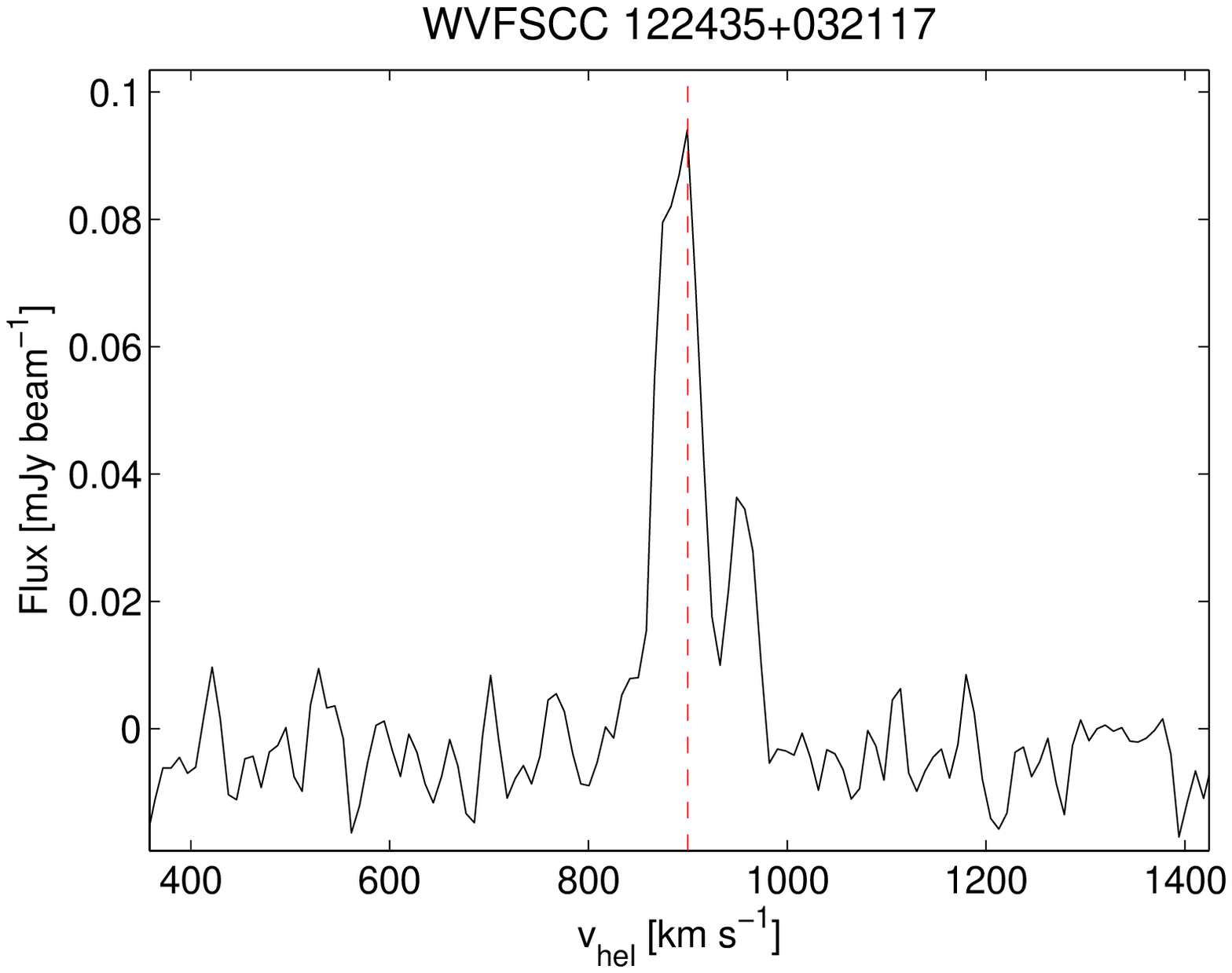}
\includegraphics[width=0.32\textwidth]{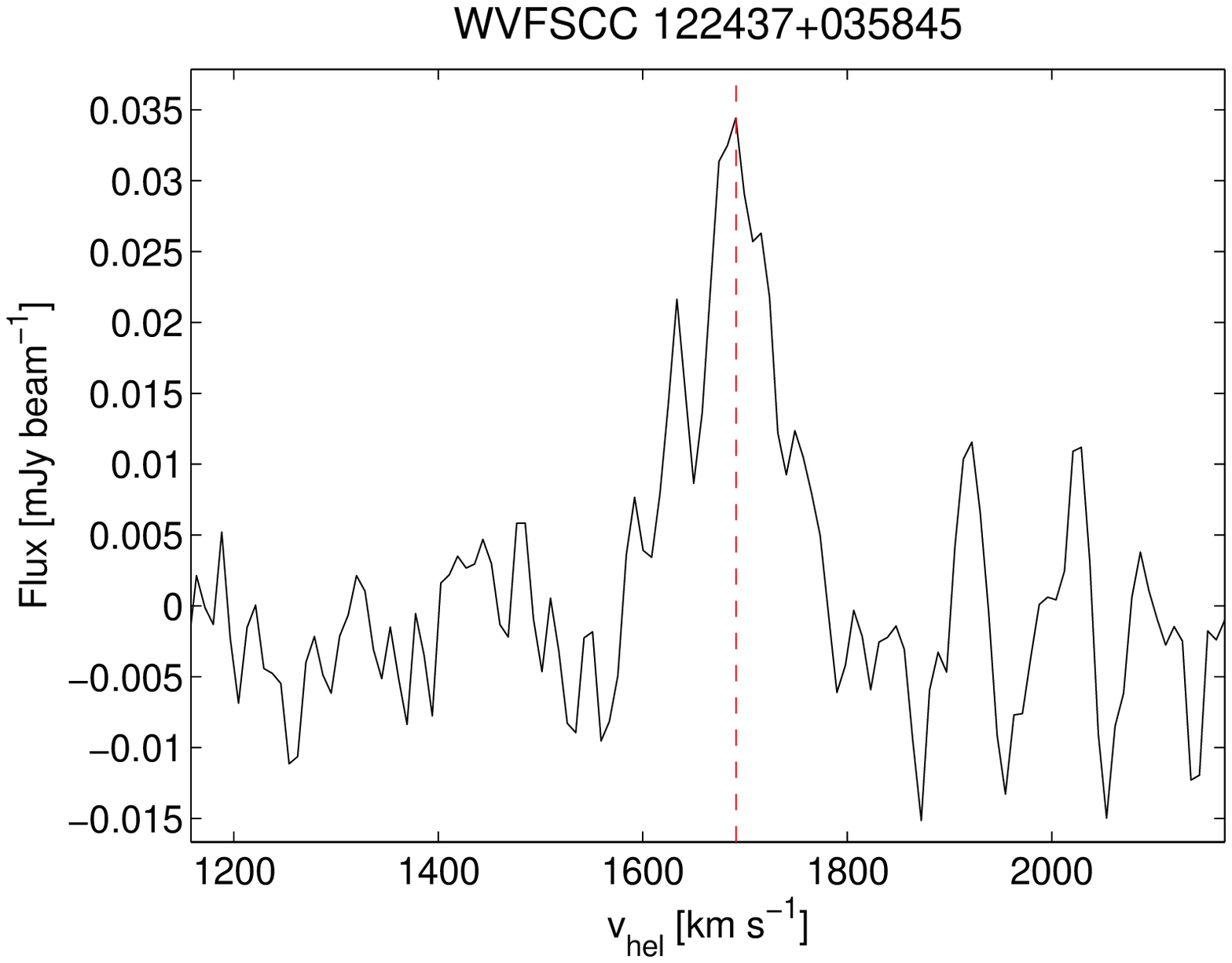}
\includegraphics[width=0.32\textwidth]{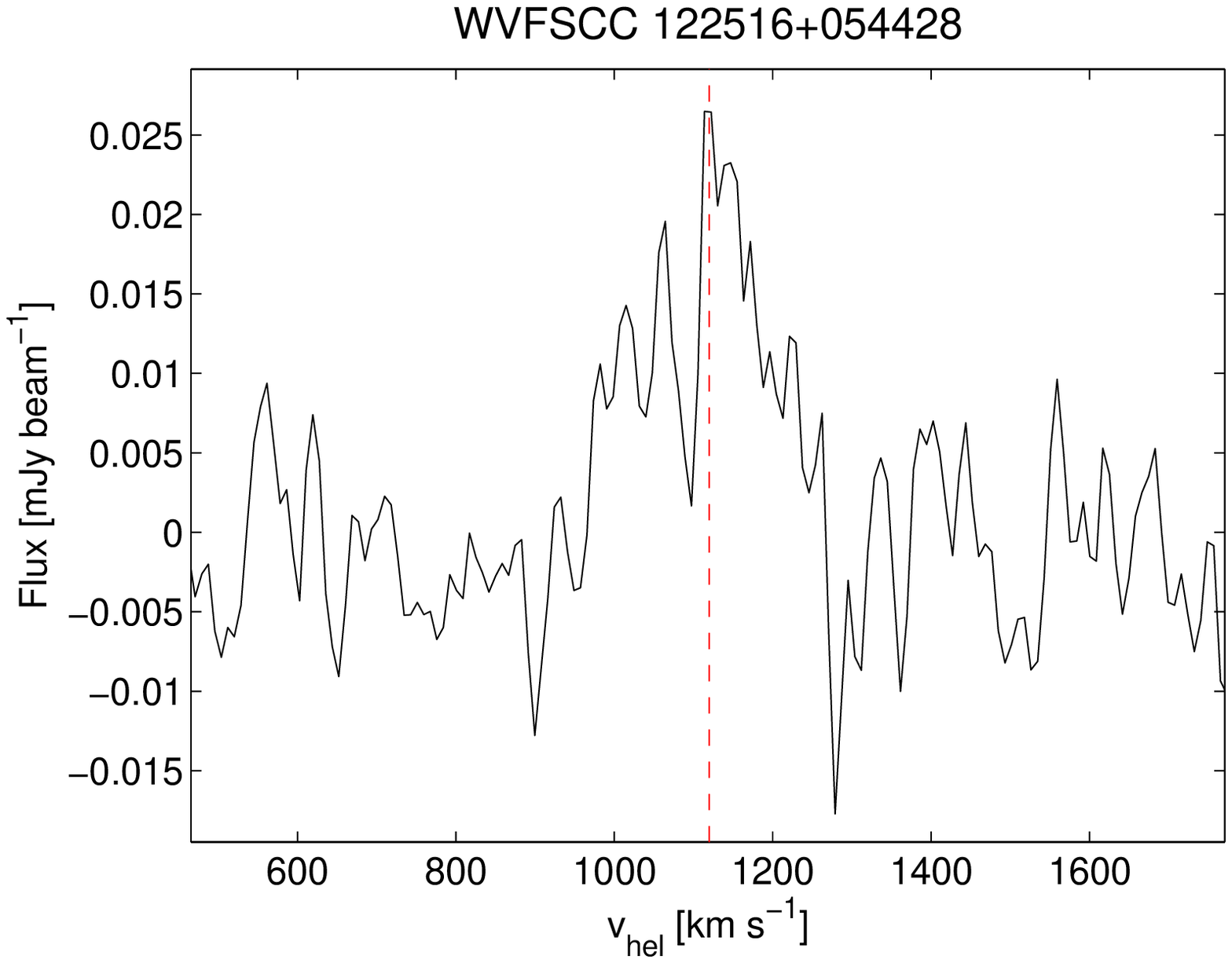}
\includegraphics[width=0.32\textwidth]{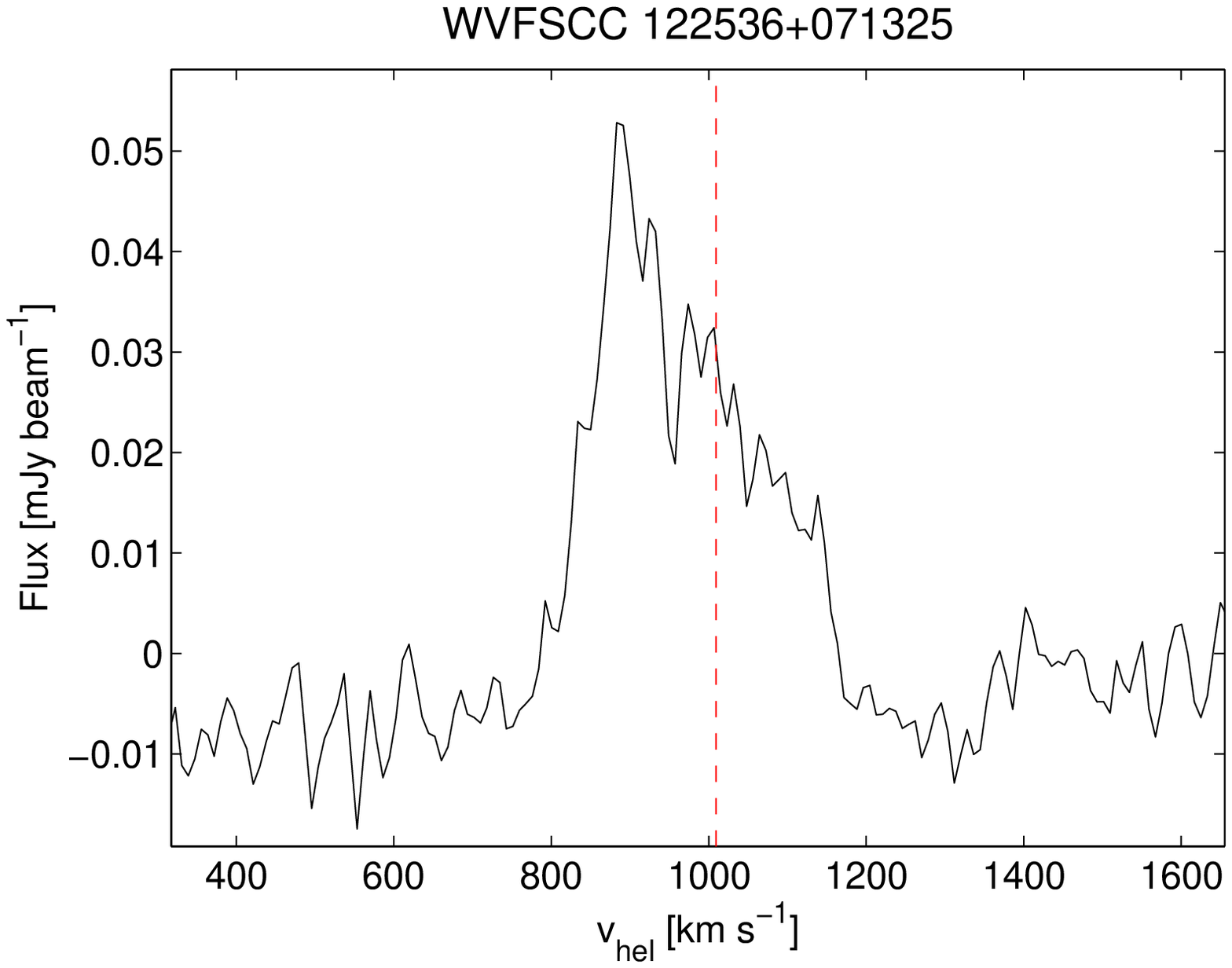} 
\includegraphics[width=0.32\textwidth]{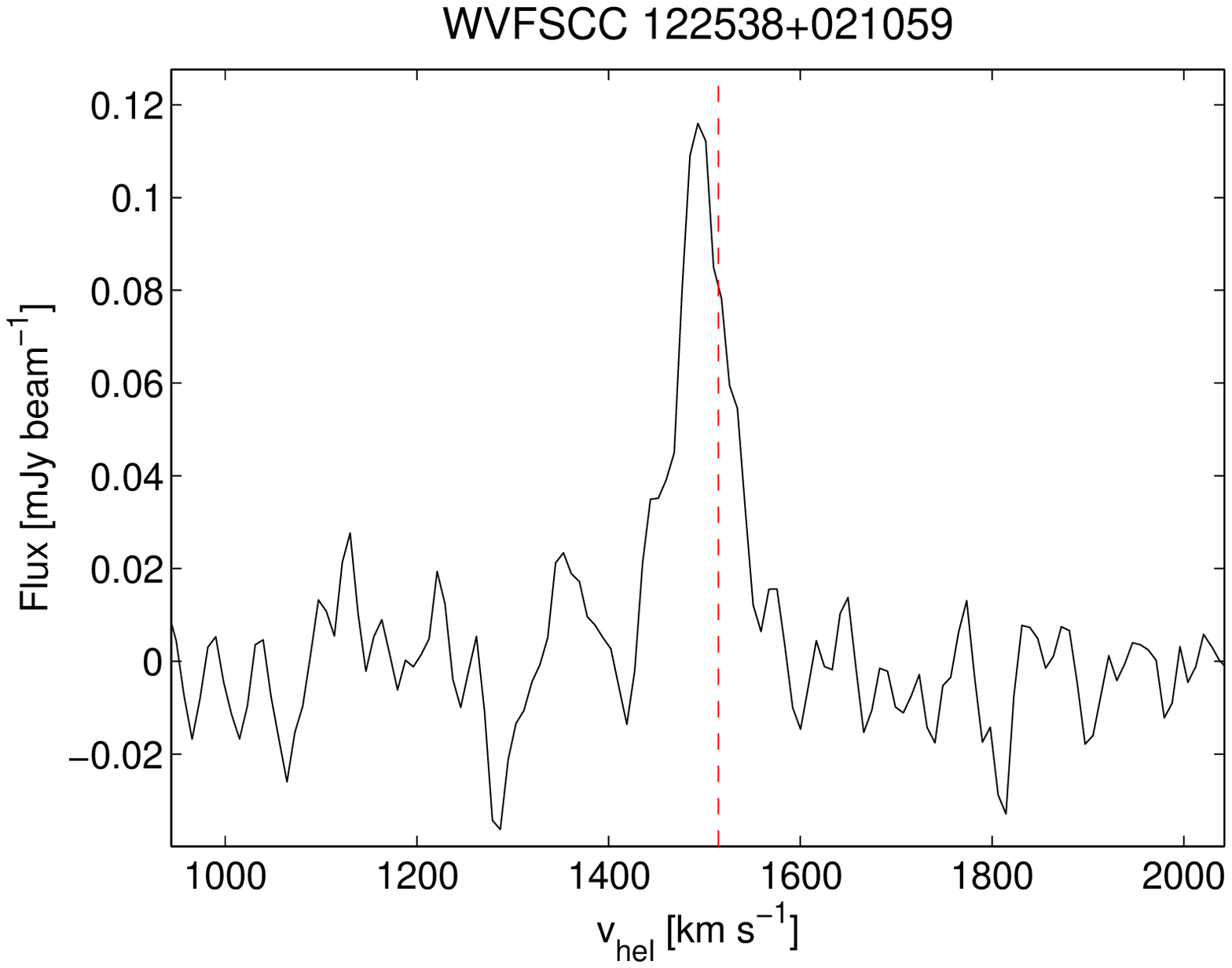}

\end{center}                                            
{\bf Fig~\ref{all_spectra2}.} (continued)                                        
 
\end{figure*}


\begin{figure*}
  \begin{center}
 
\includegraphics[width=0.32\textwidth]{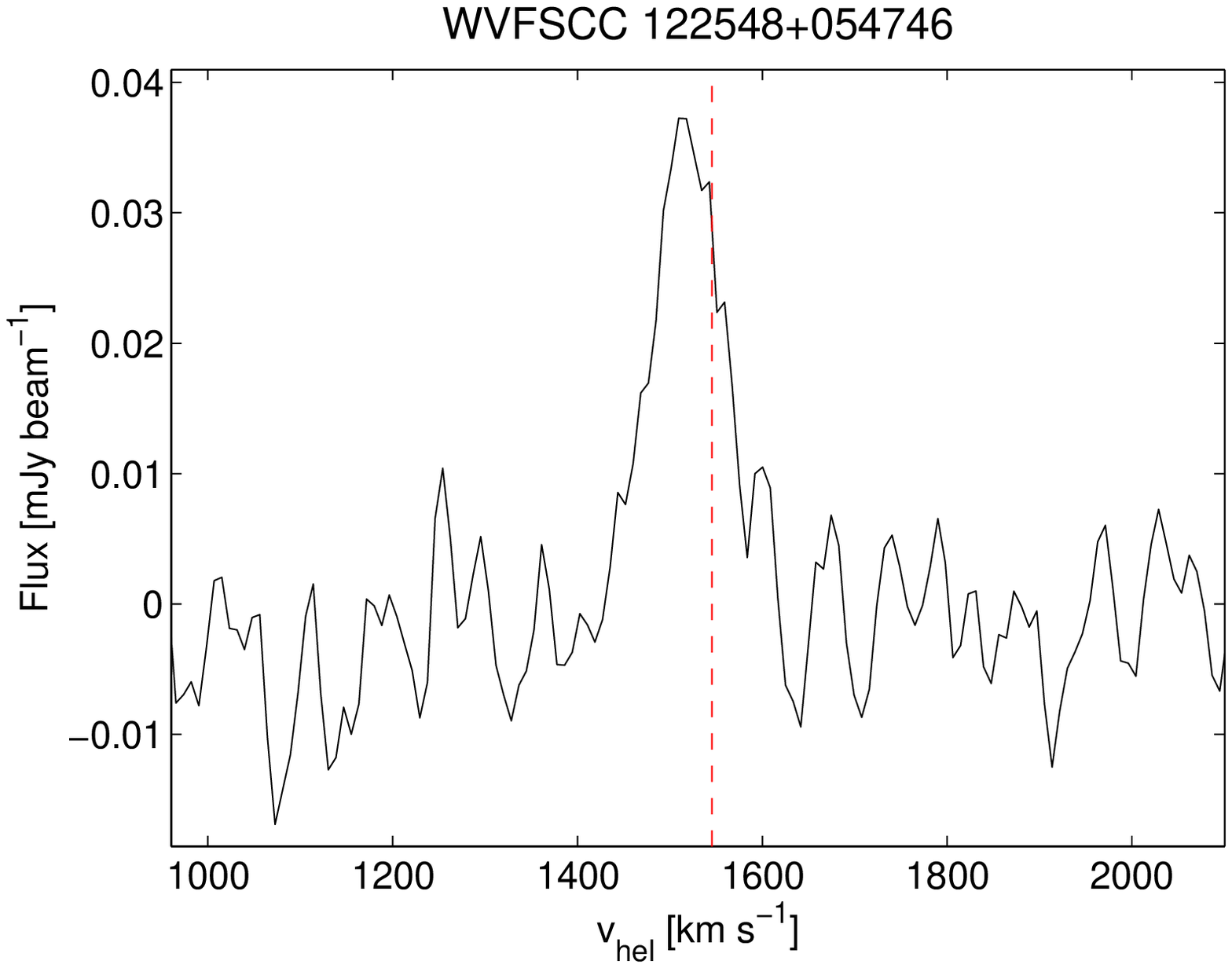}
\includegraphics[width=0.32\textwidth]{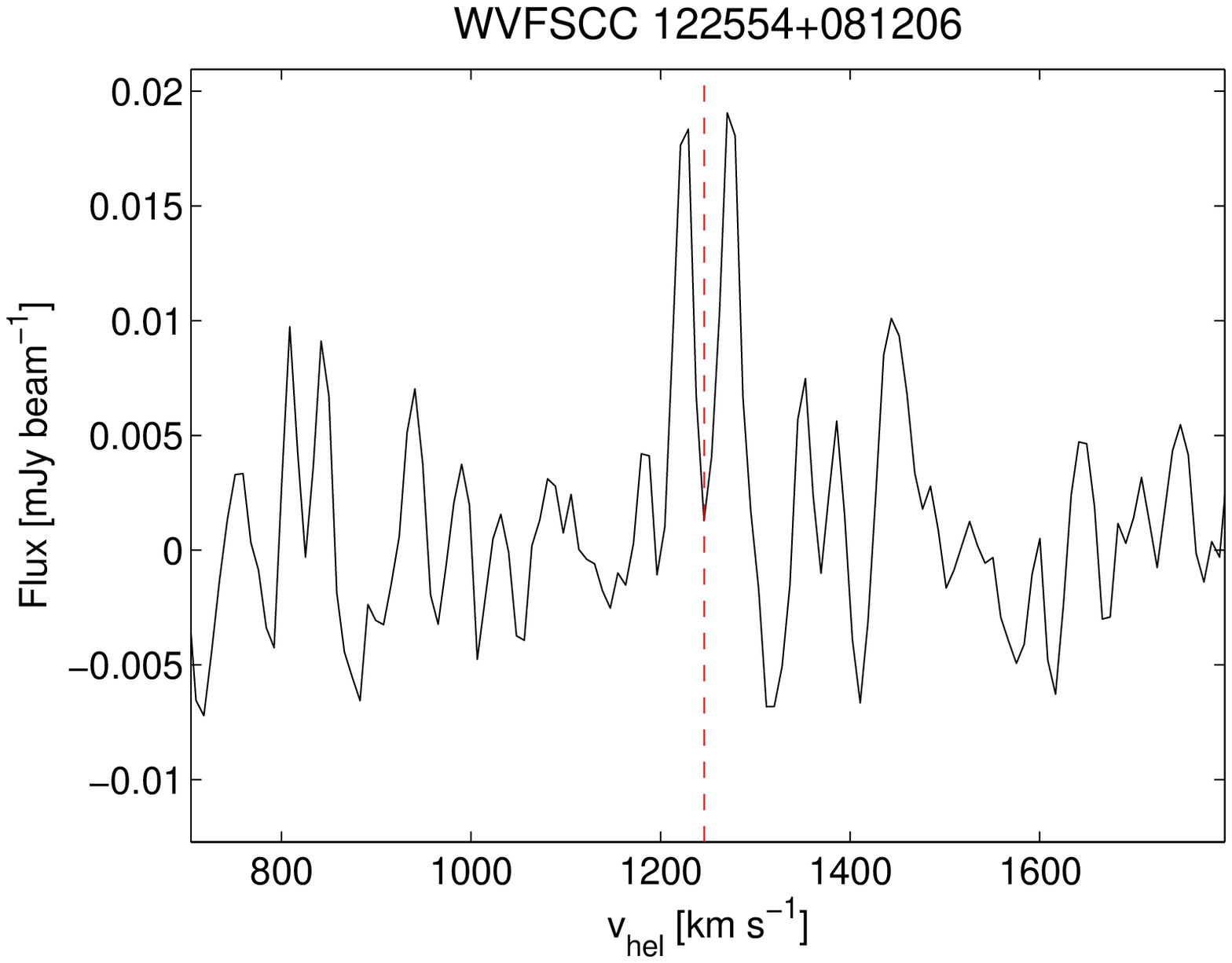}
\includegraphics[width=0.32\textwidth]{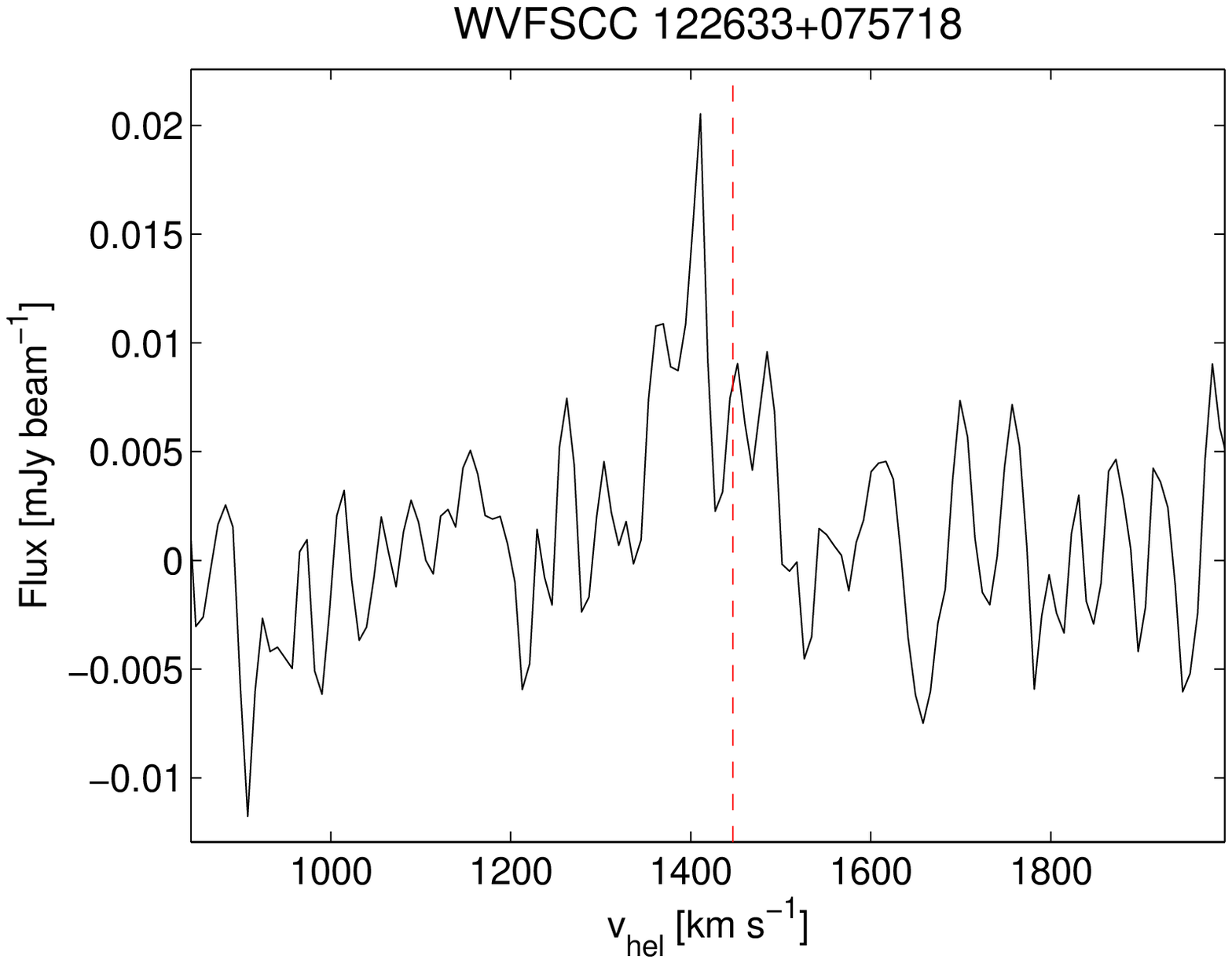}
\includegraphics[width=0.32\textwidth]{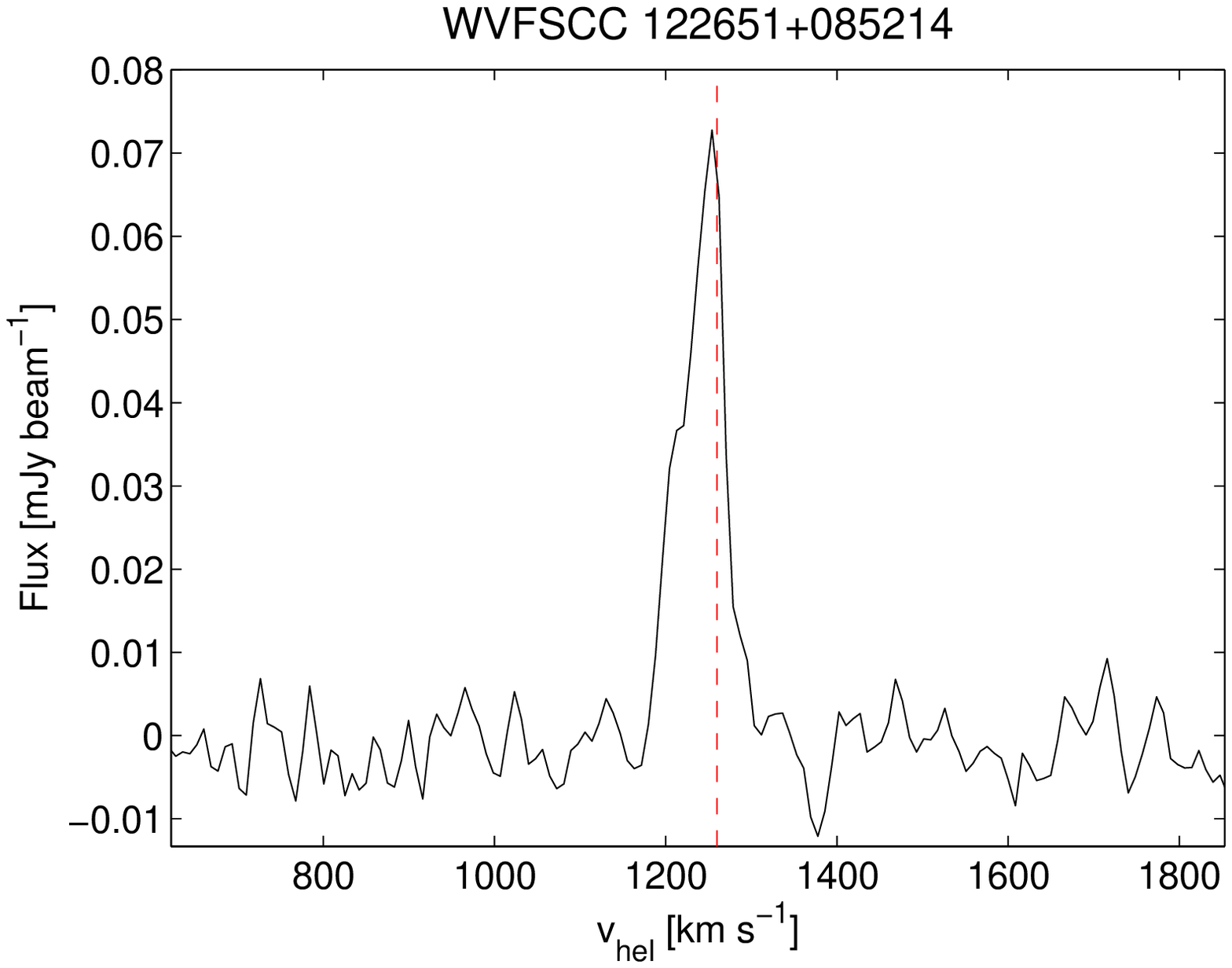}
\includegraphics[width=0.32\textwidth]{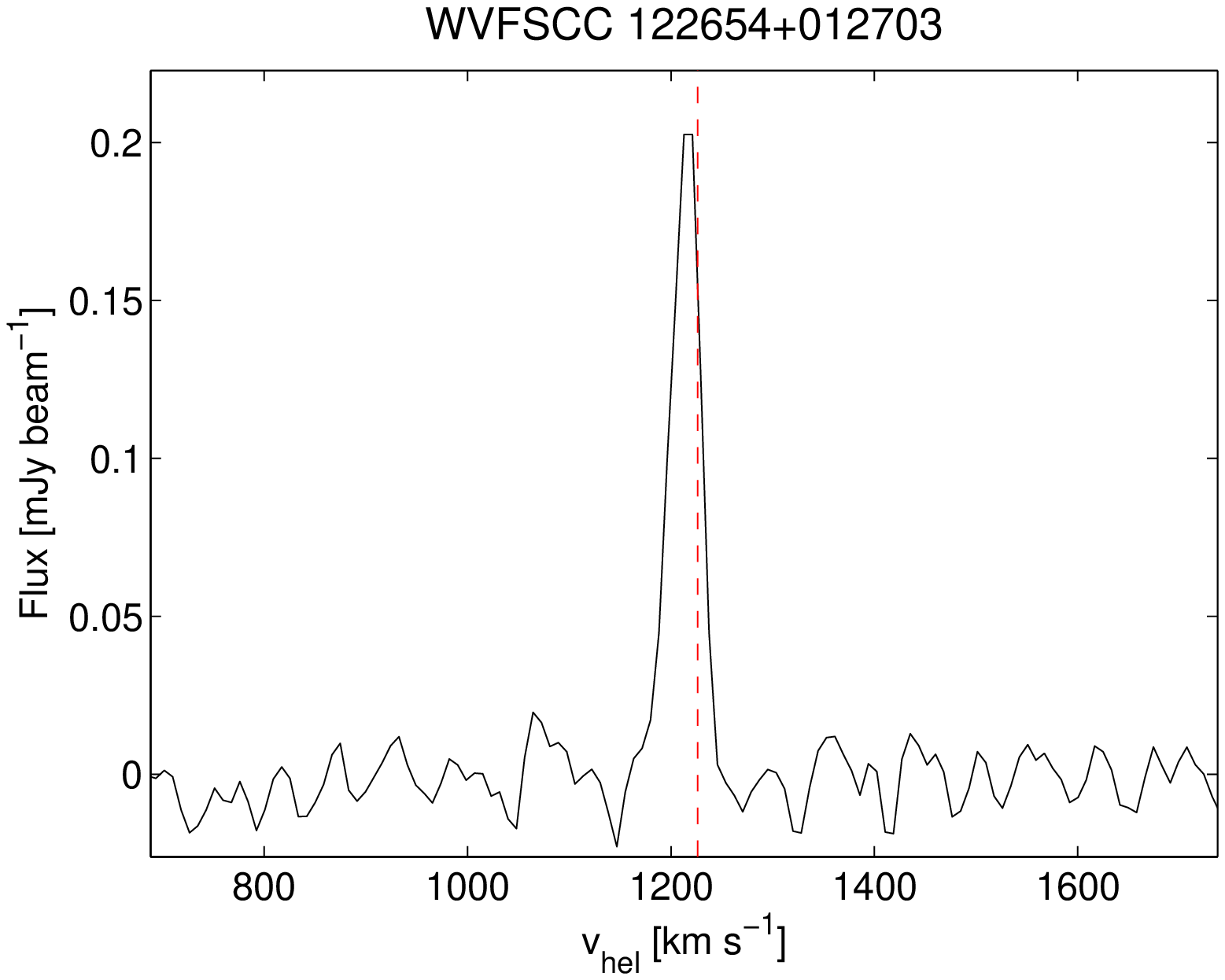}
\includegraphics[width=0.32\textwidth]{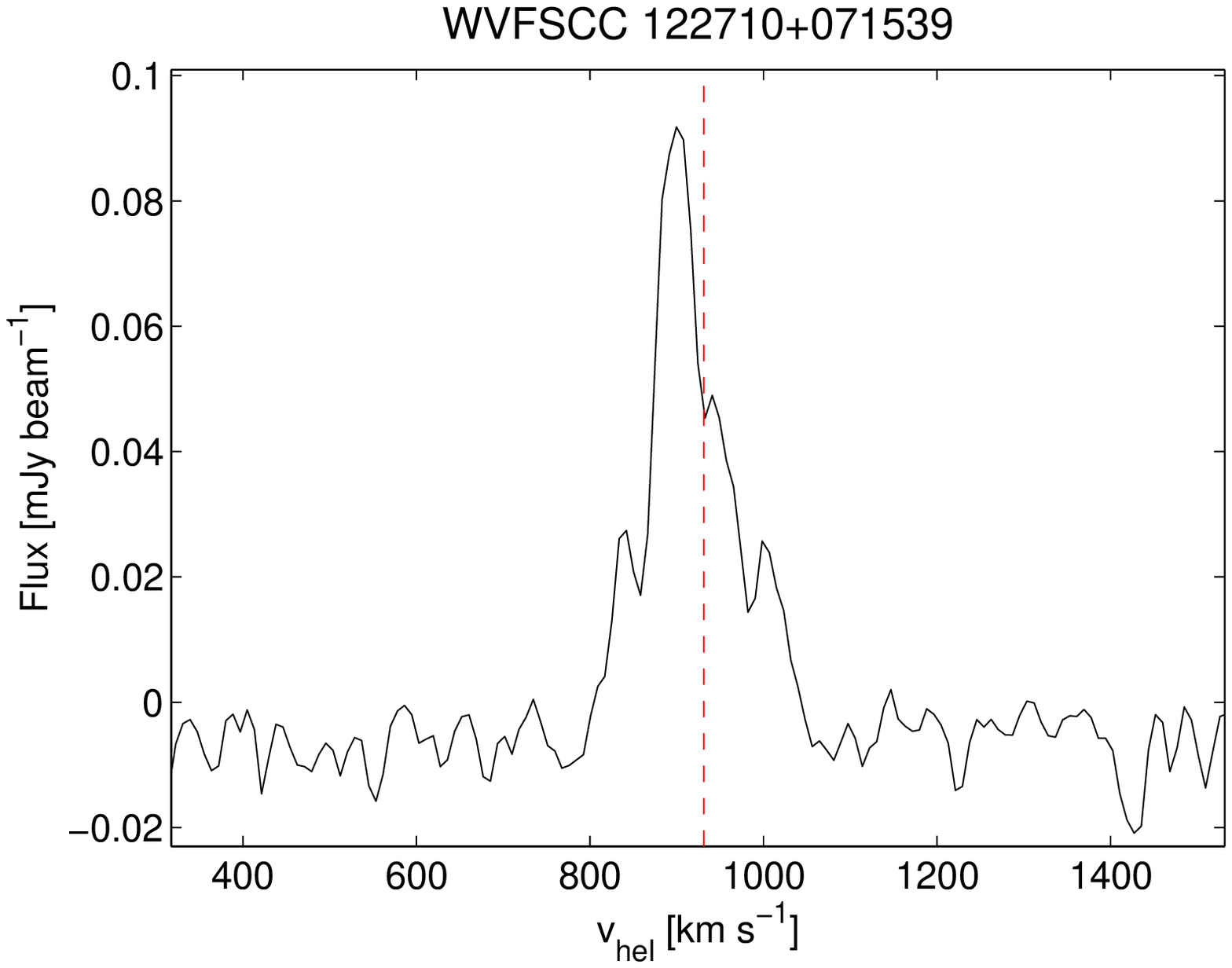}
\includegraphics[width=0.32\textwidth]{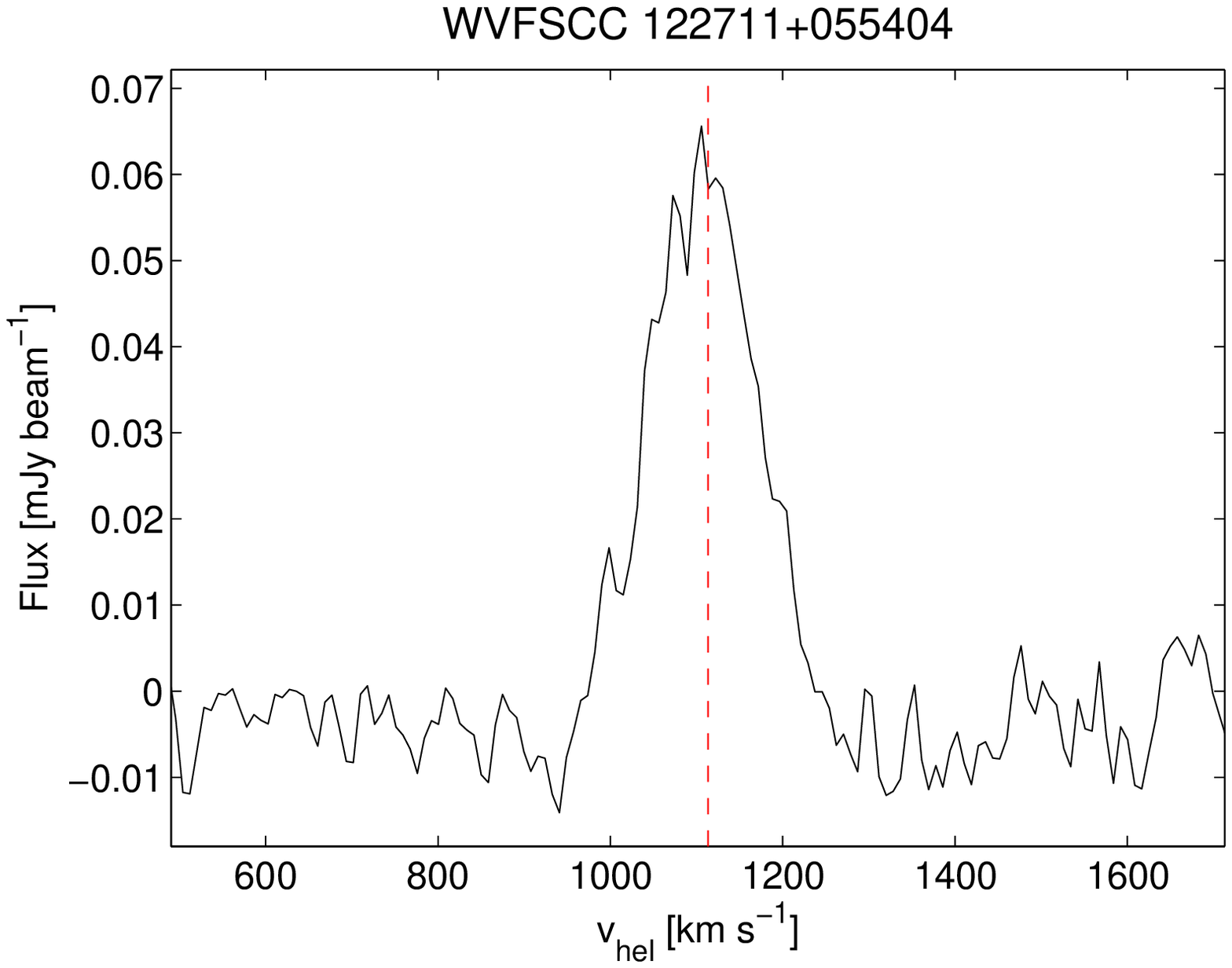}
\includegraphics[width=0.32\textwidth]{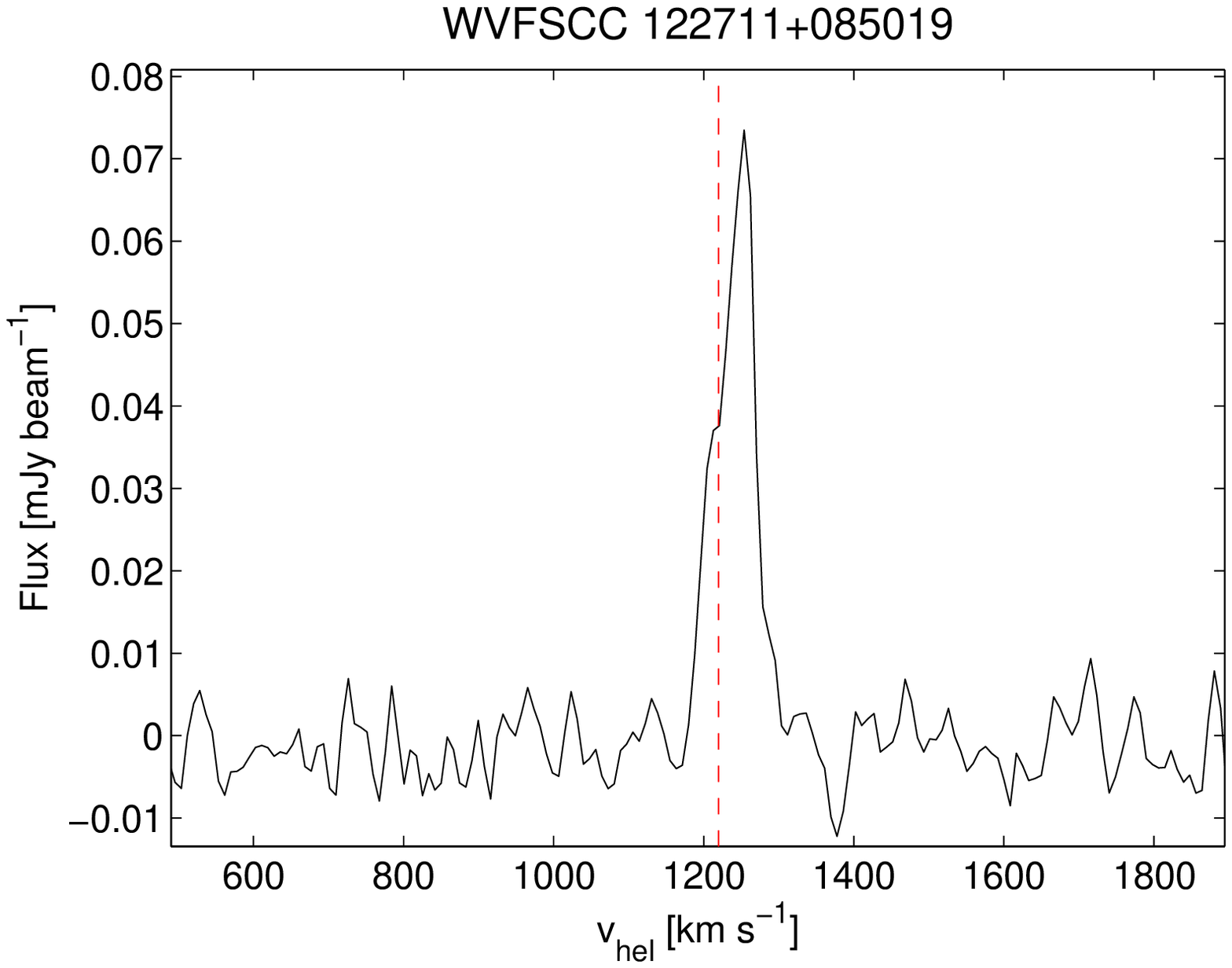}
\includegraphics[width=0.32\textwidth]{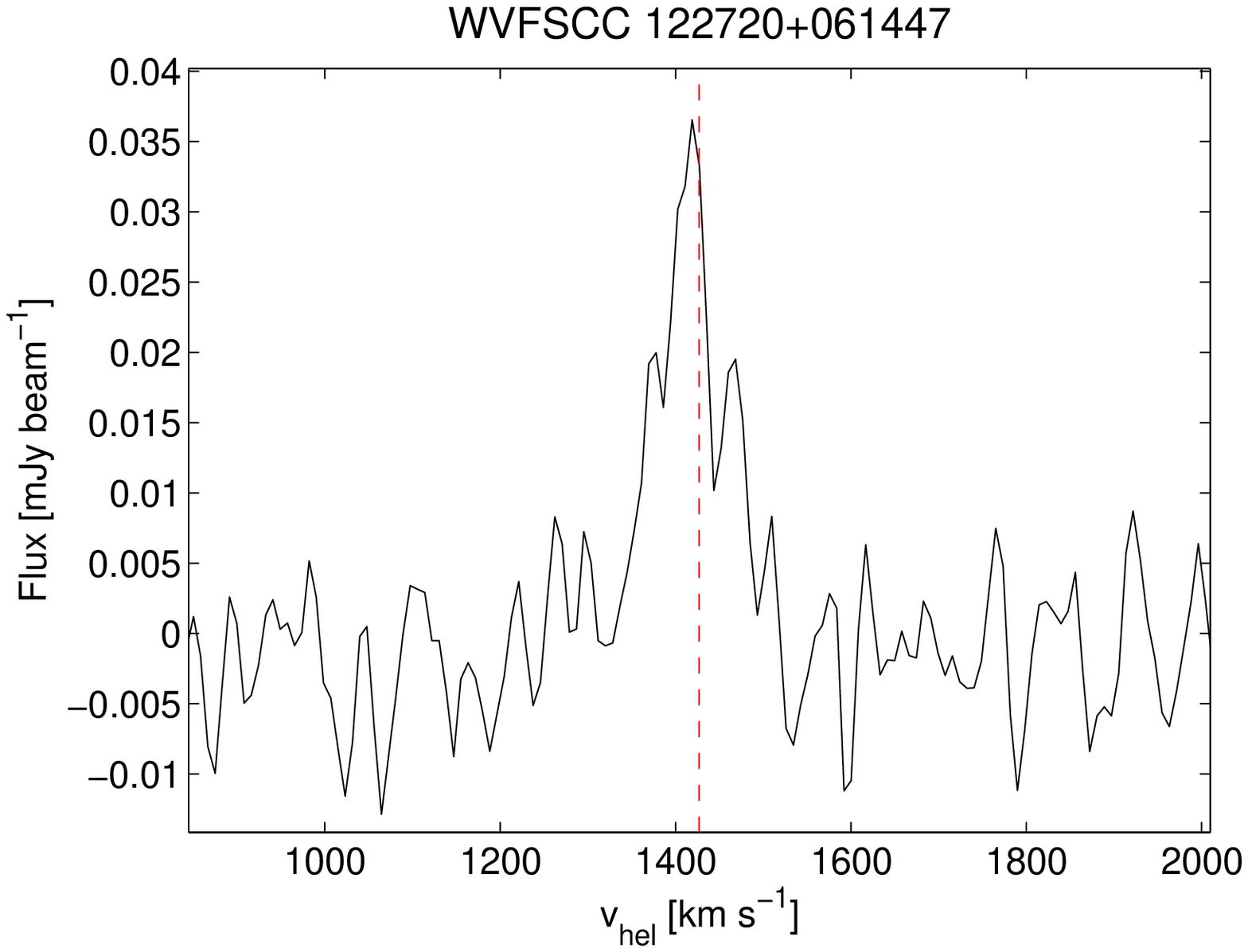}
\includegraphics[width=0.32\textwidth]{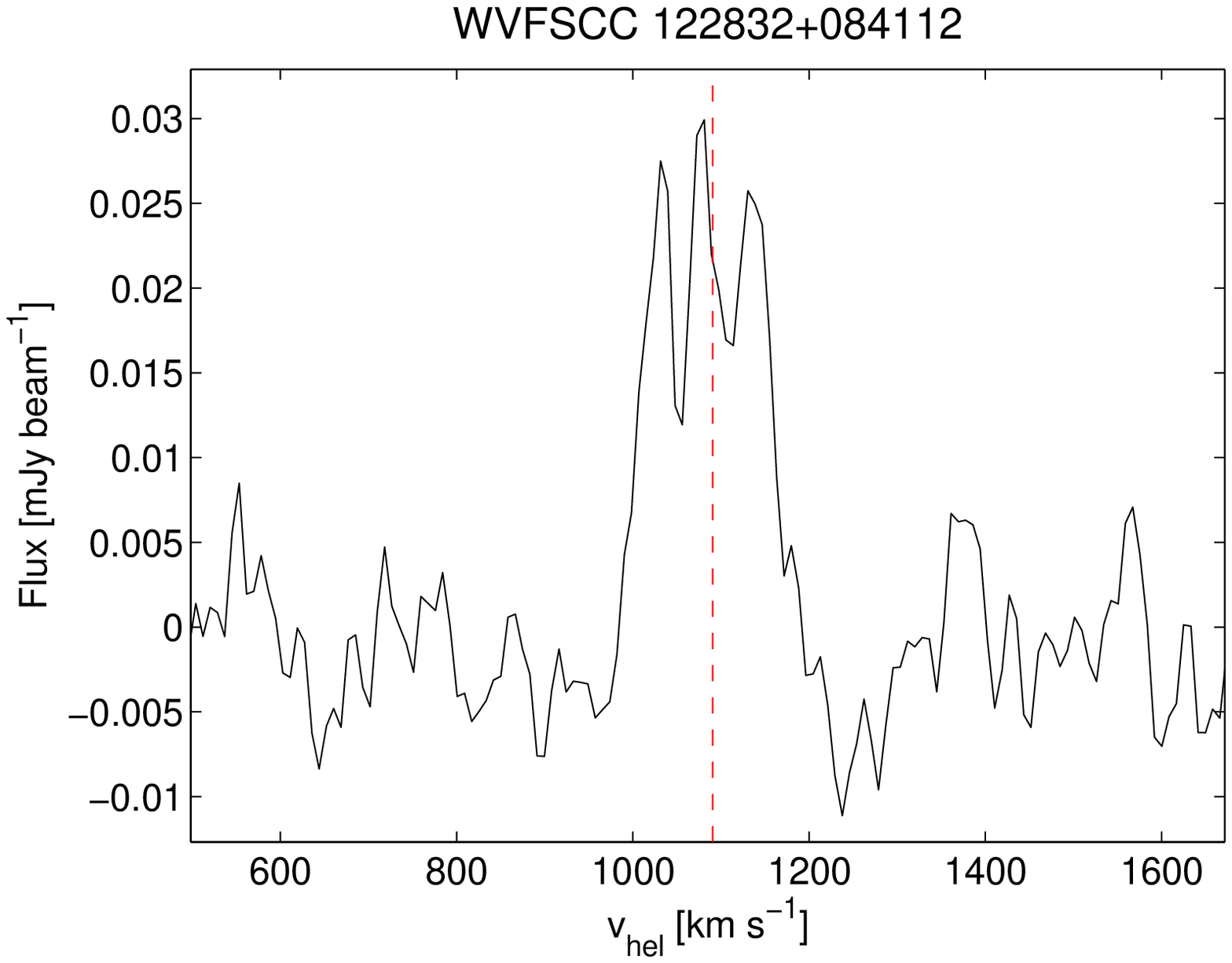}
\includegraphics[width=0.32\textwidth]{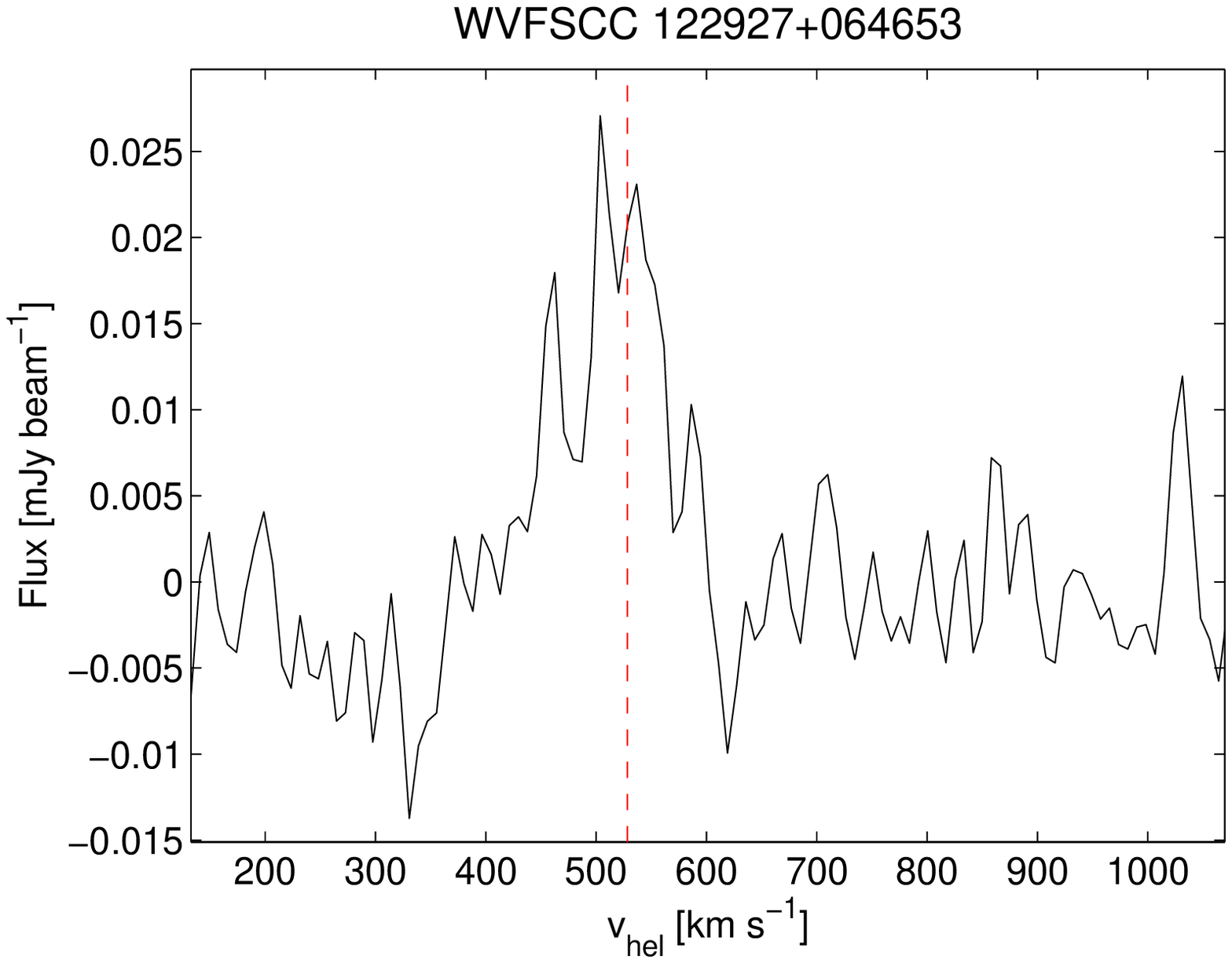}
\includegraphics[width=0.32\textwidth]{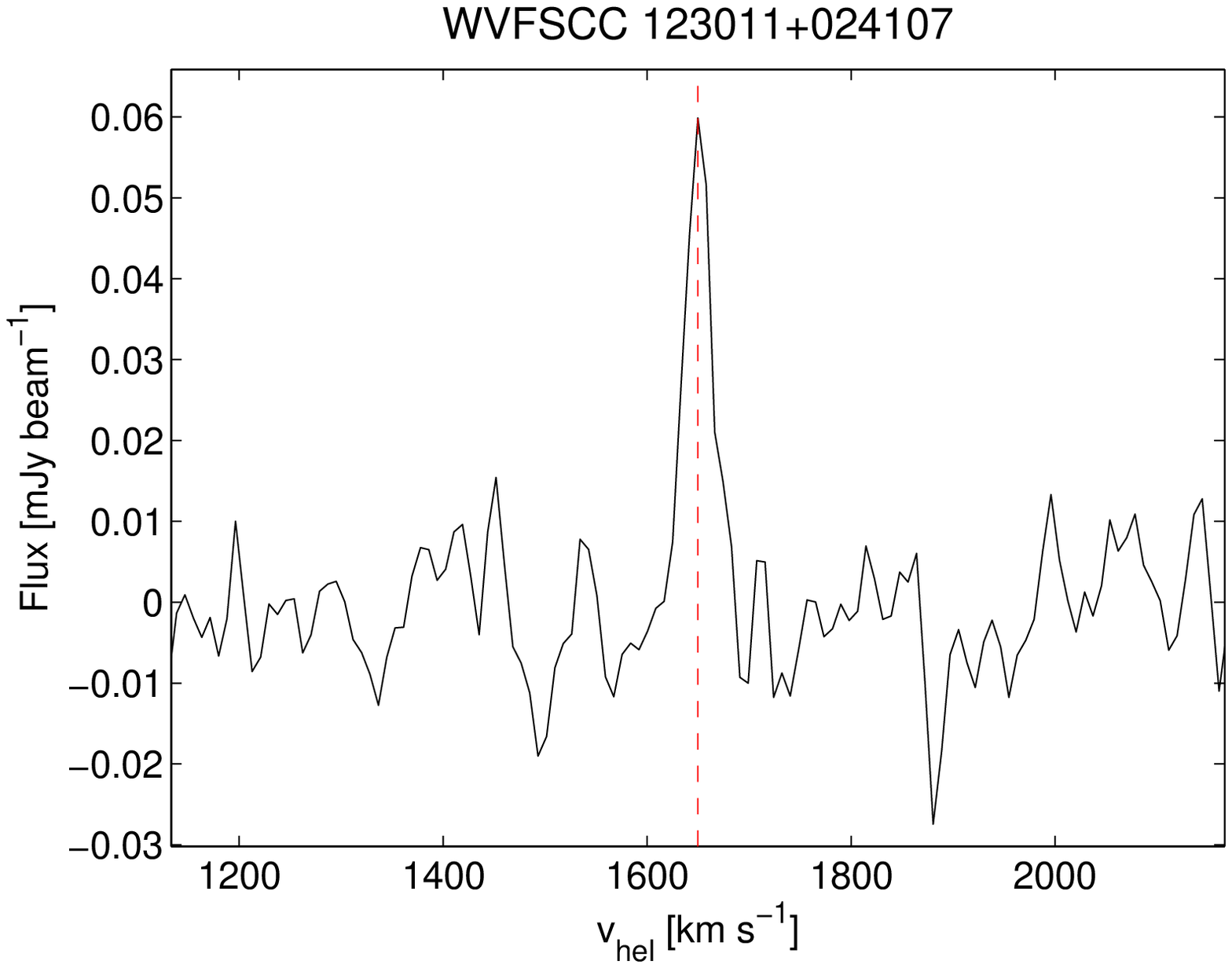}
\includegraphics[width=0.32\textwidth]{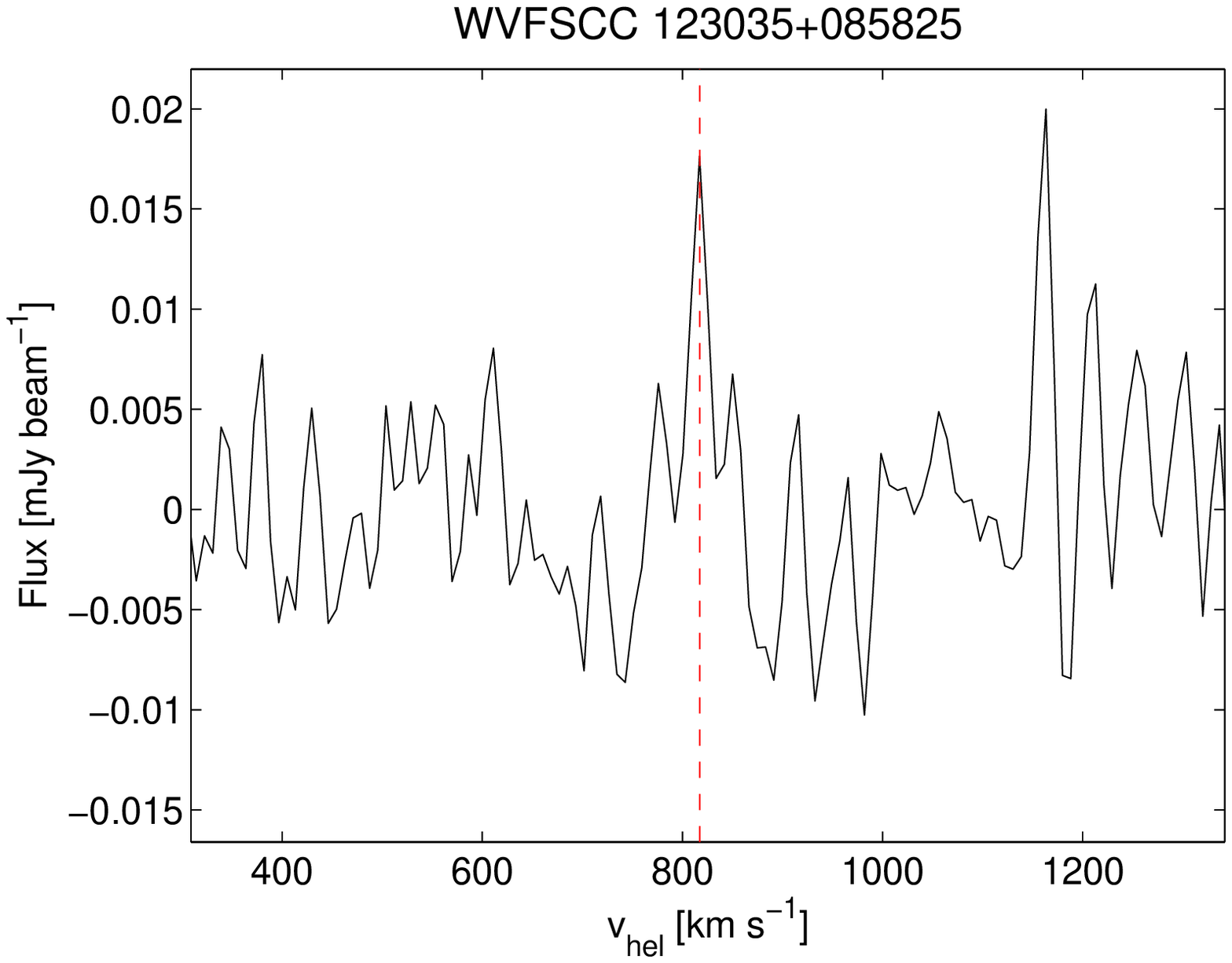}
\includegraphics[width=0.32\textwidth]{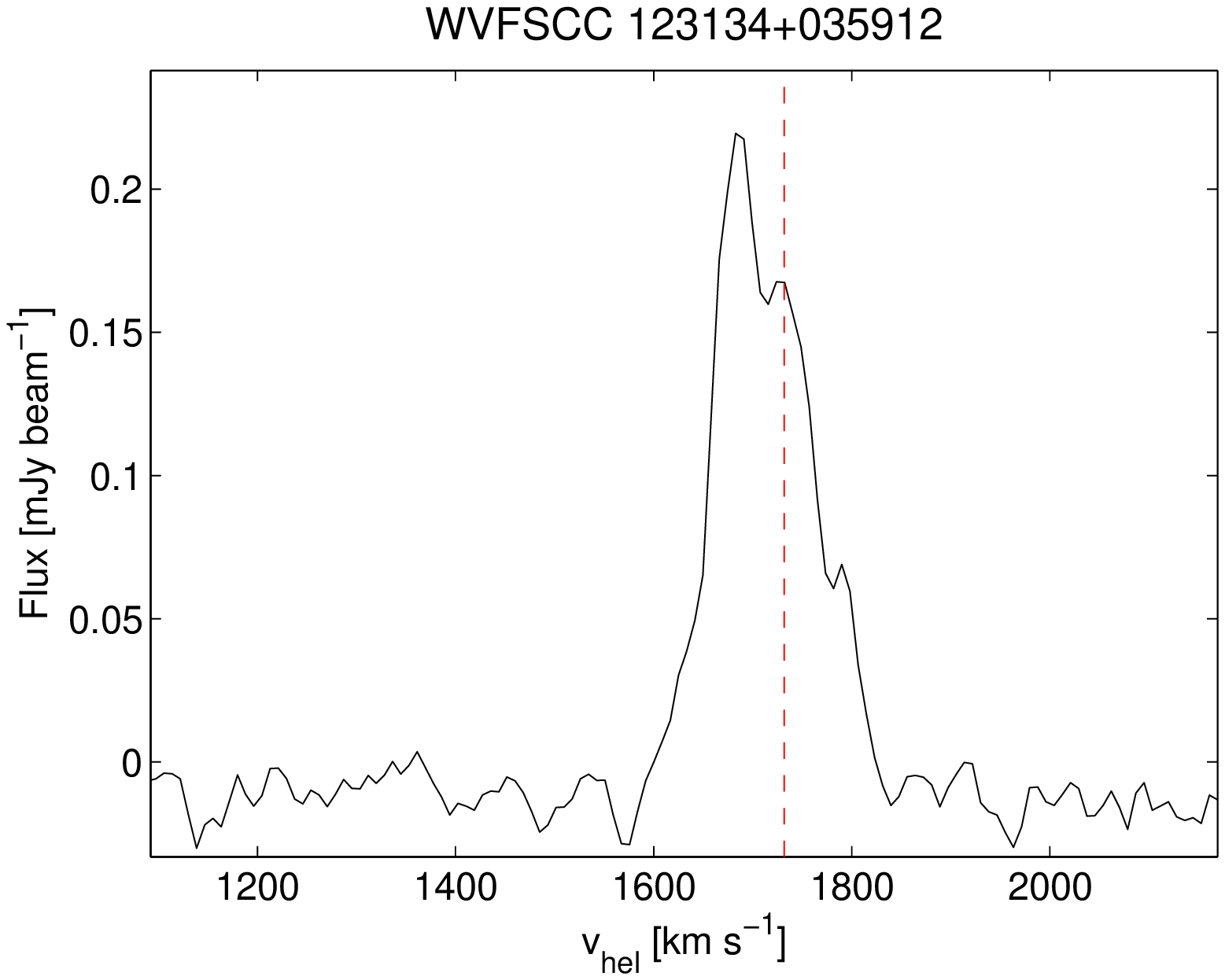}
\includegraphics[width=0.32\textwidth]{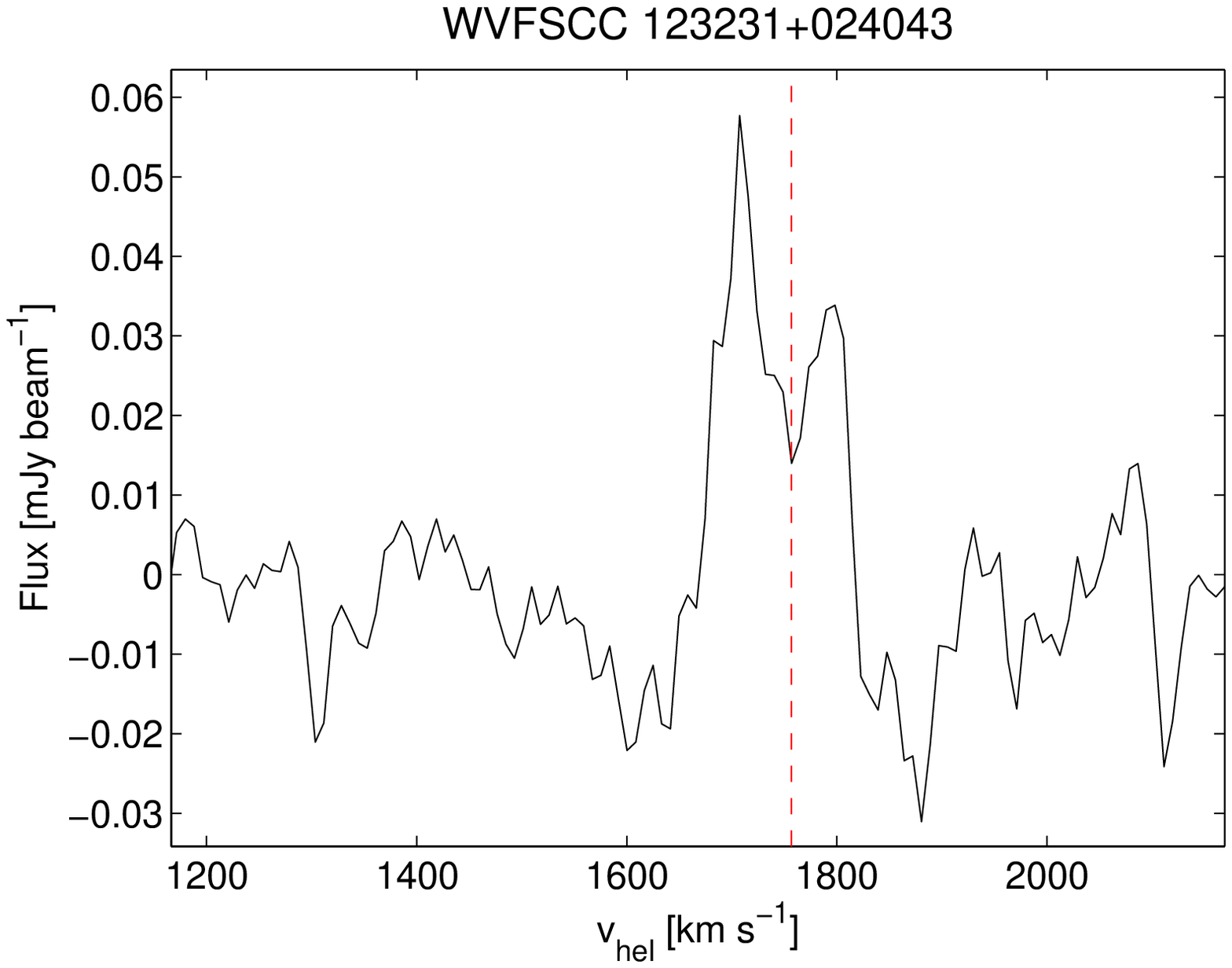}

\end{center}                                            
{\bf Fig~\ref{all_spectra2}.} (continued)                                        
 
\end{figure*}


\begin{figure*}
  \begin{center}

\includegraphics[width=0.32\textwidth]{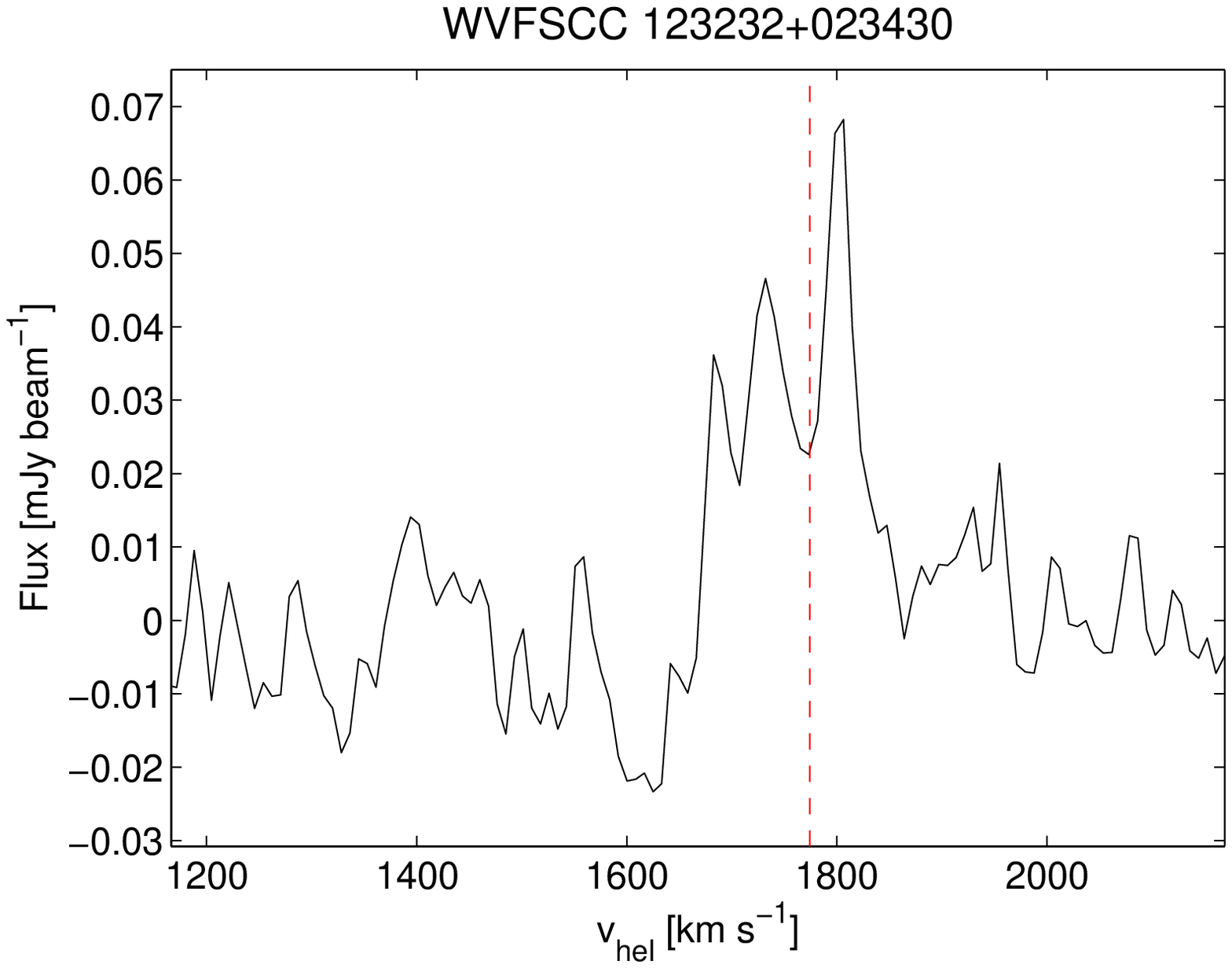}
\includegraphics[width=0.32\textwidth]{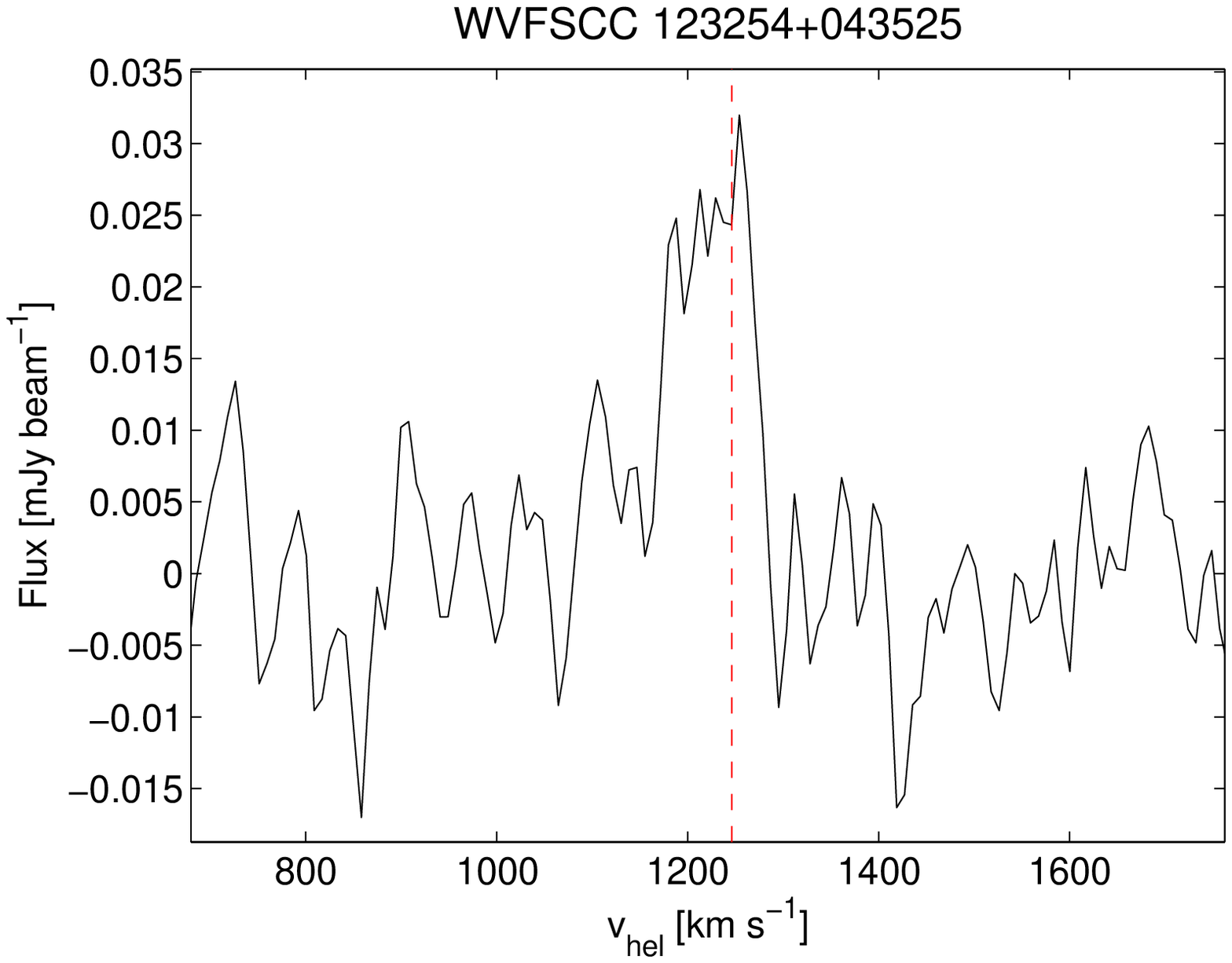} 
\includegraphics[width=0.32\textwidth]{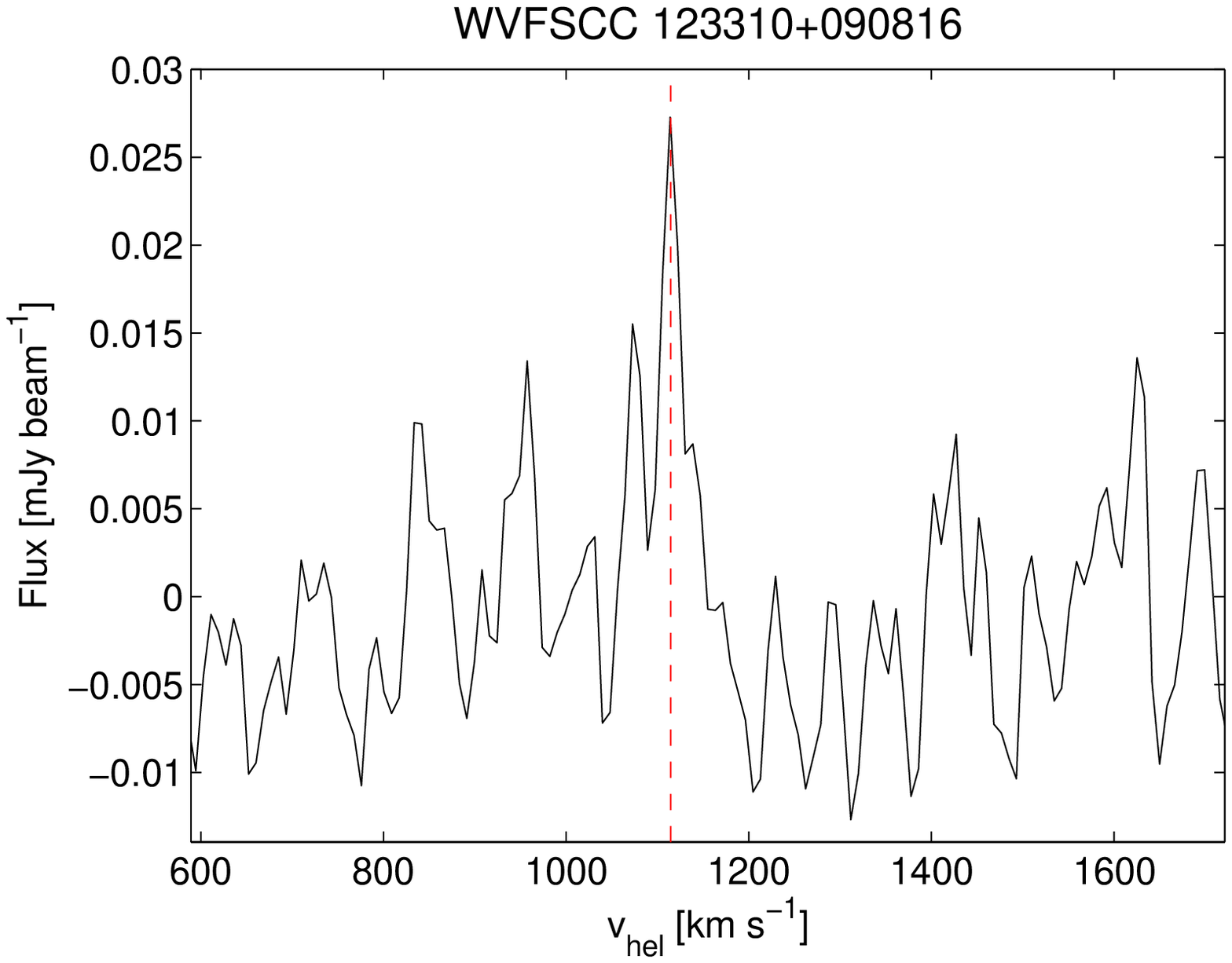}
\includegraphics[width=0.32\textwidth]{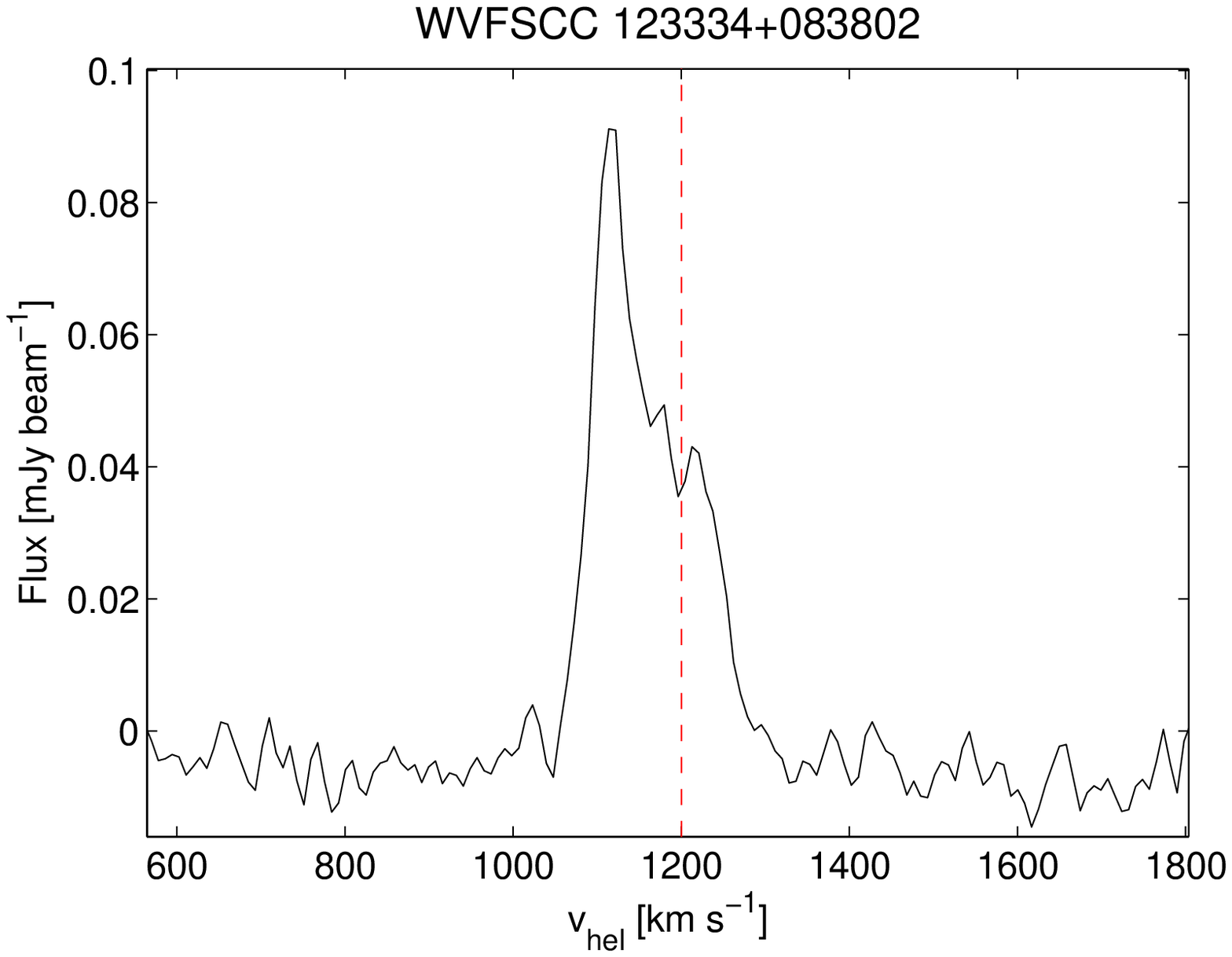}
\includegraphics[width=0.32\textwidth]{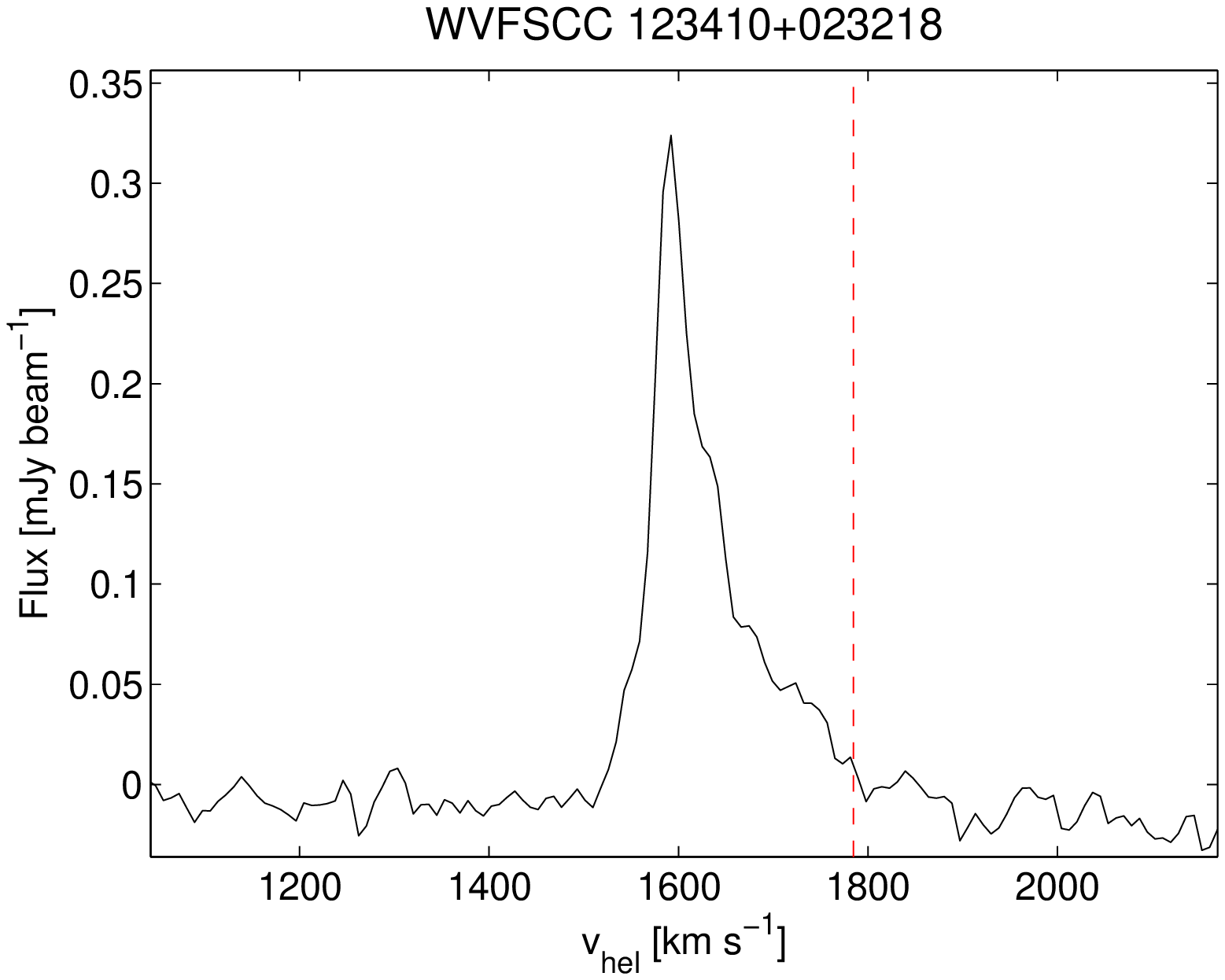}
\includegraphics[width=0.32\textwidth]{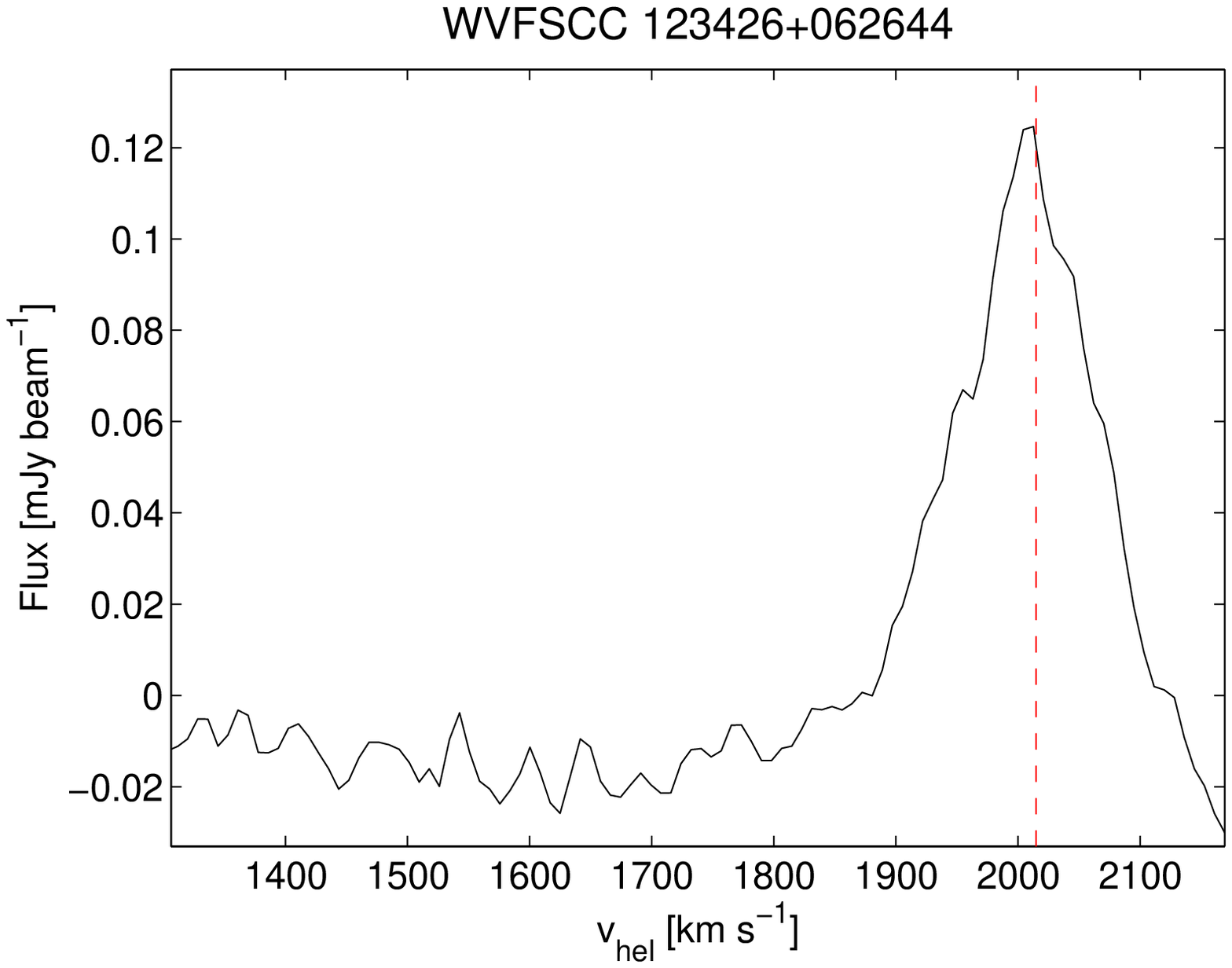}
\includegraphics[width=0.32\textwidth]{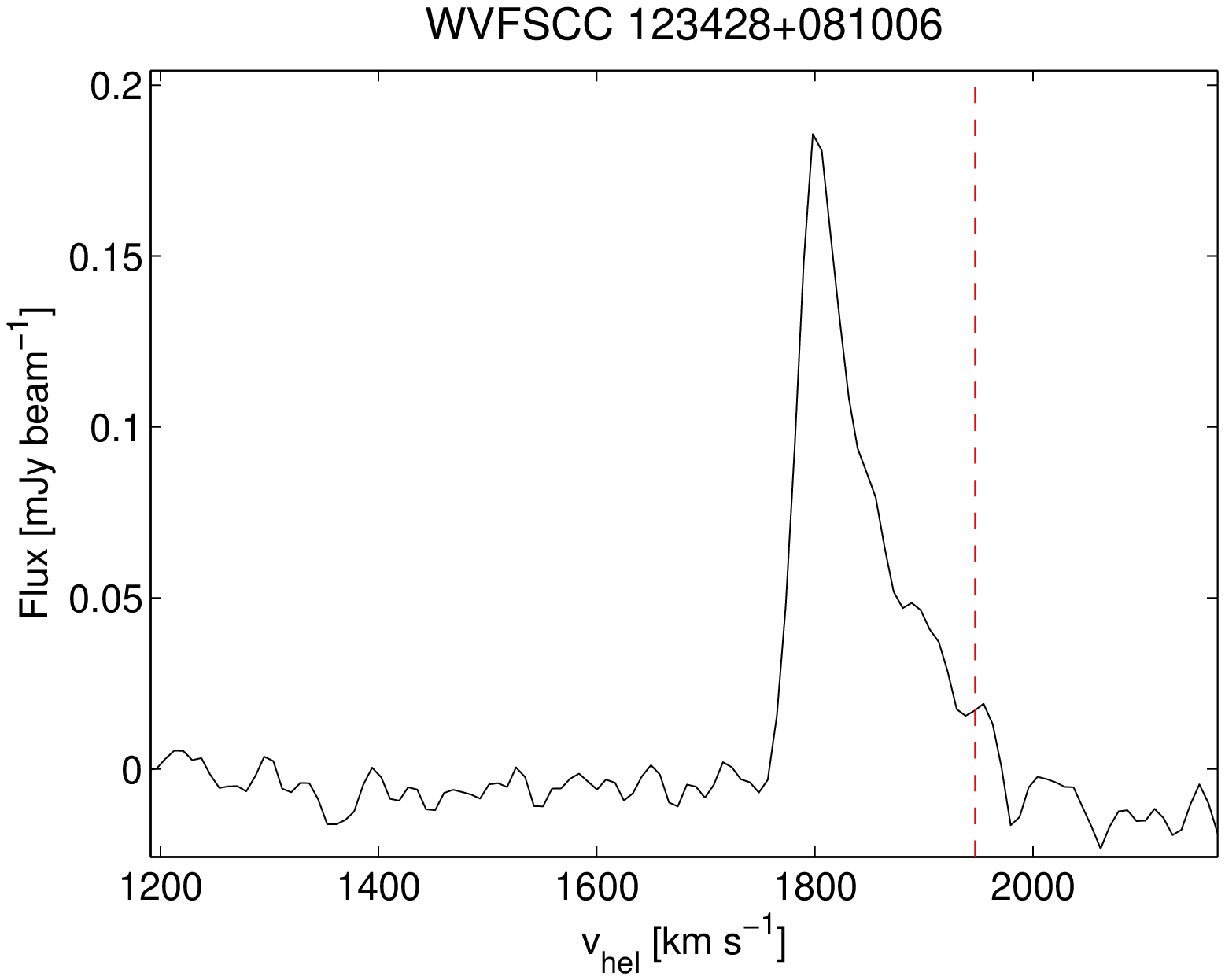}
\includegraphics[width=0.32\textwidth]{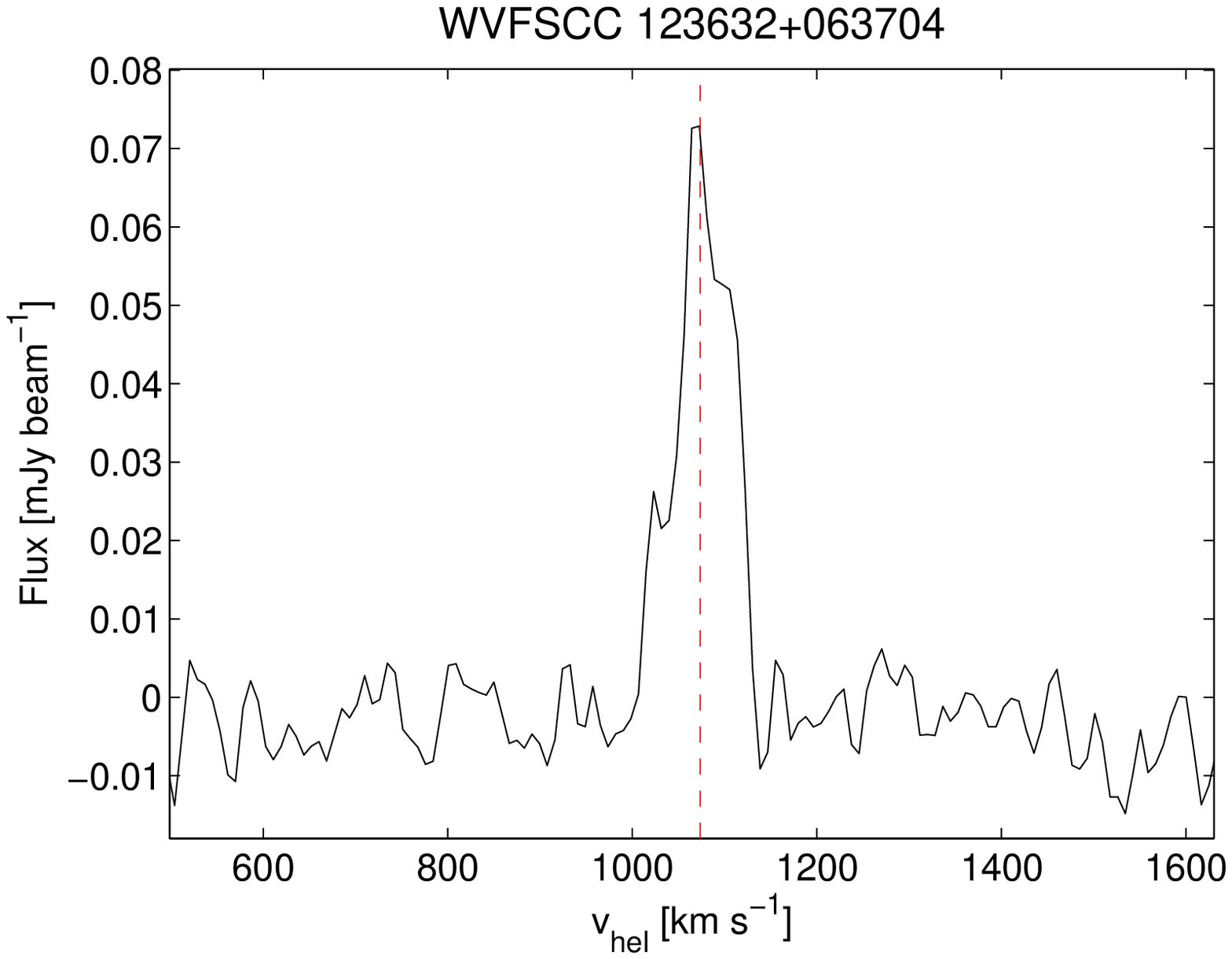}
\includegraphics[width=0.32\textwidth]{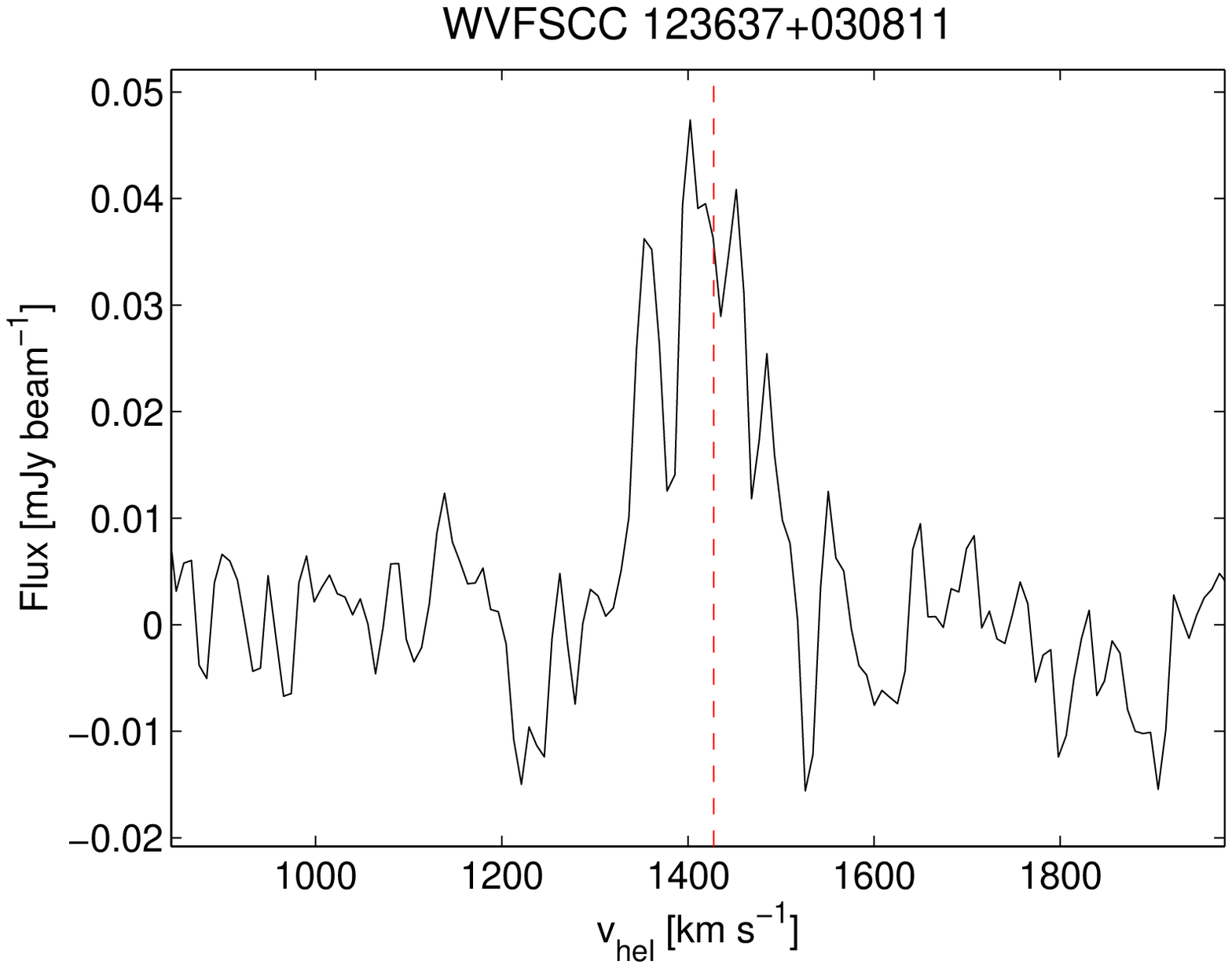}
\includegraphics[width=0.32\textwidth]{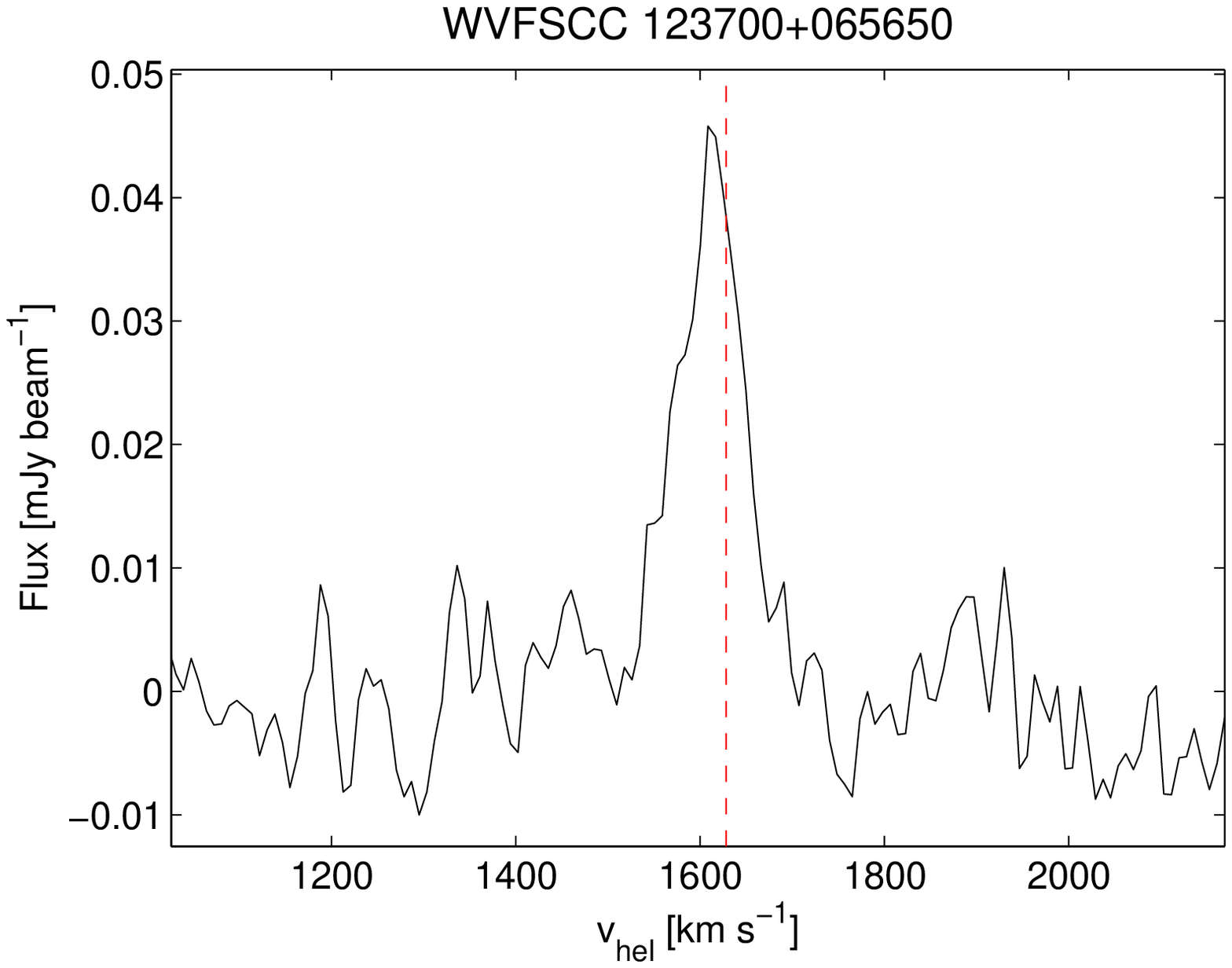}
\includegraphics[width=0.32\textwidth]{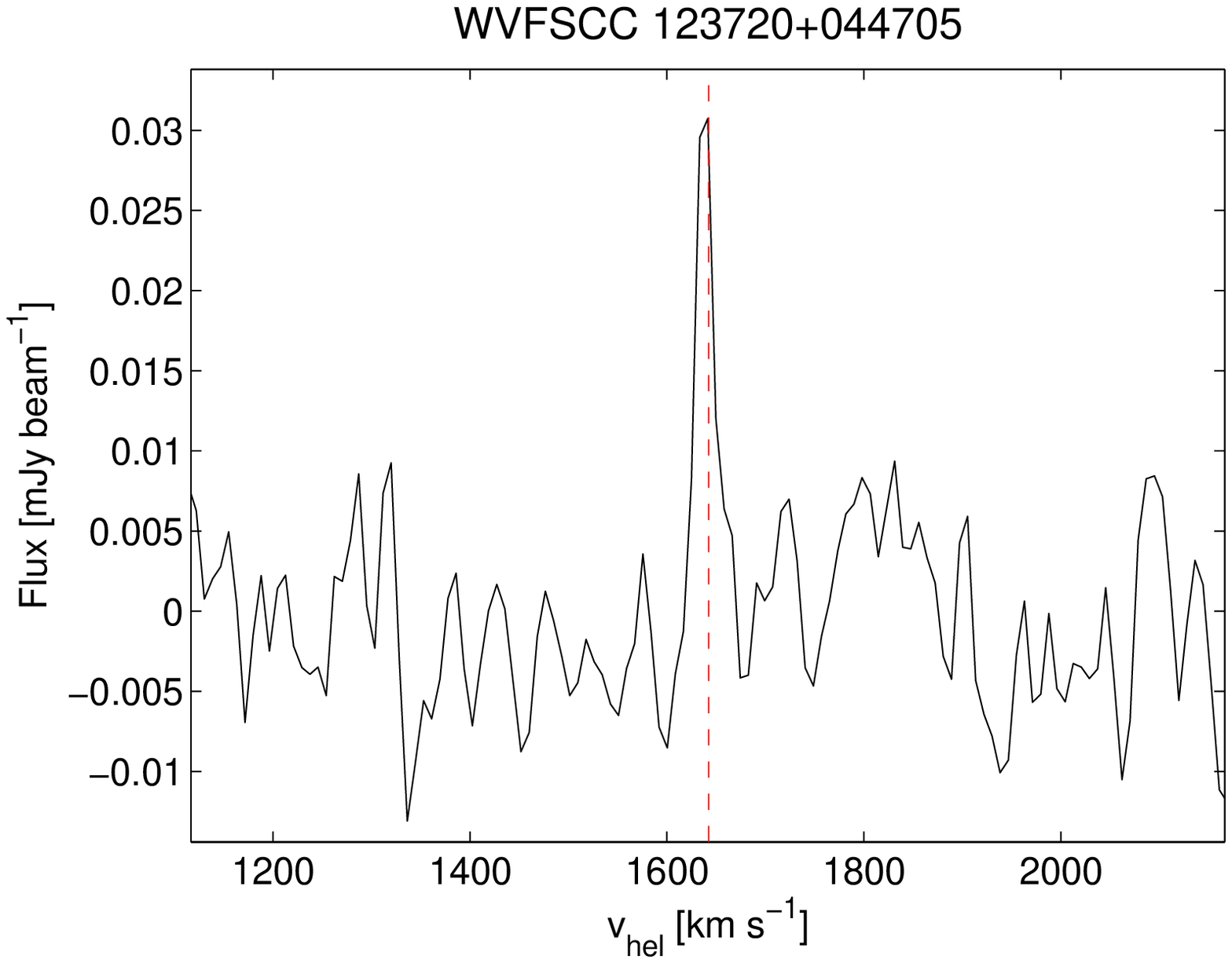}
\includegraphics[width=0.32\textwidth]{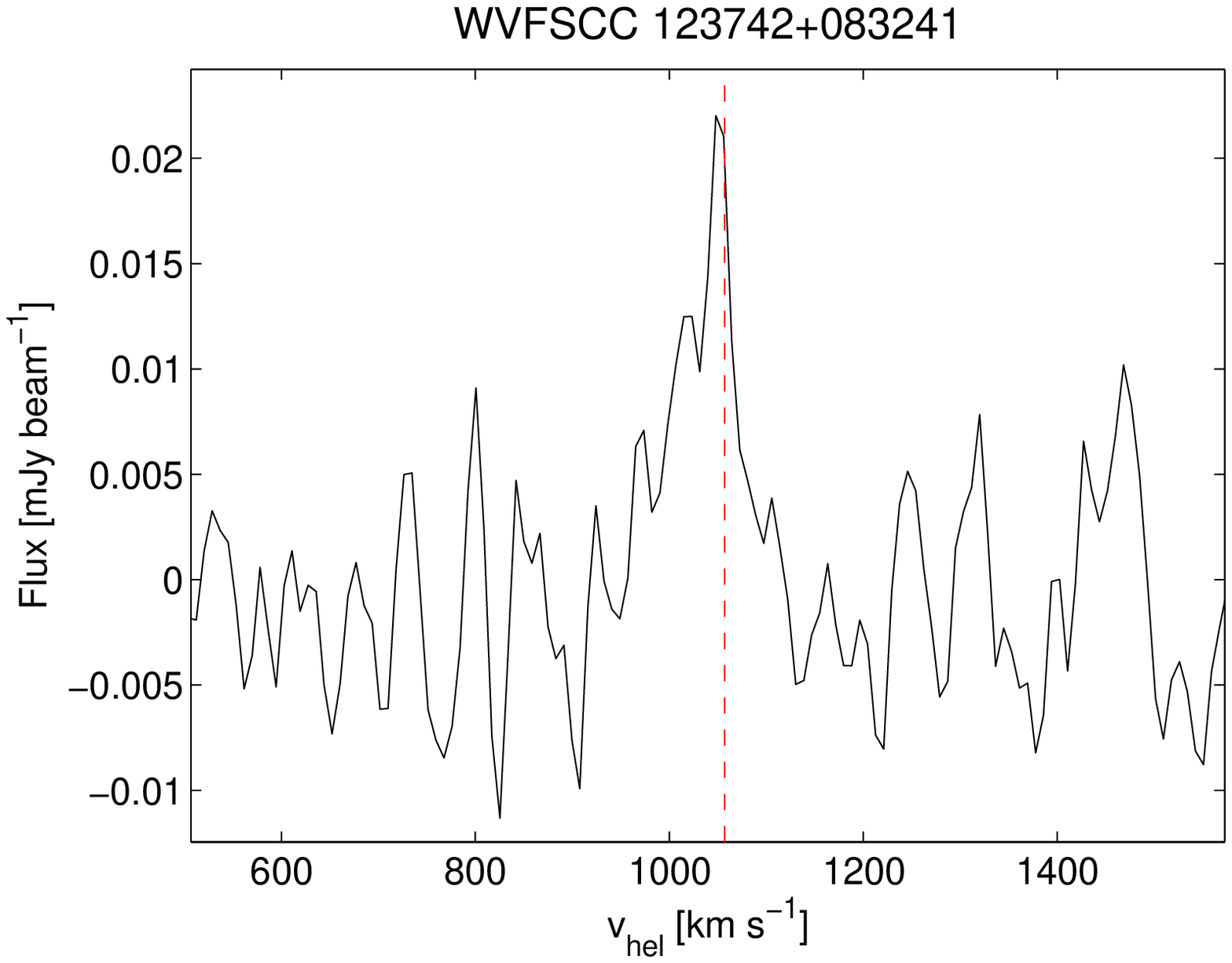}
\includegraphics[width=0.32\textwidth]{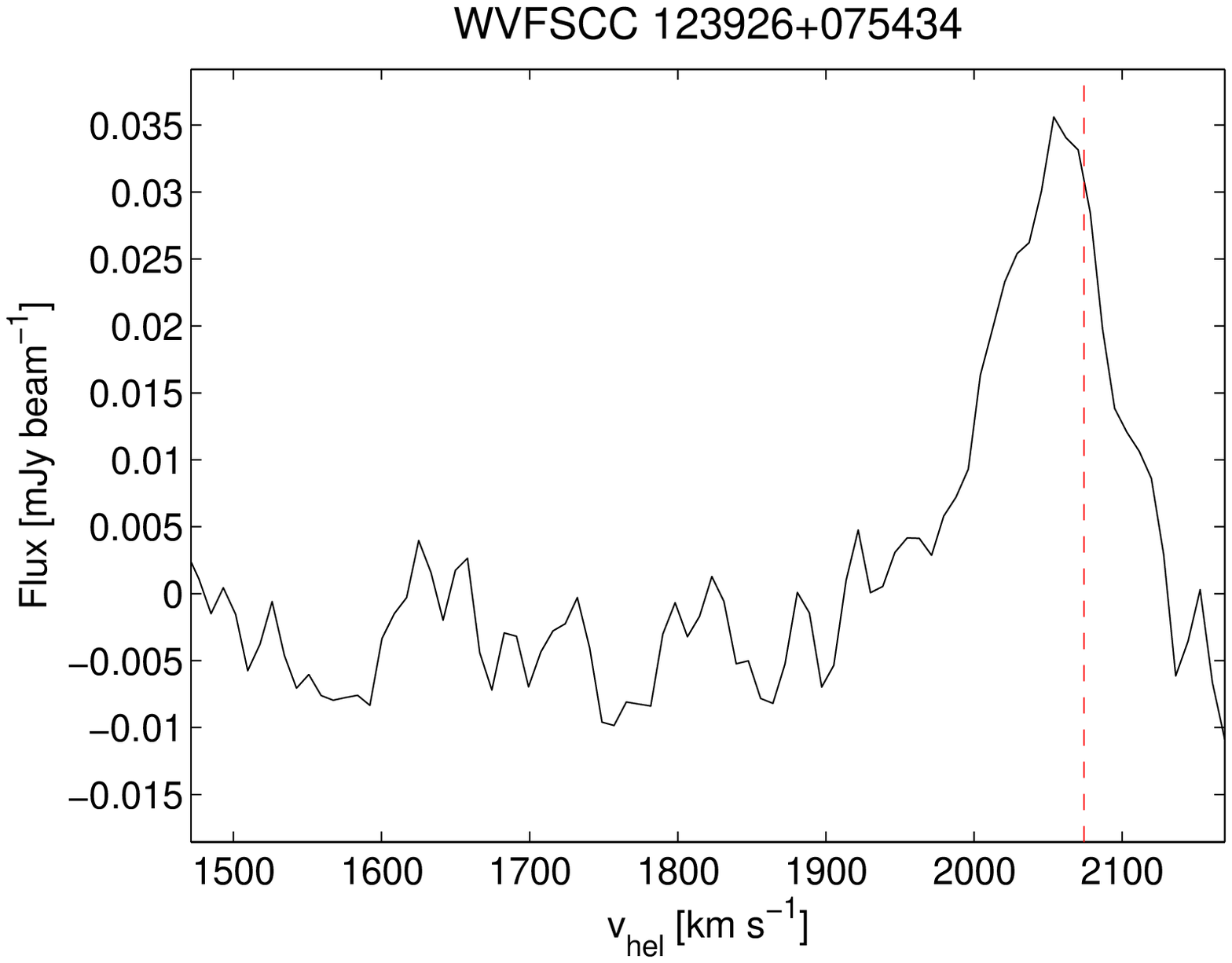}
\includegraphics[width=0.32\textwidth]{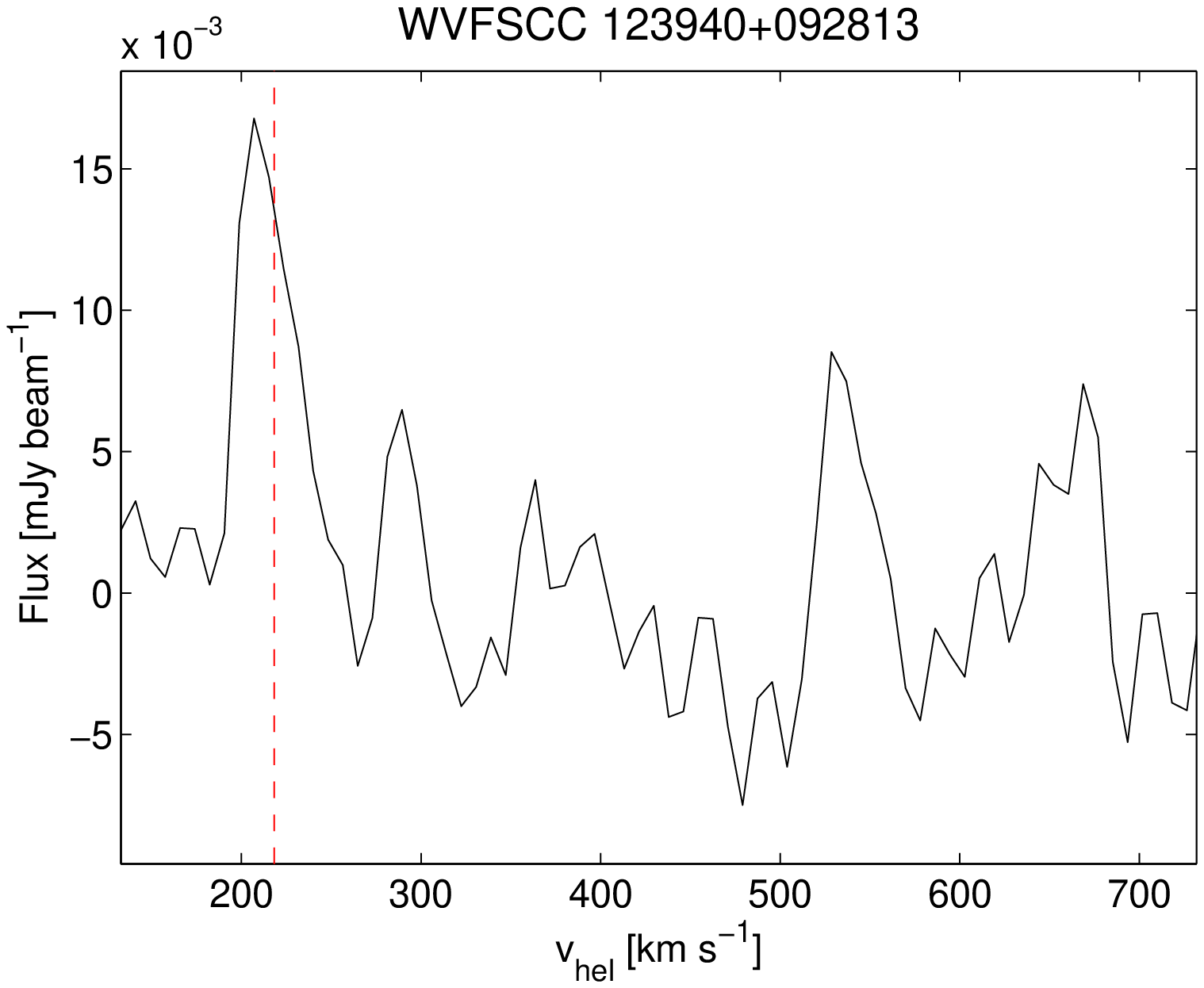}
\includegraphics[width=0.32\textwidth]{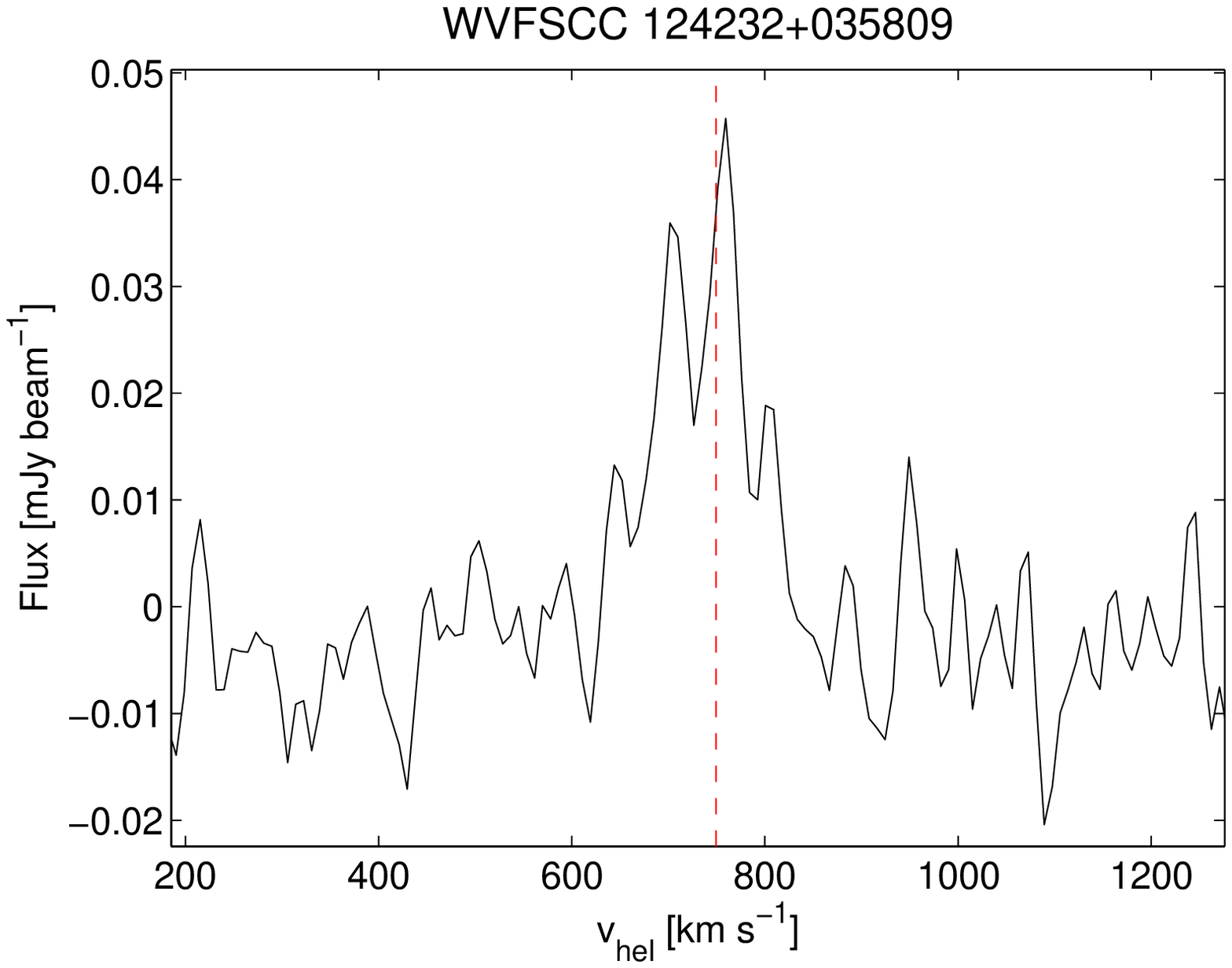}
                                                         
\end{center}                                            
{\bf Fig~\ref{all_spectra2}.} (continued)                                        
 
\end{figure*}


\begin{figure*}
  \begin{center}

\includegraphics[width=0.32\textwidth]{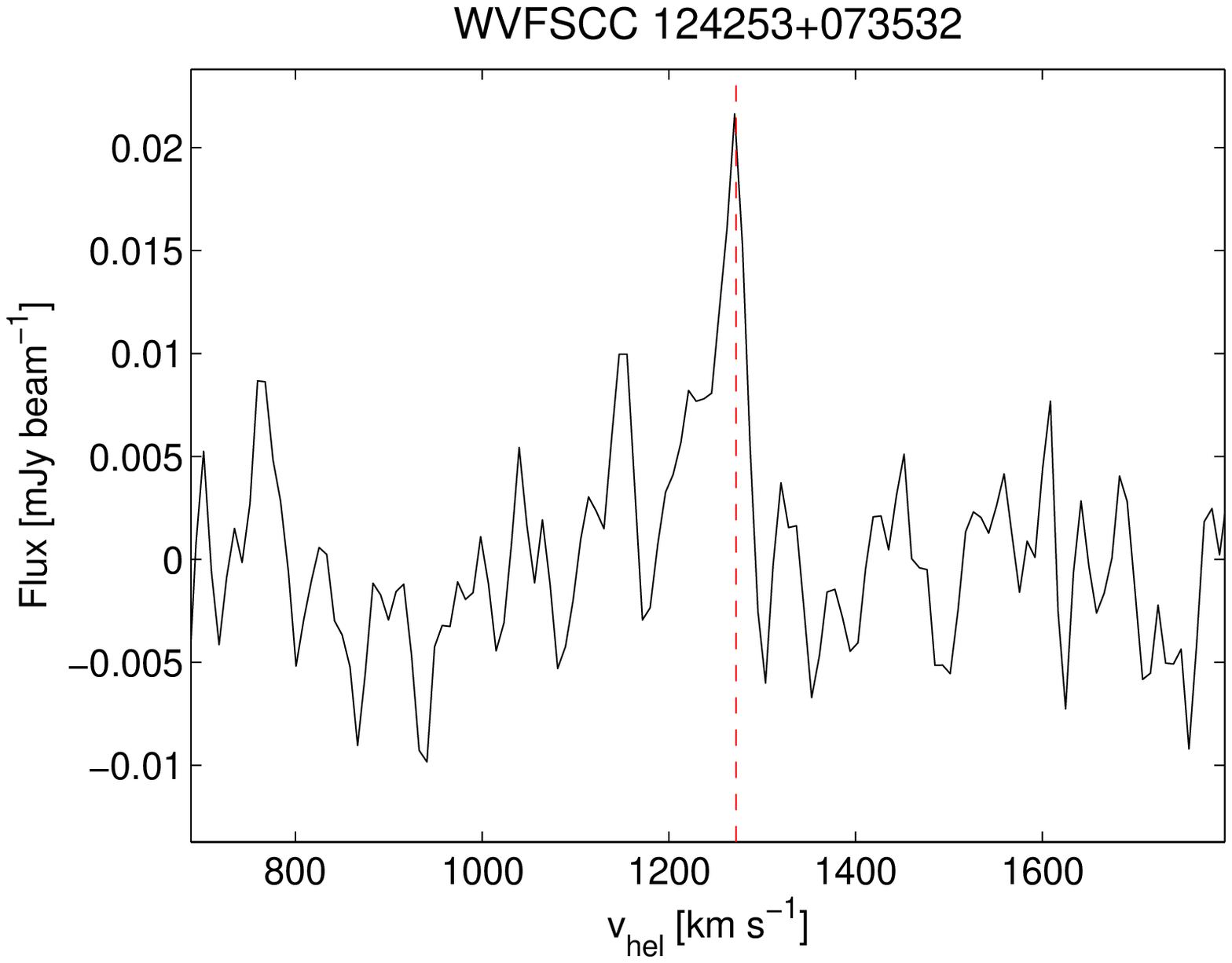} 
\includegraphics[width=0.32\textwidth]{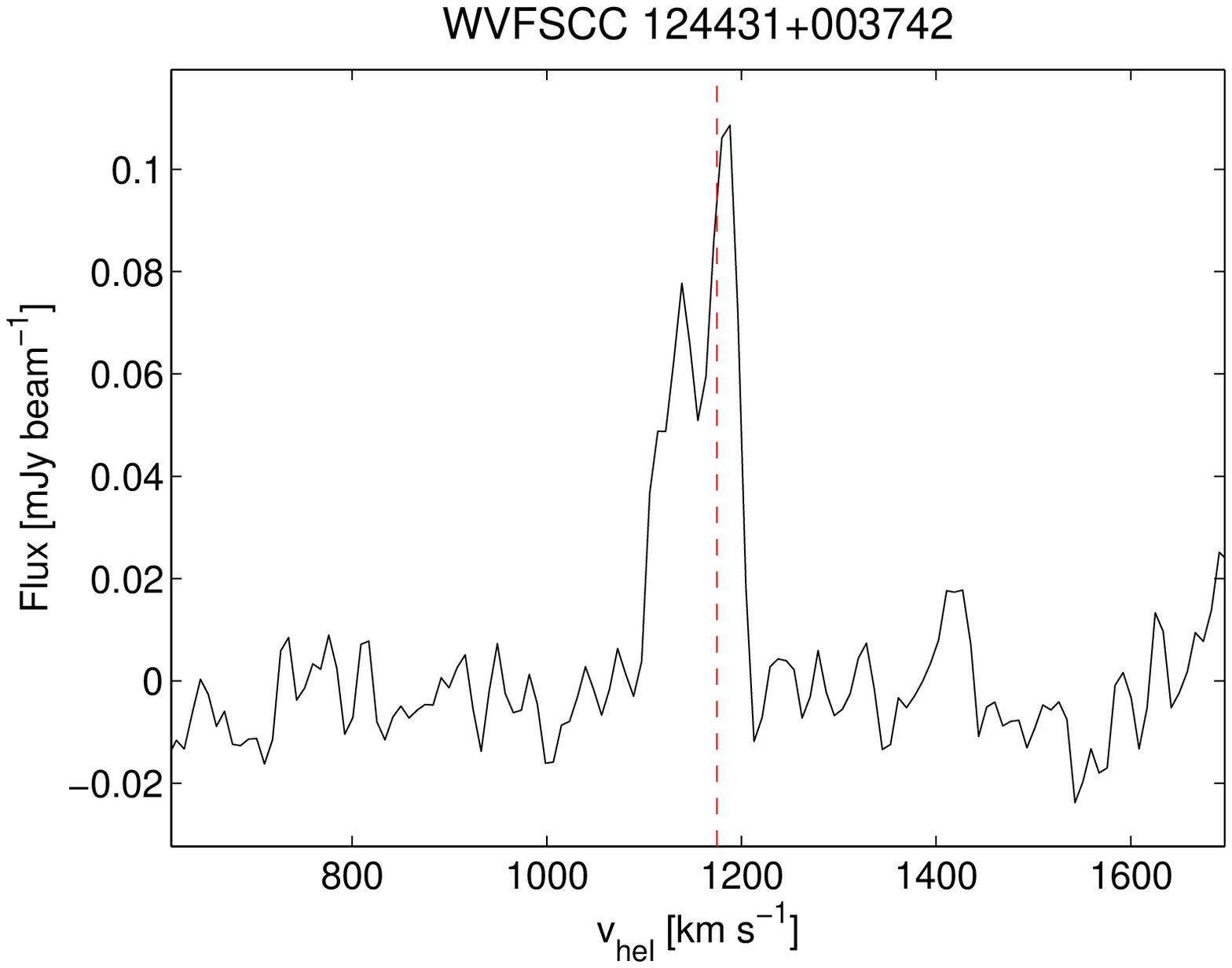}
\includegraphics[width=0.32\textwidth]{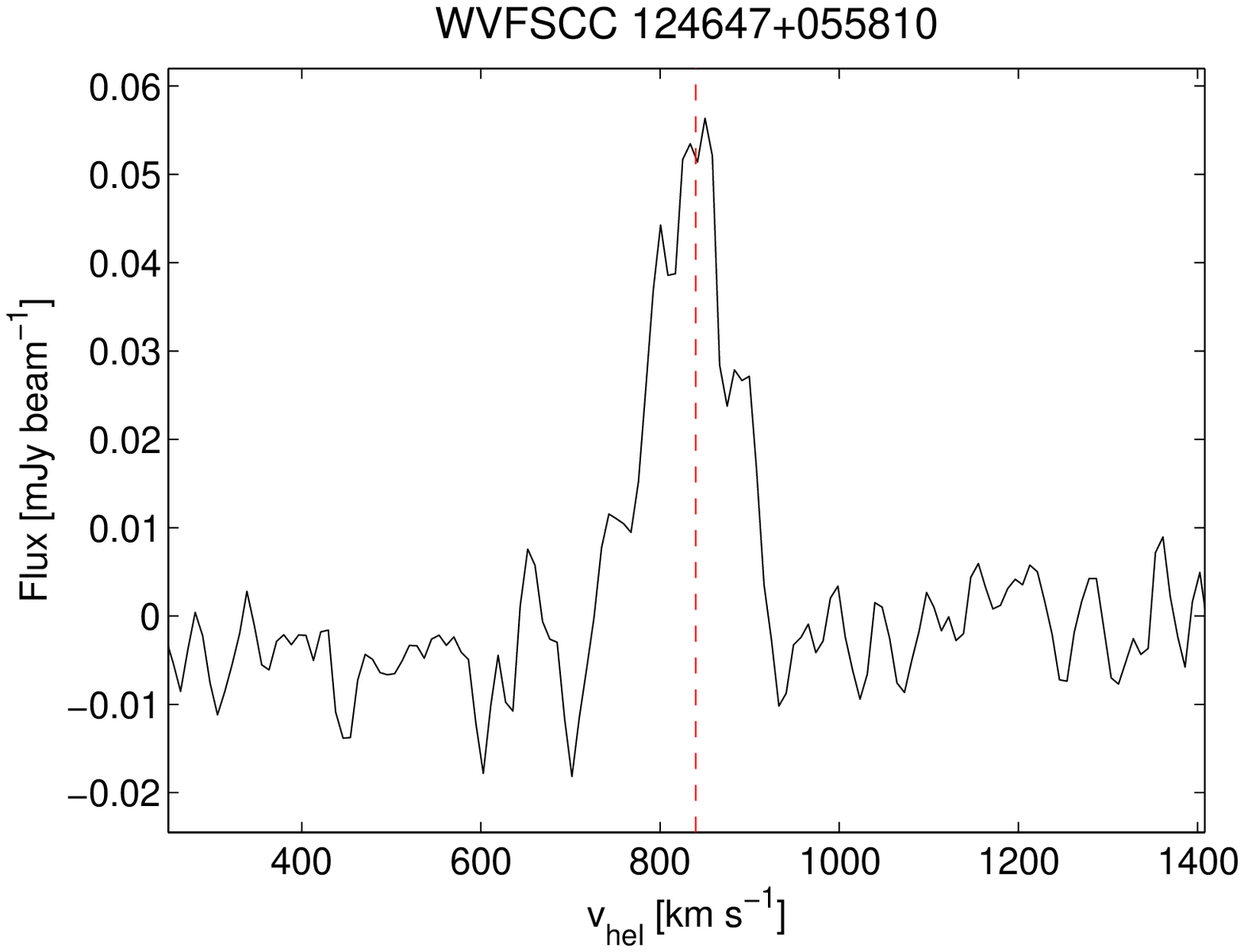}
\includegraphics[width=0.32\textwidth]{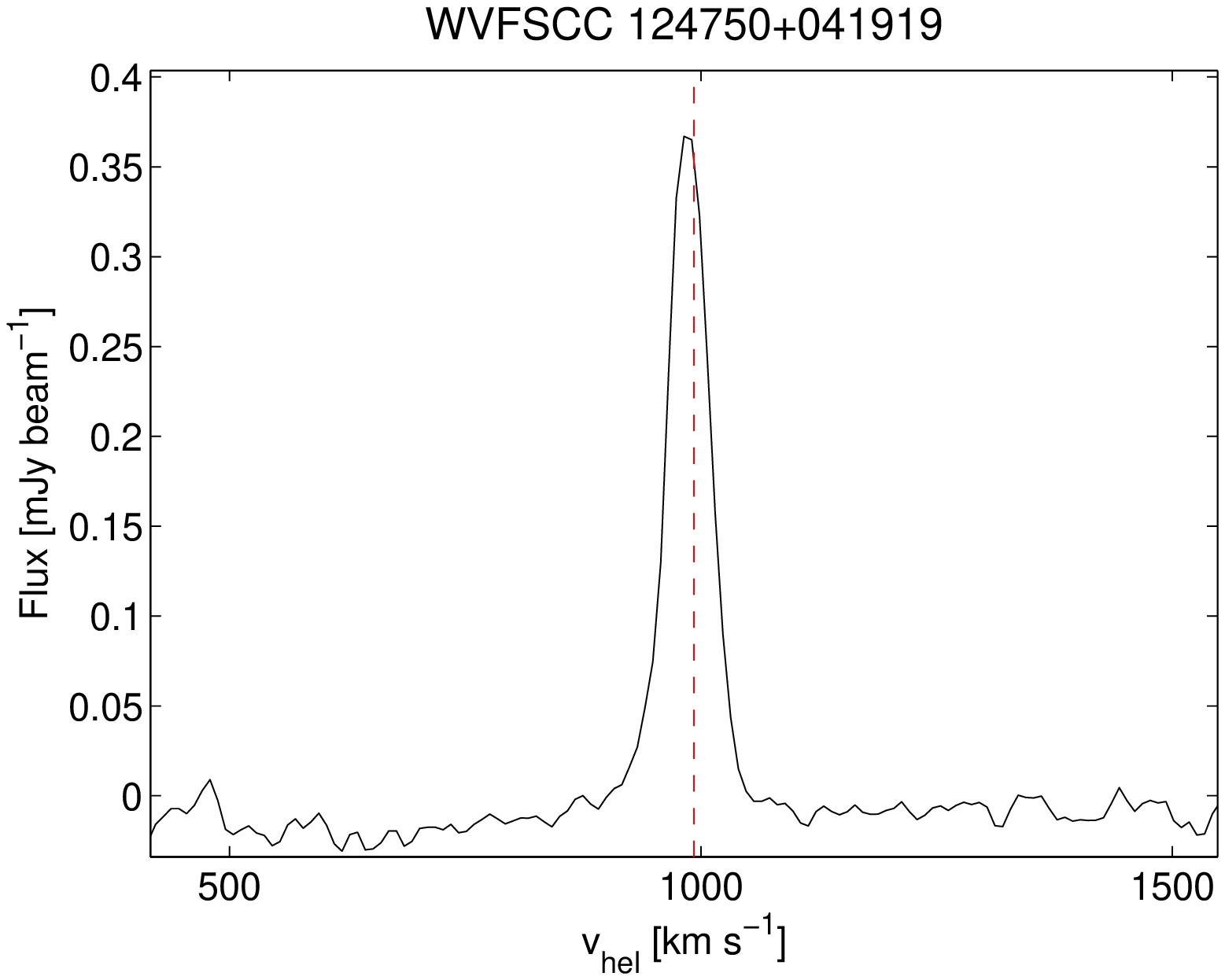}
\includegraphics[width=0.32\textwidth]{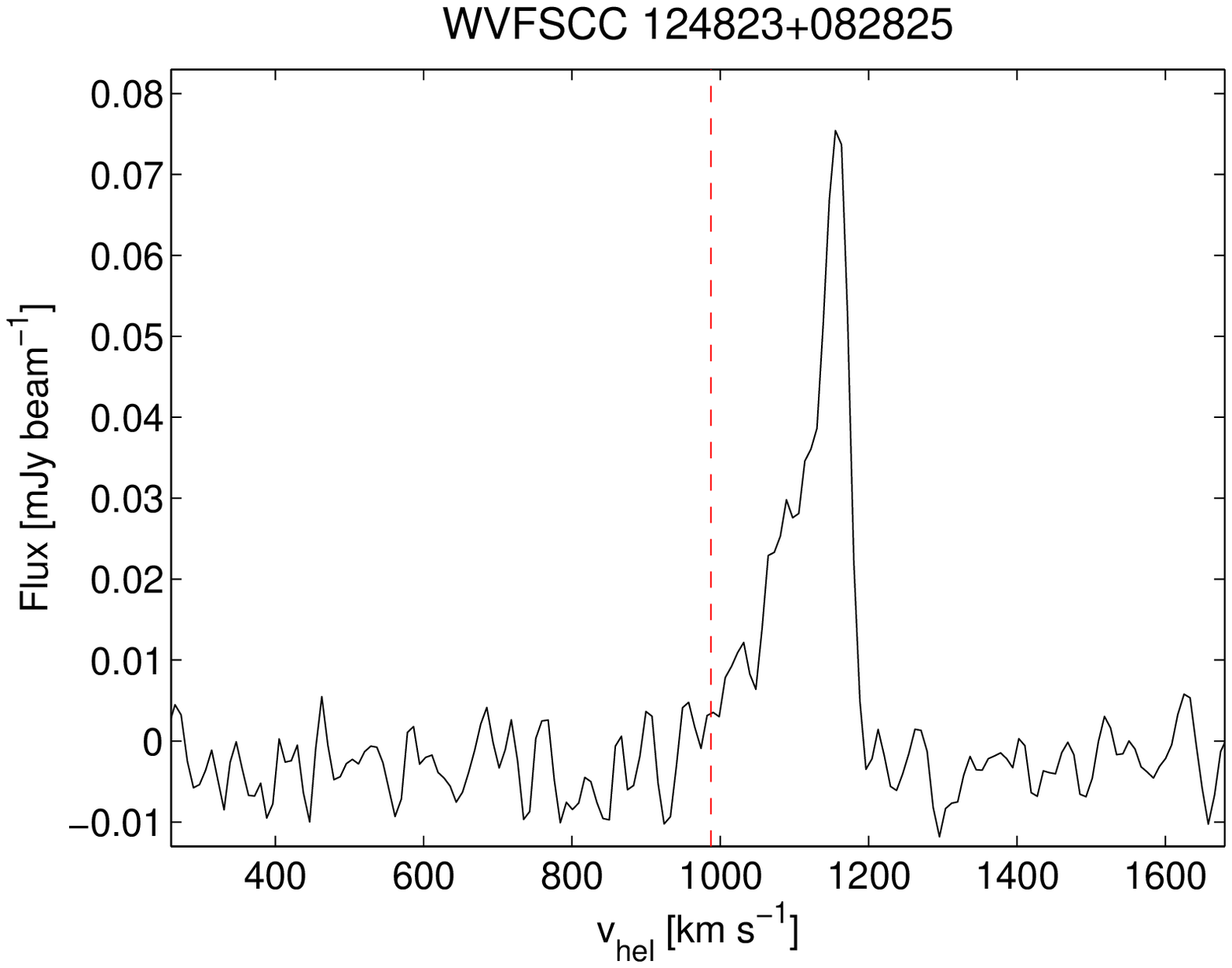}
\includegraphics[width=0.32\textwidth]{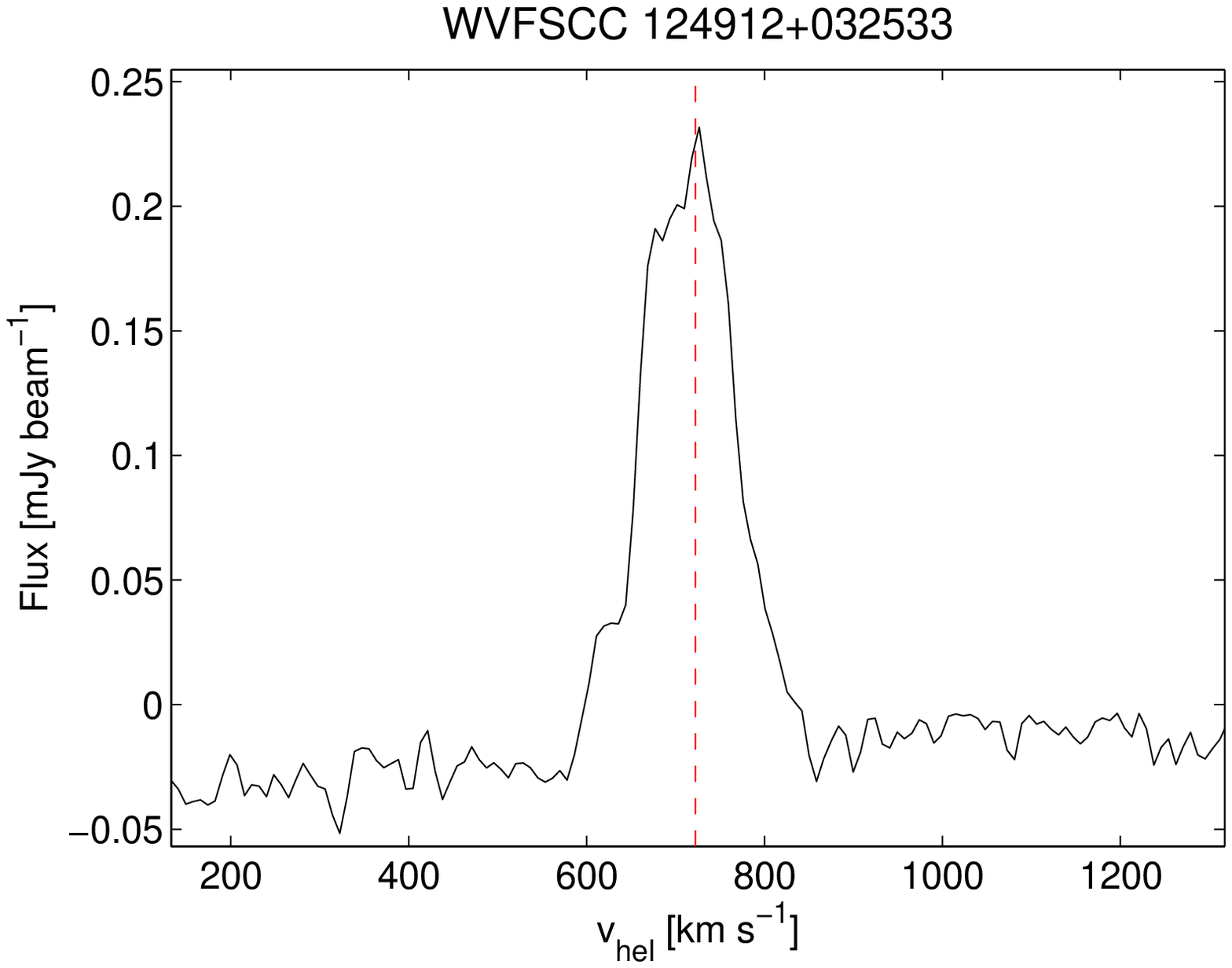}
\includegraphics[width=0.32\textwidth]{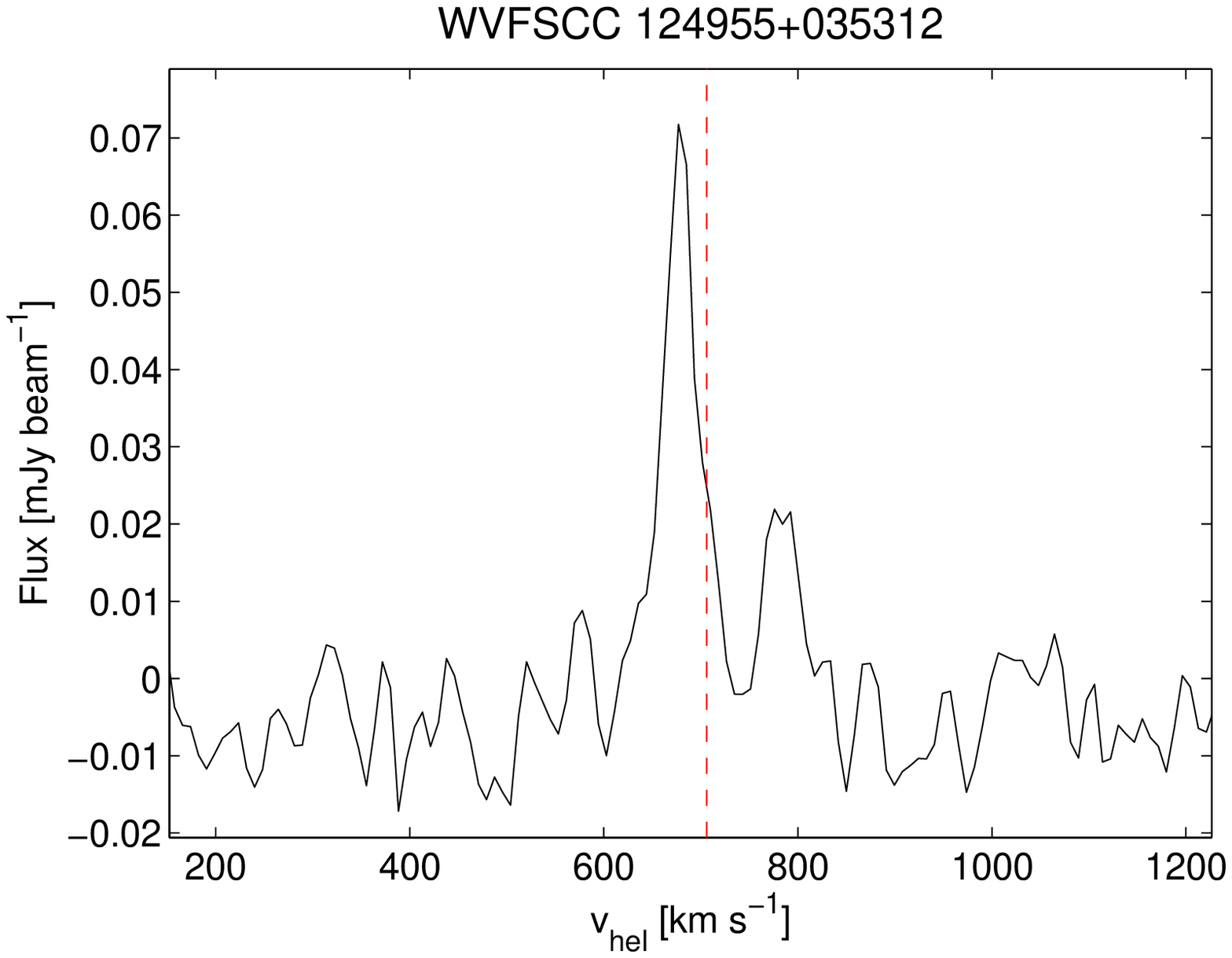}
\includegraphics[width=0.32\textwidth]{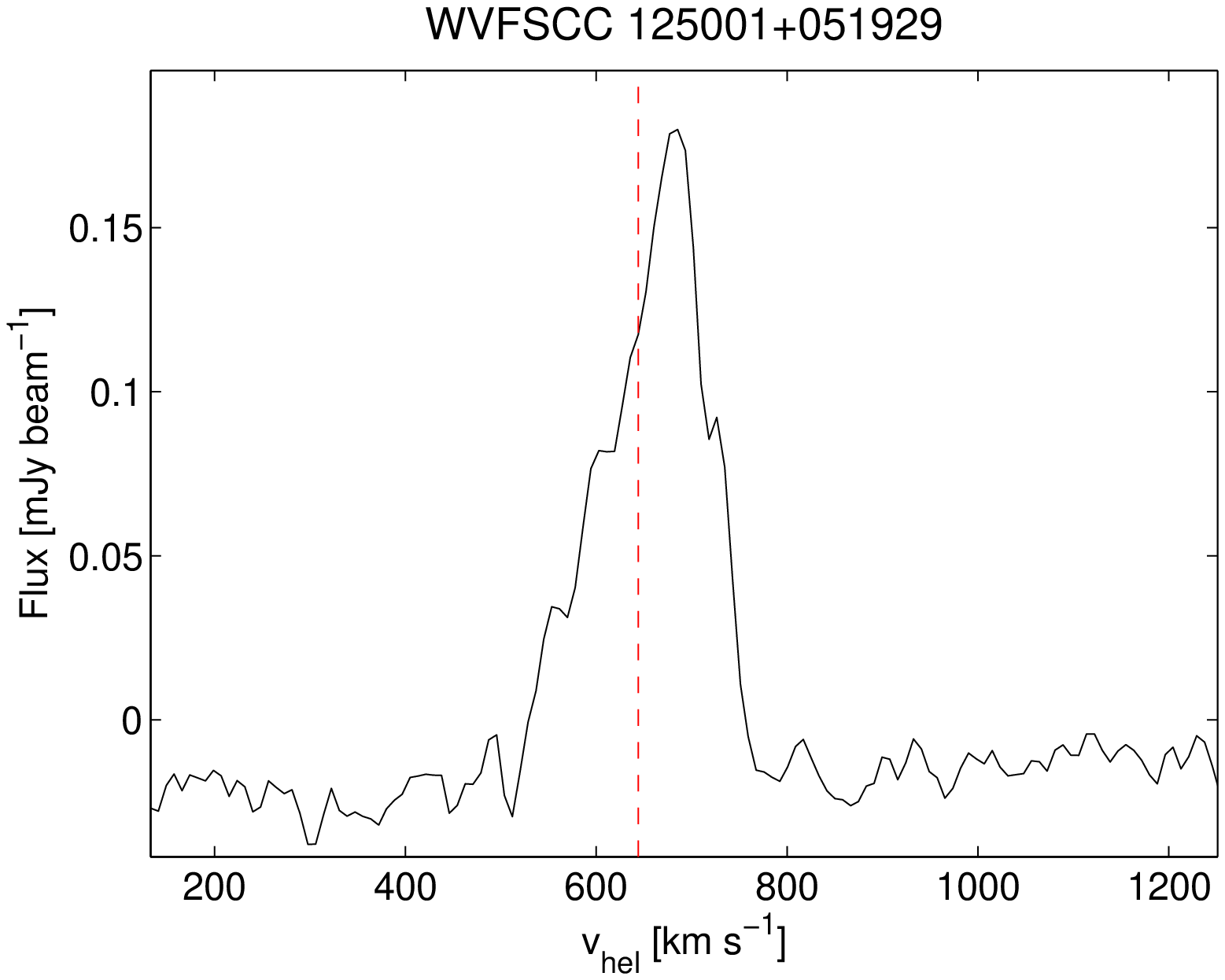}
\includegraphics[width=0.32\textwidth]{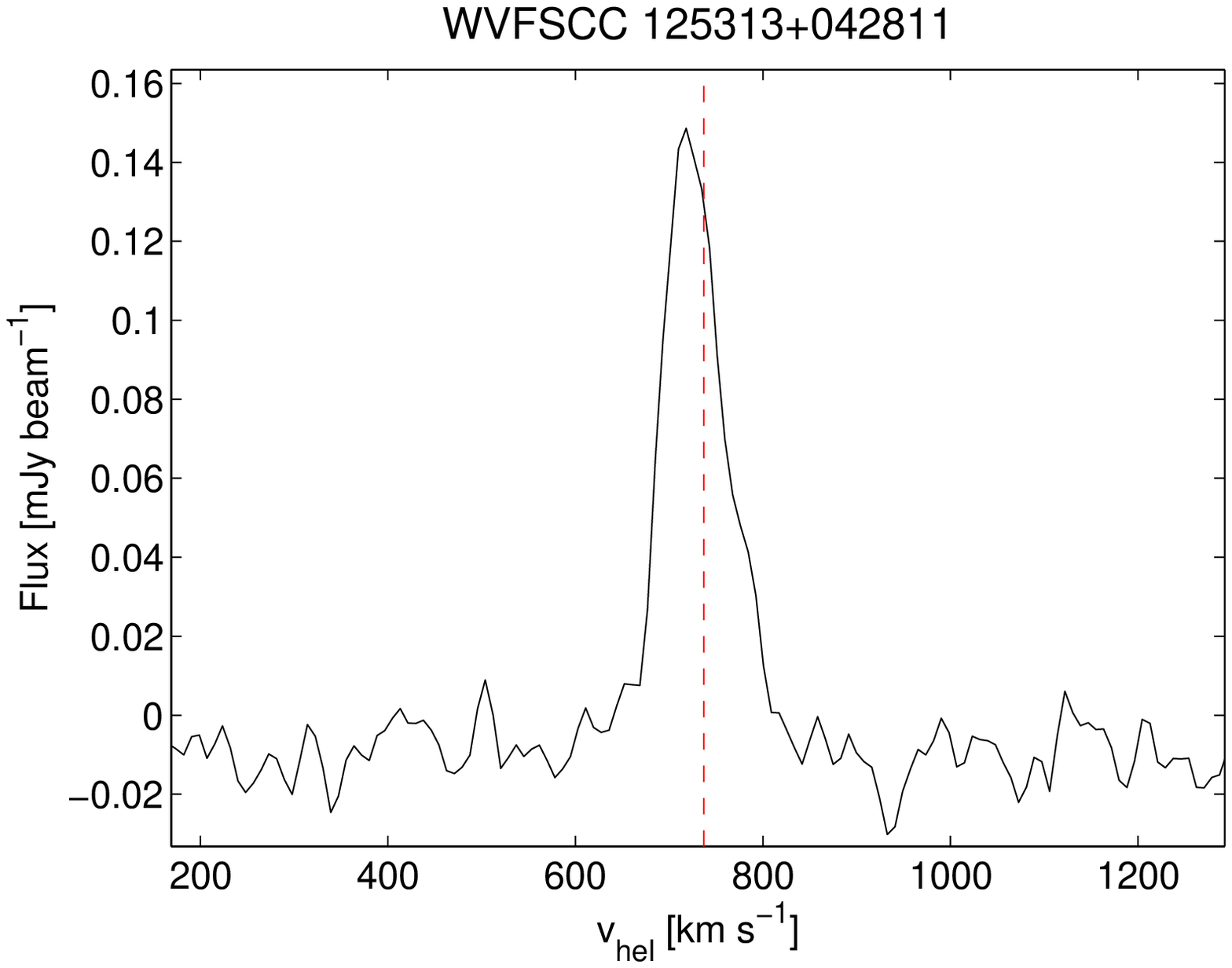}
\includegraphics[width=0.32\textwidth]{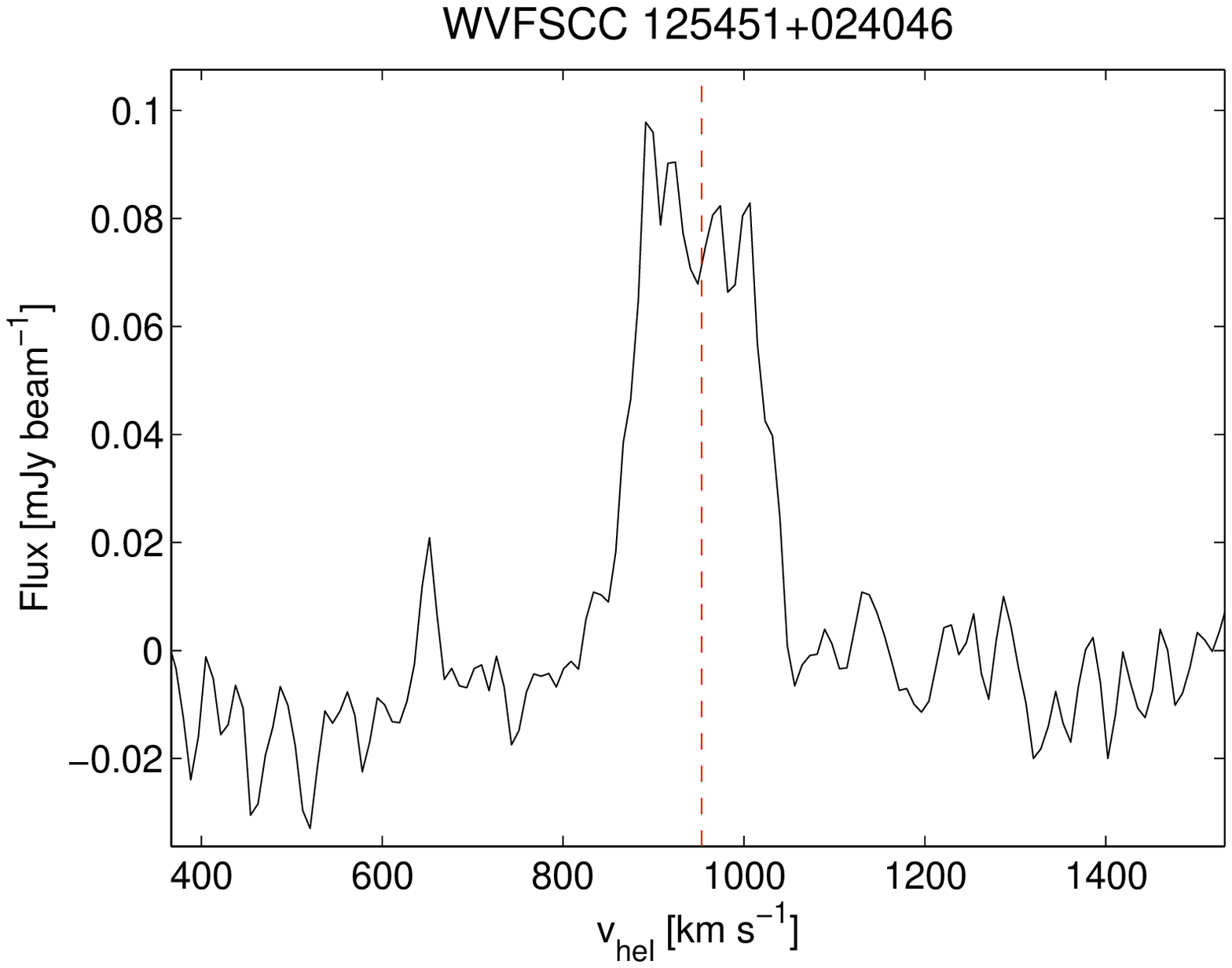}
\includegraphics[width=0.32\textwidth]{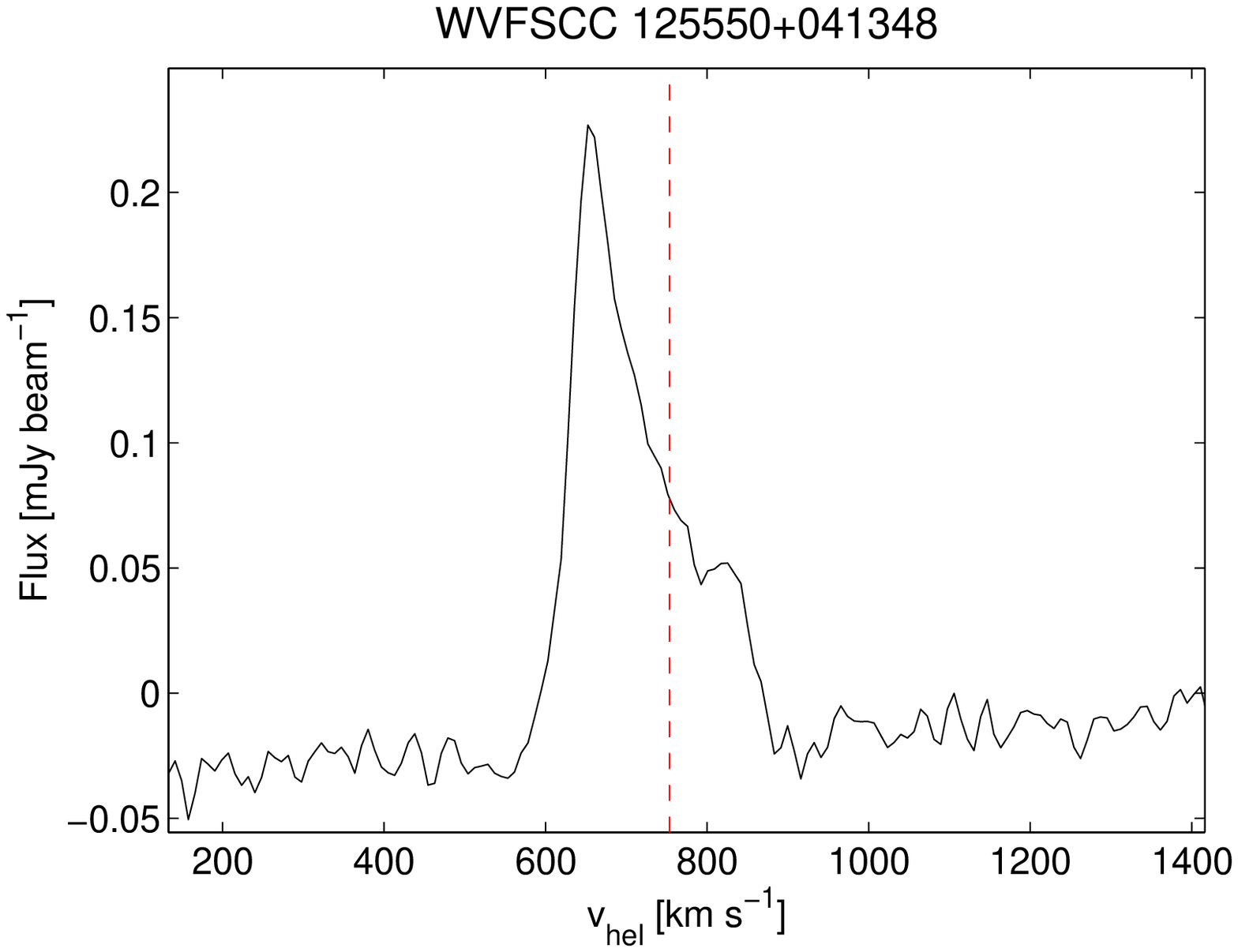}
\includegraphics[width=0.32\textwidth]{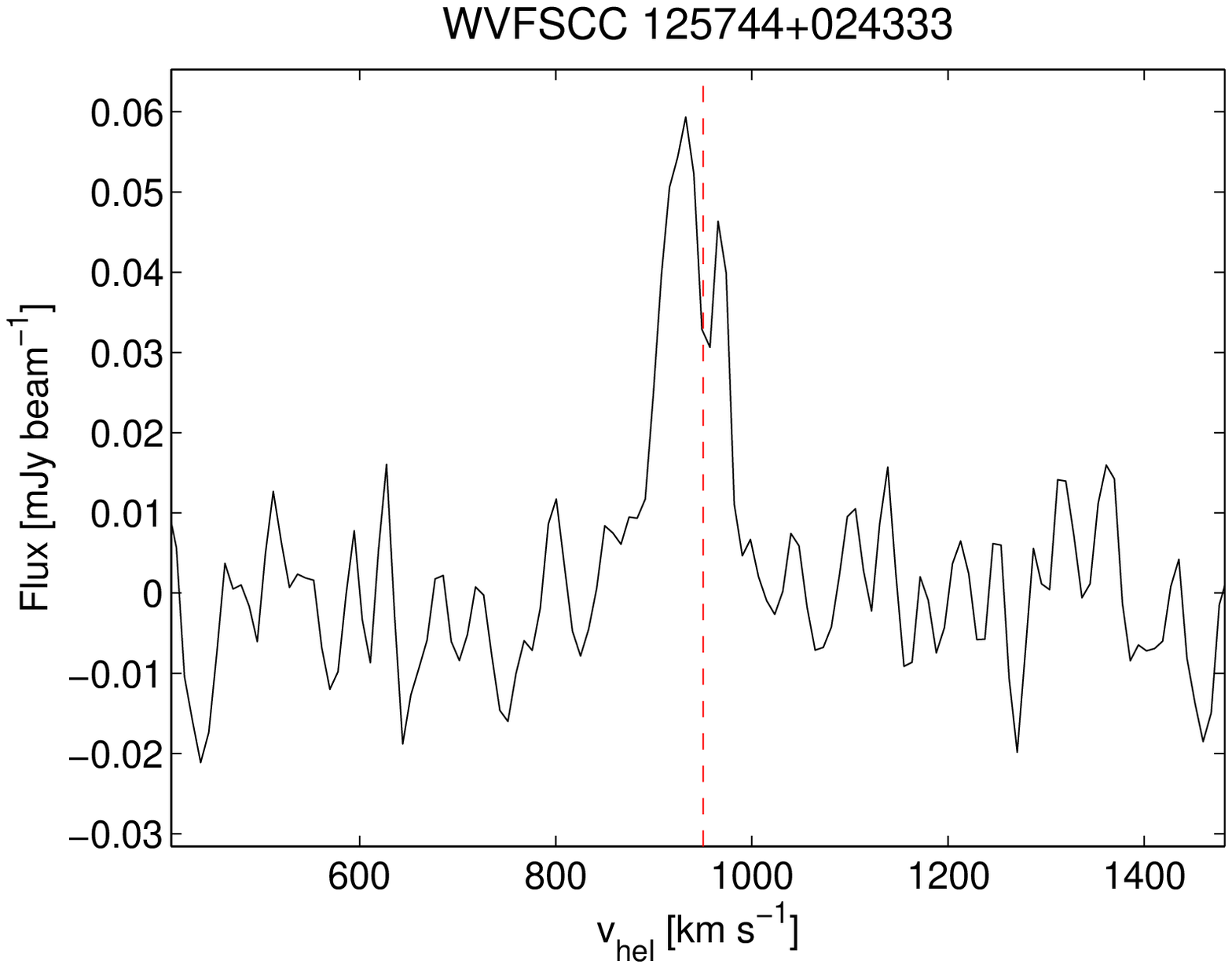}
\includegraphics[width=0.32\textwidth]{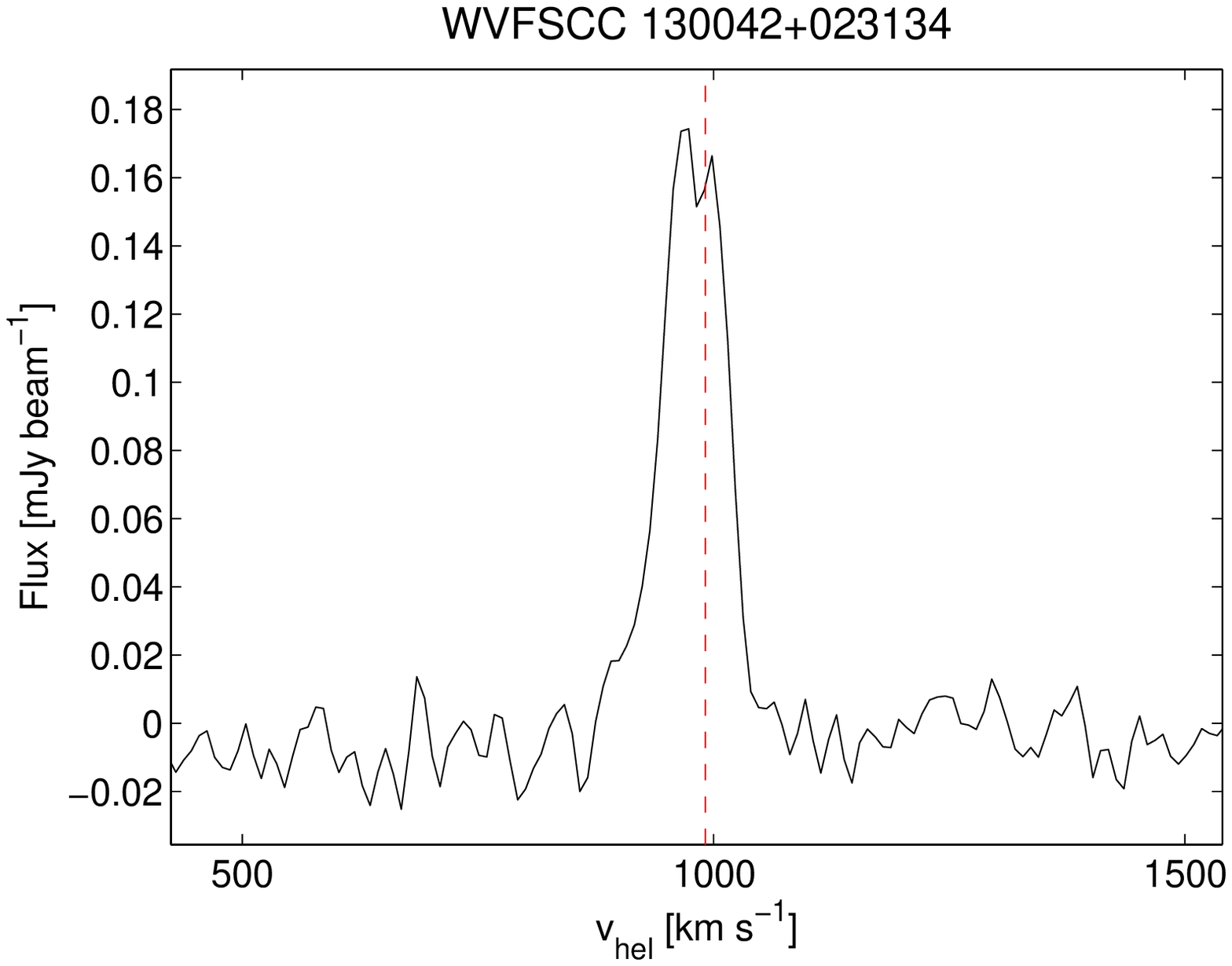}
\includegraphics[width=0.32\textwidth]{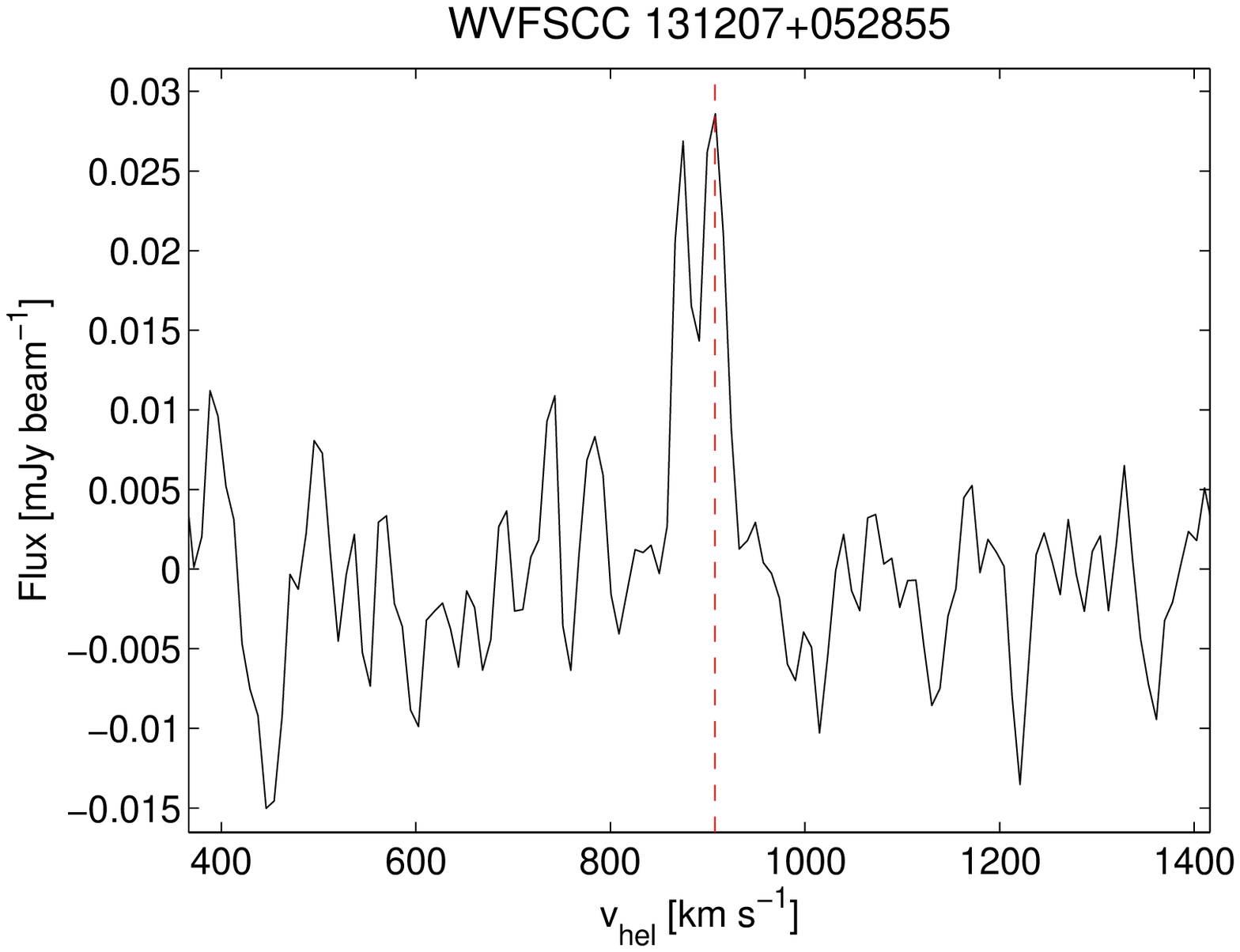}
\includegraphics[width=0.32\textwidth]{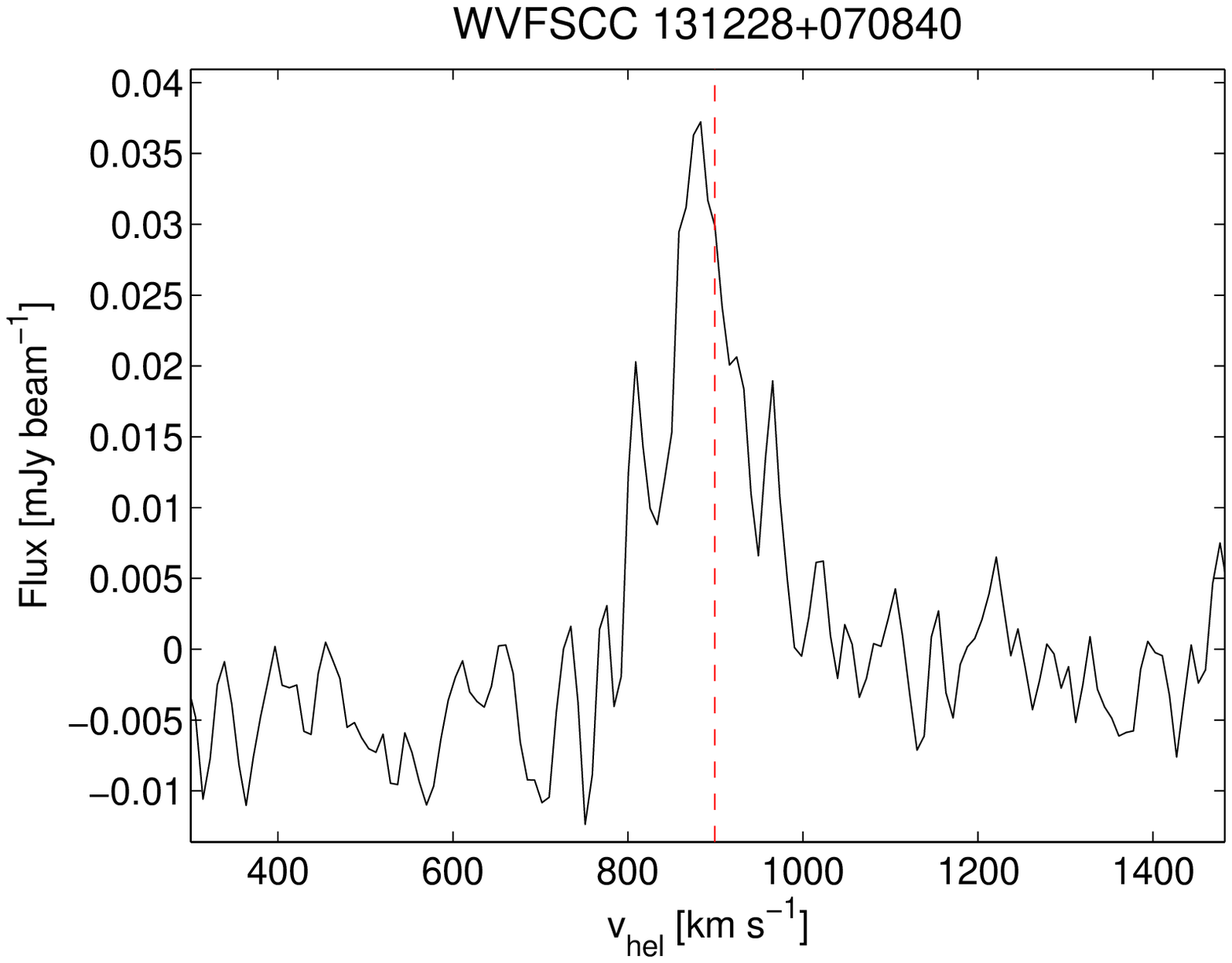}
                                                         
\end{center}                                            
{\bf Fig~\ref{all_spectra2}.} (continued)                                        
 
\end{figure*}


\begin{figure*}
  \begin{center}

\includegraphics[width=0.32\textwidth]{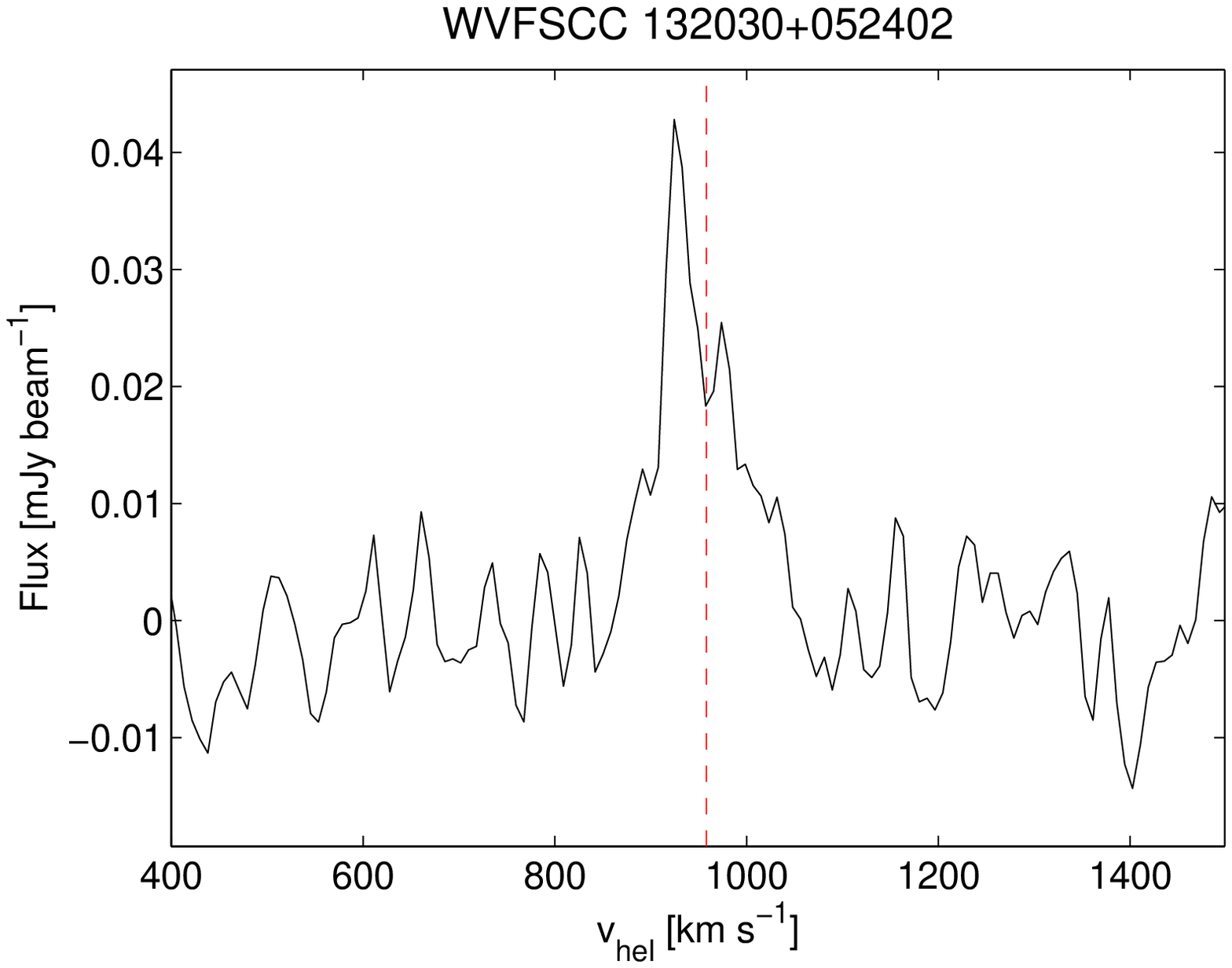}
\includegraphics[width=0.32\textwidth]{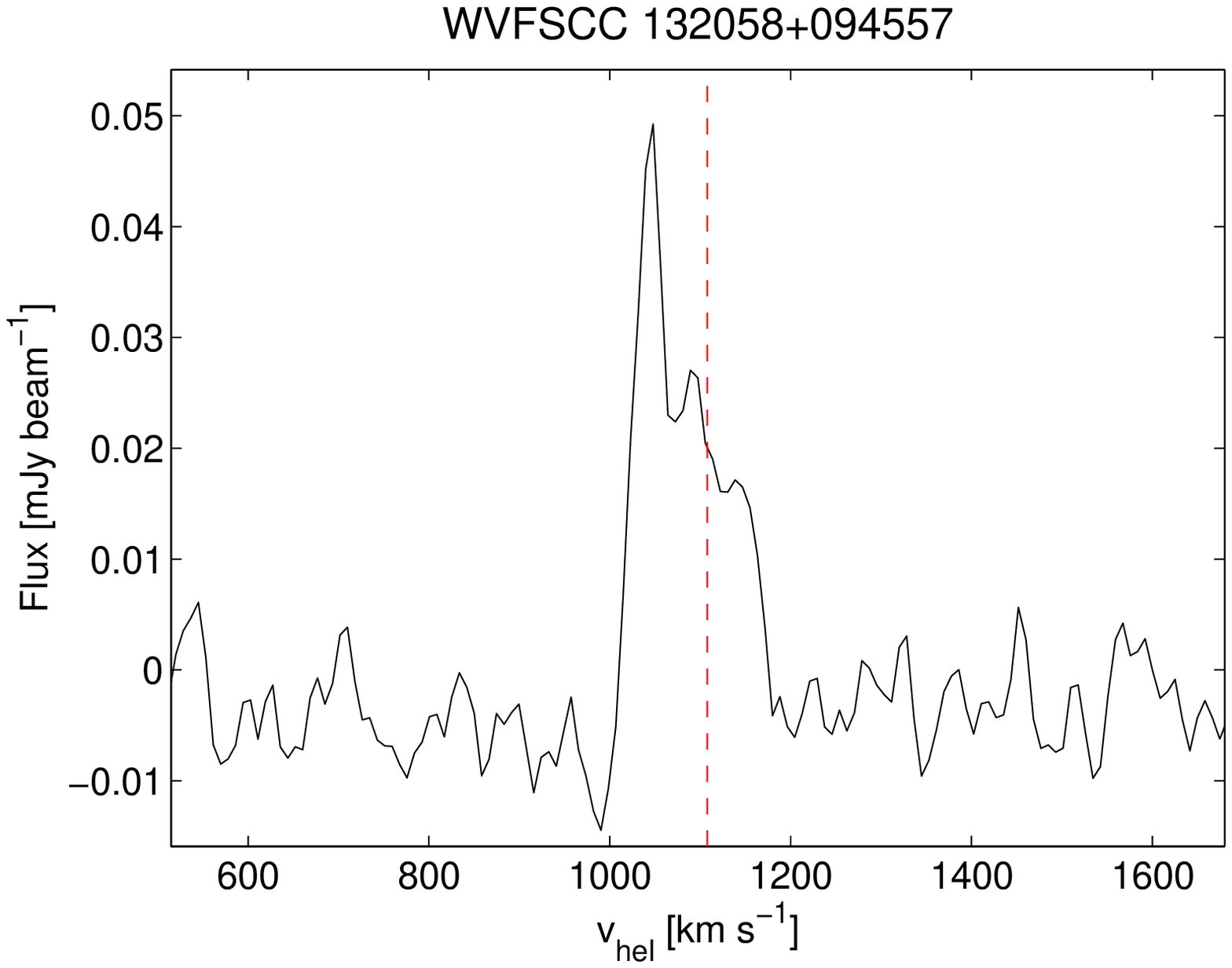}
\includegraphics[width=0.32\textwidth]{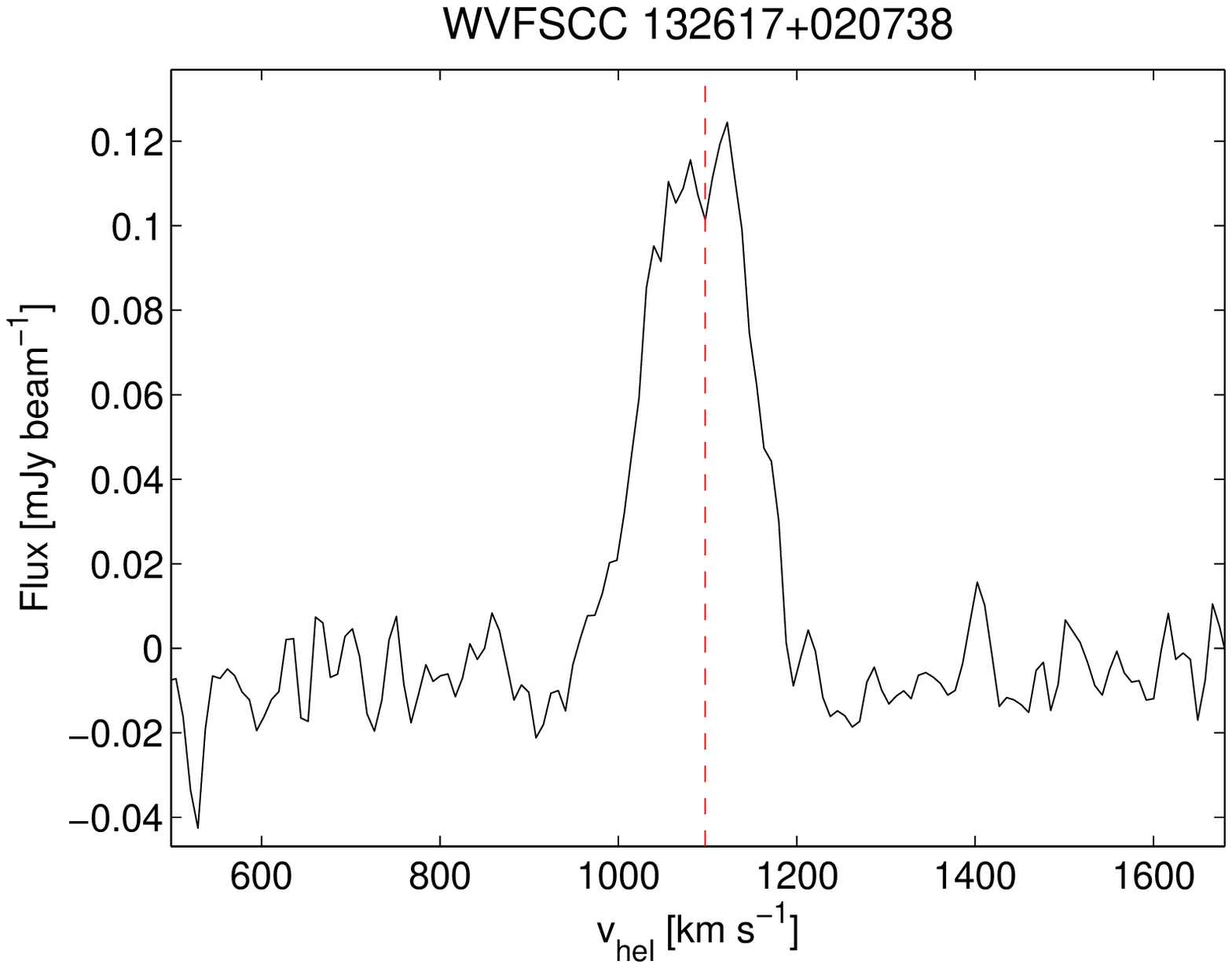}  
\includegraphics[width=0.32\textwidth]{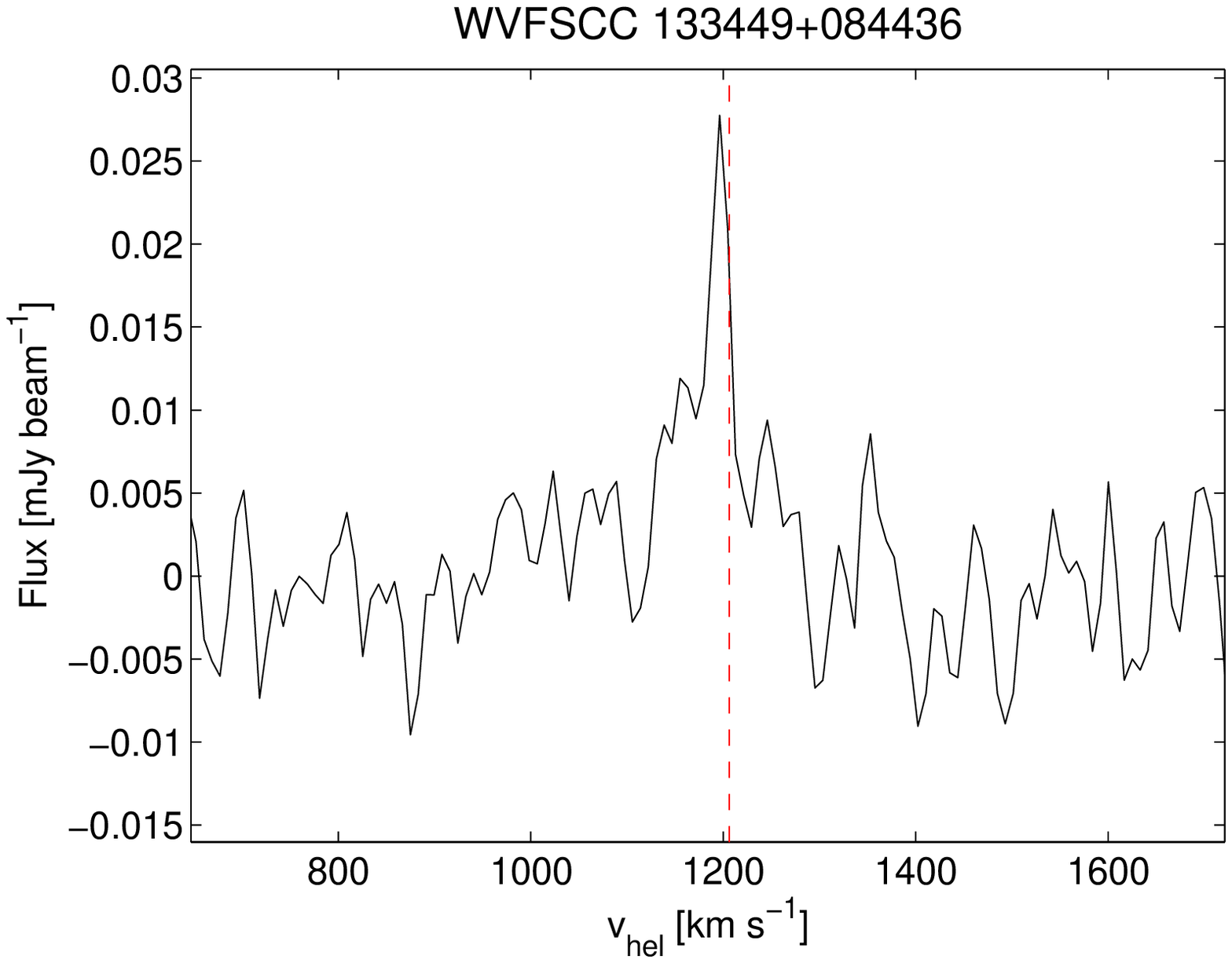}
\includegraphics[width=0.32\textwidth]{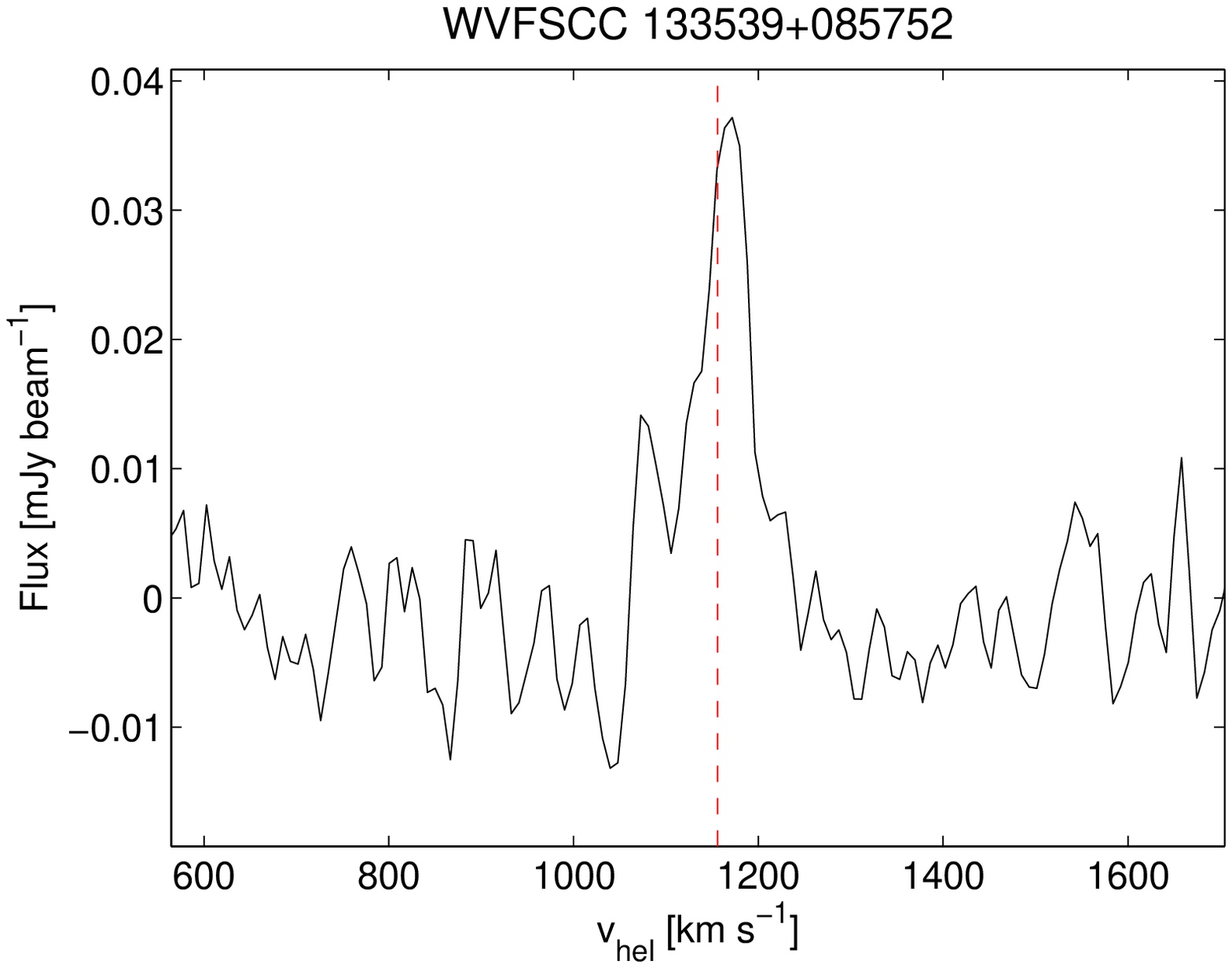}
\includegraphics[width=0.32\textwidth]{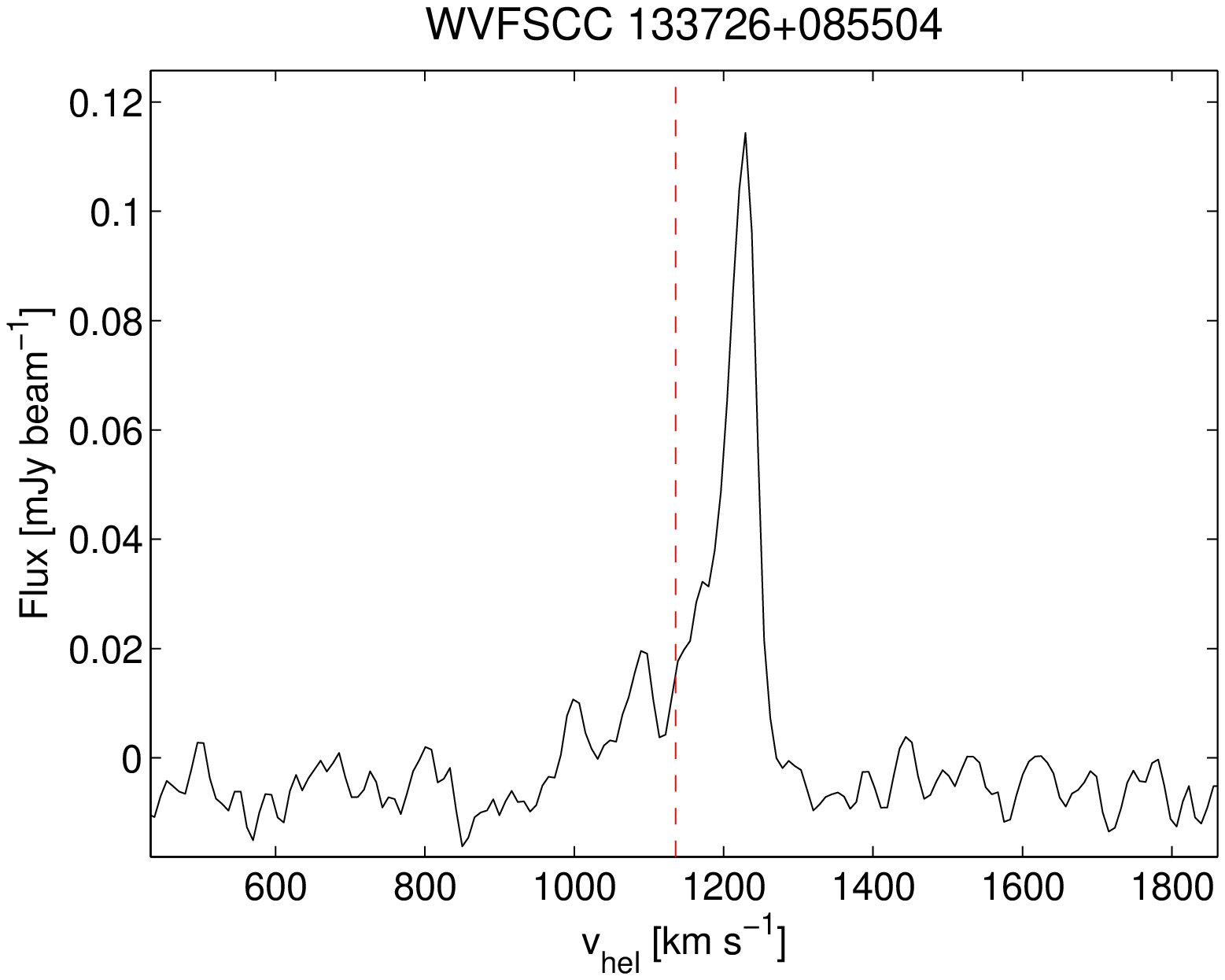}
\includegraphics[width=0.32\textwidth]{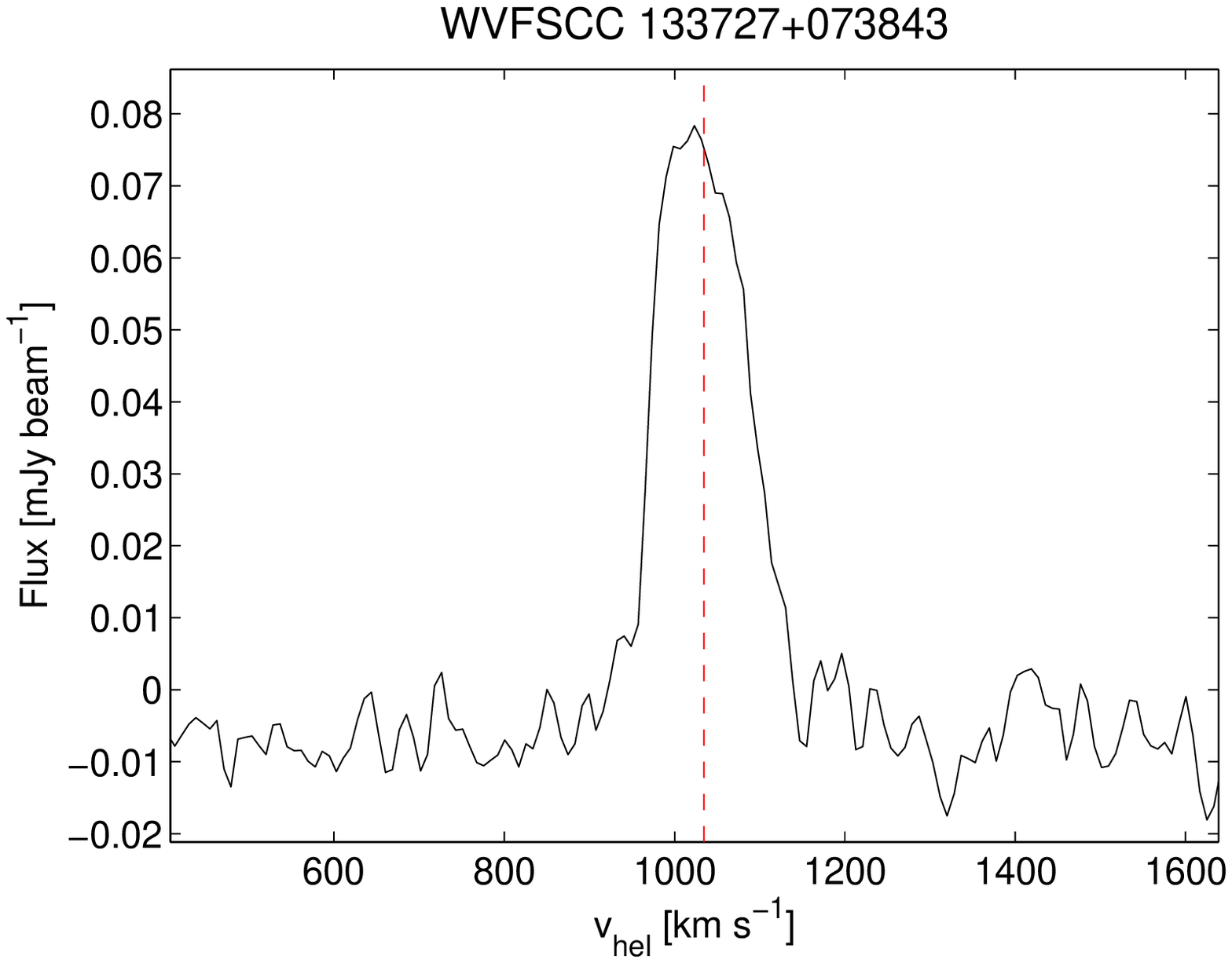}
\includegraphics[width=0.32\textwidth]{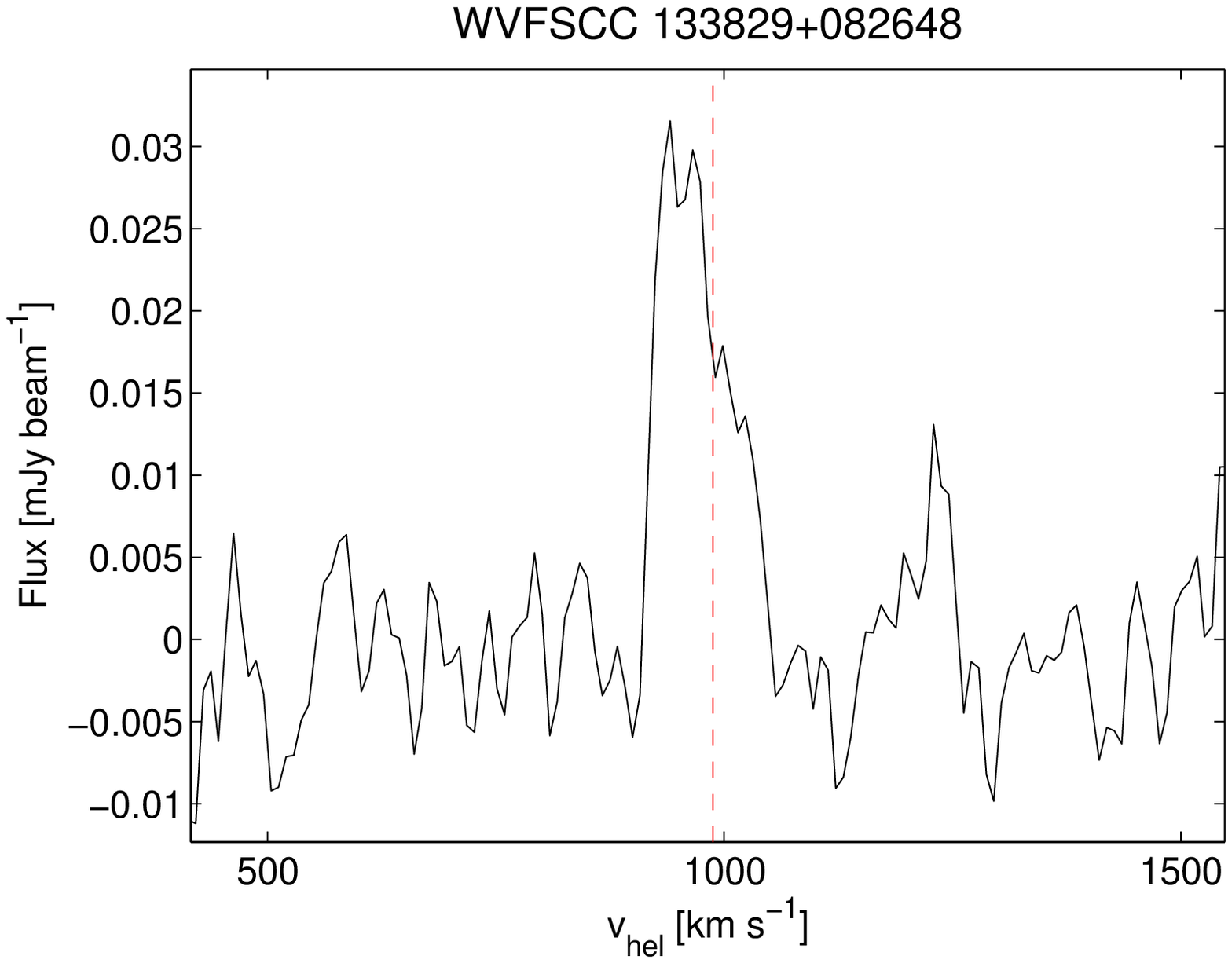}
\includegraphics[width=0.32\textwidth]{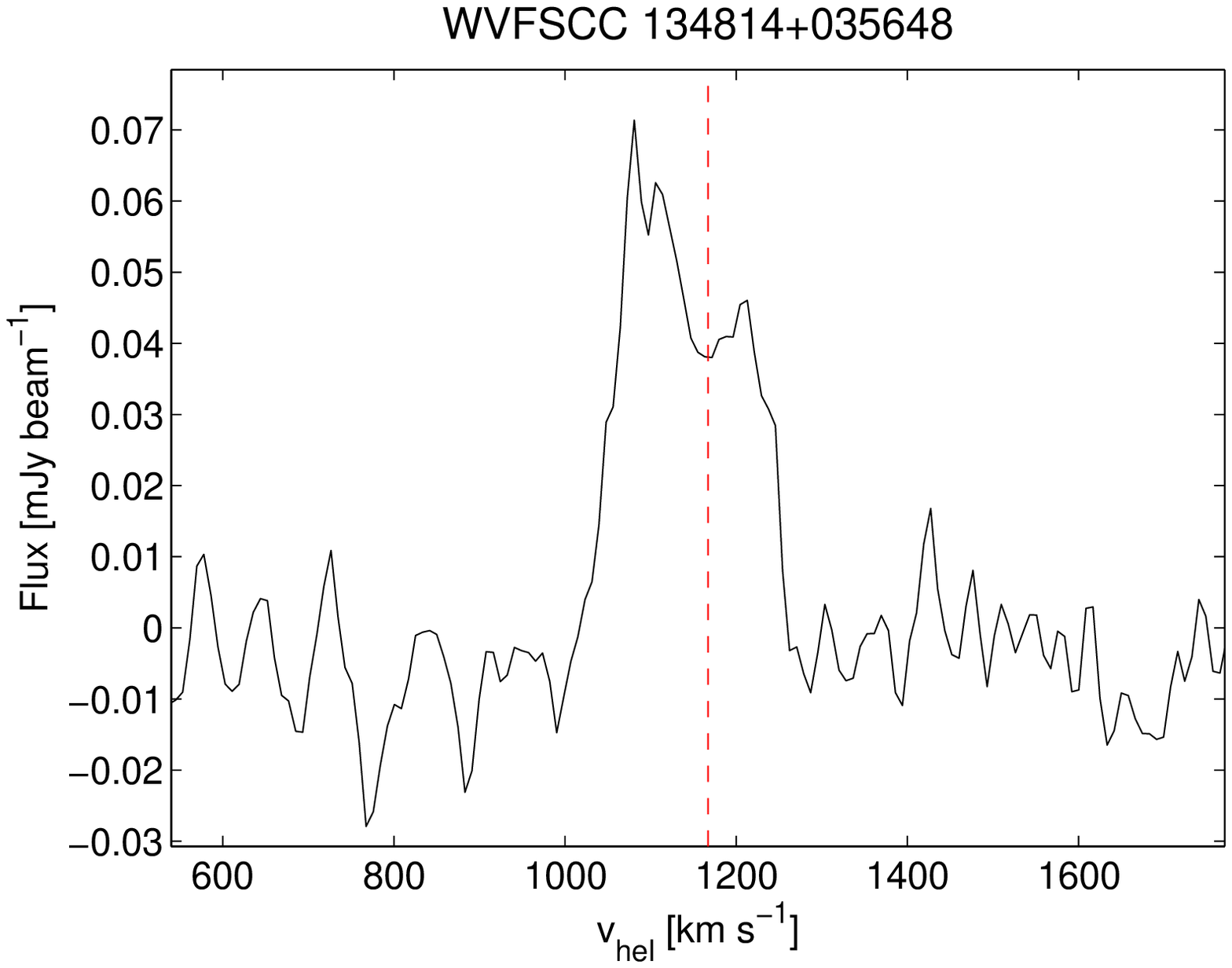}
\includegraphics[width=0.32\textwidth]{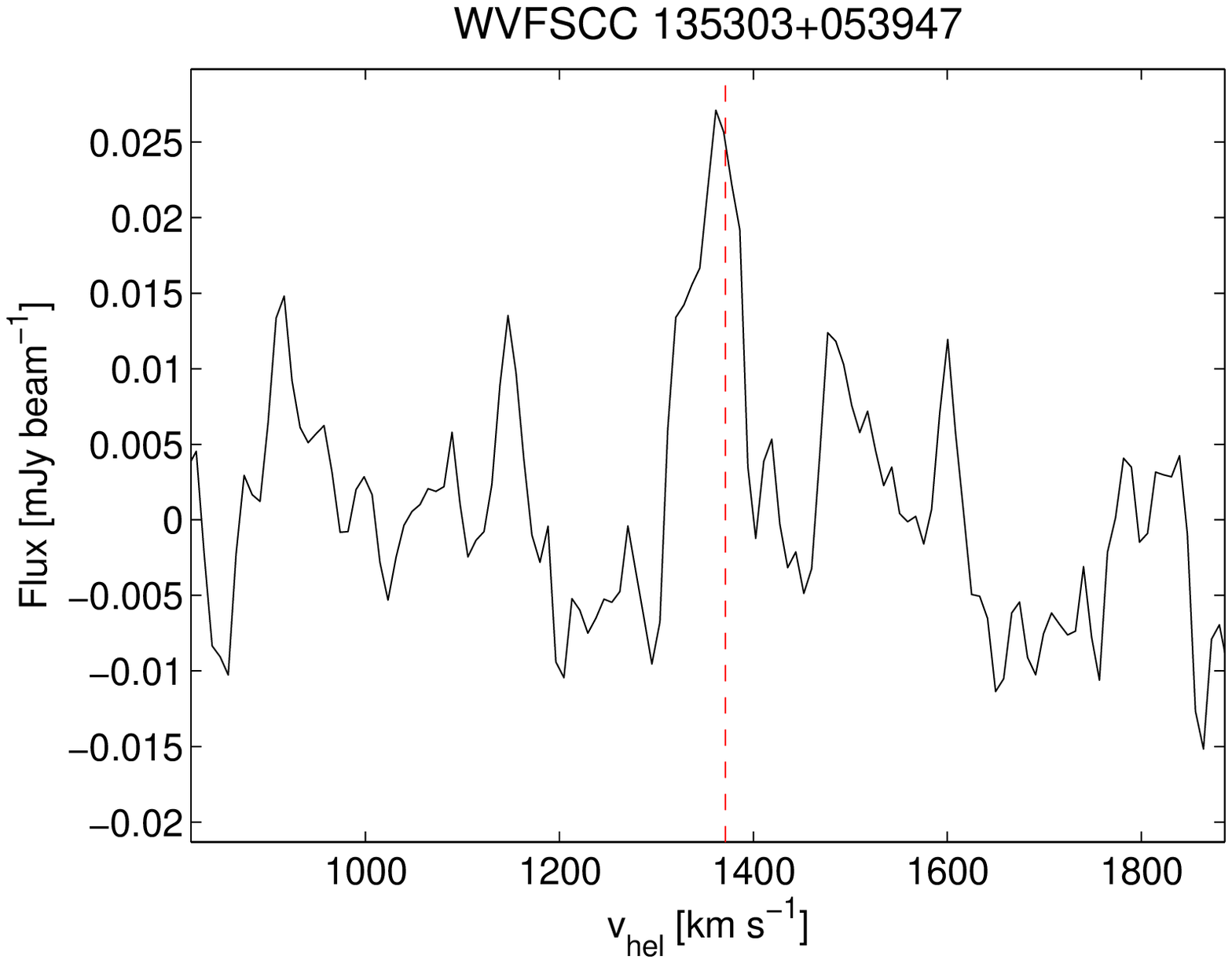}
\includegraphics[width=0.32\textwidth]{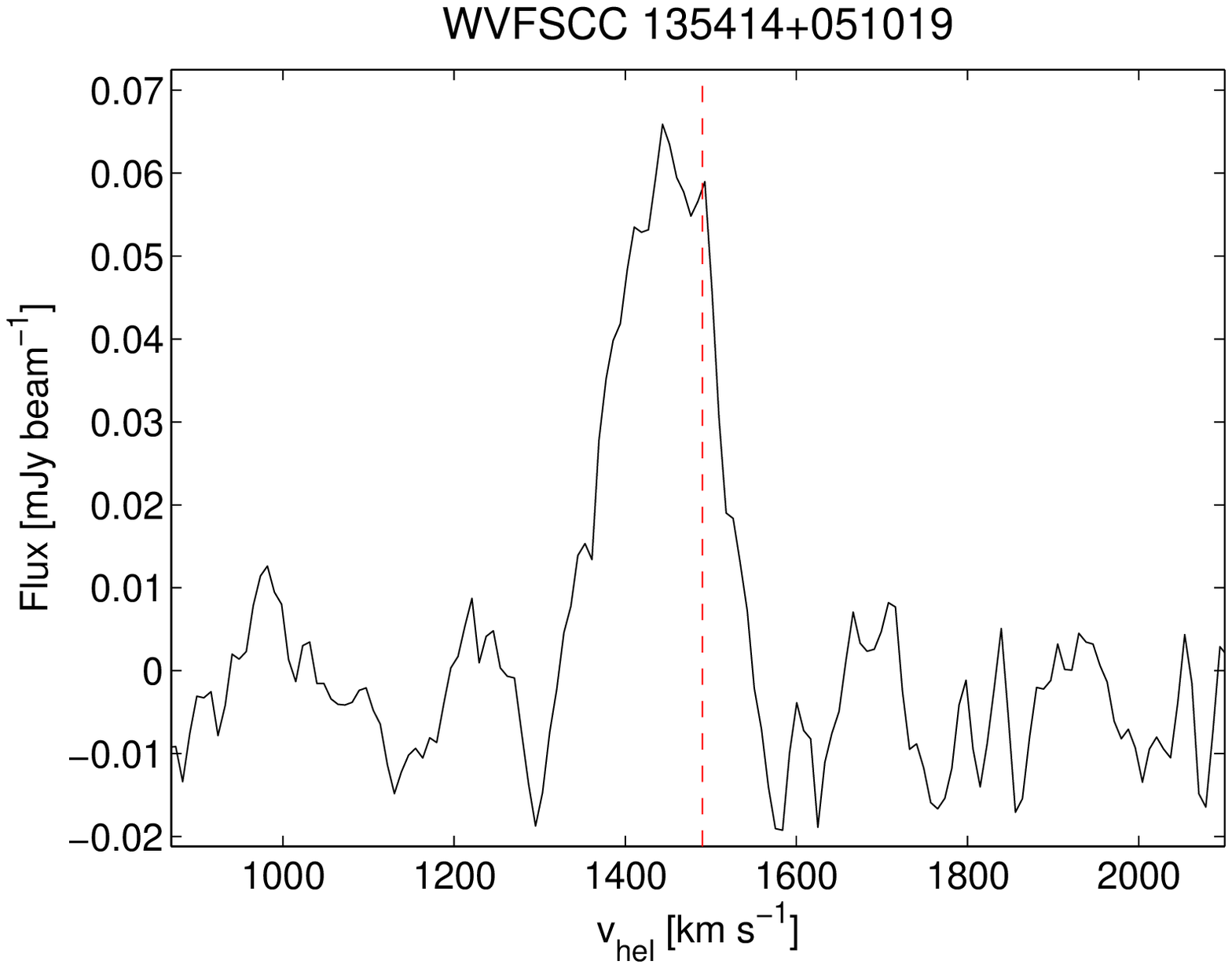}
\includegraphics[width=0.32\textwidth]{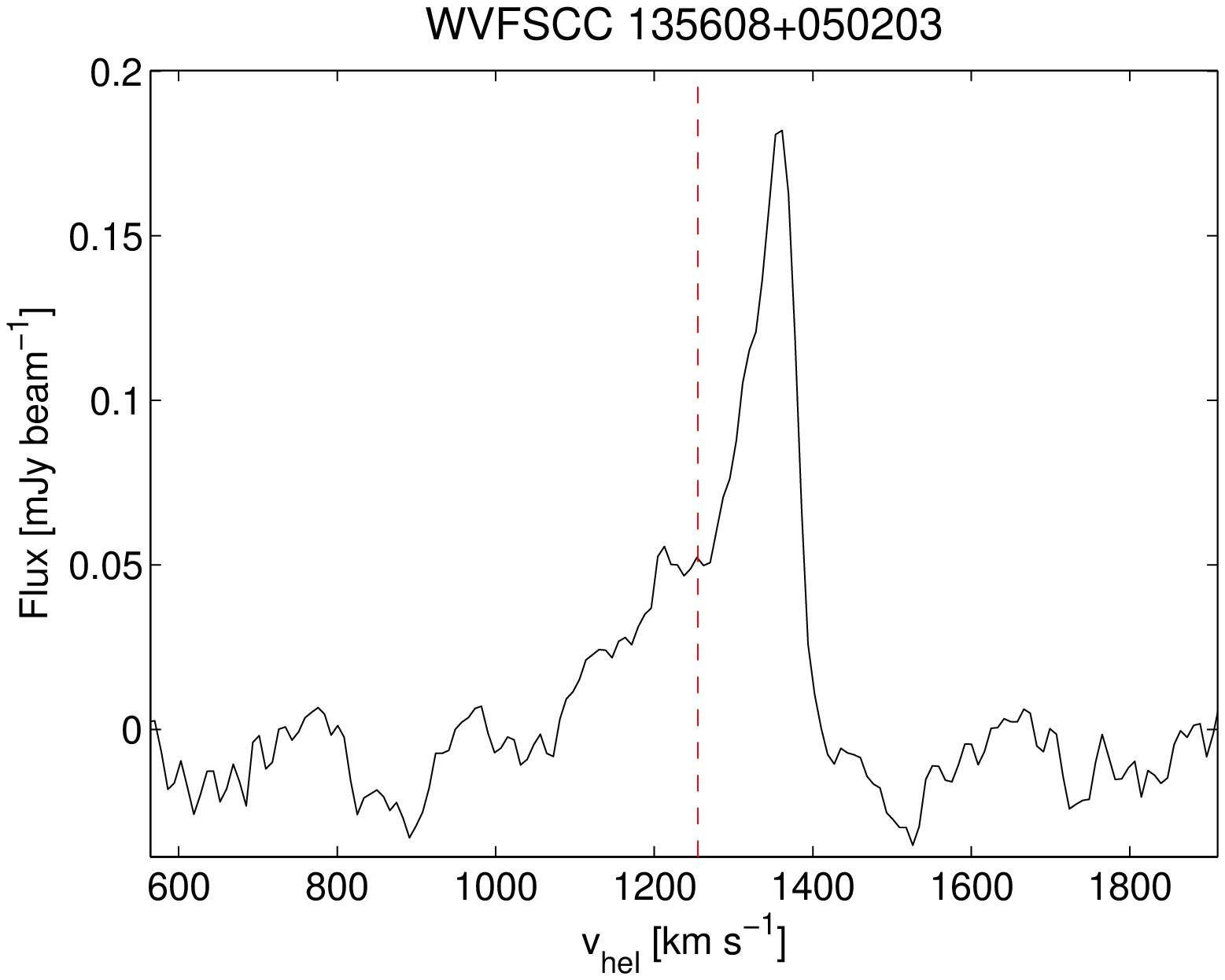}
\includegraphics[width=0.32\textwidth]{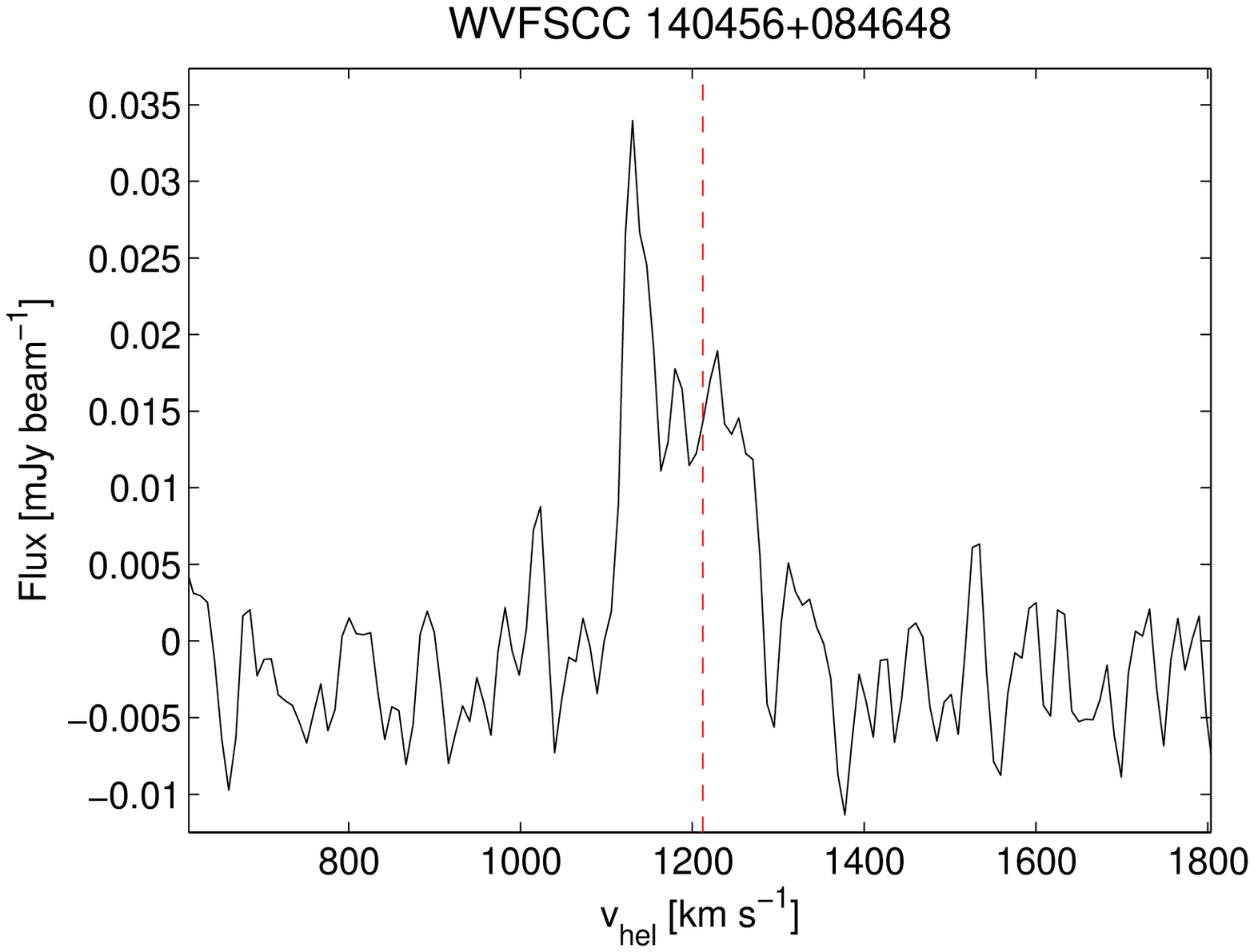}
\includegraphics[width=0.32\textwidth]{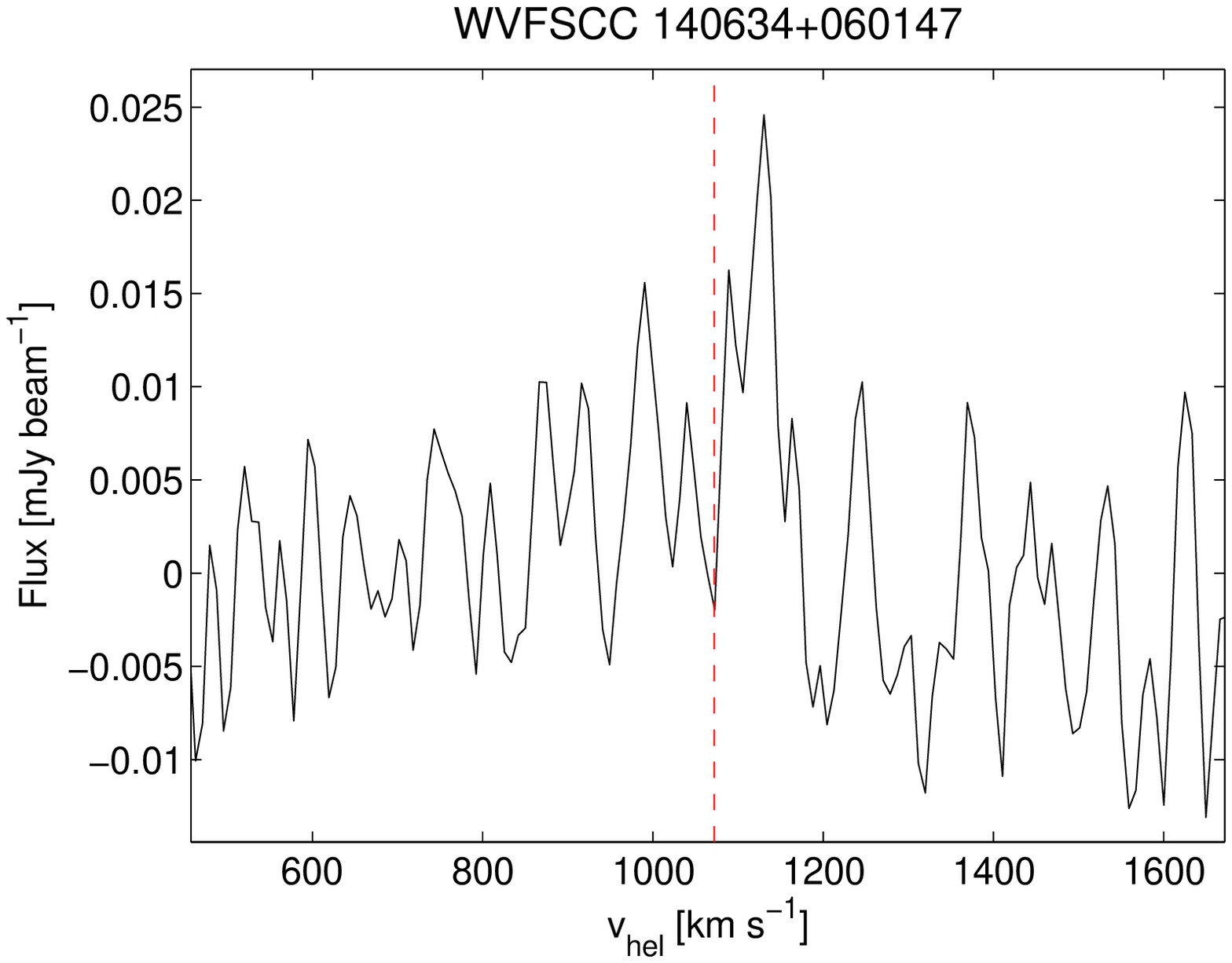}
\includegraphics[width=0.32\textwidth]{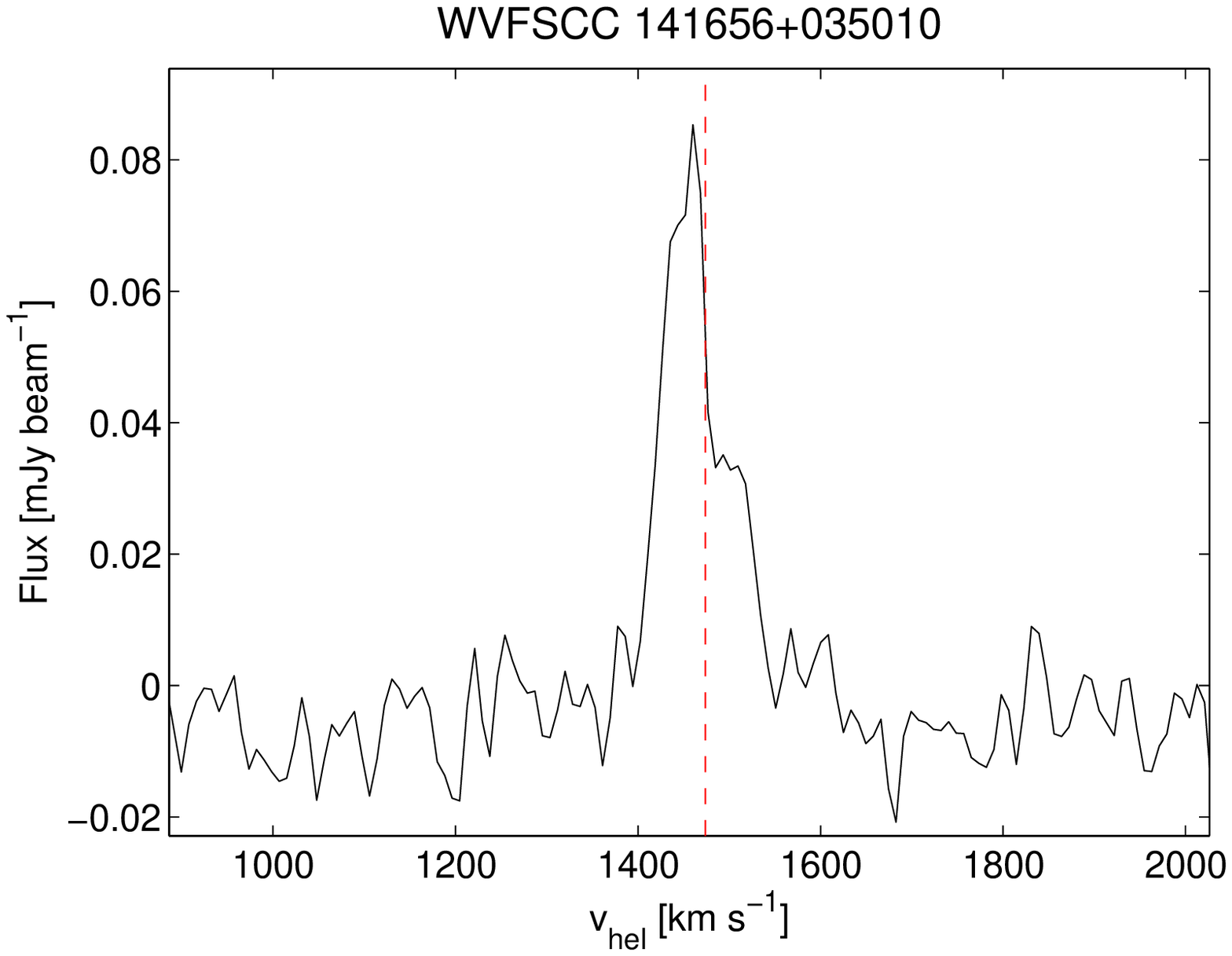}

\end{center}                                            
{\bf Fig~\ref{all_spectra2}.} (continued)                                        
 
\end{figure*}


\begin{figure*}
  \begin{center}
  
\includegraphics[width=0.32\textwidth]{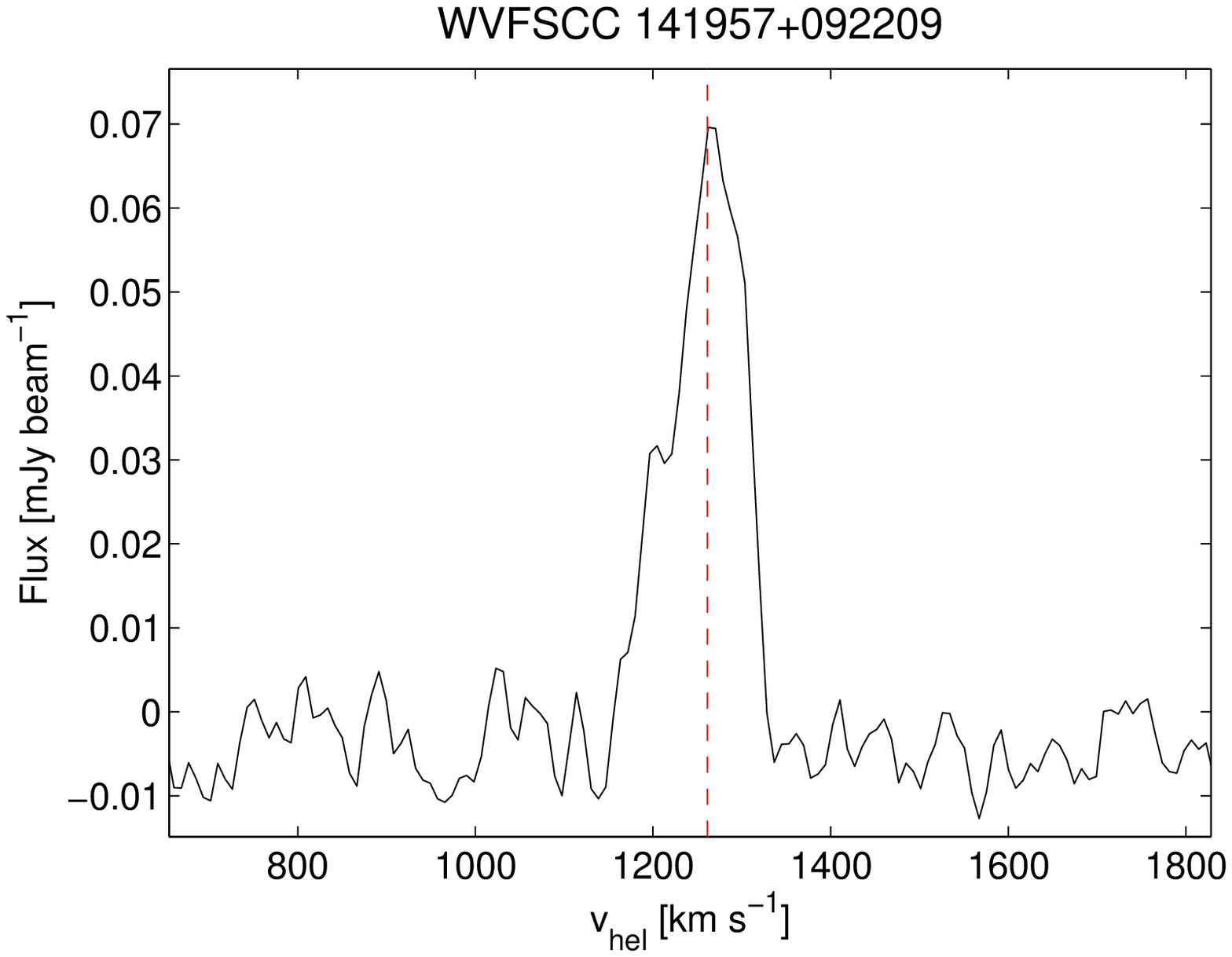}
\includegraphics[width=0.32\textwidth]{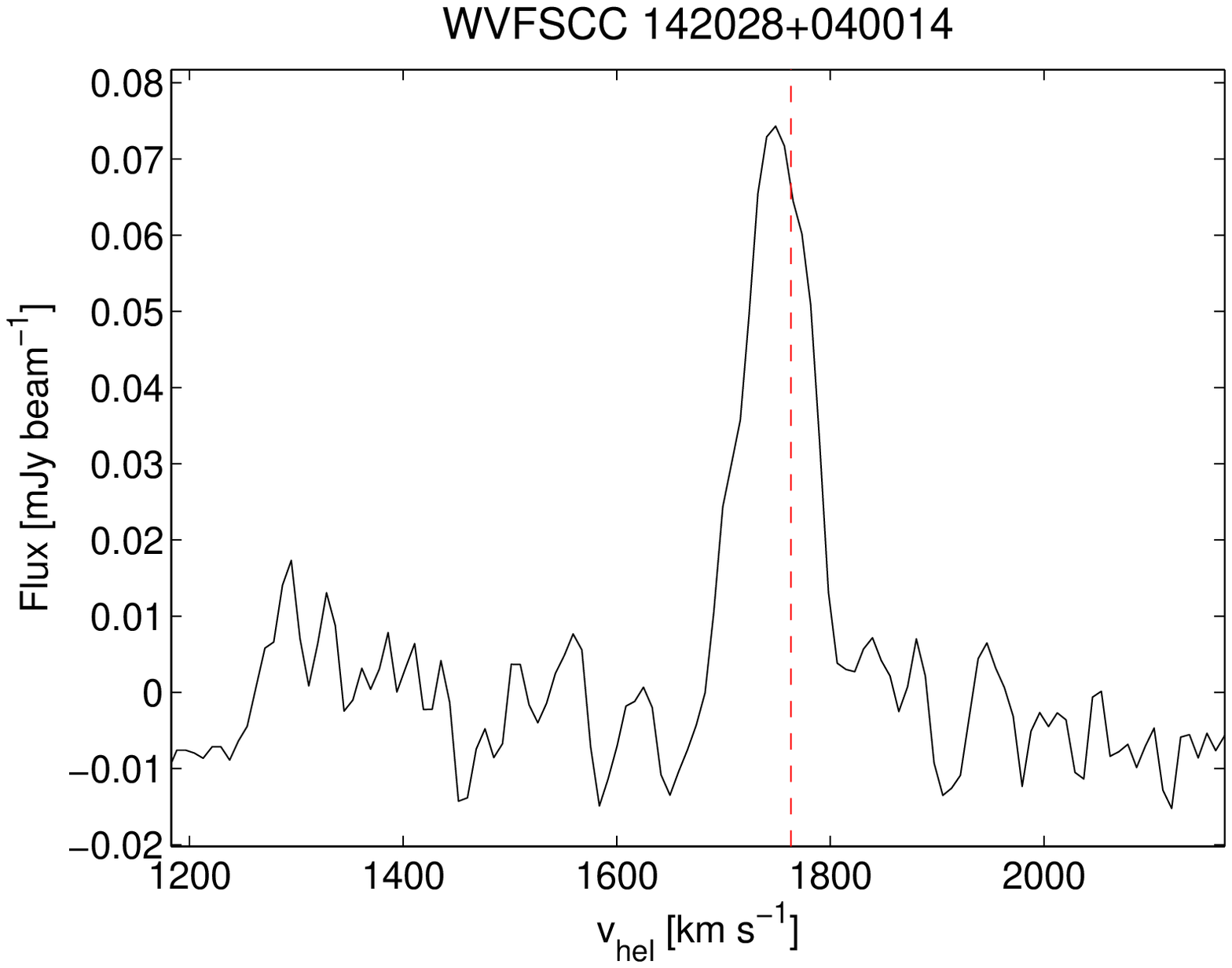}
\includegraphics[width=0.32\textwidth]{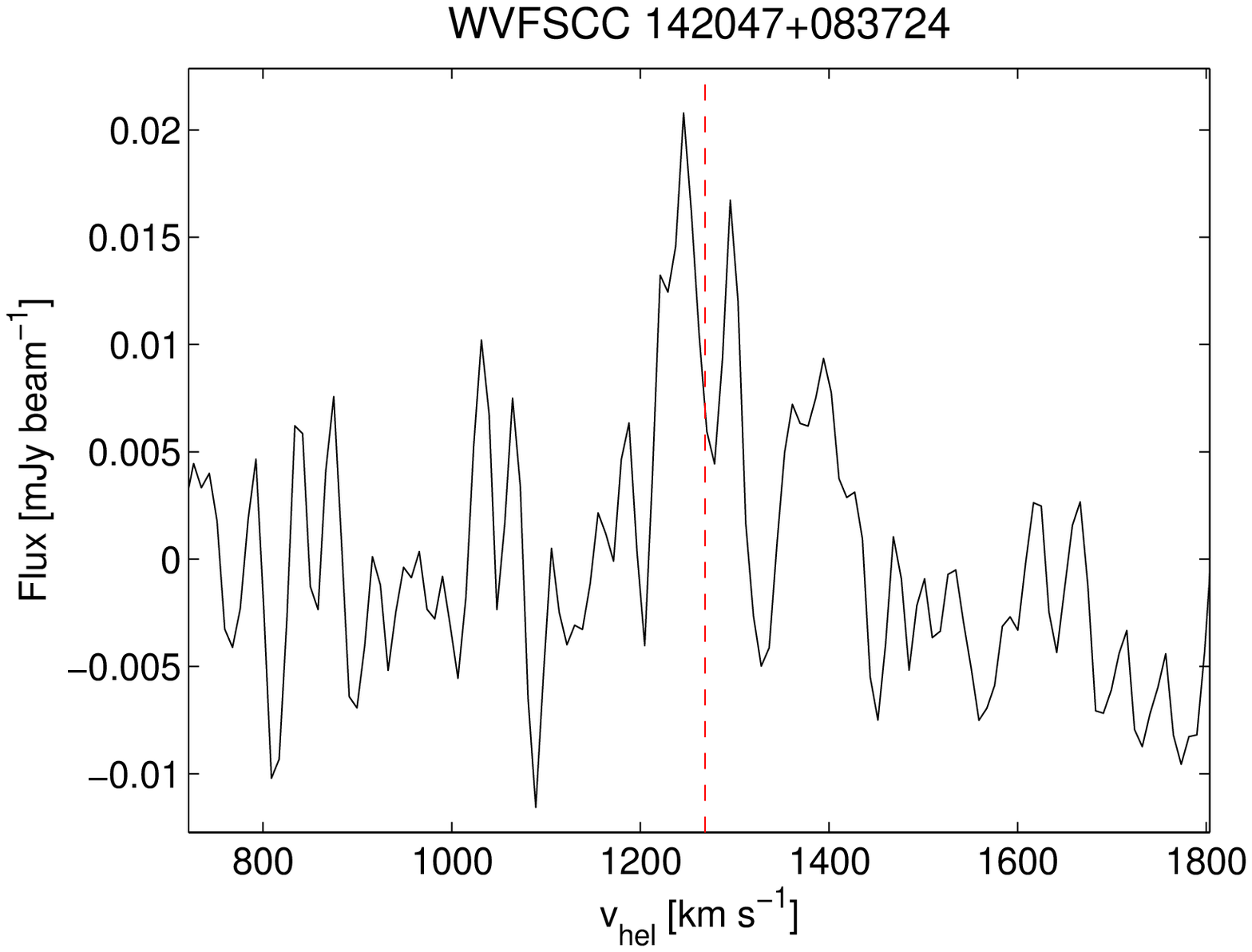}
\includegraphics[width=0.32\textwidth]{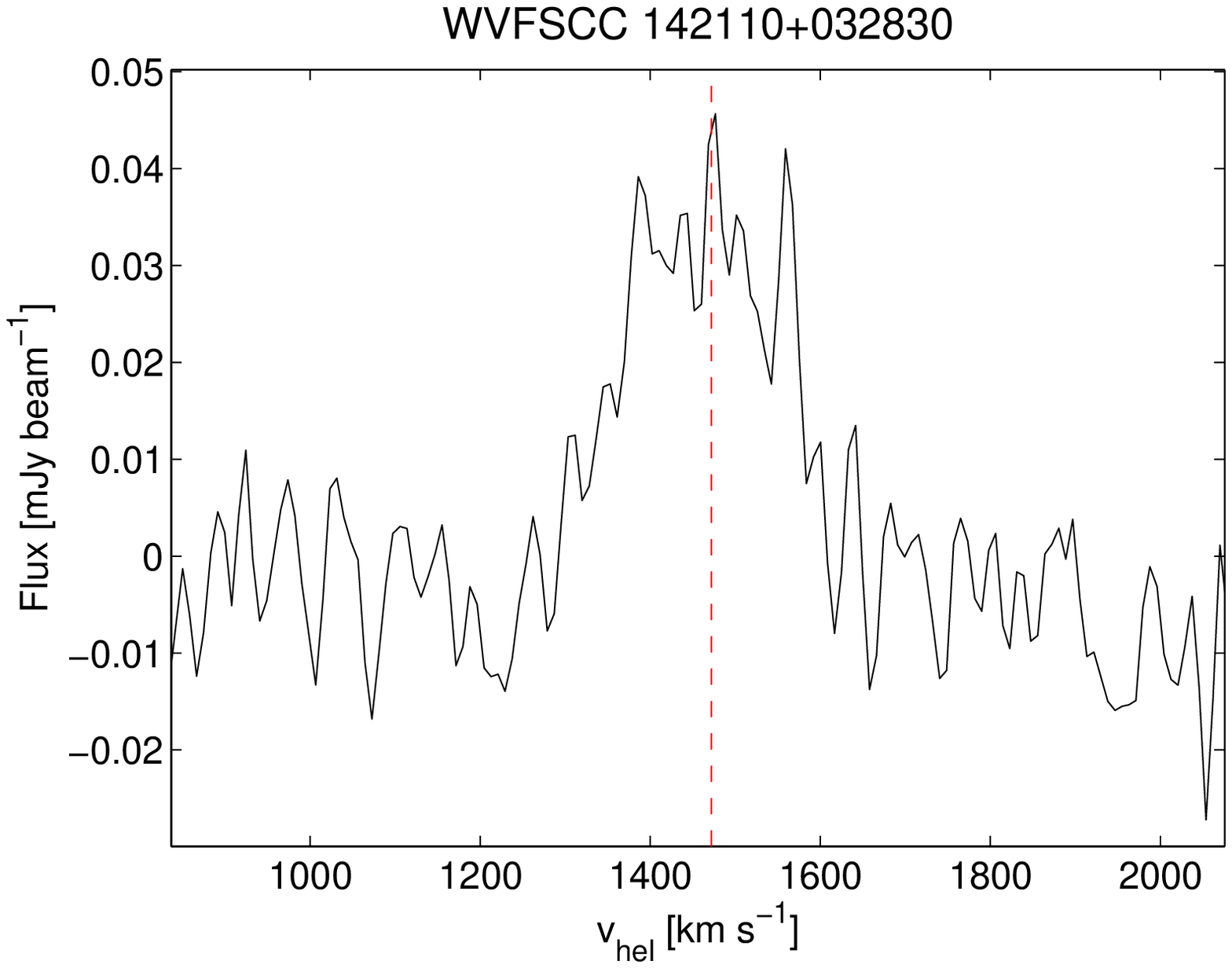}
\includegraphics[width=0.32\textwidth]{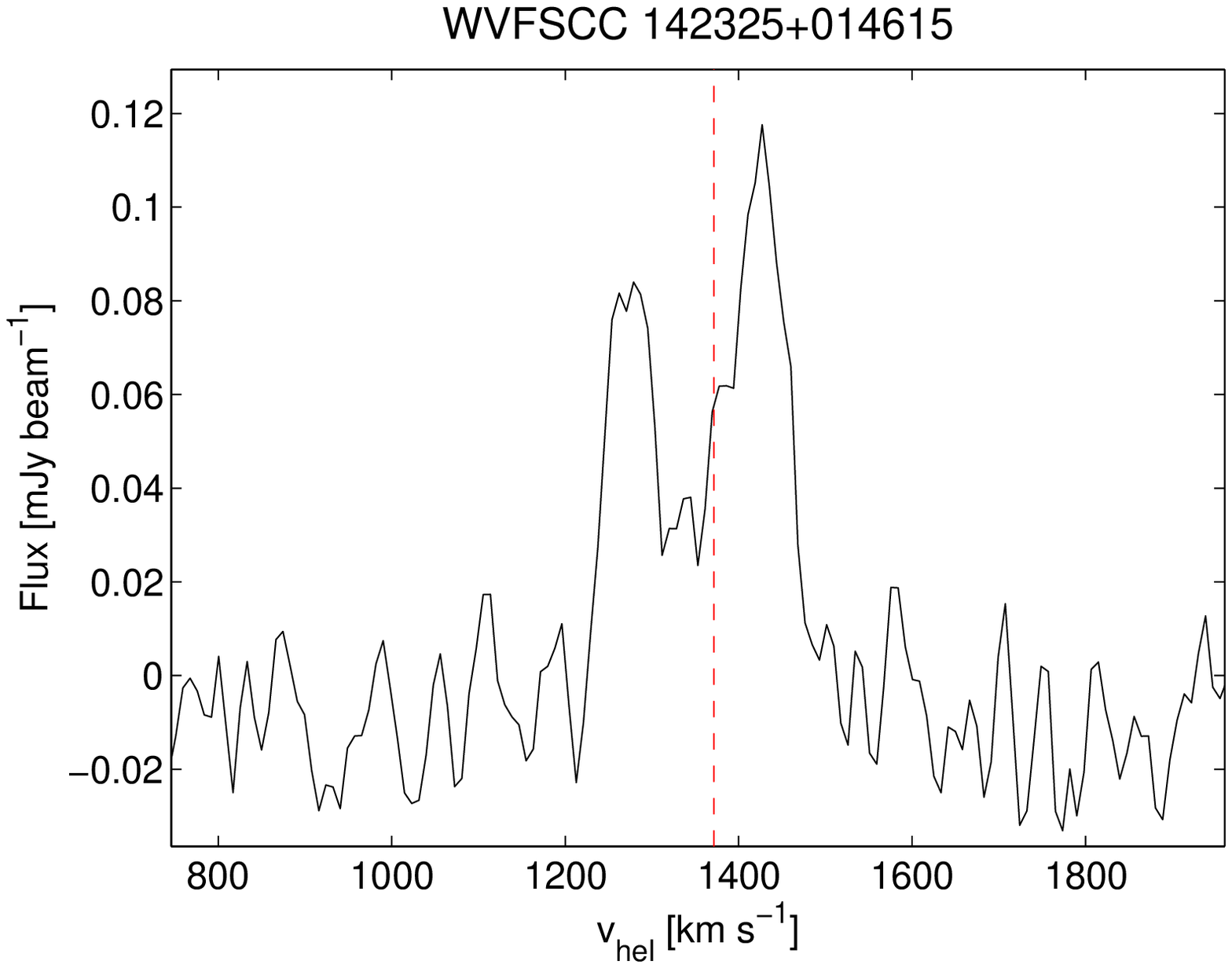}
\includegraphics[width=0.32\textwidth]{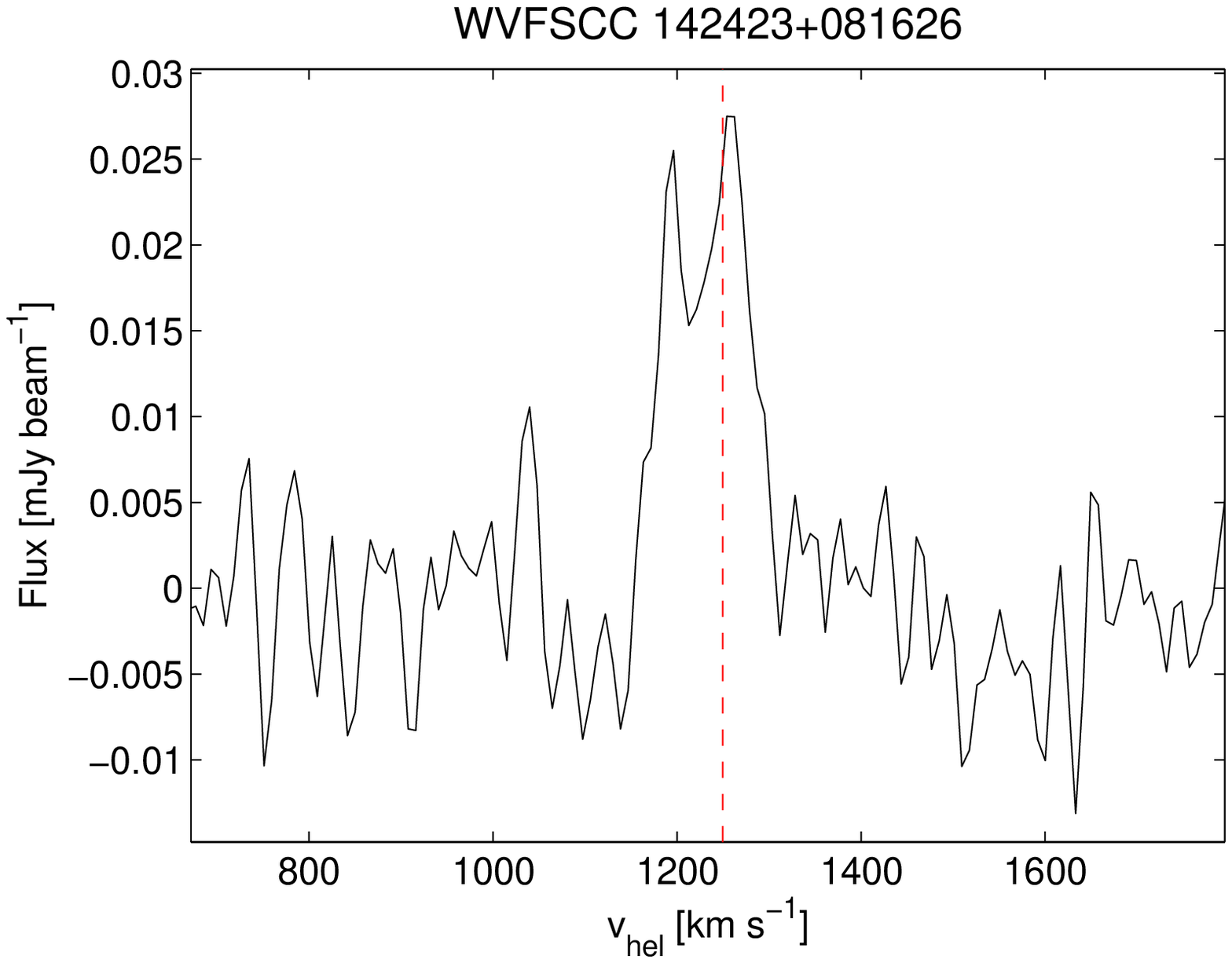}
\includegraphics[width=0.32\textwidth]{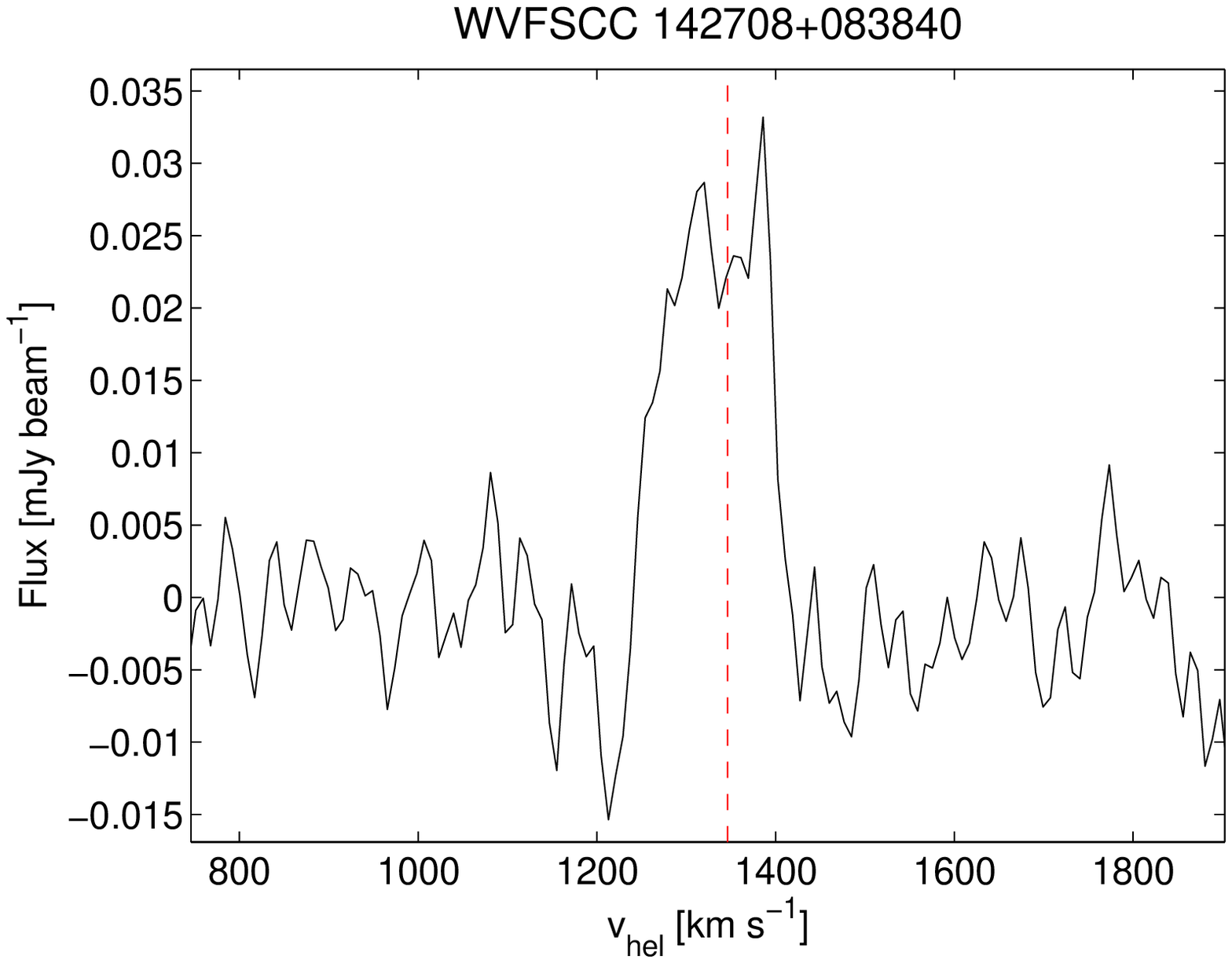}
\includegraphics[width=0.32\textwidth]{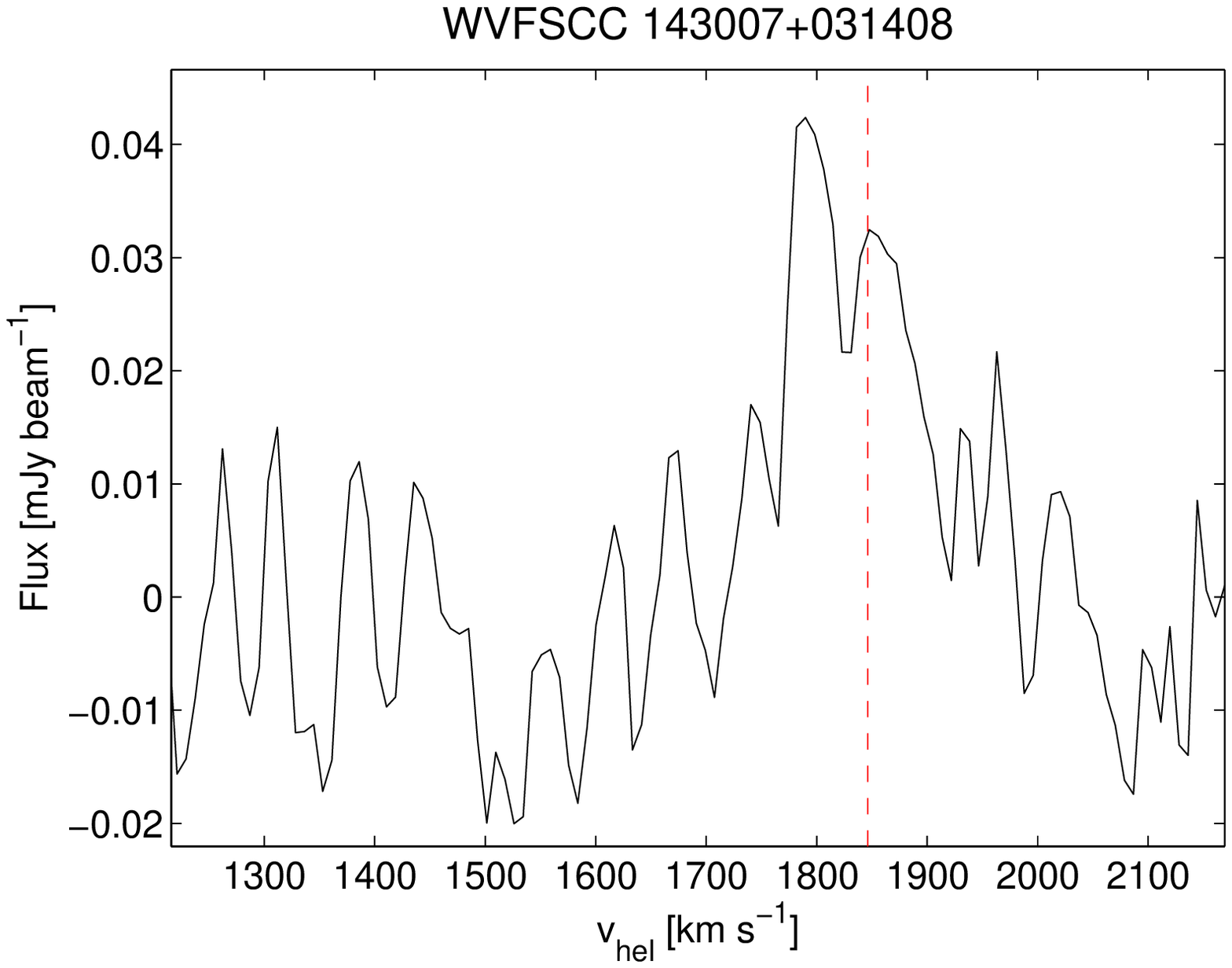}
\includegraphics[width=0.32\textwidth]{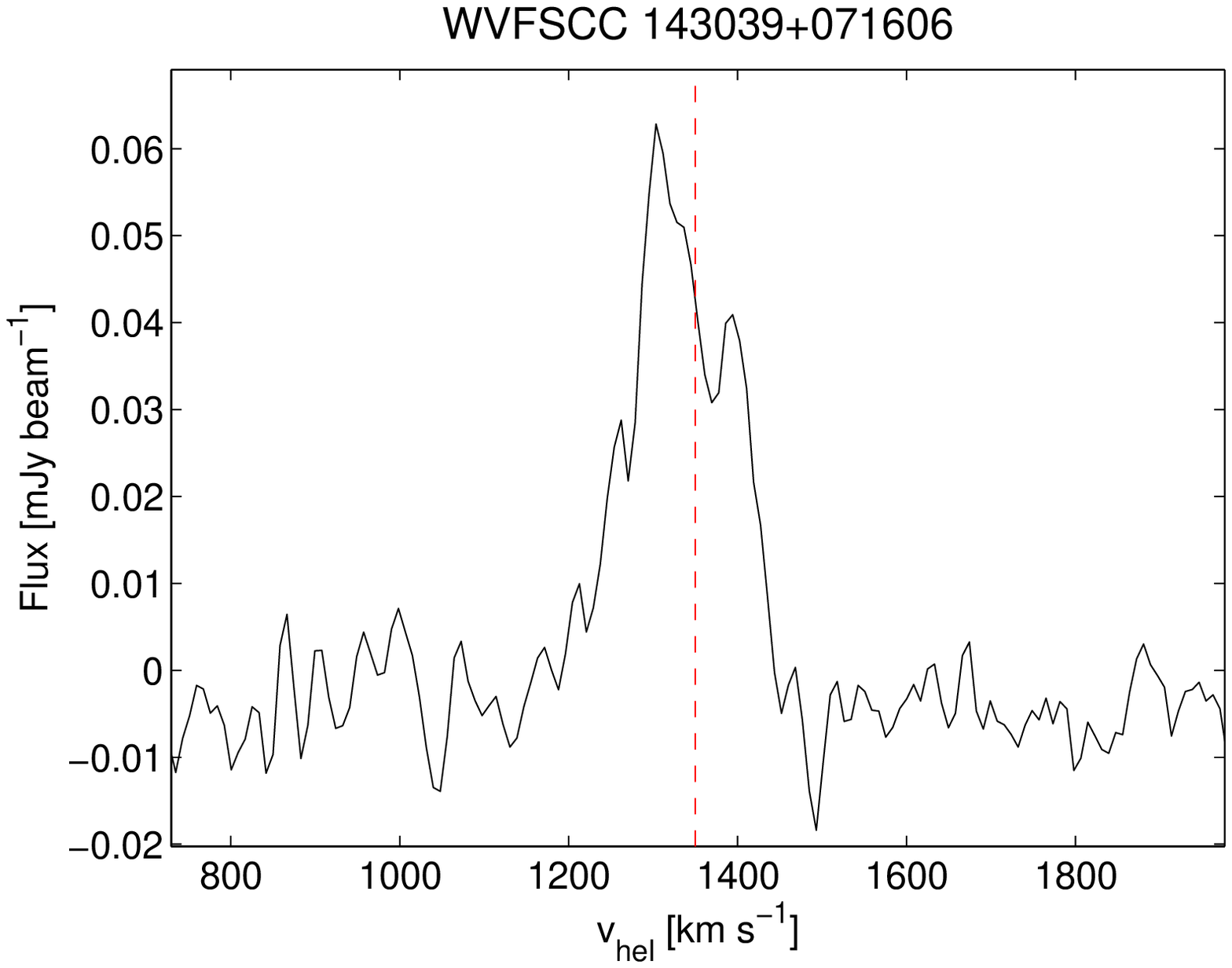}
\includegraphics[width=0.32\textwidth]{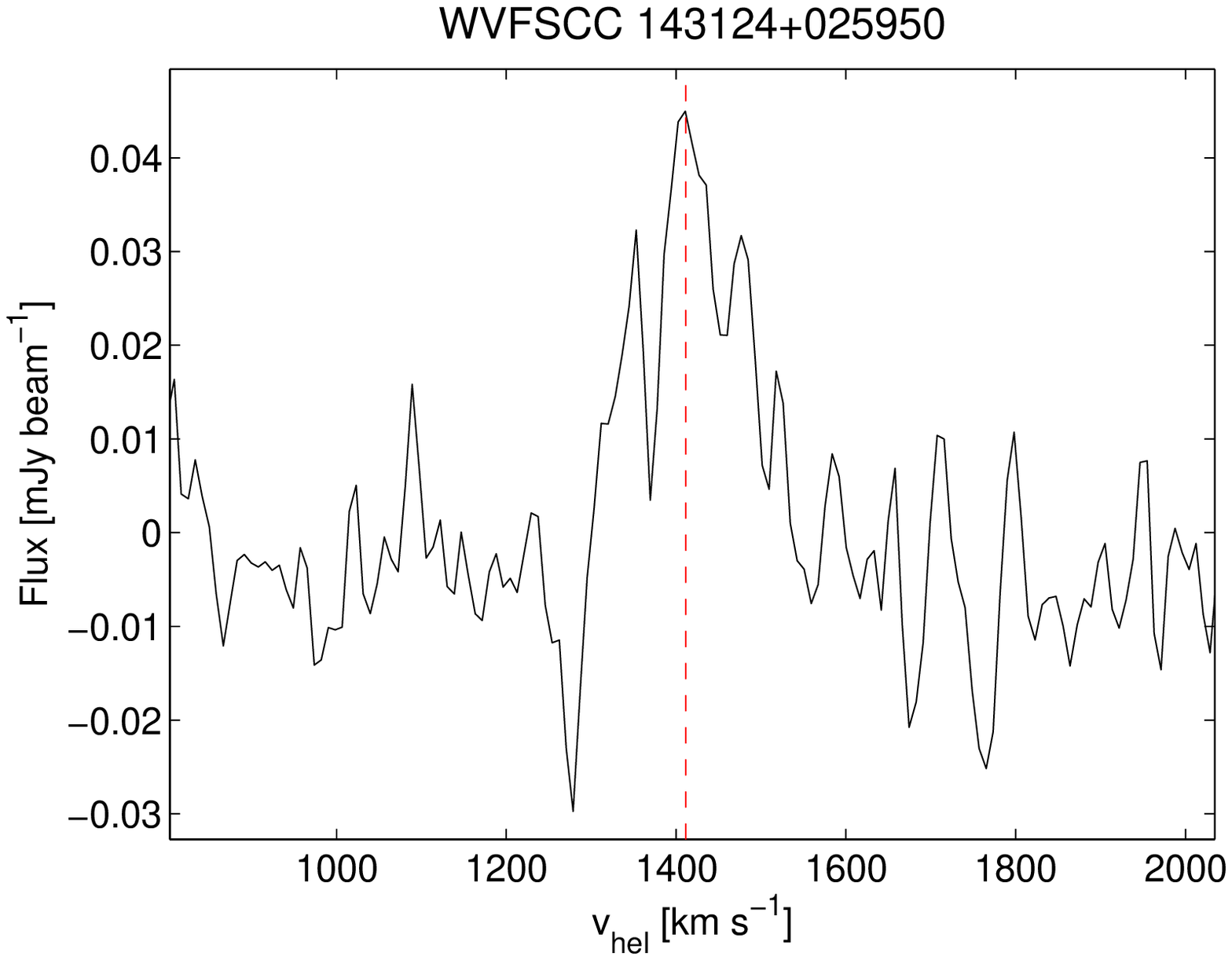}
\includegraphics[width=0.32\textwidth]{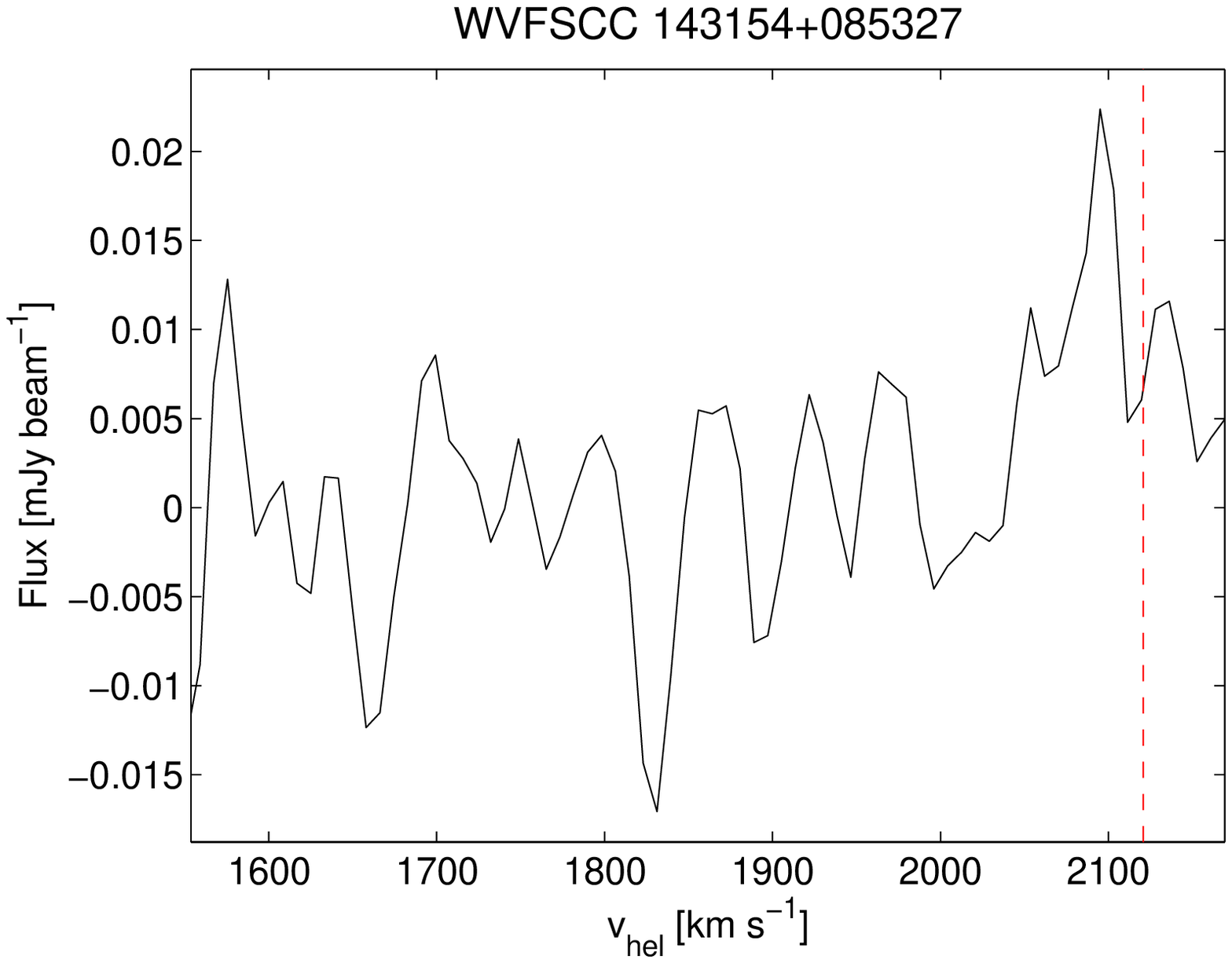}
\includegraphics[width=0.32\textwidth]{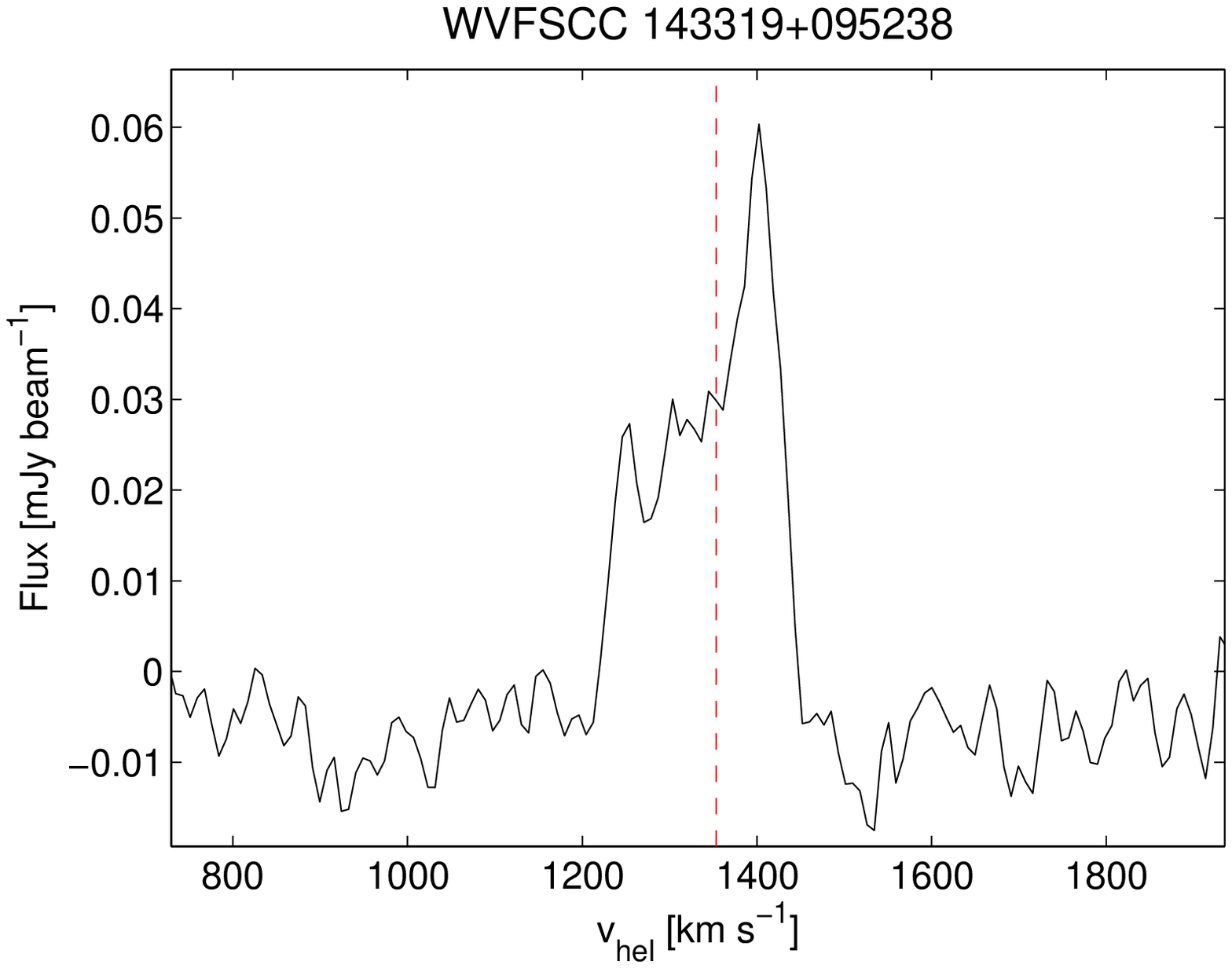}
\includegraphics[width=0.32\textwidth]{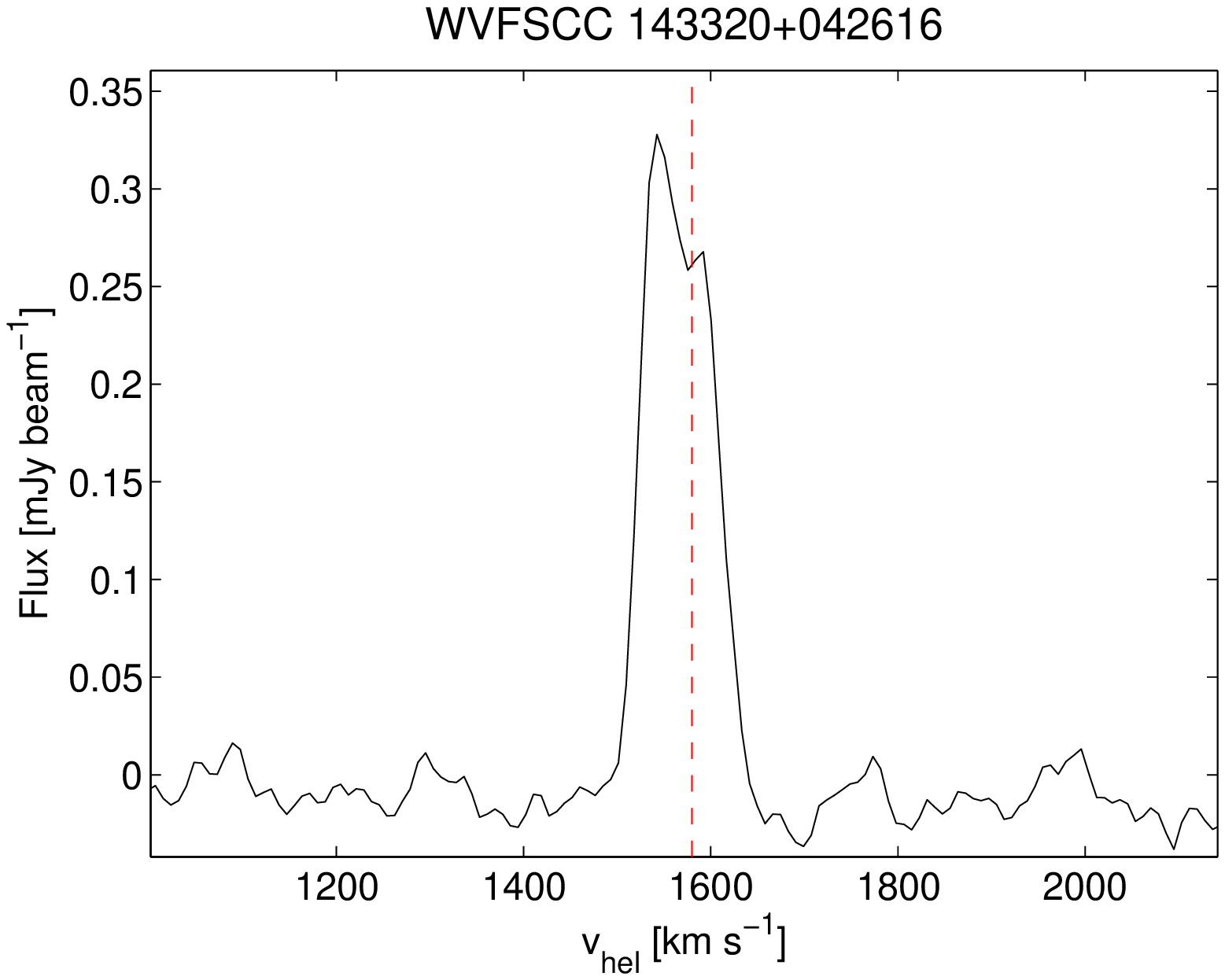}
\includegraphics[width=0.32\textwidth]{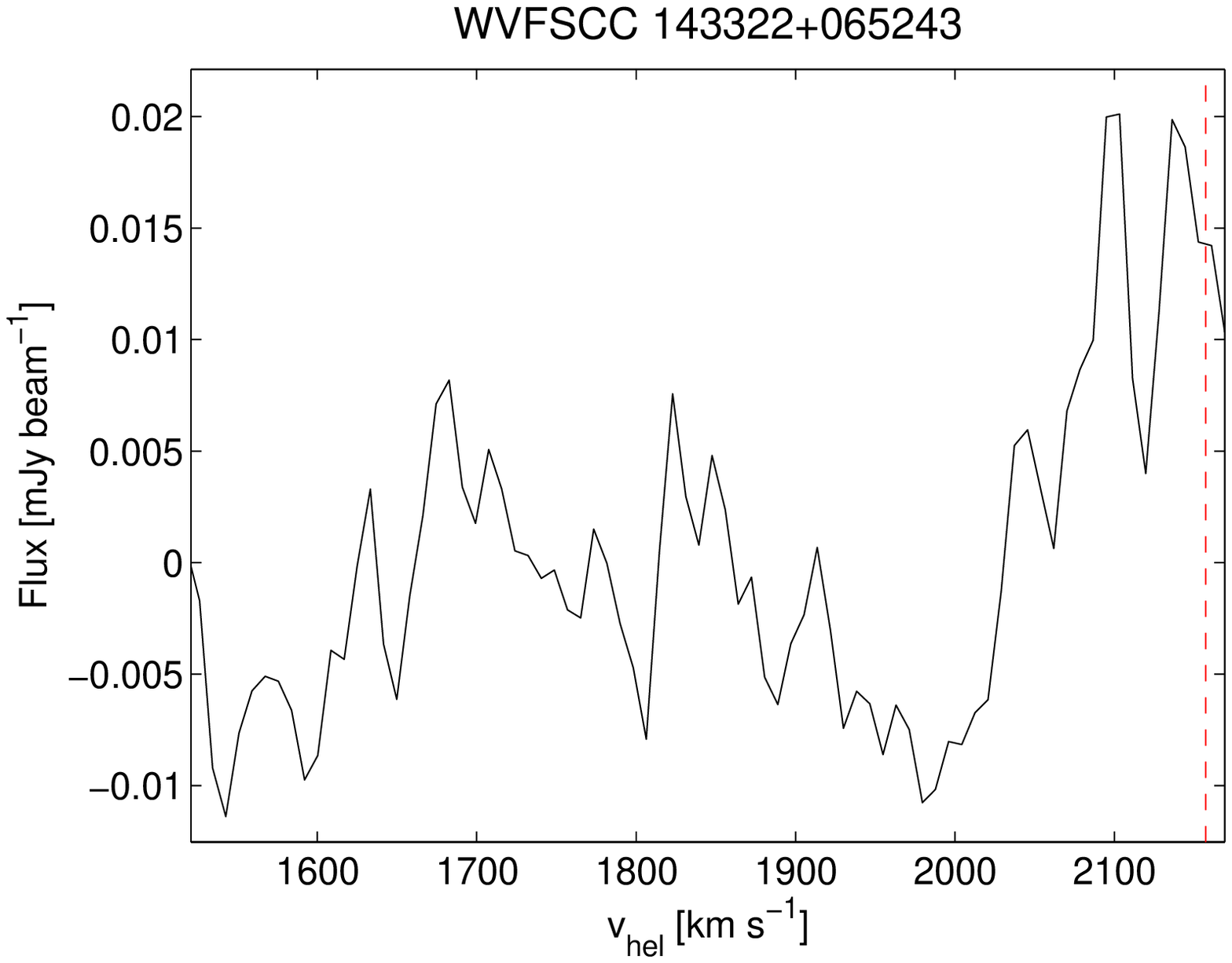}
\includegraphics[width=0.32\textwidth]{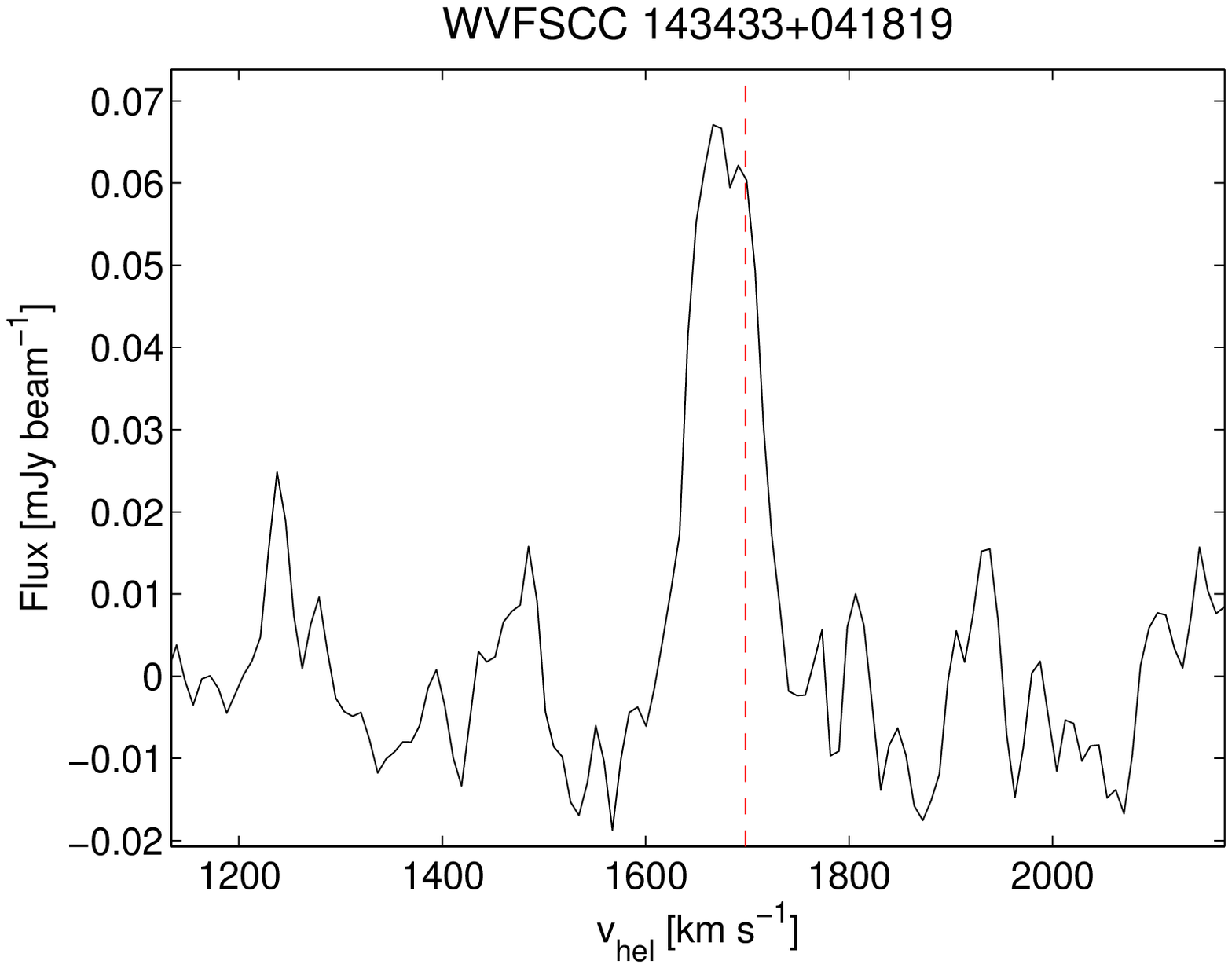}

\end{center}                                            
{\bf Fig~\ref{all_spectra2}.} (continued)                                        
 
\end{figure*}


\begin{figure*}
  \begin{center}
 
\includegraphics[width=0.32\textwidth]{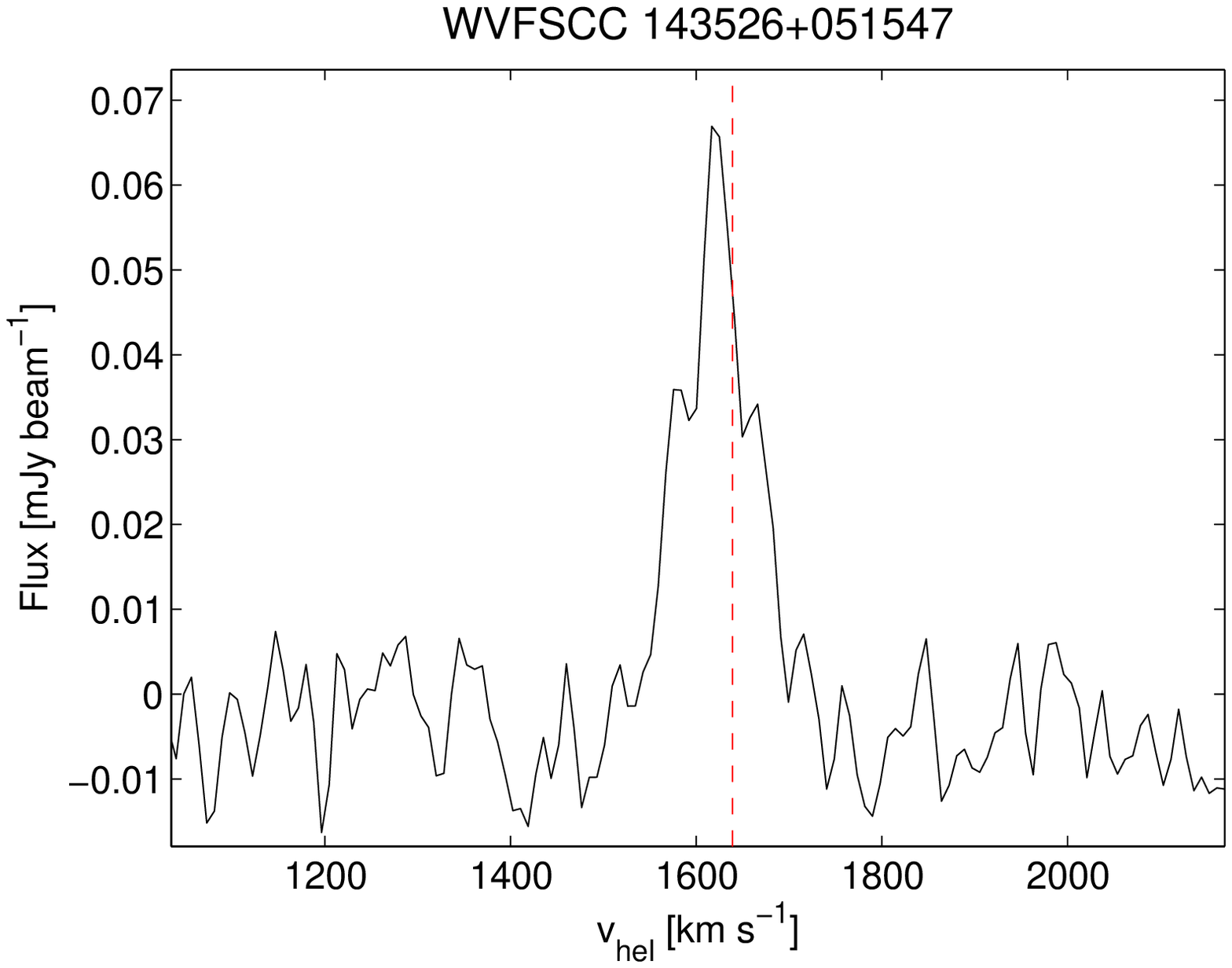}
\includegraphics[width=0.32\textwidth]{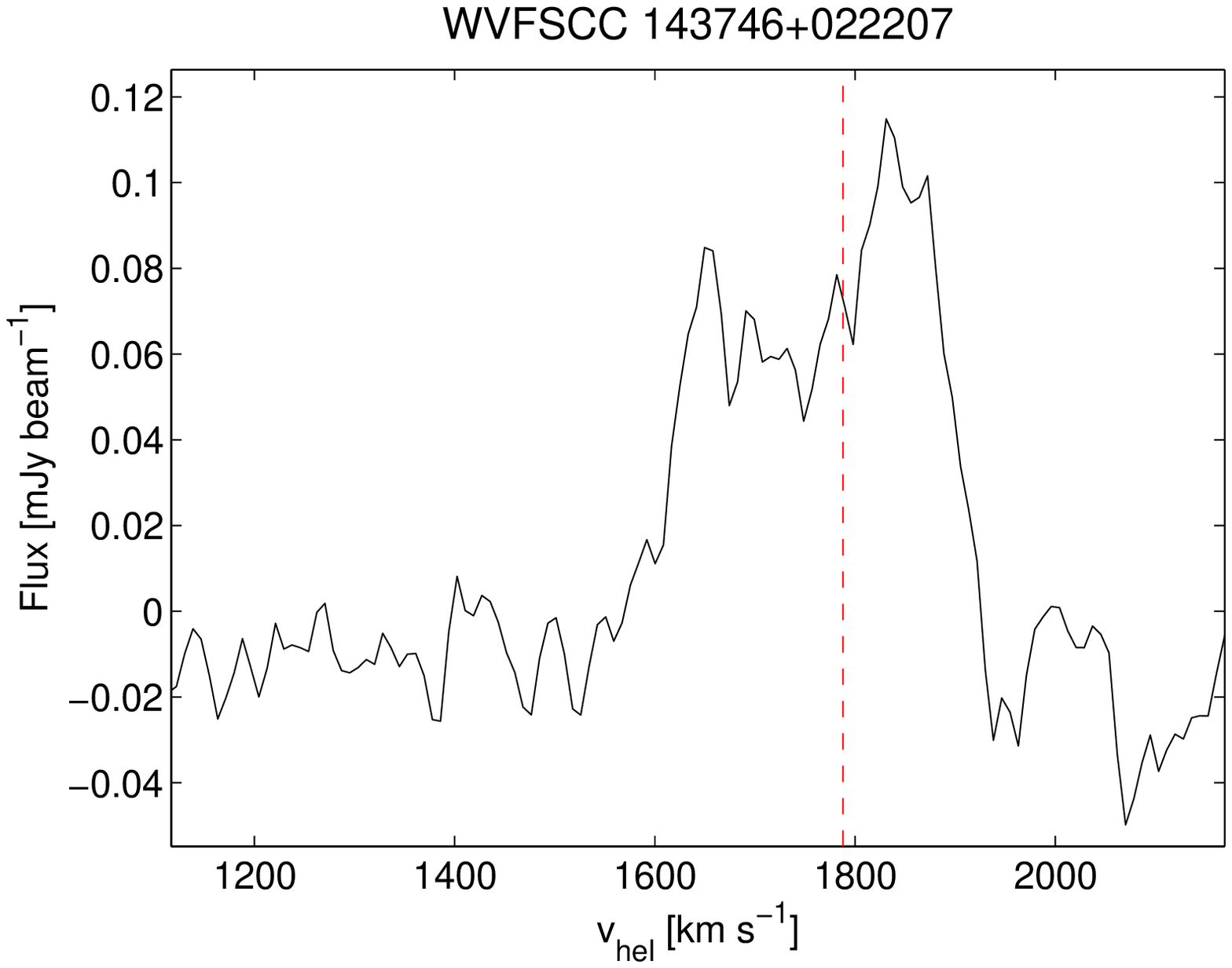}
\includegraphics[width=0.32\textwidth]{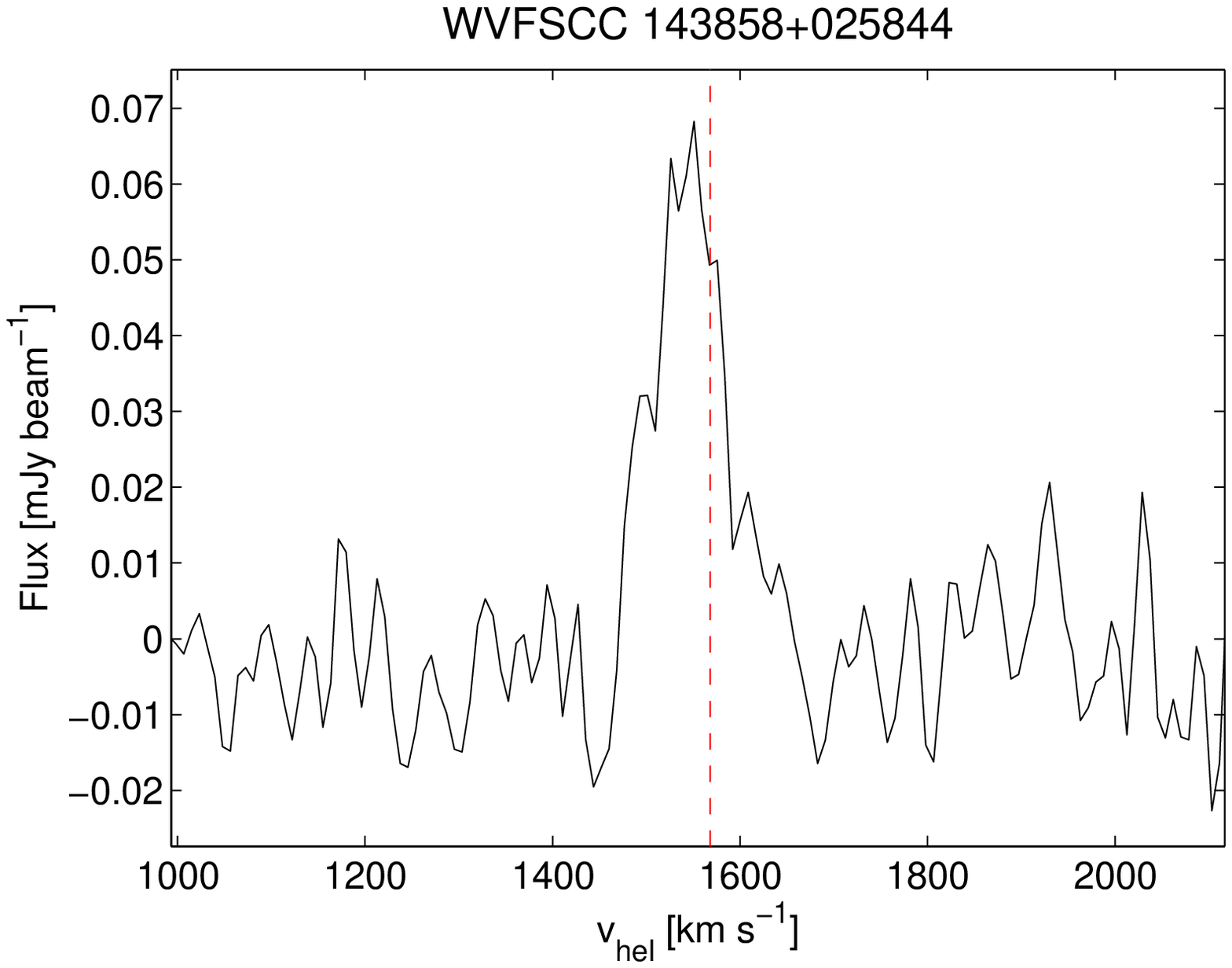}
\includegraphics[width=0.32\textwidth]{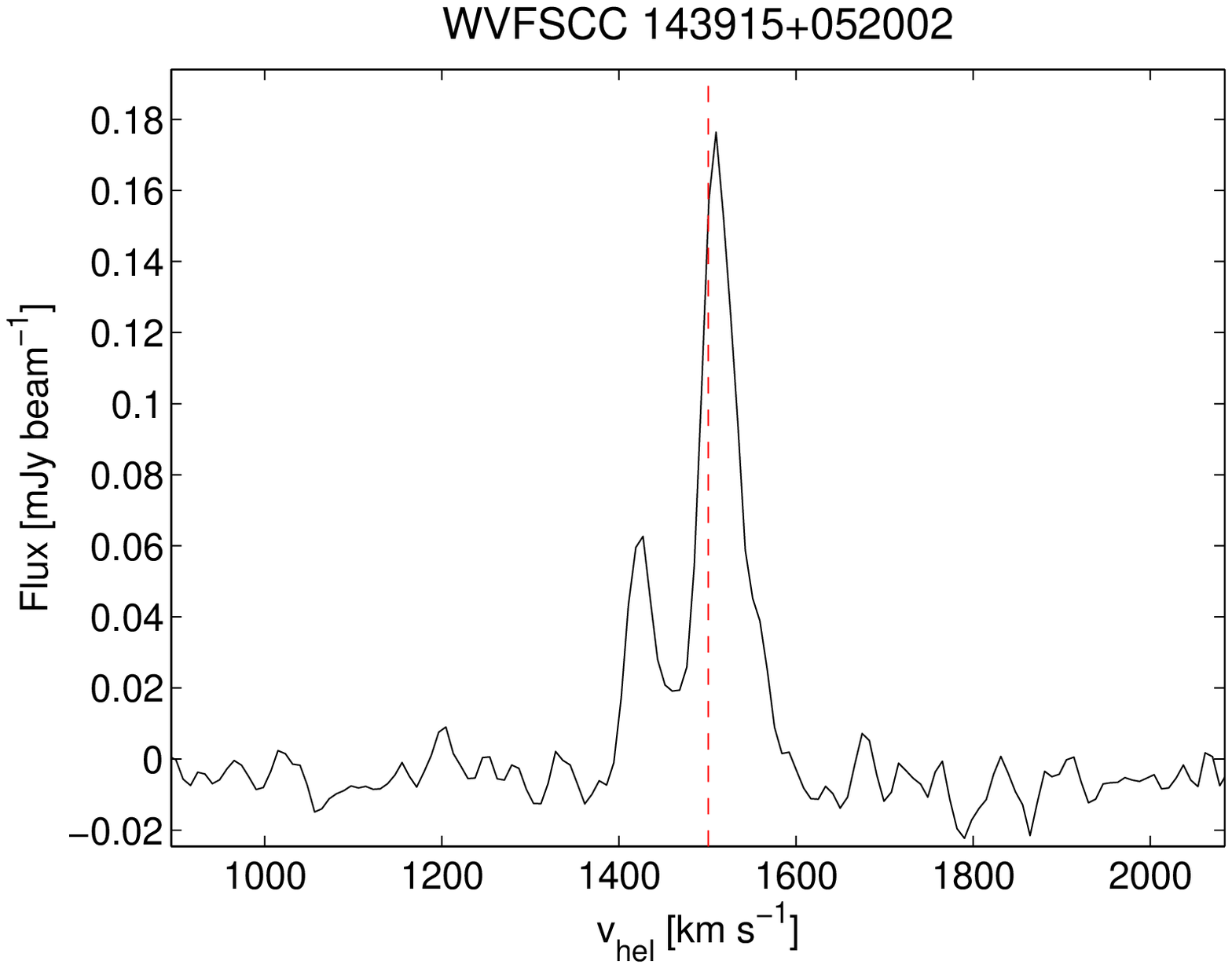}
\includegraphics[width=0.32\textwidth]{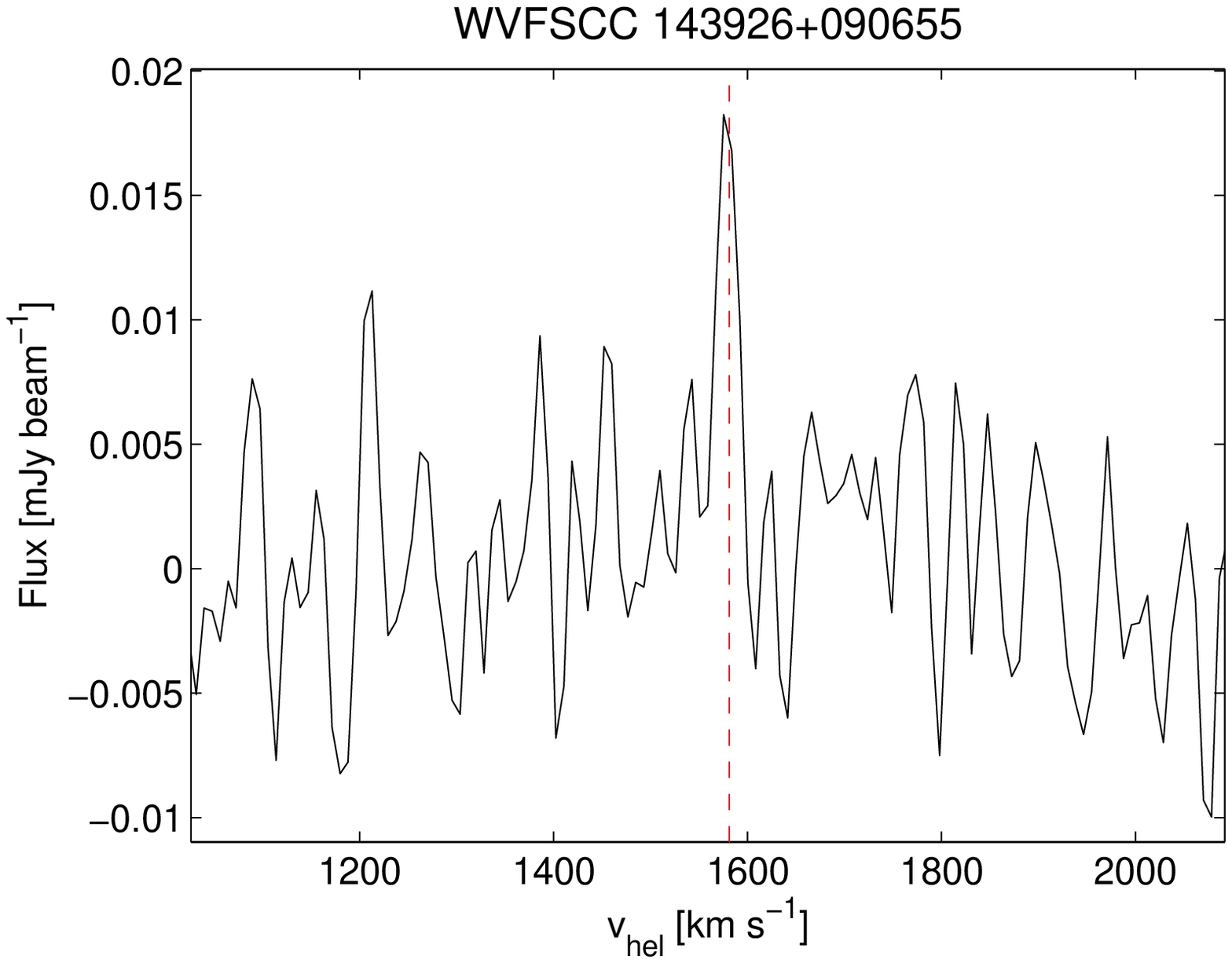} 
\includegraphics[width=0.32\textwidth]{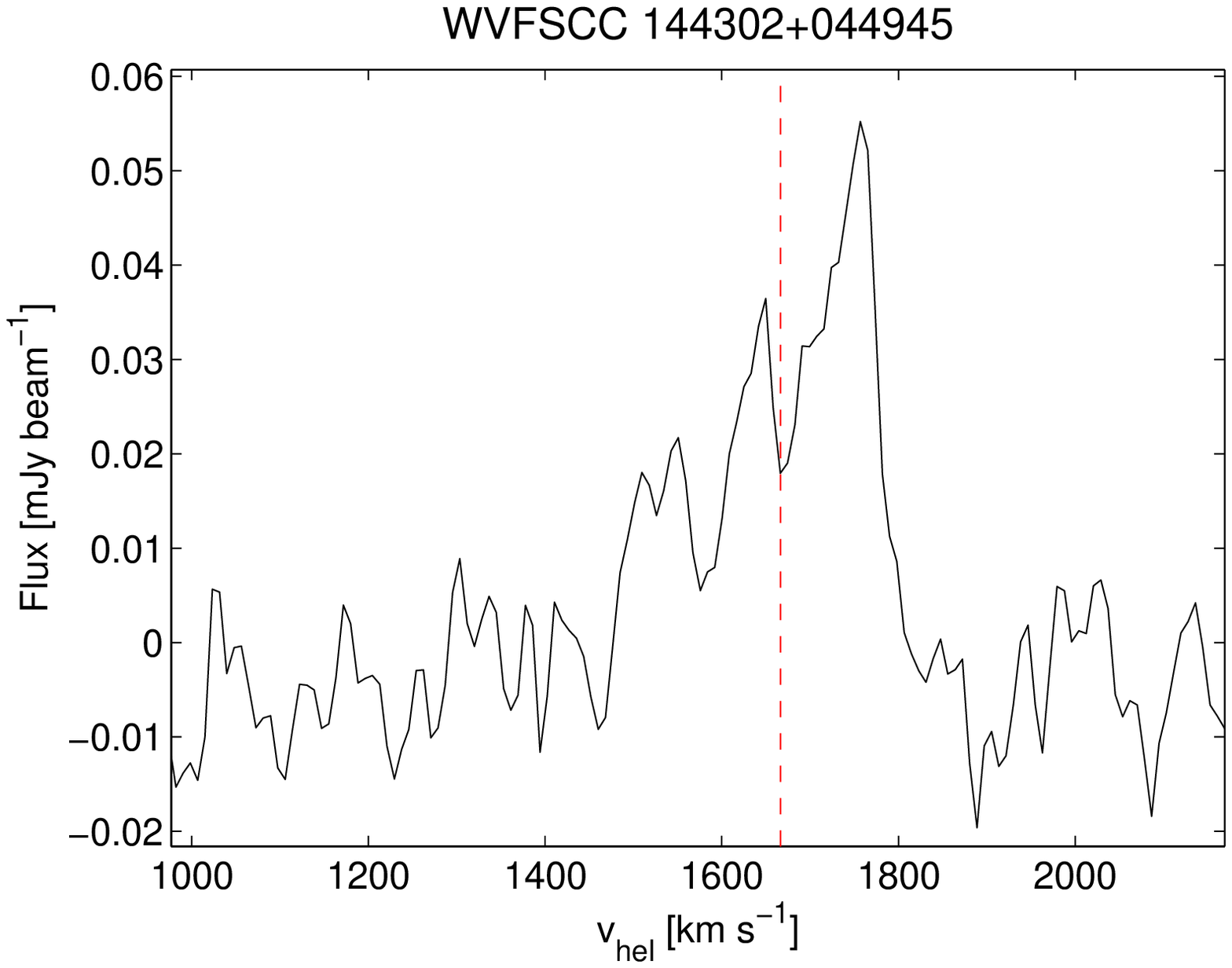}
\includegraphics[width=0.32\textwidth]{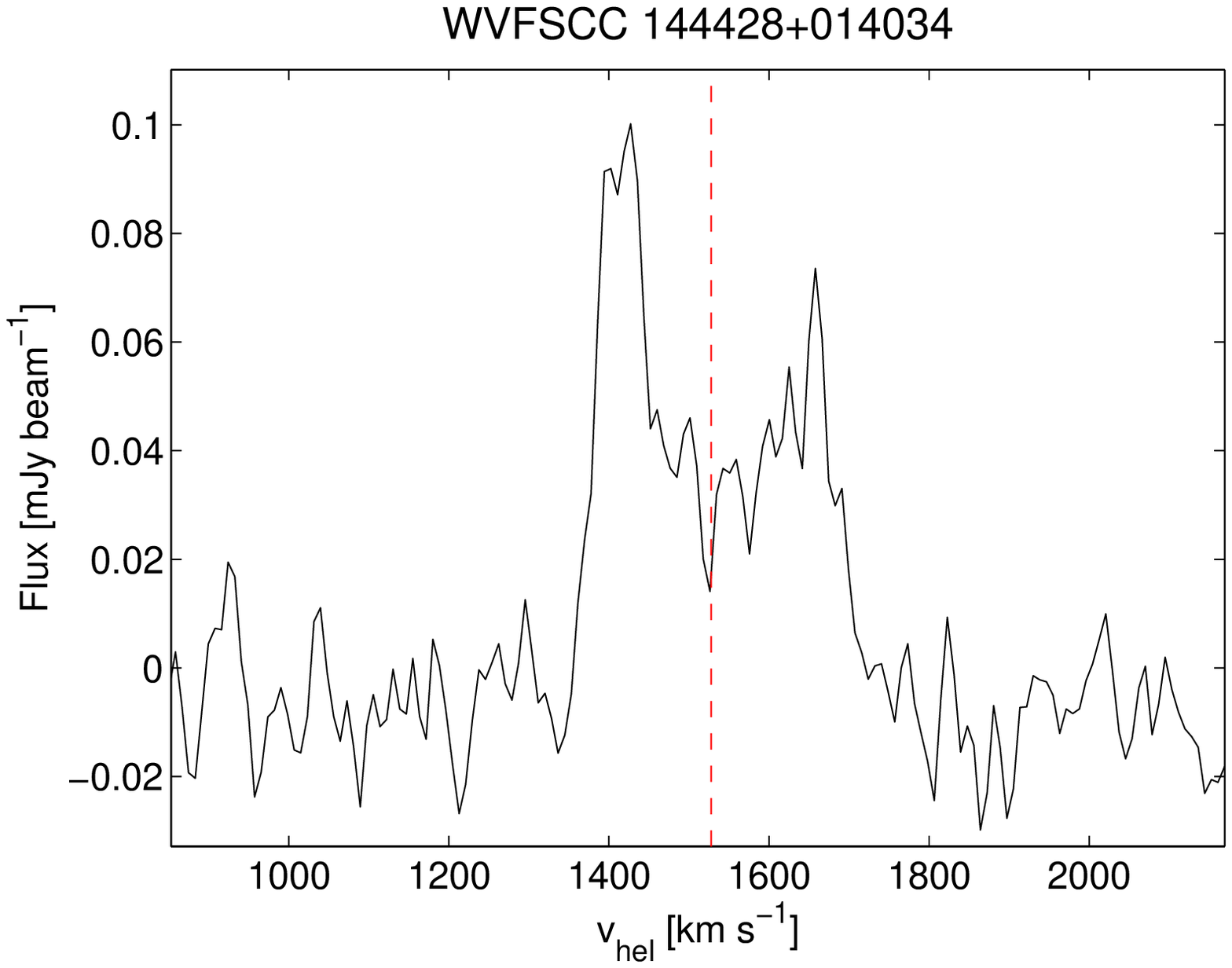}
\includegraphics[width=0.32\textwidth]{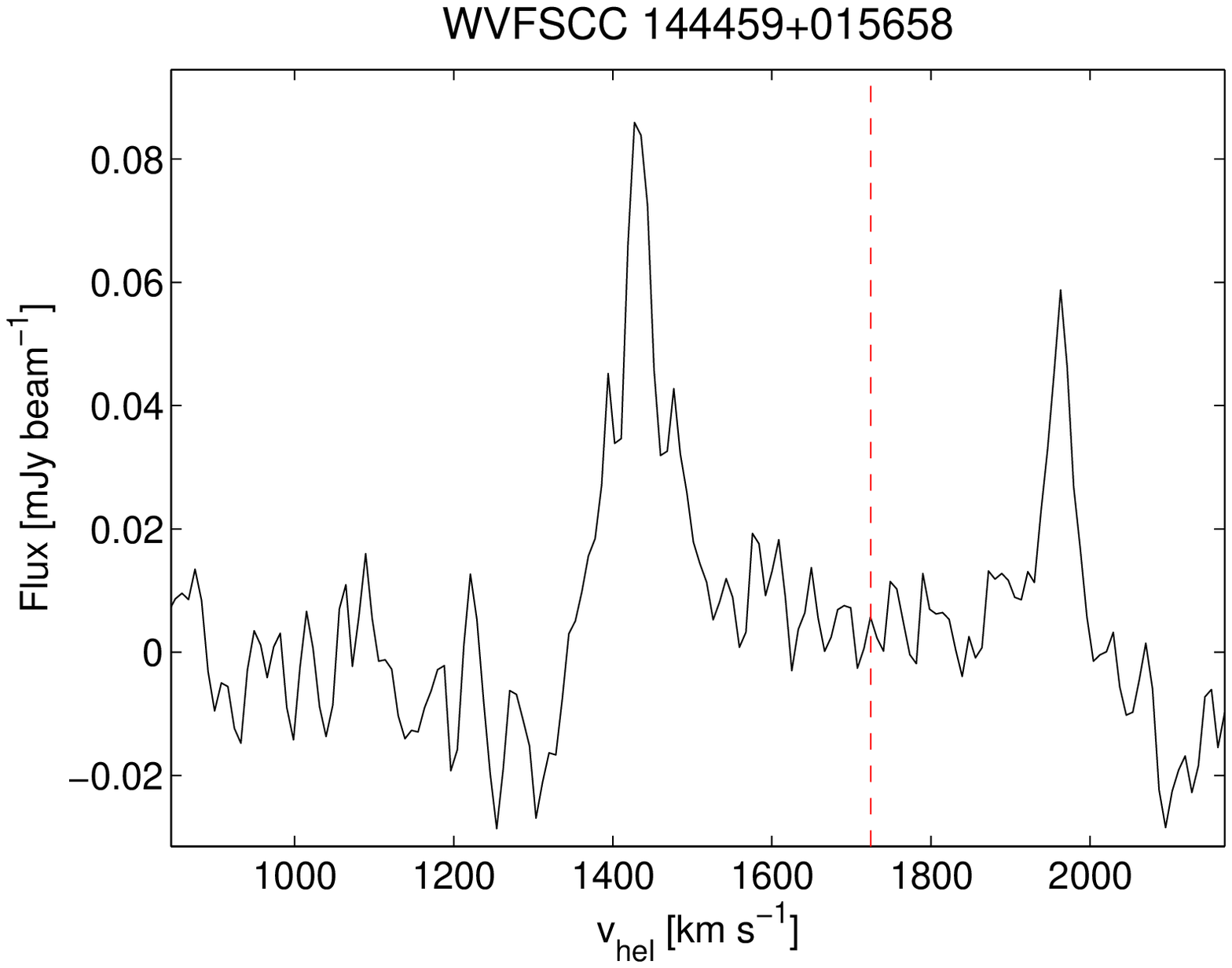}
\includegraphics[width=0.32\textwidth]{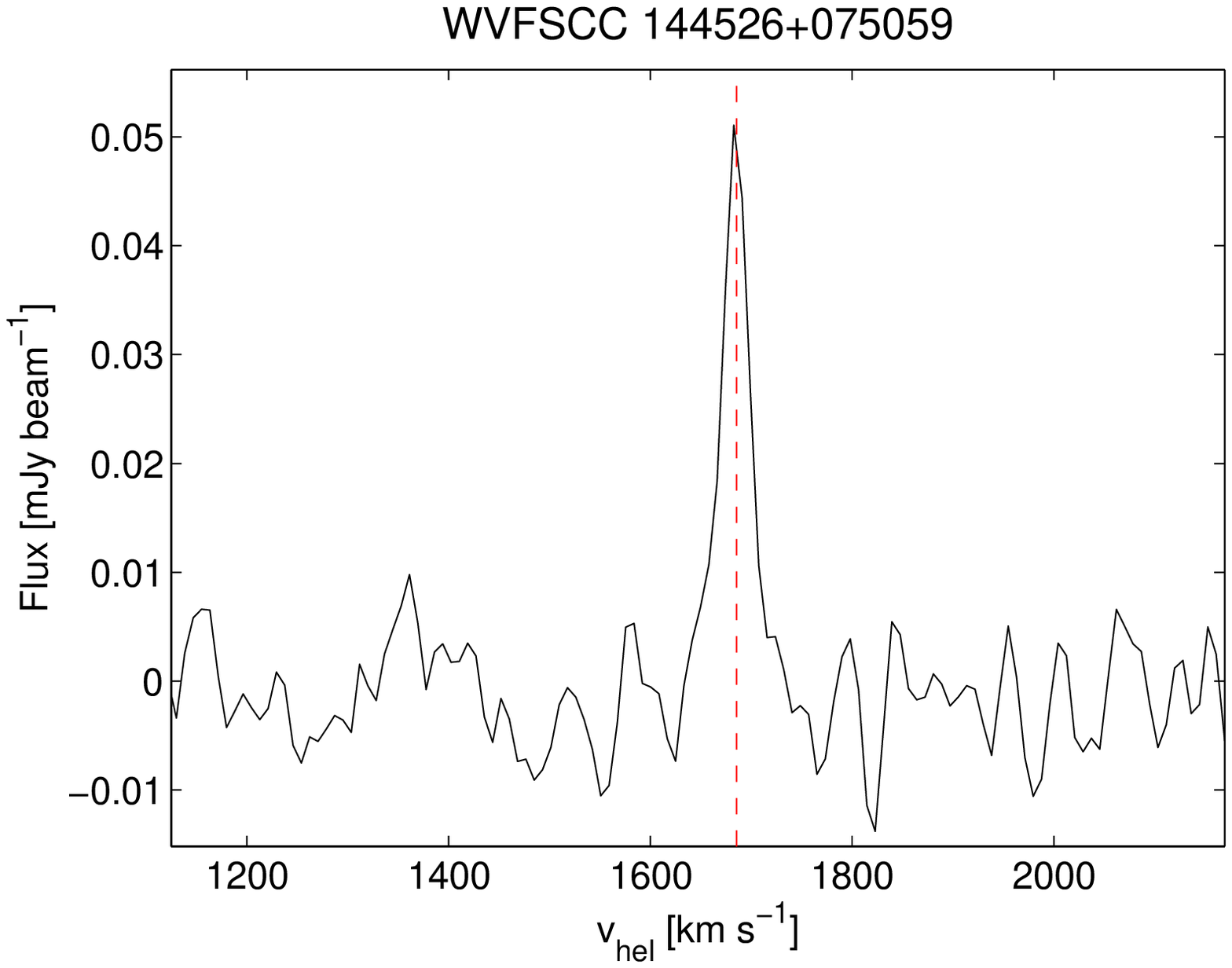}
\includegraphics[width=0.32\textwidth]{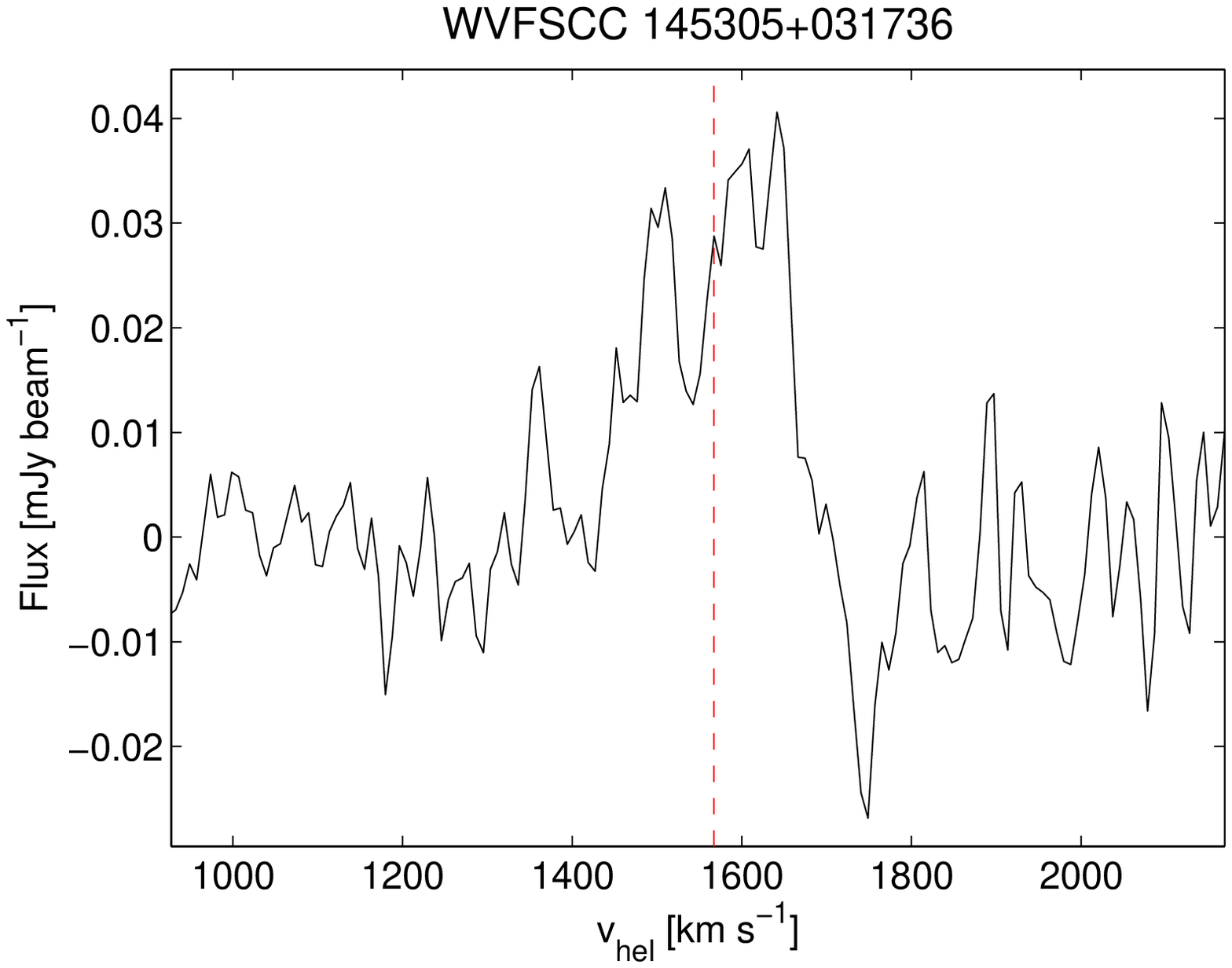}
\includegraphics[width=0.32\textwidth]{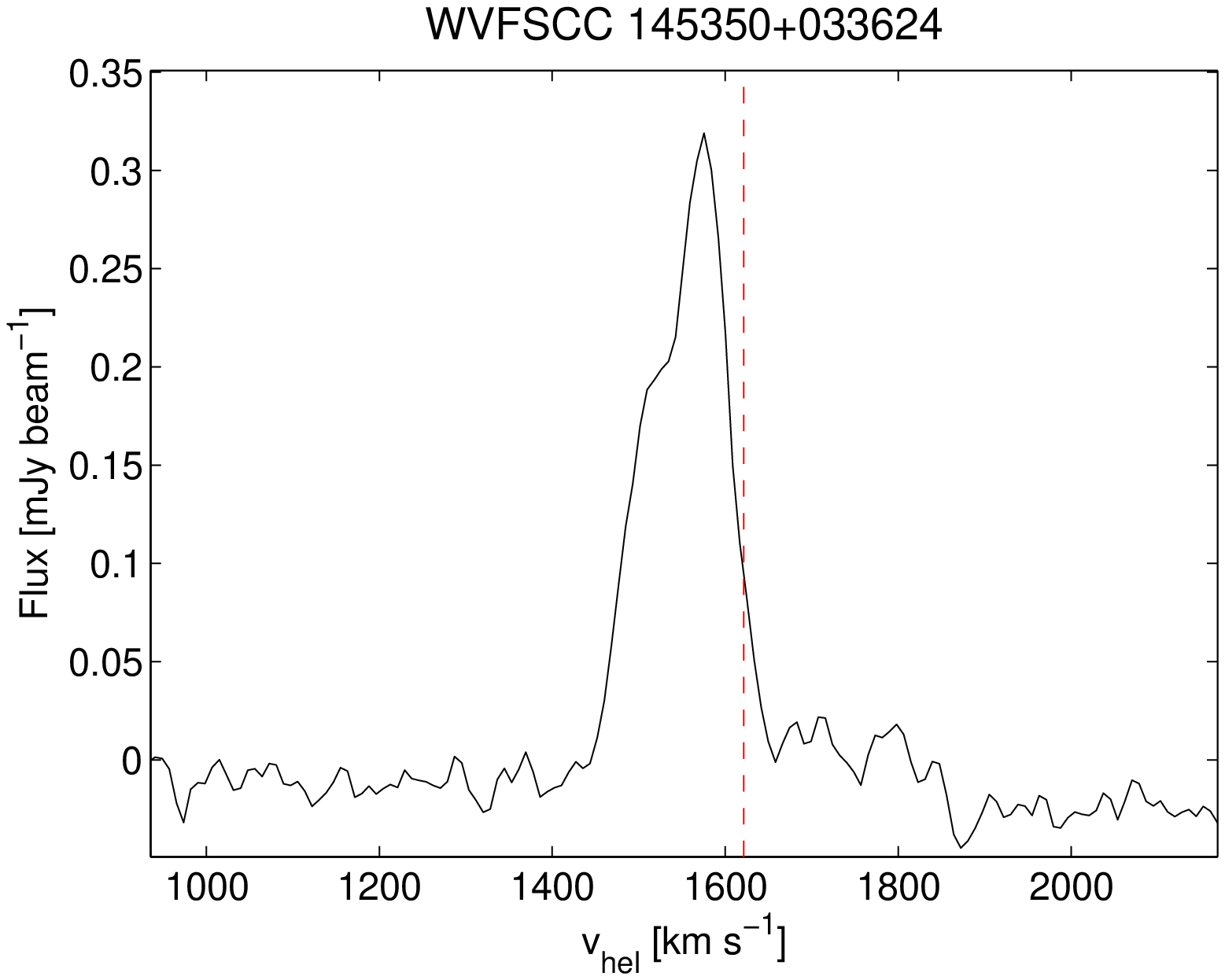}
\includegraphics[width=0.32\textwidth]{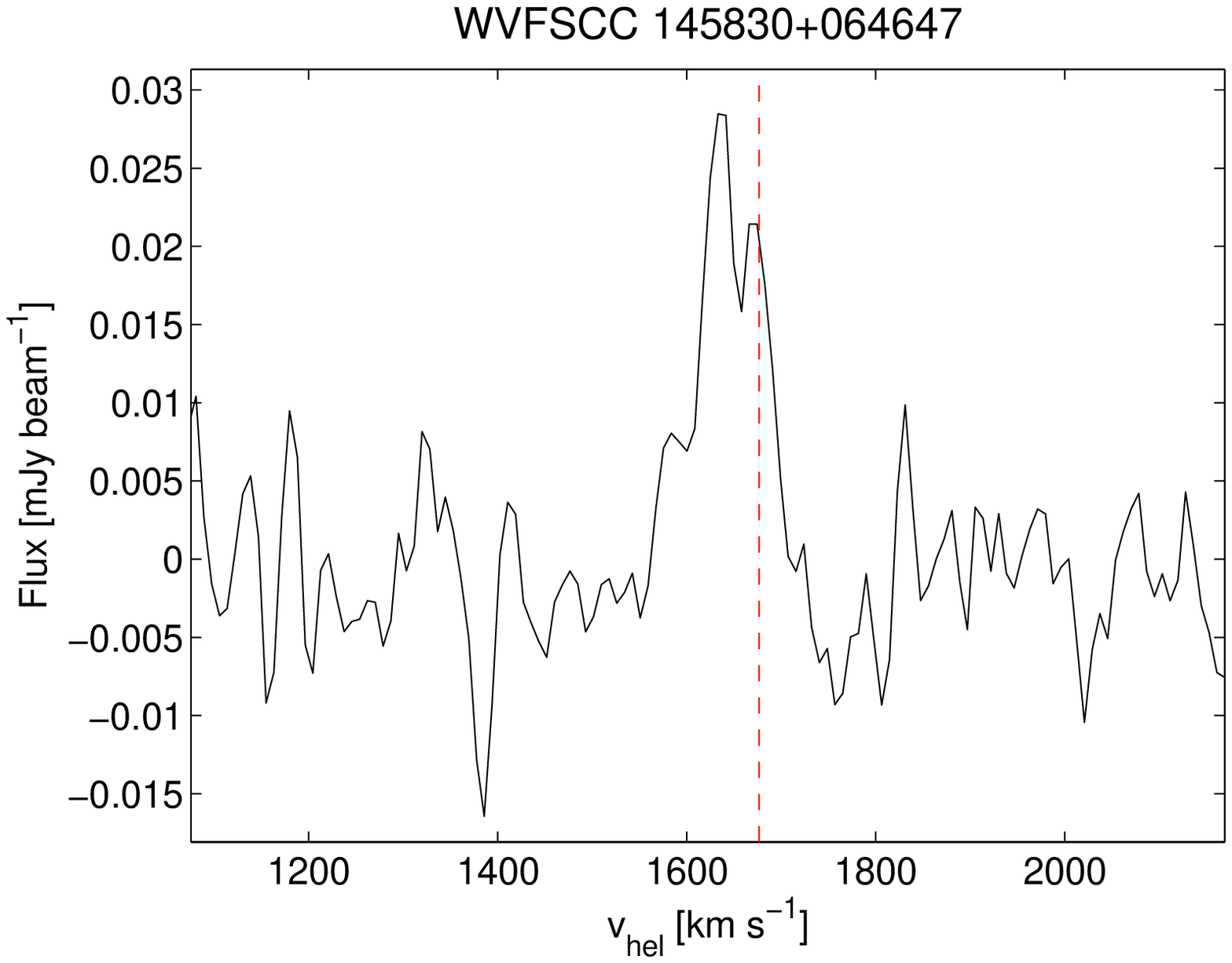}
\includegraphics[width=0.32\textwidth]{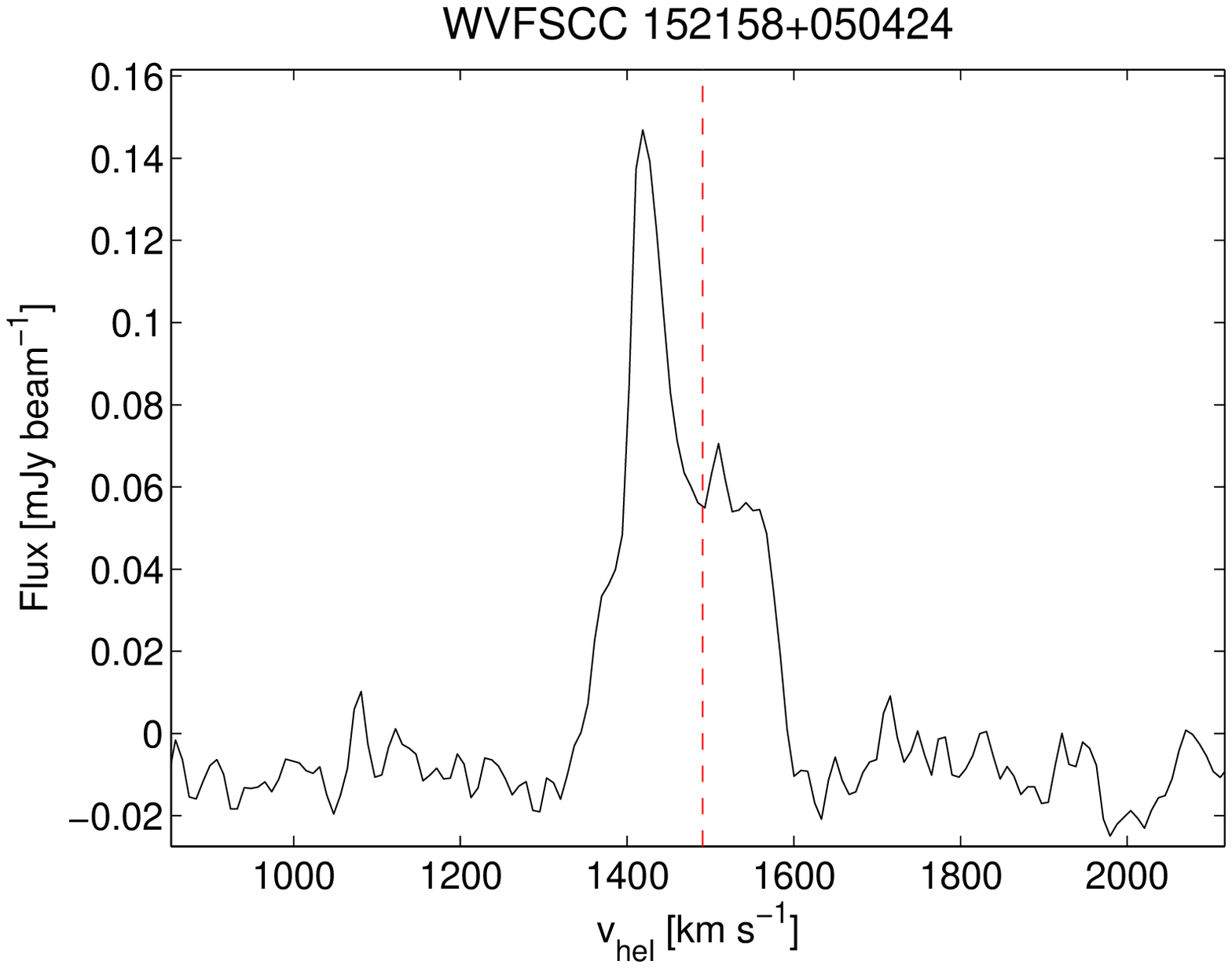}
\includegraphics[width=0.32\textwidth]{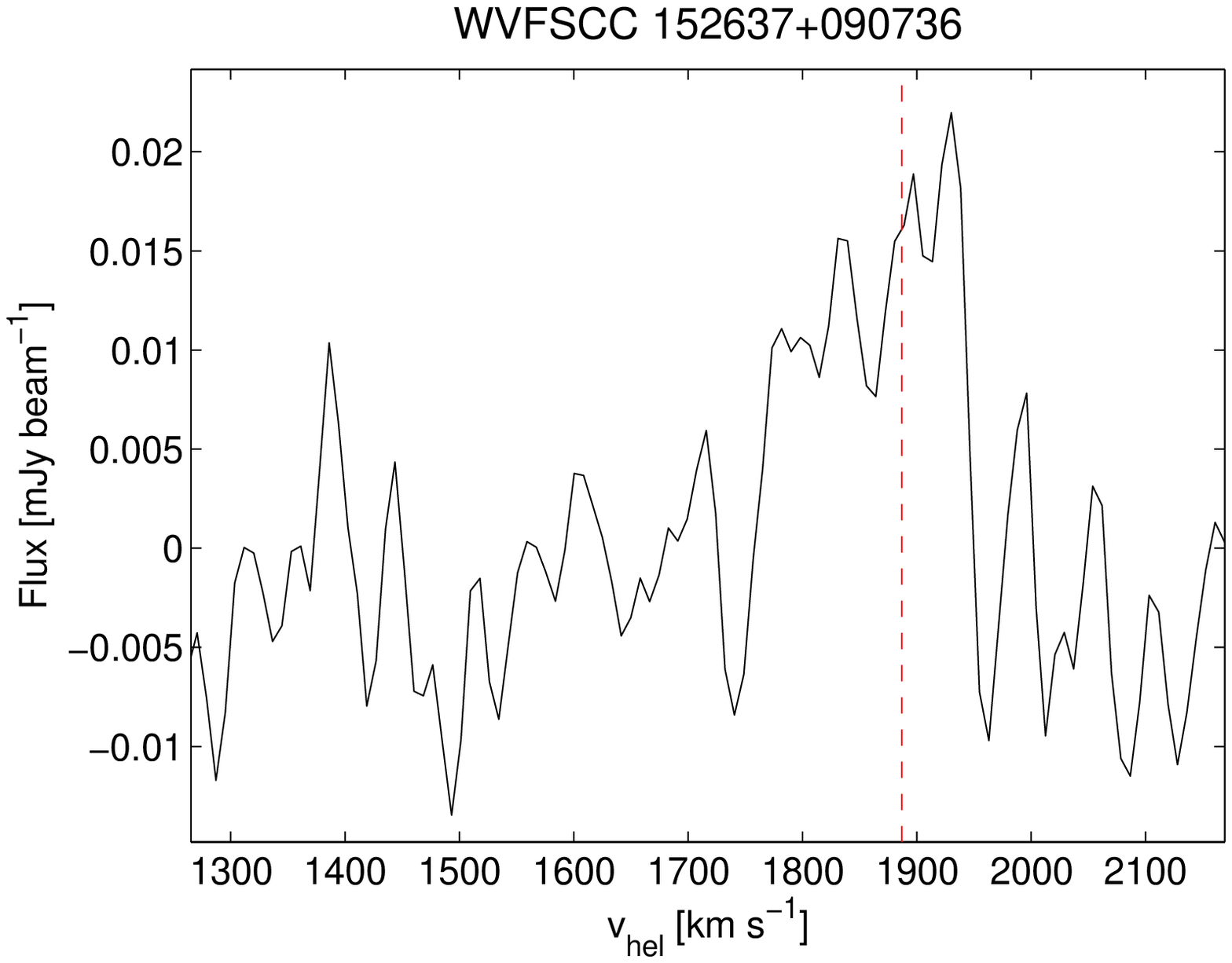}
\includegraphics[width=0.32\textwidth]{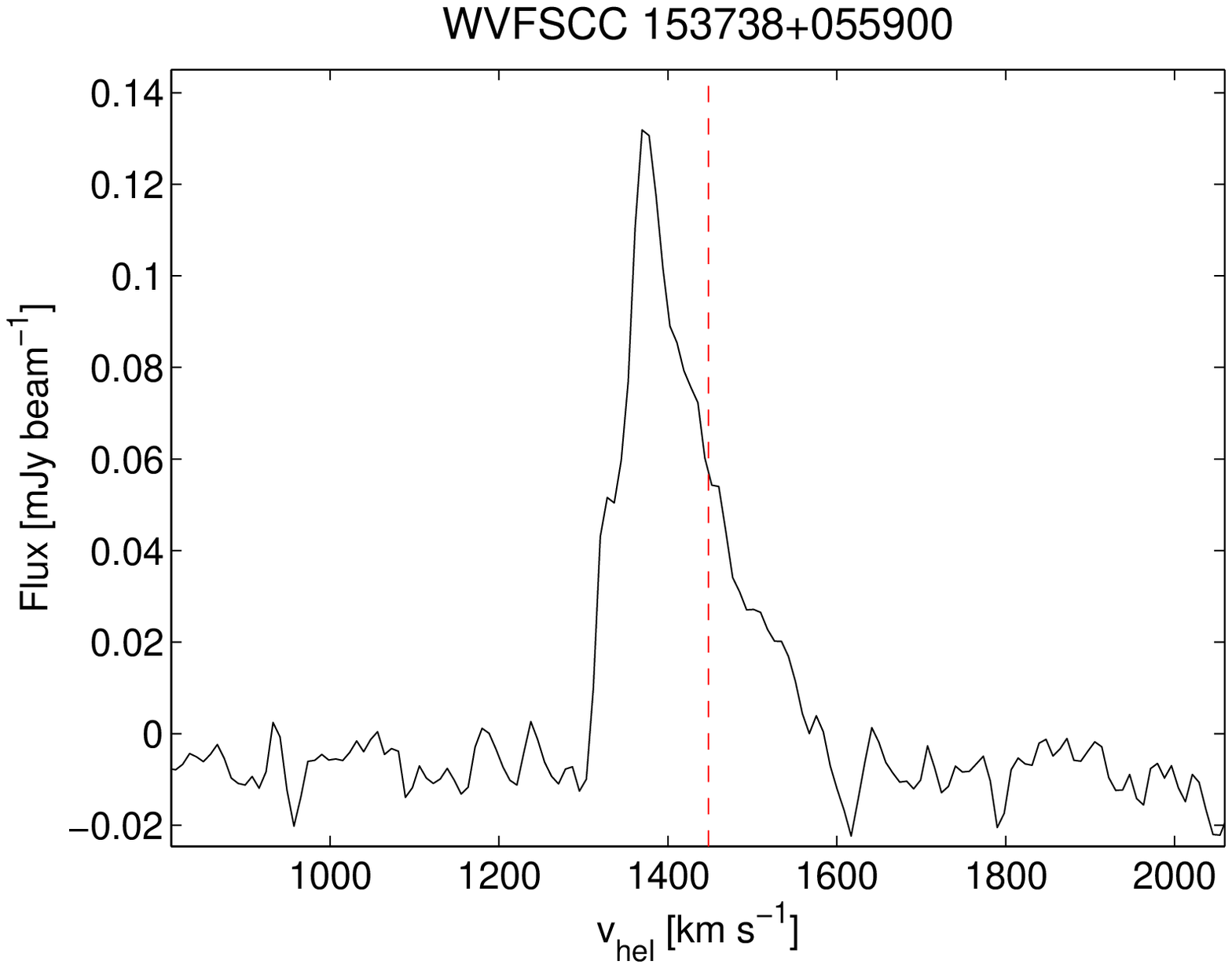}

\end{center}                                            
{\bf Fig~\ref{all_spectra2}.} (continued)                                        
 
\end{figure*}


\begin{figure*}
  \begin{center}

\includegraphics[width=0.32\textwidth]{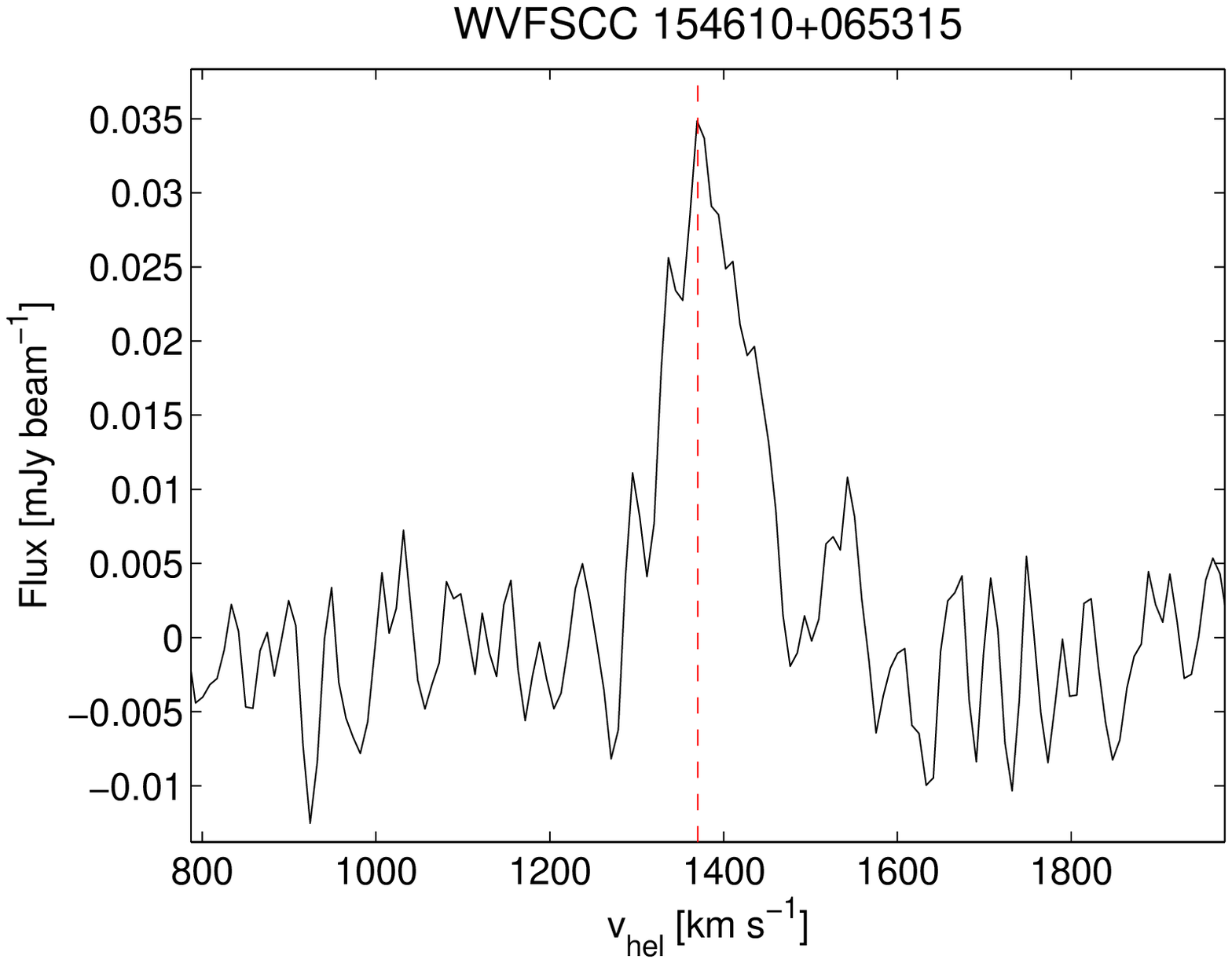}
\includegraphics[width=0.32\textwidth]{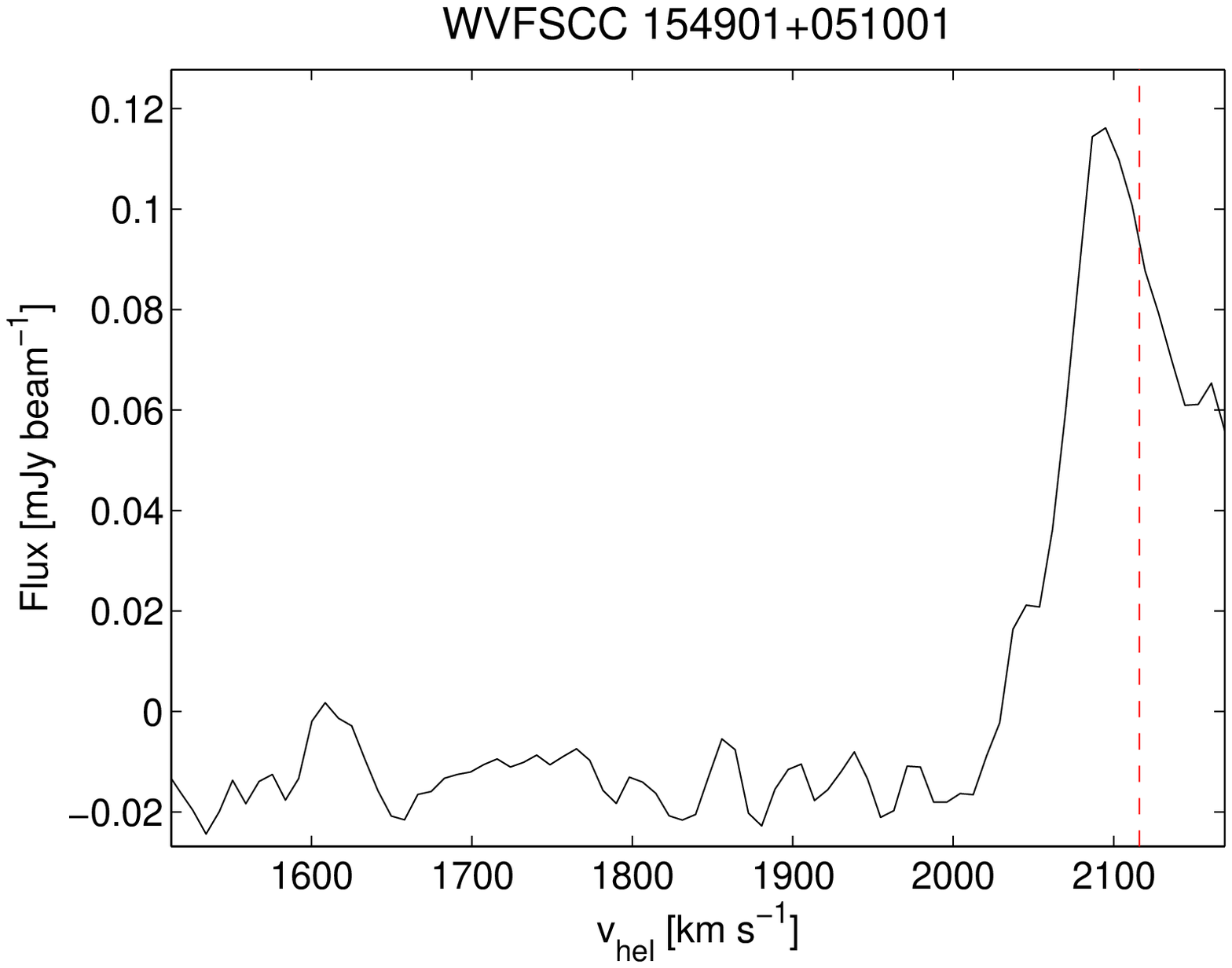}
\includegraphics[width=0.32\textwidth]{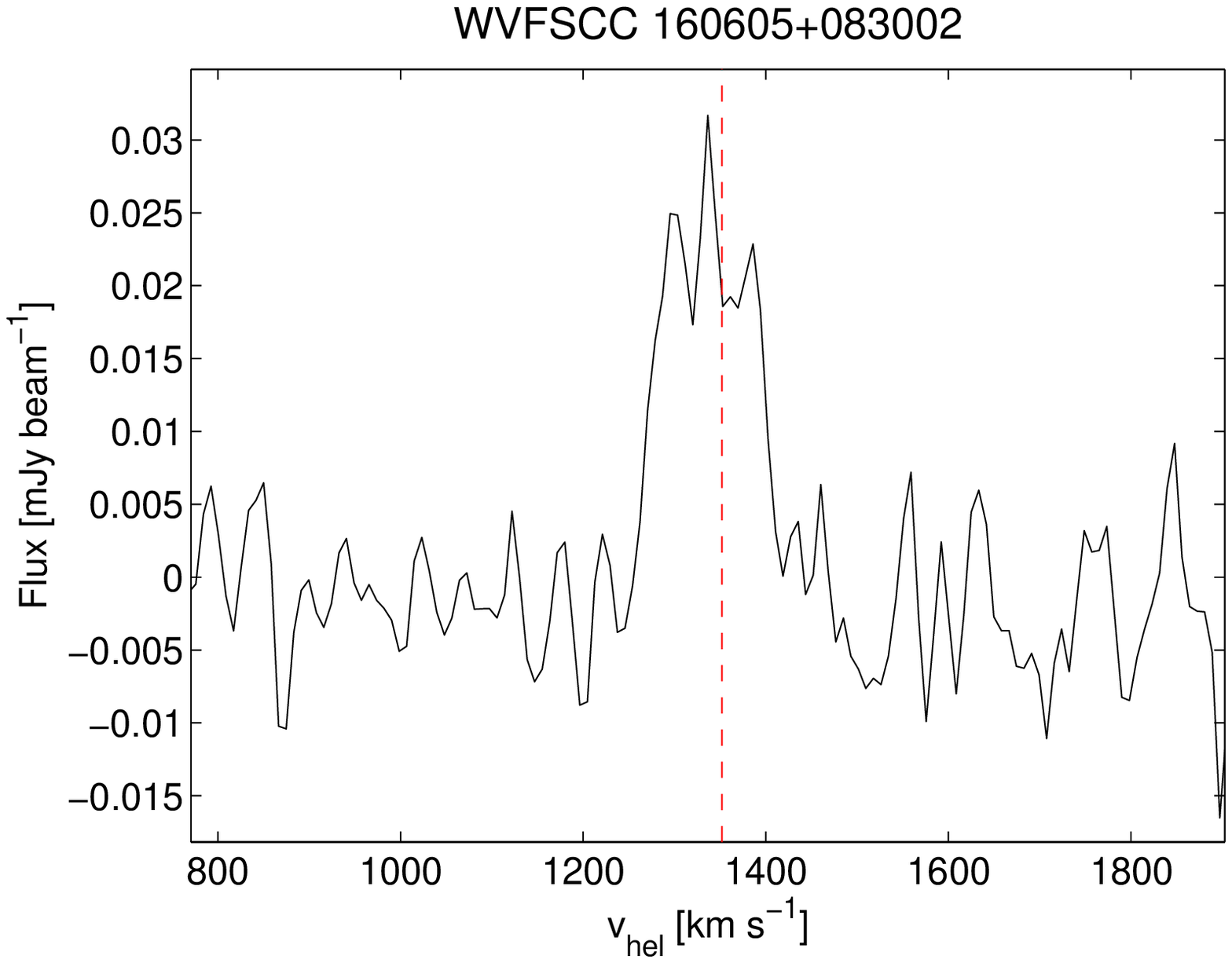}
\includegraphics[width=0.32\textwidth]{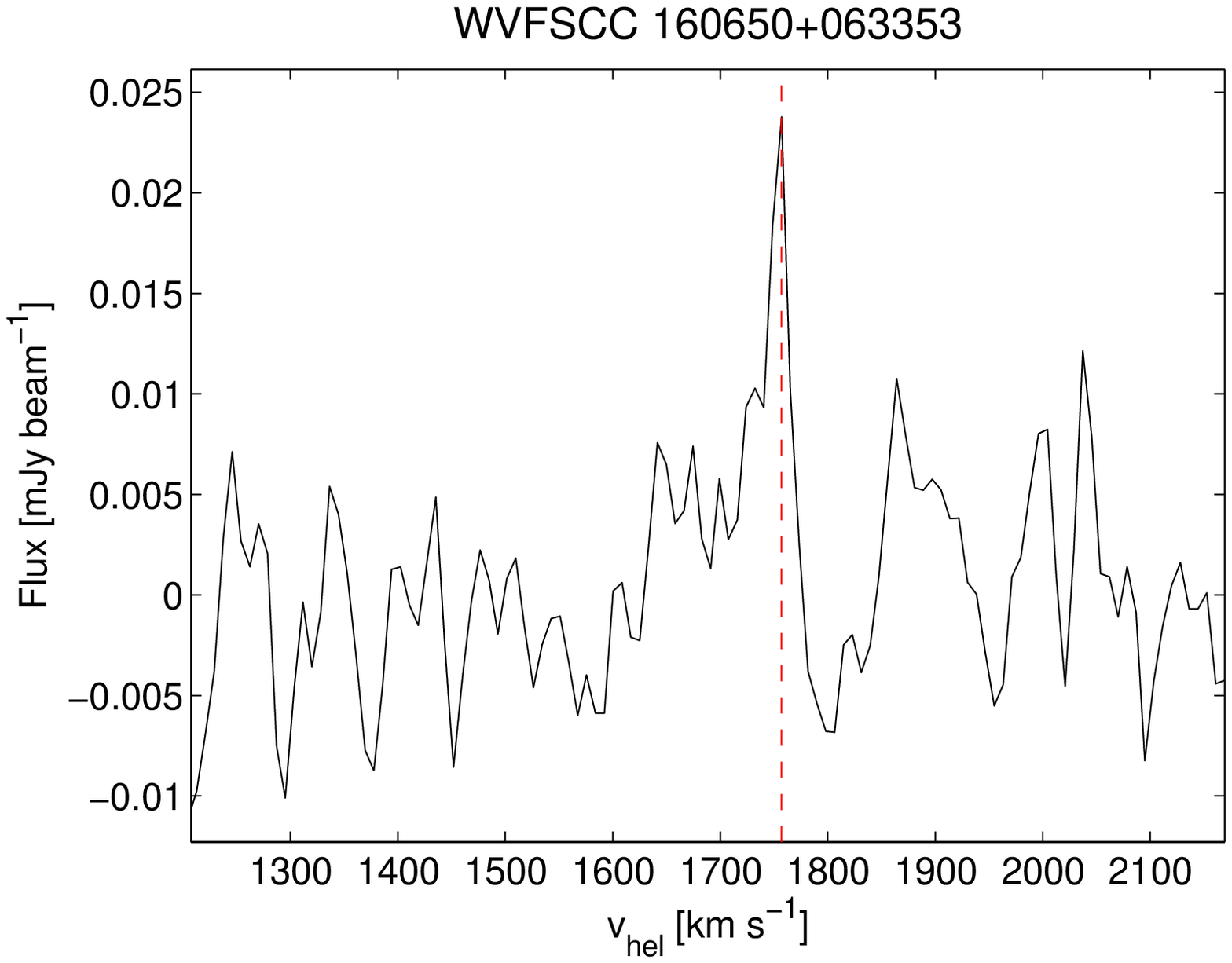}
\includegraphics[width=0.32\textwidth]{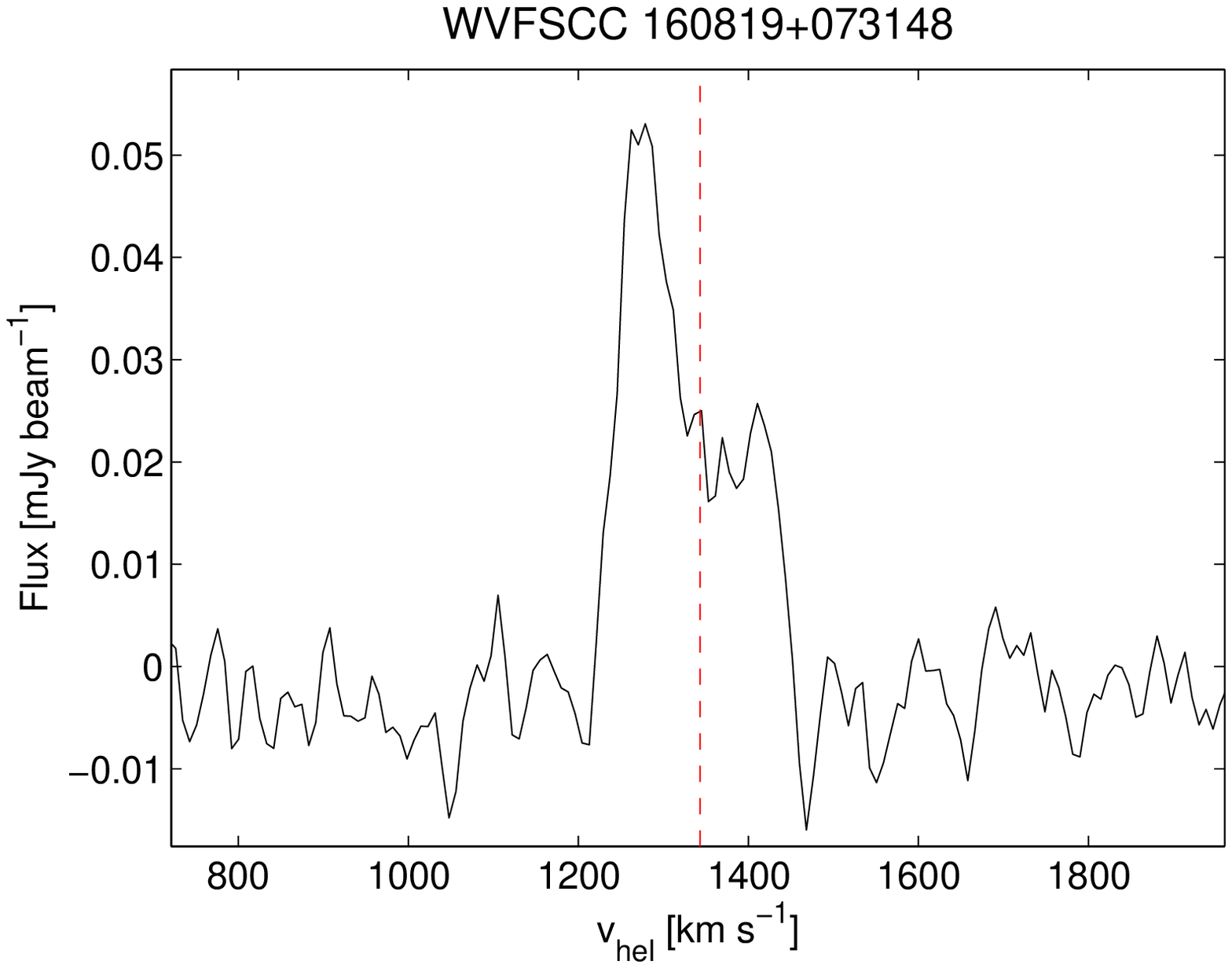}
\includegraphics[width=0.32\textwidth]{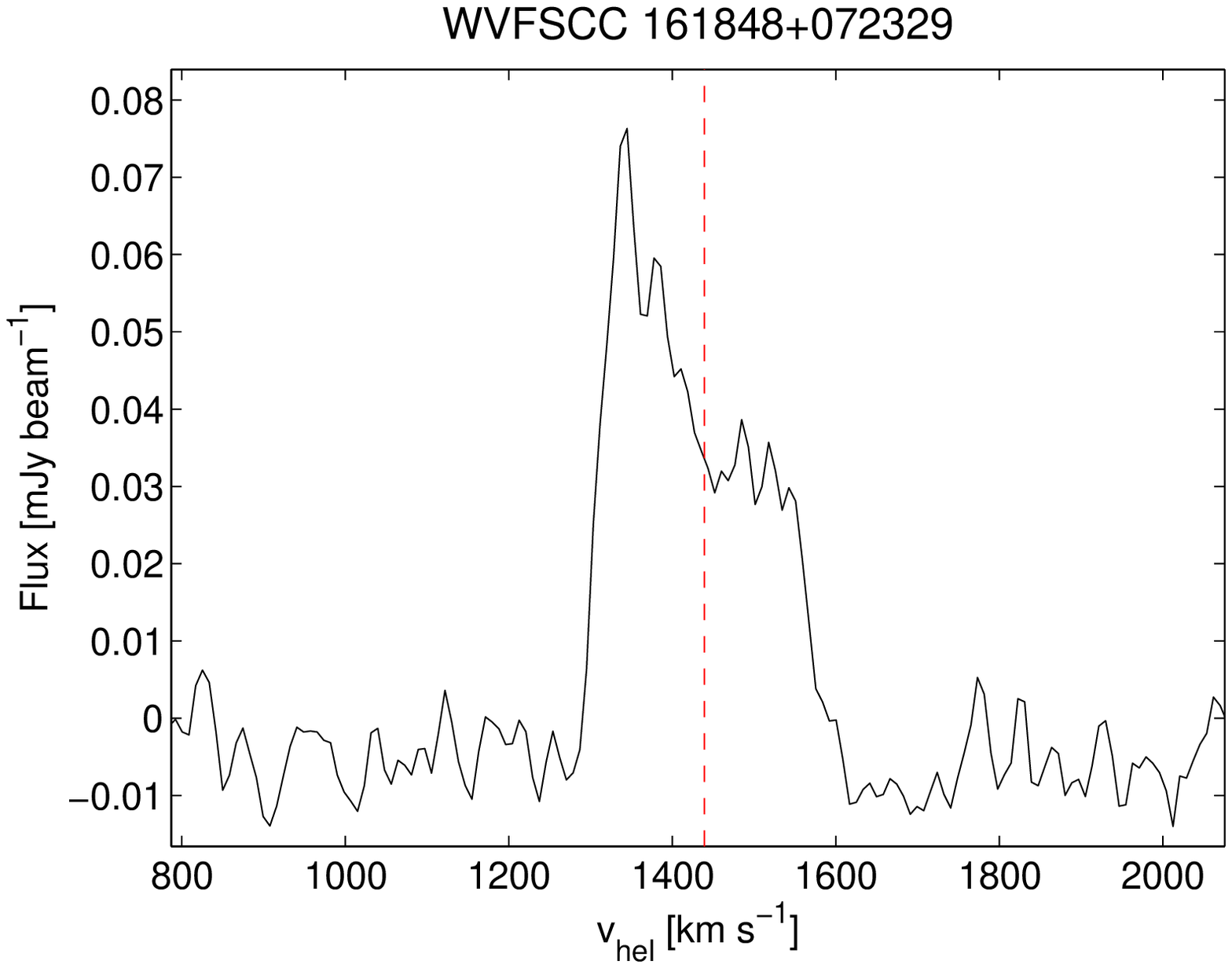}
\includegraphics[width=0.32\textwidth]{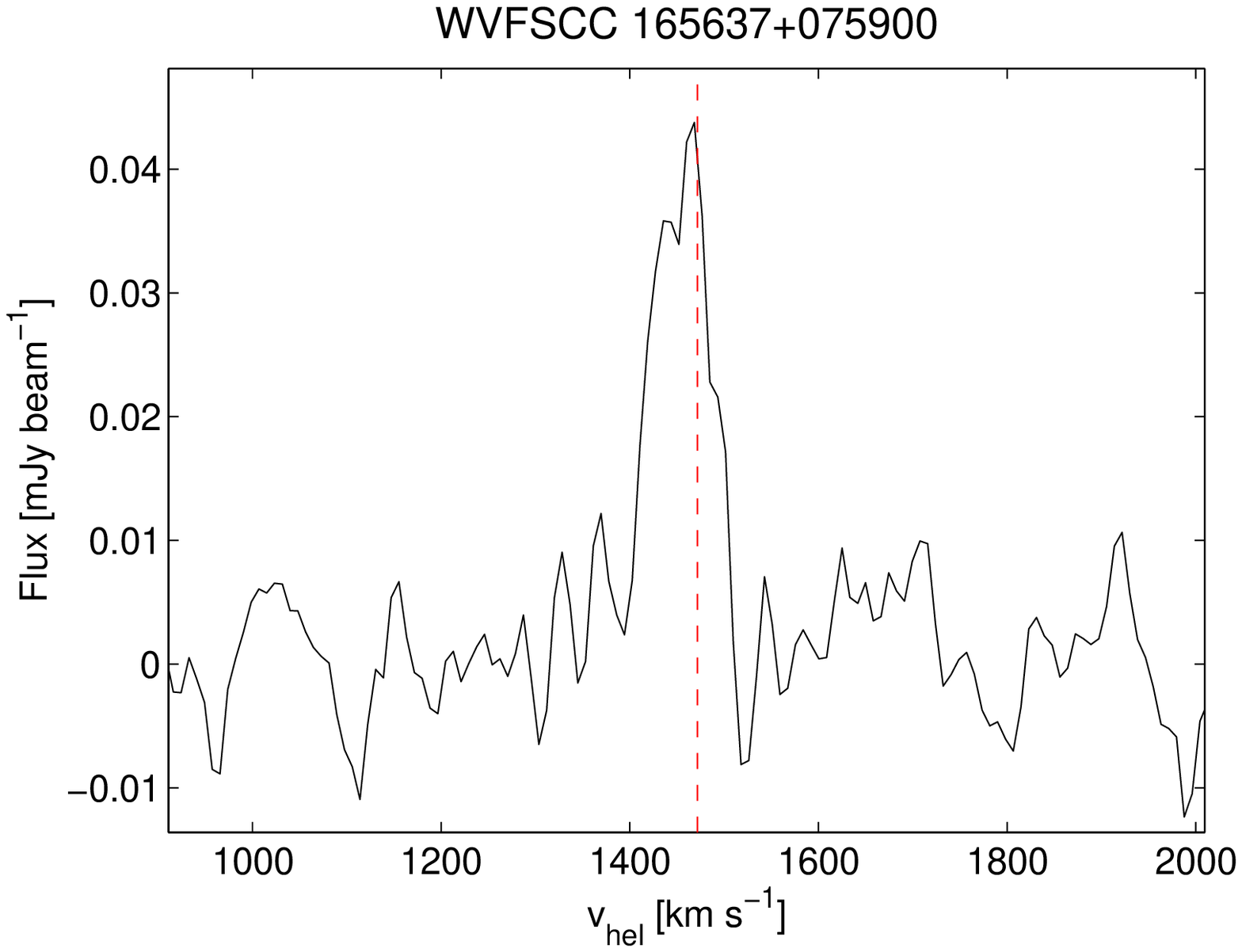}

\end{center}                                            
{\bf Fig~\ref{all_spectra2}.} (continued)                                        
 
\end{figure*}

\end{appendix}

\end{document}